\documentclass[11pt]{article}
\usepackage{amsfonts}
\usepackage{amssymb}
\usepackage{graphicx}
\usepackage{amsmath}
\usepackage{geometry}
\usepackage{appendix}

\let\dprod\prod
\let\tprod\prod
\let\tbigoplus\bigoplus
\let\tsum\sum
\DeclareMathSizes{10.95}{10}{7}{6}
\input{tcilatex}
\begin{document}

\title{Decomposition of State Spaces into Subobjects\\
in Quantum Field Theory}
\author{Pierre Gosselin\thanks{%
Pierre Gosselin : Institut Fourier, UMR 5582 CNRS-UGA, Universit\'{e}
Grenoble Alpes, BP 74, 38402 St Martin d'H\`{e}res, France.\ E-Mail:
Pierre.Gosselin@univ-grenoble-alpes.fr}}
\date{April 2025}
\maketitle

\begin{abstract}
This paper introduces a comprehensive formalism for decomposing the state
space of a quantum field into several entangled subobjects, i.e., fields
generating a subspace of states. Projecting some of the subobjects onto
degenerate background states reduces the system to an effective field theory
depending on parameters representing the degeneracies. Notably, these
parameters are not exogenous. The entanglement among subobjects in the
initial system manifests as an interrelation between parameters and
non-projected subobjects. Untangling this dependency necessitates imposing
linear first-order equations on the effective field. The geometric
characteristics of the parameter spaces depend on both the effective field
and the background of the projected subobjects. The system, governed by
arbitrary variables, has no dynamics, but the projection of some subobjects
can be interpreted as slicing the original state space according to the
lowest eigenvalues of a parameter-dependent family of operators. The slices
can be endowed with amplitudes similar to some transitions between each
other, contingent upon these eigenvalues. Averaging over all possible
transitions shows that the amplitudes are higher for maps with increased
eigenvalue than for maps with decreasing eigenvalue.
\end{abstract}

\subparagraph*{Acknowledgements}

\emph{The Author is deeply grateful to A\"{\i}leen Lotz for her
contributions to this research. Her insightful discussions, thoughtful
feedback, and constant involvement have been determining factors in the
development of this work.}

\section{Introduction}

Composite objects in quantum field theory are typically treated as
combinations of various states or fields through perturbative computations
(see $\left[ 1\right] $-$\left[ 9\right] $ for the context of Bethe-Salpeter
equation and $\left[ 9\right] $-$\left[ 20\right] $ for QCD) or
non-perturbative methods such as operator product expansion (see $\left[ 21%
\right] $-$\left[ 32\right] $). This work takes a different approach,
starting with an arbitrary field theory and considering states that
decompose into constrained substates, built from subfields referred to as
subobjects throughout this paper. These subobjects define a field theory
themselves, but due to the constraints arising from the decomposition, these
fields become entangled with each other. Thus, the decomposition describes
interacting tensor products of fields defined by subobjects.

For each decomposition, we consider the projection of the states of one or
several subobjects onto some background or some operator eigenspace. Doing
so, the projected subobjects (denoted PS) keeps track of some degrees of
freedom of the field we started with and of the interactions with the
remaining subobjects, also refered as non projected subobjects (written
NPS). The PS constitute a background on which the NPS will be redefined in
an apparent effective field theory. The degeneracy of the background state
induces the emergence of effective parameter-dependent fields, reducing the
initial field theory to an effective field theory depending on these
parameters. The effective fields are the initial NPS dressed by the
parameters. These parameters are exogenous in the first approximation, and
this effective field theory may be considered to be defined on an a priori
parameter space.

However, this parameter space keeps track of the characteristics of the
initial system and the PS as well as their interactions with NPS. Thus, the
parameter space depends directly on the remaining degrees of freedom, that
is on the NPS. In other words, due to the entanglement of the subobjects,
the NPS's degrees of freedom globally condition the geometry of the
parameter space. Imposing independence to the first order between the
effective field variation and parameters leads to some field equations.
Moreover, both the NPS degrees of freedom and the projected background
determine the metric of the parameter space. This dependency in the
background is the trace of the states from which the parameter space has
emerged.

Alternatively, this approach can be described in terms of operators and
states formalism. Choosing an operator acting on one or several subobjects,
the state space of intertwined subobjects can be projected onto the
operator's lowest eigenvalue's eigenspace with degeneracy. Since this
projection acts on the state space of one or several subobjects, the
projected states are similar to the subobject background previously
described.

Due to the subobjects' entanglement, the space of lowest eigenvalues depends
on the NPS degrees of freedom, that is on the states of the NPS. Then, for
every state of the NPS, we obtain a projected space defined by the lowest
eigenvalue of the operator. As the non projected state varies, so do the
projected spaces. We can consider the resulting state-space defined as the
union of all these projected spaces, which corresponds to some fibration
above the space of non projected states. Then, we can consider a partition
of the resulting state-space, decomposed into "slices," each of them
consisting of states of this fibration such that the lowest eigenvalue for
these states has a given value. These slices are not orthogonal with each
other, as the state-dependency of the eigenvalue implies that two different
states correspond to different operators: the eigenvalues are not the
eigenvalues of a single operator, but all of them are the minimum
eigenvalues of a state-dependent set of operators.

These projections result in describing the effective system in terms of
states and operators depending on some degeneracy parameters, one of these
parameters being distinguished from others. The field equations resulting
from the independence between field variations and the parameter space are
recovered. This approach allows computing amplitudes between states with
different distinguished eigenvalues: for each eigenvalue, we consider the
corresponding slice. Then, we can compute transitions between slices.
Indeed, the states for a given eigenvalue generate a state subspace, and we
can define a map between these subspaces. They are defined by summing
infinitesimal transitions maps between close slices. The form of the
transitions is similar to some path integrals between slices. These maps and
the corresponding amplitudes depend on the states and on the background
constituted by the PS. The geometry underlying the transition is local,
since the projection defining the effective states varies with the states
themselves. It implies that the transition between several states depends
both the PS and the NPS. That is, the apparent geometry underlying
transitions depends on the states but also on some background space that can
be considered as inert in first approximation only.

The field theory presented here is defined on an abstract parameter space,
and no dynamics occur. However, after projection on the background, the set
of lowest eigenvalues defining the slices in the state space along with
their amplitudes, allows defining an analog of such dynamics. Considering
the averaged transitions between slices, we can show that under some
conditions about the number of states in each slice, the amplitudes are
weighted for transitions corresponding to an increase in eigenvalues. This
asymmetry results from the characteristics of the slices. The number of
transition maps between these spaces increases with the eigenvalue, which
induces a bias of transition amplitude towards an increase in this variable.
Translated in an effective field perspective, the state space can be
decomposed in slices describing some dynamics for an irreversible parameter,
this parameter being in fact the track of the states on which the system can
be projected.

This work is organized into four parts. The first part presents the
formalism for the particular case of a state space defined by functionals of
one field. Section 2 presents the initial field theory, the full space of
states, and some notations. The states are functionals of a field defined on
some parameter spaces, including some constraints on these parameters. We
consider independently the tensor products of the field arising in the
states. The states are thus functional of the field tensor power with some
constraints on the parameters.

In section 3, the state space is decomposed into two particular subobjects,
i.e., entangled fields together with their state spaces. We present the
resulting decomposition of the initial field theory. Section 4 derives the
projection on the background of one of the subobjects. It results in an
effective field theory for the NPS. The degeneracy of the background
translates into an effective field depending on parameters describing the
symmetries of the background. We describe the effective projected field
theory. The NPS absorbs these parameters and becomes a field theory defined
on this parameter space. However, due to the constraints between initial
subobjects, these parameters are not globally independent from the field.

Then, section 5 studies the consequences of this decomposition. Dependency
of parameters in the field translates into joint variations of these two
variables. We show that these variables can be considered as independent if
the field satisfies some first-order equations. These equations involve the
field over the entire parameter space. Averaging over this space yields
equations for one value of the parameters, similar to some local equations.

Section 6 focuses on the constraints of the parameter space and the geometry
of this space. The mutual dependency between parameters and field translates
into a description similar to some set of metric spaces depending
functionally on the field and the projected background, which implies that
the field dependence of metric depends also on some inert quantity. These
inert quantities are a parameters space that keep track of the NPS degres of
freedom of the hidden projcted states..

In section 7, we develop an equivalent approach. Rather than considering
solutions to saddle point equations for one of the subobjects, we rather
project the space of states onto the minimal eigenstates of an operator
acting on this subobject. Due to the constraint between the two subobjects,
the eigenvalues and the projected state space depend on the states for the
remaining subobject. This approach allows defining subspaces of effective
states, depending on the eigenvalues of the operator. For each eigenvalue,
we consider the subspace and parameter values corresponding to this
eigenvalue. In section 8, we define the transitions between such subspaces,
or slices. Integrating over infinitesimal variation of slices, corresponding
to a variation in eigenvalues, yields a path integral formulation for
transitions between subspaces of states. These transitions are also defined
for operators.

Section 9 builds on this approach to define average transition between
slices in the state space. We show that for a given slice, the average
amplitudes of transitions towards other slices arise with an increase in
eigenvalues.

The second part develops a general formalism for decomposing the state space
of a field into several subobjects. Section 10 presents the setup and the
main elements. Defining formally a subobject independently from any
decomposition, as a composite of states, we consider the sequences of
inclusion between subobjects, that is, composed subobjects. In section 11,
we introduce all possible decompositions of the initial field and state
space. We include the possibility that in a given decomposition, any
subobject itself decomposes in subobjects of the subobjects defining the
decomposition. The formalism results in describing the system as sets of
maps between various constrained parameter spaces. These maps describe
sequences of inclusions between subobjects.

Sections 12, 13, 14 and 15 focus on the case of a projection of all
subobjects except one. We recover the description of the first part: the
effective field, the field equations corresponding to independent
parameters, and the geometric aspects for the constraints. In section 16, we
also study the transformation properties of the field under
reparametrization. In section 17, we consider the general case, where an
arbitrary number of subobjects remain unprojected. This allows presenting
the emergence of several fields and parameter spaces, arising from an
arbitrary initial formalism. Section 18 presents the alternative method of
projecting over some eigenspace of an operator and derives the transitions
in the general formalism.

The third part presents an alternative and less general formalism that is
closer to the usual formalism of field theory. We also consider
decomposition of states into several types of objects, but the formalism
starts directly from states in section 19 and with operators in section 20.
We recover the main characteristics of the formalism developed in the two
first parts. Transitions are considered in section 21. Due to our choice to
start our description with states, these transitions are first derived in a
context similar to a first quantized system and then reconsidered in a field
second quantization context.

The fourth part concludes this work by outlining several potential
developments. Section 22 interprets the preceding formalism as a field
theory on a singular fibred space of states, with the dimension of the fiber
contingent upon the states in the basis space. In section 23, we focus on
the constraints that define the system. We examine their continuous
variations as the basis state defining the system undergoes changes, that
is, when a displacement is considered in the basis. These variations in the
constraints manifest as differential equation in the state space, akin to
some sort of dynamical equations. Section 24 considers discrete, i.e. non
continuous, modifications of the contraints. Imposing some consistency
conditions between different modifications imply some commutation relations
for degeneracy operators. Thus modifying the basis state of the system may,
at the effective level, translates by the creation or destruction of
additional fields and states, along some local parameters, or points, or
clouds of points, that accompany these new states and flds.

In section 25, we consider more closely the effective action functional for
projected fields in a given states. The constraints impose some general
form, including an action for the cloud of points defining the parameter
space. Section 26 gathers the results of the last three sections to connect
effective actions and basis state variations. We argue that our formalism
can be considerd in an effective perspective as a landscape of effective
theories, one for global states. These theories are connected through
variations of global states. At this effective level, these variations
correspond to transition from one theory to an other. A transition is
characterized by the possible extension of the theory with new stats, points
and field. They are labelled by the variation of a parameter measuring the
variation in state. We recover a global view of the apparent dynamical
system presented in the previous parts.

\part*{Part I Decomposition of state space in two subobjects and resulting
parameter space.}

We present the decomposition of the functional space for a field into two
subobjects. The realizations of the field decompose into sums of products of
realizations, with each product constrained by relations between the
parameters defining the subobjects. Projecting onto a degenerate background
state for one of the subobjects provides an effective description of the
resulting system. The degeneracy of the background state implies that the
system can be described as an effective field theory for a field depending
on certain parameters. These parameters are exogenous only at first glance.
In reality, the initial entanglement between subobjects results in
intertwining between the NPS and the parameters. The interrelation between
the two initial subobjects implies that the parameter space itself is a
dynamic object in the effective theory. Untangling the constraints between
this parameter space and the effective field implies an analogue of
dynamical equations for the effective field. The dynamical characteristic of
the parameter space translates into a relation between the metric
characteristics of the space and the effective field. These geometric
characteristics also includes an inert part due to the background created by
the PS.

\section{General set up and fields functional description}

We describe the system as space of functional of an arbitrary field $\Psi $.
This field is defined over some parameter space $U$. To a field is
associated its set of realizations, i.e. the infinite set of values this
field can take. Each of these values is a function defined over $U$. To such
a realization $\Psi _{\alpha }$ we associate usually a weight given by $\exp
\left( iS\left( \Psi _{\alpha }\right) \right) $ where $S$ is a given action
functional. In the sequel, since we consider decomposition of a field into
other fields, called subobjects of the field, we will work at some points
with the realizations of the field rather\ than with the field itself. For
this reason, we will consider the tensor product of the field $\Psi
^{\otimes m}$ as an independent field defined on $U^{m}$ including some
constraint on this set. This amounts to consider multiple states as
themselves as basic elements. Actually, a realization of the product $\Psi
^{\otimes m}$ is a is an infinite sum of products $\Psi _{\alpha _{1}}...$ $%
\Psi _{\alpha _{m}}$, an intricate relation between the realization of $\Psi
^{\otimes m}$ and those of $\Psi $, which justifies our choice.

Moreover, considering a functional for $\Psi $ as a sum of linear functional
of $\Psi ^{\otimes m}$, we will restrict the functionals in the $\Psi
^{\otimes m}$ to be linear. Any product of functional of the $\Psi ^{\otimes
m}$ can be itself considered as a series of linear functionals of the $\Psi
^{\otimes m^{\prime }}$.

\subsection{States basis:}

We consider some parameters spaces $U,U^{k}$,... with $U^{k}$ is given by $k$
copies of $U$. We introduce implicit relations:%
\begin{equation*}
U/c\left( U\right) ,...,U^{k}/c\left( U^{k}\right)
\end{equation*}%
where the $c\left( U^{k}\right) $ are constraints.

This is leading to states combinations of:

\begin{equation*}
\left\vert u_{i}\right\rangle ,...\left\vert u_{i_{1}}\right\rangle \otimes
...\otimes \left\vert u_{i_{k}}\right\rangle ,...
\end{equation*}%
plus some implicit constraint:%
\begin{eqnarray*}
&&\left\vert u_{i_{1}}\right\rangle \otimes ...\otimes \left\vert
u_{i_{k}}\right\rangle /c\left( u_{i_{1}},...,u_{i_{k}}\right) \\
&\rightarrow &\left\vert u_{i_{1},}...,u_{i_{k}}\right\rangle
\end{eqnarray*}%
so that states are:%
\begin{eqnarray*}
&&\sum_{k,i_{1}...i_{k}}\left\vert u_{i_{1}}\right\rangle \otimes ...\otimes
\left\vert u_{i_{k}}\right\rangle /c\left( \left(
u_{i_{1}},...,u_{i_{k}}\right) _{k}\right) \\
&\rightarrow &\sum_{k,i_{1}...i_{k}}\left\vert
u_{i_{1},}...,u_{i_{k}}\right\rangle
\end{eqnarray*}

\subsection{States and functionals}

If we fix a basis $\left\vert u_{i}\right\rangle $, general states in this
set up have form:%
\begin{equation*}
\sum \otimes _{s}\Psi \left( u_{i_{s}}\right) \left\vert
u_{i_{s}}\right\rangle =\sum_{k,i_{1}...i_{k}}\tprod\limits_{s}\Psi \left(
u_{i_{s}}\right) \left\vert u_{i_{1}}\right\rangle \otimes ...\otimes
\left\vert u_{i_{k}}\right\rangle /c\left( \left(
u_{i_{1}},...,u_{i_{k}}\right) _{k}\right)
\end{equation*}%
They are considered as particular realization $\left\vert \Psi \right\rangle 
$ of a field $\Psi $:%
\begin{equation*}
\left\vert \Psi \right\rangle =\sum \Psi ^{\otimes k}\left(
u_{i_{1},}...,u_{i_{k}}\right) \left\vert
u_{i_{1},}...,u_{i_{k}}\right\rangle
\end{equation*}%
with $\Psi ^{\otimes k}\left( u_{i_{1},}...,u_{i_{k}}\right) $ represents
the "wave function" of the state in the basis $\left\vert
u_{i_{1},}...,u_{i_{k}}\right\rangle $.

and more genrally, products write:%
\begin{equation*}
\sum \otimes _{l}\Psi ^{\otimes k_{l}}\left( u_{i_{1},}^{\left( l\right)
}...,u_{i_{k_{l}}}^{\left( l\right) }\right) \left\vert u_{i_{1},}^{\left(
l\right) }...,u_{i_{k_{l}}}^{\left( l\right) }\right\rangle /c\left( \left(
u_{i_{1},}^{\left( l\right) }...,u_{i_{k_{l}}}^{\left( l\right) }\right)
_{l}\right) =\sum_{k,\left( i_{1}...i_{k}\right) _{l}}\Psi \left( \left(
u_{i_{1},}^{\left( l\right) }...,u_{i_{k}}^{\left( l\right) }\right)
_{l}\right) \left\vert \left( u_{i_{1},}^{\left( l\right)
}...,u_{i_{k}}^{\left( l\right) }\right) _{l}\right\rangle
\end{equation*}%
In a perspective of second quantized formalism, we consider functionals:%
\begin{eqnarray*}
&&\sum \left\langle u_{i_{1},}^{\left( l^{\prime }\right)
}...,u_{i_{k_{l^{\prime }}^{\prime }}}^{\left( l^{\prime }\right)
}\right\vert \tprod \Psi ^{\dag \otimes k_{l^{\prime }}^{\prime }}\left(
u_{i_{1},}^{\left( l^{\prime }\right) }...,u_{i_{k_{l^{\prime }}^{\prime
}}}^{\left( l^{\prime }\right) }\right) F\left( \left( u_{i_{1},}^{\left(
l^{\prime }\right) }...,u_{i_{k_{l^{\prime }}^{\prime }}}^{\left( l^{\prime
}\right) }\right) _{l^{\prime }},\left( u_{i_{1},}^{\left( l\right)
}...,u_{i_{k_{l}}}^{\left( l\right) }\right) _{l}\right) \tprod \Psi
^{\otimes k_{l}}\left( u_{i_{1},}^{\left( l\right)
}...,u_{i_{k_{l}}}^{\left( l\right) }\right) \left\vert u_{i_{1},}^{\left(
l\right) }...,u_{i_{k_{l}}}^{\left( l\right) }\right\rangle \\
&=&\sum \tprod \Psi ^{\dag \otimes k_{l^{\prime }}^{\prime }}\left(
u_{i_{1},}^{\left( l^{\prime }\right) }...,u_{i_{k_{l^{\prime }}^{\prime
}}}^{\left( l^{\prime }\right) }\right) F\left( \left( u_{i_{1},}^{\left(
l^{\prime }\right) }...,u_{i_{k_{l^{\prime }}^{\prime }}}^{\left( l^{\prime
}\right) }\right) _{l^{\prime }},\left( u_{i_{1},}^{\left( l\right)
}...,u_{i_{k_{l}}}^{\left( l\right) }\right) _{l}\right) \tprod \Psi
^{\otimes k_{l}}\left( u_{i_{1},}^{\left( l\right)
}...,u_{i_{k_{l}}}^{\left( l\right) }\right)
\end{eqnarray*}%
that are thus linear combinations of:%
\begin{equation*}
\Psi ^{\dag \otimes \sum k_{l^{\prime }}^{\prime }}\left( \left(
u_{i_{1},}^{\left( l^{\prime }\right) }...,u_{i_{k^{\prime }}}^{\left(
l^{\prime }\right) }\right) _{l^{\prime }}\right) F\left( \left(
u_{i_{1},}^{\left( l^{\prime }\right) }...,u_{i_{k^{\prime }}}^{\left(
l^{\prime }\right) }\right) _{l^{\prime }},\left( u_{i_{1},}^{\left(
l\right) }...,u_{i_{k}}^{\left( l\right) }\right) _{l}\right) \Psi ^{\otimes
\sum k_{l}}\left( \left( u_{i_{1},}^{\left( l\right) }...,u_{i_{k}}^{\left(
l\right) }\right) _{l}\right)
\end{equation*}%
This expression will be compacted as:%
\begin{equation*}
F\left( \left( u_{i_{1},}^{\left( l^{\prime }\right) }...,u_{i_{k^{\prime
}}}^{\left( l^{\prime }\right) }\right) _{l^{\prime }},\left(
u_{i_{1},}^{\left( l\right) }...,u_{i_{k}}^{\left( l\right) }\right)
_{l}\right) \Psi ^{\otimes \left[ \sum k_{l}+\sum k_{l^{\prime }}^{\prime }%
\right] }\left( \underline{\left( u_{i_{1},}^{\left( l^{\prime }\right)
}...,u_{i_{k^{\prime }}}^{\left( l^{\prime }\right) }\right) _{l^{\prime }}}%
,\left( u_{i_{1},}^{\left( l\right) }...,u_{i_{k}}^{\left( l\right) }\right)
_{l}\right)
\end{equation*}%
so that we will consider a single field:%
\begin{equation*}
\Psi ^{\otimes \left[ kl+k^{\prime }l^{\prime }\right] }\left( \underline{%
\left( u_{i_{1},}^{\left( l^{\prime }\right) }...,u_{i_{k_{l^{\prime
}}^{\prime }}}^{\left( l^{\prime }\right) }\right) _{l^{\prime }}},\left(
u_{i_{1},}^{\left( l\right) }...,u_{i_{k_{l}}}^{\left( l\right) }\right)
_{l}\right) \rightarrow \Psi ^{\otimes k}\left( U^{k}\right)
\end{equation*}%
so that $U$ stands for a product $\underline{U}\times U$.

The fields $\Psi ^{\otimes k}\left( U^{k}\right) $ will be dealt with as if
they were independent quantity. Actually, up to the constraints, a
realization $\Psi ^{\otimes k+l}\left( U^{k+l}\right) $ is not the product
of two realizations of $\Psi ^{\otimes k}\left( U^{k}\right) $ and $\Psi
^{\otimes l}\left( U^{l}\right) $, but rather a sum of products of such
realizations:%
\begin{equation}
\sum_{\alpha }\Psi _{\alpha }^{\otimes k}\left( U^{k}\right) \Psi _{\alpha
}^{\otimes l}\left( U^{l}\right)  \label{DC}
\end{equation}%
In coordinates it means that a realization $\Psi ^{\otimes k+l}\left(
u_{i_{1}},...,u_{i_{k+l}}\right) $ can decomposed as sum of products of
realization:%
\begin{equation*}
\Psi ^{\otimes k+l}\left( u_{i_{1}},...,u_{i_{k+l}}\right) \rightarrow
\sum_{\alpha }\Psi _{\alpha }^{\otimes k}\left(
u_{i_{1}},...,u_{i_{k}}\right) \Psi _{\alpha }^{\otimes l}\left(
u_{i_{1+k}},...,u_{i_{k+l}}\right)
\end{equation*}%
The constraints are included by identifying some parameters:%
\begin{equation*}
\Psi \left( U^{l}\right) =\Psi ^{\otimes l}\left( U^{l}\right) /f
\end{equation*}%
and the tensor product of realizations of $\Psi \left( U\right) $ is a
particular realization of $\Psi \left( U^{l}\right) $. In coordinates: 
\begin{equation*}
\Psi _{1}\left( u_{i_{1}}\right) \otimes ...\otimes \Psi _{k}\left(
u_{i_{k}}\right) /f\left( \left( u_{i_{1}},...,u_{i_{k}}\right) \right)
\end{equation*}%
is a realization of $\Psi \left( U^{l}\right) =\Psi \left(
u_{i_{1}},...u_{i_{k}}\right) $ where implicitly $\left(
u_{i_{1}},...u_{i_{k}}\right) $ stands for $\left(
u_{i_{1}},...u_{i_{k}}\right) /f\left( \left( u_{i_{1}},...u_{i_{k}}\right)
\right) $. All realizations of the field $\Psi \left( U^{l}\right) $ are
given by linear combinations of products of realizations $\Psi \left(
u_{i_{l}}\right) $.

Using the all sequence $\left\{ \Psi ^{\otimes k}\left( U^{k}\right)
\right\} $ as variables allow to restrict the functionals that define the
states to linear combinations:%
\begin{equation*}
F\left( \Psi \left( U\right) \right) =\sum_{k}F_{lin}\left( \Psi ^{\otimes
k}\left( U^{k}\right) \right)
\end{equation*}

\subsection{Remarks}

1. This decomposition will be generalized later to several collections of
sets $\left\{ U_{j}\right\} _{j}$ to write the decomposition in fields.

2. Since a realization of: 
\begin{equation*}
\Psi ^{\otimes k+l}\left( U^{k+l}\right)
\end{equation*}%
writes as a sum: 
\begin{equation*}
\sum_{\alpha }\Psi _{\alpha }^{\otimes k}\left( U^{k}\right) \Psi _{\alpha
}^{\otimes l}\left( U^{l}\right)
\end{equation*}%
the product of the realizations of two fields:%
\begin{equation*}
\Psi ^{\otimes k}\left( U^{k}\right) \Psi ^{\otimes l}\left( U^{l}\right)
\end{equation*}%
is a particular realization of $\Psi ^{\otimes k+l}\left( U^{k+l}\right) $.

\bigskip

3. In terms of of states, i.e. functional, decomposition (\ref{DC}) becomes
sums of terms of the type:%
\begin{equation*}
\sum F_{k,lin}\left( \Psi ^{\otimes k}\left( U^{k}\right) \right)
F_{l,lin}\left( \Psi ^{\otimes k}\left( U^{k}\right) \right)
\end{equation*}

\section{Decompostion in two types of fields}

\subsection{Decomposition and states}

We consider the decomposition of the previous system for field $\Psi
^{\otimes k}\left( U^{k}\right) $ into two subsystems or subobjects, each of
them defined by a field written $\Psi _{J}^{\otimes l}\left(
U_{j}^{l}\right) $ and $\Psi _{I}^{\otimes k}\left( U_{i}^{k}\right) $
respectively along with their associated state space. The states generated
by these two fields, modulo some constraints entangling subobjects, span the
entire states' space of the system. This description generalizes the tensor
product decomposition (\ref{DC}).

Assume a system described by fields:%
\begin{equation*}
\oplus \Psi ^{\otimes m}\left( U^{m}/f_{m}\right)
\end{equation*}%
with some constraint $f_{m}$ on $U^{m}$ and $\mathcal{H}\left( \left(
U\right) \right) $ the spaces of linear functionals of $\oplus \Psi
^{\otimes m}\left( U^{m}/f_{m}\right) $. We consider that the parameters:%
\begin{equation*}
\sum U^{m}/f_{m}
\end{equation*}%
decompose into two different families in parameters by the following map:%
\begin{equation*}
\sum U^{m}/f_{m}\overset{{\Large g}}{\hookrightarrow }V\left(
\sum_{l_{j},k_{i}}\left( U_{j}^{l_{j}}/f_{l_{j}}\times
U_{i}^{k_{i}}/f_{k_{i}}\right) /f_{l_{j},k_{i}}\right)
\end{equation*}%
where:%
\begin{equation*}
\mathbf{V}=V\left( \sum_{l_{j},k_{i}}\left( U_{j}^{l_{j}}/f_{l_{j}}\times
U_{i}^{k_{i}}/f_{k_{i}}\right) /f_{l_{j},k_{i}}\right)
\end{equation*}%
denotes the set of subvarieties of:%
\begin{equation*}
U_{ij}=\sum_{l_{j},k_{i}}\left( U_{j}^{l_{j}}/f_{l_{j}}\times
U_{i}^{k_{i}}/f_{k_{i}}\right) /f_{l_{j},k_{i}}
\end{equation*}%
with implicit constraints. The previous decomposition is not a decomposition
into the powers of subsets of $U$. Due to the arbitrary form of the
constraint, we assume that the full series may enter in the decomposition of
a given $U^{m}/f_{m}$. Associating to an element of $U^{m}/f_{m}$
subvarieties of $U_{ij}$ is analog to the description of some bound state in
field theory in which such state is described by an infinite series of
products of states, involving an infinite number of variable, later
integrated to produce a state, the infinite series coming from some
perturbative expansion. We will write for each $V\subset \mathbf{V}$, we
write the decomposition:%
\begin{equation*}
V=\cup V_{l_{i}k_{j}}
\end{equation*}%
with:%
\begin{equation*}
V_{l_{i}k_{j}}=V\cap \left( U_{j}^{l_{j}}/f_{l_{j}}\times
U_{i}^{k_{i}}/f_{k_{i}}\right) /f_{l_{j},k_{i}}
\end{equation*}

Previous formula leads thus to assume at the field level the corresponding
decomposition in subobjects: 
\begin{equation*}
\Psi ^{\otimes m}\left( U^{m}/f_{m}\right) \rightarrow \sum_{V\subset
g\left( U^{m}/f_{m}\right) }\sum_{V=\cup V_{i,j}}\int_{V_{i,j}}h_{\left(
k,l\right) }^{m}\left( U^{m},U_{j}^{l},U_{i}^{k},V_{i,j}\right) \Psi
_{J}^{\otimes l}\left( U_{j}^{l}\right) \underset{k,l}{\otimes }\Psi
_{I}^{\otimes k}\left( U_{i}^{k}\right) \delta \left( f_{lk}\left(
U_{j}^{l},U_{i}^{k}\right) \right) dU_{j}^{l}dU_{i}^{k}
\end{equation*}%
That formula has to be read is in terms of realization. The realizations $%
\Psi ^{\otimes m}\left( U^{m}/f_{m}\right) $ write as series over products
of realizations of fields $\Psi _{J}\left( U_{j}^{l}\right) $, $\Psi
_{I}\left( U_{i}^{k}\right) $. Including the constraints, the tensor
products represent series expansions:%
\begin{eqnarray*}
&&\Psi _{J}\left( U_{j}^{l}\right) \underset{k,l}{\otimes }\Psi _{I}\left(
U_{i}^{k}\right) \delta \left( f_{lk}\left( U_{j}^{l},U_{i}^{k}\right)
\right) \\
&\equiv &\sum_{\alpha }\Psi _{J,\alpha }\left( U_{j}^{l}\right) \Psi
_{I,\alpha }\left( U_{i}^{k}\right) \delta \left( f_{lk}\left(
U_{j}^{l},U_{i}^{k}\right) \right)
\end{eqnarray*}%
similar to that arising in the tensor product (\ref{DC}).

At the level of states, the functionals:%
\begin{equation}
\sum_{m}\int a_{m}\left( U^{m}\right) \Psi ^{\otimes m}\left( U^{m}\right)
\label{FCN}
\end{equation}%
expand as:%
\begin{equation*}
\sum_{m}\int a_{m}\left( U^{m}\right) \sum_{V\subset g\left(
U^{m}/f_{m}\right) }\sum_{V=\cup V_{i,j}}\int h_{\left( k,l\right)
}^{m}\left( U^{m},U_{j}^{l},U_{i}^{k},V_{i,j}\right) \Psi _{J}^{\otimes
l}\left( U_{j}^{l}\right) \underset{k,l}{\otimes }\Psi _{I}^{\otimes
k}\left( U_{i}^{k}\right) \delta \left( f_{lk}\left(
U_{j}^{l},U_{i}^{k}\right) \right) dU_{j}^{l}dU_{i}^{k}
\end{equation*}%
The sum:%
\begin{equation*}
\sum_{V\subset g\left( U^{m}/f_{m}\right) }\sum_{V=\cup
V_{i,j}}\int_{V_{i,j}}h_{\left( k,l\right) }^{m}\left(
U^{m},U_{j}^{l},U_{i}^{k},V_{i,j}\right)
\end{equation*}%
can be replaced by an unconstrained integral:%
\begin{equation*}
\int h_{\left( k,l\right) }^{m}\left(
U^{m},U_{j}^{l},U_{i}^{k},V_{i,j}\right) \delta \left( f\left(
U^{m},U_{j}^{l},U_{i}^{k}\right) \right)
\end{equation*}%
where the $\delta $ functions:%
\begin{equation*}
\delta \left( f\left( U^{m},U_{j}^{l},U_{i}^{k}\right) \right)
\end{equation*}%
implement the condition $\cup V_{i,j}\subset g\left( U^{m}/f_{m}\right) $.
The integral:%
\begin{equation*}
\sum_{m}\int a_{m}\left( U^{m}\right) h_{\left( k,l\right) }^{m}\left(
U^{m},U_{j}^{l},U_{i}^{k},V_{i,j}\right) \delta \left( f\left(
U^{m},U_{j}^{l},U_{i}^{k}\right) \right)
\end{equation*}%
is a function $a_{l,k}\left( U_{j}^{l},U_{i}^{k}\right) $ and the functional
(\ref{FCN}) writes: 
\begin{equation*}
\sum_{\alpha }\int a_{l,k}\left( U_{j}^{l},U_{i}^{k}\right) \Psi _{J,\alpha
}^{\otimes l}\left( U_{j}^{l}\right) \Psi _{I,\alpha }^{\otimes k}\left(
U_{i}^{k}\right) \delta \left( f_{lk}\left( U_{j}^{l},U_{i}^{k}\right)
\right) dU_{j}^{l}dU_{i}^{k}
\end{equation*}%
which leads to consider a decomposition of spaces of state:

\begin{equation*}
\mathcal{H}\left( \left( U\right) \right) \subset \mathcal{H}\left( \left\{
\left( U_{j}\right) ,\left( U_{i}\right) \right\} \right)
\end{equation*}%
Given that the space $\mathcal{H}\left( \left( U\right) \right) $ is an
infinite series of tensor products, we assume in the sequel that:%
\begin{equation*}
\mathcal{H}\left( \left( U\right) \right) =\mathcal{H}\left( \left\{ \left(
U_{j}\right) ,\left( U_{i}\right) \right\} \right)
\end{equation*}

Note that the decomposition presented above represents the reverse path
compared to the one leading to the formation of a composed states. We
decompose a given state into a series of products of functionals associated
with different subobjects. In the next sections, we will work directly with
the products:%
\begin{equation*}
\Psi _{J,\alpha }^{\otimes l}\left( U_{j}^{l}\right) \otimes \Psi _{I,\alpha
}^{\otimes k}\left( U_{i}^{k}\right) \delta \left( f_{lk}\left(
U_{j}^{l},U_{i}^{k}\right) \right)
\end{equation*}

\subsection{Partial states}

Usually, a state for one single field, say $\left\{ \Psi _{J}^{\otimes
l}\right\} _{l}$ is defined by some functional of this field:%
\begin{equation}
\sum \int s\left( U_{j}^{l}\right) \Psi _{J}^{\otimes l}\left(
U_{j}^{l}\right) dU_{j}^{l}  \label{SND}
\end{equation}%
that is, by the set of functions $\left\{ s\left( U_{j}^{l}\right) \right\} $%
.

However, in the present context, both fields arising in the decomposition
are subject to some constraints $\left\{ f_{lk}\left(
U_{j}^{l},U_{i}^{k}\right) \right\} $. Thus, we have to consider that a
state for one field is subject to some constraints in its integration
variables.

We begin by defining the evaluation functional at $\left(
U_{j}^{l},U_{i}^{k}\right) $ for two realizations of the field $\Psi
_{J}^{\otimes l}$ and $\Psi _{I}^{\otimes k}$:%
\begin{equation*}
ev_{U_{j}^{l},U_{i}^{k}}\left( \Psi _{J}^{\otimes l}\underset{l,k}{\otimes }%
\Psi _{I}^{\otimes k}\right) \rightarrow \Psi _{J}^{\otimes l}\left(
U_{j}^{l}\right) \underset{l,k}{\otimes }\Psi _{I}^{\otimes k}\left(
U_{i}^{k}\right)
\end{equation*}%
and impose the constraints $\delta \left( f_{lk}\left(
U_{j}^{l},U_{i}^{k}\right) \right) $ between these two subobjects: 
\begin{equation}
ev_{U_{j}^{l},U_{i}^{k}}\left( \Psi _{J}^{\otimes l}\underset{l,k}{\otimes }%
\Psi _{I}^{\otimes k}\right) \delta \left( f_{lk}\left(
U_{j}^{l},U_{i}^{k}\right) \right) =\Psi _{J}^{\otimes l}\left(
U_{j}^{l}\right) \underset{l,k}{\otimes }\Psi _{I}^{\otimes k}\left(
U_{i}^{k}\right) \delta \left( f_{lk}\left( U_{j}^{l},U_{i}^{k}\right)
\right)  \label{VL}
\end{equation}%
The definition of a partial state for $\Psi _{J}^{\otimes l}$ or $\Psi
_{I}^{\otimes k}$ should thus respect the constraints. We define a state for 
$\Psi _{J}^{\otimes l}$ by a set of functions:%
\begin{equation*}
s\equiv \left\{ s_{U_{i}^{k}}\left( U_{j}^{l}\right) \right\} _{\left(
U_{j}^{l},U_{i}^{k}\right) }
\end{equation*}%
and a functional of $\Psi _{J}^{\otimes l}\underset{l,k}{\otimes }\Psi
_{I}^{\otimes k}$ defined by the combinations of evaluations (\ref{VL}):%
\begin{eqnarray*}
&&\sum \int s_{U_{i}^{k}}\left( U_{j}^{l}\right)
ev_{U_{j}^{l},U_{i}^{k}}\left( \Psi _{J}^{\otimes l}\underset{l,k}{\otimes }%
\Psi _{I}^{\otimes k}\right) \delta \left( f_{lk}\left(
U_{j}^{l},U_{i}^{k}\right) \right) dU_{j}^{l} \\
&=&\sum \int s_{U_{i}^{k}}\left( U_{j}^{l}\right) \Psi _{J}^{\otimes
l}\left( U_{j}^{l}\right) \underset{l,k}{\otimes }\Psi _{I}^{\otimes
k}\left( U_{i}^{k}\right) \delta \left( f_{lk}\left(
U_{j}^{l},U_{i}^{k}\right) \right) \delta \left( f_{lk}\left(
U_{j}^{l},U_{i}^{k}\right) \right) dU_{j}^{l}
\end{eqnarray*}

Note that we will alternatively write $s\equiv \left\{ s_{U_{i}^{k}}\left(
\Psi _{J}^{\otimes l}\right) \right\} $ or $\left\{ s_{U_{j}^{l}}\left( \Psi
_{I}^{\otimes k}\right) \right\} $ to define a partial functional of $\Psi
_{J}^{\otimes l}$ or $\Psi _{I}^{\otimes k}$ respectively.

\section{Projection over functional minima and effective field}

Once the system has been decomposed into subobjects $\Psi _{J}^{\otimes
l}\left( U_{j}^{l}\right) $ and $\Psi _{I}^{\otimes k}\left(
U_{i}^{k}\right) $, we examine the projection of $\Psi _{I}^{\otimes
k}\left( U_{i}^{k}\right) $ onto the space generated by the saddle-point
solutions of a functional $S$, analog to the background states in a typical
action functional. Given the interdependence of subobjects through
constraints and the initial system's dependence on products such as $\Psi
_{J}\left( U_{j}^{l}\right) \underset{k,l}{\otimes }\Psi _{I}\left(
U_{i}^{k}\right) $, we posit that the functional whose saddle points
determine the projection depends on both fields $\Psi _{J}^{\otimes l}\left(
U_{j}^{l}\right) $ and $\Psi _{I}^{\otimes k}\left( U_{i}^{k}\right) $, \
i.e. on both sets of realizations $\Psi _{J,\alpha }^{\otimes l}\left(
U_{j}^{l}\right) $ and $\Psi _{I,\alpha }^{\otimes k}\left( U_{i}^{k}\right) 
$. The projection that defines the states minimizing $S$ will consequently
depend on $\Psi _{J}\left( U_{j}^{l}\right) \underset{k,l}{\otimes }\Psi
_{I}\left( U_{i}^{k}\right) $ and potentially on some functional $v$ of $%
\Psi _{J,\alpha }^{\otimes l}\left( U_{j}^{l}\right) $ that characterizes a
state for $\Psi _{J,\alpha }^{\otimes l}\left( U_{j}^{l}\right) $. We thus
write $S\left( \Psi _{J}\left( U_{j}^{l}\right) \underset{k,l}{\otimes }\Psi
_{I}\left( U_{i}^{k}\right) ,v\right) $ for this functional or $S\left(
v\right) $ for short.

\subsection{States and projections}

We consider the projection of states onto a background of the subspace
defined by field $\Psi _{I}^{\otimes k}\left( U_{i}^{k}\right) $.
Additionally, we assume the existence of a basis of states $v$ for $\Psi
_{J}^{\otimes l}$, i.e. functionals of the fields $\Psi _{J}^{\otimes l}${}
so that:%
\begin{equation*}
\sum_{v}\tprod\limits_{v}
\end{equation*}%
represents the identity. As a consequence, the states are projected onto
spaces:%
\begin{equation*}
\sum_{v}\tprod\limits_{v}\otimes \tprod_{\min S\left( v\right) }
\end{equation*}%
by projecting the $\Psi _{I}^{\otimes k}$ onto the minima of $S\left(
v\right) $ which depend on $\Psi _{J}^{\otimes l}$ through states $v$. This
reflects that the decomposition is carried out for entangled states, that
is, interacting fields.

To write a functional in the basis $\left\{ v\right\} $, recall that each
element of this basis is defined by a collection: $\left\{
v_{U_{i}^{k}}\left( \Psi _{J}^{\otimes l}\right) \right\} $. Starting with
an arbitrary functional respecting the conservation $\delta \left(
f_{lk}\left( U_{j}^{l},U_{i}^{k}\right) \right) $:%
\begin{equation}
\int g\left( U_{j}^{l},U_{i}^{k}\right) \Psi _{J}^{\otimes l}\left(
U_{j}^{l}\right) \Psi _{I}^{\otimes k}\left( U_{i}^{k}\right) \delta \left(
f_{lk}\left( U_{j}^{l},U_{i}^{k}\right) \right)  \label{FNL}
\end{equation}%
that can be rewritten in the basis $\left\{ v\right\} $ as: 
\begin{equation}
\sum_{v,v_{U_{i}^{k}}}\int g\left( v,U_{i}^{k}\right) v_{U_{i}^{k}}\left(
\Psi _{J}^{\otimes l}\right) ev_{U_{i}^{k}}\left( \Psi _{I}^{\otimes
k}\left( U_{i}^{k}\right) \right) \delta \left( f_{lk}\left(
U_{j}^{l},U_{i}^{k}\right) \right)  \label{FNV}
\end{equation}%
where:%
\begin{equation*}
ev_{U_{i}^{k}}\left( \Psi _{I}^{\otimes k}\left( U_{i}^{k}\right) \right)
=\Psi _{I}^{\otimes k}\left( U_{i}^{k}\right)
\end{equation*}%
is the evaluation functional for $\Psi _{I}^{\otimes k}$. The states $%
v_{U_{i}^{k}}\left( \Psi _{J}^{\otimes l}\right) $ on which the projection
arise depend on the constraint $\delta \left( f_{lk}\left(
U_{j}^{l},U_{i}^{k}\right) \right) $.

In coordinates, we can solve the constraints and $v_{U_{i}^{k}}\left( \Psi
_{J}^{\otimes l}\right) $ writes:%
\begin{equation*}
v_{U_{i}^{k}}\left( \Psi _{J}^{\otimes l}\right) =\int v\left(
U_{j}^{l}/f_{lk},\left\{ U_{i}^{k}\right\} \right) \Psi _{J}^{\otimes
l}\left( U_{j}^{l}/f_{lk},\left\{ U_{i}^{k}\right\} \right) d\left(
U_{j}^{l}/f_{lk}\right)
\end{equation*}%
and $v\left( U_{j}^{l}/f_{lk},\left\{ U_{i}^{k}\right\} \right) $ is the
functional density associated to $v_{U_{i}^{k}}\left( \Psi _{J}^{\otimes
l}\right) $.

Given the infinite number of realizations, i.e. components involved in the
products, we consider the multiple functional and replace: 
\begin{equation*}
v_{U_{i}^{k}}\left( \Psi _{J}^{\otimes l}\right) ev_{U_{i}^{k}}\left( \Psi
_{I}^{\otimes k}\right) \rightarrow v_{U_{i}^{k}}\left( \Psi _{J,\alpha
}^{\otimes l}\right) ev_{U_{i}^{k}}\left( \Psi _{I,\alpha }^{\otimes
k}\right)
\end{equation*}%
for each realization $\alpha $.

Note that we can consider these functionals:%
\begin{equation*}
v_{\left\{ U_{i}^{k}\right\} }\left( \left\{ \Psi _{J}^{\otimes l}\right\}
\right)
\end{equation*}%
as eigenstates of some operator:%
\begin{equation*}
\Phi \left( \Psi _{J}^{\otimes k_{l}},\frac{\delta }{\delta \Psi
_{J}^{\otimes k_{l}}},U_{i}^{k},\Psi _{I}^{\otimes k}\right)
\end{equation*}%
that depend on the $U_{i}^{k}$ and $\Psi _{I}^{\otimes k}$ through the
constraint and some operators involving interactions between both subobjects.

\subsubsection{Remark}

We will assume in general that fields are already chosen as eigenstates of $%
\Lambda $ so that the $\sum_{v}$ is performed for a collection:%
\begin{equation*}
\sum_{U_{j}^{l}/f_{lk}}v_{\left\{ U_{i}^{k}\right\}
}^{U_{j}^{l}/f_{lk}}\left\{ \Psi _{J}^{\otimes l}\right\}
\end{equation*}%
\begin{equation*}
v_{\left\{ U_{i}^{k}\right\} }^{U_{j}^{l}/f_{lk}}\left\{ \Psi _{J}^{\otimes
l}\right\} =\Psi _{J}^{\otimes l}\left( U_{j}^{l}\right) =\Psi _{J}^{\otimes
l}\left( U_{j}^{l}/f_{lk},\left\{ U_{i}^{k}\right\} \right)
\end{equation*}%
which corresponds to sum over the evaluation at points $U_{j}^{l}/f_{lk}$
and the decomposition is performd with respect to the values of $%
U_{j}^{l}/f_{kl}$. We can also consider restrict the projection to a
subspace $V$ of $U_{j}^{l}/f_{lk}$:%
\begin{equation*}
v_{\left\{ U_{i}^{k}\right\} }^{V}\left\{ \Psi _{J}^{\otimes l}\right\}
=\Psi _{J}^{\otimes l}\left( V,\left\{ U_{i}^{k}\right\} \right)
\end{equation*}%
which reduces the evaluation to $\left( V,\left\{ U_{i}^{k}\right\} \right) $%
.

\subsection{Action functional saddle point and projection}

We assume that the projection of $\Psi _{I,\alpha }^{\otimes k}\left(
U_{i}^{k}\right) $ comes from the minimization:%
\begin{equation*}
\exp \left( -S\left( \left\{ \Psi _{J,\alpha }^{\otimes l}\left(
U_{j}^{l}\right) \underset{l,k}{\otimes }\Psi _{I,\alpha }^{\otimes k}\left(
U_{i}^{k}\right) \right\} _{\alpha },\left\{ v_{\left\{ U_{i}^{k}\right\}
}\left( \Psi _{J,\alpha }^{\otimes k}\right) \otimes \Psi _{I,\alpha
}^{\otimes k}\right\} _{\left\{ U_{i}^{k}\right\} }\right) \right)
\end{equation*}

This form of $S$ emphasizes that the primary object is the initial field $%
\Psi ^{\otimes m}\left( U^{m}/f_{m}\right) $. Multiple realizations of the
fields are implicated, as a consequence of our assumption about
decomposition, indicating that realizations of the field $\Psi ^{\otimes
m}\left( U^{m}/f_{m}\right) $ involve multiple realizations $\Psi _{J,\alpha
}^{\otimes l}$, $\Psi _{I,\alpha }^{\otimes k}$. We have assumed the
functional depends on some specific intertwined states $v_{\left\{
U_{i}^{k}\right\} }$ on which the field $\Psi _{J,\alpha }^{\otimes k}$ is
projected to model that the projection of $\Psi _{I,\alpha }^{\otimes k}$
depends on a state for $\Psi _{J,\alpha }^{\otimes l}$. This reminiscent of
current-current interaction involving some functionals (usually local) of
the fields.

In appendix 1, assuming that $S$ is a series in the variables:%
\begin{equation*}
\left\{ \Psi _{J,\alpha }^{\otimes l}\left( U_{j}^{l}\right) \underset{l,k}{%
\otimes }\Psi _{I,\alpha }^{\otimes k}\left( U_{i}^{k}\right) \right\}
_{\alpha }
\end{equation*}%
including a quadratic term for the $\Psi _{I,\alpha }^{\otimes k}\left(
U_{i}^{k}\right) $, we decompose the solutions of the saddle point equation
as:%
\begin{equation*}
\left\{ \Psi _{J}^{\otimes l}\left( U_{j}^{l}\right) \tbigoplus \Psi
_{I,0}^{\otimes k}\left( U_{i}^{k}\right) \right\} _{\substack{ l\leqslant m 
\\ l^{\prime }\leqslant m^{\prime }}}\equiv \left\{ \Psi _{I,0}^{\otimes
k}\left( U_{i}^{k}\right) \right\} _{\Psi _{J}^{\otimes l}\left(
U_{j}^{l}\right) }
\end{equation*}%
and write a series expansion for $\Psi _{I,0}^{\otimes k}\left(
U_{i}^{k}\right) $. We show that:%
\begin{eqnarray*}
&&\Psi _{I,0,\alpha }^{\otimes k}\left( U_{i}^{k},\left\{ \Psi _{J}^{\otimes
l}\right\} \right) \\
&=&\sum_{s,l_{i},...,l_{s}}\int d\left( \left\{ U_{j}^{l_{i}}\right\}
_{l_{i}}\right) d\left( \left\{ U_{i}^{k_{i}}/f_{k_{i}l_{i}}\right\}
_{k_{i},l_{i}}\right) \mathcal{K}_{0}^{v}\left( U_{i}^{k},\left\{
U_{j}^{l_{i}}\right\} _{l_{i}},\left\{ U_{i}^{k_{i}}/f_{k_{i}l_{i}}\right\}
_{i}\right) \Psi _{J,\alpha }^{\otimes \sum_{i\leqslant n}l_{i}}\left(
\left( U_{j}^{l_{i}}\right) \right)
\end{eqnarray*}%
where:%
\begin{equation*}
\Psi _{J,\alpha }^{\otimes \sum_{i\leqslant n}l_{i}}\left( \left(
U_{j}^{l_{i}}\right) \right) =\sum_{\left\{ \alpha _{i}^{\prime }\right\}
_{_{i\leqslant n}}}\tprod_{i}\Psi _{J,\alpha _{i}^{\prime }}^{\otimes
l_{i}}\left( U_{j}^{l_{i}}\right)
\end{equation*}%
and:%
\begin{equation*}
\mathcal{K}_{0}^{v}\left( U_{i}^{k},\left\{ U_{j}^{l_{i}}\right\}
_{l_{i}},\left\{ U_{i}^{k_{i}}/f_{k_{i}l_{i}}\right\} _{i}\right) =\mathcal{K%
}_{0}\left( U_{i}^{k},\left\{ U_{j}^{l_{i}}\right\} _{l_{i}},\left\{
U_{i}^{k_{i}}/f_{k_{i}l_{i}}\right\} _{i}\right) \tprod v\left(
U_{i}^{k_{i}},U_{j}^{l_{i}}/f_{k_{i}l_{i}}\right)
\end{equation*}%
The dependency in $\alpha $ is justified by the fact that an element of $%
\left\{ \alpha _{i}^{\prime }\right\} _{_{i\leqslant n}}$ is $\alpha $ or
arises in the action in products involving $\alpha $.

\subsection{Degeneracy of saddle points}

So far the formulas have been obtained for one saddle point. However,
considering the presence of multiple realizations of the\ field in the
functional, we can expect some symmetry permuting or combining these
realizations to arise. We will assume certain symmetry groups that imply the
degeneracy of the saddle point.

\subsubsection{General formula}

Consider the action:%
\begin{equation*}
S\left( \left\{ \Psi _{J,\alpha }^{\otimes l}\left( U_{j}^{l}\right) 
\underset{l,k}{\otimes }\Psi _{I,\alpha }^{\otimes k}\left( U_{i}^{k}\right)
\right\} _{\alpha },\left\{ v_{\left\{ U_{i}^{k}\right\} }\left( \Psi
_{J,\alpha }^{\otimes k}\right) \otimes \Psi _{I,\alpha }^{\otimes
k}\right\} _{\left\{ U_{i}^{k}\right\} }\right)
\end{equation*}%
depending on a sequence of realizations:%
\begin{equation*}
\left\{ \Psi _{J,\alpha }^{\otimes l}\left( U_{j}^{l}\right) \underset{l,k}{%
\otimes }\Psi _{I,\alpha }^{\otimes k}\left( U_{i}^{k}\right) \right\}
_{\alpha }
\end{equation*}%
that can also be written:%
\begin{equation*}
\left\{ \Psi _{J,\alpha _{i}}^{\otimes l}\left( U_{j}^{l}\right) \underset{%
l,k}{\otimes }\Psi _{I,\alpha _{i}}^{\otimes k}\left( U_{i}^{k}\right)
\right\} _{i=1,...,n...}
\end{equation*}

We look for the invariance of $S$ as a fnction of a given sequence $\left\{
\Psi _{J,\alpha _{i}}^{\otimes l}\left( U_{j}^{l}\right) \right\} _{i}$

We assume the existence of sequences of groups of transformations $G_{\alpha
_{1},...\alpha _{n}}^{k_{i},...,k_{n}}\left( \left\{ \Psi _{J,\alpha
_{i}}^{\otimes l}\left( U_{j}^{l}\right) \right\} _{i=1,...n}\right) $ with $%
n\geqslant 1$, written $G^{k_{i},...,k_{n}}$ fr short, such that there is an
invariance of $S$ by action of $G_{k_{i},...,k_{n}}$ on:%
\begin{equation*}
\left\{ \Psi _{I,\alpha _{i}}^{\otimes k_{i}}\left( U_{i}^{k_{i}}\right)
\right\} _{i\leqslant n}
\end{equation*}%
The transformations for a set of realizations are parametrized by $\mathbf{%
\hat{\Lambda}}_{\alpha _{1},...\alpha _{n}}^{\left[ \left\{ k_{i\leqslant
n}\right\} \right] }\left[ \Psi _{J},\nu \right] $ with $\left[ \left\{
k_{i\leqslant n}\right\} \right] =\left( k_{1},...k_{n}\right) $ and $\alpha
_{1},...\alpha _{n}$ is a $n-$uplet of realizations. These parameters span
for example the symmetry groups of a series of the form:%
\begin{eqnarray}
&&S\left( \left( \Psi _{I,\alpha _{i}}^{\otimes k_{i}}\left(
U_{i}^{k_{i}}\right) \right) \right)  \label{SRS} \\
&=&\sum_{m}\sum_{\left\{ k_{i\leqslant m}\right\} \subset \left\{
k_{i\leqslant n}\right\} }\int f\left( \left\{
U_{i}^{k_{i}},U_{j}^{l_{i}}/f_{k_{i}l_{i}}\right\} _{i}\right)
\tprod_{i}\Psi _{J,\alpha _{i}^{\prime }}^{\otimes l_{i}}\left(
U_{j}^{l_{i}}/f_{k_{i}l_{i}},U_{i}^{k_{i}}\right) \tprod_{i=1}^{m}\Psi
_{I,\alpha _{i}}^{\otimes k_{i}}\left( U_{i}^{k_{i}}\right) d\left( \left\{
U_{i}^{k_{i}},U_{j}^{l_{i}}/f_{k_{i}l_{i}}\right\} _{i}\right)  \notag
\end{eqnarray}%
where the transformation acts on $n$ realizations:

\begin{equation*}
G^{k_{i},...,k_{n}}\left( \mathbf{\hat{\Lambda}}_{\alpha _{1},...\alpha
_{n}}^{\left[ \left\{ k_{i\leqslant n}\right\} \right] }\right) .\left(
\tprod\limits_{i=1...n}\left\{ \Psi _{I,\alpha _{i}}^{\otimes k_{i}}\left(
U_{i}^{k_{i}}\right) \right\} _{i\leqslant n}\right) =\left(
\tprod\limits_{i=1...n}R_{\alpha _{1},...\alpha _{n}}^{\left[ \left\{
k_{i\leqslant n}\right\} \right] }\left( \mathbf{\hat{\Lambda}}_{\alpha
_{1},...\alpha _{n}}^{\left[ \left\{ k_{i\leqslant n}\right\} \right]
}\right) \Psi _{I,\alpha _{i}}^{\otimes k_{i}}\left(
g_{k_{i},...,k_{n}}^{k_{i}}U_{i}^{k_{i}}\right) \right)
\end{equation*}%
with $R_{\left\{ \alpha _{i\leqslant n}\right\} }^{\left\{ k_{i\leqslant
n}\right\} }$ acting on indices $\alpha _{1},...\alpha _{n}$:%
\begin{equation*}
R_{\alpha _{1},...\alpha _{n}}^{\left\{ k_{i\leqslant n}\right\} }\left( 
\mathbf{\hat{\Lambda}}_{\alpha _{1},...\alpha _{n}}^{\left[ \left\{
k_{i\leqslant n}\right\} \right] }\right) \Psi _{I,\alpha _{i}}^{\otimes
k_{i}}\left( g_{k_{i},...,k_{n}}^{k_{i}}U_{i}^{k_{i}}\right) =\sum_{\alpha
_{i}^{\prime }}\left[ R_{\alpha _{1},...\alpha _{n}}^{\left[ \left\{
k_{i\leqslant n}\right\} \right] }\left( \mathbf{\hat{\Lambda}}_{\alpha
_{1},...\alpha _{n}}^{\left[ \left\{ k_{i\leqslant n}\right\} \right]
}\right) \right] _{\alpha _{i}}^{\alpha _{i}^{\prime }}\Psi _{I,\alpha
_{i}^{\prime }}^{\otimes k_{i}}\left(
g_{k_{i},...,k_{n}}^{k_{i}}U_{i}^{k_{i}}\right)
\end{equation*}%
and:%
\begin{equation*}
g_{k_{i},...,k_{n}}\left( \left( U_{i}^{k_{i}}\right) \right) =\left(
g_{k_{i},...,k_{n}}^{k_{i}}U_{i}^{k_{i}}\right)
\end{equation*}%
acts on the coordinates $\left( U_{i}^{k_{i}}\right) $. The parameters $%
\left\{ \mathbf{\hat{\Lambda}}_{\alpha _{1},...,\alpha _{n}}^{\left[ \left\{
k_{i\leqslant n}\right\} \right] }\right\} $ describe the group elements.
The transformation depends on the $\left\{ \Psi _{J,\alpha _{i}}^{\otimes
l}\left( U_{j}^{l}\right) \right\} _{i=1,...n}$ so that in the sequel,we
will write:%
\begin{equation*}
\left\{ \mathbf{\hat{\Lambda}}_{\alpha _{1},...,\alpha _{n}}^{\left[ \left\{
k_{i\leqslant n}\right\} \right] }\left[ \Psi _{J},\nu \right] \right\}
\end{equation*}%
for the set of parameters of the group $G^{k_{i},...,k_{n}}$ acting on
several realizations $\alpha _{1},...,\alpha _{n}$. The $k_{i\leqslant n}$
are integer variables.

More generally, we can assume that the transformations mix the elements of a
collection of $n-$uplet of realizations. We write $\left\{ \alpha
_{i\leqslant n}\right\} $ for an arbitrary collection $\left\{ \alpha
_{1},...,\alpha _{n}\right\} $ of such $n-$uplets. In this case, the
transformation writes:%
\begin{eqnarray}
&&G_{\left\{ \alpha _{i\leqslant n}\right\} }^{k_{i},...,k_{n}}\left( 
\mathbf{\hat{\Lambda}}_{\left\{ \alpha _{i\leqslant n}\right\} }^{\left[
\left\{ k_{i\leqslant n}\right\} \right] }\right) .\left(
\tprod\limits_{i=1...n}\left\{ \Psi _{I,\alpha _{i}}^{\otimes k_{i}}\left(
U_{i}^{k_{i}}\right) \right\} _{i\leqslant n}\right)  \label{GRT} \\
&=&\sum_{\left( \alpha _{1}^{\prime },...,\alpha _{n}^{\prime }\right) \in
\left\{ \alpha _{i\leqslant n}\right\} }\left( \tprod\limits_{i=1...n}\left[
R_{\left\{ \alpha _{i\leqslant n}\right\} }^{\left\{ k_{i\leqslant
n}\right\} }\left( \mathbf{\hat{\Lambda}}_{\left\{ \alpha _{i\leqslant
n}\right\} }^{\left[ \left\{ k_{i\leqslant n}\right\} \right] }\right) %
\right] _{\left( \alpha _{1},...,\alpha _{n}\right) }^{\left( \alpha
_{1}^{\prime },...,\alpha _{n}^{\prime }\right) }\Psi _{I,\alpha
_{i}^{\prime }}^{\otimes k_{i}}\left(
g_{k_{i},...,k_{n}}^{k_{i}}U_{i}^{k_{i}}\right) \right)  \notag
\end{eqnarray}%
where the sum is over the collection considered, and with $\left[ R_{\left\{
\alpha _{i\leqslant n}\right\} }^{\left\{ k_{i\leqslant n}\right\} }\left( 
\mathbf{\hat{\Lambda}}_{\left\{ \alpha _{i\leqslant n}\right\} }^{\left[
\left\{ k_{i\leqslant n}\right\} \right] }\right) \right] _{\left( \alpha
_{1},...,\alpha _{n}\right) }^{\left( \alpha _{1}^{\prime },...,\alpha
_{n}^{\prime }\right) }$ acting on indices $\alpha _{1},...,\alpha _{n}$:%
\begin{eqnarray}
&&R_{\left\{ \alpha _{i\leqslant n}\right\} }^{\left\{ k_{i\leqslant
n}\right\} }\left( \mathbf{\hat{\Lambda}}_{\left\{ \alpha _{i\leqslant
n}\right\} }^{\left[ \left\{ k_{i\leqslant n}\right\} \right] }\right) \Psi
_{I,\alpha _{i}}^{\otimes k_{i}}\left(
g_{k_{i},...,k_{n}}^{k_{i}}U_{i}^{k_{i}}\right)  \label{GRD} \\
&=&\sum_{\left( \alpha _{1}^{\prime },...,\alpha _{n}^{\prime }\right) \in
\left\{ \alpha _{i\leqslant n}\right\} }\sum_{\alpha _{i}^{\prime }\in
\left( \alpha _{1}^{\prime },...,\alpha _{n}^{\prime }\right) }\left[ \left[
R_{\left\{ \alpha _{i\leqslant n}\right\} }^{\left\{ k_{i\leqslant
n}\right\} }\left( \mathbf{\hat{\Lambda}}_{\left\{ \alpha _{i\leqslant
n}\right\} }^{\left[ \left\{ k_{i\leqslant n}\right\} \right] }\right) %
\right] _{\left( \alpha _{1},...,\alpha _{n}\right) }^{\left( \alpha
_{1}^{\prime },...,\alpha _{n}^{\prime }\right) }\right] _{\alpha
_{i}}^{\alpha _{i}^{\prime }}\Psi _{I,\alpha _{i}^{\prime }}^{\otimes
k_{i}}\left( g_{k_{i},...,k_{n}}^{k_{i}}U_{i}^{k_{i}}\right)  \notag
\end{eqnarray}%
the parameters $\mathbf{\hat{\Lambda}}_{\left\{ \alpha _{i\leqslant
n}\right\} }^{\left[ \left\{ k_{i\leqslant n}\right\} \right] }\left[ \Psi
_{J},\nu \right] $ are local coordinates for the transformation $G_{\left\{
\alpha _{i\leqslant n}\right\} }^{k_{i},...,k_{n}}\left( \mathbf{\hat{\Lambda%
}}_{\left\{ \alpha _{i\leqslant n}\right\} }^{\left[ \left\{ k_{i\leqslant
n}\right\} \right] }\right) $.

Considering all transformations, the full set of parameters is thus:%
\begin{equation*}
\left( \left\{ \mathbf{\hat{\Lambda}}_{\left\{ \alpha _{i\leqslant
n}\right\} }^{\left[ \left\{ k_{i\leqslant n}\right\} \right] }\left[ \Psi
_{J},\nu \right] \right\} \right) _{n}
\end{equation*}%
where $\left\{ \mathbf{\hat{\Lambda}}_{\left\{ \alpha _{i\leqslant
n}\right\} }^{\left[ \left\{ k_{i\leqslant n}\right\} \right] }\left[ \Psi
_{J},\nu \right] \right\} $ describes all set of parameters for a given
collection $\left\{ \alpha _{i\leqslant n}\right\} $.

When all realizations $\left\{ \alpha _{i\leqslant n}\right\} $ arise, the
variables $\left\{ \mathbf{\hat{\Lambda}}_{\left\{ \alpha _{i\leqslant
n}\right\} }^{\left[ \left\{ k_{i\leqslant n}\right\} \right] }\left[ \Psi
_{J},\nu \right] \right\} $ is independent from a given set $\alpha
_{1},...,\alpha _{n}$. As a consequence, considering the symetry groups for $%
\left\{ \Psi _{I,\alpha _{i}}^{\otimes k_{i}}\left( U_{i}^{k_{i}}\right)
\right\} _{i\leqslant n}$ and assume that one of the index is a given
realization $\alpha $, the set $\left\{ \mathbf{\hat{\Lambda}}_{\left\{
\alpha _{i\leqslant n}\right\} }^{\left[ \left\{ k_{i\leqslant n}\right\} %
\right] }\left[ \Psi _{J},\nu \right] \right\} $ only depends on $\alpha $
and we write:%
\begin{equation*}
\left\{ \mathbf{\hat{\Lambda}}_{\alpha }^{\left[ \left\{ k,k_{i\leqslant
n-1}\right\} \right] }\left[ \Psi _{J},\nu \right] \right\}
\end{equation*}%
and:%
\begin{equation*}
\left( \left\{ \mathbf{\hat{\Lambda}}_{\alpha }^{\left[ \left\{
k,k_{i\leqslant n-1}\right\} \right] }\left[ \Psi _{J},\nu \right] \right\}
\right) _{n}
\end{equation*}%
for $n$ running from $1$ to $\infty $ when all sets of parameters are taken
into account. We show in appendix 2 that the fields solving saddle point
equation write:%
\begin{equation*}
\Psi _{I,0,\alpha }^{\otimes k}\left( U_{i}^{k},\left( \left\{ \mathbf{\hat{%
\Lambda}}_{\alpha }^{\left[ \left\{ k,k_{i\leqslant n-1}\right\} \right] }%
\left[ \Psi _{J},\nu \right] \right\} \right) _{n},\left\{ \Psi
_{J}^{\otimes l}\right\} \right)
\end{equation*}

We also show in appendix 2 that for transformation groups that satisfisy:%
\begin{equation*}
G_{\alpha _{1},...\alpha _{n}}^{k_{i},...,k_{n}}\subset G_{\alpha
_{1},...\alpha _{n+1}}^{k_{i},...,k_{n+1}}
\end{equation*}%
the whole set of parameters:%
\begin{equation*}
\left( \left\{ \mathbf{\hat{\Lambda}}_{\alpha }^{\left[ \left\{
k,k_{i\leqslant n-1}\right\} \right] }\left[ \Psi _{J},\nu \right] \right\}
\right) _{n}
\end{equation*}%
describes set of infinite dimensional flag manifolds, all starting with $%
\mathbf{\hat{\Lambda}}_{\alpha }^{\left[ k\right] }\left[ \Psi _{J},\nu %
\right] $.

In the sequel we note:%
\begin{eqnarray*}
\mathbf{\hat{\Lambda}}_{n,\alpha }\left[ \Psi _{J},\nu \right] &=&\left\{ 
\mathbf{\hat{\Lambda}}_{\alpha }^{\left[ \left\{ k_{i\leqslant n}\right\} %
\right] }\left[ \Psi _{J},\nu \right] \right\} \\
\mathbf{\hat{\Lambda}}_{\infty ,\alpha }\left[ \Psi _{J},\nu \right]
&=&\left( \left\{ \mathbf{\hat{\Lambda}}_{\alpha }^{\left[ \left\{
k_{i\leqslant n}\right\} \right] }\left[ \Psi _{J},\nu \right] \right\}
\right) _{n}
\end{eqnarray*}%
and:%
\begin{eqnarray*}
\mathbf{\hat{\Lambda}}_{n,\alpha }^{\left[ k\right] }\left[ \Psi _{J},\nu %
\right] &=&\left\{ \mathbf{\hat{\Lambda}}_{\alpha }^{\left[ \left\{
k,k_{i\leqslant n-1}\right\} \right] }\left[ \Psi _{J},\nu \right] \right\}
\\
\mathbf{\hat{\Lambda}}_{\infty ,\alpha }^{\left[ k\right] }\left[ \Psi
_{J},\nu \right] &=&\left( \left\{ \mathbf{\hat{\Lambda}}_{\alpha }^{\left[
\left\{ k,k_{i\leqslant n-1}\right\} \right] }\left[ \Psi _{J},\nu \right]
\right\} \right) _{n}
\end{eqnarray*}

When these states are independent from $\alpha $, that index will be removed.

\subsubsection{Example: Case of reparametrization invariance for original
field}

We start by considering general terms:%
\begin{equation}
\sum_{\alpha _{i}^{\left( 1\right) },...,\alpha _{i}^{\left( l\right) }}\int
f\left( \left\{ U_{i}^{k_{i}},U_{j}^{l_{i}}/f_{k_{i}l_{i}}\right\}
_{1,...,l},\left[ \Psi _{J}\right] \right) \tprod\limits_{t=1,...,l}\Psi
_{J,\alpha _{i}^{\left( t\right) }}^{\otimes l_{i}}\left( \left(
U_{j}^{l_{i}}/f_{k_{i}l_{i}},U_{i}^{k_{i}}\right) _{t}\right) \Psi
_{I,\alpha _{i}^{\left( t\right) }}^{\otimes k_{i}}\left( \left(
U_{i}^{k_{i}}\right) _{t}\right) d\left( \left\{ \left(
U_{i}^{k_{i}},U_{j}^{l_{i}}/f_{k_{i}l_{i}}\right) _{t}\right\} \right)
\label{GTV}
\end{equation}%
arising from the power functions in the initial field theory:%
\begin{equation}
\int f\left( \left\{ U_{i}^{k_{i}},U_{j}^{l_{i}}/f_{k_{i}l_{i}}\right\}
_{i=1,...,l},\left[ \Psi _{J}\right] \right) \tprod\limits_{i=1,...,l}\Psi
^{\otimes \left( k_{i}+l_{i}\right) }\left( \left\{
U_{i}^{k_{i}},U_{j}^{l_{i}}/f_{k_{i}l_{i}}\right\} _{i=1,...,l}\right)
\label{GPR}
\end{equation}%
or equivalently, from:%
\begin{equation}
\int f\left( \left\{ U_{i}^{k_{i}},U_{j}^{l_{i}}/f_{k_{i}l_{i}}\right\}
_{i=1,...,l},\left[ \Psi _{J}\right] \right) \Psi ^{\otimes l\left(
k_{i}+l_{i}\right) }\left( \left\{
U_{i}^{k_{i}},U_{j}^{l_{i}}/f_{k_{i}l_{i}}\right\} _{i=1,...,l}\right)
\label{GPS}
\end{equation}%
Starting with (\ref{GPS}), we can consider some reparametrization invariance:%
\begin{eqnarray*}
&&g^{\left\{ U_{i}^{k_{i}}\right\} }.\Psi ^{\otimes l\left(
k_{i}+l_{i}\right) }\left( \left\{
U_{i}^{k_{i}},U_{j}^{l_{i}}/f_{k_{i}l_{i}}\right\} _{i=1,...,l}\right) \\
&=&\Psi ^{\otimes l\left( k_{i}+l_{i}\right) }\left( g^{\left\{
U_{i}^{k_{i}}\right\} }.\left\{
U_{i}^{k_{i}},U_{j}^{l_{i}}/f_{k_{i}l_{i}}\right\} _{i=1,...,l}\right) \\
&=&R\left( g^{\left\{ U_{i}^{k_{i}}\right\} }\right) \Psi ^{\otimes l\left(
k_{i}+l_{i}\right) }\left( \left\{
U_{i}^{k_{i}},U_{j}^{l_{i}}/f_{k_{i}l_{i}}\right\} _{i=1,...,l}\right)
\end{eqnarray*}%
depending on the symmetries of the function%
\begin{equation*}
f\left( \left\{ U_{i}^{k_{i}},U_{j}^{l_{i}}/f_{k_{i}l_{i}}\right\}
_{i=1,...,l},\left[ \Psi _{J}\right] \right)
\end{equation*}%
where $R\left( g^{\left\{ U_{i}^{k_{i}}\right\} }\right) $ is a differential
operator acting on $\left\{
U_{i}^{k_{i}},U_{j}^{l_{i}}/f_{k_{i}l_{i}}\right\} _{i=1,...,l}$.

In this case, the transformation for (\ref{GTV}) is:%
\begin{eqnarray*}
&&G^{k_{i},...,k_{i}}\left( \mathbf{\hat{\Lambda}}^{\left[ k_{i},...,k_{i}%
\right] }\left[ \Psi _{J}\right] \right) .\left(
\tprod\limits_{t=1,...,l}\Psi _{I,\alpha _{i}^{\left( t\right) }}^{\otimes
k_{i}}\left( \left( U_{i}^{k_{i}}\right) _{t}\right) \right) \\
&=&R\left( g^{\left\{ U_{i}^{k_{i}}\right\} }\right) \left\{
\tprod\limits_{t=1,...,l}\Psi _{I,\alpha _{i}^{\left( t\right) }}^{\otimes
k_{i}}\left( g^{\left\{ U_{i}^{k_{i}}\right\} }\left( \left(
U_{i}^{k_{i}}\right) _{t}\right) \right) \right\}
\end{eqnarray*}%
Here, the parameters $\mathbf{\hat{\Lambda}}^{\left[ k_{i},...,k_{i}\right] }%
\left[ \Psi _{J}\right] $ for transformation $g^{\left\{
U_{i}^{k_{i}}\right\} }$\ are independent from the copies. The inclusion of
groups $G_{\alpha _{1},...\alpha _{n}}^{k_{i},...,k_{i}}$%
\begin{equation*}
G^{\overset{l^{\prime }}{\overbrace{k_{i},...,k_{i}}}}\left( \mathbf{\hat{%
\Lambda}}^{\left[ \overset{l^{\prime }}{\overbrace{k_{i},...,k_{i}}}\right] }%
\left[ \Psi _{J}\right] \right) \subset G^{\overset{l}{\overbrace{%
k_{i},...,k_{i}}}}\left( \mathbf{\hat{\Lambda}}^{\left[ \overset{l}{%
\overbrace{k_{i},...,k_{i}}}\right] }\left[ \Psi _{J}\right] \right)
\end{equation*}%
comes from inclusion of symmetry groups of $f\left( \left\{
U_{i}^{k_{i}},U_{j}^{l_{i}}/f_{k_{i}l_{i}}\right\} _{i=1,...,l^{\prime }},%
\left[ \Psi _{J}\right] \right) $ and $f\left( \left\{
U_{i}^{k_{i}},U_{j}^{l_{i}}/f_{k_{i}l_{i}}\right\} _{i=1,...,l},\left[ \Psi
_{J}\right] \right) $.

\subsubsection{Example 2: partial reparametrization symmetry}

As an example we consider reparametrizations that do not originate from
initial field. In (\ref{SRS}) we look at the following second order term in $%
\Psi _{I,\alpha _{i}}^{\otimes k_{i}}\left( U_{i}^{k_{i}}\right) $:%
\begin{eqnarray}
&&\sum_{\alpha _{i},\alpha _{i}^{\prime }}\int \Psi _{J,\alpha
_{i}}^{\otimes l_{i}}\left( \left(
U_{j}^{l_{i}}/f_{k_{i}l_{i}},U_{i}^{k_{i}}\right) ^{\prime }\right) \Psi
_{I,\alpha _{i}}^{\otimes k_{i}}\left( \left( U_{i}^{k_{i}}\right) ^{\prime
}\right)  \label{QSR} \\
&&\times f\left( \left( U_{j}^{l_{i}}/f_{k_{i}l_{i}},U_{i}^{k_{i}}\right)
^{\prime },\left( U_{j}^{l_{i}}/f_{k_{i}l_{i}},U_{i}^{k_{i}}\right) ,\left[
\Psi _{J}\right] \right) \Psi _{J,\alpha _{i}^{\prime }}^{\otimes
l_{i}}\left( U_{j}^{l_{i}}/f_{k_{i}l_{i}},U_{i}^{k_{i}}\right) \Psi
_{I,\alpha _{i}^{\prime }}^{\otimes k_{i}}\left( U_{i}^{k_{i}}\right)
d\left( \left\{ U_{i}^{k_{i}},U_{j}^{l_{i}}/f_{k_{i}l_{i}}\right\}
_{i}\right)  \notag
\end{eqnarray}%
which arises from the decomposition of quadratic terms in the original field
theory:%
\begin{equation}
\int \left( \Psi ^{\otimes k_{i}+l_{i}}\left(
U_{i}^{k_{i}},U_{j}^{l_{i}}/f_{k_{i}l_{i}}\right) \right) f\left( \left(
U_{j}^{l_{i}}/f_{k_{i}l_{i}},U_{i}^{k_{i}}\right) ^{\prime },\left(
U_{j}^{l_{i}}/f_{k_{i}l_{i}},U_{i}^{k_{i}}\right) ,\left[ \Psi _{J}\right]
\right) \Psi ^{\otimes k_{i}+l_{i}}\left(
U_{i}^{k_{i}},U_{j}^{l_{i}}/f_{k_{i}l_{i}}\right)  \label{PTW}
\end{equation}%
The dependence in $\left[ \Psi _{J}\right] $ being functional, i.e. arising
from integrals of function in $\Psi _{J}$.

The term (\ref{QSR}) is invariant for any transformation $G_{k_{i},k_{i}}%
\left[ \Psi _{J,\alpha _{i}}^{\otimes l_{i}},\Psi _{J,\alpha _{i}^{\prime
}}^{\otimes l_{i}}\right] $ acting on $\left( \Psi _{I,\alpha _{i}}^{\otimes
k_{i}}\left( U_{i}^{k_{i}}\right) ,\Psi _{I,\alpha _{i}^{\prime }}^{\otimes
k_{i}}\left( U_{i}^{k_{i}}\right) \right) $, preserving the form (\ref{QSR}%
). Writing this action: 
\begin{equation*}
G_{\alpha _{i},\alpha _{i}^{\prime }}^{k_{i},k_{i}}\left( \mathbf{\hat{%
\Lambda}}_{\alpha _{i},\alpha _{i}^{\prime }}^{\left[ k_{i},k_{i}\right] }%
\left[ \Psi _{J}\right] \right) .\left( \Psi _{I,\alpha _{i}}^{\otimes
k_{i}}\left( U_{i}^{k_{i}}\right) \Psi _{I,\alpha _{i}^{\prime }}^{\otimes
k_{i}}\left( U_{i}^{k_{i}}\right) \right) =\left( \tprod\limits_{\alpha
_{i},\alpha _{i}^{\prime }}\sum_{\substack{ \alpha =\alpha _{i},\alpha
_{i}^{\prime }  \\ \alpha ^{\prime }=\alpha _{i},\alpha _{i}^{\prime }}}%
\left[ R_{\alpha _{i},\alpha _{i}^{\prime }}^{\left[ k_{i},k_{i}\right]
}\left( \mathbf{\hat{\Lambda}}_{\alpha _{i},\alpha _{i}^{\prime }}^{\left[
k_{i},k_{i}\right] }\left[ \Psi _{J}\right] \right) \right] _{\alpha
_{i}}^{\alpha ^{\prime }}\Psi _{I,\alpha ^{\prime }}^{\otimes k_{i}}\left(
g_{k_{i},k_{i}}^{k_{i}}U_{i}^{k_{i}}\right) \right)
\end{equation*}%
The dependency in $\left[ \Psi _{J}\right] $ arises from its realizations $%
\Psi _{J,\alpha _{i}}^{\otimes l_{i}}$, $\Psi _{J,\alpha _{i}^{\prime
}}^{\otimes l_{i}}$.

This symmetry is still present, if we assume a series expansion:%
\begin{eqnarray*}
&&S\left( \left( \Psi _{I,\alpha _{i}}^{\otimes k_{i}}\left(
U_{i}^{k_{i}}\right) \right) \right) \\
&=&\sum_{n}a_{n}\left( \int \left( \Psi ^{\otimes k_{i}+l_{i}}\left(
U_{i}^{k_{i}},U_{j}^{l_{i}}/f_{k_{i}l_{i}}\right) \right) f\left( \left(
U_{j}^{l_{i}}/f_{k_{i}l_{i}},U_{i}^{k_{i}}\right) ^{\prime },\left(
U_{j}^{l_{i}}/f_{k_{i}l_{i}},U_{i}^{k_{i}}\right) ,\left[ \Psi _{J}\right]
\right) \Psi ^{\otimes k_{i}+l_{i}}\left(
U_{i}^{k_{i}},U_{j}^{l_{i}}/f_{k_{i}l_{i}}\right) \right) ^{n}
\end{eqnarray*}%
In this case the symmetry groups are:%
\begin{equation*}
G_{k_{i},k_{i}}\left[ \Psi _{J}\right] \subset G_{k_{i},k_{i}}\left[ \Psi
_{J}\right] \times G_{k_{i},k_{i}}\left[ \Psi _{J}\right] \subset ...
\end{equation*}%
More generally, considering expression (\ref{GTV}), it is preserved for
transformations $G_{\alpha _{i},...,\alpha _{l}}^{k_{i},...,k_{i}}$ for one
set of realizations:%
\begin{eqnarray*}
&&G_{\alpha _{i},...,\alpha _{l}}^{k_{i},...,k_{i}}\left( \mathbf{\hat{%
\Lambda}}_{\alpha _{i},...,\alpha _{l}}^{k_{i},...,k_{i}}\left[ \Psi _{J}%
\right] \right) .\left( \tprod\limits_{t=1,...,l}\Psi _{I,\alpha
_{t}}^{\otimes k_{i}}\left( \left( U_{i}^{k_{i}}\right) _{t}\right) \right)
\\
&=&\tprod\limits_{t=1,...,l}\left\{ \sum_{t^{\prime }}\left[ R_{\alpha
_{1},...\alpha _{n}}^{k_{i},...,k_{i}}\left( \mathbf{\hat{\Lambda}}_{\alpha
_{i},...,\alpha _{l}}^{k_{i},...,k_{i}}\left[ \Psi _{J}\right] \right) %
\right] _{\alpha _{i}^{\left( t\right) }}^{\alpha _{i}^{\left( t^{\prime
}\right) }}\Psi _{I,\alpha _{i}^{\left( t^{\prime }\right) }}^{\otimes
k_{i}}\left( g^{\left\{ U_{i}^{k_{i}}\right\} }\left( \left(
U_{i}^{k_{i}}\right) _{t}\right) \right) \right\}
\end{eqnarray*}%
compatible with the symmetries of functions $f\left( \left(
U_{j}^{l_{i}}/f_{k_{i}l_{i}},U_{i}^{k_{i}}\right) ^{\prime },\left(
U_{j}^{l_{i}}/f_{k_{i}l_{i}},U_{i}^{k_{i}}\right) ,\left[ \Psi _{J}\right]
\right) $. If a collection of realizations is involved in the
transformation, the transformation formula is rather given by (\ref{GRT})
and (\ref{GRD}) with $R_{\left\{ \alpha _{i\leqslant n}\right\} }^{\left\{
k_{i\leqslant n}\right\} }\left( \mathbf{\hat{\Lambda}}_{\left\{ \alpha
_{i\leqslant n}\right\} }^{\left[ \left\{ k_{i\leqslant n}\right\} \right]
}\right) \rightarrow R_{\left\{ \alpha _{i\leqslant n}\right\}
}^{k_{1}...k_{n}}\left( \mathbf{\hat{\Lambda}}_{\left\{ \alpha _{i\leqslant
n}\right\} }^{\left[ \left\{ k_{i\leqslant n}\right\} \right] }\right) $.

The inclusion between groups arises if $G_{\left\{ \alpha _{i\leqslant
n}\right\} }^{k_{i},...,k_{i}}\left( \mathbf{\hat{\Lambda}}_{\left\{ \alpha
_{i\leqslant n}\right\} }^{\left[ \left\{ k_{i\leqslant n}\right\} \right]
}\right) $ contains some subgroups:%
\begin{equation*}
\left( G_{\left\{ \alpha _{i\leqslant l}\right\} }^{k_{i},...,k_{i}}\left( 
\mathbf{\hat{\Lambda}}_{\left\{ \alpha _{i\leqslant l}\right\} }^{\left[
\left\{ k_{i\leqslant l}\right\} \right] }\right) \right) ^{\prime }\times
\left( G_{\left\{ \alpha _{i\leqslant n-l}\right\} }^{k_{i},...,k_{i}}\left( 
\mathbf{\hat{\Lambda}}_{\left\{ \alpha _{i\leqslant n-l}\right\} }^{\left[
\left\{ k_{i\leqslant n-l}\right\} \right] }\right) \right) ^{\prime
}\subset G^{k_{i},...,k_{i}}\left( \mathbf{\hat{\Lambda}}_{\left\{ \alpha
_{i\leqslant n}\right\} }^{\left[ \left\{ k_{i\leqslant n}\right\} \right]
}\right)
\end{equation*}%
with the inclusion of subgroups:%
\begin{eqnarray*}
\left( G_{\left\{ \alpha _{i\leqslant l}\right\} }^{k_{i},...,k_{i}}\left( 
\mathbf{\hat{\Lambda}}_{\left\{ \alpha _{i\leqslant l}\right\} }^{\left[
\left\{ k_{i\leqslant l}\right\} \right] }\right) \right) ^{\prime }
&\subset &G_{\left\{ \alpha _{i\leqslant l}\right\}
}^{k_{i},...,k_{i}}\left( \mathbf{\hat{\Lambda}}_{\left\{ \alpha
_{i\leqslant l}\right\} }^{\left[ \left\{ k_{i\leqslant l}\right\} \right]
}\right) \\
\left( G_{\left\{ \alpha _{i\leqslant n-l}\right\} }^{k_{i},...,k_{i}}\left( 
\mathbf{\hat{\Lambda}}_{\left\{ \alpha _{i\leqslant n-l}\right\} }^{\left[
\left\{ k_{i\leqslant n-l}\right\} \right] }\right) \right) ^{\prime }
&\subset &G_{\left\{ \alpha _{i\leqslant n-l}\right\}
}^{k_{i},...,k_{i}}\left( \mathbf{\hat{\Lambda}}_{\left\{ \alpha
_{i\leqslant n-l}\right\} }^{\left[ \left\{ k_{i\leqslant n-l}\right\} %
\right] }\right)
\end{eqnarray*}

\subsubsection{Remarks}

1. Except\ for the case of reparametrization invariance of the original
theory, the parameters $\mathbf{\hat{\Lambda}}_{\left\{ \alpha _{i}^{\left(
1\right) },...,\alpha _{i}^{\left( p\right) }\right\} }^{\left[
k_{i},...,k_{i}\right] }\left[ \Psi _{J}\right] $ depend on realizations $%
\left\{ \alpha _{i}^{\left( 1\right) },...,\alpha _{i}^{\left( p\right)
}\right\} $ of $\Psi _{J}$ in general. In functionals, by change of
variables in integrals, we may expect they can be identified with parameters
independent from the realizations, i.e. that the various actions are copies
of the same groups.

2. If $S\left( \left( \Psi _{I,\alpha _{i}}^{\otimes k_{i}}\left(
U_{i}^{k_{i}}\right) \right) \right) $ depends only on the state $\nu $,
then: 
\begin{eqnarray*}
&&S\left( \left( \Psi _{I,\alpha _{i}}^{\otimes k_{i}}\left(
U_{i}^{k_{i}}\right) \right) \right) \\
&=&\sum_{l}a_{l}\sum_{\alpha _{i}^{\left( 1\right) },...,\alpha _{i}^{\left(
l\right) }}\int f\left( \left\{ U_{i}^{k_{i}},\right\} _{1,...,l}\right) \\
&&\times \tprod\limits_{t=1,...,l}\nu _{\left\{ \left( U_{i}^{k_{i}}\right)
_{t}\right\} }\left( \Psi _{J,\alpha _{i}^{\left( t\right) }}^{\otimes
l_{i}}\left( \left( U_{j}^{l_{i}}/f_{k_{i}l_{i}},U_{i}^{k_{i}}\right)
_{t}\right) \right) \Psi _{I,\alpha _{i}^{\left( t\right) }}^{\otimes
k_{i}}\left( \left( U_{i}^{k_{i}}\right) _{t}\right) d\left( \left\{ \left(
U_{i}^{k_{i}},U_{j}^{l_{i}}/f_{k_{i}l_{i}}\right) _{t}\right\} \right)
\end{eqnarray*}%
and the dependence of the symmetry parameters writes: 
\begin{equation*}
\mathbf{\hat{\Lambda}}_{\left\{ \alpha _{i}^{\left( 1\right) },...,\alpha
_{i}^{\left( l\right) }\right\} }^{\left[ k_{i},...,k_{i}\right] }\left[
\Psi _{J}\right] \rightarrow \mathbf{\hat{\Lambda}}_{\left\{ \alpha
_{i}^{\left( 1\right) },...,\alpha _{i}^{\left( l\right) }\right\} }^{\left[
k_{i},...,k_{i}\right] }\left[ \Psi _{J},\nu \right]
\end{equation*}

\subsubsection{Projection over minima}

Minimization of the action functional including degeneracy can be performed
and yields the projected field:%
\begin{eqnarray}
&&\Psi _{I,\alpha ,0}^{\otimes k}\left( U_{i}^{k},\mathbf{\hat{\Lambda}}%
_{\infty }^{\left[ k\right] }\left[ \Psi _{J},\nu \right] ,\left\{ \Psi
_{J}^{\otimes l}\right\} _{l}\right)  \label{PS} \\
&=&\sum_{s}\sum_{s,l_{1},...,l_{s},\left\{ \alpha _{i}^{\prime }\right\}
_{_{i\leqslant s}}}d\left( \left\{ U_{j}^{l_{i}}\right\} _{l_{i}}\right)
d\left( \left\{ U_{i}^{k_{i}}/f_{k_{i}l_{i}}\right\} _{k_{i},l_{i}}\right) 
\notag \\
&&\times \tprod_{i}\Psi _{J,\alpha _{i}^{\prime }}^{\otimes l_{i}}\left(
U_{j}^{l_{i}}\right) \mathcal{K}_{0}^{v}\left( U_{i}^{k},\left\{
U_{j}^{l_{i}}\right\} _{l_{i}},\left\{ U_{i}^{k_{i}}/f_{k_{i}l_{i}}\right\}
_{i},\mathbf{\hat{\Lambda}}_{\infty }^{\left[ k\right] }\left[ \Psi _{J},\nu %
\right] \right)  \notag
\end{eqnarray}%
where the kernel:%
\begin{eqnarray*}
&&\mathcal{K}_{0}^{v}\left( U_{i}^{k},\left\{ U_{j}^{l_{i}}\right\} _{l_{i}},%
\mathbf{\hat{\Lambda}}_{\infty }^{\left[ k\right] }\left[ \Psi _{J},\nu %
\right] \right) \\
&=&\mathcal{K}_{0}\left( U_{i}^{k},\left\{ U_{j}^{l_{i}}\right\} _{l_{i}},%
\mathbf{\hat{\Lambda}}_{\infty }^{\left[ k\right] }\left[ \Psi _{J},\nu %
\right] \right) \tprod_{i}v\left(
U_{i}^{k_{i}},U_{j}^{l_{i}}/f_{k_{i}l_{i}}\right)
\end{eqnarray*}%
includes possibly derivatives in the various variables.

\subsubsection{Restriction including constraints}

Up until now, the derivation has been performed without taking into account
the constraints between the two fields. Accounting for these constraints
will reduce the symmetry group. As we need to consider the following terms
in the functionals:%
\begin{equation*}
\Psi _{J}^{\otimes l}\left( U_{j}^{l}\right) \Psi _{I}^{\otimes k}\left(
U_{i}^{k}\right) \delta \left( f_{lk}\left( U_{j}^{l},U_{i}^{k}\right)
\right)
\end{equation*}%
we have to restrict saddle point solutions $\Psi _{I,0}^{\otimes k}\left(
U_{i}^{k}\right) $ by including a factor $\delta \left( f_{lk}\left(
U_{j}^{l},U_{i}^{k}\right) \right) $ and replace:%
\begin{equation*}
\Psi _{I}^{\otimes k}\left( U_{i}^{k}\right) \rightarrow \Psi _{I}^{\otimes
k}\left( U_{i}^{k}\right) \delta \left( f_{lk}\left(
U_{j}^{l},U_{i}^{k}\right) \right)
\end{equation*}%
This restriction reduces the symetry group to a subgroup preserving the
constraint. This also applies to the terms including $\tprod_{i\leqslant
s}\Psi _{J,\alpha _{i}^{\prime }}^{\otimes l_{i}}\left( U_{j}^{l_{i}}\right) 
$ in (\ref{PS}) which imposes the constraints $\delta \left(
f_{l_{i}k_{i}}\left( U_{i}^{k_{i}},U_{j}^{l_{i}}\right) \right) $. We show
in appendix 2 that locally, this amounts to reduce $\mathbf{\hat{\Lambda}}%
_{\infty }^{\left[ k\right] }\left[ \Psi _{J},\nu \right] $ to a set
depending on $\left\{ U_{j}^{l_{i\leqslant s}}\right\} \equiv \left\{
U_{j}^{l_{i}}\right\} $ and rewrite the variables $\mathbf{\hat{\Lambda}}%
_{\infty ,\alpha }^{\left[ k\right] }\left[ \Psi _{J},\nu \right] $ as
functions of the set $\left\{ U_{j}^{l_{i}}\right\} $: 
\begin{equation}
\mathbf{\hat{\Lambda}}_{\infty ,\alpha }^{\left[ k\right] }\left[ \Psi
_{J},\nu \right] \rightarrow \mathbf{\hat{\Lambda}}_{\infty ,\alpha }^{\left[
k\right] }\left[ \Psi _{J},\nu ,\left\{ U_{j}^{l_{i}}\right\} \right]
=\left( \left\{ \mathbf{\hat{\Lambda}}_{\alpha }^{\left[ \left\{
k,k_{i\leqslant n-1}\right\} \right] }\left[ \Psi _{J},\nu ,\left\{
U_{j}^{l_{i}}\right\} \right] \right\} \right) _{n}  \label{VR}
\end{equation}%
and:%
\begin{equation*}
\mathbf{\hat{\Lambda}}_{n,\alpha }^{\left[ k\right] }\left[ \Psi _{J},\nu %
\right] \rightarrow \mathbf{\hat{\Lambda}}_{n,\alpha }^{\left[ k\right] }%
\left[ \Psi _{J},\nu ,\left\{ U_{j}^{l_{i}}\right\} \right] =\left\{ \mathbf{%
\hat{\Lambda}}_{\alpha }^{\left[ \left\{ k,k_{i\leqslant n-1}\right\} \right]
}\left[ \Psi _{J},\nu ,\left\{ U_{j}^{l_{i}}\right\} \right] \right\}
\end{equation*}

\subsection{Projected functional and effective field}

Once the saddle point has been found, we can compute the projected
functional (\ref{FNV}):%
\begin{equation}
\sum_{v_{U_{i}^{k}}}\int g\left( v,U_{i}^{k}\right) v_{U_{i}^{k}}\left( \Psi
_{J}^{\otimes l}\right) \tprod_{\min S\left( v\right) }ev_{U_{i}^{k}}\left(
\Psi _{I}^{\otimes k}\left( U_{i}^{k}\right) \right) \delta \left(
f_{lk}\left( U_{j}^{l},U_{i}^{k}\right) \right)  \label{FCP}
\end{equation}%
by replacing $\Psi _{I}^{\otimes k}\left( U_{i}^{k}\right) $ with its saddle
point value. This will enable to describe the effective theory after
projection, as states of an effective field including the degrees of freedom
of $\Psi _{I}^{\otimes k}\left( U_{i}^{k}\right) $.

The computation is presented in appendix 3 under the assumption:%
\begin{equation*}
v_{U_{i}^{k}}\left( U_{i}^{k_{i}},U_{j}^{l_{i}}/f_{k_{i}l_{i}}\right)
=v\left( U_{j}^{l_{i}}\right) \delta \left( f_{lk}\left(
U_{j}^{l},U_{i}^{k}\right) \right)
\end{equation*}%
corresponding roughly to current-current interaction. As before, defining
the realization $\alpha $ as a sum of products of realizations:%
\begin{equation*}
\Psi _{J,\alpha }^{\otimes \sum_{i\leqslant n}l_{i}}\left( \left(
U_{j}^{l_{i}}\right) \right) =\sum_{\left\{ \alpha _{i}^{\prime }\right\}
}\tprod_{i}\Psi _{J,\left\{ \alpha _{i}^{\prime }\right\} ,\alpha }^{\otimes
l_{i}}\left( U_{j}^{l_{i}}\right)
\end{equation*}%
Defining also:%
\begin{eqnarray*}
&&\bar{g}\left( v,\left\{ U_{j}^{l_{i}}\right\} _{i\leqslant n},\mathbf{\hat{%
\Lambda}}_{\infty ,\alpha }^{\left[ k\right] }\left[ \Psi _{J},\nu ,\left\{
U_{j}^{l_{i}}\right\} \right] \right) \\
&=&\int g\left( v,U_{i}^{k}\right) \mathcal{K}_{0}\left( \left\{
U_{j}^{l_{i}}\right\} _{l_{i}},\left\{ U_{i}^{k_{i}}/f_{k_{i}l_{i}}\right\}
_{i\leqslant n},\mathbf{\hat{\Lambda}}_{\infty ,\alpha }^{\left[ k\right] }%
\left[ \Psi _{J},\nu ,\left\{ U_{j}^{l_{i}}\right\} \right] \right)
dU_{i}^{k}/f_{kl}\tprod dU_{i}^{k_{i}}/f_{k_{i}l_{i}}
\end{eqnarray*}%
formula (\ref{FCP}) writes:%
\begin{equation}
\sum_{\alpha }\int \bar{g}\left( v,\left\{ U_{j}^{l_{i}}\right\} _{i},%
\mathbf{\hat{\Lambda}}_{\infty ,\alpha }^{\left[ k\right] }\left[ \Psi
_{J},\nu ,\left\{ U_{j}^{l_{i}}\right\} \right] ,v\left( \left(
U_{j}^{l_{i}}\right) \right) \right) \left( \Psi _{J,\alpha }^{\otimes
\left( \sum_{i}l_{i}\right) }\left( \left\{ U_{j}^{l_{i}}\right\} _{i},%
\mathbf{\hat{\Lambda}}_{\infty ,\alpha }^{\left[ k\right] }\left[ \Psi
_{J},\nu ,\left\{ U_{j}^{l_{i}}\right\} \right] ,v\right) \right)
\label{Fcp}
\end{equation}%
with:%
\begin{equation*}
\Psi _{J,\alpha }^{\otimes \left( \sum_{i}l_{i}\right) }\left( \left\{
U_{j}^{l_{i}}\right\} _{i},\mathbf{\hat{\Lambda}}_{\infty ,\alpha }^{\left[ k%
\right] }\left[ \Psi _{J},\nu ,\left\{ U_{j}^{l_{i}}\right\} \right]
,v\right) =\Psi _{J,\alpha }^{\otimes \left( \sum_{i}l_{i}\right) }\left(
\left\{ U_{j}^{l_{i}}\right\} _{i}\right) \Psi _{\alpha }\left( \mathbf{\hat{%
\Lambda}}_{\infty ,\alpha }^{\left[ k\right] }\left[ \Psi _{J},\nu ,\left\{
U_{j}^{l_{i}}\right\} \right] ,v\right)
\end{equation*}%
and:%
\begin{equation*}
v\left( \left( U_{j}^{l_{i}}\right) \right) =\sum_{S}\tprod_{i\in S}v\left(
U_{i}^{k_{i}}\right)
\end{equation*}%
Setting $\bar{g}_{v}\rightarrow \bar{g}$ since the fnctnls are arbitrary and
the $v$ dependency can be absorbed in the definition of $\bar{g}$. If the: 
\begin{equation*}
\mathbf{\hat{\Lambda}}_{\infty ,\alpha }^{\left[ k\right] }\left[ \Psi
_{J},\nu ,\left\{ U_{j}^{l_{i}}\right\} \right] =\mathbf{\hat{\Lambda}}%
_{\infty }^{\left[ k\right] }\left[ \Psi _{J},\nu ,\left\{
U_{j}^{l_{i}}\right\} \right]
\end{equation*}%
are independent from $\alpha $, we find the projected functional:%
\begin{equation}
\int \bar{g}\left( \left\{ U_{j}^{l_{i}}\right\} _{i},\mathbf{\hat{\Lambda}}%
_{\infty }^{\left[ k\right] }\left[ \Psi _{J},\nu ,\left\{
U_{j}^{l_{i}}\right\} \right] \right) \Psi _{J}^{\otimes \left(
\sum_{i}l_{i}\right) }\left( \left\{ U_{j}^{l_{i}}\right\} _{i},\mathbf{\hat{%
\Lambda}}_{\infty }^{\left[ k\right] }\left[ \Psi _{J},\nu ,\left\{
U_{j}^{l_{i}}\right\} \right] ,v\right)  \label{PFT}
\end{equation}%
wth the field given by a sum over realizations:%
\begin{eqnarray}
&&\Psi _{J}^{\otimes \left( \sum_{i}l_{i}\right) }\left( \left\{
U_{j}^{l_{i}}\right\} _{i},\mathbf{\hat{\Lambda}}_{\infty }^{\left[ k\right]
}\left[ \Psi _{J},\nu ,\left\{ U_{j}^{l_{i}}\right\} \right] ,v\right)
\label{Sr} \\
&=&\sum_{\alpha }\Psi _{J,\alpha }^{\otimes \left( \sum_{i}l_{i}\right)
}\left( \left\{ U_{j}^{l_{i}}\right\} _{i},\mathbf{\hat{\Lambda}}_{\infty }^{%
\left[ k\right] }\left[ \Psi _{J},\nu ,\left\{ U_{j}^{l_{i}}\right\} \right]
,v\right)  \notag
\end{eqnarray}

If the $\mathbf{\hat{\Lambda}}_{\infty ,\alpha }^{\left[ k\right] }\left[
\Psi _{J},\nu ,\left\{ U_{j}^{l_{i}}\right\} \right] $ are not independent
of the copies $\Psi _{J,\alpha }^{\otimes k_{p_{0}}}$, the identification is
local. The fields $\Psi _{J}^{\otimes \left( \sum_{i}l_{i}\right) }$ are
effective fields after projection of the field $\Psi _{I}$. They depend on
some parameters that both keep track of the projcted states, and the
remaining object, through the field $\Psi _{J}$ and the state $\nu $. This
dependency arises from the fact that the saddle point equations, and the
symmetry groups depend on those quantities. We will detail below the form of
the constraints and their dependency on the parameters $\mathbf{\hat{\Lambda}%
}_{\infty ,\alpha }^{\left[ k\right] }\left[ \Psi _{J},\nu ,\left\{
U_{j}^{l_{i}}\right\} \right] $, the field $\Psi _{J}$ and state $\nu $.

\subsection{Averaging over cloud substratum and local field}

We aim at describing the effective field:%
\begin{equation*}
\Psi _{J}^{\otimes \left( \sum_{i}l_{i}\right) }\left( \left\{
U_{j}^{l_{i}}\right\} _{i},\mathbf{\hat{\Lambda}}_{\infty }^{\left[ k\right]
}\left[ \Psi _{J},\nu ,\left\{ U_{j}^{l_{i}}\right\} \right] ,v\right)
\end{equation*}%
involved in the functionals (\ref{PFT}) as a more usual field defined at
some given points. This is achieved by showing that $\mathbf{\hat{\Lambda}}%
_{\infty }^{\left[ k\right] }\left[ \Psi _{J},\nu ,\left\{
U_{j}^{l_{i}}\right\} \right] $ can be depicted as cloud of points
surrounding these particular points.

\subsubsection{Cloud of points}

To explain this point, we assume, at least in first approximation that:%
\begin{equation*}
\mathbf{\hat{\Lambda}}_{\infty }^{\left[ k\right] }\left[ \Psi _{J},\nu
,\left\{ U_{j}^{l_{i}}\right\} \right] =\left( \mathbf{\hat{\Lambda}}^{\left[
k\right] }\left[ \Psi _{J},\nu ,\left\{ U_{j}^{l_{i}}\right\} \right] ,%
\mathbf{\hat{\Lambda}}_{n}\left[ \Psi _{J},\nu ,\left\{
U_{j}^{l_{i}}\right\} \right] \right) _{n}
\end{equation*}%
where $\left\{ U_{j}^{l_{i}}\right\} =\left\{ U_{j}^{l_{i\leqslant
s}}\right\} $ when insertd in products $\tprod_{i\leqslant s}\Psi _{J,\alpha
_{i}^{\prime }}^{\otimes l_{i}}\left( U_{j}^{l_{i}}\right) $.

We have given above the description of $\mathbf{\hat{\Lambda}}^{\left[
\left\{ k,k_{i\leqslant n-1}\right\} \right] }\left[ \Psi _{J},\nu ,U_{j}^{l}%
\right] $ in terms of set of flag manifolds starting with $\mathbf{\hat{%
\Lambda}}^{\left[ k\right] }\left[ \Psi _{J},\nu ,U_{j}^{l}\right] $. As a
consequence, the field:%
\begin{equation*}
\Psi _{J}^{\otimes \left( \sum_{i}l_{i}\right) }\left( \left\{
U_{j}^{l_{i}}\right\} _{i},\left( \mathbf{\hat{\Lambda}}^{\left[ k\right] }%
\left[ \Psi _{J},\nu ,\left\{ U_{j}^{l_{i}}\right\} \right] ,\left\{ \mathbf{%
\hat{\Lambda}}^{\left[ \left\{ k_{i\leqslant n-1}\right\} \right] }\left[
\Psi _{J},\nu ,\left\{ U_{j}^{l_{i}}\right\} \right] \right\} \right)
_{n},v\right)
\end{equation*}%
dscribes both the integratd presenc of a cloud, that is the points of a
space, with distinguished points $\mathbf{\hat{\Lambda}}^{\left[ k\right] }%
\left[ \Psi _{J},\nu ,U_{j}^{l}\right] $. \ These points bear the physical
quantities $\left\{ U_{j}^{l_{i}}\right\} _{i}$.

\subsubsection{Series expnsion of the field in cloud variables}

To describe more precisely $\left\{ \mathbf{\hat{\Lambda}}_{\alpha }^{\left[
\left\{ k_{i\leqslant n-1}\right\} \right] }\left[ \Psi _{J},\nu ,U_{j}^{l}%
\right] \right\} $, we use our two previous assmptions:%
\begin{equation*}
\mathbf{\hat{\Lambda}}^{\left[ \left\{ k_{i\leqslant n}\right\} \right] }%
\left[ \Psi _{J},\nu ,\left\{ U_{j}^{l_{i}}\right\} \right] =\left\{ \mathbf{%
\hat{\Lambda}}^{\left[ k_{i}\right] }\left[ \Psi _{J},\nu ,\left\{
U_{j}^{l_{i}}\right\} \right] \right\} _{k_{1},..k_{n}}
\end{equation*}%
and:%
\begin{equation*}
\mathbf{\hat{\Lambda}}^{\left[ k_{i}\right] }\left[ \Psi _{J},\nu ,\left\{
U_{j}^{l_{i\leqslant n}}\right\} \right] =\left( \mathbf{\hat{\Lambda}}%
^{k_{i}}\left[ \Psi _{J},\nu ,\left\{ U_{j}^{l_{i\leqslant n}}\right\} %
\right] \right)
\end{equation*}%
Thus, the set $\left\{ \mathbf{\hat{\Lambda}}^{\left[ \left\{ k_{i\leqslant
n-1}\right\} \right] }\left[ \Psi _{J},\nu ,\left\{ U_{j}^{l_{i}}\right\} %
\right] \right\} $ consists of sequences:%
\begin{eqnarray*}
&&\mathbf{\hat{\Lambda}}^{k_{1}}\left[ \Psi _{J},\nu ,\left\{
U_{j}^{l_{i}}\right\} \right] ,\left( \mathbf{\hat{\Lambda}}^{k_{1,1}}\left[
\Psi _{J},\nu ,\left\{ U_{j}^{l_{i}}\right\} \right] ,\mathbf{\hat{\Lambda}}%
^{k_{1,2}}\left[ \Psi _{J},\nu ,\left\{ U_{j}^{l_{i}}\right\} \right]
\right) , \\
&&\left( \mathbf{\hat{\Lambda}}^{k_{2,1}}\left[ \Psi _{J},\nu ,\left\{
U_{j}^{l_{i}}\right\} \right] ,\mathbf{\hat{\Lambda}}_{\alpha }^{k_{2,2}}%
\left[ \Psi _{J},\nu ,\left\{ U_{j}^{l_{i}}\right\} \right] ,\mathbf{\hat{%
\Lambda}}_{\alpha }^{k_{2,3}}\left[ \Psi _{J},\nu ,\left\{
U_{j}^{l_{i}}\right\} \right] \right) ...
\end{eqnarray*}%
Each $\mathbf{\hat{\Lambda}}^{k_{i}}\left[ \Psi _{J},\nu \right] $ is
described by an infinite number of representants. Consequently, if we
consider the expansion of $\Psi _{J}^{\otimes \left( \sum_{i}l_{i}\right) }$
in terms of representations of the $G_{k_{i},...,k_{n}}$:%
\begin{eqnarray*}
&&\Psi _{J}^{\otimes \left( \sum_{i}l_{i}\right) }\left( \left\{
U_{j}^{l_{i}}\right\} _{i},\left( \mathbf{\hat{\Lambda}}^{\left[ k\right] }%
\left[ \Psi _{J},\nu ,\left\{ U_{j}^{l_{i}}\right\} \right] ,\left\{ \mathbf{%
\hat{\Lambda}}^{\left[ \left\{ k_{i\leqslant n-1}\right\} \right] }\left[
\Psi _{J},\nu ,\left\{ U_{j}^{l_{i}}\right\} \right] \right\} \right)
_{n},v\right) \\
&=&\sum_{k_{i}}\Psi _{J}^{\otimes \left( \sum_{i}l_{i}\right) }\left(
\left\{ U_{j}^{l_{i}}\right\} _{i},\mathbf{\hat{\Lambda}}^{\left[ k\right] }%
\left[ \Psi _{J},\nu ,\left\{ U_{j}^{l_{i}}\right\} \right] ,\left\{ \mathbf{%
\hat{\Lambda}}^{k_{i}}\left[ \Psi _{J},\nu ,\left\{ U_{j}^{l_{i}}\right\} %
\right] \right\} ,v\right) \\
&&+\sum_{k_{i,1/2}}\Psi _{J}^{\otimes \left( \sum_{i}l_{i}\right) }\left(
\left\{ U_{j}^{l_{i}}\right\} _{i},\mathbf{\hat{\Lambda}}^{\left[ k\right] }%
\left[ \Psi _{J},\nu ,\left\{ U_{j}^{l_{i}}\right\} \right] ,\left\{ \mathbf{%
\hat{\Lambda}}^{k_{i,1}}\left[ \Psi _{J},\nu ,\left\{ U_{j}^{l_{i}}\right\} %
\right] ,\mathbf{\hat{\Lambda}}^{k_{i,2}}\left[ \Psi _{J},\nu ,\left\{
U_{j}^{l_{i}}\right\} \right] \right\} ,v\right) +...
\end{eqnarray*}%
where $\left\{ \mathbf{\hat{\Lambda}}^{k_{i}}\left[ \Psi _{J},\nu ,\left\{
U_{j}^{l_{i}}\right\} \right] \right\} $ is the set of all the $\mathbf{\hat{%
\Lambda}}^{k_{i}}\left[ \Psi _{J},\nu \right] $ for given $k_{i}$, and $%
\left\{ \mathbf{\hat{\Lambda}}^{k_{i,1}}\left[ \Psi _{J},\nu ,\left\{
U_{j}^{l_{i}}\right\} \right] ,\mathbf{\hat{\Lambda}}^{k_{i,2}}\left[ \Psi
_{J},\nu ,\left\{ U_{j}^{l_{i}}\right\} \right] \right\} $ the set of the $%
\left( \mathbf{\hat{\Lambda}}^{k_{i,1}}\left[ \Psi _{J},\nu ,\left\{
U_{j}^{l_{i}}\right\} \right] ,\mathbf{\hat{\Lambda}}^{k_{i,2}}\left[ \Psi
_{J},\nu ,\left\{ U_{j}^{l_{i}}\right\} \right] \right) $ for given $k_{i,1}$%
, $k_{i,2}$.

The field is thus a series:%
\begin{eqnarray}
&&\Psi _{J}^{\otimes \left( \sum_{i}l_{i}\right) }\left( \left\{
U_{j}^{l_{i}}\right\} _{i},\mathbf{\hat{\Lambda}}_{\infty }^{\left[ k\right]
}\left[ \Psi _{J},\nu ,\left\{ U_{j}^{l_{i}}\right\} \right] ,v\right)
\label{SR} \\
&=&\Psi _{J}^{\otimes \left( \sum_{i}l_{i}\right) }\left( \left\{
U_{j}^{l_{i}}\right\} _{i},\mathbf{\hat{\Lambda}}^{\left[ k\right] }\left[
\Psi _{J},\nu ,\left\{ U_{j}^{l_{i}}\right\} \right] ,V,v\right) +\Psi
_{J}^{\otimes \left( \sum_{i}l_{i}\right) }\left( \left\{
U_{j}^{l_{i}}\right\} _{i},\mathbf{\hat{\Lambda}}^{\left[ k\right] }\left[
\Psi _{J},\nu ,\left\{ U_{j}^{l_{i}}\right\} \right] ,V^{2},v\right) +... 
\notag
\end{eqnarray}%
for:%
\begin{equation*}
V=\left\{ \left\{ \mathbf{\hat{\Lambda}}^{k_{i}}\left[ \Psi _{J},\nu
,\left\{ U_{j}^{l_{i}}\right\} \right] \right\} \right\} _{k_{i}}
\end{equation*}%
gathering all the states $\left\{ \mathbf{\hat{\Lambda}}^{k_{i}}\left[ \Psi
_{J},\nu \right] \right\} $.

\subsubsection{Averaging over cloud variables}

Given that we have considered independently the tensor power $\Psi
_{J}^{\otimes l}\left( U_{j}^{l}\right) $, we can decompose the effective
field into series of products of realizations of several fields:%
\begin{eqnarray*}
&&\sum_{k_{i}}\Psi _{J,k_{i}}^{\otimes \left( \sum_{i}l_{i}\right) }\left(
\left\{ U_{j}^{l_{i}}\right\} _{i},\mathbf{\hat{\Lambda}}^{\left[ k\right] }%
\left[ \Psi _{J},\nu ,\left\{ U_{j}^{l_{i}}\right\} \right]
,V_{k_{i}},v\right) \\
&&+\sum_{k_{i,1/2}}\Psi _{J,k_{i,1},\alpha }^{\otimes \left(
\sum_{i}l_{i}^{\prime }\right) }\left( \left\{ U_{j}^{l_{i}}\right\} _{i},%
\mathbf{\hat{\Lambda}}^{\left[ k\right] }\left[ \Psi _{J},\nu ,\left\{
U_{j}^{l_{i}}\right\} \right] ,V_{k_{i,1}},v\right) \Psi _{J,k_{i,2},\alpha
}^{\otimes \left( \sum_{i}l_{i}-l_{i}^{\prime }\right) }\left( \left\{
U_{j}^{l_{i}}\right\} _{i},\mathbf{\hat{\Lambda}}^{\left[ k\right] }\left[
\Psi _{J},\nu ,\left\{ U_{j}^{l_{i}}\right\} \right] ,V_{k_{i,2}},v\right)
+...
\end{eqnarray*}%
where:%
\begin{equation*}
V_{k_{i}}=\left\{ \mathbf{\hat{\Lambda}}^{k_{i}}\left[ \Psi _{J},\nu
,\left\{ U_{j}^{l_{i}}\right\} \right] \right\}
\end{equation*}%
By expanding the field as functional of one point, two points and so on:%
\begin{eqnarray*}
&&\Psi _{J,k_{i}}^{\otimes \left( \sum_{i}l_{i}\right) }\left( \left\{
U_{j}^{l_{i}}\right\} _{i},\mathbf{\hat{\Lambda}}^{\left[ k\right] }\left[
\Psi _{J},\nu ,U_{j}^{l}\right] ,V_{k_{i}},v\right) \\
&=&\sum_{\Lambda _{k_{i}}}\Psi _{J,k_{i}}^{\otimes \left(
\sum_{i}l_{i}\right) }\left( \left\{ U_{j}^{l_{i}}\right\} _{i},\mathbf{\hat{%
\Lambda}}^{\left[ k\right] }\left[ \Psi _{J},\nu ,U_{j}^{l}\right] ,\Lambda
_{k_{i}},v\right) \\
&&+\sum_{\Lambda _{k_{i}}^{\left( 1\right) },\Lambda _{k_{i}}^{\left(
2\right) }}\Psi _{J,k_{i}}^{\otimes \left( \sum_{i}l_{i}\right) }\left(
\left\{ U_{j}^{l_{i}}\right\} _{i},\mathbf{\hat{\Lambda}}^{\left[ k\right] }%
\left[ \Psi _{J},\nu ,U_{j}^{l}\right] ,\Lambda _{k_{i}}^{\left( 1\right)
},\Lambda _{k_{i}}^{\left( 2\right) },v\right) +...
\end{eqnarray*}%
where $\Lambda _{k_{i}}$ are coordinates in $V_{k_{i}}$. Consequently the
field can be seen as an average and expands in series:%
\begin{eqnarray}
&&\Psi _{J,k_{i}}^{\otimes \left( \sum_{i}l_{i}\right) }\left( \left\{
U_{j}^{l_{i}}\right\} _{i},\mathbf{\hat{\Lambda}}^{\left[ k\right] }\left[
\Psi _{J},\nu ,U_{j}^{l}\right] ,V_{k_{i}},v\right)  \label{NGF} \\
&=&\sum_{r}\int_{\left( V_{k_{i}}\right) ^{r}}\Psi _{J,k_{i}}^{\otimes
\left( \sum_{i}l_{i}\right) }\left( \left\{ U_{j}^{l_{i}}\right\} _{i},%
\mathbf{\hat{\Lambda}}^{\left[ k\right] }\left[ \Psi _{J},\nu ,\left\{
U_{j}^{l_{i}}\right\} \right] ,\left( \Lambda _{k_{i}}\right) ^{r},v\right)
d\left( \Lambda _{k_{i}}\right) ^{r}  \notag
\end{eqnarray}%
The fields at stake are thus fields depending on one set of parameter
variables $\mathbf{\hat{\Lambda}}^{\left[ k\right] }\left[ \Psi _{J},\nu
,\left\{ U_{j}^{l_{i}}\right\} \right] $, but also include in an integrated
manner, the internal space of points.

\section{Variations of fields}

\subsection{Infinitesimal variation of the field}

The field:%
\begin{equation*}
\Psi _{J}^{\otimes \sum_{i}l_{i}}\left( \left\{ U_{j}^{l_{i}}\right\} _{i},%
\mathbf{\hat{\Lambda}}_{\infty ,\alpha }^{\left[ k\right] }\left[ \Psi
_{J},\nu ,\left\{ U_{j}^{l_{i}}\right\} \right] ,v\right)
\end{equation*}%
encompasses the parameters $\mathbf{\hat{\Lambda}}_{\infty ,\alpha }^{\left[
k\right] }\left[ \Psi _{J},\nu ,\left\{ U_{j}^{l_{i}}\right\} \right] $.
Since $\Psi _{J}^{\otimes \sum_{i}l_{i}}$ \ and $\Psi _{I}^{\otimes
\sum_{i}k_{i}}$ are intertwined through the constraints, $\mathbf{\hat{%
\Lambda}}_{\infty ,\alpha }^{\left[ k\right] }\left[ \Psi _{J},\nu ,\left\{
U_{j}^{l_{i}}\right\} \right] $ depends on $\Psi _{J}$ and $\left\{
U_{j}^{l_{i}}\right\} $. A variation of $\Psi _{J}$ modifies $\mathbf{\hat{%
\Lambda}}_{\infty ,\alpha }^{\left[ k\right] }$. We aim at finding the
conditions for parameters $\mathbf{\hat{\Lambda}}_{\infty ,\alpha }^{\left[ k%
\right] }$ to be locally independent from $\Psi _{J}$, so that the field can
be considered as function of the parameter space and that a field variation
does not affect this parameter space.

We begin with the field decomposed as a sum over product of components:%
\begin{eqnarray*}
&&\Psi _{J}^{\otimes \sum_{i}l_{i}}\left( \left\{ U_{j}^{l_{i}}\right\} _{i},%
\mathbf{\hat{\Lambda}}_{\infty ,\alpha }^{\left[ k\right] }\left[ \Psi
_{J},\nu ,\left\{ U_{j}^{l_{i}}\right\} \right] ,v\right) \\
&=&\sum_{\alpha }\Psi _{J,\alpha }^{\otimes \sum_{i}l_{i}}\left( \left\{
U_{j}^{l_{i}}\right\} _{i},\mathbf{\hat{\Lambda}}_{\infty ,\alpha }^{\left[ k%
\right] }\left[ \Psi _{J},\nu ,\left\{ U_{j}^{l_{i}}\right\} \right]
,v\right) =\sum_{\alpha }\Psi _{J,\alpha }^{\otimes \sum l_{i}}\left(
U_{j}^{l_{i}}\right) \Psi _{\alpha }\left( \mathbf{\hat{\Lambda}}_{\infty
,\alpha }^{\left[ k\right] }\left[ \Psi _{J},\nu ,\left\{
U_{j}^{l_{i}}\right\} \right] ,v\right)
\end{eqnarray*}%
This field will be inserted in the state-functionals for one realization by
introducing:%
\begin{equation*}
\delta \left( h_{i}\left( \mathbf{\hat{\Lambda}}_{\infty ,\alpha }^{\left[ k%
\right] }\left[ \Psi _{J},\nu ,\left\{ U_{j}^{l_{i}}\right\} \right] ,\left(
\Psi _{J,\alpha }^{\otimes l_{i}},v\right) \right) \right)
\end{equation*}%
which represents the constraint defining $\mathbf{\hat{\Lambda}}_{\infty
,\alpha }^{\left[ k\right] }\left[ \Psi _{J},\nu ,\left\{
U_{j}^{l_{i}}\right\} \right] $. The field can thus also be written by
including this constraint: 
\begin{eqnarray*}
&&\Psi _{J,\alpha }^{\otimes \sum_{i}l_{i}}\left( \left\{
U_{j}^{l_{i}}\right\} _{i},\mathbf{\hat{\Lambda}}_{\infty ,\alpha }^{\left[ k%
\right] }\left[ \Psi _{J},\nu ,\left\{ U_{j}^{l_{i}}\right\} \right]
,v\right) \\
&=&\Psi _{J,\alpha }^{\otimes \sum_{i}l_{i}}\left( \left\{
U_{j}^{l_{i}}\right\} _{i},\mathbf{\hat{\Lambda}}_{\infty ,\alpha }^{\left[ k%
\right] }\left[ \Psi _{J},\nu ,\left\{ U_{j}^{l_{i}}\right\} \right]
,v\right) \delta \left( h_{i}\left( \mathbf{\hat{\Lambda}}_{\alpha }^{\left[
\left\{ k_{i\leqslant n}\right\} \right] }\left[ \Psi _{J},\nu ,U_{j}^{l}%
\right] ,\left( \Psi _{J,\alpha }^{\otimes l_{i}},v\right) \right) \right)
\end{eqnarray*}%
Since $\Psi _{J,\alpha }^{\otimes \sum_{i}l_{i}}$ depends both directly in $%
\Psi _{J,\alpha }^{\otimes l_{i}}\left( U_{j}^{l_{i}}\right) $ and
indirectly via $\mathbf{\hat{\Lambda}}_{\alpha }^{\left[ k_{i}\right] }\left[
\Psi _{J},\nu ,U_{j}^{l}\right] $, a set of variations $\left\{ \delta \Psi
_{J,\alpha }^{\otimes l_{i}}\left( U_{j}^{l_{i}}\right) \right\} $ for each $%
U_{j}^{l_{i}}$, induces a variation:%
\begin{eqnarray*}
&&\frac{\delta \Psi _{J,\alpha }^{\otimes \sum_{i}l_{i}}\left( \left\{
U_{j}^{l_{i}}\right\} _{i},\mathbf{\hat{\Lambda}}_{\infty ,\alpha }^{\left[ k%
\right] }\left[ \Psi _{J},\nu ,\left\{ U_{j}^{l_{i}}\right\} \right]
,v\right) }{\delta \Psi _{J,\alpha }^{\otimes l_{i}}\left( \left(
U_{j}^{l_{i}}\right) ^{\prime }\right) } \\
&=&\frac{\delta ^{\prime }\Psi _{J,\alpha }^{\otimes \sum_{i}l_{i}}\left(
\left\{ U_{j}^{l_{i}}\right\} _{i},\mathbf{\hat{\Lambda}}_{\infty ,\alpha }^{%
\left[ k\right] }\left[ \Psi _{J},\nu ,\left\{ U_{j}^{l_{i}}\right\} \right]
,v\right) }{\delta \Psi _{J,\alpha }^{\otimes l_{i}}\left( \left(
U_{j}^{l_{i}}\right) ^{\prime }\right) } \\
&&+\frac{\delta \mathbf{\hat{\Lambda}}_{\infty ,\alpha }^{\left[ k\right] }%
\left[ \Psi _{J},\nu ,\left\{ U_{j}^{l_{i}}\right\} \right] }{\delta \Psi
_{J,\alpha }^{\otimes l_{i}}\left( \left( U_{j}^{l_{i}}\right) ^{\prime
}\right) }\nabla _{\mathbf{\hat{\Lambda}}_{\infty ,\alpha }^{\left[ k\right]
}}\Psi _{J,\alpha }^{\otimes \sum_{i}l_{i}}\left( \left\{
U_{j}^{l_{i}}\right\} _{i},\mathbf{\hat{\Lambda}}_{\infty ,\alpha }^{\left[ k%
\right] }\left[ \Psi _{J},\nu ,\left\{ U_{j}^{l_{i}}\right\} \right]
,v\right)
\end{eqnarray*}%
with:%
\begin{equation*}
\frac{\delta \mathbf{\hat{\Lambda}}_{\infty ,\alpha }^{\left[ k\right] }}{%
\delta \Psi _{J,\alpha }^{\otimes l_{i}}\left( \left( U_{j}^{l_{i}}\right)
^{\prime }\right) }\nabla _{\mathbf{\hat{\Lambda}}_{\infty ,\alpha }^{\left[
k\right] }}=\sum_{n,\left\{ k,k_{i\leqslant n-1}\right\} }\frac{\delta 
\mathbf{\hat{\Lambda}}_{\alpha }^{\left[ \left\{ k,k_{i\leqslant
n-1}\right\} \right] }}{\delta \Psi _{J,\alpha }^{\otimes l_{i}}\left(
\left( U_{j}^{l_{i}}\right) ^{\prime }\right) }\nabla _{\mathbf{\hat{\Lambda}%
}_{\alpha }^{\left[ \left\{ k,k_{i\leqslant n-1}\right\} \right] }}
\end{equation*}%
and:%
\begin{equation*}
\frac{\delta ^{\prime }\Psi _{J,\alpha }^{\otimes \sum_{i}l_{i}}\left(
\left\{ U_{j}^{l_{i}}\right\} _{i},\mathbf{\hat{\Lambda}}_{\infty ,\alpha }^{%
\left[ k\right] }\left[ \Psi _{J},\nu ,\left\{ U_{j}^{l_{i}}\right\} \right]
,v\right) }{\delta \Psi _{J,\alpha }^{\otimes l_{i}}\left( \left(
U_{j}^{l_{i}}\right) ^{\prime }\right) }
\end{equation*}%
represents the variation of $\Psi _{J,\alpha }^{\otimes \sum_{i}l_{i}}$ for
constant $\mathbf{\hat{\Lambda}}$.

In turn, this associates the variation of the state $v$:%
\begin{eqnarray*}
&&\int v\left( \left( U_{j}^{l_{i}}\right) \right) \frac{\delta \Psi
_{J,\alpha }^{\otimes \sum_{i}l_{i}}\left( \left\{ U_{j}^{l_{i}}\right\}
_{i},\mathbf{\hat{\Lambda}}_{\infty ,\alpha }^{\left[ k\right] }\left[ \Psi
_{J},\nu ,\left\{ U_{j}^{l_{i}}\right\} \right] ,v\right) }{\delta \Psi
_{J,\alpha }^{\otimes l_{i}}\left( \left( U_{j}^{l_{i}}\right) ^{\prime
}\right) } \\
&=&\int v\left( \left( U_{i}^{l}\right) \right) \frac{\delta ^{\prime }\Psi
_{J,\alpha }^{\otimes \sum_{i}l_{i}}\left( \left\{ U_{j}^{l_{i}}\right\}
_{i},\mathbf{\hat{\Lambda}}_{\infty ,\alpha }^{\left[ k\right] }\left[ \Psi
_{J},\nu ,\left\{ U_{j}^{l_{i}}\right\} \right] ,v\right) }{\delta \Psi
_{J,\alpha }^{\otimes l_{i}}\left( \left( U_{j}^{l_{i}}\right) ^{\prime
}\right) } \\
&&+\int v\left( \left( U_{j}^{l_{i}}\right) \right) \frac{\delta \mathbf{%
\hat{\Lambda}}_{\infty ,\alpha }^{\left[ k\right] }}{\delta \Psi _{J,\alpha
}^{\otimes l_{i}}\left( \left( U_{j}^{l_{i}}\right) ^{\prime }\right) }%
\nabla _{\mathbf{\hat{\Lambda}}_{\infty ,\alpha }^{\left[ k\right] }}\Psi
_{J,\alpha }^{\otimes \sum_{i}l_{i}}\left( \left\{ U_{j}^{l_{i}}\right\}
_{i},\mathbf{\hat{\Lambda}}_{\infty ,\alpha }^{\left[ k\right] }\left[ \Psi
_{J},\nu ,\left\{ U_{j}^{l_{i}}\right\} \right] ,v\right)
\end{eqnarray*}%
and invariance writes for this state:%
\begin{equation}
\int v\left( \left( U_{j}^{l_{i}}\right) \right) \frac{\delta \mathbf{\hat{%
\Lambda}}_{\infty ,\alpha }^{\left[ k\right] }}{\delta \Psi _{J,\alpha
}^{\otimes l_{i}}\left( \left( U_{j}^{l_{i}}\right) ^{\prime }\right) }%
\nabla _{\mathbf{\hat{\Lambda}}_{\infty ,\alpha }^{\left[ k\right] }}\Psi
_{J,\alpha }^{\otimes \sum_{i}l_{i}}\left( \left\{ U_{j}^{l_{i}}\right\}
_{i},\mathbf{\hat{\Lambda}}_{\infty ,\alpha }^{\left[ k\right] }\left[ \Psi
_{J},\nu ,\left\{ U_{j}^{l_{i}}\right\} \right] ,v\right) =0  \label{NV}
\end{equation}%
If $\mathbf{\hat{\Lambda}}_{\alpha }^{\left[ k_{i}\right] }$ independent
from copy, and if in first approximation $\frac{\delta \mathbf{\hat{\Lambda}}%
_{\infty ,\alpha }^{\left[ k\right] }}{\delta \Psi _{J,\alpha }^{\otimes
l_{i}}\left( \left( U_{j}^{l_{i}}\right) ^{\prime }\right) }$ is independent
from $\alpha $, then writing:%
\begin{equation*}
\frac{\delta \mathbf{\hat{\Lambda}}_{\infty ,\alpha }^{\left[ k\right] }%
\left[ \Psi _{J},\nu ,\left\{ U_{j}^{l_{i}}\right\} \right] }{\delta \Psi
_{J,\alpha }^{\otimes l_{i}}\left( \left( U_{j}^{l_{i}}\right) ^{\prime
}\right) }=\frac{\delta \mathbf{\hat{\Lambda}}_{\infty }^{\left[ k\right] }%
\left[ \Psi _{J},\nu ,\left\{ U_{j}^{l_{i}}\right\} \right] }{\delta \Psi
_{J}^{\otimes l_{i}}\left( \left( U_{j}^{l_{i}}\right) ^{\prime }\right) }
\end{equation*}%
the variations can be summed over realizations and the invariance of
parameters space with respect to field variations implies the effective
field equation:%
\begin{equation}
0=\int v\left( U_{i}^{l}\right) \frac{\delta \mathbf{\hat{\Lambda}}_{\infty
}^{\left[ k\right] }\left[ \Psi _{J},\nu ,\left\{ U_{j}^{l_{i}}\right\} %
\right] }{\delta \Psi _{J}^{\otimes l_{i}}\left( \left( U_{j}^{l_{i}}\right)
^{\prime }\right) }\nabla _{\mathbf{\hat{\Lambda}}_{\infty }^{\left[ k\right]
}}\Psi _{J}^{\otimes \left( \sum_{i}l_{i}\right) }\left( \left\{
U_{j}^{l_{i}}\right\} _{i},\mathbf{\hat{\Lambda}}_{\infty }^{\left[ k\right]
}\left[ \Psi _{J},\nu ,\left\{ U_{j}^{l_{i}}\right\} \right] \right)
\label{NR}
\end{equation}

\subsection{Fields in local coordinates}

When equation (\ref{NR}) is satisfied, the field can be rewritten:%
\begin{equation}
\hat{\Psi}_{J}^{\otimes \sum_{i}l_{i}}\left( \left\{ U_{j}^{l_{i}}\right\}
_{i},\mathbf{\hat{\Lambda}}_{\infty }^{\left[ k\right] },v\right) \delta
\left( f\left( \mathbf{\hat{\Lambda}}^{\left[ \left\{ k,k_{i\leqslant
n-1}\right\} \right] },\left( \Psi _{J,0,\alpha }^{\otimes l_{i}},v\right)
\right) \right)  \label{RF}
\end{equation}%
with the same $\mathbf{\hat{\Lambda}}^{\left[ k_{i}\right] }$ for all fields 
$\hat{\Psi}_{J}^{\otimes \sum_{i}l_{i}}$. The field $\Psi _{J,0,\alpha
}^{\otimes l_{i}},v$ \ is some\ reference realization from which the $%
\mathbf{\hat{\Lambda}}^{\left[ \left\{ k,k_{i\leqslant n-1}\right\} \right]
} $ are defined and:%
\begin{equation*}
\hat{\Psi}_{J}^{\otimes \sum_{i}l_{i}}\left( \left\{ U_{j}^{l_{i}}\right\}
_{i},\mathbf{\hat{\Lambda}}_{\infty }^{\left[ k\right] },v,\alpha \right)
=\Psi _{J}^{\otimes \left( \sum_{i}l_{i}\right) }\left( \left\{
U_{j}^{l_{i}}\right\} _{i},\mathbf{\hat{\Lambda}}_{\infty }^{\left[ k\right]
}\left[ \Psi _{J},\nu ,\left\{ U_{j}^{l_{i}}\right\} \right] \right)
\end{equation*}%
is obtained by change of variable. This description implements locally the
independence between field variation and parameter space, since in (\ref{RF}%
) the parameters depend on a fixed reference $\Psi _{J,0,\alpha }^{\otimes
l_{i}}$ and remains inert with respect to the variations of $\hat{\Psi}%
_{J}^{\otimes \sum_{i}l_{i}}$.

The equation for $\hat{\Psi}_{J}^{\otimes \sum_{i}l_{i}}\left( \left\{
U_{j}^{l_{i}}\right\} _{i},\mathbf{\hat{\Lambda}}_{\infty }^{\left[ k\right]
},v,\alpha \right) $ is:%
\begin{equation}
0=\int v\left( U_{i}^{l}\right) \Gamma \left( \left[ \Psi _{J},\nu ,U_{j}^{l}%
\right] ,\left( U_{j}^{l_{i}}\right) ^{\prime }\right) \nabla _{\mathbf{\hat{%
\Lambda}}_{\infty }^{\left[ k\right] }}\hat{\Psi}_{J}^{\otimes
\sum_{i}l_{i}}\left( \left\{ U_{j}^{l_{i}}\right\} _{i},\mathbf{\hat{\Lambda}%
}_{\infty }^{\left[ k\right] },v,\alpha \right)   \label{CGR}
\end{equation}%
where:%
\begin{equation*}
\Gamma \left( \left[ \Psi _{J},\nu ,U_{j}^{l}\right] ,\left(
U_{j}^{l_{i}}\right) ^{\prime }\right) =\frac{\delta \mathbf{\hat{\Lambda}}%
_{\infty }^{\left[ k\right] }\left[ \Psi _{J},\nu ,\left\{
U_{j}^{l_{i}}\right\} \right] }{\delta \Psi _{J}^{\otimes l_{i}}\left(
\left( U_{j}^{l_{i}}\right) ^{\prime }\right) }\left( \frac{\delta \mathbf{%
\hat{\Lambda}}_{\infty }^{\left[ k\right] }\left[ \Psi _{J},\nu ,\left\{
U_{j}^{l_{i}}\right\} \right] }{\delta \mathbf{\hat{\Lambda}}_{\infty }^{%
\left[ k\right] }}\right) ^{-1}
\end{equation*}%
Equation (\ref{CGR}), is a first order differential equation similar to a
spinor equation, where the equivalent of the $\gamma $ matrices are:%
\begin{equation}
\hat{\Gamma}\left( \left( U_{j}^{l_{i}}\right) ^{\prime },U_{j}^{l}\right)
=v\left( U_{i}^{l}\right) \Gamma \left( \left[ \Psi _{J},\nu ,U_{j}^{l}%
\right] ,\left( U_{j}^{l_{i}}\right) ^{\prime }\right)   \label{GMM}
\end{equation}%
In (\ref{GMM}); the variables $U_{i}^{l}$ and $\left( U_{j}^{l_{i}}\right)
^{\prime }$ act as spinor indices. In the case where these variable are
discrete, a sum replaces the integral and we can rewrite (\ref{CGR}):%
\begin{equation}
0=\sum_{U_{j}^{l}}\hat{\Gamma}\left( \left( U_{j}^{l_{i}}\right) ^{\prime
},U_{j}^{l}\right) \nabla _{\mathbf{\hat{\Lambda}}_{\infty }^{\left[ k\right]
}}\hat{\Psi}_{J}^{\otimes \sum_{i}l_{i}}\left( \left\{ U_{j}^{l_{i}}\right\}
_{i},\mathbf{\hat{\Lambda}}_{\infty }^{\left[ k\right] },v,\alpha \right) 
\label{MLN}
\end{equation}%
which looks like massless spinor equation. Such representation is relatively
general since (see $\left[ 32\right] $-$\left[ 41\right] $), at least
Maxwell equation and $SU(2)$ gauge fields can be described, at least
formally, by spinor formalism.

The derivates:%
\begin{equation*}
\frac{\delta \mathbf{\hat{\Lambda}}_{\infty }^{\left[ k\right] }\left[ \Psi
_{J},\nu ,\left\{ U_{j}^{l_{i}}\right\} \right] }{\delta \mathbf{\hat{\Lambda%
}}_{\infty }^{\left[ k\right] }}
\end{equation*}%
reflect that the identification of the coordinates $\mathbf{\hat{\Lambda}}%
_{\infty }^{\left[ k\right] }\left[ \Psi _{J},\nu ,\left\{
U_{j}^{l_{i}}\right\} \right] $ with some referece local set of coordinates $%
\mathbf{\hat{\Lambda}}_{\infty }^{\left[ k\right] }$ depends on the context
framed by the fiel $\Psi _{J}$ and the stat $\nu $.

However, in (\ref{MLN}), the parameters $\mathbf{\hat{\Lambda}}_{\infty }^{%
\left[ k\right] }$ represents the entire cloud of points, rather than single
coordinates. Local variation of $\hat{\Psi}_{J}^{\otimes
\sum_{i}l_{i}}\left( \left\{ U_{j}^{l_{i}}\right\} _{i}\right) $ may locally
modify the points of $\mathbf{\hat{\Lambda}}_{\infty }^{\left[ k\right] }$
without modifying globally the cloud. This point is studied in the next
paragraph by considering the averaged field.

\subsection{Averaged field variation}

\subsubsection{General form}

Equation (\ref{NR}) can be rewritten for the averaged field (\ref{SR}).
Considering only the first term in (\ref{SR}):%
\begin{equation*}
\Psi _{J}^{\otimes \left( \sum_{i}l_{i}\right) }\left( \left\{
U_{j}^{l_{i}}\right\} _{i},\mathbf{\hat{\Lambda}}^{\left[ k\right] }\left[
\Psi _{J},\nu ,\left\{ U_{j}^{l_{i}}\right\} \right] ,V,v\right)
\end{equation*}%
where $V$ is a $V_{k_{i}}$, the variation equation becomes:%
\begin{eqnarray}
0 &=&\int v\left( \left( U_{j}^{l_{i}}\right) \right) \frac{\delta \mathbf{%
\hat{\Lambda}}^{\left[ k\right] }\left[ \Psi _{J},\nu ,\left\{
U_{j}^{l_{i}}\right\} \right] }{\delta \Psi _{J}^{\otimes l_{i}}\left(
\left( U_{j}^{l_{i}}\right) ^{\prime }\right) }\nabla _{\mathbf{\hat{\Lambda}%
}^{\left[ k\right] }}\Psi _{J}^{\otimes \left( \sum_{i}l_{i}\right) }\left(
\left\{ U_{j}^{l_{i}}\right\} _{i},\mathbf{\hat{\Lambda}}^{\left[ k\right] }%
\left[ \Psi _{J},\nu ,\left\{ U_{j}^{l_{i}}\right\} \right] ,V,v\right)
\label{VT} \\
&&+\int v\left( \left( U_{j}^{l_{i}}\right) \right) \frac{\delta V}{\delta
\Psi _{J}^{\otimes l_{i}}\left( \left( U_{j}^{l_{i}}\right) ^{\prime
}\right) }\nabla _{\mathbf{V}}\Psi _{J}^{\otimes \left( \sum_{i}l_{i}\right)
}\left( \left\{ U_{j}^{l_{i}}\right\} _{i},\mathbf{\hat{\Lambda}}^{\left[ k%
\right] }\left[ \Psi _{J},\nu ,\left\{ U_{j}^{l_{i}}\right\} \right]
,V,v\right)  \notag
\end{eqnarray}%
The computation of the variation is similar for other terms of (\ref{SR}).
In (\ref{VT}), the first term corresponds to a local variation, while the
second one describes global effect through a deformation of the manifold $V$.

Using the integral form (\ref{NGF}), allows to rewrite (\ref{VT}):%
\begin{eqnarray}
&&\int v\left( \left( U_{j}^{l_{i}}\right) \right) \frac{\delta V}{\delta
\Psi _{J}^{\otimes l_{i}}\left( \left( U_{j}^{l_{i}}\right) ^{\prime
}\right) }\nabla _{\mathbf{V}}\Psi _{J}^{\otimes \left( \sum_{i}l_{i}\right)
}\left( \left\{ U_{j}^{l_{i}}\right\} _{i},\mathbf{\hat{\Lambda}}^{\left[ k%
\right] }\left[ \Psi _{J},\nu ,\left\{ U_{j}^{l_{i}}\right\} \right]
,V,v\right)  \label{FRM} \\
&=&\int v\left( \left( U_{j}^{l_{i}}\right) \right) \sum_{r}\int_{\left(
V_{k_{i}}\right) ^{r}}\frac{\delta \left( \Lambda _{k_{i}}\right) ^{r}}{%
\delta \Psi _{J}^{\otimes l_{i}}\left( \left( U_{j}^{l_{i}}\right) ^{\prime
}\right) }\nabla _{\left( \Lambda _{k_{i}}\right) ^{r}}\Psi
_{J,k_{i}}^{\otimes \left( \sum_{i}l_{i}\right) }\left( \left\{
U_{j}^{l_{i}}\right\} _{i},\mathbf{\hat{\Lambda}}^{\left[ k\right] }\left[
\Psi _{J},\nu ,\left\{ U_{j}^{l_{i}}\right\} \right] ,\left( \Lambda
_{k_{i}}\right) ^{r},v\right) d\left( \Lambda _{k_{i}}\right) ^{r}  \notag \\
&=&\int v\left( \left( U_{j}^{l_{i}}\right) \right) \sum_{r}\int_{B\left[
\left( V_{k_{i}}\right) ^{r}\right] }\frac{\delta \left( \Lambda
_{k_{i}}\right) ^{r}}{\delta \Psi _{J}^{\otimes l_{i}}\left( \left(
U_{j}^{l_{i}}\right) ^{\prime }\right) }\Psi _{J,k_{i}}^{\otimes \left(
\sum_{i}l_{i}\right) }\left( \left\{ U_{j}^{l_{i}}\right\} _{i},\mathbf{\hat{%
\Lambda}}^{\left[ k\right] }\left[ \Psi _{J},\nu ,\left\{
U_{j}^{l_{i}}\right\} \right] ,\left( \Lambda _{k_{i}}\right) ^{r},v\right)
d\left( \Lambda _{k_{i}}\right) ^{r}  \notag \\
&&-\int v\left( \left( U_{j}^{l_{i}}\right) \right) \sum_{r}\int_{\left(
V_{k_{i}}\right) ^{r}}\nabla _{\left( \Lambda _{k_{i}}\right) ^{r}}h\left(
\left( \Lambda _{k_{i}}\right) ^{r},\Psi _{J}^{\otimes l_{i}}\left( \left(
U_{j}^{l_{i}}\right) ^{\prime }\right) \right)  \notag \\
&&\times \Psi _{J,k_{i}}^{\otimes \left( \sum_{i}l_{i}\right) }\left(
\left\{ U_{j}^{l_{i}}\right\} _{i},\mathbf{\hat{\Lambda}}^{\left[ k\right] }%
\left[ \Psi _{J},\nu ,\left\{ U_{j}^{l_{i}}\right\} \right] ,\left( \Lambda
_{k_{i}}\right) ^{r},v\right) d\left( \Lambda _{k_{i}}\right) ^{r}  \notag
\end{eqnarray}%
with $B\left[ \left( V_{k_{i}}\right) ^{r}\right] $ the boundary of $\left(
V_{k_{i}}\right) ^{r}$ and:%
\begin{equation*}
h\left( \left( \Lambda _{k_{i}}\right) ^{r},\Psi _{J}^{\otimes l_{i}}\left(
\left( U_{j}^{l_{i}}\right) ^{\prime }\right) \right) =\frac{\delta \left(
\Lambda _{k_{i}}\right) ^{r}}{\delta \Psi _{J}^{\otimes l_{i}}\left( \left(
U_{j}^{l_{i}}\right) ^{\prime }\right) }
\end{equation*}%
This last equation depends on $\left( \Lambda _{k_{i}}\right) ^{r}$. Formula
(\ref{FRM}) can be rewritten as:%
\begin{eqnarray*}
&&\int v\left( \left( U_{j}^{l_{i}}\right) \right) \frac{\delta V}{\delta
\Psi _{J}^{\otimes l_{i}}\left( \left( U_{j}^{l_{i}}\right) ^{\prime
}\right) }\nabla _{\mathbf{V}}\Psi _{J}^{\otimes \left( \sum_{i}l_{i}\right)
}\left( \left\{ U_{j}^{l_{i}}\right\} _{i},\mathbf{\hat{\Lambda}}^{\left[ k%
\right] }\left[ \Psi _{J},\nu ,\left\{ U_{j}^{l_{i}}\right\} \right]
,V,v\right) \\
&=&\int v\left( \left( U_{j}^{l_{i}}\right) \right) \frac{\delta \left(
B\left( V\right) \right) }{\delta \Psi _{J}^{\otimes l_{i}}\left( \left(
U_{j}^{l_{i}}\right) ^{\prime }\right) }\Psi _{J}^{\otimes \left(
\sum_{i}l_{i}\right) }\left( \left\{ U_{j}^{l_{i}}\right\} _{i},\mathbf{\hat{%
\Lambda}}^{\left[ k\right] }\left[ \Psi _{J},\nu ,\left\{
U_{j}^{l_{i}}\right\} \right] ,B\left( V\right) ,v\right) \\
&&-\int v\left( \left( U_{j}^{l_{i}}\right) \right) h\left( V,\Psi
_{J}^{\otimes l_{i}}\left( \left( U_{j}^{l_{i}}\right) ^{\prime }\right)
\right) \Psi _{J}^{\otimes \left( \sum_{i}l_{i}\right) }\left( \left\{
U_{j}^{l_{i}}\right\} _{i},\mathbf{\hat{\Lambda}}^{\left[ k\right] }\left[
\Psi _{J},\nu ,\left\{ U_{j}^{l_{i}}\right\} \right] ,V,v\right)
\end{eqnarray*}%
If we assume in average a proportionality factor $\epsilon $ between the
field on the boundary $B\left[ \left( V_{k_{i}}\right) ^{r}\right] $ and the
field on the entire $\left( V_{k_{i}}\right) ^{r}$, we can write:%
\begin{eqnarray*}
&&\frac{\delta \left( B\left( V\right) \right) }{\delta \Psi _{J}^{\otimes
l_{i}}\left( \left( U_{j}^{l_{i}}\right) ^{\prime }\right) }\Psi
_{J}^{\otimes \left( \sum_{i}l_{i}\right) }\left( \left\{
U_{j}^{l_{i}}\right\} _{i},\mathbf{\hat{\Lambda}}^{\left[ k\right] }\left[
\Psi _{J},\nu ,\left\{ U_{j}^{l_{i}}\right\} \right] ,B\left( V\right)
,v\right) \\
&=&\epsilon \frac{\delta V}{\delta \Psi _{J}^{\otimes l_{i}}\left( \left(
U_{j}^{l_{i}}\right) ^{\prime }\right) }\Psi _{J}^{\otimes \left(
\sum_{i}l_{i}\right) }\left( \left\{ U_{j}^{l_{i}}\right\} _{i},\mathbf{\hat{%
\Lambda}}^{\left[ k\right] }\left[ \Psi _{J},\nu ,\left\{
U_{j}^{l_{i}}\right\} \right] ,V,v\right)
\end{eqnarray*}%
So that, we have in average:%
\begin{eqnarray}
&&\int v\left( \left( U_{j}^{l_{i}}\right) \right) \frac{\delta V}{\delta
\Psi _{J}^{\otimes l_{i}}\left( \left( U_{j}^{l_{i}}\right) ^{\prime
}\right) }\nabla _{\mathbf{V}}\Psi _{J}^{\otimes \left( \sum_{i}l_{i}\right)
}\left( \left\{ U_{j}^{l_{i}}\right\} _{i},\mathbf{\hat{\Lambda}}^{\left[ k%
\right] }\left[ \Psi _{J},\nu ,\left\{ U_{j}^{l_{i}}\right\} \right]
,V,v\right)  \label{BND} \\
&=&\int v\left( \left( U_{j}^{l_{i}}\right) \right) \left( \epsilon \frac{%
\delta V}{\delta \Psi _{J}^{\otimes l_{i}}\left( \left( U_{j}^{l_{i}}\right)
^{\prime }\right) }-h\left( V,\Psi _{J}^{\otimes l_{i}}\left( \left(
U_{j}^{l_{i}}\right) ^{\prime }\right) \right) \right) \Psi _{J}^{\otimes
\left( \sum_{i}l_{i}\right) }\left( \left\{ U_{j}^{l_{i}}\right\} _{i},%
\mathbf{\hat{\Lambda}}^{\left[ k\right] }\left[ \Psi _{J},\nu ,\left\{
U_{j}^{l_{i}}\right\} \right] ,V,v\right)  \notag
\end{eqnarray}%
where $h\left( V,\Psi _{J}^{\otimes l_{i}}\left( \left( U_{j}^{l_{i}}\right)
^{\prime }\right) \right) $ is the average of $h\left( \left( \Lambda
_{k_{i}}\right) ^{r},\Psi _{J}^{\otimes l_{i}}\left( \left(
U_{j}^{l_{i}}\right) ^{\prime }\right) \right) $\ over $V$.

Given (\ref{BND}), equation (\ref{VT}) writes:%
\begin{eqnarray}
0 &=&\int v\left( \left( U_{j}^{l_{i}}\right) \right) \frac{\delta \mathbf{%
\hat{\Lambda}}^{\left[ k\right] }\left[ \Psi _{J},\nu ,U_{j}^{l}\right] }{%
\delta \Psi _{J}^{\otimes l_{i}}\left( \left( U_{j}^{l_{i}}\right) ^{\prime
}\right) }\nabla _{\mathbf{\hat{\Lambda}}^{\left[ k\right] }}\Psi
_{J}^{\otimes \left( \sum_{i}l_{i}\right) }\left( \left\{
U_{j}^{l_{i}}\right\} _{i},\mathbf{\hat{\Lambda}}^{\left[ k\right] }\left[
\Psi _{J},\nu ,\left\{ U_{j}^{l_{i}}\right\} \right] ,V,v\right)  \label{BNF}
\\
&&+\int v\left( \left( U_{j}^{l_{i}}\right) \right) \left( \epsilon \frac{%
\delta V}{\delta \Psi _{J}^{\otimes l_{i}}\left( \left( U_{j}^{l_{i}}\right)
^{\prime }\right) }-h\left( V,\Psi _{J}^{\otimes l_{i}}\left( \left(
U_{j}^{l_{i}}\right) ^{\prime }\right) \right) \right) \Psi _{J}^{\otimes
\left( \sum_{i}l_{i}\right) }\left( \left\{ U_{j}^{l_{i}}\right\} _{i},%
\mathbf{\hat{\Lambda}}^{\left[ k\right] }\left[ \Psi _{J},\nu ,\left\{
U_{j}^{l_{i}}\right\} \right] ,V,v\right)  \notag
\end{eqnarray}%
Considered as an equation where the parameter space is integrated, this
looks like an equation for a local field. The variation in $\Psi
_{J}^{\otimes \left( \sum_{i}l_{i}\right) }$ induces a modification of this
parameter space at its border. This induces an analog of a mass term in (\ref%
{BNF}), through its last contribution.

\subsubsection{Field expansion and coherent states}

Formula (\ref{BNF}) can be computed if we expand the field $\Psi
_{J}^{\otimes \left( \sum_{i}l_{i}\right) }$ as a function of $V$: 
\begin{eqnarray}
&&\Psi _{J}^{\otimes \left( \sum_{i}l_{i}\right) }\left( \left\{
U_{j}^{l_{i}}\right\} _{i},\mathbf{\hat{\Lambda}}^{\left[ k\right] }\left[
\Psi _{J},\nu ,\left\{ U_{j}^{l_{i}}\right\} \right] ,V,v\right)  \label{PNS}
\\
&=&\sum_{r}\int_{\left( V\right) ^{r}}\Psi _{J}^{\otimes \left(
\sum_{i}l_{i}\right) }\left( \left\{ U_{j}^{l_{i}}\right\} _{i},\mathbf{\hat{%
\Lambda}}^{\left[ k\right] }\left[ \Psi _{J},\nu ,\left\{
U_{j}^{l_{i}}\right\} \right] ,\left( V\right) ^{r},v\right) d\left(
V\right) ^{r}  \notag
\end{eqnarray}%
For coherent states, formula (\ref{PNS}) reduces to:%
\begin{equation}
\Psi _{J}^{\otimes \left( \sum_{i}l_{i}\right) }\left( \left\{
U_{j}^{l_{i}}\right\} _{i},\mathbf{\hat{\Lambda}}^{\left[ k\right] }\left[
\Psi _{J},\nu ,\left\{ U_{j}^{l_{i}}\right\} \right] ,v\right) \sum_{r}\frac{%
1}{r!}\tprod \int_{V^{r}}\bar{\Psi}\left( V^{\left( r\right) },v\right)
d\left( V^{\left( r\right) }\right)  \label{CN}
\end{equation}%
The computation leading to (\ref{BND}) can be performed directly in this
case. Neglecting $h\left( V,\Psi _{J}^{\otimes l_{i}}\left( \left(
U_{j}^{l_{i}}\right) ^{\prime }\right) \right) $, the variation of the
second part in (\ref{CN}) is:%
\begin{eqnarray*}
&&\frac{\delta V}{\delta \Psi _{J}^{\otimes l_{i}}\left( \left(
U_{j}^{l_{i}}\right) ^{\prime }\right) }\nabla _{\mathbf{V}}\sum_{r}\frac{1}{%
r!}\tprod \int_{V^{r}}\bar{\Psi}\left( V^{\left( r\right) },v\right) d\left(
V^{\left( r\right) }\right) \\
&=&\sum_{r}\int_{B\left( V\right) }\frac{\delta V}{\delta \Psi _{J}^{\otimes
l_{i}}\left( \left( U_{j}^{l_{i}}\right) ^{\prime }\right) }\bar{\Psi}\left(
B\left( V\right) ,v\right) \frac{1}{\left( r-1\right) !}\tprod \int_{V^{r-1}}%
\bar{\Psi}\left( V^{\left( r-1\right) },v\right) d\left( V^{\left(
r-1\right) }\right) \\
&=&\int_{B\left( V\right) }c\left( B\left( V\right) ,\Psi _{J}^{\otimes
l_{i}}\left( \left( U_{j}^{l_{i}}\right) ^{\prime }\right) \right) \bar{\Psi}%
\left( B\left( V\right) ,v\right) \sum_{r}\frac{1}{r!}\tprod \int_{V^{\left(
r\right) }}\bar{\Psi}\left( V^{\left( r\right) },v\right) d\left( V^{\left(
r\right) }\right)
\end{eqnarray*}%
Now, defining:%
\begin{eqnarray*}
&&\int_{B\left( V\right) }c\left( B\left( V\right) ,\Psi _{J}^{\otimes
l_{i}}\left( \left( U_{j}^{l_{i}}\right) ^{\prime }\right) \right) \bar{\Psi}%
\left( B\left( V\right) ,v\right) \\
&=&C\left( \Psi _{J}^{\otimes \left( \sum_{i}l_{i}\right) }\left( \left(
U_{j}^{l_{i}}\right) ^{\prime },\mathbf{\hat{\Lambda}}^{\left[ k\right] }%
\left[ \Psi _{J},\nu ,\left\{ U_{j}^{l_{i}}\right\} \right] ,v\right) \right)
\end{eqnarray*}%
the invariance equation (\ref{BNF}) becomes:%
\begin{eqnarray*}
0 &=&\int v\left( \left( U_{j}^{l_{i}}\right) \right) \frac{\delta \mathbf{%
\hat{\Lambda}}^{\left[ k\right] }\left[ \Psi _{J},\nu ,\left\{
U_{j}^{l_{i}}\right\} \right] }{\delta \Psi _{J}^{\otimes l_{i}}\left(
\left( U_{j}^{l_{i}}\right) ^{\prime }\right) }\nabla _{\mathbf{\hat{\Lambda}%
}^{\left[ k\right] }}\Psi _{J}^{\otimes \left( \sum_{i}l_{i}\right) }\left(
\left\{ U_{j}^{l_{i}}\right\} _{i},\mathbf{\hat{\Lambda}}^{\left[ k\right] }%
\left[ \Psi _{J},\nu ,\left\{ U_{j}^{l_{i}}\right\} \right] ,V,v\right) \\
&&+C\left( \Psi _{J}^{\otimes \left( \sum_{i}l_{i}\right) }\left( \left(
U_{j}^{l_{i}}\right) ^{\prime },\mathbf{\hat{\Lambda}}^{\left[ k\right] }%
\left[ \Psi _{J},\nu ,\left\{ U_{j}^{l_{i}}\right\} \right] ,v\right)
\right) \Psi _{J}^{\otimes \left( \sum_{i}l_{i}\right) }\left( \left\{
U_{j}^{l_{i}}\right\} _{i},\mathbf{\hat{\Lambda}}^{\left[ k\right] }\left[
\Psi _{J},\nu ,\left\{ U_{j}^{l_{i}}\right\} \right] ,V,v\right)
\end{eqnarray*}%
After changing of variable for local coordinats, it also rewrites:%
\begin{eqnarray}
0 &=&\int v\left( U_{i}^{l}\right) \Gamma \left( \left[ \Psi _{J},\nu
,U_{j}^{l}\right] ,\left( U_{j}^{l_{i}}\right) ^{\prime }\right) \nabla _{%
\mathbf{\hat{\Lambda}}_{\alpha }^{\left[ \left\{ k,k_{i\leqslant
n-1}\right\} \right] }}\hat{\Psi}_{J}^{\otimes \sum_{i}l_{i}}\left( \left\{
U_{j}^{l_{i}}\right\} _{i},\left( \left\{ \mathbf{\hat{\Lambda}}^{\left[
\left\{ k,k_{i\leqslant n-1}\right\} \right] }\right\} \right) _{n},v,\alpha
\right)  \label{CT} \\
&&+C\left( \Psi _{J}^{\otimes \left( \sum_{i}l_{i}\right) }\left( \left(
U_{j}^{l_{i}}\right) ^{\prime },\mathbf{\hat{\Lambda}}^{\left[ k\right]
},v\right) \right) \hat{\Psi}_{J}^{\otimes \sum_{i}l_{i}}\left( \left\{
U_{j}^{l_{i}}\right\} _{i},\left( \left\{ \mathbf{\hat{\Lambda}}^{\left[
\left\{ k,k_{i\leqslant n-1}\right\} \right] }\right\} \right) _{n},v,\alpha
\right)  \notag
\end{eqnarray}%
which is analogous to the dynamic equation of a massive field.

\subsection{Generalization: projective invariance}

More generaly, rather than considering full invariance, we can consider only
projective invariance with local additional contribution. In that case,
equation (\ref{NR}) is not satisfied anymore, but includes an additional
term:%
\begin{eqnarray*}
&&\int v\left( \left( U_{j}^{l_{i}}\right) \right) \frac{\delta \mathbf{\hat{%
\Lambda}}_{\alpha }^{\left[ \left\{ k,k_{i\leqslant n-1}\right\} \right] }%
\left[ \Psi _{J},\nu ,\left\{ U_{j}^{l_{i}}\right\} \right] }{\delta \Psi
_{J,\alpha }^{\otimes l_{i}}\left( \left( U_{j}^{l_{i}}\right) ^{\prime
}\right) }\nabla _{\mathbf{\hat{\Lambda}}_{\alpha }^{\left[ \left\{
k,k_{i\leqslant n-1}\right\} \right] }}\Psi _{J,\alpha }^{\otimes
\sum_{i}l_{i}}\left( \left\{ U_{j}^{l_{i}},\right\} _{i},\mathbf{\hat{\Lambda%
}}_{\infty ,\alpha }^{\left[ k\right] }\left[ \Psi _{J},\nu ,\left\{
U_{j}^{l_{i}}\right\} \right] \right) \\
&=&\int v\left( \left\{ \left( U_{j}^{l_{i}}\right) \right\} \right) \frac{%
\delta ^{\prime }V\left( \left\{ \Psi _{J,\alpha }^{\otimes
\sum_{i}l_{i}}\left( \left\{ U_{j}^{l_{i}},\right\} _{i},\mathbf{\hat{\Lambda%
}}_{\infty }^{\left[ k\right] }\left[ \Psi _{J},\nu ,\left\{
U_{j}^{l_{i}}\right\} \right] \right) \right\} \right) }{\delta \Psi
_{J,\alpha }^{\otimes l_{i}}\left( \left( U_{j}^{l_{i}}\right) ^{\prime },%
\mathbf{\hat{\Lambda}}^{\left[ \left\{ k,k_{i\leqslant n-1}\right\} \right] }%
\left[ \Psi _{J},\nu ,\left\{ U_{j}^{l_{i}}\right\} \right] \right) }\Psi
_{J,\alpha }^{\otimes \sum_{i}l_{i}}\left( \mathbf{\hat{\Lambda}}_{\infty
,\alpha }^{\left[ k\right] }\left[ \Psi _{J},\nu ,\left\{
U_{j}^{l_{i}}\right\} \right] \right)
\end{eqnarray*}%
Appendix 5 shows that in this case, we can still define, at least locally a
field depending on fixed set of parameters. In this case, the initial
variation of the field rewrites:%
\begin{equation*}
\delta \Psi _{J,\alpha }^{\otimes \sum_{i}l_{i}}\left( \left\{
U_{j}^{l_{i}}\right\} _{i},\mathbf{\hat{\Lambda}}_{\infty ,\alpha }^{\left[ k%
\right] }\left[ \Psi _{J},\nu ,\left\{ U_{j}^{l_{i}}\right\} \right] \right)
=\delta ^{\prime }\hat{\Psi}_{J,\alpha }^{\otimes \sum_{i}l_{i}}\left(
\left( U_{j}^{l_{i}}\right) \right) \Psi _{J,\alpha }^{\otimes
\sum_{i}l_{i}}\left( \mathbf{\hat{\Lambda}}_{\infty ,\alpha }^{\left[ k%
\right] }\left[ \Psi _{J},\nu ,\left\{ U_{j}^{l_{i}}\right\} \right] \right)
\end{equation*}%
which is performed at fixed $\mathbf{\hat{\Lambda}}_{\infty ,\alpha }^{\left[
k\right] }\left[ \Psi _{J},\nu ,\left\{ U_{j}^{l_{i}}\right\} \right] $. As
a consquence, even if the effective field $\Psi _{J,\alpha }^{\otimes
\sum_{i}l_{i}}$\ is not invariant, there is a related field $\hat{\Psi}%
_{J,\alpha }^{\otimes \sum_{i}l_{i}}$ that can be defined locally as a
function of an invariant family of parametrs $\mathbf{\hat{\Lambda}}_{\infty
,\alpha }^{\left[ k\right] }\left[ \Psi _{J},\nu ,\left\{
U_{j}^{l_{i}}\right\} \right] $.

The fields $\hat{\Psi}_{J,\alpha }^{\otimes \sum_{i}l_{i}}$ writes:%
\begin{eqnarray*}
&&\hat{\Psi}_{J,\alpha }^{\otimes \sum_{i}l_{i}}\left( \left\{
U_{j}^{l_{i}}\right\} _{i},\mathbf{\hat{\Lambda}}_{\infty ,\alpha }^{\left[ k%
\right] }\left[ \Psi _{J},\nu ,\left\{ U_{j}^{l_{i}}\right\} \right] \right)
\\
&=&\hat{\Psi}_{J,\alpha }^{\otimes \sum_{i}l_{i}}\left( \left\{
U_{j}^{l_{i}}\right\} _{i},\mathbf{\hat{\Lambda}}_{\infty ,\alpha }^{\left[ k%
\right] },v,\alpha \right) \delta \left( f\left( \mathbf{\hat{\Lambda}}%
_{\infty ,\alpha }^{\left[ k\right] },\left( \Psi _{J,0,\alpha }^{\otimes
l_{i}},v\right) \right) \right)
\end{eqnarray*}%
If, as before, $\frac{\delta \mathbf{\hat{\Lambda}}_{\alpha }^{\left[
\left\{ k,k_{i\leqslant n-1}\right\} \right] }}{\delta \Psi _{J,\alpha
}^{\otimes l_{i}}\left( \left( U_{j}^{l_{i}}\right) ^{\prime }\right) }$ is
independent from realizations, $\hat{\Psi}_{J,\alpha }^{\otimes
\sum_{i}l_{i}}$ satisfies:%
\begin{eqnarray*}
&&\int \hat{v}\left( \left( U_{j}^{l_{i}}\right) \right) \frac{\delta 
\mathbf{\hat{\Lambda}}^{\left[ \left\{ k,k_{i\leqslant n-1}\right\} \right] }%
}{\delta \Psi _{J}^{\otimes l_{i}}\left( \left( U_{j}^{l_{i}}\right)
^{\prime }\right) }\nabla _{\mathbf{\hat{\Lambda}}^{\left[ \left\{
k,k_{i\leqslant n-1}\right\} \right] }}\hat{\Psi}_{J}^{\otimes
\sum_{i}l_{i}}\left( \left\{ U_{j}^{l_{i}}\right\} _{i},\mathbf{\hat{\Lambda}%
}_{\infty }^{\left[ k\right] }\left[ \Psi _{J},\nu ,\left\{
U_{j}^{l_{i}}\right\} \right] \right) \\
&\simeq &v\left( \left\{ \left( U_{j}^{l_{i}}\right) \right\} \right) \frac{%
\delta ^{\prime }\hat{V}\left( \hat{\Psi}_{J}^{\otimes \sum_{i}l_{i}}\left(
\left\{ U_{j}^{l_{i}}\right\} _{i},\mathbf{\hat{\Lambda}}_{\infty }^{\left[ k%
\right] }\left[ \Psi _{J},\nu ,\left\{ U_{j}^{l_{i}}\right\} \right] \right)
\right) }{\delta \Psi _{J}^{\otimes l_{i}}\left( \left\{ \left(
U_{j}^{l_{i}}\right) ^{\prime },\mathbf{\hat{\Lambda}}_{\infty }^{\left[ k%
\right] }\left[ \Psi _{J},\nu ,\left\{ U_{j}^{l_{i}}\right\} \right]
\right\} \right) }
\end{eqnarray*}%
where:%
\begin{equation*}
\hat{v}\left( \left( U_{j}^{l_{i}}\right) \right) =\int v\left( \left(
U_{j}^{l_{i}}\right) ^{\prime \prime }\right) \frac{\delta \Psi
_{J}^{\otimes \sum_{i}l_{i}}\left( \left( U_{j}^{l_{i}}\right) ^{\prime
\prime }\right) }{\delta \hat{\Psi}_{J}^{\otimes \sum_{i}l_{i}}\left(
U_{j}^{l_{i}}\right) }
\end{equation*}

\section{Constraints and geometry of parameter space}

So far, we have set aside the functional relation between the field $\Psi
_{J}^{\otimes l}$ and the parameter space. We assumed that this space
describes the parameters of some transformation groups among the
realizations of the field $\Psi _{I}^{\otimes k}$. Considering that the
generators of these groups satisfy some algebraic relations conditions the
parameter space through some defining equatns. The equations defining ths
space depend on the fields $\Psi _{J}^{\otimes l}$ since the action to
minimize depends on these fields. This implies that the metric on the
parameter space depends on $\Psi _{J}^{\otimes l}$, but also on the
background for the fld $\Psi _{I}^{\otimes k}$.

\subsection{Form of the constraint and metric}

We assume that The constraint describes relations on the symmetry operators
for $n\in 
\mathbb{N}
$:%
\begin{equation}
h_{k_{n}}\left( \left\{ \mathbf{L}_{\left\{ \alpha _{i}\right\} }\left( \Psi
_{J}^{\otimes l}\right) \right\} _{i\leqslant n},\left\{ \mathbf{U}%
_{i}^{k}\right\} _{k},h_{p}\left( \left( \Psi _{J}\right) ,U_{j}^{l},\nu
\right) \right) =0  \label{CNS}
\end{equation}%
where the functionals $h_{l}\left( \left( \Psi _{J}\right) ,U_{j}^{l},\nu
\right) $ have the form: 
\begin{equation}
h_{p}\left( \left( \Psi _{J}\right) ,\left\{ U_{j}^{l_{i}}\right\} ,\nu
\right) =\left[ \int h_{p}\left( \left( U_{j}^{l_{i}^{\prime }}\right) ,\nu
\right) \left( \tprod\limits_{i}\Psi _{J}^{\otimes l_{i}^{\prime }}\left(
U_{j}^{l_{i}^{\prime }}\right) \right) d\left( U_{j}^{l_{i}^{\prime
}}\right) \right] \tprod\limits_{i}\Psi _{J}^{\otimes l_{i}}\left(
U_{j}^{l_{i}}\right)  \label{HFT}
\end{equation}%
and $\mathbf{U}_{i}^{k}$ acts on functions of $U_{i}^{k}$ as the
multiplication operator by $U_{i}^{k}$. Expression (\ref{HFT}) corresponds
to consider states $F\left( \Psi _{J,\alpha }^{\otimes \sum_{i}l_{i}}\right) 
$ localizd at $U_{j}^{l_{i}}$.

We then evaluate (\ref{CNS}) the projected states:%
\begin{equation*}
\tprod\limits_{k}\Psi _{I,\alpha ,0}^{\otimes k}\left( U_{i}^{k},\mathbf{%
\hat{\Lambda}}_{\infty ,\alpha }^{\left[ k\right] }\left[ \Psi _{J},\nu
,\left\{ U_{j}^{l_{i}}\right\} \right] ,\left\{ \Psi _{J}^{\otimes
l}\right\} _{l}\right)
\end{equation*}%
defined by the saddle point solutions $\Psi _{I,\alpha ,0}^{\otimes k}$:%
\begin{eqnarray}
&&\Psi _{I,\alpha ,0}^{\otimes k}\left( U_{i}^{k},\mathbf{\hat{\Lambda}}%
_{\infty ,\alpha }^{\left[ k\right] }\left[ \Psi _{J},\nu ,\left\{
U_{j}^{l_{i}}\right\} \right] ,\left\{ \Psi _{J}^{\otimes l}\right\}
_{l}\right)  \label{VRN} \\
&=&\sum_{l_{i}}\Psi _{J,\alpha }^{\otimes \sum_{i}l_{i}}\left( \left\{
U_{j}^{l_{i}}\right\} ,\mathbf{\hat{\Lambda}}_{\infty ,\alpha }^{\left[ k%
\right] }\left[ \Psi _{J},\nu ,\left\{ U_{j}^{l_{i}}\right\} \right] \right) 
\mathcal{K}_{0}\left( U_{i}^{k},\left\{ U_{j}^{l_{i}}\right\} _{l_{i}},%
\mathbf{\hat{\Lambda}}_{\infty ,\alpha }^{\left[ k\right] }\left[ \Psi
_{J},\nu \right] \right)  \notag
\end{eqnarray}%
and equation (\ref{CNS}) becomes a relation for each realization $\alpha $:%
\begin{eqnarray}
0 &=&\int h_{k_{n}}\left( \left\{ \mathbf{\hat{\Lambda}}_{n,\alpha }^{\left[
k\right] }\left[ \Psi _{J},\nu ,\left\{ U_{j}^{l_{i}}\right\} \right]
\right\} _{k},\left\{ U_{i}^{k}\right\} ,h_{p}\left( \left( \Psi _{J}\right)
,U_{j}^{l},\nu \right) \right)  \label{CNT} \\
&&\tprod\limits_{k}\left\vert \Psi _{I,\alpha ,0}^{\otimes k}\left(
U_{i}^{k},\mathbf{\hat{\Lambda}}_{\infty ,\alpha }^{\left[ k\right] }\left[
\Psi _{J},\nu ,\left\{ U_{j}^{l_{i}}\right\} \right] ,\left\{ \Psi
_{J}^{\otimes l}\right\} _{l}\right) \right\vert ^{2}dU_{i}^{k}  \notag
\end{eqnarray}%
If we assume the independence of parameters from the realizations: 
\begin{equation*}
\mathbf{\hat{\Lambda}}_{n,\alpha }^{\left[ k\right] }\left[ \Psi _{J},\nu
,\left\{ U_{j}^{l_{i}}\right\} \right] =\mathbf{\hat{\Lambda}}_{n}^{\left[ k%
\right] }\left[ \Psi _{J},\nu ,\left\{ U_{j}^{l_{i}}\right\} \right]
\end{equation*}%
and:%
\begin{equation*}
\mathbf{\hat{\Lambda}}_{\infty ,\alpha }^{\left[ k\right] }\left[ \Psi
_{J},\nu ,\left\{ U_{j}^{l_{i}}\right\} \right] =\mathbf{\hat{\Lambda}}%
_{\infty }^{\left[ k\right] }\left[ \Psi _{J},\nu ,\left\{
U_{j}^{l_{i}}\right\} \right]
\end{equation*}%
we can sum over realizations before evaluation of (\ref{CNS}) on the
projected states. Defining:%
\begin{equation}
\Psi _{I,0}^{\otimes k}\left( U_{i}^{k},\mathbf{\hat{\Lambda}}_{\infty }^{%
\left[ k\right] }\left[ \Psi _{J},\nu ,\left\{ U_{j}^{l_{i}}\right\} \right]
,\left\{ \Psi _{J}^{\otimes l}\right\} _{l}\right) =\sum_{\alpha }\Psi
_{I,\alpha ,0}^{\otimes k}\left( U_{i}^{k},\mathbf{\hat{\Lambda}}_{\infty
,\alpha }^{\left[ k\right] }\left[ \Psi _{J},\nu ,\left\{
U_{j}^{l_{i}}\right\} \right] ,\left\{ \Psi _{J}^{\otimes l}\right\}
_{l}\right)  \label{SRN}
\end{equation}%
equation (\ref{CNS}) becomes a relation:%
\begin{equation}
\int h_{k_{n}}\left( \mathbf{\hat{\Lambda}}_{n}\left[ \Psi _{J},\nu ,\left\{
U_{j}^{l_{i}}\right\} \right] ,\left\{ U_{i}^{k}\right\} ,h_{p}\left( \left(
\Psi _{J}\right) ,U_{j}^{l},\nu \right) \right) \tprod\limits_{k}\left\vert
\Psi _{I,0}^{\otimes k}\left( U_{i}^{k},\mathbf{\hat{\Lambda}}_{\infty
,\alpha }^{\left[ k\right] }\left[ \Psi _{J},\nu ,\left\{
U_{j}^{l_{i}}\right\} \right] ,\left\{ \Psi _{J}^{\otimes l}\right\}
_{l}\right) \right\vert ^{2}dU_{i}^{k}=0  \label{CTQ}
\end{equation}

We can expand $h_{k_{n}}$ as a series of products: 
\begin{eqnarray*}
&&h_{k_{n}}\left( \mathbf{\hat{\Lambda}}_{n}\left[ \Psi _{J},\nu ,\left\{
U_{j}^{l_{i}}\right\} \right] ,\left\{ U_{i}^{k}\right\} ,h_{p}\left( \left(
\Psi _{J}\right) ,U_{j}^{l},\nu \right) \right) \\
&=&\sum_{m}h_{k_{n}}^{\left( m\right) }\left( \mathbf{\hat{\Lambda}}_{n}%
\left[ \Psi _{J},\nu ,\left\{ U_{j}^{l_{i}}\right\} \right] ,\left\{
U_{i}^{k}\right\} ,h_{p}\left( \left( \Psi _{J}\right) ,U_{j}^{l},\nu
\right) \right) \bar{h}_{m}\left( \left\{ U_{i}^{k}\right\} \right)
\end{eqnarray*}%
and equation (\ref{CTQ}) writes:%
\begin{equation}
0=\sum_{m}h_{k_{n}}^{\left( m\right) }\left( \mathbf{\hat{\Lambda}}_{n}\left[
\Psi _{J},\nu ,\left\{ U_{j}^{l_{i}}\right\} \right] ,\left\{
U_{i}^{k}\right\} ,h_{p}\left( \left( \Psi _{J}\right) ,U_{j}^{l},\nu
\right) \right) \bar{H}_{m}\left( \Psi _{I,0}^{\otimes k}\right)  \label{CTB}
\end{equation}%
with:%
\begin{equation}
\bar{H}_{m}\left( \Psi _{I,0}^{\otimes k}\right) =\int \bar{h}_{m}\left(
\left\{ U_{i}^{k}\right\} \right) \tprod\limits_{k}\left\vert \Psi
_{I,0}^{\otimes k}\left( U_{i}^{k},\mathbf{\hat{\Lambda}}_{\infty }^{\left[ k%
\right] }\left[ \Psi _{J},\nu ,\left\{ U_{j}^{l_{i}}\right\} \right]
,\left\{ \Psi _{J}^{\otimes l}\right\} _{l}\right) \right\vert ^{2}dU_{i}^{k}
\label{H}
\end{equation}

Given the form of $h_{p}\left( \left( \Psi _{J}\right) ,U_{j}^{l},\nu
\right) $, the constraint (\ref{CTB}) depends on: 
\begin{equation*}
\left( \Psi _{J}\right) ,\nu ,\left\{ \Psi _{J}^{\otimes l_{i}}\left(
U_{j}^{l_{i}}\right) ,\bar{H}_{m}\left( \Psi _{I,0}^{\otimes k}\right)
\right\}
\end{equation*}%
The form of $\Psi _{I,\alpha ,0}^{\otimes k}$ ((\ref{SRN}) and (\ref{VRN}))
and $\bar{H}_{m}\left( \Psi _{I,\alpha ,0}^{\otimes k}\right) $ defined in (%
\ref{H}) omply that the dependence of the constraint in $\left\{ \Psi
_{J}^{\otimes l_{i}}\left( U_{j}^{l_{i}}\right) \right\} ,\bar{H}_{m}\left(
\Psi _{I,\alpha ,0}^{\otimes k}\right) $ can be described as function
dependence in $\Psi _{J}^{\otimes \sum_{i}l_{i}}\mathcal{K}_{0}$ through: 
\begin{equation*}
\int \bar{h}_{m}\left( \left\{ U_{i}^{k}\right\} \right)
\tprod\limits_{k}\left\vert \Psi _{J}^{\otimes \sum_{i}l_{i}}\left( \left\{
U_{j}^{l_{i}}\right\} ,\mathbf{\hat{\Lambda}}_{\infty }^{\left[ k\right] }%
\left[ \Psi _{J},\nu ,\left\{ U_{j}^{l_{i}}\right\} \right] \right)
\right\vert ^{2}\left\vert \mathcal{K}_{0}\left( U_{i}^{k},\left\{
U_{j}^{l_{i}}\right\} _{l_{i}},\mathbf{\hat{\Lambda}}_{\infty }^{\left[ k%
\right] }\left[ \Psi _{J},\nu \right] \right) \right\vert ^{2}dU_{i}^{k}
\end{equation*}%
Remark that we can consider non localized states by replacing:%
\begin{eqnarray*}
&&\Psi _{J}^{\otimes \sum_{i}l_{i}}\left( \left\{ U_{j}^{l_{i}}\right\} ,%
\mathbf{\hat{\Lambda}}_{\infty }^{\left[ k\right] }\left[ \Psi _{J},\nu
,\left\{ U_{j}^{l_{i}}\right\} \right] \right) \\
&\rightarrow &\nu \left( \Psi _{J}^{\otimes \sum_{i}l_{i}}\left( \mathbf{%
\hat{\Lambda}}_{\infty }^{\left[ k\right] }\left[ \Psi _{J},\nu \right]
\right) \right) =\int \nu \left( \left\{ U_{j}^{l_{i}}\right\} \right) \Psi
_{J}^{\otimes \sum_{i}l_{i}}\left( \left\{ U_{j}^{l_{i}}\right\} ,\mathbf{%
\hat{\Lambda}}_{\infty }^{\left[ k\right] }\left[ \Psi _{J},\nu ,\left\{
U_{j}^{l_{i}}\right\} \right] \right) d\left\{ U_{j}^{l_{i}}\right\}
\end{eqnarray*}

\subsection{Lowest order expansion and metrics on projected states}

A second order expansion of the previous constraint enables to define a
metric on the parameter space. It depends both on the \ field $\Psi
_{J}^{\otimes \sum_{i}l_{i}}$ and the background $\Psi _{I,0}^{\otimes k}$.

\subsubsection{Local functional}

For a local functional at the lowest order:%
\begin{eqnarray*}
0 &=&\gamma \left( \Psi _{J},\nu \right) +N\left( \Psi _{J},\nu ,\left\{
\left( \Psi _{J}^{\otimes l_{i}}\left( U_{j}^{l_{i}}\right) ,\bar{H}%
_{m}\left( \Psi _{I,0}^{\otimes k}\right) \right) \right\} \right) .\left(
\left\{ \mathbf{\hat{\Lambda}}_{n}\left[ \Psi _{J},\nu ,\left\{
U_{j}^{l_{i}}\right\} \right] \right\} \right) \\
&&+\left\{ \mathbf{\hat{\Lambda}}_{n}\left[ \Psi _{J},\nu ,\left\{
U_{j}^{l_{i}}\right\} \right] \right\} M\left( \Psi _{J},\nu ,\left\{ \left(
\Psi _{J}^{\otimes l_{i}}\left( U_{j}^{l_{i}}\right) ,\bar{H}_{m}\left( \Psi
_{I,0}^{\otimes k}\right) \right) \right\} \right) \left\{ \mathbf{\hat{%
\Lambda}}_{n}\left[ \Psi _{J},\nu ,\left\{ U_{j}^{l_{i}}\right\} \right]
\right\}
\end{eqnarray*}%
and redefining operators by translation, we find:%
\begin{equation}
0=\gamma \left( \Psi _{J},\nu \right) +\left\{ \mathbf{\hat{\Lambda}}%
_{n_{1}}^{\left[ \mathbf{k}_{1}\right] }\left[ \Psi _{J},\nu ,\left\{
U_{j}^{l_{i}}\right\} \right] \right\} M\left( \left\{ \left[ \mathbf{k}_{1}%
\right] ,\left[ \mathbf{k}_{2}\right] \right\} ,\Psi _{J},\nu ,\left\{
\left( \Psi _{J}^{\otimes l_{i}}\left( U_{j}^{l_{i}}\right) ,\bar{H}%
_{m}\left( \Psi _{I,0}^{\otimes k}\right) \right) \right\} \right) \left\{ 
\mathbf{\hat{\Lambda}}_{n_{2}}^{\left[ \mathbf{k}_{2}\right] }\left[ \Psi
_{J},\nu ,\left\{ U_{j}^{l_{i}}\right\} \right] \right\}  \label{MF}
\end{equation}%
with:%
\begin{eqnarray*}
\left[ \mathbf{k}_{1}\right] &=&\left[ k_{1,1}...k_{1,n_{1}}\right] \\
\left[ \mathbf{k}_{2}\right] &=&\left[ k_{2,1},...,k_{2,n_{2}}\right]
\end{eqnarray*}

The scalar functional has the form (we reintroduce the notation $\Psi _{J}$, 
$\Psi _{J}^{\dag }$): 
\begin{equation*}
\alpha \left( \Psi _{J}\right) =\int \Psi _{J}^{\dag }A\Psi _{J}+\int \Psi
_{J}^{\dag }\Psi _{J}^{\dag }B\Psi _{J}\Psi _{J}+...
\end{equation*}%
At the quadratic order the quadratic form defined in (\ref{MF}) becomes:%
\begin{equation*}
M\left( \left\{ \left[ \mathbf{k}_{1}\right] ,\left[ \mathbf{k}_{2}\right]
\right\} ,\Psi _{J},\nu ,\left\{ \left( \Psi _{J}^{\otimes l_{i}}\left(
U_{j}^{l_{i}}\right) ,\bar{H}_{m}\left( \Psi _{I,0}^{\otimes k}\right)
\right) \right\} \right)
\end{equation*}%
or equivalently:%
\begin{equation*}
M\left( \left\{ \left[ \mathbf{k}_{1}\right] ,\left[ \mathbf{k}_{2}\right]
\right\} ,\nu ,\left\{ \Psi _{J}^{\otimes \sum_{i}l_{i}}\mathcal{K}%
_{0}\right\} \right)
\end{equation*}%
that is, in the projected space, the metric is dynamicaly an object defined
by the field: 
\begin{equation*}
\Psi _{I,\alpha ,0}^{\otimes k}\left( U_{i}^{k},\mathbf{\hat{\Lambda}}%
_{\infty }^{\left[ k\right] }\left[ \Psi _{J},\nu ,\left\{
U_{j}^{l_{i}}\right\} \right] ,\left\{ \Psi _{J}^{\otimes l}\right\}
_{l}\right)
\end{equation*}

Ultimately, if (\ref{NR}), or its versions (\ref{BNF}) of (\ref{CT}) are
satisfied, the sytem is described by dynamical quantities $\Psi
_{J}^{\otimes \sum_{i}l_{i}}$, $M\left( \left\{ \left[ \mathbf{k}_{1}\right]
,\left[ \mathbf{k}_{2}\right] \right\} ,\Psi _{J},\nu ,\left( \Psi
_{J}^{\otimes l_{i}}\left( U_{j}^{l_{i}}\right) \right) ,\bar{H}_{m}\left(
\Psi _{I,\alpha ,0}^{\otimes k}\right) \right) $ and is built on apparently
exogeneous parameter space $\left\{ \mathbf{\hat{\Lambda}}_{\infty ,\alpha
}^{\left[ k\right] }\right\} $.

For one variable $\mathbf{\hat{\Lambda}}^{\left[ k\right] }\left[ \Psi
_{J},\nu ,\left\{ U_{j}^{l_{i}}\right\} \right] $ the metric tensor deriving
from: 
\begin{equation*}
M\left( \left\{ \left[ \mathbf{k}_{1}\right] ,\left[ \mathbf{k}_{2}\right]
\right\} ,\nu ,\left\{ \Psi _{J}^{\otimes \sum_{i}l_{i}}\mathcal{K}%
_{0}\right\} \right)
\end{equation*}%
is derived by chosing some ntrinsc coordnates $\widetilde{\mathbf{\hat{%
\Lambda}}^{\left[ k\right] }\left[ \Psi _{J},\nu ,\left\{
U_{j}^{l_{i}}\right\} \right] }$ satisng (\ref{MF}) nd btnd b vltng:%
\begin{equation*}
\frac{\partial \mathbf{\hat{\Lambda}}^{\left[ k\right] }\left[ \Psi _{J},\nu
,\left\{ U_{j}^{l_{i}}\right\} \right] }{\nabla \widetilde{\mathbf{\hat{%
\Lambda}}^{\left[ k\right] }\left[ \Psi _{J},\nu ,\left\{
U_{j}^{l_{i}}\right\} \right] }}.\frac{\partial \mathbf{\hat{\Lambda}}^{%
\left[ k\right] }\left[ \Psi _{J},\nu ,\left\{ U_{j}^{l_{i}}\right\} \right] 
}{\nabla \widetilde{\mathbf{\hat{\Lambda}}^{\left[ k\right] }\left[ \Psi
_{J},\nu ,\left\{ U_{j}^{l_{i}}\right\} \right] }}
\end{equation*}%
whr th sclr prdct is on th coordntes $\mathbf{\hat{\Lambda}}^{\left[ k\right]
}\left[ \Psi _{J},\nu ,\left\{ U_{j}^{l_{i}}\right\} \right] $

It is wrtten: 
\begin{equation*}
g\left( \left\{ \left[ k\right] ,\left[ k\right] \right\} ,\Psi _{J},\nu
,\left\{ \left( \Psi _{J}^{\otimes l_{i}}\left( U_{j}^{l_{i}}\right) ,\bar{H}%
_{m}\left( \Psi _{I,0}^{\otimes k}\right) \right) \right\} \right)
\end{equation*}%
and depends on the projected field. This ones includes $\sum \Psi
_{J}^{\otimes \sum_{i}l_{i}}\left( \left\{ U_{j}^{l_{i}}\right\} ,\mathbf{%
\hat{\Lambda}}_{\infty }^{\left[ k\right] }\left[ \Psi _{J},\nu ,\left\{
U_{j}^{l_{i}}\right\} \right] \right) $ as needed and the background. $%
\mathcal{K}_{0}\left( U_{i}^{k},\left\{ U_{j}^{l_{i}}\right\} _{l_{i}},%
\mathbf{\hat{\Lambda}}_{\infty }^{\left[ k\right] }\left[ \Psi _{J},\nu %
\right] \right) $. The metrics depends on the integrated background.

Considering the constraint for one variable $\mathbf{\hat{\Lambda}}^{\left[ k%
\right] }\left[ \Psi _{J},\nu ,\left\{ U_{j}^{l_{i}}\right\} \right] $:%
\begin{equation*}
\left\{ \mathbf{\hat{\Lambda}}^{\left[ k\right] }\left[ \Psi _{J},\nu
,\left\{ U_{j}^{l_{i}}\right\} \right] \right\} M\left( \left\{ \left[ k%
\right] ,\left[ k\right] \right\} ,\Psi _{J},\nu ,\left\{ \left( \Psi
_{J}^{\otimes l_{i}}\left( U_{j}^{l_{i}}\right) ,\bar{H}_{m}\left( \Psi
_{I,0}^{\otimes k}\right) \right) \right\} \right) .\left\{ \mathbf{\hat{%
\Lambda}}^{\left[ k\right] }\left[ \Psi _{J},\nu ,\left\{
U_{j}^{l_{i}}\right\} \right] \right\} +\gamma \left( \Psi _{J}\right) =0
\end{equation*}%
we recover some "usual" metric tensor: 
\begin{equation*}
g\left( \left\{ \left[ k\right] ,\left[ k\right] \right\} ,\Psi _{J},\nu
,\left\{ \left( \Psi _{J}^{\otimes l_{i}}\left( U_{j}^{l_{i}}\right) ,\bar{H}%
_{m}\left( \Psi _{I,0}^{\otimes k}\right) \right) \right\} \right)
\end{equation*}%
that also writes:%
\begin{equation*}
g\left( \left\{ \left[ k\right] ,\left[ k\right] \right\} ,\nu ,\left\{ \Psi
_{J}^{\otimes \sum_{i}l_{i}}\mathcal{K}_{0}\right\} \right)
\end{equation*}%
The metric depends functionally of $\Psi _{J}^{\otimes \sum_{i}l_{i}}$, as
needed, but also on the integrated field $\mathcal{K}_{0}$. The metric
component depends on physical characteristics of space.

More that the metric for the whole set of parameters $\mathbf{\hat{\Lambda}}%
_{\infty }^{\left[ k\right] }\left[ \Psi _{J},\nu ,\left\{
U_{j}^{l_{i}}\right\} \right] $,can be described given that the set of
parameters is defined by flag manifold:%
\begin{equation*}
\mathbf{\hat{\Lambda}}^{\left[ k\right] }\left[ \Psi _{J},\nu ,\left\{
U_{j}^{l_{i}}\right\} \right] ,\left\{ \mathbf{\hat{\Lambda}}^{\left[
\left\{ k_{i\leqslant n-1}\right\} \right] }\left[ \Psi _{J},\nu ,\left\{
U_{j}^{l_{i}}\right\} \right] \right\}
\end{equation*}%
If the intrinsic coordinates $\widetilde{\mathbf{\hat{\Lambda}}_{\infty }^{%
\left[ k\right] }\left[ \Psi _{J},\nu ,\left\{ U_{j}^{l_{i}}\right\} \right] 
}$ respects these sequence of inclusion, the metric decomposes as sequences:%
\begin{eqnarray*}
&&\left\{ g\left( \left\{ \left[ \left\{ k_{i\leqslant n-1}\right\} \right] ,%
\left[ \left\{ k_{i\leqslant n-1}\right\} \right] \right\} ,\nu ,\left\{
\Psi _{J}^{\otimes \sum_{i}l_{i}}\mathcal{K}_{0}\right\} \right) \right\} \\
&\rightarrow &\left\{ \frac{\partial \mathbf{\hat{\Lambda}}^{\left[ \left\{
k_{i\leqslant n-1}\right\} \right] }\left[ \Psi _{J},\nu ,\left\{
U_{j}^{l_{i}}\right\} \right] }{\nabla \widetilde{\mathbf{\hat{\Lambda}}^{%
\left[ \left\{ k_{i\leqslant n-1}\right\} \right] }\left[ \Psi _{J},\nu
,\left\{ U_{j}^{l_{i}}\right\} \right] }}.\frac{\partial \mathbf{\hat{\Lambda%
}}^{\left[ \left\{ k_{i\leqslant n-1}\right\} \right] }\left[ \Psi _{J},\nu
,\left\{ U_{j}^{l_{i}}\right\} \right] }{\nabla \widetilde{\mathbf{\hat{%
\Lambda}}^{\left[ \left\{ k_{i\leqslant n-1}\right\} \right] }\left[ \Psi
_{J},\nu ,\left\{ U_{j}^{l_{i}}\right\} \right] }}\right\}
\end{eqnarray*}%
This describes some metrics on the increasing sequences of set of point.

\subsubsection{Non local functional}

For non local functional, the derivation is similar, where:%
\begin{equation*}
\nu \left( \Psi _{J}^{\otimes \sum_{i}l_{i}}\left( \mathbf{\hat{\Lambda}}%
_{\infty }^{\left[ k\right] }\left[ \Psi _{J},\nu \right] \right) \right)
=\nu \left( \Psi _{J}^{\otimes \sum_{i}l_{i}}\right)
\end{equation*}%
replaces $\Psi _{J}^{\otimes l_{i}}\left( U_{j}^{l_{i}}\right) $. We have
for the quadratic form:%
\begin{equation*}
M\left( \left\{ \left[ \mathbf{k}_{1}\right] ,\left[ \mathbf{k}_{2}\right]
\right\} ,\Psi _{J},\left\{ \left( \nu \left( \Psi _{J}^{\otimes
\sum_{i}l_{i}}\right) ,\bar{H}_{m}\left( \Psi _{I,0}^{\otimes k}\right)
\right) \right\} \right) =M\left( \left\{ \left[ \mathbf{k}_{1}\right] ,%
\left[ \mathbf{k}_{2}\right] \right\} ,\left\{ \left( \nu \left( \Psi
_{J}^{\otimes \sum_{i}l_{i}}\right) \mathcal{K}_{0}\right) \right\} \right)
\end{equation*}%
and for any variable, the form of the associated metric:%
\begin{equation*}
g\left( \left\{ \left[ k\right] ,\left[ k\right] \right\} ,\Psi _{J},\left\{
\left( \nu \left( \Psi _{J}^{\otimes \sum_{i}l_{i}}\right) ,\bar{H}%
_{m}\left( \Psi _{I,0}^{\otimes k}\right) \right) \right\} \right) =g\left(
\left\{ \left[ k\right] ,\left[ k\right] \right\} ,\left\{ \left( \nu \left(
\Psi _{J}^{\otimes \sum_{i}l_{i}}\right) \mathcal{K}_{0}\right) \right\}
\right)
\end{equation*}

\subsubsection{Expression in terms of averaged field}

The constraints can also be rewritten in terms of the averaged fields. They
were defined in (\ref{SR}) as:%
\begin{eqnarray*}
&&\Psi _{J}^{\otimes \left( \sum_{i}l_{i}\right) }\left( \left\{
U_{j}^{l_{i}}\right\} _{i},\mathbf{\hat{\Lambda}}_{\infty }^{\left[ k\right]
}\left[ \Psi _{J},\nu ,\left\{ U_{j}^{l_{i}}\right\} \right] ,v\right) \\
&\equiv &\Psi _{J}^{\otimes \left( \sum_{i}l_{i}\right) }\left( \left\{
U_{j}^{l_{i}}\right\} _{i},\mathbf{\hat{\Lambda}}^{\left[ k\right] }\left[
\Psi _{J},\nu ,\left\{ U_{j}^{l_{i}}\right\} \right] ,\left\{ V^{k}\right\}
,v\right) \\
&=&\Psi _{J}^{\otimes \left( \sum_{i}l_{i}\right) }\left( \left\{
U_{j}^{l_{i}}\right\} _{i},\mathbf{\hat{\Lambda}}^{\left[ k\right] }\left[
\Psi _{J},\nu ,\left\{ U_{j}^{l_{i}}\right\} \right] ,V,v\right) +\Psi
_{J}^{\otimes \left( \sum_{i}l_{i}\right) }\left( \left\{
U_{j}^{l_{i}}\right\} _{i},\mathbf{\hat{\Lambda}}^{\left[ k\right] }\left[
\Psi _{J},\nu ,\left\{ U_{j}^{l_{i}}\right\} \right] ,V^{2},v\right) +...
\end{eqnarray*}%
where:%
\begin{equation*}
V=\left\{ \left\{ \mathbf{\hat{\Lambda}}^{k_{i}}\left[ \Psi _{J},\nu
,\left\{ U_{j}^{l_{i}}\right\} \right] \right\} \right\} _{k_{i}}
\end{equation*}%
depicts the entire set of parameters $\left\{ \mathbf{\hat{\Lambda}}^{k_{i}}%
\left[ \Psi _{J},\nu \right] \right\} $ and $\mathbf{\hat{\Lambda}}^{\left[ k%
\right] }\left[ \Psi _{J},\nu ,\left\{ U_{j}^{l_{i}}\right\} \right] $ is a
given parameter. Using the same decomposition for the constraints:%
\begin{equation*}
h_{k_{n}}\left( \left\{ \mathbf{L}_{\left\{ \alpha _{i}\right\} }\left( \Psi
_{J}^{\otimes l}\right) \right\} _{i\leqslant n},\left\{ \mathbf{U}%
_{i}^{k}\right\} _{k},h_{p}\left( \left( \Psi _{J}\right) ,U_{j}^{l},\nu
\right) \right) =0
\end{equation*}%
evaluated on a state of the field (\ref{VRN}):%
\begin{equation*}
\int \nu \left( \left\{ U_{j}^{l_{i}}\right\} _{i}\right) \Psi _{J}^{\otimes
\left( \sum_{i}l_{i}\right) }\left( \left\{ U_{j}^{l_{i}}\right\} _{i},%
\mathbf{\hat{\Lambda}}^{\left[ k\right] }\left[ \Psi _{J},\nu ,\left\{
U_{j}^{l_{i}}\right\} \right] ,\left\{ V^{k}\right\} ,v\right) \mathcal{K}%
_{0}\left( U_{i}^{k},\left\{ U_{j}^{l_{i}}\right\} _{l_{i}},\mathbf{\hat{%
\Lambda}}_{\infty ,\alpha }^{\left[ k\right] }\left[ \Psi _{J},\nu \right]
\right) d\left\{ U_{j}^{l_{i}}\right\} _{i}
\end{equation*}%
yields:%
\begin{eqnarray*}
0 &=&\int h_{k_{n}}\left( \mathbf{\hat{\Lambda}}^{\left[ k\right] }\left[
\Psi _{J},\nu ,\left\{ U_{j}^{l_{i}}\right\} \right] ,\left\{ V^{k}\right\}
,\left\{ U_{i}^{k}\right\} _{k},h_{p}\left( \left( \Psi _{J}\right)
,U_{j}^{l},\nu \right) \right) \\
&&\times \left\vert \nu \left( \left\{ U_{j}^{l_{i}}\right\} _{i}\right)
\right\vert ^{2}\left\vert \Psi _{J}^{\otimes \left( \sum_{i}l_{i}\right)
}\left( \left\{ U_{j}^{l_{i}}\right\} _{i},\mathbf{\hat{\Lambda}}^{\left[ k%
\right] }\left[ \Psi _{J},\nu ,\left\{ U_{j}^{l_{i}}\right\} \right]
,\left\{ V^{k}\right\} ,v\right) \right\vert ^{2} \\
&&\times \left\vert \mathcal{K}_{0}\left( \left\{ U_{i}^{k}\right\}
_{k},\left\{ U_{j}^{l_{i}}\right\} _{l_{i}},\mathbf{\hat{\Lambda}}_{\infty
,\alpha }^{\left[ k\right] }\left[ \Psi _{J},\nu \right] \right) \right\vert
^{2}d\left\{ V^{k}\right\} d\left\{ U_{i}^{k}\right\} _{k}
\end{eqnarray*}%
and this become an equation:%
\begin{equation*}
0=H\left( \mathbf{\hat{\Lambda}}^{\left[ k\right] }\left[ \Psi _{J},\nu %
\right] ,\nu \left( \Psi _{J}^{\otimes \left( \sum_{i}l_{i}\right) }\right) ,%
\mathcal{K}_{0}\right)
\end{equation*}%
This a manifold equation depending on the state of the system and the
background kernel $\mathcal{K}_{0}$. This means that the metric of this
manifold:%
\begin{equation*}
g\left( \mathbf{\hat{\Lambda}}^{\left[ k\right] }\left[ \Psi _{J},\nu \right]
,\nu \left( \Psi _{J}^{\otimes \left( \sum_{i}l_{i}\right) }\right) ,%
\mathcal{K}_{0}\right)
\end{equation*}%
is a series expansion:%
\begin{eqnarray}
&&g_{0}\left( \mathbf{\hat{\Lambda}}^{\left[ k\right] }\left[ \Psi _{J},\nu %
\right] ,\Psi _{J}^{\otimes \left( \sum_{i}l_{i}\right) }\left( \mathbf{\hat{%
\Lambda}}^{\left[ k\right] }\left[ \Psi _{J},\nu \right] \right) ,\mathcal{K}%
_{0}\right)  \label{GPN} \\
&&+\int g_{1}\left( \mathbf{\hat{\Lambda}}^{\left[ k\right] }\left[ \Psi
_{J},\nu \right] ,\Psi _{J}^{\otimes \left( \sum_{i}l_{i}\right) }\left( 
\mathbf{\hat{\Lambda}}^{\left[ k\right] }\left[ \Psi _{J},\nu \right]
\right) ,\mathcal{K}_{0},\left( \mathbf{\hat{\Lambda}}^{\left[ k\right] }%
\left[ \Psi _{J},\nu \right] \right) ^{\prime }\right) \int \Psi
_{J}^{\otimes \left( \sum_{i}l_{i}\right) }\left( \left( \mathbf{\hat{\Lambda%
}}^{\left[ k\right] }\left[ \Psi _{J},\nu \right] \right) ^{\prime }\right)
d\left( \mathbf{\hat{\Lambda}}^{\left[ k\right] }\left[ \Psi _{J},\nu \right]
\right) ^{\prime }  \notag \\
&&+\int g_{1}\left( \mathbf{\hat{\Lambda}}^{\left[ k\right] }\left[ \Psi
_{J},\nu \right] ,\Psi _{J}^{\otimes \left( \sum_{i}l_{i}\right) }\left( 
\mathbf{\hat{\Lambda}}^{\left[ k\right] }\left[ \Psi _{J},\nu \right]
\right) ,\mathcal{K}_{0},\left( \mathbf{\hat{\Lambda}}^{\left[ k\right] }%
\left[ \Psi _{J},\nu \right] \right) ^{\prime },\left( \mathbf{\hat{\Lambda}}%
^{\left[ k\right] }\left[ \Psi _{J},\nu \right] \right) ^{\prime \prime
}\right)  \notag \\
&&\times \Psi _{J}^{\otimes \left( \sum_{i}l_{i}\right) }\left( \left( 
\mathbf{\hat{\Lambda}}^{\left[ k\right] }\left[ \Psi _{J},\nu \right]
\right) ^{\prime }\right) d\left( \mathbf{\hat{\Lambda}}^{\left[ k\right] }%
\left[ \Psi _{J},\nu \right] \right) ^{\prime }\Psi _{J}^{\otimes \left(
\sum_{i}l_{i}\right) }\left( \left( \mathbf{\hat{\Lambda}}^{\left[ k\right] }%
\left[ \Psi _{J},\nu \right] \right) ^{\prime \prime }\right) d\left( 
\mathbf{\hat{\Lambda}}^{\left[ k\right] }\left[ \Psi _{J},\nu \right]
\right) ^{\prime \prime }  \notag
\end{eqnarray}%
where the $\Psi _{J}^{\otimes \left( \sum_{i}l_{i}\right) }\left( \mathbf{%
\hat{\Lambda}}^{\left[ k\right] }\left[ \Psi _{J},\nu \right] \right) $ are
some apparent local fields arising from expansions of:%
\begin{equation*}
\Psi _{J}^{\otimes \left( \sum_{i}l_{i}\right) }\left( \left\{
U_{j}^{l_{i}}\right\} _{i},\mathbf{\hat{\Lambda}}^{\left[ k\right] }\left[
\Psi _{J},\nu ,\left\{ U_{j}^{l_{i}}\right\} \right] ,\left\{ V^{k}\right\}
,v\right)
\end{equation*}%
The series (\ref{GPN}) is similar to the one we would obtain by considering
perturbative expansion for a system where metric and field interact, execept
that here, we have not tried to derive some equations relating those
quantitites. Our formula is only a general result without considerations
about the way these quantities should interact. The main difference comes
from the background kernel:%
\begin{equation*}
\mathcal{K}_{0}\left( U_{i}^{k},\left\{ U_{j}^{l_{i}}\right\} _{l_{i}},%
\mathbf{\hat{\Lambda}}_{\infty ,\alpha }^{\left[ k\right] }\left[ \Psi
_{J},\nu \right] \right)
\end{equation*}%
that does not appear directly, since the corresponding subobject has been
proected and integrated in the fields, but this kernel should take part to
the metric equations.

\subsection{Change ef variable}

We write for short:%
\begin{equation*}
\Psi _{I,\alpha ,0}^{\otimes k}\left( U_{i}^{k},\mathbf{\hat{\Lambda}}%
_{\infty }^{\left[ k\right] }\left[ \Psi _{J},\nu ,\left\{
U_{j}^{l_{i}}\right\} \right] ,\left\{ \Psi _{J}^{\otimes l}\right\}
_{l}\right) \rightarrow \Psi _{J}^{\otimes \sum_{i}l_{i}}\mathcal{K}_{0}
\end{equation*}%
where:%
\begin{equation*}
\Psi _{J}^{\otimes \sum_{i}l_{i}}=\Psi _{J}^{\otimes \sum_{i}l_{i}}\left(
\left\{ U_{j}^{l_{i}}\right\} ,\mathbf{\hat{\Lambda}}_{\infty ,\alpha }^{%
\left[ k\right] }\left[ \Psi _{J},\nu ,\left\{ U_{j}^{l_{i}}\right\} \right]
\right)
\end{equation*}%
We can then consider a change of variable:%
\begin{eqnarray*}
&&M\left( \left\{ \left[ \mathbf{k}_{1}\right] ,\left[ \mathbf{k}_{2}\right]
\right\} ,\left\{ \Psi _{J}^{\otimes \sum_{i}l_{i}}\mathcal{K}_{0}\right\}
\right) \\
&=&U\left( \left\{ \left[ \mathbf{k}_{1}\right] ,\left[ \mathbf{k}%
_{1}^{\prime }\right] \right\} \right) N\left( \left\{ \left[ \mathbf{k}%
_{1}^{\prime }\right] ,\left[ \mathbf{k}_{1}^{\prime }\right] \right\}
,\left\{ \Psi _{J}^{\otimes \sum_{i}l_{i}}\mathcal{K}_{0}\right\} \right)
U^{-1}\left( \left\{ \left[ \mathbf{k}_{1}^{\prime }\right] ,\left[ \mathbf{k%
}_{2}\right] \right\} \right)
\end{eqnarray*}%
where the vector of parameters is: 
\begin{equation*}
\left[ \mathbf{k}_{1}^{\prime }\right] =\left[ k_{1,1}^{\prime
}...k_{1,n_{1}}^{\prime }\right]
\end{equation*}%
Under this change of variable:%
\begin{equation*}
\mathbf{\hat{\Lambda}}_{n_{1}}^{\left[ \mathbf{k}_{1}^{\prime }\right] }%
\left[ \Psi _{J},\nu ,\left\{ U_{j}^{l_{i}}\right\} \right] =U\left( \left\{ %
\left[ \mathbf{k}_{1}^{\prime }\right] ,\left[ \mathbf{k}_{1}\right]
\right\} \right) \sqrt{N\left( \left\{ \left[ \mathbf{k}_{1}\right] ,\left[ 
\mathbf{k}_{1}\right] \right\} ,\left\{ \Psi _{J}^{\otimes \sum_{i}l_{i}}%
\mathcal{K}_{0}\right\} \right) }\mathbf{\hat{\Lambda}}_{n_{1}}^{\left[ 
\mathbf{k}_{1}\right] }\left[ \Psi _{J},\nu ,\left\{ U_{j}^{l_{i}}\right\} %
\right]
\end{equation*}%
with:%
\begin{equation*}
\hat{U}\left( \left\{ \left[ \mathbf{k}_{1}^{\prime }\right] ,\left[ \mathbf{%
k}_{1}\right] \right\} \right) =U\left( \left\{ \left[ \mathbf{k}%
_{1}^{\prime }\right] ,\left[ \mathbf{k}_{1}\right] \right\} \right) \sqrt{%
N\left( \left\{ \left[ \mathbf{k}_{1}\right] ,\left[ \mathbf{k}_{1}\right]
\right\} ,\left\{ \Psi _{J}^{\otimes \sum_{i}l_{i}}\mathcal{K}_{0}\right\}
\right) }
\end{equation*}%
The variation of parameters in this change of variable is given by:%
\begin{eqnarray*}
\frac{\partial \mathbf{\hat{\Lambda}}_{n_{1}}^{\left[ \mathbf{k}_{1}^{\prime
}\right] }\left[ \Psi _{J},\nu ,\left\{ U_{j}^{l_{i}}\right\} \right] }{%
\partial \mathbf{\hat{\Lambda}}_{n_{2}}^{\left[ \mathbf{k}_{2}\right] }\left[
\Psi _{J},\nu ,\left\{ U_{j}^{l_{i}}\right\} \right] } &=&\frac{\partial 
\hat{U}\left( \left\{ \left[ \mathbf{k}_{1}^{\prime }\right] ,\left[ \mathbf{%
k}_{1}\right] \right\} \right) }{\partial \mathbf{\hat{\Lambda}}_{n_{2}}^{%
\left[ \mathbf{k}_{2}\right] }\left[ \Psi _{J},\nu ,\left\{
U_{j}^{l_{i}}\right\} \right] }\mathbf{\hat{\Lambda}}_{n_{1}}^{\left[ 
\mathbf{k}_{1}\right] }\left[ \Psi _{J},\nu ,\left\{ U_{j}^{l_{i}}\right\} %
\right] +\hat{U}\left( \left\{ \left[ \mathbf{k}_{1}^{\prime }\right] ,\left[
\mathbf{k}_{2}\right] \right\} \right) \\
&=&\frac{\partial \left( \Psi _{J}^{\otimes \sum_{i}l_{i}}\mathcal{K}%
_{0}\right) }{\partial \mathbf{\hat{\Lambda}}_{n_{2}}^{\left[ \mathbf{k}_{2}%
\right] }\left[ \Psi _{J},\nu ,\left\{ U_{j}^{l_{i}}\right\} \right] }\frac{%
\partial \hat{U}\left( \left\{ \left[ \mathbf{k}_{1}^{\prime }\right] ,\left[
\mathbf{k}_{1}\right] \right\} \right) }{\partial \left( \Psi _{J}^{\otimes
\sum_{i}l_{i}}\mathcal{K}_{0}\right) }\mathbf{\hat{\Lambda}}_{n_{1}}^{\left[ 
\mathbf{k}_{1}\right] }\left[ \Psi _{J},\nu ,\left\{ U_{j}^{l_{i}}\right\} %
\right] +\hat{U}\left( \left\{ \left[ \mathbf{k}_{1}^{\prime }\right] ,\left[
\mathbf{k}_{2}\right] \right\} \right)
\end{eqnarray*}%
Any associated tensor transforms as%
\begin{eqnarray*}
R_{\left[ \mathbf{k}_{1}\right] ,\left[ \mathbf{k}_{2}\right] } &=&\frac{%
\partial \left( \Psi _{J}^{\otimes \sum_{i}l_{i}}\mathcal{K}_{0}\right) }{%
\partial \mathbf{\hat{\Lambda}}_{n_{2}}^{\left[ \mathbf{k}_{2}\right] }\left[
\Psi _{J},\nu ,\left\{ U_{j}^{l_{i}}\right\} \right] }\frac{\partial \hat{U}%
\left( \left\{ \left[ \mathbf{k}_{2}^{\prime }\right] ,\left[ \mathbf{k}_{3}%
\right] \right\} \right) }{\partial \left( \Psi _{J}^{\otimes \sum_{i}l_{i}}%
\mathcal{K}_{0}\right) }\mathbf{\hat{\Lambda}}_{n_{3}}^{\left[ \mathbf{k}%
_{n_{3}}\right] }R_{\left[ \mathbf{k}_{1}^{\prime }\right] ,\left[ \mathbf{k}%
_{2}^{\prime }\right] }^{\prime }\frac{\partial \hat{U}^{-1}\left( \left\{ %
\left[ \mathbf{k}_{4}\right] ,\left[ \mathbf{k}_{1}^{\prime }\right]
\right\} \right) }{\partial \left( \Psi _{J}^{\otimes \sum_{i}l_{i}}\mathcal{%
K}_{0}\right) }\mathbf{\hat{\Lambda}}_{n_{4}}^{\left[ \mathbf{k}_{n_{4}}%
\right] }\frac{\partial \left( \Psi _{J}^{\otimes \sum_{i}l_{i}}\mathcal{K}%
_{0}\right) }{\partial \mathbf{\hat{\Lambda}}_{n_{1}}^{\left[ \mathbf{k}_{1}%
\right] }\left[ \Psi _{J},\nu ,\left\{ U_{j}^{l_{i}}\right\} \right] } \\
&&+\hat{U}\left( \left\{ \left[ \mathbf{k}_{1}^{\prime }\right] ,\left[ 
\mathbf{k}_{1}\right] \right\} \right) R_{\left[ \mathbf{k}_{1}^{\prime }%
\right] ,\left[ \mathbf{k}_{2}^{\prime }\right] }^{\prime }\hat{U}%
^{-1}\left( \left\{ \left[ \mathbf{k}_{2}\right] ,\left[ \mathbf{k}%
_{2}^{\prime }\right] \right\} \right)
\end{eqnarray*}%
it includes the inert contributions that depends on $\mathcal{K}_{0}$
through $\Psi _{J}^{\otimes \sum_{i}l_{i}}\mathcal{K}_{0}$. The tensor $R_{%
\left[ \mathbf{k}_{1}\right] ,\left[ \mathbf{k}_{2}\right] }$ can be seen as
a generalized curvature, involving any set of points.

\subsection{Invariance of constraint by change of parametrization}

We consider the quadratic form $M$ and assume the general form:%
\begin{eqnarray*}
&&M\left( \left\{ \left[ k\right] ,\left[ k\right] \right\} ,\Psi _{J},\nu
,\left\{ \left( \Psi _{J}^{\otimes l_{i}}\left( U_{j}^{l_{i}}\right) ,\bar{H}%
_{m}\left( \Psi _{I,0}^{\otimes k}\right) \right) \right\} \right) \\
&=&\sum_{h_{l_{i}}}h_{l_{i}}\left( \left( U_{j}^{l}\right) ,\left\{ \left[ 
\mathbf{k}_{1}\right] ,\left[ \mathbf{k}_{2}\right] \right\} ,\left\{ \int
G\left( U_{j}^{l^{\prime }}\right) \Psi _{J}^{\otimes l^{\prime }}\left(
U_{j}^{l^{\prime }}\right) dU_{j}^{l^{\prime }}\right\} \right) \Psi
_{J}^{\otimes l_{i}}\left( U_{j}^{l_{i}}\right)
\end{eqnarray*}%
where $\left\{ \int G\left( U_{j}^{l^{\prime }}\right) \Psi _{J}^{\otimes
l^{\prime }}\left( U_{j}^{l^{\prime }}\right) dU_{j}^{l^{\prime }}\right\} $
is a set of aribitrary functionals of the $\Psi _{J}^{\otimes l^{\prime
}}\left( U_{j}^{l^{\prime }}\right) $

If this quadratic form is invariant through:%
\begin{equation*}
\Psi _{J}\rightarrow \mathbf{g.}\Psi _{J}=\Psi _{J}\left( \mathbf{g.}%
U_{j}^{l}\right)
\end{equation*}%
then globally the entire set of parameters $\left\{ \mathbf{\hat{\Lambda}}%
_{\infty ,\alpha }^{\left[ k_{1}\right] }\left[ \Psi _{J},\nu ,\left\{
U_{j}^{l_{i}}\right\} \right] \right\} $ is invariant under :%
\begin{equation*}
\mathbf{g.}\left\{ \mathbf{\hat{\Lambda}}_{\infty ,\alpha }^{\left[ k_{1}%
\right] }\left[ \Psi _{J},\nu ,\left\{ U_{j}^{l_{i}}\right\} \right]
\right\} =\left\{ \mathbf{\hat{\Lambda}}_{\infty ,\alpha }^{\left[ k_{1}%
\right] }\left[ \Psi _{J},\nu ,\left\{ \mathbf{g.}U_{j}^{l_{i}}\right\} %
\right] \right\} =\left\{ \mathbf{\hat{\Lambda}}^{\left[ k_{1}\right] }\left[
\Psi _{J},\nu ,U_{j}^{l}\right] \right\}
\end{equation*}

Locally, assume that there is an action:%
\begin{equation*}
\Psi _{J}\rightarrow \mathbf{g.}\Psi _{J}=\Psi _{J}\left( \mathbf{g.}%
U_{j}^{l}\right)
\end{equation*}%
such that:%
\begin{equation*}
\gamma \left( \Psi _{J}\right)
\end{equation*}%
and the:%
\begin{equation*}
\left\{ \int G\left( U_{j}^{l^{\prime }}\right) \Psi _{J}^{\otimes l^{\prime
}}\left( U_{j}^{l^{\prime }}\right) dU_{j}^{l^{\prime }}\right\}
\end{equation*}%
If there are representations $\emph{R}_{\left[ \mathbf{k}_{1}^{\prime }%
\right] }^{\left[ \mathbf{k}_{1}\right] }$, $\emph{R}_{\left[ \mathbf{k}%
_{2}^{\prime }\right] }^{\left[ \mathbf{k}_{2}\right] }$, such that:%
\begin{eqnarray*}
&&h_{l_{i}}\left( \mathbf{g.}\left( U_{j}^{l}\right) ,\left\{ \left[ \mathbf{%
k}_{1}\right] ,\left[ \mathbf{k}_{2}\right] \right\} ,\left\{ \int G\left(
U_{j}^{l^{\prime }}\right) \Psi _{J}^{\otimes l^{\prime }}\left(
U_{j}^{l^{\prime }}\right) dU_{j}^{l^{\prime }}\right\} \right) \mathbf{g.}%
\Psi _{J}^{\otimes l_{i}}\left( U_{j}^{l_{i}}\right) \\
&=&\sum_{\left[ \mathbf{k}_{1}^{\prime }\right] ,\left[ \mathbf{k}%
_{2}^{\prime }\right] }\emph{R}_{\left[ \mathbf{k}_{1}^{\prime }\right] }^{%
\left[ \mathbf{k}_{1}\right] }\left( \mathbf{g}\right) \mathbf{\times }\emph{%
R}_{\left[ \mathbf{k}_{2}^{\prime }\right] }^{\left[ \mathbf{k}_{2}\right]
}\left( \mathbf{g}\right) h_{l_{i}}\left( \left( U_{j}^{l}\right) ,\left\{ %
\left[ \mathbf{k}_{1}\right] ,\left[ \mathbf{k}_{2}\right] \right\} ,\left\{
\int G\left( U_{j}^{l^{\prime }}\right) \Psi _{J}^{\otimes l^{\prime
}}\left( U_{j}^{l^{\prime }}\right) dU_{j}^{l^{\prime }}\right\} \right)
\Psi _{J}^{\otimes l_{i}}\left( U_{j}^{l_{i}}\right)
\end{eqnarray*}%
we can rewrite the transformed constraints:%
\begin{equation*}
0=\gamma \left( \Psi _{J},\nu \right) +\left\{ \emph{R}_{\left[ \mathbf{k}%
_{1}^{\prime }\right] }^{\left[ \mathbf{k}_{1}\right] }\mathbf{\hat{\Lambda}}%
_{n_{1}}^{\left[ \mathbf{k}_{1}^{\prime }\right] }\left[ \Psi _{J},\nu
,\left\{ U_{j}^{l_{i}}\right\} \right] \right\} G\left( \left\{ \left[ 
\mathbf{k}_{1}\right] ,\left[ \mathbf{k}_{2}\right] \right\} ,\Psi _{J},\Psi
_{J}^{\otimes l}\left( U_{j}^{l}\right) \right) \left\{ \emph{R}_{\left[ 
\mathbf{k}_{2}^{\prime }\right] }^{\left[ \mathbf{k}_{2}\right] }\mathbf{%
\hat{\Lambda}}_{n_{2}}^{\left[ \mathbf{k}_{2}^{\prime }\right] }\left[ \Psi
_{J},\nu ,\left\{ U_{j}^{l_{i}}\right\} \right] \right\}
\end{equation*}%
The representation $\emph{R}_{\left[ \mathbf{k}_{1}^{\prime }\right] }^{%
\left[ \mathbf{k}_{1}\right] }$ is a symmetry of parameter space:%
\begin{equation*}
\mathbf{\hat{\Lambda}}_{n_{1}}^{\left[ \mathbf{k}_{1}^{\prime }\right] }%
\left[ \Psi _{J},\nu ,\left\{ U_{j}^{l_{i}}\right\} \right] \rightarrow 
\emph{R}_{\left[ \mathbf{k}_{1}^{\prime }\right] }^{\left[ \mathbf{k}_{1}%
\right] }\mathbf{\hat{\Lambda}}_{n_{1}}^{\left[ \mathbf{k}_{1}^{\prime }%
\right] }\left[ \Psi _{J},\nu ,\left\{ U_{j}^{l_{i}}\right\} \right]
\end{equation*}
and the function $G\left( \left\{ \left[ \mathbf{k}_{1}\right] ,\left[ 
\mathbf{k}_{2}\right] \right\} ,\Psi _{J},\Psi _{J}^{\otimes l}\left(
U_{j}^{l}\right) \right) $ transforms as:%
\begin{equation*}
G\left( \left\{ \left[ \mathbf{k}_{1}\right] ,\left[ \mathbf{k}_{2}\right]
\right\} ,\Psi _{J},\Psi _{J}^{\otimes l}\left( U_{j}^{l}\right) \right)
\rightarrow \emph{R}_{\left[ \mathbf{k}_{1}^{\prime }\right] }^{\left[ 
\mathbf{k}_{1}\right] }G\left( \left\{ \left[ \mathbf{k}_{1}\right] ,\left[ 
\mathbf{k}_{2}\right] \right\} ,\Psi _{J},\Psi _{J}^{\otimes l}\left(
U_{j}^{l}\right) \right) \emph{R}_{\left[ \mathbf{k}_{2}^{\prime }\right] }^{%
\left[ \mathbf{k}_{2}\right] }
\end{equation*}

\section{Operators perspective: Average values on eigenstates of operators}

We present an equivalent description to the one developed in the previous
section. We begin with functional states and project onto the lowest
eigenspace of operators acting on the state space of $\Psi _{I,\alpha
}^{\otimes k}\left( U_{i}^{k}\right) $. This operator depends on the states
of $\Psi _{J}^{\otimes l}\left( U_{j}^{l}\right) $ and we denote $\lambda
\left( \Psi _{J}^{\otimes k}\right) $ the eigenvalues. Degeneracies induce
as before parameters $\left\{ \underline{\mathbf{\hat{\Lambda}}}_{\infty
,\alpha }^{\left[ k_{i}\right] }\left[ \Psi _{J},\nu ,\left\{
U_{j}^{l_{i}}\right\} \right] \right\} _{i}$, and fields depending on these
parameters. The projection amounts to comput averages over these
eigenstates. The projection involves computing averages over these
eigenstates. This approach has the particularity of dividing the state
spaces into slices with respect to the eigenvalues. These spaces are not
orthogonal to each other, as different eigenvalues correspond to different
operators depending on different states from $\Psi _{J,\alpha }^{\otimes
l}\left( U_{j}^{l}\right) $.

Then, we can define states and operators acting on these subspaces. For a
given eigenvalue, we recover a similar formalism to the states and operators
Hamiltonian formalism. For a given $\lambda \left( \Psi _{J}^{\otimes
k}\right) $, we can locally describe the degrees of freedom for a state and
an operator by some functional of $\Psi _{J}^{\otimes k}$ and some $\left\{ 
\underline{\mathbf{\hat{\Lambda}}}_{\infty ,\alpha }^{\left[ k_{i}\right] }%
\left[ \Psi _{J},\nu ,\left\{ U_{j}^{l_{i}}\right\} \right] \right\} _{i}$.
States and operators have internal degrees of freedom defined on some
parameter space at $\lambda \left( \Psi _{J}^{\otimes k}\right) $. We can
then define transitions between states for infinitesimally different
eigenvalues $\lambda \left( \Psi _{J}^{\otimes k}\right) $ and $\lambda
\left( \Psi _{J}^{\otimes k}\right) +\delta \lambda \left( \Psi
_{J}^{\otimes k}\right) $, that is, transition operators between the
different eigenspaces. Summing such transformation defines transitions
between state spaces with different eigenvalues.

\subsection{Principle}

We start with the functionals without projection:

\begin{equation}
\sum_{\alpha }\int a_{l,k}\left( U_{j}^{l},U_{i}^{k}\right) \Psi _{J,\alpha
}^{\otimes l}\left( U_{j}^{l}\right) \Psi _{I,\alpha }^{\otimes k}\left(
U_{i}^{k}\right) \delta \left( f_{lk}\left( U_{j}^{l},U_{i}^{k}\right)
\right) dU_{j}^{l}dU_{i}^{k}  \label{FCL}
\end{equation}%
and as above replace:%
\begin{equation*}
\Psi _{J,\alpha }^{\otimes l}\left( U_{j}^{l}\right)
\end{equation*}%
by:%
\begin{equation*}
\Psi _{J,\alpha }^{\otimes l}\left( U_{j}^{l}/f_{lk},\left\{
U_{i}^{k}\right\} \right)
\end{equation*}%
to parametrize the functional by the $\left\{ U_{i}^{k}\right\} $ nd rsdual
parameters. Then, we will replace $\Psi _{I,\alpha }^{\otimes k}\left(
U_{i}^{k}\right) $ in (\ref{FCL}) by its the average on the minimal
eigenstates of an operator:%
\begin{equation*}
\mathcal{H}\left( \left\{ \Psi _{I,\alpha }^{\otimes k}\left(
U_{i}^{k}\right) ,\Pi _{\Psi _{I,\alpha }^{\otimes k}\left( U_{i}^{k}\right)
}\right\} _{k},\left\{ \Psi _{J,\alpha }^{\otimes l}\left(
U_{j}^{l}/f_{lk},\left\{ U_{i}^{k}\right\} \right) \right\} _{l}\right)
\end{equation*}%
with:%
\begin{equation*}
\Pi _{\Psi _{I,\alpha }^{\otimes k}\left( U_{i}^{k}\right) }=\frac{\delta }{%
\delta \Psi _{I,\alpha }^{\otimes k}\left( U_{i}^{k}\right) }
\end{equation*}%
This is a similar approach to the previous one, but it replaces the
formalism of saddle point by some projection over background eigenstates.

\subsection{Average over eigenspaces}

If the operator considered has only one eigenstate, this one has the form:%
\begin{equation}
F_{0}\left[ \left\{ \Psi _{I,\alpha }^{\otimes k_{i}}\right\} ,\left\{ \Psi
_{J,\alpha }^{\otimes l_{i}}\right\} \right]  \label{FT}
\end{equation}%
state functional of $\Psi _{I,\alpha }^{\otimes k}$ depending on a given $%
\Psi _{J,\alpha }^{\otimes l}$. The functional includes the contraints
between $\Psi _{I,\alpha }^{\otimes k}$ and $\Psi _{J,\alpha }^{\otimes
l_{i}}$.

In general including explicitly the constraint :%
\begin{equation*}
\Psi _{I}^{\otimes k}\left( U_{i}^{k}\right) \delta \left( f_{lk}\left(
U_{j}^{l_{i}},U_{i}^{k}\right) \right)
\end{equation*}%
state (\ref{FT}) writes:%
\begin{equation*}
F_{0}\left[ \left\{ \Psi _{I,\alpha }^{\otimes k_{i}}\right\} ,\left\{ \Psi
_{J,\alpha }^{\otimes l_{i}}\right\} \right] \equiv \int F_{0}\left[ \left(
\left\{ U_{i}^{k_{i}}\right\} \right) ,\left\{ \Psi _{J,\alpha }^{\otimes
l_{i}}\right\} ,U_{i}^{k_{i}}\right] \otimes _{i}\Psi _{I,\alpha }^{\otimes
k_{i}}\left( U_{i}^{k_{i}}\right) \delta \left( f_{lk_{i}}\left(
U_{j}^{l},U_{i}^{k_{i}}\right) \right) dU_{i}^{k_{i}}
\end{equation*}%
As before, the tensor $\otimes $ stands for series of products of identical
copies of fields, or in terms of states series of products of realizations.
The dependency in $U_{i}^{k_{i}}$ depicts the constraints between $\left\{
\Psi _{J,\alpha }^{\otimes l_{i}}\right\} $ and $\Psi _{I,\alpha }^{\otimes
k_{i}}\left( U_{i}^{k_{i}}\right) $. This rewrites as series expansion:%
\begin{eqnarray*}
&&F_{0}\left[ \left\{ \Psi _{I,\alpha }^{\otimes k_{i}}\right\} ,\left\{
\Psi _{J,\alpha }^{\otimes l_{i}}\right\} \right] \\
&=&\sum_{s,l_{1},...,l_{s}}d\left( \left\{ U_{j}^{l_{i}}\right\}
_{l_{i}}\right) d\left( \left\{ U_{i}^{k_{i}}/f_{k_{i}l_{i}}\right\}
_{k_{i},l_{i}}\right) F_{0}^{\left( s\right) }\left[ \left\{
U_{j}^{l_{i}}\right\} ,\left\{ U_{i}^{k_{i}}/f_{k_{i}l_{i}}\right\} \right]
\tprod_{i}\Psi _{J}^{\otimes l_{i}}\left( U_{j}^{l_{i}}\right) \otimes
_{i}\Psi _{I,\alpha }^{\otimes k_{i}}\left( U_{i}^{k_{i}}\right)
\end{eqnarray*}

Including symetries and degeneracy amounts to introdce parameters $\left\{ 
\underline{\mathbf{\hat{\Lambda}}}_{\infty ,\alpha }^{\left[ k_{i}\right] }%
\left[ \Psi _{J},\nu ,\left\{ U_{j}^{l_{i}}\right\} \right] \right\} _{i}$
and given a state $\nu $ for $\Psi _{J,\alpha }^{\otimes l}$, th states $%
F_{0}\left[ \left\{ \Psi _{I,\alpha }^{\otimes k_{i}}\right\} ,\left\{ \Psi
_{J,\alpha }^{\otimes l_{i}}\right\} \right] $ becomes:%
\begin{eqnarray}
&&F_{0}^{\underline{\mathbf{\hat{\Lambda}}}}\left[ \left\{ \Psi _{I,\alpha
}^{\otimes k_{i}}\right\} ,\left\{ \Psi _{J,\alpha }^{\otimes l_{i}}\right\}
,v\right]  \label{FR} \\
&\equiv &\sum_{s,l_{1},...,l_{s}}d\left( \left\{ U_{j}^{l_{i}}\right\}
_{l_{i}}\right) d\left( \left\{ U_{i}^{k_{i}}/f_{k_{i}l_{i}}\right\}
_{k_{i},l_{i}}\right) F_{0}^{\left( s\right) }\left[ \left\{
U_{j}^{l_{i}}\right\} ,\left\{ U_{i}^{k_{i}}/f_{k_{i}l_{i}}\right\} \right]
\tprod_{i}\Psi _{J}^{\otimes l_{i}}\left( U_{j}^{l_{i}}\right) \otimes
_{i}\Psi _{I,\alpha }^{\mathbf{\hat{\Lambda}}^{-1}\otimes k_{i}}\left(
U_{i}^{k_{i}}\right)  \notag
\end{eqnarray}%
where $\Psi _{I,\alpha }^{\mathbf{\hat{\Lambda}}^{-1}\otimes k_{i}}$ are the
transformd fields induced by the inverse transformation $\mathbf{\hat{\Lambda%
}}^{-1}$. The average values of $\Psi _{I}^{\otimes k}\left(
U_{i}^{k}\right) $ in these states are:%
\begin{equation*}
\left\langle \Psi _{I}^{\otimes k}\left( U_{i}^{k}\right) \right\rangle _{%
\mathbf{\hat{\Lambda}}}=\int \Psi _{I}^{\otimes k}\left( U_{i}^{k}\right)
\left\vert F_{0}^{\mathbf{\hat{\Lambda}}}\left[ \left\{ \Psi _{I,\alpha
}^{\otimes k_{i}}\right\} ,\left\{ \Psi _{J,\alpha }^{\otimes l_{i}}\right\}
,v\right] \right\vert ^{2}\tprod \mathcal{D}\left\{ \Psi _{I,\alpha
}^{\otimes k_{i}}\right\}
\end{equation*}%
Then, using a change of variabl:%
\begin{equation*}
\Psi _{I,\alpha }^{\mathbf{\hat{\Lambda}}^{-1}\otimes k_{i}}\rightarrow \Psi
_{I,\alpha }^{\otimes k_{i}}
\end{equation*}%
we find ultimately the averages:%
\begin{equation}
\left\langle \Psi _{I}^{\otimes k}\left( U_{i}^{k}\right) \right\rangle _{%
\mathbf{\hat{\Lambda}}}=\left\langle \Psi _{I}^{\mathbf{\hat{\Lambda}}%
\otimes k}\left( U_{i}^{k}\right) \right\rangle _{\mathbf{\hat{\Lambda}}%
}\equiv v_{I}^{\otimes k}\left( U_{i}^{k},\left\{ \Psi _{J,\alpha }^{\otimes
l_{i}}\right\} ,\underline{\mathbf{\hat{\Lambda}}}_{\infty ,\alpha }^{\left[
k\right] }\left[ \Psi _{J},\nu ,\left\{ U_{j}^{l_{i}}\right\} \right] \right)
\label{VRF}
\end{equation}

We rewrite the functional (\ref{FCL}) using (\ref{VRF}) in appendix 5. As
before if the set $\underline{\mathbf{\hat{\Lambda}}}_{\infty ,\alpha }^{%
\left[ k\right] }\left[ \Psi _{J},\nu ,\left\{ U_{j}^{l_{i}}\right\} \right] 
$ are independent of the realization, it becomes:%
\begin{eqnarray}
&&F_{f,lin}\left( \left\{ \Psi _{J}^{\otimes l}\left( U_{j}^{l}\right)
\right\} _{l}\right)  \label{PTF} \\
&=&\int \bar{g}\left( \left\{ U_{j}^{l_{i}}\right\} ,\underline{\mathbf{\hat{%
\Lambda}}}_{\infty }^{\left[ k\right] }\left[ \Psi _{J},\nu ,\left\{
U_{j}^{l_{i}}\right\} \right] ,\lambda \left( \Psi _{J}^{\otimes k}\right)
\right) \Psi _{J}^{\otimes \sum l_{i}}\left( \left\{ U_{j}^{l_{i}}\right\}
_{i},\underline{\mathbf{\hat{\Lambda}}}_{\infty }^{\left[ k\right] }\left[
\Psi _{J},\nu ,\left\{ U_{j}^{l_{i}}\right\} \right] ,\lambda \left( \Psi
_{J}^{\otimes k}\right) ,v\right)  \notag
\end{eqnarray}%
where the expression for $\Psi _{J}^{\otimes \sum l_{i}}$ and $\bar{g}$ are
given in appendix 5. The parametr $\lambda \left( \Psi _{J}^{\otimes
k}\right) $ is the genval considered. The function $\bar{g}$ is an averaged
functional.

Appendix 5 also provides expressions for projected functionals over space of
vector-valued eigenvalues. Functional (\ref{PTF}) is similar to previous
section and involves the effective field:%
\begin{equation}
\Psi _{J}^{\otimes \sum l_{i}}\left( \left\{ U_{j}^{l_{i}}\right\} _{i},%
\underline{\mathbf{\hat{\Lambda}}}_{\infty }^{\left[ k\right] }\left[ \Psi
_{J},\nu ,\left\{ U_{j}^{l_{i}}\right\} \right] ,\lambda \left( \Psi
_{J}^{\otimes k}\right) ,v\right)  \label{FCV}
\end{equation}%
the difference with the previous sections is that parameters $\mathbf{\hat{%
\Lambda}}_{\infty }^{\left[ k\right] }\left[ \Psi _{J},\nu ,\left\{
U_{j}^{l_{i}}\right\} \right] $ have been decomposed in $\underline{\mathbf{%
\hat{\Lambda}}}_{\infty }^{\left[ k\right] }\left[ \Psi _{J},\nu ,\left\{
U_{j}^{l_{i}}\right\} \right] $ and $\lambda \left( \Psi _{J}^{\otimes
k}\right) $.

\subsection{\textbf{Variations}}

Once the effective field (\ref{FCV}) is obtained, the same principles as in
the first approach apply. The independenc t frst rdr f parameters with
respect to $\Psi _{J}^{\otimes \sum l_{i}}$ yields the following equation,
similar to (\ref{NR}) where the role of $\lambda \left( \Psi _{J}^{\otimes
k}\right) $ is distinguished:%
\begin{eqnarray}
0 &=&\int v\left( U_{i}^{l}\right) \frac{\delta \mathbf{\hat{\Lambda}}%
_{\infty }^{\left[ k\right] }\left[ \Psi _{J},\nu ,\left\{
U_{j}^{l_{i}}\right\} \right] }{\delta \Psi _{J}^{\otimes l_{i}}\left(
\left( U_{j}^{l_{i}}\right) ^{\prime }\right) }\nabla _{\mathbf{\hat{\Lambda}%
}_{\infty }^{\left[ k\right] }}\Psi _{J}^{\otimes \left(
\sum_{i}l_{i}\right) }\left( \left\{ U_{j}^{l_{i}}\right\} _{i},\underline{%
\mathbf{\hat{\Lambda}}}_{\infty }^{\left[ k\right] }\left[ \Psi _{J},\nu
,\left\{ U_{j}^{l_{i}}\right\} \right] ,\lambda \left( \Psi _{J}^{\otimes
k}\right) ,v\right)  \label{NS} \\
&&+\int v\left( U_{i}^{l}\right) \frac{\delta \mathbf{\hat{\Lambda}}_{\infty
}^{\left[ k\right] }\left[ \Psi _{J},\nu ,\left\{ U_{j}^{l_{i}}\right\} %
\right] }{\delta \Psi _{J}^{\otimes l_{i}}\left( \left( U_{j}^{l_{i}}\right)
^{\prime }\right) }\nabla _{\lambda \left( \Psi _{J}^{\otimes k}\right)
}\Psi _{J}^{\otimes \left( \sum_{i}l_{i}\right) }\left( \left\{
U_{j}^{l_{i}}\right\} _{i},\underline{\mathbf{\hat{\Lambda}}}_{\infty }^{%
\left[ k\right] }\left[ \Psi _{J},\nu ,\left\{ U_{j}^{l_{i}}\right\} \right]
,\lambda \left( \Psi _{J}^{\otimes k}\right) ,v\right)  \notag
\end{eqnarray}%
and if averaged fields are considered, we find an equation similar to (\ref%
{BNF}):%
\begin{eqnarray}
0 &=&\int v\left( \left( U_{j}^{l_{i}}\right) \right) \frac{\delta \mathbf{%
\hat{\Lambda}}^{\left[ k\right] }\left[ \Psi _{J},\nu ,U_{j}^{l}\right] }{%
\delta \Psi _{J}^{\otimes l_{i}}\left( \left( U_{j}^{l_{i}}\right) ^{\prime
}\right) }\nabla _{\mathbf{\hat{\Lambda}}^{\left[ k\right] }}\Psi
_{J}^{\otimes \left( \sum_{i}l_{i}\right) }\left( \left\{
U_{j}^{l_{i}}\right\} _{i},\underline{\mathbf{\hat{\Lambda}}}^{\left[ k%
\right] }\left[ \Psi _{J},\nu ,\left\{ U_{j}^{l_{i}}\right\} \right]
,\lambda \left( \Psi _{J}^{\otimes k}\right) ,\underline{V},v\right)
\label{BNG} \\
&&+\int v\left( \left( U_{j}^{l_{i}}\right) \right) \frac{\delta \mathbf{%
\hat{\Lambda}}^{\left[ k\right] }\left[ \Psi _{J},\nu ,U_{j}^{l}\right] }{%
\delta \Psi _{J}^{\otimes l_{i}}\left( \left( U_{j}^{l_{i}}\right) ^{\prime
}\right) }\nabla _{\mathbf{\hat{\Lambda}}^{\left[ k\right] }}\Psi
_{J}^{\otimes \left( \sum_{i}l_{i}\right) }\left( \left\{
U_{j}^{l_{i}}\right\} _{i},\underline{\mathbf{\hat{\Lambda}}}^{\left[ k%
\right] }\left[ \Psi _{J},\nu ,\left\{ U_{j}^{l_{i}}\right\} \right]
,\lambda \left( \Psi _{J}^{\otimes k}\right) ,\underline{V},v\right)  \notag
\\
&&+\int v\left( \left( U_{j}^{l_{i}}\right) \right) \left( \epsilon \frac{%
\delta \underline{V}}{\delta \Psi _{J}^{\otimes l_{i}}\left( \left(
U_{j}^{l_{i}}\right) ^{\prime }\right) }-h\left( \underline{V},\Psi
_{J}^{\otimes l_{i}}\left( \left( U_{j}^{l_{i}}\right) ^{\prime }\right)
\right) \right) \Psi _{J}^{\otimes \left( \sum_{i}l_{i}\right) }\left(
\left\{ U_{j}^{l_{i}}\right\} _{i},\underline{\mathbf{\hat{\Lambda}}}^{\left[
k\right] }\left[ \Psi _{J},\nu ,\left\{ U_{j}^{l_{i}}\right\} \right] ,%
\underline{V},v\right)  \notag
\end{eqnarray}%
where $\underline{V}$ is similar to $V$ in the derivation of (\ref{BNF}) but
with the parameters restricted to $,\underline{\mathbf{\hat{\Lambda}}}%
_{\infty }^{\left[ k\right] }\left[ \Psi _{J},\nu ,\left\{
U_{j}^{l_{i}}\right\} \right] $.

\subsection{Fixed $\protect\lambda \left( \Psi _{J}\right) $ slices}

We consider the projection of the constraint on background states such that,
at the lowest order: 
\begin{eqnarray*}
0 &=&\left( \left\{ \underline{\mathbf{\hat{\Lambda}}}_{n_{1}}^{\left[ 
\mathbf{k}_{1}\right] }\left[ \Psi _{J},\nu ,\left\{ U_{j}^{l_{i}}\right\} %
\right] \right\} M\left( \left\{ \left[ \mathbf{k}_{1}\right] ,\left[ 
\mathbf{k}_{2}\right] \right\} ,\Psi _{J},\Psi _{J}^{\otimes l}\left(
U_{j}^{l}\right) \right) .\left\{ \mathbf{\hat{\Lambda}}_{n_{2}}^{\left[ 
\mathbf{k}_{2}\right] }\left[ \Psi _{J},\nu ,\left\{ U_{j}^{l_{i}}\right\} %
\right] \right\} \right. \\
&&+N\left\{ \left[ \mathbf{k}_{1}\right] \right\} ,\Psi _{J},\Psi
_{J}^{\otimes l}\left( U_{j}^{l}\right) .\left\{ \underline{\mathbf{\hat{%
\Lambda}}}_{n_{1}}^{\left[ \mathbf{k}_{1}\right] }\left[ \Psi _{J},\nu
,\left\{ U_{j}^{l_{i}}\right\} \right] \right\} +\alpha \left( \Psi
_{J}\right) -\lambda \left( \Psi _{J}\right)
\end{eqnarray*}%
implements $\lambda \left( \Psi _{J}\right) $ bounded from below. A change
of variable leads to:

\begin{equation*}
\left\{ \underline{\mathbf{\hat{\Lambda}}}_{n_{1}}^{\left[ \mathbf{k}_{1}%
\right] }\left[ \Psi _{J},\nu ,\left\{ U_{j}^{l_{i}}\right\} \right]
\right\} M\left( \left\{ \left[ \mathbf{k}_{1}\right] ,\left[ \mathbf{k}_{2}%
\right] \right\} ,\Psi _{J},\Psi _{J}^{\otimes l}\left( U_{j}^{l}\right)
\right) .\left\{ \underline{\mathbf{\hat{\Lambda}}}_{n_{2}}^{\left[ \mathbf{k%
}_{2}\right] }\left[ \Psi _{J},\nu ,\left\{ U_{j}^{l_{i}}\right\} \right]
\right\} +\alpha \left( \Psi _{J}\right) -\lambda \left( \Psi _{J}\right) =0
\end{equation*}%
This can be also written in a form where the parameters are not
distinguished:%
\begin{equation*}
\left( \underline{\mathbf{\hat{\Lambda}}}_{n_{1}}^{\left[ \mathbf{k}_{1}%
\right] },\lambda \left( \Psi _{J}\right) \right)
\end{equation*}%
by a change of variable:%
\begin{eqnarray*}
V_{\left[ \mathbf{k}_{1}^{\prime }\right] }^{\left[ k_{0}\right] }\mathbf{%
\hat{\Lambda}}_{n_{1}}^{\left[ \mathbf{k}_{1}^{\prime }\right] }\left[ \Psi
_{J},\nu ,\left\{ U_{j}^{l_{i}}\right\} \right] &\rightarrow &\lambda \left(
\Psi _{J}\right) \\
V_{\left[ \mathbf{k}_{1}^{\prime }\right] }^{\left[ \mathbf{k}_{1}\right]
/k_{0}}\mathbf{\hat{\Lambda}}_{n_{1}}^{\left[ \mathbf{k}_{1}^{\prime }\right]
}\left[ \Psi _{J},\nu ,\left\{ U_{j}^{l_{i}}\right\} \right] &\rightarrow &%
\underline{\mathbf{\hat{\Lambda}}}_{n_{1}}^{\left[ \mathbf{k}_{1}\right] }%
\left[ \Psi _{J},\nu ,\left\{ U_{j}^{l_{i}}\right\} \right]
\end{eqnarray*}%
leading to:%
\begin{equation}
\underline{\mathbf{\hat{\Lambda}}}_{n_{1}}^{\left[ \mathbf{k}_{1}\right] }%
\left[ \Psi _{J},\nu ,\left\{ U_{j}^{l_{i}}\right\} \right] M\left( \left\{ %
\left[ \mathbf{k}_{1}\right] ,\left[ \mathbf{k}_{2}\right] \right\} ,\Psi
_{J},\Psi _{J}^{\otimes l}\left( U_{j}^{l}\right) \right) .\underline{%
\mathbf{\hat{\Lambda}}}_{n_{2}}^{\left[ \mathbf{k}_{2}\right] }\left[ \Psi
_{J},\nu ,\left\{ U_{j}^{l_{i}}\right\} \right] +\alpha \left( \Psi
_{J}\right) -\lambda \left( \Psi _{J}\right) =0  \label{CTR}
\end{equation}

Recall that $\alpha \left( \Psi _{J}\right) $ is a functional of $\Psi _{J}$%
: 
\begin{equation*}
\alpha \left( \Psi _{J}\right) =\int \Psi _{J}^{\dag }A\Psi _{J}+\int \Psi
_{J}^{\dag }\Psi _{J}^{\dag }B\Psi _{J}\Psi _{J}+...
\end{equation*}%
and note for later purpose that the number of states saisfying the
constraint (\ref{CTR}) increases with $\lambda $. since the quadratic
relation involves tensor products of states. The higher the quadratic
quantity, the higher the number of states satifying the equation: 
\begin{equation*}
\sharp \left\{ \underline{\mathbf{\hat{\Lambda}}}_{n_{1}}^{\left[ \mathbf{k}%
_{1}\right] }\left[ \Psi _{J},\nu ,\left\{ U_{j}^{l_{i}}\right\} \right]
M\left( \left\{ \left[ \mathbf{k}_{1}\right] ,\left[ \mathbf{k}_{2}\right]
\right\} ,\Psi _{J},\Psi _{J}^{\otimes l}\left( U_{j}^{l}\right) \right) .%
\underline{\mathbf{\hat{\Lambda}}}_{n_{2}}^{\left[ \mathbf{k}_{2}\right] }%
\left[ \Psi _{J},\nu ,\left\{ U_{j}^{l_{i}}\right\} \right] +\alpha \left(
\Psi _{J}\right) =\lambda \right\} \nearrow \text{ fr }\alpha \left( \Psi
_{J}\right) \nearrow
\end{equation*}

\section{States and operators}

We detail the states on the subspaces defined by parameters $\left( 
\underline{\mathbf{\hat{\Lambda}}}_{\infty }^{\left[ k\right] }\left[ \Psi
_{J},\nu ,\left\{ U_{j}^{l_{i}}\right\} \right] ,\lambda \left( \Psi
_{J}^{\otimes k}\right) \right) $. This will allow to compute transitions
between those spaces.

\subsection{Functionals for projected fields}

The states after projections are described by fields of the type:%
\begin{equation*}
\int v\left( \left( U_{j}^{l_{i}}\right) \right) \Psi _{J}^{\otimes \sum
l_{i}}\left( \left\{ U_{j}^{l_{i}}\right\} ,\underline{\mathbf{\hat{\Lambda}}%
}_{\infty }^{\left[ k\right] }\left[ \Psi _{J},\nu ,\left\{
U_{j}^{l_{i}}\right\} \right] ,\lambda \left( \Psi _{J}^{\otimes k}\right)
,v\right) d\left( U_{j}^{l_{i}}\right)
\end{equation*}%
and the values $\underline{\mathbf{\hat{\Lambda}}}^{\left[ k_{i}\right] }$
nd $\lambda \left( \Psi _{J}^{\otimes k}\right) $ are functionals of the
field and state $\nu $. Note that: 
\begin{equation*}
\left( \underline{\mathbf{\hat{\Lambda}}}_{\infty }^{\left[ k\right] }\left[
\Psi _{J},\nu ,\left\{ U_{j}^{l_{i}}\right\} \right] ,\lambda \left( \Psi
_{J}^{\otimes k}\right) \right) \equiv \mathbf{\hat{\Lambda}}_{\infty }^{%
\left[ k\right] }\left[ \Psi _{J},\nu ,\left\{ U_{j}^{l_{i}}\right\} \right]
\end{equation*}%
For functionals, we replace $v\left( \left( U_{j}^{l_{i}}\right) \right) $:%
\begin{equation*}
v\left( \left( U_{j}^{l_{i}}\right) \right) \rightarrow v\left( \left(
U_{j}^{l_{i}},\underline{\mathbf{\hat{\Lambda}}}_{\infty }^{\left[ k\right] }%
\left[ \Psi _{J},\nu ,\left\{ U_{j}^{l_{i}}\right\} \right] \right) ,\lambda
\left( \Psi _{J}^{\otimes k}\right) \right)
\end{equation*}%
We assume that a change of variable in intgrals leaves unchanged the
boundary condition and we can replace:%
\begin{equation*}
\underline{\mathbf{\hat{\Lambda}}}_{\infty }^{\left[ k\right] }\left[ \Psi
_{J},\nu ,\left\{ U_{j}^{l_{i}}\right\} \right] \rightarrow \underline{%
\mathbf{\hat{\Lambda}}}_{\infty }^{\left[ k\right] }\left[ \Psi _{J},\nu %
\right]
\end{equation*}%
so that a general functional writes: 
\begin{eqnarray}
&&\sum_{v}\int v\left( \left( U_{j}^{l_{i}},\right) ,\underline{\mathbf{\hat{%
\Lambda}}}_{\infty }^{\left[ k\right] }\left[ \Psi _{J},\nu ,\left\{
U_{j}^{l_{i}}\right\} \right] ,\lambda \left( \Psi _{J}^{\otimes k}\right)
\right)  \label{fnc} \\
&&\times g\left( v,\underline{\mathbf{\hat{\Lambda}}}_{\infty }^{\left[ k%
\right] }\left[ \Psi _{J},\nu ,\left\{ U_{j}^{l_{i}}\right\} \right]
,\lambda \left( \Psi _{J}^{\otimes k}\right) \right)  \notag \\
&&\Psi _{J}^{\otimes \sum l_{i}}\left( \left\{ U_{j}^{l_{i}}\right\} ,%
\underline{\mathbf{\hat{\Lambda}}}_{\infty }^{\left[ k\right] }\left[ \Psi
_{J},\nu ,\left\{ U_{j}^{l_{i}}\right\} \right] ,\lambda \left( \Psi
_{J}^{\otimes k}\right) ,v\right) \times d\underline{\mathbf{\hat{\Lambda}}}%
_{\infty }^{\left[ k\right] }\left[ \Psi _{J},\nu ,\left\{
U_{j}^{l_{i}}\right\} \right] d\left( U_{j}^{l_{i}}\right)  \notag \\
&=&\sum_{v}\int v\left( \left( U_{j}^{l_{i}},\right) ,\underline{\mathbf{%
\hat{\Lambda}}}_{\infty }^{\left[ k\right] }\left[ \Psi _{J},\nu \right]
,\lambda \left( \Psi _{J}^{\otimes k}\right) \right)  \notag \\
&&g\left( v,\underline{\mathbf{\hat{\Lambda}}}_{\infty }^{\left[ k\right] }%
\left[ \Psi _{J},\nu \right] ,\lambda \left( \Psi _{J}^{\otimes k}\right)
\right) \Psi _{J}^{\otimes \sum l_{i}}\left( \left\{ U_{j}^{l_{i}}\right\} ,%
\underline{\mathbf{\hat{\Lambda}}}_{\infty }^{\left[ k\right] }\left[ \Psi
_{J},\nu \right] ,\lambda \left( \Psi _{J}^{\otimes k}\right) ,v\right) d%
\underline{\mathbf{\hat{\Lambda}}}_{\infty }^{\left[ k\right] }\left[ \Psi
_{J},\nu \right] d\left( U_{j}^{l_{i}}\right)  \notag
\end{eqnarray}%
This formula describes functionals of the field:%
\begin{eqnarray}
&&\Psi _{J}^{\otimes \sum l_{i}}\left( \underline{\mathbf{\hat{\Lambda}}}%
_{\infty }^{\left[ k\right] }\left[ \Psi _{J},\nu \right] ,\lambda \left(
\Psi _{J}^{\otimes k}\right) ,v\right)  \label{FV} \\
&=&\int v\left( \left( U_{j}^{l_{i}},\right) ,\underline{\mathbf{\hat{\Lambda%
}}}_{\infty }^{\left[ k\right] }\left[ \Psi _{J},\nu \right] ,\lambda \left(
\Psi _{J}^{\otimes k}\right) \right) \Psi _{J}^{\otimes \sum l_{i}}\left(
\left\{ U_{j}^{l_{i}}\right\} ,\underline{\mathbf{\hat{\Lambda}}}_{\infty }^{%
\left[ k\right] }\left[ \Psi _{J},\nu \right] ,\lambda \left( \Psi
_{J}^{\otimes k}\right) ,v\right) d\left( U_{j}^{l_{i}}\right)  \notag
\end{eqnarray}%
This effective field will be the bss to describe the states on subspaces
defined by $\lambda \left( \Psi _{J}^{\otimes k}\right) $.

\subsection{States associated to functionals}

This description can be translated in terms of states for the field (\ref{FV}%
). Functionals (\ref{fnc}) are equivalent to states:%
\begin{equation*}
\left\vert \nu ,\left\{ \underline{\mathbf{\hat{\Lambda}}}_{\infty }^{\left[
k\right] }\left[ \Psi _{J},\nu \right] \right\} _{k},\lambda \left( \nu
\right) \right\rangle
\end{equation*}%
with $v$ is a functional state of $\Psi _{J}$. in the sequel, we omit the
subscript $k$ in $\left\{ \underline{\mathbf{\hat{\Lambda}}}_{\infty }^{%
\left[ k\right] }\left[ \Psi _{J},\nu \right] \right\} _{k}$.

Assume such state $v$ can be parametrized through some parameters space $%
U^{\left( j\right) }$. The projected states become:%
\begin{equation*}
\left\vert U^{\left( j\right) },\left\{ \underline{\mathbf{\hat{\Lambda}}}%
_{\infty }^{\left[ k\right] }\left[ U^{\left( j\right) }\right] \right\}
,\lambda \left( U^{\left( j\right) }\right) \right\rangle
\end{equation*}%
where the parameters $U^{\left( j\right) }$ are analogous to some $\cup
_{j}U_{j}^{l_{j}}$. For example, any functional:%
\begin{equation*}
v\left( \Psi _{J}\right) =\sum_{t}v_{t}\left( \Psi _{J}\right) =\sum_{t}\int
a_{t}\left( U_{j_{1}},...,U_{j_{t}}\right) \Psi _{J}\left(
U_{j_{1}},...,U_{j_{t}}\right)
\end{equation*}%
where the $U_{j_{k}}$ are the components of $U_{j}$, can be parametrized by $%
\left\{ a_{t}\left( U_{j_{1}},...,U_{j_{t}}\right) \right\} _{t}$ \ and thus
we can choose:%
\begin{equation*}
U^{\left( j\right) }=\left\{ a_{t}\left( U_{j_{1}},...,U_{j_{t}}\right)
\right\} _{t}
\end{equation*}

Practically:%
\begin{equation*}
\left\vert U^{\left( j\right) },\left\{ \underline{\mathbf{\hat{\Lambda}}}%
_{\infty }^{\left[ k\right] }\left[ U^{\left( j\right) }\right] \right\}
,\lambda \left( U^{\left( j\right) }\right) \right\rangle
\end{equation*}%
are combination of products of states:%
\begin{equation*}
\sum \left\vert U^{\left( j\right) }\right\rangle _{\gamma }\left\vert
\left\{ \underline{\mathbf{\hat{\Lambda}}}_{\infty }^{\left[ k\right] }\left[
U^{\left( j\right) }\right] \right\} ,\lambda \left( U^{\left( j\right)
}\right) \right\rangle _{\gamma }
\end{equation*}%
Since $U^{\left( j\right) }$ and $\left\{ \underline{\mathbf{\hat{\Lambda}}}%
_{\infty }^{\left[ k\right] }\left[ U^{\left( j\right) }\right] \right\} $
are related through the constraints, we will use these constraints in the
sequel to reduce locally the number of parameters.

\subsection{Constrained states}

If some components $\left[ U^{\left( j\right) }\right] $ of $U^{\left(
j\right) }$ can be replaced by functions of $\underline{\mathbf{\hat{\Lambda}%
}}_{n}^{\left[ k\right] }\left[ U^{\left( j\right) }\right] $ through the
constraints, the states can be rewritten in the follwing manner. We start by
writing the constraints:%
\begin{equation*}
h_{k_{n}}\left( \left\{ \underline{\mathbf{\hat{\Lambda}}}_{n}^{\left[ k%
\right] }\left[ U^{\left( j\right) }\right] \right\} ,h_{p}\left( \left(
\Psi _{J}\right) ,U^{\left( j\right) }\right) \right) =0
\end{equation*}%
so that if $h_{p}\left( \left( \Psi _{J}\right) ,U_{j}^{l},\nu \right) $
includes combinations of the $a_{t}\left( \Psi _{J}\right) $, some of them
may be replaced by:%
\begin{equation*}
a_{t^{\prime }}\left( \Psi _{J}\right) =f_{t}\left( \left\{ \underline{%
\mathbf{\hat{\Lambda}}}_{\infty }^{\left[ k\right] }\left[ U^{\left(
j\right) }\right] \right\} \right)
\end{equation*}%
with:%
\begin{equation*}
\left\{ t^{\prime }\right\} \subset \left\{ t\right\}
\end{equation*}%
and the states write in a reduced form:%
\begin{equation*}
\sum \left\vert U^{\left( j\right) }/\left[ U^{\left( j\right) }\right] ,%
\left[ U^{\left( j\right) }\right] \left( \left\{ \underline{\mathbf{\hat{%
\Lambda}}}_{\infty }^{\left[ k\right] }\left[ U^{\left( j\right) }\right]
\right\} ,\lambda \left( U^{\left( j\right) }\right) \right) \right\rangle
_{\gamma }\left\vert \left\{ \underline{\mathbf{\hat{\Lambda}}}_{\infty }^{%
\left[ k\right] }\left[ U^{\left( j\right) }\right] \right\} ,\lambda \left(
U^{\left( j\right) }\right) \right\rangle _{\gamma }
\end{equation*}%
We will omit the subscript $\gamma $\ in the sequel. Moreover, if the $%
\underline{\mathbf{\hat{\Lambda}}}^{\left[ k\right] }\left[ U^{\left(
j\right) }\right] $ can be locally identified with fixed parameters $%
\underline{\mathbf{\hat{\Lambda}}}^{\left[ k\right] }$, we find the states:%
\begin{equation}
\sum \left\vert U^{\left( j\right) }/\left[ U^{\left( j\right) }\right] ,%
\left[ U^{\left( j\right) }\right] \left( \left\{ \underline{\mathbf{\hat{%
\Lambda}}}_{\infty }^{\left[ k\right] }\right\} ,\lambda \left( U^{\left(
j\right) }\right) \right) \right\rangle \left\vert \left\{ \underline{%
\mathbf{\hat{\Lambda}}}_{\infty }^{\left[ k\right] }\right\} ,\lambda \left(
U^{\left( j\right) }\right) \right\rangle  \label{ST}
\end{equation}%
In the sequel we replace $\left\{ \underline{\mathbf{\hat{\Lambda}}}_{\infty
}^{\left[ k\right] }\right\} \rightarrow \left\{ \underline{\mathbf{\hat{%
\Lambda}}}_{\infty }\right\} $, that stands for all realization of the $%
\underline{\mathbf{\hat{\Lambda}}}_{\infty }^{\left[ k\right] }$.

\subsection{State space for a given $\protect\lambda $}

Until now, we have considered the state as depending on the states
parametrized by $U^{\left( j\right) }$. We can reverse this point of view
and regroup states such that $\lambda \left( U^{\left( j\right) }\right)
=\lambda $, for each value of $\lambda $.

We first consider linear combinations of states (\ref{ST}) rewritten as:%
\begin{equation}
\sum \left\vert \left\{ \underline{\mathbf{\hat{\Lambda}}}_{\infty }\right\}
,\lambda \left( U^{\left( j\right) }\right) \right\rangle \left\vert h\left(
\left\{ \underline{\mathbf{\hat{\Lambda}}}_{\infty }\right\} ,\lambda \left(
U^{\left( j\right) }\right) \right) \right\rangle _{U^{\left( j\right) }}
\label{TS}
\end{equation}%
with:%
\begin{equation*}
\left\vert h\left( \left\{ \underline{\mathbf{\hat{\Lambda}}}_{\infty
}\right\} ,\lambda \left( U^{\left( j\right) }\right) \right) \right\rangle
_{U^{\left( j\right) }}=\left\vert U^{\left( j\right) }/\left[ U^{\left(
j\right) }\right] ,\left[ U^{\left( j\right) }\right] \left( \left\{ 
\underline{\mathbf{\hat{\Lambda}}}_{\infty }\right\} ,\lambda \left(
U^{\left( j\right) }\right) \right) \right\rangle
\end{equation*}%
The subscript $U^{\left( j\right) }$ reminds that the identification is
local, since th $\left\{ \underline{\mathbf{\hat{\Lambda}}}_{\infty
}\right\} $ are functionals of $U^{\left( j\right) }$.

Then, we restrict states $U^{\left( j\right) }$ to subspaces $U_{\lambda
}^{\left( j\right) }$ such.that. $\lambda \left( U_{\lambda }^{\left(
j\right) }\right) =\lambda $, and\ we write states (\ref{TS}) locally:%
\begin{equation*}
\left\vert \left\{ \underline{\mathbf{\hat{\Lambda}}}_{\infty }\right\}
,\lambda \right\rangle \left\vert h\left( \left\{ \underline{\mathbf{\hat{%
\Lambda}}}_{\infty }\right\} ,\lambda \right) \right\rangle _{U_{\lambda
}^{\left( j\right) }}\equiv \left\vert \left\{ \underline{\mathbf{\hat{%
\Lambda}}}_{\infty }\right\} ,\lambda \right\rangle \left\vert h\left(
\left\{ \underline{\mathbf{\hat{\Lambda}}}_{\infty }\right\} \right)
\right\rangle _{U_{\lambda }^{\left( j\right) }}
\end{equation*}%
where the relation for $\lambda $ is satisfied:%
\begin{equation*}
\lambda \left( h\left( \left\{ \underline{\mathbf{\hat{\Lambda}}}_{\infty
}\right\} ,\lambda \right) ,U_{\lambda }^{\left( j\right) }\right) =\lambda
\end{equation*}%
and the constraint writes:%
\begin{equation*}
h_{k_{i}}\left( \left\{ \mathbf{\hat{\Lambda}}^{\left[ k_{i}\right] }\left[
\Psi _{J},\left( h\left( \left\{ \underline{\mathbf{\hat{\Lambda}}}_{\infty
}\right\} ,\lambda \right) ,U_{\lambda }^{\left( j\right) }\right) \right]
\right\} _{i},h_{p}\left( \left( \Psi _{J}\right) ,\left( h\left( \left\{ 
\underline{\mathbf{\hat{\Lambda}}}_{\infty }\right\} ,\lambda \right)
,U_{\lambda }^{\left( j\right) }\right) \right) \right) =0
\end{equation*}

We also define the space spanned by states:%
\begin{equation*}
\mathcal{H}_{\left( \lambda ,\left\{ \left\{ \underline{\mathbf{\hat{\Lambda}%
}}_{\infty }\left( U_{\lambda }^{\left( j\right) }\right) \right\}
,U_{\lambda }^{\left( j\right) }\right\} \right) }=\left\{ \left\vert
\left\{ \underline{\mathbf{\hat{\Lambda}}}_{\infty }\right\} ,\lambda
\right\rangle \left\vert h\left( \left\{ \underline{\mathbf{\hat{\Lambda}}}%
_{\infty }\right\} \right) \right\rangle _{U_{\lambda }^{\left( j\right)
}}\right\} _{U_{\lambda }^{\left( j\right) }}
\end{equation*}%
which is also written $\mathcal{H}_{\lambda }$ for short. States belonging
to this space are series expansions:%
\begin{equation*}
\sum_{n,\hat{U}_{n}^{\left( j\right) }}\alpha _{n}\left( \left\{ \underline{%
\mathbf{\hat{\Lambda}}}_{\infty }\right\} ,\lambda ,U_{n}^{\left( j\right)
}\right) \delta \left( \lambda \left( h\left( \left\{ \underline{\mathbf{%
\hat{\Lambda}}}_{\infty }\right\} ,\lambda \right) ,U_{n}^{\left( j\right)
}\right) -\lambda \right) \left\vert \left\{ \underline{\mathbf{\hat{\Lambda}%
}}_{\infty }\right\} ,\lambda \right\rangle \left\vert h_{n}\left( \left\{ 
\underline{\mathbf{\hat{\Lambda}}}_{\infty }\right\} \right) \right\rangle
_{U_{n}^{\left( j\right) }}
\end{equation*}

Remark:

1. locally, these states can be generated by a field:%
\begin{equation*}
\Psi _{J}\left( U_{j}/\left[ U_{j}\right] ,\left\{ \underline{\mathbf{\hat{%
\Lambda}}}_{\infty }\right\} ,\lambda \right)
\end{equation*}

2. Covariantly the states are functionals $\alpha \left( \Sigma \right) $
(standing for $\alpha \left( \Sigma \left( U^{\left( j\right) }\right)
\right) $.

\subsection{Operators}

In this context, using a basis of states defined by some parameters $A$,
operators writes:%
\begin{equation*}
\left[ A\right] \Xi \left[ A^{\prime }\right]
\end{equation*}%
with the notation:%
\begin{equation*}
\left[ A\right] \Xi \left[ A^{\prime }\right] =\int \left\vert
A\right\rangle \Xi \left( A,A^{\prime }\right) \left\langle A^{\prime
}\right\vert
\end{equation*}%
where:%
\begin{eqnarray*}
\left[ A\right] &=&\left[ U^{\left( j\right) }/\left[ U^{\left( j\right) }%
\right] ,\left[ U^{\left( j\right) }\right] \left( \left\{ \underline{%
\mathbf{\hat{\Lambda}}}_{\infty }^{\left[ k_{i}\right] }\right\} ,\lambda
\left( U^{\left( j\right) }\right) \right) ,\left\{ \underline{\mathbf{\hat{%
\Lambda}}}_{\infty }^{\left[ k_{i}\right] }\right\} ,\lambda \left(
U^{\left( j\right) }\right) \right] \\
&=&\left[ h\left( \left\{ \underline{\mathbf{\hat{\Lambda}}}_{\infty }^{%
\left[ k_{i}\right] }\right\} ,\lambda \left( U^{\left( j\right) }\right)
\right) _{U^{\left( j\right) }},\left\{ \underline{\mathbf{\hat{\Lambda}}}%
_{\infty }^{\left[ k_{i}\right] }\right\} ,\lambda \left( U^{\left( j\right)
}\right) \right]
\end{eqnarray*}%
and:%
\begin{eqnarray*}
\left[ A^{\prime }\right] &=&\left[ U^{\left( j^{\prime }\right) }/\left[
U^{\left( j^{\prime }\right) }\right] ,\left[ U^{\left( j^{\prime }\right) }%
\right] \left( \left\{ \underline{\mathbf{\hat{\Lambda}}}_{\infty }^{\left[
k_{i}\right] }\right\} ^{\prime },\lambda \left( U^{\left( j^{\prime
}\right) }\right) \right) ,\left\{ \underline{\mathbf{\hat{\Lambda}}}%
_{\infty }^{\left[ k_{i}\right] }\right\} ^{\prime },\lambda \left(
U^{\left( j^{\prime }\right) }\right) \right] \\
&=&\left[ h\left( \left\{ \underline{\mathbf{\hat{\Lambda}}}_{\infty }^{%
\left[ k_{i}\right] }\right\} ^{\prime },\lambda \left( U^{\left( j^{\prime
}\right) }\right) \right) _{U^{\left( j^{\prime }\right) }},\left\{ 
\underline{\mathbf{\hat{\Lambda}}}_{\infty }^{\left[ k_{i}\right] }\right\}
^{\prime },\lambda \left( U^{\left( j^{\prime }\right) }\right) \right]
\end{eqnarray*}%
The $\left\{ \underline{\mathbf{\hat{\Lambda}}}_{\infty }^{\left[ k_{i}%
\right] }\right\} $ stand for any collection of $\left\{ \underline{\mathbf{%
\hat{\Lambda}}}_{\infty }^{\left[ k_{i}\right] }\right\} $.

In terms of eignvalues $\lambda $ and $\lambda ^{\prime }$ and with the
decompsition in terms f $U_{\lambda }^{\left( j\right) }$\ this also reduces
to:%
\begin{equation*}
\left[ U_{j}/\left[ U_{j}\right] ,h\left( \left\{ \underline{\mathbf{\hat{%
\Lambda}}}_{\infty }^{\left[ k_{i}\right] }\right\} ,\lambda \right)
_{U_{\lambda }^{\left( j\right) }},\left\{ \underline{\mathbf{\hat{\Lambda}}}%
_{\infty }^{\left[ k_{i}\right] }\right\} ,\lambda \right] \Xi \left[ \left(
U_{j}/\left[ U_{j}\right] \right) ^{\prime },h^{\prime }\left( \left\{ 
\underline{\mathbf{\hat{\Lambda}}}_{\infty }^{\left[ k_{i}\right] }\right\}
_{U_{\lambda ^{\prime }}^{\left( j^{\prime }\right) }}^{\prime },\lambda
^{\prime }\right) ,\left\{ \underline{\mathbf{\hat{\Lambda}}}_{\infty }^{%
\left[ k_{i}\right] }\right\} ^{\prime },\lambda ^{\prime }\right]
\end{equation*}%
The same forms are recovered in part 3 starting directly from states and
operators formalism. In the sequel, we write $\left\{ \underline{\mathbf{%
\hat{\Lambda}}}^{\left[ k_{i}\right] }\right\} $ fr$\left\{ \underline{%
\mathbf{\hat{\Lambda}}}_{\infty }^{\left[ k_{i}\right] }\right\} $. Conclude
by noting that locally, these states and operator are states build from
fields:%
\begin{equation*}
\Psi _{J}\left( U_{j}/\left[ U_{j}\right] ,\left\{ \underline{\mathbf{\hat{%
\Lambda}}}^{\left[ k_{i}\right] }\right\} ,\lambda \right)
\end{equation*}

\section{\textbf{Transitions between spaces }$\mathcal{H}_{\protect\lambda }$%
}

In this section we consider the state transitions betwn states due to
transition operator betwen spaces $\mathcal{H}_{\lambda _{0}}\rightarrow 
\mathcal{H}_{\lambda }$.

\subsection{Transformation $\mathcal{H}_{\protect\lambda _{0}}\rightarrow 
\mathcal{H}_{\protect\lambda }$}

Consider two eigenvalues $\lambda _{0}$ and $\lambda $. We assume that there
are isomorphism of spaces:%
\begin{equation*}
\mathcal{H}_{\left( \lambda _{0},\left\{ \underline{\mathbf{\hat{\Lambda}}}^{%
\left[ k_{i}\right] }\left( U_{\lambda _{0}}^{\left( j\right) }\right)
,U_{\lambda _{0}}^{\left( j\right) }\right\} \right) }\mathcal{\simeq H}%
_{\left( \lambda ,\left\{ \underline{\mathbf{\hat{\Lambda}}}^{\left[ k_{i}%
\right] }\left( U_{\lambda }^{\left( j\right) }\right) ,U_{\lambda }^{\left(
j\right) }\right\} \right) }
\end{equation*}%
where:%
\begin{equation*}
\mathcal{H}_{\left( \lambda ,\left\{ \underline{\mathbf{\hat{\Lambda}}}^{%
\left[ k_{i}\right] }\left( U_{\lambda }^{\left( j\right) }\right)
,U_{\lambda }^{\left( j\right) }\right\} \right) }=\left\{ \left\vert
\lambda ,\left\{ \underline{\mathbf{\hat{\Lambda}}}^{\left[ k_{i}\right]
}\right\} \right\rangle \left\vert h\left( \left\{ \underline{\mathbf{\hat{%
\Lambda}}}^{\left[ k_{i}\right] }\right\} \right) \right\rangle _{U_{\lambda
}^{\left( j\right) }}\right\} _{U_{\lambda }^{\left( j\right) }}
\end{equation*}%
given by:%
\begin{equation*}
T_{\lambda _{0}\lambda }:\mathcal{H}_{\left( \lambda _{0},\left\{ \underline{%
\mathbf{\hat{\Lambda}}}^{\left[ k_{i}\right] }\left( U_{\lambda
_{0}}^{\left( j\right) }\right) ,U_{\lambda _{0}}^{\left( j\right) }\right\}
\right) }\rightarrow \mathcal{H}_{\left( \lambda ,\left\{ \underline{\mathbf{%
\hat{\Lambda}}}^{\left[ k_{i}\right] }\left( U_{\lambda }^{\left( j\right)
}\right) ,U_{\lambda }^{\left( j\right) }\right\} \right) }
\end{equation*}%
so that:%
\begin{equation*}
T_{\lambda _{0}\lambda }\left\vert \lambda _{0},\left\{ \underline{\mathbf{%
\hat{\Lambda}}}^{\left[ k_{i}\right] }\right\} _{0}\right\rangle \left\vert
h\left( \left\{ \underline{\mathbf{\hat{\Lambda}}}^{\left[ k_{i}\right]
}\right\} _{0}\right) \right\rangle _{U_{0}^{\left( j\right) }}\in \left\{
\left\vert \lambda ,\left\{ \underline{\mathbf{\hat{\Lambda}}}^{\left[ k_{i}%
\right] }\right\} \right\rangle \left\vert h\left( \left\{ \underline{%
\mathbf{\hat{\Lambda}}}^{\left[ k_{i}\right] }\right\} \right) \right\rangle
_{U^{\left( j\right) }}\right\}
\end{equation*}

\subsubsection*{\textbf{Remark}}

The transformation can be considered covariantly by replacing:%
\begin{equation*}
\left\vert \lambda _{0},\left\{ \mathbf{\hat{\Lambda}}^{\left[ k_{i}\right]
}\right\} \right\rangle \left\vert H\left( \Sigma \left\{ \mathbf{\hat{%
\Lambda}}^{\left[ k_{i}\right] }\right\} \right) \right\rangle
_{U_{0}^{\left( j\right) }}\rightarrow \left\vert \Sigma \left\{ \mathbf{%
\hat{\Lambda}}^{\left[ k_{i}\right] }\right\} _{0}\right\rangle \left\vert
H\left( \Sigma \left\{ \mathbf{\hat{\Lambda}}^{\left[ k_{i}\right] }\right\}
\right) \right\rangle _{U_{0}^{\left( j\right) }}
\end{equation*}%
where $\Sigma \left\{ \mathbf{\hat{\Lambda}}^{\left[ k_{i}\right] }\right\} $
are hypersurface of $\left\{ \mathbf{\hat{\Lambda}}^{\left[ k_{i}\right]
}\right\} $. The transformation rewrites: 
\begin{equation*}
T_{\Sigma \left\{ \mathbf{\hat{\Lambda}}^{\left[ k_{i}\right] }\right\}
_{0}\Sigma \left\{ \mathbf{\hat{\Lambda}}^{\left[ k_{i}\right] }\right\}
}\left\vert \Sigma \left\{ \mathbf{\hat{\Lambda}}^{\left[ k_{i}\right]
}\right\} \right\rangle \left\vert h\left( \left\{ \underline{\mathbf{\hat{%
\Lambda}}}^{\left[ k_{i}\right] }\right\} \right) \right\rangle _{U^{\left(
j\right) }}\in \left\{ \left\vert \Sigma \left\{ \mathbf{\hat{\Lambda}}^{%
\left[ k_{i}\right] }\right\} _{0}\right\rangle \left\vert h\left( \Sigma
\left\{ \mathbf{\hat{\Lambda}}^{\left[ k_{i}\right] }\right\} \right)
\right\rangle _{U_{0}^{\left( j\right) }}\right\}
\end{equation*}

\subsection{Amplitudes}

To describe the amplitudes of transitions between states, we assume that
infinitesimally, the transformation has the form:%
\begin{equation*}
\delta T_{\lambda \lambda +\delta \lambda }=\delta \lambda \bar{V}\left(
U_{\lambda }^{\left( j\right) },U_{\lambda +\delta \lambda }^{\left(
j\right) }\right)
\end{equation*}%
where operator $\delta T_{\lambda \lambda +\delta \lambda }$ is a
transformation:%
\begin{equation*}
\delta T_{\lambda \lambda +\delta \lambda }:\mathcal{H}_{\left( \lambda
,\left\{ \underline{\mathbf{\hat{\Lambda}}}^{\left[ k_{i}\right] }\left(
U_{\lambda }^{\left( j\right) }\right) ,U_{\lambda }^{\left( j\right)
}\right\} \right) }\rightarrow \mathcal{H}_{\left( \lambda +\delta \lambda
,\left\{ \underline{\mathbf{\hat{\Lambda}}}^{\left[ k_{i}\right] }\left(
U_{\lambda +\delta \lambda }^{\left( j\right) }\right) ,U_{\lambda +\delta
\lambda }^{\left( j\right) }\right\} \right) }
\end{equation*}

To find the transformation $\delta T_{\lambda \lambda +\delta \lambda }$
recall that the states of $\mathcal{H}_{\left( \lambda ,\left\{ \underline{%
\mathbf{\hat{\Lambda}}}^{\left[ k_{i}\right] }\left( U_{\lambda }^{\left(
j\right) }\right) ,U_{\lambda }^{\left( j\right) }\right\} \right) }$ can be
built from fields:%
\begin{equation*}
\Psi _{J}\left( U_{j}/\left[ U_{j}\right] ,\left\{ \underline{\mathbf{\hat{%
\Lambda}}}^{\left[ k_{i}\right] }\right\} ,\lambda \right) 
\end{equation*}%
The amplitudes:%
\begin{equation*}
_{U_{j}}\left\langle \Psi _{J}\left( U_{j}/\left[ U_{j}\right] ,\left\{ 
\underline{\mathbf{\hat{\Lambda}}}^{\left[ k_{i}\right] }\right\} ,\lambda
+\delta \lambda \right) \right\vert \delta T_{\lambda \lambda +\delta
\lambda }\left\vert \Psi _{J}\left( \left( U_{j}\right) ^{\prime }/\left[
\left( U_{j}\right) ^{\prime }\right] ,\left\{ \underline{\mathbf{\hat{%
\Lambda}}}^{\left[ k_{i}\right] }\right\} ,\lambda \right) \right\rangle
_{\left( U_{j}\right) ^{\prime }}
\end{equation*}%
are computed by pulling back the field state:%
\begin{equation*}
\left\langle \Psi _{J}\left( U_{j}/\left[ U_{j}\right] ,\left\{ \underline{%
\mathbf{\hat{\Lambda}}}^{\left[ k_{i}\right] }\right\} ,\lambda +\delta
\lambda \right) \right\vert \rightarrow \left\langle \Psi _{J}\left( U_{j}/%
\left[ U_{j}\right] ,\left\{ \underline{\mathbf{\hat{\Lambda}}}^{\left[ k_{i}%
\right] }\right\} ,\lambda \right) \right\vert 
\end{equation*}%
from $\mathcal{H}_{\left( \lambda +\delta \lambda ,\left\{ \underline{%
\mathbf{\hat{\Lambda}}}^{\left[ k_{i}\right] }\left( U_{\lambda +\delta
\lambda }^{\left( j\right) }\right) ,U_{\lambda +\delta \lambda }^{\left(
j\right) }\right\} \right) }$ to $\mathcal{H}_{\left( \lambda ,\left\{ 
\underline{\mathbf{\hat{\Lambda}}}^{\left[ k_{i}\right] }\left( U_{\lambda
}^{\left( j\right) }\right) ,U_{\lambda }^{\left( j\right) }\right\} \right)
}$ by some parallel transport operator $P_{\lambda \lambda +\delta \lambda }$
and then computing the matrix elements of the pulled-backed transformation:%
\begin{equation}
T_{\lambda }:\mathcal{H}_{\left( \lambda ,\left\{ \underline{\mathbf{\hat{%
\Lambda}}}^{\left[ k_{i}\right] }\left( U_{\lambda }^{\left( j\right)
}\right) ,U_{\lambda }^{\left( j\right) }\right\} \right) }\rightarrow 
\mathcal{H}_{\left( \lambda ,\left\{ \underline{\mathbf{\hat{\Lambda}}}^{%
\left[ k_{i}\right] }\left( U_{\lambda }^{\left( j\right) }\right)
,U_{\lambda }^{\left( j\right) }\right\} \right) }  \label{Tp}
\end{equation}%
so that the transition is the composition:%
\begin{equation*}
\delta T_{\lambda \lambda +\delta \lambda }=P_{\lambda \lambda +\delta
\lambda }T_{\lambda }
\end{equation*}

\subsubsection{Global identification}

We assume in first approximation that the parameters $\left\{ \underline{%
\mathbf{\hat{\Lambda}}}^{\left[ k_{i}\right] }\right\} $ are global. The
pull back of the fld state $P_{\lambda \lambda +\delta \lambda }$ is
trivial: we can identify the various spaces, the transport involves standard
"derivatives". Formally, it is generated by the operator:%
\begin{eqnarray*}
&&\left( \Psi _{J}\left( U_{j}/\left[ U_{j}\right] ,\left\{ \underline{%
\mathbf{\hat{\Lambda}}}^{\left[ k_{i}\right] }\right\} ,\lambda +\delta
\lambda \right) -\Psi _{J}\left( U_{j}/\left[ U_{j}\right] ,\left\{ 
\underline{\mathbf{\hat{\Lambda}}}^{\left[ k_{i}\right] }\right\} ,\lambda
\right) \right) \frac{\delta }{\delta \Psi _{J}\left( U_{j}/\left[ U_{j}%
\right] ,\left\{ \underline{\mathbf{\hat{\Lambda}}}^{\left[ k_{i}\right]
}\right\} ,\lambda \right) } \\
&=&\frac{d}{d\lambda }\Psi _{J}\left( U_{j}/\left[ U_{j}\right] ,\left\{ 
\underline{\mathbf{\hat{\Lambda}}}^{\left[ k_{i}\right] }\right\} ,\lambda
\right) \frac{\delta }{\delta \Psi _{J}\left( U_{j}/\left[ U_{j}\right]
,\left\{ \underline{\mathbf{\hat{\Lambda}}}^{\left[ k_{i}\right] }\right\}
,\lambda \right) }
\end{eqnarray*}%
that can be exponentiated:%
\begin{equation*}
P_{\lambda \lambda +\delta \lambda }=\exp \left( \int i\delta \lambda \frac{d%
}{d\lambda }\Psi _{J}\left( U_{j}/\left[ U_{j}\right] ,\left\{ \underline{%
\mathbf{\hat{\Lambda}}}^{\left[ k_{i}\right] }\right\} ,\lambda \right) 
\frac{\delta }{\delta \Psi _{J}\left( U_{j}/\left[ U_{j}\right] ,\left\{ 
\underline{\mathbf{\hat{\Lambda}}}^{\left[ k_{i}\right] }\right\} ,\lambda
\right) }\mathcal{D}\Psi _{J}\left( U_{j}/\left[ U_{j}\right] ,\left\{ 
\underline{\mathbf{\hat{\Lambda}}}^{\left[ k_{i}\right] }\right\} ,\lambda
\right) \right) 
\end{equation*}%
Assuming a standard form for the transformation part $\delta T_{\lambda
\lambda +\delta \lambda }$ \ with $\frac{\delta }{\delta \Psi _{J}\left(
U_{j}/\left[ U_{j}\right] ,\left\{ \underline{\mathbf{\hat{\Lambda}}}^{\left[
k_{i}\right] }\right\} ,\lambda \right) }$ separable from other variables:%
\begin{eqnarray*}
\delta T_{\lambda \lambda +\delta \lambda } &=&V\left( \Psi _{J}\left( U_{j}/
\left[ U_{j}\right] ,\left\{ \underline{\mathbf{\hat{\Lambda}}}^{\left[ k_{i}%
\right] }\right\} ,\lambda \right) ,\nabla _{\left\{ \underline{\mathbf{\hat{%
\Lambda}}}^{\left[ k_{i}\right] }\right\} }\Psi _{J}\left( U_{j}/\left[ U_{j}%
\right] ,\left\{ \underline{\mathbf{\hat{\Lambda}}}^{\left[ k_{i}\right]
}\right\} ,\lambda \right) \right)  \\
&&+V_{2}\left( \frac{\delta }{\delta \Psi _{J}\left( U_{j}/\left[ U_{j}%
\right] ,\left\{ \underline{\mathbf{\hat{\Lambda}}}^{\left[ k_{i}\right]
}\right\} ,\lambda \right) }\right) 
\end{eqnarray*}%
the composition:%
\begin{equation*}
P_{\lambda \lambda +\delta \lambda }\delta T_{\lambda \lambda +\delta
\lambda }
\end{equation*}%
can be simplified by the computation of the matrices element of:%
\begin{eqnarray*}
&&\exp \left( \int i\delta \lambda \frac{d}{d\lambda }\Psi _{J}\left( U_{j}/%
\left[ U_{j}\right] ,\left\{ \underline{\mathbf{\hat{\Lambda}}}^{\left[ k_{i}%
\right] }\right\} ,\lambda \right) \frac{\delta }{\delta \Psi _{J}\left(
U_{j}/\left[ U_{j}\right] ,\left\{ \underline{\mathbf{\hat{\Lambda}}}^{\left[
k_{i}\right] }\right\} ,\lambda \right) }\mathcal{D}\Psi _{J}\left( U_{j}/%
\left[ U_{j}\right] ,\left\{ \underline{\mathbf{\hat{\Lambda}}}^{\left[ k_{i}%
\right] }\right\} ,\lambda \right) \right)  \\
&&\times \exp \left( iV_{2}\left( \frac{\delta }{\delta \Psi _{J}\left(
U_{j}/\left[ U_{j}\right] ,\left\{ \underline{\mathbf{\hat{\Lambda}}}^{\left[
k_{i}\right] }\right\} ,\lambda \right) }\right) \right) 
\end{eqnarray*}%
so that the transition is generated by: 
\begin{eqnarray*}
&&S\left( \Psi _{J}\right)  \\
&=&S\left( \frac{d}{d\lambda }\Psi _{J}\left( U_{j}/\left[ U_{j}\right]
,\left\{ \underline{\mathbf{\hat{\Lambda}}}^{\left[ k_{i}\right] }\right\}
,\lambda \right) ,\nabla _{\left\{ \underline{\mathbf{\hat{\Lambda}}}^{\left[
k_{i}\right] }\right\} }\Psi _{J}\left( U_{j}/\left[ U_{j}\right] ,\left\{ 
\underline{\mathbf{\hat{\Lambda}}}^{\left[ k_{i}\right] }\right\} ,\lambda
\right) ,\Psi _{J}\left( U_{j}/\left[ U_{j}\right] ,\left\{ \underline{%
\mathbf{\hat{\Lambda}}}^{\left[ k_{i}\right] }\right\} ,\lambda \right)
\right) 
\end{eqnarray*}

\subsubsection{Local identification}

When the identifications are local, the transport is non trivial. Changing
the variables, modifies the states. The transport is generated by some
covariant derivatives. We show in appendix 6 that the general form for the
amplitudes of this operator are generated by some functional:%
\begin{eqnarray*}
&&S\left( \Psi _{J}\right) \\
&=&S\left( \underline{\nabla }_{\lambda }\Psi _{J}\left( U_{j}/\left[ U_{j}%
\right] ,\left\{ \underline{\mathbf{\hat{\Lambda}}}^{\left[ k_{i}\right]
}\right\} ,\lambda \right) ,\underline{\nabla }_{\left\{ \underline{\mathbf{%
\hat{\Lambda}}}^{\left[ k_{i}\right] }\right\} }\Psi _{J}\left( U_{j}/\left[
U_{j}\right] ,\left\{ \underline{\mathbf{\hat{\Lambda}}}^{\left[ k_{i}\right]
}\right\} ,\lambda \right) ,\Psi _{J}\left( U_{j}/\left[ U_{j}\right]
,\left\{ \underline{\mathbf{\hat{\Lambda}}}^{\left[ k_{i}\right] }\right\}
,\lambda \right) \right)
\end{eqnarray*}%
with:

\begin{eqnarray*}
&&\underline{\nabla }_{\left\{ \underline{\mathbf{\hat{\Lambda}}}^{\left[
k_{i}\right] }\right\} }\Psi _{J}\left( U_{j}/\left[ U_{j}\right] ,\left\{ 
\underline{\mathbf{\hat{\Lambda}}}^{\left[ k_{i}\right] }\right\} ,\lambda
\right) \\
&=&\nabla _{\left\{ \underline{\mathbf{\hat{\Lambda}}}^{\left[ k_{i}\right]
}\right\} }\Psi _{J}\left( U_{j}/\left[ U_{j}\right] ,\left\{ \underline{%
\mathbf{\hat{\Lambda}}}^{\left[ k_{i}\right] }\right\} ,\lambda \right)
+\left( \left( A_{\left\{ \underline{\mathbf{\hat{\Lambda}}}^{\left[ k_{i}%
\right] }\right\} }\right) _{k}^{k^{\prime }}\left( U_{j}/\left[ U_{j}\right]
\right) \Psi _{J}\left( \left( U_{j}/\left[ U_{j}\right] \right) _{k^{\prime
}},\left\{ \underline{\mathbf{\hat{\Lambda}}}^{\left[ k_{i}\right] }\right\}
,\lambda \right) \right) _{k}
\end{eqnarray*}%
and:%
\begin{eqnarray*}
&&\delta \lambda \underline{\nabla }_{\lambda }\Psi _{J}\left( U_{j}/\left[
U_{j}\right] ,\left\{ \underline{\mathbf{\hat{\Lambda}}}^{\left[ k_{i}\right]
}\right\} ,\lambda \right) \\
&=&\frac{\partial \Psi _{J}\left( U_{j}/\left[ U_{j}\right] ,\left\{ 
\underline{\mathbf{\hat{\Lambda}}}^{\left[ k_{i}\right] }\right\} ,\lambda
\right) }{\partial \lambda }+\left( \left( A_{\lambda }\right)
_{k}^{k^{\prime }}\left( U_{j}/\left[ U_{j}\right] \right) \Psi _{J}\left(
\left( U_{j}/\left[ U_{j}\right] \right) _{k^{\prime }},\left\{ \underline{%
\mathbf{\hat{\Lambda}}}^{\left[ k_{i}\right] }\right\} ,\lambda \right)
\right) _{k}
\end{eqnarray*}%
are some covariant derivatives with connection matrices:%
\begin{equation*}
\left( A_{\left\{ \underline{\mathbf{\hat{\Lambda}}}^{\left[ k_{i}\right]
}\right\} },\left( A_{\lambda }\right) \right)
\end{equation*}%
connecting the states with different values of $\lambda $.

The matrix elements of transitions $T_{\lambda \lambda ^{\prime }}$ are
obtained by exponentiation of $S$. Actually:%
\begin{equation}
T_{\lambda \lambda ^{\prime }}=\tprod \left( 1+\delta T_{\lambda \lambda
+\delta \lambda }\right)  \label{CMTRS}
\end{equation}%
so that:%
\begin{eqnarray}
&&\left\langle \Psi _{J}\left( U_{j}/\left[ U_{j}\right] ,\left\{ \underline{%
\mathbf{\hat{\Lambda}}}^{\left[ k_{i}\right] }\right\} ,\lambda ^{\prime
}\right) \right\vert T_{\lambda \lambda ^{\prime }}\left\vert \Psi
_{J}\left( U_{j}/\left[ U_{j}\right] ,\left\{ \underline{\mathbf{\hat{\Lambda%
}}}^{\left[ k_{i}\right] }\right\} ,\lambda \right) \right\rangle
\label{FRMTRST} \\
&=&\left\langle \Psi _{J}\left( U_{j}/\left[ U_{j}\right] ,\left\{ 
\underline{\mathbf{\hat{\Lambda}}}^{\left[ k_{i}\right] }\right\} ,\lambda
^{\prime }\right) \right\vert \exp \left( i\int \delta \lambda S\left( \Psi
_{J}\right) \delta \lambda \right) \left\vert \Psi _{J}\left( U_{j}/\left[
U_{j}\right] ,\left\{ \underline{\mathbf{\hat{\Lambda}}}^{\left[ k_{i}\right]
}\right\} ,\lambda \right) \right\rangle  \notag
\end{eqnarray}%
\bigskip

\subsubsection{Remark 1}

These amplitudes between states can also be rewritten covariantly:%
\begin{equation*}
\left\langle \Sigma \left\{ \mathbf{\hat{\Lambda}}^{\left[ k_{i}\right]
}\right\} \right\vert \left\langle H\left( \Sigma \left\{ \mathbf{\hat{%
\Lambda}}^{\left[ k_{i}\right] }\right\} \right) \right\vert _{U_{0}^{\left(
j\right) }}\exp \left( iS\left( \Psi _{J}\left( U_{j}/\left[ U_{j}\right]
,\left\{ \mathbf{\hat{\Lambda}}^{\left[ k_{i}\right] }\right\} \right)
\right) \right) \left\vert \Sigma \left\{ \mathbf{\hat{\Lambda}}^{\left[
k_{i}\right] }\right\} \right\rangle \left\vert H\left( \Sigma \left\{ 
\mathbf{\hat{\Lambda}}^{\left[ k_{i}\right] }\right\} \right) \right\rangle
_{U_{0}^{\left( j\right) }}
\end{equation*}%
with local invariance with respect to transformation of $\left\{ \mathbf{%
\hat{\Lambda}}^{\left[ k_{i}\right] }\right\} $.

\subsubsection{Remark 2}

The connection 
\begin{equation*}
\Gamma \equiv \left( \left( A_{\left\{ \underline{\mathbf{\hat{\Lambda}}}^{%
\left[ k_{i}\right] }\right\} }\right) \left( U_{j}/\left[ U_{j}\right]
\right) ,\left( A_{\lambda }\right) \left( U_{j}/\left[ U_{j}\right] \right)
\right)
\end{equation*}%
is itself a field dependent object, since it is derived from the constraint.
Given the form of the constraint, it has the form:%
\begin{equation*}
R\left( \Gamma \right) =F\left( \Psi _{J}\left( U_{j}/\left[ U_{j}\right]
,\left\{ \underline{\mathbf{\hat{\Lambda}}}^{\left[ k_{i}\right] }\right\}
,\lambda \right) ,\nabla _{\mathbf{\hat{\Lambda}}^{\left[ k_{i}\right]
}}^{\Gamma }\Psi _{J}\left( U_{j}/\left[ U_{j}\right] ,\left\{ \underline{%
\mathbf{\hat{\Lambda}}}^{\left[ k_{i}\right] }\right\} ,\lambda \right)
\right)
\end{equation*}%
and some transition amplitude should be associated to this quantity.

\subsubsection{Non uniqueness of connection and modification of amplitude}

Even if connections are considered as inert, there should be several
possible connections corresponding to the transport:%
\begin{equation*}
\left\langle \Psi _{J}\left( U_{j}/\left[ U_{j}\right] ,\left\{ \underline{%
\mathbf{\hat{\Lambda}}}^{\left[ k_{i}\right] }\right\} ,\lambda +\delta
\lambda \right) \right\vert \rightarrow \left\langle \Psi _{J}\left( U_{j}/%
\left[ U_{j}\right] ,\left\{ \underline{\mathbf{\hat{\Lambda}}}^{\left[ k_{i}%
\right] }\right\} ,\lambda \right) \right\vert
\end{equation*}%
since there should be multiple way to send a state with parameter $\left(
U_{j}/\left[ U_{j}\right] ,\left\{ \underline{\mathbf{\hat{\Lambda}}}^{\left[
k_{i}\right] }\right\} ,\lambda +\delta \lambda \right) $ to $\left( U_{j}/%
\left[ U_{j}\right] ,\left\{ \underline{\mathbf{\hat{\Lambda}}}^{\left[ k_{i}%
\right] }\right\} ,\lambda \right) $. This corresponds to the fact that some
relative dimension may arise between these two spaces of parameters:%
\begin{equation*}
\dim \left( \left( U_{j}/\left[ U_{j}\right] ,\left\{ \underline{\mathbf{%
\hat{\Lambda}}}^{\left[ k_{i}\right] }\right\} ,\lambda +\delta \lambda
\right) /\left( U_{j}/\left[ U_{j}\right] ,\left\{ \underline{\mathbf{\hat{%
\Lambda}}}^{\left[ k_{i}\right] }\right\} ,\lambda \right) \right) >0
\end{equation*}%
and that there are many maps sending $\left( U_{j}/\left[ U_{j}\right]
,\left\{ \underline{\mathbf{\hat{\Lambda}}}^{\left[ k_{i}\right] }\right\}
,\lambda \right) $ to $\left( U_{j}/\left[ U_{j}\right] ,\left\{ \underline{%
\mathbf{\hat{\Lambda}}}^{\left[ k_{i}\right] }\right\} ,\lambda +\delta
\lambda \right) $.

The amplitudes thus should modify (\ref{FRMTRST}). Actually, we first
consider that $\delta T_{\lambda \lambda +\delta \lambda }$ is replaced by
the set:%
\begin{equation*}
\left( \delta T_{\lambda \lambda +\delta \lambda }^{\left( \left(
U_{j}\right) _{\lambda }\hookrightarrow \left( U_{j}\right) _{\lambda
+d\lambda }\right) }\right) _{\left( U_{j}\right) _{\lambda }\hookrightarrow
\left( U_{j}\right) _{\lambda +d\lambda }}
\end{equation*}%
accounting for all possible maps from one space to the other. Then (\ref%
{CMTRS}) is modified by taking into account of these multiple paths, so that
we replace:%
\begin{equation}
1+\delta T_{\lambda \lambda +\delta \lambda }\rightarrow \sum_{\left(
U_{j}\right) _{\lambda }\hookrightarrow \left( U_{j}\right) _{\lambda
+d\lambda }}1+\delta T_{\lambda \lambda +\delta \lambda }^{\left( \left(
U_{j}\right) _{\lambda }\hookrightarrow \left( U_{j}\right) _{\lambda
+d\lambda }\right) }
\end{equation}%
and the transitions are:%
\begin{equation*}
T_{\lambda \lambda ^{\prime }}=\tprod \left( 1+\sum_{\left( U_{j}\right)
_{\lambda }\hookrightarrow \left( U_{j}\right) _{\lambda +d\lambda }}\delta
T_{\lambda \lambda +\delta \lambda }^{\left( \left( U_{j}\right) _{\lambda
}\hookrightarrow \left( U_{j}\right) _{\lambda +d\lambda }\right) }\right)
\end{equation*}%
In terms of amplitudes this becomes:%
\begin{eqnarray}
&&\left\langle \Psi _{J}\left( U_{j}/\left[ U_{j}\right] ,\left\{ \underline{%
\mathbf{\hat{\Lambda}}}^{\left[ k_{i}\right] }\right\} ,\lambda ^{\prime
}\right) \right\vert T_{\lambda \lambda ^{\prime }}\left\vert \Psi
_{J}\left( U_{j}/\left[ U_{j}\right] ,\left\{ \underline{\mathbf{\hat{\Lambda%
}}}^{\left[ k_{i}\right] }\right\} ,\lambda \right) \right\rangle
\label{NBC} \\
&=&\left\langle \Psi _{J}\left( U_{j}/\left[ U_{j}\right] ,\left\{ 
\underline{\mathbf{\hat{\Lambda}}}^{\left[ k_{i}\right] }\right\} ,\lambda
+\delta \lambda \right) \right\vert  \notag \\
&&\times \tprod_{\delta \lambda }\tprod_{\left( U_{j}\right) _{\lambda
}\hookrightarrow \left( U_{j}\right) _{\lambda +d\lambda }}\exp \left(
i\delta \lambda S\left( \Psi _{J},\left( A_{\left\{ \underline{\mathbf{\hat{%
\Lambda}}}^{\left[ k_{i}\right] }\right\} },\left( A_{\lambda }\right)
\right) ^{\left( \left( U_{j}\right) _{\lambda }\hookrightarrow \left(
U_{j}\right) _{\lambda +d\lambda }\right) }\right) \right) \left\vert \Psi
_{J}\left( U_{j}/\left[ U_{j}\right] ,\left\{ \underline{\mathbf{\hat{\Lambda%
}}}^{\left[ k_{i}\right] }\right\} ,\lambda \right) \right\rangle  \notag
\end{eqnarray}%
The upperscript $\left( \left( U_{j}\right) _{\lambda }\hookrightarrow
\left( U_{j}\right) _{\lambda +d\lambda }\right) $ reminds that the
connections $\left( A_{\left\{ \underline{\mathbf{\hat{\Lambda}}}^{\left[
k_{i}\right] }\right\} },\left( A_{\lambda }\right) \right) $ depend on the
path chosen. Assuming that we can replace in average:%
\begin{equation*}
\left( A_{\left\{ \underline{\mathbf{\hat{\Lambda}}}^{\left[ k_{i}\right]
}\right\} },\left( A_{\lambda }\right) \right) ^{\left( \left( U_{j}\right)
_{\lambda }\hookrightarrow \left( U_{j}\right) _{\lambda +d\lambda }\right)
}\rightarrow \left( A_{\left\{ \underline{\mathbf{\hat{\Lambda}}}^{\left[
k_{i}\right] }\right\} },\left( A_{\lambda }\right) \right)
\end{equation*}%
the transition becomes:%
\begin{eqnarray*}
&&\left\langle \Psi _{J}\left( U_{j}/\left[ U_{j}\right] ,\left\{ \underline{%
\mathbf{\hat{\Lambda}}}^{\left[ k_{i}\right] }\right\} ,\lambda ^{\prime
}\right) \right\vert T_{\lambda \lambda ^{\prime }}\left\vert \Psi
_{J}\left( U_{j}/\left[ U_{j}\right] ,\left\{ \underline{\mathbf{\hat{\Lambda%
}}}^{\left[ k_{i}\right] }\right\} ,\lambda \right) \right\rangle \\
&=&\left\langle \Psi _{J}\left( U_{j}/\left[ U_{j}\right] ,\left\{ 
\underline{\mathbf{\hat{\Lambda}}}^{\left[ k_{i}\right] }\right\} ,\lambda
+\delta \lambda \right) \right\vert \\
&&\times N\left( \left( U_{j}\right) _{\lambda }\hookrightarrow \left(
U_{j}\right) _{\lambda +d\lambda }\right) \tprod_{\delta \lambda }\exp
\left( i\delta \lambda S\left( \Psi _{J},\left( A_{\left\{ \underline{%
\mathbf{\hat{\Lambda}}}^{\left[ k_{i}\right] }\right\} },\left( A_{\lambda
}\right) \right) \right) \right) \left\vert \Psi _{J}\left( U_{j}/\left[
U_{j}\right] ,\left\{ \underline{\mathbf{\hat{\Lambda}}}^{\left[ k_{i}\right]
}\right\} ,\lambda \right) \right\rangle
\end{eqnarray*}%
where $N\left( \left( U_{j}\right) _{\lambda }\hookrightarrow \left(
U_{j}\right) _{\lambda +d\lambda }\right) $ is the number, or the volume of
set of maps $\left( \left( U_{j}\right) _{\lambda }\hookrightarrow \left(
U_{j}\right) _{\lambda +d\lambda }\right) $.

Ultimately, the transitions become:%
\begin{eqnarray}
&&\left\langle \Psi _{J}\left( U_{j}/\left[ U_{j}\right] ,\left\{ \underline{%
\mathbf{\hat{\Lambda}}}^{\left[ k_{i}\right] }\right\} ,\lambda ^{\prime
}\right) \right\vert T_{\lambda \lambda ^{\prime }}\left\vert \Psi
_{J}\left( U_{j}/\left[ U_{j}\right] ,\left\{ \underline{\mathbf{\hat{\Lambda%
}}}^{\left[ k_{i}\right] }\right\} ,\lambda \right) \right\rangle
\label{TRM} \\
&=&\left\langle \Psi _{J}\left( U_{j}/\left[ U_{j}\right] ,\left\{ 
\underline{\mathbf{\hat{\Lambda}}}^{\left[ k_{i}\right] }\right\} ,\lambda
+\delta \lambda \right) \right\vert  \notag \\
&&\times N\left( \left( U_{j}\right) _{\lambda }\hookrightarrow \left(
U_{j}\right) _{\lambda ^{\prime }}\right) \exp \left( i\int \delta \lambda
S\left( \Psi _{J},\left( A_{\left\{ \underline{\mathbf{\hat{\Lambda}}}^{%
\left[ k_{i}\right] }\right\} },\left( A_{\lambda }\right) \right) \right)
\right) \left\vert \Psi _{J}\left( U_{j}/\left[ U_{j}\right] ,\left\{ 
\underline{\mathbf{\hat{\Lambda}}}^{\left[ k_{i}\right] }\right\} ,\lambda
\right) \right\rangle  \notag
\end{eqnarray}%
with $N\left( \left( U_{j}\right) _{\lambda }\hookrightarrow \left(
U_{j}\right) _{\lambda ^{\prime }}\right) $ the number, or the volume of set
of maps $\left( \left( U_{j}\right) _{\lambda }\hookrightarrow \left(
U_{j}\right) _{\lambda \prime }\right) $.

\subsection{Particular states}

We consider a particular case in which a state can be decomposed into two
separate sets:%
\begin{eqnarray*}
U_{0}^{\left( j/p\right) } &=&\left( U_{0}^{\left( j/p\right)
},U_{0}^{\left( jp\right) }\right) \\
U^{\left( j/p\right) } &=&\left( U^{\left( j/p\right) },U^{\left( jp\right)
}\right)
\end{eqnarray*}%
where $\lambda \left( U_{0}^{\left( j\right) }\right) =\lambda \left(
U_{0}^{\left( j/p\right) }\right) $ and $\lambda \left( U^{\left( j\right)
}\right) =\lambda \left( U^{\left( j/p\right) }\right) $. It describes a
decomposition in which $U_{0}^{\left( j/p\right) }$ is much larger than $%
U^{\left( jp\right) }$ and determines the parameter space, while the stes
defined by $U^{\left( jp\right) }$ consists in some system evolving in these
parameter space. We consider the states: \ 
\begin{equation*}
\left\vert \left\{ \underline{\mathbf{\hat{\Lambda}}}^{\left[ k_{i}\right]
}\right\} ,\lambda \right\rangle \left\vert h\left( \left\{ \underline{%
\mathbf{\hat{\Lambda}}}^{\left[ k_{i}\right] }\right\} ,\lambda \right) ,%
\hat{U}^{\left( j\right) }\right\rangle _{U^{\left( j\right) }}=\left\vert
\left\{ \underline{\mathbf{\hat{\Lambda}}}^{\left[ k_{i}\right] }\right\}
_{/p},\lambda \right\rangle \left\vert \left\{ \underline{\mathbf{\hat{%
\Lambda}}}^{\left[ k_{i}\right] }\right\} _{p}\right\rangle \left\vert
h\left( \left\{ \underline{\mathbf{\hat{\Lambda}}}^{\left[ k_{i}\right]
}\right\} _{p},\lambda \right) \right\rangle _{U^{\left( j\right) }}
\end{equation*}%
This corresponds to sum tensor products of fields states. The parameter
space decomposes into a large set $\left\{ \underline{\mathbf{\hat{\Lambda}}}%
^{\left[ k_{i}\right] }\right\} _{/p}$ and a small set $\left\{ \underline{%
\mathbf{\hat{\Lambda}}}^{\left[ k_{i}\right] }\right\} _{p}$ describing the
system studied.

\begin{eqnarray*}
\left\vert \lambda ,\left\{ \underline{\mathbf{\hat{\Lambda}}}^{\left[ k_{i}%
\right] }\right\} \right\rangle \left\vert h\left( \left\{ \underline{%
\mathbf{\hat{\Lambda}}}^{\left[ k_{i}\right] }\right\} \right) \right\rangle
_{U^{\left( j\right) }} &=&\left\vert \lambda ,\left\{ \underline{\mathbf{%
\hat{\Lambda}}}^{\left[ k_{i}\right] }\right\} _{/p}\right\rangle \left\vert
h\left( \left\{ \underline{\mathbf{\hat{\Lambda}}}^{\left[ k_{i}\right]
}\right\} _{/p}\right) \right\rangle _{U^{\left( j\right) }}\left\vert
\lambda ,\left\{ \underline{\mathbf{\hat{\Lambda}}}^{\left[ k_{i}\right]
}\right\} _{p}\right\rangle \left\vert h\left( \left\{ \underline{\mathbf{%
\hat{\Lambda}}}^{\left[ k_{i}\right] }\right\} _{p}\right) \right\rangle
_{U^{\left( j\right) }} \\
&=&\left\vert \lambda ,\left\{ \underline{\mathbf{\hat{\Lambda}}}^{\left[
k_{i}\right] }\right\} \right\rangle \left\vert h\left( \left\{ \underline{%
\mathbf{\hat{\Lambda}}}^{\left[ k_{i}\right] }\right\} _{/p}\right)
\right\rangle _{U^{\left( j\right) }}\left\vert h\left( \left\{ \underline{%
\mathbf{\hat{\Lambda}}}^{\left[ k_{i}\right] }\right\} _{p}\right)
\right\rangle _{U^{\left( j\right) }}
\end{eqnarray*}%
whr: 
\begin{equation*}
V\left( \left\{ \underline{\mathbf{\hat{\Lambda}}}^{\left[ k_{i}\right]
}\right\} _{p}\right) <<V\left( \left\{ \underline{\mathbf{\hat{\Lambda}}}^{%
\left[ k_{i}\right] }\right\} _{/p}\right)
\end{equation*}%
and:%
\begin{equation*}
\Psi _{J}\left( U_{j}/\left[ U_{j}\right] ,\left\{ \underline{\mathbf{\hat{%
\Lambda}}}^{\left[ k_{i}\right] }\right\} ,\lambda \right) =\left\{ \Psi
_{J}\left( U_{j}/\left[ U_{j}\right] ,\left\{ \underline{\mathbf{\hat{\Lambda%
}}}^{\left[ k_{i}\right] }\right\} _{p},\lambda \right) ,\Psi _{J}\left(
U_{j}/\left[ U_{j}\right] ,\left\{ \underline{\mathbf{\hat{\Lambda}}}^{\left[
k_{i}\right] }\right\} _{/p},\lambda \right) \right\}
\end{equation*}%
and:%
\begin{equation*}
S\left( \Psi _{J}\left( U_{j}/\left[ U_{j}\right] ,\left\{ \underline{%
\mathbf{\hat{\Lambda}}}^{\left[ k_{i}\right] }\right\} ,\lambda \right)
\right) =S_{p}\left( \Psi _{J}\left( U_{j}/\left[ U_{j}\right] ,\left\{ 
\underline{\mathbf{\hat{\Lambda}}}^{\left[ k_{i}\right] }\right\}
_{p},\lambda \right) \right) +S_{/p}\left( \Psi _{J}\left( U_{j}/\left[ U_{j}%
\right] ,\left\{ \underline{\mathbf{\hat{\Lambda}}}^{\left[ k_{i}\right]
}\right\} _{/p},\lambda \right) \right)
\end{equation*}

Then we can factor:%
\begin{equation*}
\left\langle \Sigma \left\{ \mathbf{\hat{\Lambda}}^{\left[ k_{i}\right]
}\right\} _{p}\right\vert \left\langle H\left( \Sigma \left\{ \mathbf{\hat{%
\Lambda}}^{\left[ k_{i}\right] }\right\} _{p}\right) \right\vert
_{U_{0}^{\left( j\right) }}\exp \left( iS_{p}\left( \Psi _{J}\left( U_{j}/%
\left[ U_{j}\right] ,\left\{ \underline{\mathbf{\hat{\Lambda}}}^{\left[ k_{i}%
\right] }\right\} _{p},\lambda \right) \right) \right) \left\vert \Sigma
\left\{ \mathbf{\hat{\Lambda}}^{\left[ k_{i}\right] }\right\}
_{p}\right\rangle \left\vert H\left( \Sigma \left\{ \mathbf{\hat{\Lambda}}^{%
\left[ k_{i}\right] }\right\} _{p}\right) \right\rangle _{U_{0}^{\left(
j\right) }}
\end{equation*}%
\bigskip

and interaction with full field arises through $\Gamma $.

\subsection{\textbf{Transitions for operators}:}

We consider some operator:%
\begin{equation*}
\mathcal{H}_{\left( \lambda ,\left\{ \underline{\mathbf{\hat{\Lambda}}}^{%
\left[ k_{i}\right] }\left( U_{\lambda }^{\left( j\right) }\right)
,U_{\lambda }^{\left( j\right) }\right\} \right) }\mathcal{\simeq H}_{\left(
\lambda ^{\prime },\left\{ \underline{\mathbf{\hat{\Lambda}}}^{\left[ k_{i}%
\right] }\left( U_{\lambda ^{\prime }}^{\left( j\right) }\right) ,U_{\lambda
^{\prime }}^{\left( j\right) }\right\} \right) }
\end{equation*}%
with matrices elements:%
\begin{equation*}
\left[ \lambda ^{\prime },\left\{ \underline{\mathbf{\hat{\Lambda}}}^{\prime %
\left[ k_{i}\right] }\right\} ,h\left( \left\{ \underline{\mathbf{\hat{%
\Lambda}}}^{\prime \left[ k_{i}\right] }\right\} \right) \right] _{U^{\left(
j\right) \prime }}\Phi \left[ \lambda ,\left\{ \underline{\mathbf{\hat{%
\Lambda}}}^{\left[ k_{i}\right] }\right\} ,h\left( \left\{ \underline{%
\mathbf{\hat{\Lambda}}}^{\left[ k_{i}\right] }\right\} \right) \right]
_{U^{\left( j\right) }}
\end{equation*}%
We consider that these matrices elements can be decomposed between $/p$ and $%
p$ states: 
\begin{equation*}
\left[ \lambda ^{\prime },\left\{ \underline{\mathbf{\hat{\Lambda}}}^{\prime %
\left[ k_{i}\right] }\right\} ,h\left( \left\{ \underline{\mathbf{\hat{%
\Lambda}}}^{\prime \left[ k_{i}\right] }\right\} _{/p}\right) ,h\left(
\left\{ \underline{\mathbf{\hat{\Lambda}}^{\prime }}^{\left[ k_{i}\right]
}\right\} _{p}\right) \right] _{U^{\left( j\right) \prime }}\Phi \left[
\lambda ,\left\{ \underline{\mathbf{\hat{\Lambda}}}^{\left[ k_{i}\right]
}\right\} ,h\left( \left\{ \underline{\mathbf{\hat{\Lambda}}}^{\left[ k_{i}%
\right] }\right\} _{/p}\right) ,h\left( \left\{ \underline{\mathbf{\hat{%
\Lambda}}}^{\left[ k_{i}\right] }\right\} _{p}\right) \right] _{U^{\left(
j\right) }}
\end{equation*}%
for $\lambda \left( U^{\left( j\right) }\right) \simeq \lambda \left( \left(
U^{\left( j\right) }\right) ^{\prime }\right) $. Here $\left\vert U^{\left(
j\right) p}\right\rangle $ is at scale of some system. We assume that $\dim
U^{\left( j\right) p}<<\dim U^{\left( j\right) }$, so that $\lambda ^{\left(
i\right) }\left( U^{\left( j\right) }\right) $ and $\left\{ \underline{%
\mathbf{\hat{\Lambda}}}^{\left[ k_{i}\right] }\right\} $ can be considered
independent from $U^{\left( j\right) p}$ and $\lambda ^{\left( i\right)
}\left( U^{\left( j\right) }\right) =\lambda ^{\left( i\right) }$. For
operators depending only on the $U^{\left( j\right) p}$, \ we can discard
the $U^{\left( j\right) /p}$, $\left( U^{\left( j\right) /p}\right) ^{\prime
}$

Restricting to:%
\begin{equation*}
\left[ \lambda ^{\prime },\left\{ \underline{\mathbf{\hat{\Lambda}}}^{\prime %
\left[ k_{i}\right] }\right\} ,h\left( \left\{ \underline{\mathbf{\hat{%
\Lambda}}^{\prime }}^{\left[ k_{i}\right] }\right\} _{p}\right) \right]
_{U^{\left( j\right) \prime }}\Phi \left[ \lambda ,\left\{ \underline{%
\mathbf{\hat{\Lambda}}}^{\left[ k_{i}\right] }\right\} ,h\left( \left\{ 
\underline{\mathbf{\hat{\Lambda}}}^{\left[ k_{i}\right] }\right\}
_{p}\right) \right] _{U^{\left( j\right) }}
\end{equation*}%
\bigskip 

One can also write the states as function of some state $\left\vert
U^{\left( j\right) p}\right\rangle $ by using:%
\begin{equation*}
\left\vert h\left( \left\{ \underline{\mathbf{\hat{\Lambda}}}^{\left[ k_{i}%
\right] }\right\} _{p}\right) \right\rangle =\int g^{\left( j\right) }\left(
U^{\left( j\right) p},h\left( \left\{ \underline{\mathbf{\hat{\Lambda}}}^{%
\left[ k_{i}\right] }\right\} _{p}\right) \right) \left\vert U^{\left(
j\right) p}\right\rangle dU^{\left( j\right) p}
\end{equation*}%
and:%
\begin{equation*}
\left[ \lambda ^{\prime },\left\{ \underline{\mathbf{\hat{\Lambda}}}^{\prime %
\left[ k_{i}\right] }\right\} ,U^{\left( j\right) p\prime }\right]
_{U^{\left( j\right) \prime }}\Phi \left[ \lambda ,\left\{ \underline{%
\mathbf{\hat{\Lambda}}}^{\left[ k_{i}\right] }\right\} ,U^{\left( j\right) p}%
\right] _{U^{\left( j\right) }}
\end{equation*}%
\begin{eqnarray*}
&&\left\vert \left\{ \underline{\mathbf{\hat{\Lambda}}}^{\prime \left[ k_{i}%
\right] }\right\} \right\rangle \left[ \lambda ^{\prime },U^{\left( j\right)
p\prime }\right] _{U^{\left( j\right) \prime }}\Phi \left[ \lambda
,U^{\left( j\right) p}\right] _{U^{\left( j\right) }}\left\langle \left\{ 
\underline{\mathbf{\hat{\Lambda}}}^{\left[ k_{i}\right] }\right\} \right\vert
\\
&=&\left\vert \left\{ \underline{\mathbf{\hat{\Lambda}}}^{\prime \left[ k_{i}%
\right] }\right\} \right\rangle \int g^{\left( j\right) \dag }\left( \left(
U^{\left( j\right) p}\right) ^{\prime },h\left( \left\{ \underline{\mathbf{%
\hat{\Lambda}}^{\prime }}^{\left[ k_{i}\right] }\right\} _{p}\right) \right)
\\
&&\times \left[ \lambda ^{\prime },U^{\left( j\right) p\prime }\right]
_{U^{\left( j\right) \prime }}\Phi \left[ \lambda ,U^{\left( j\right) p}%
\right] _{U^{\left( j\right) }}g^{\left( j\right) }\left( U^{\left( j\right)
p},h\left( \left\{ \underline{\mathbf{\hat{\Lambda}}}^{\left[ k_{i}\right]
}\right\} _{p}\right) \right) dU^{\left( j\right) p}d\left( U^{\left(
j\right) p}\right) ^{\prime }\left\langle \left\{ \underline{\mathbf{\hat{%
\Lambda}}}^{\left[ k_{i}\right] }\right\} \right\vert
\end{eqnarray*}%
writing:%
\begin{equation*}
\left[ \lambda _{0},U^{\left( j\right) p\prime }\right] _{U^{\left( j\right)
\prime }}\Phi \left[ \lambda _{0},U^{\left( j\right) p}\right] _{U^{\left(
j\right) }}=\left[ U^{\left( j\right) p\prime }\right] _{U^{\left( j\right)
\prime }}\Phi \left[ U^{\left( j\right) p}\right] _{U^{\left( j\right) }}
\end{equation*}%
\begin{eqnarray*}
&=&\left\vert \left\{ \underline{\mathbf{\hat{\Lambda}}^{\prime }}^{\left[
k_{i}\right] }\right\} \right\rangle \int g^{\dag }\left( \left( U^{\left(
j\right) p}\right) ^{\prime },h\left( \left\{ \underline{\mathbf{\hat{\Lambda%
}}^{\prime }}^{\left[ k_{i}\right] }\right\} _{p}\right) \right) \\
&&\exp \left( -i\left( \lambda \left( U^{\left( j\right) \prime }\right)
-\lambda _{0}\right) V\right) \left[ U^{\left( j\right) p\prime }\right]
_{U^{\left( j\right) \prime }}\Phi \left[ U^{\left( j\right) p}\right]
_{U^{\left( j\right) }}\exp \left( i\left( \lambda \left( U^{\left( j\right)
}\right) -\lambda _{0}\right) V\right) \\
&&\times g\left( U^{\left( j\right) p},h\left( \left\{ \underline{\mathbf{%
\hat{\Lambda}}}^{\left[ k_{i}\right] }\right\} _{p}\right) \right)
dU^{\left( j\right) p}d\left( U^{\left( j\right) p}\right) ^{\prime
}\left\langle \left\{ \underline{\mathbf{\hat{\Lambda}}}^{\left[ k_{i}\right]
}\right\} \right\vert
\end{eqnarray*}%
The usual transformation corresponds to the case where $V$ is diagonal in
the basis $U^{\left( j\right) p}$ and $\lambda \left( U^{\left( j\right)
\prime }\right) =\lambda \left( U^{\left( j\right) }\right) $:%
\begin{eqnarray*}
&&\exp \left( -i\left( \lambda \left( U^{\left( j\right) \prime }\right)
-\lambda _{0}\right) V\right) \left[ U^{\left( j\right) p\prime }\right]
_{U^{\left( j\right) \prime }}\Phi \left[ U^{\left( j\right) p}\right]
_{U^{\left( j\right) }}\exp \left( i\left( \lambda \left( U^{\left( j\right)
}\right) -\lambda _{0}\right) V\right) \\
&=&\exp \left( -i\left( \lambda \left( U^{\left( j\right) }\right) -\lambda
_{0}\right) \left( V\left( \left( U^{\left( j\right) p}\right) ^{\prime
}\right) -V\left( U^{\left( j\right) p}\right) \right) \right) \left[
U^{\left( j\right) p\prime }\right] _{U^{\left( j\right) \prime }}\Phi \left[
U^{\left( j\right) p}\right] _{U^{\left( j\right) }}
\end{eqnarray*}%
and:%
\begin{eqnarray}
&=&\left\vert \left\{ \underline{\mathbf{\hat{\Lambda}}^{\prime }}^{\left[
k_{i}\right] }\right\} \right\rangle \int g^{\dag }\left( \left( U^{\left(
j\right) p}\right) ^{\prime },h\left( \left\{ \underline{\mathbf{\hat{\Lambda%
}}^{\prime }}^{\left[ k_{i}\right] }\right\} _{p}\right) \right) g\left(
U^{\left( j\right) p},h\left( \left\{ \underline{\mathbf{\hat{\Lambda}}}^{%
\left[ k_{i}\right] }\right\} _{p}\right) \right)  \label{TRPRS} \\
&&\exp \left( -i\left( \lambda \left( U^{\left( j\right) }\right) -\lambda
_{0}\right) \left( V\left( \left( U^{\left( j\right) p}\right) ^{\prime
}\right) -V\left( U^{\left( j\right) p}\right) \right) \right) \left[
U^{\left( j\right) p\prime }\right] _{U^{\left( j\right) \prime }}\Phi \left[
U^{\left( j\right) p}\right] _{U^{\left( j\right) }}  \notag \\
&&\times dU^{\left( j\right) p}d\left( U^{\left( j\right) p}\right) ^{\prime
}\left\langle \left\{ \underline{\mathbf{\hat{\Lambda}}}^{\left[ k_{i}\right]
}\right\} \right\vert  \notag
\end{eqnarray}

\subsubsection{Example}

For operators:%
\begin{equation*}
\left[ \lambda ^{\left( i\right) },U_{j}^{p\prime }\right] \Psi \left[
\lambda ^{\left( i\right) },U_{j}^{p}\right] =\left\vert
U_{j}^{p+k}\right\rangle \left\langle U_{j}^{p}\right\vert
\end{equation*}%
formula (\ref{TRPRS}) writes:%
\begin{equation*}
\left\vert \left\{ \underline{\mathbf{\hat{\Lambda}}^{\prime }}^{\left[ k_{i}%
\right] }\right\} \right\rangle \int g^{\dag }\left( U_{j}^{p+k},h\left(
\left\{ \underline{\mathbf{\hat{\Lambda}}}^{\left[ k_{i}\right] }\right\}
_{p+k}\right) \right) \left\vert U_{j}^{p+k}\right\rangle \left\langle
U_{j}^{p}\right\vert g\left( U_{j}^{p},h\left( \left\{ \underline{\mathbf{%
\hat{\Lambda}}}^{\left[ k_{i}\right] }\right\} _{p}\right) \right)
dU_{j}^{p}dU_{j}^{p+k}\left\langle \left\{ \underline{\mathbf{\hat{\Lambda}}}%
^{\left[ k_{i}\right] }\right\} \right\vert
\end{equation*}%
In the usual set up, we can consider translation invariance: 
\begin{equation*}
g^{\dag }\left( U_{j}^{p+k},h\left( \left\{ \underline{\mathbf{\hat{\Lambda}}%
}^{\left[ k_{i}\right] }\right\} _{p+k}\right) \right) g\left(
U_{j}^{p},h\left( \left\{ \underline{\mathbf{\hat{\Lambda}}}^{\left[ k_{i}%
\right] }\right\} _{p}\right) \right) =\exp \left( iU_{j}^{p+k/p}.\left\{ 
\underline{\mathbf{\hat{\Lambda}}}^{\left[ k_{i}\right] }\right\}
_{p+k/p}\right)
\end{equation*}%
where $p+k/p$ stands for the set $\left\{ \underline{\mathbf{\hat{\Lambda}}}%
^{\left[ k_{i}\right] }\right\} _{p+k}$ , parameters $\left\{ \underline{%
\mathbf{\hat{\Lambda}}}^{\left[ k_{i}\right] }\right\} _{p}$ excluded, so
that the operator is given by:%
\begin{equation*}
\left\vert \left\{ \underline{\mathbf{\hat{\Lambda}}^{\prime }}^{\left[ k_{i}%
\right] }\right\} \right\rangle \int \exp \left( iU_{j}^{p+k/p}.\left\{ 
\underline{\mathbf{\hat{\Lambda}}}^{\left[ k_{i}\right] }\right\}
_{p+k/p}\right) \left\vert U_{j}^{p+k}\right\rangle \left\langle
U_{j}^{p}\right\vert dU_{j}^{p}dU_{j}^{p+k/p}\left\langle \left\{ \underline{%
\mathbf{\hat{\Lambda}}}^{\left[ k_{i}\right] }\right\} \right\vert
\end{equation*}%
This is similar to a usual field description in terms of creation and
annihilation operators.

\section{Number of states and connections amplitude}

\subsection{Number of states}

We write the constraint:%
\begin{equation*}
\left\{ \underline{\mathbf{\hat{\Lambda}}}^{\left[ k_{i}\right] }\left[ \Psi
_{J},\nu ,U_{j}^{l}\right] \right\} _{i}M\left( \left\{ \left[ k_{i}\right] ,%
\left[ k_{i^{\prime }}\right] \right\} ,\Psi _{J},\Psi _{J}^{\otimes
l}\left( U_{j}^{l}\right) \right) .\left\{ \underline{\mathbf{\hat{\Lambda}}}%
^{\left[ k_{i^{\prime }}\right] }\left[ \Psi _{J},\nu ,U_{j}^{l}\right]
\right\} _{i}+\alpha \left( \Psi _{J}\right) -\lambda \left( \Psi
_{J}\right) =0
\end{equation*}

where:%
\begin{equation*}
\alpha \left( \Psi _{J}\right) =\int \Psi _{J}^{\dag }A\Psi _{J}+\int \Psi
_{J}^{\dag }\Psi _{J}^{\dag }B\Psi _{J}\Psi _{J}+...
\end{equation*}%
We assume that the number of states satisfying $\alpha \left( \Psi
_{J}\right) =\lambda $ increases with $\lambda $: 
\begin{equation*}
\sharp N\left( U_{\lambda }^{\left( j\right) }\right) =\sharp \left\{
S\left( \Psi _{J}\right) ,\alpha \left( S\left( \Psi _{J}\right) \right)
=\lambda \right\} \nearrow \text{ fr }\alpha \left( \Psi _{J}\right) \nearrow
\end{equation*}%
with $S\left( \Psi _{J}\right) $ denoting any state of $\Psi _{J}$. This is
justified since the constraint:%
\begin{equation*}
\lambda \left( \Psi _{J}\right) =\left\{ \underline{\mathbf{\hat{\Lambda}}}^{%
\left[ k_{i}\right] }\left[ \Psi _{J},\nu ,U_{j}^{l}\right] \right\}
_{i}M\left( \left\{ \left[ k_{i}\right] ,\left[ k_{i^{\prime }}\right]
\right\} ,\Psi _{J},\Psi _{J}^{\otimes l}\left( U_{j}^{l}\right) \right)
.\left\{ \underline{\mathbf{\hat{\Lambda}}}^{\left[ k_{i^{\prime }}\right] }%
\left[ \Psi _{J},\nu ,U_{j}^{l}\right] \right\} _{i}+\alpha \left( \Psi
_{J}\right)
\end{equation*}%
involves a quadratic quantity depending on the states. Since this is
computed on tensor products of states, increasing $\lambda \left( \Psi
_{J}\right) $ implies a larger number of states satisfying this equation.

\subsection{\textbf{States combinations}:}

We consider some smeared transitions combining the transitions defined in
the previous sections.

For a set decomposed into global system and local one through decomposition $%
\left( U^{\left( j\right) /p},U^{\left( j\right) p}\right) $, we define the
smeared state by $\alpha $ as: 
\begin{eqnarray*}
&&\sum_{U^{\left( j\right) },\lambda \left( U^{\left( j\right) }\right)
=\lambda }\alpha \left( U^{\left( j\right) }\right) \left\vert U^{\left(
j\right) /p},U^{\left( j\right) p},\lambda \left( U^{\left( j\right)
}\right) ,\left\{ \underline{\mathbf{\hat{\Lambda}}}^{\left[ k_{i}\right]
}\left( U^{\left( j\right) }\right) \right\} \right\rangle \\
&=&\sum_{U^{\left( j\right) },MU^{\left( j\right) }=\alpha ^{-1}\left(
\lambda \right) }\alpha \left( U^{\left( j\right) }\right) \left\vert
U^{\left( j\right) /p},U^{\left( j\right) p},\lambda \left( U^{\left(
j\right) }\right) ,\left\{ \underline{\mathbf{\hat{\Lambda}}}^{\left[ k_{i}%
\right] }\left( U^{\left( j\right) }\right) \right\} \right\rangle \\
&\equiv &\left\vert \alpha ,U^{\left( j\right) p},\lambda ,\left\{ 
\underline{\mathbf{\hat{\Lambda}}}^{\left[ k_{i}\right] }\right\}
\right\rangle
\end{eqnarray*}%
We can consider the map:%
\begin{equation*}
\left\vert \alpha ,U^{\left( j\right) p},\lambda ,\left\{ \underline{\mathbf{%
\hat{\Lambda}}}^{\left[ k_{i}\right] }\right\} \right\rangle _{\left\{
\alpha \right\} }\overset{\delta }{\rightarrow }\left\vert \alpha ^{\prime
},\left( U^{\left( j\right) p}\right) ^{\prime },\lambda ^{\prime },\left\{ 
\underline{\mathbf{\hat{\Lambda}}}^{\left[ k_{i}\right] }\right\}
\right\rangle _{\left\{ \alpha ^{\prime }\right\} }
\end{equation*}%
for $\lambda ^{\prime }>\lambda $ since $\sharp N\left( U_{\lambda ^{\prime
}}^{\left( j\right) }\right) >\sharp N\left( U_{\lambda }^{\left( j\right)
}\right) $ with $\lambda ^{\prime }>\lambda $. The corresponding amplitude
are computed through (\ref{TRM}).

\subsection{\textbf{Connection}}

We consider the set of \ infinitesimal maps 
\begin{equation*}
\delta _{k}:\mathcal{H}_{\left( \lambda ,\left\{ \underline{\mathbf{\hat{%
\Lambda}}}^{\left[ k_{i}\right] }\left( U_{\lambda }^{\left( j\right)
}\right) ,U_{\lambda }^{\left( j\right) }\right\} \right) }\rightarrow 
\mathcal{H}_{\left( \lambda +\delta _{k}\lambda ,\left\{ \underline{\mathbf{%
\hat{\Lambda}}}^{\left[ k_{i}\right] }\left( U_{\lambda }^{\left( j\right)
}\right) ,U_{\lambda }^{\left( j\right) }\right\} \right) }
\end{equation*}%
summed up by following graph:%
\begin{equation*}
\underset{\searrow }{\left\vert \lambda \right\rangle +\delta _{2}\left\vert
\lambda \right\rangle }\leftarrow \underset{\left\vert \lambda \right\rangle
+\delta _{k}\left\vert \lambda \right\rangle \nearrow }{\overset{\left\vert
\lambda \right\rangle +\delta _{1}\left\vert \lambda \right\rangle }{%
\underset{\downarrow }{\overset{\uparrow }{\left\vert \lambda \right\rangle }%
}}}\rightarrow \left\vert \lambda \right\rangle +\delta _{n}\left\vert
\lambda \right\rangle
\end{equation*}%
where $\left\vert \lambda \right\rangle $ represents $\mathcal{H}_{\left(
\lambda ,\left\{ \underline{\mathbf{\hat{\Lambda}}}^{\left[ k_{i}\right]
}\left( U_{\lambda }^{\left( j\right) }\right) ,U_{\lambda }^{\left(
j\right) }\right\} \right) }$ and $\left\vert \lambda \right\rangle +\delta
_{k}\left\vert \lambda \right\rangle $ stands for $\mathcal{H}_{\left(
\lambda +\delta _{k},\left\{ \underline{\mathbf{\hat{\Lambda}}}^{\left[ k_{i}%
\right] }\left( U_{\lambda }^{\left( j\right) }\right) ,U_{\lambda }^{\left(
j\right) }\right\} \right) }$. Our aim is to sum transitions over all these
maps and compute an average $\delta \lambda $. In the sequel, we assume that
the parameter $\lambda \left( U^{\left( j\right) }\right) $ is defined by
the "surrounding" part of the system: 
\begin{equation*}
\lambda \left( U^{\left( j\right) }\right) \simeq \lambda \left( U^{\left(
j\right) /p}\right)
\end{equation*}%
We also consider the number of maps between $\left\vert \lambda
\right\rangle $ and $\left\vert \lambda \right\rangle +\delta \left\vert
\lambda \right\rangle $:%
\begin{equation*}
N\left( \left( U_{j}\right) _{\lambda }\hookrightarrow \left( U_{j}\right)
_{\lambda +\delta \lambda }\right)
\end{equation*}%
counting the number of maps from $\left( U_{j}\right) _{\lambda }$ to $%
\left( U_{j}\right) _{\lambda +\delta \lambda }$.%
\begin{equation*}
\delta T\left\vert \lambda ,\left\{ \underline{\mathbf{\hat{\Lambda}}}^{%
\left[ k_{i}\right] }\right\} _{\lambda },U^{\left( j\right) p}\right\rangle
=\sum_{\left( U_{\lambda +\delta \lambda }^{\left( j\right) }\right)
^{\prime }}\frac{N\left( U_{\lambda }^{\left( j\right) }\rightarrow \left(
U_{\lambda +\delta \lambda }^{\left( j\right) }\right) ^{\prime }\right) }{%
\sum N\left( U_{\lambda }^{\left( j\right) }\rightarrow \left( U_{\lambda
+\delta \lambda }^{\left( j\right) }\right) ^{\prime }\right) }T_{U_{\lambda
}^{\left( j\right) },\left( U_{\lambda +\delta \lambda }^{\left( j\right)
}\right) ^{\prime }}\left\vert \lambda ,\left\{ \underline{\mathbf{\hat{%
\Lambda}}}^{\left[ k_{i}\right] }\right\} _{\lambda },U^{\left( j\right)
p}\right\rangle
\end{equation*}%
\begin{eqnarray*}
N_{k_{0}} &>&>N_{k},k\neq k_{0} \\
\delta T &=&\sum_{\left( U_{\lambda +\delta \lambda }^{\left( j\right)
}\right) ^{\prime }}\frac{N\left( U_{\lambda }^{\left( j\right) }\rightarrow
\left( U_{\lambda +\delta \lambda }^{\left( j\right) }\right) ^{\prime
}\right) }{\sum N\left( U_{\lambda }^{\left( j\right) }\rightarrow \left(
U_{\lambda +\delta \lambda }^{\left( j\right) }\right) ^{\prime }\right) }%
T_{U_{\lambda }^{\left( j\right) },\left( U_{\lambda +\delta \lambda
}^{\left( j\right) }\right) ^{\prime }}
\end{eqnarray*}

We will consider the transitions for states composed from:%
\begin{equation*}
\left\vert U^{\left( j\right) }\right\rangle =\left\vert U^{\left( j\right)
/p}\right\rangle \left\vert U^{\left( j\right) p}\right\rangle
\end{equation*}%
and smeared by coefficients $\alpha $:%
\begin{equation*}
\left\vert \alpha ,U^{\left( j\right) p},\lambda _{0},\left\{ \underline{%
\mathbf{\hat{\Lambda}}}^{\left[ k_{i}\right] }\right\} _{\lambda
_{0}}\right\rangle =\sum_{U^{\left( j\right) },\lambda \left( U^{\left(
j\right) }\right) =\lambda _{0}}\alpha \left( U^{\left( j\right) }\right)
\left\vert U^{\left( j\right) /p},U^{\left( j\right) p},\lambda _{0},\left\{ 
\underline{\mathbf{\hat{\Lambda}}}^{\left[ k_{i}\right] }\right\} _{\lambda
_{0}}\right\rangle
\end{equation*}%
And a sequence of infinitesimal transitions will write:%
\begin{equation*}
\left\vert \alpha ^{\left( 1\right) },\lambda _{0}^{\left( 1\right)
},U^{\left( j\right) p},\left\{ \underline{\mathbf{\hat{\Lambda}}}^{\left[
k_{i}\right] }\right\} _{\lambda _{0}^{\left( 1\right) }}\right\rangle
\rightarrow \left\vert \alpha ^{\left( 2\right) },\lambda _{0}^{\left(
2\right) },U^{\left( j\right) p},\left\{ \underline{\mathbf{\hat{\Lambda}}}^{%
\left[ k_{i}\right] }\right\} _{\lambda _{0}^{\left( 2\right)
}}\right\rangle \rightarrow ...\rightarrow \left\vert \alpha ^{\left(
n\right) },\lambda _{0}^{\left( n\right) },U^{\left( j\right) p},\left\{ 
\underline{\mathbf{\hat{\Lambda}}}^{\left[ k_{i}\right] }\right\} _{\lambda
_{0}^{\left( n\right) }}\right\rangle
\end{equation*}%
with:%
\begin{equation*}
\lambda _{0}^{\left( l\right) }>\lambda _{0}^{\left( l-1\right) }
\end{equation*}%
Covariantly, this sequence writes also:%
\begin{equation*}
\left\vert \Sigma _{1}\left( U_{1}^{\left( j\right) }\right) \right\rangle
\rightarrow \left\vert \Sigma _{2}\left( U_{2}^{\left( j\right) }\right)
\right\rangle \rightarrow \left\vert \Sigma _{3}\left( U_{3}^{\left(
j\right) }\right) \right\rangle \rightarrow ...
\end{equation*}%
with:%
\begin{equation*}
\lambda \left( U_{n}^{\left( j\right) }\right) >...>\lambda \left(
U_{1}^{\left( j\right) }\right)
\end{equation*}

\subsection{\textbf{Mechanism:}}

\subsubsection{Global principle}

Recall that:%
\begin{equation*}
U_{\lambda }^{\left( j\right) }=U^{\left( j\right) }\text{ s.t. }\lambda
\left( U^{\left( j\right) }\right) =\lambda
\end{equation*}

The transition for a state $\left( U^{\left( j\right) /p},U^{\left( j\right)
p}\right) $ writes: 
\begin{eqnarray*}
&&\delta _{\lambda _{0}\rightarrow \lambda }\left\vert \left( U^{\left(
j\right) /p},U^{\left( j\right) p}\right) _{\lambda _{0}},\lambda
_{0},\left\{ \underline{\mathbf{\hat{\Lambda}}}^{\left[ k_{i}\right]
}\right\} _{\lambda _{0}}\right\rangle \\
&=&\sum_{U_{\lambda }^{\left( j\right) },\lambda \left( U_{\lambda }^{\left(
j\right) }\right) =\lambda }\frac{N\left( U_{\lambda _{0}}^{\left( j\right)
}\rightarrow U_{\lambda }^{\left( j\right) }\right) \exp \left( i\left(
\lambda \left( U_{\lambda _{0}}^{\left( j\right) }\right) -\lambda \left(
U_{\lambda }^{\left( j\right) }\right) \right) V\left( U_{\lambda }^{\left(
j\right) },U_{\lambda _{0}}^{\left( j\right) }\right) \right) }{%
\sum_{U_{\lambda }^{\left( j\right) },\lambda \left( U_{\lambda }^{\left(
j\right) }\right) =\lambda }N\left( U_{\lambda _{0}}^{\left( j\right)
}\rightarrow U_{\lambda }^{\left( j\right) }\right) }\left\vert \left(
U^{\left( j\right) /p},U^{\left( j\right) p}\right) _{\lambda _{0}},\lambda
_{0},\left\{ \underline{\mathbf{\hat{\Lambda}}}^{\left[ k_{i}\right]
}\right\} _{\lambda _{0}}\right\rangle
\end{eqnarray*}%
which rewrites after smearing:%
\begin{eqnarray*}
&&\delta _{\lambda _{0}\rightarrow \lambda }\left\vert \alpha ,\left(
U^{\left( j\right) p}\right) _{0},\lambda _{0},\left\{ \underline{\mathbf{%
\hat{\Lambda}}}^{\left[ k_{i}\right] }\right\} _{\lambda _{0}}\right\rangle
\\
&=&\sum_{\substack{ U_{\lambda }^{\left( j\right) }  \\ \lambda \left(
U_{\lambda }^{\left( j\right) }\right) =\lambda }}\sum_{\substack{ %
U_{\lambda _{0}}^{\left( j\right) }  \\ \lambda \left( U_{\lambda
_{0}}^{\left( j\right) }\right) =\lambda _{0}}}\frac{N\left( U_{\lambda
_{0}}^{\left( j\right) }\rightarrow U_{\lambda }^{\left( j\right) }\right)
\exp \left( i\left( \lambda _{0}-\lambda \right) V\left( U_{\lambda
}^{\left( j\right) },U_{\lambda _{0}}^{\left( j\right) }\right) \right)
\alpha \left( U_{\lambda _{0}}^{\left( j\right) }\right) }{\sum_{\substack{ %
U_{\lambda }^{\left( j\right) },\lambda \left( U_{\lambda }^{\left( j\right)
}\right) =\lambda ,  \\ U_{\lambda _{0}}^{\left( j\right) },\lambda \left(
U_{\lambda _{0}}^{\left( j\right) }\right) =\lambda _{0}}}N\left( U_{\lambda
_{0}}^{\left( j\right) }\rightarrow U_{\lambda }^{\left( j\right) }\right) }%
\left\vert \left( U^{\left( j\right) /p},U^{\left( j\right) p}\right)
_{\lambda _{0}},\lambda _{0},\left\{ \underline{\mathbf{\hat{\Lambda}}}^{%
\left[ k_{i}\right] }\right\} _{\lambda _{0}}\right\rangle
\end{eqnarray*}%
As before, we assume that we can decompose the degrees of freedom $%
U_{\lambda }^{\left( j\right) p}$ and $U_{\lambda }^{\left( j\right) /p}$:%
\begin{equation*}
V\left( U_{\lambda }^{\left( j\right) },U_{\lambda _{0}}^{\left( j\right)
}\right) =V\left( U_{\lambda }^{\left( j\right) p},U_{\lambda _{0}}^{\left(
j\right) p}\right) +V\left( U_{\lambda }^{\left( j\right) /p},U_{\lambda
_{0}}^{\left( j\right) /p}\right)
\end{equation*}%
and:%
\begin{eqnarray*}
&&\exp \left( i\left( \lambda _{0}-\lambda \right) V\left( U_{\lambda
}^{\left( j\right) },U_{\lambda _{0}}^{\left( j\right) }\right) \right)
\left\vert \left( U^{\left( j\right) /p},U^{\left( j\right) p}\right)
_{\lambda _{0}},\lambda _{0},\left\{ \underline{\mathbf{\hat{\Lambda}}}^{%
\left[ k_{i}\right] }\right\} _{\lambda _{0}}\right\rangle \\
&=&\sum_{U^{\left( j\right) p}}\gamma \left( \left( U^{\left( j\right)
p}\right) _{\lambda _{0}},\left( U^{\left( j\right) p}\right) \right)
\left\vert \left( U^{\left( j/p\right) },\left( U^{\left( j\right) p}\right)
\right) _{\lambda },\lambda ,\left\{ \underline{\mathbf{\hat{\Lambda}}}^{%
\left[ k_{i}\right] }\right\} _{\lambda }\right\rangle
\end{eqnarray*}%
As a consequence, the statistic transition is:%
\begin{eqnarray*}
&&\delta _{\lambda _{0}\rightarrow \lambda }\left\vert \alpha ,U_{\lambda
_{0}}^{\left( j\right) p},\lambda _{0},\left\{ \underline{\mathbf{\hat{%
\Lambda}}}^{\left[ k_{i}\right] }\right\} _{\lambda _{0}}\right\rangle \\
&=&\sum_{U_{\lambda }^{\left( j\right) },\lambda \left( U_{\lambda }^{\left(
j\right) }\right) =\lambda }\left( \frac{N\left( U_{\lambda _{0}}^{\left(
j\right) }\rightarrow U_{\lambda }^{\left( j\right) }\right) }{\sum 
_{\substack{ U_{\lambda }^{\left( j\right) },\lambda \left( U_{\lambda
}^{\left( j\right) }\right) =\lambda ,  \\ U_{\lambda _{0}}^{\left( j\right)
},\lambda \left( U_{\lambda _{0}}^{\left( j\right) }\right) =\lambda _{0}}}%
N\left( U_{\lambda _{0}}^{\left( j\right) }\rightarrow U_{\lambda }^{\left(
j\right) }\right) }\right. \\
&&\left. \times \sum_{\substack{ U_{\lambda _{0}}^{\left( j\right) },\lambda
\left( U_{\lambda _{0}}^{\left( j\right) }\right) =\lambda _{0},  \\ \delta
_{\lambda _{0}\rightarrow \lambda }U_{\lambda _{0}}^{\left( j\right)
}=U_{\lambda }^{\left( j\right) }}}\alpha \left( U_{\lambda _{0}}^{\left(
j\right) }\right) \gamma \left( \left( U^{\left( j\right) p}\right)
_{\lambda _{0}},\left( U^{\left( j\right) p}\right) \right) \right)
\left\vert \left( U^{\left( j/p\right) },\left( U^{\left( j\right) p}\right)
\right) _{\lambda },\lambda ,\left\{ \underline{\mathbf{\hat{\Lambda}}}^{%
\left[ k_{i}\right] }\right\} _{\lambda }\right\rangle \\
&\equiv &\sum_{U_{\lambda }^{\left( j\right) },\lambda \left( U_{\lambda
}^{\left( j\right) }\right) =\lambda }\alpha ^{\prime }\left( U_{\lambda
}^{\left( j\right) },\alpha ,\left( U^{\left( j\right) p}\right) _{\lambda
_{0}}\right) \left\vert \left( U^{\left( j/p\right) },\left( U^{\left(
j\right) p}\right) \right) _{\lambda },\lambda ,\left\{ \underline{\mathbf{%
\hat{\Lambda}}}^{\left[ k_{i}\right] }\right\} _{\lambda }\right\rangle
=\left\vert \alpha ^{\prime },U_{\lambda }^{\left( j\right) p},\lambda
,\left\{ \underline{\mathbf{\hat{\Lambda}}}^{\left[ k_{i}\right] }\right\}
_{\lambda }\right\rangle
\end{eqnarray*}

which writes in a compact form:%
\begin{equation*}
\delta _{\lambda _{0}\rightarrow \lambda }\left\vert \alpha ,U_{\lambda
_{0}}^{\left( j\right) p},\lambda _{0},\left\{ \underline{\mathbf{\hat{%
\Lambda}}}^{\left[ k_{i}\right] }\right\} _{\lambda _{0}}\right\rangle
_{\left\{ \alpha \right\} }=\left\vert \alpha ^{\prime },U_{\lambda
}^{\left( j\right) p},\lambda ,\left\{ \underline{\mathbf{\hat{\Lambda}}}^{%
\left[ k_{i}\right] }\right\} _{\lambda }\right\rangle _{\left\{ \alpha
^{\prime }\right\} }
\end{equation*}%
As before, locally, we can assume that:%
\begin{equation*}
\left\{ \underline{\mathbf{\hat{\Lambda}}}^{\left[ k_{i}\right] }\right\}
_{\lambda _{0}}=\left\{ \underline{\mathbf{\hat{\Lambda}}}^{\left[ k_{i}%
\right] }\right\} _{\lambda }=\left\{ \underline{\mathbf{\hat{\Lambda}}}^{%
\left[ k_{i}\right] }\right\}
\end{equation*}%
and that the change in manifold:%
\begin{equation*}
\left\{ \underline{\mathbf{\hat{\Lambda}}}^{\left[ k_{i}\right] }\right\}
_{\lambda },\lambda
\end{equation*}%
is included in the transition $V$ through the sum over maps.

\subsubsection{Infinitesimal transitions}

\begin{eqnarray*}
&&\delta \left\vert \left( U^{\left( j\right) /p},U^{\left( j\right)
p}\right) _{\lambda _{0}},\lambda _{0},\left\{ \underline{\mathbf{\hat{%
\Lambda}}}^{\left[ k_{i}\right] }\right\} _{\lambda _{0}}\right\rangle \\
&=&\sum_{\delta \lambda }\sum_{U_{\lambda _{0}}^{\left( j\right) }+\delta
U_{\lambda _{0}}^{\left( j\right) }\mid \delta \lambda \left( U_{\lambda
_{0}}^{\left( j\right) }\right) =\delta \lambda }\frac{N\left( U_{\lambda
_{0}}^{\left( j\right) }\rightarrow U_{\lambda _{0}}^{\left( j\right)
}+\delta U_{\lambda _{0}}^{\left( j\right) }\mid \delta \lambda \left(
\delta U_{\lambda _{0}}^{\left( j\right) }\right) =\delta \lambda \right) }{%
\sum_{\delta \lambda }\sum_{U_{\lambda _{0}}^{\left( j\right) }+\delta
U_{\lambda _{0}}^{\left( j\right) }}N\left( U_{\lambda _{0}}^{\left(
j\right) }\rightarrow U_{\lambda _{0}}^{\left( j\right) }+\delta U_{\lambda
_{0}}^{\left( j\right) }\right) } \\
&&\times \exp \left( i\delta \lambda V\left( U_{\lambda }^{\left( j\right)
},U_{\lambda _{0}}^{\left( j\right) }\right) \right) \left\vert \left(
U^{\left( j\right) /p},U^{\left( j\right) p}\right) _{\lambda _{0}},\lambda
_{0},\left\{ \underline{\mathbf{\hat{\Lambda}}}^{\left[ k_{i}\right]
}\right\} _{\lambda _{0}}\right\rangle
\end{eqnarray*}%
with $\delta U_{\lambda _{0}}^{\left( j\right) }$ an elementary modification
tht is defnd by:%
\begin{equation*}
\nexists \left( \delta U_{\lambda _{0}}^{\left( j\right) }\right) ^{\prime
}\neq \delta U_{\lambda _{0}}^{\left( j\right) }:U_{\lambda _{0}}^{\left(
j\right) }\rightarrow \left( \delta U_{\lambda _{0}}^{\left( j\right)
}\right) ^{\prime }\rightarrow \delta U_{\lambda _{0}}^{\left( j\right) }
\end{equation*}%
\begin{eqnarray*}
&&\delta \left\vert \alpha ,\left( U^{\left( j\right) p}\right) _{0},\lambda
_{0},\left\{ \underline{\mathbf{\hat{\Lambda}}}^{\left[ k_{i}\right]
}\right\} _{\lambda _{0}}\right\rangle \\
&=&\sum_{U_{\lambda _{0}}^{\left( j\right) },\lambda \left( U_{\lambda
_{0}}^{\left( j\right) }\right) =\lambda _{0}}\sum_{\delta \lambda
}\sum_{U_{\lambda _{0}}^{\left( j\right) }+\delta U_{\lambda _{0}}^{\left(
j\right) }\mid \delta \lambda \left( \delta U_{\lambda _{0}}^{\left(
j\right) }\right) =\delta \lambda } \\
&&\frac{N\left( U_{\lambda _{0}}^{\left( j\right) }\rightarrow U_{\lambda
_{0}+\delta \lambda }^{\left( j\right) }\right) \exp \left( i\delta \lambda
V\left( U_{\lambda _{0}+\delta \lambda }^{\left( j\right) },U_{\lambda
_{0}}^{\left( j\right) }\right) \right) }{\sum_{\delta \lambda }\sum 
_{\substack{ U_{\lambda }^{\left( j\right) },\lambda \left( U_{\lambda
}^{\left( j\right) }\right) =\lambda _{0}+\delta \lambda ,  \\ U_{\lambda
_{0}}^{\left( j\right) },\lambda \left( U_{\lambda _{0}}^{\left( j\right)
}\right) =\lambda _{0}}}N\left( U_{\lambda _{0}}^{\left( j\right)
}\rightarrow U_{\lambda _{0}+\delta \lambda }^{\left( j\right) }\right) }%
\alpha \left( U_{\lambda _{0}}^{\left( j\right) }\right) \left\vert \left(
U^{\left( j\right) /p},U^{\left( j\right) p}\right) _{\lambda _{0}},\lambda
_{0},\left\{ \underline{\mathbf{\hat{\Lambda}}}^{\left[ k_{i}\right]
}\right\} _{\lambda _{0}}\right\rangle \\
&\equiv &\sum_{\delta \lambda }\sum_{U_{\lambda _{0}}^{\left( j\right)
}+\delta U_{\lambda _{0}}^{\left( j\right) }\mid \delta \lambda \left(
\delta U_{\lambda _{0}}^{\left( j\right) }\right) =\delta \lambda }\alpha
^{\prime }\left( U_{\lambda }^{\left( j\right) },\alpha ,\delta \lambda
,U_{\lambda _{0}}^{\left( j\right) p}\right) \left\vert U_{\lambda
_{0}+\delta \lambda }^{\left( j\right) /p},U_{\lambda }^{\left( j\right)
p},\lambda _{0}+\delta \lambda ,\left\{ \underline{\mathbf{\hat{\Lambda}}}^{%
\left[ k_{i}\right] }\right\} _{\lambda _{0}+\delta \lambda }\right\rangle \\
&\equiv &\left\vert \bar{\alpha}^{\prime },\left( U^{\left( j\right)
p}\right) ,\lambda _{0}+\delta \bar{\lambda},\left\{ \underline{\mathbf{\hat{%
\Lambda}}}^{\left[ k_{i}\right] }\right\} _{\lambda _{0}+\delta \bar{\lambda}%
}\right\rangle
\end{eqnarray*}%
If we consider the $U^{\left( j\right) p}$ states and assuming that locally
we can identify the $\left\{ \underline{\mathbf{\hat{\Lambda}}}^{\left[ k_{i}%
\right] }\right\} _{\lambda }$ with a constant set of parametrs:%
\begin{equation*}
\left\{ \underline{\mathbf{\hat{\Lambda}}}^{\left[ k_{i}\right] }\right\}
_{\lambda }\rightarrow \left\{ \underline{\mathbf{\hat{\Lambda}}}^{\left[
k_{i}\right] }\right\}
\end{equation*}%
then the transition becomes%
\begin{eqnarray*}
\delta \left\vert \left( U^{\left( j\right) p}\right) _{0},\lambda
_{0},\left\{ \underline{\mathbf{\hat{\Lambda}}}^{\left[ k_{i}\right]
}\right\} \right\rangle &=&\left\vert \left( U^{\left( j\right) p}\right)
_{0},\lambda _{0}+\delta \bar{\lambda},\left\{ \underline{\mathbf{\hat{%
\Lambda}}}^{\left[ k_{i}\right] }\right\} \right\rangle \\
&=&\exp \left( i\delta \bar{\lambda}V\right) \left\vert \left( U^{\left(
j\right) p}\right) _{0},\lambda _{0}+\delta \bar{\lambda},\left\{ \underline{%
\mathbf{\hat{\Lambda}}}^{\left[ k_{i}\right] }\right\} \right\rangle
\end{eqnarray*}%
where:%
\begin{equation*}
\exp \left( i\delta \bar{\lambda}V\right) =\sum_{\delta \lambda
}\sum_{U_{\lambda _{0}}^{\left( j\right) },\lambda \left( U_{\lambda
_{0}}^{\left( j\right) }\right) =\lambda _{0}}\sum_{U_{\lambda _{0}}^{\left(
j\right) }+\delta U_{\lambda _{0}}^{\left( j\right) }\mid \delta \lambda
\left( \delta U_{\lambda _{0}}^{\left( j\right) }\right) =\delta \lambda }%
\frac{N\left( U_{\lambda _{0}}^{\left( j\right) }\rightarrow U_{\lambda
_{0}+\delta \lambda }^{\left( j\right) }\right) \exp \left( i\delta \lambda
V\left( U_{\lambda _{0}+\delta \lambda }^{\left( j\right) },U_{\lambda
_{0}}^{\left( j\right) }\right) \right) }{\sum_{\delta \lambda }\sum 
_{\substack{ U_{\lambda }^{\left( j\right) },\lambda \left( U_{\lambda
}^{\left( j\right) }\right) =\lambda _{0}+\delta \lambda ,  \\ U_{\lambda
_{0}}^{\left( j\right) },\lambda \left( U_{\lambda _{0}}^{\left( j\right)
}\right) =\lambda _{0}}}N\left( U_{\lambda _{0}}^{\left( j\right)
}\rightarrow U_{\lambda _{0}+\delta \lambda }^{\left( j\right) }\right) }%
\alpha \left( U_{\lambda _{0}}^{\left( j\right) }\right)
\end{equation*}%
and $\delta \bar{\lambda}$ is given by a first order expansion:%
\begin{equation*}
\delta \bar{\lambda}=\sum_{U_{\lambda _{0}}^{\left( j\right) },\lambda
\left( U_{\lambda _{0}}^{\left( j\right) }\right) =\lambda _{0}}\sum_{\delta
\lambda }\sum_{U_{\lambda _{0}}^{\left( j\right) }+\delta U_{\lambda
_{0}}^{\left( j\right) }\mid \delta \lambda \left( \delta U_{\lambda
_{0}}^{\left( j\right) }\right) =\delta \lambda }N\left( U_{\lambda
_{0}}^{\left( j\right) }\rightarrow U_{\lambda _{0}+\delta \lambda }^{\left(
j\right) }\right) \delta \lambda
\end{equation*}%
under our previous hypothesis $N\left( U_{\lambda _{0}}^{\left( j\right)
}\rightarrow U_{\lambda _{0}+\delta \lambda }^{\left( j\right) }\right)
\nearrow $ for $\delta \lambda \nearrow $. Then the averaged parameter is
positive: 
\begin{equation*}
\delta \bar{\lambda}>0
\end{equation*}%
As a consequence the connection between the different spaces $\mathcal{H}%
_{\left( \lambda ,\left\{ \underline{\mathbf{\hat{\Lambda}}}^{\left[ k_{i}%
\right] }\left( U_{\lambda }^{\left( j\right) }\right) ,U_{\lambda }^{\left(
j\right) }\right\} \right) }$ are weighted such that in average $\delta
\lambda >0$. In other word, the amplitudes:%
\begin{equation*}
\mathcal{H}_{\left( \lambda ,\left\{ \underline{\mathbf{\hat{\Lambda}}}^{%
\left[ k_{i}\right] }\left( U_{\lambda }^{\left( j\right) }\right)
,U_{\lambda }^{\left( j\right) }\right\} \right) }\rightarrow \mathcal{H}%
_{\left( \lambda ^{\prime },\left\{ \underline{\mathbf{\hat{\Lambda}}}^{%
\left[ k_{i}\right] }\left( U_{\lambda ^{\prime }}^{\left( j\right) }\right)
,U_{\lambda ^{\prime }}^{\left( j\right) }\right\} \right) }
\end{equation*}%
are greater for $\lambda ^{\prime }>\lambda $.

\part*{Part II General formalism}

We extend the previous formalism in the following manner. We assume that
decompositions can involve an arbitrary set of subobjects, and they can be
multiple; that is, several possible decompositions may be considered
simultaneously, akin to a superposition of states. Additionally,
decompositions can have an impact on subobjects. To achieve this, we
consider that each state space involved may itself include subobjects,
representing some bound states that concentrate a series of states in the
considered space. Such subobjects are themselves described by their field,
considered as independent quantities, eventhough these fields are connected
to the initial field by restriction maps. Furthermore, any subobject may
itself be decomposed by the decomposition of the space to which it belongs.
In other words, an initial subobject may be decomposed into subobjects of
this subobject. This is achieved by introducing products and compositions of
decompositions.

\section{Fields formulation}

\subsection{General set up}

We keep the formalism and notations of the first part and consider some
parameters spaces $U,U^{k}$,... with $U^{k}$ is given by $k$ copies of $U$.
We also consider implicit relations:%
\begin{equation*}
U/c\left( U\right) ,...,U^{k}/c\left( U^{k}\right)
\end{equation*}%
where the $c\left( U^{k}\right) $ are constraints and states on $\cup
U^{k}/c\left( U^{k}\right) $ are tensor products plus constraints:%
\begin{eqnarray*}
&&\sum_{k,i_{1}...i_{k}}\left\vert u_{i_{1}}\right\rangle \otimes ...\otimes
\left\vert u_{i_{k}}\right\rangle /c\left( \left(
u_{i_{1}},...,u_{i_{k}}\right) _{k}\right) \\
&\rightarrow &\sum_{k,i_{1}...i_{k}}\left\vert
u_{i_{1},}...,u_{i_{k}}\right\rangle
\end{eqnarray*}%
As before, the states are built from fields $\Psi ^{\otimes k}\left(
U^{k}\right) $ and up to the constraints, a realization $\Psi ^{\otimes
k+l}\left( U^{k+l}\right) $ is a sum of products of such realizations:%
\begin{equation*}
\sum_{\alpha }\Psi _{\alpha }^{\otimes k}\left( U^{k}\right) \Psi _{\alpha
}^{\otimes l}\left( U^{l}\right)
\end{equation*}%
so that a realization $\Psi ^{\otimes k+l}\left(
u_{i_{1}},...,u_{i_{k+l}}\right) $ can decomposed as sum of products of
realization:%
\begin{equation*}
\Psi ^{\otimes k+l}\left( u_{i_{1}},...,u_{i_{k+l}}\right) \rightarrow
\sum_{\alpha }\Psi _{\alpha }^{\otimes k}\left(
u_{i_{1}},...,u_{i_{k}}\right) \Psi _{\alpha }^{\otimes l}\left(
u_{i_{1+k}},...,u_{i_{k+l}}\right)
\end{equation*}%
The constraints are included by identifying some parameters:%
\begin{equation*}
\Psi \left( U^{l}\right) =\Psi ^{\otimes l}\left( U^{l}\right) /f
\end{equation*}%
ndrstd tht $U_{j}^{k}$ stnds fr $U_{j}^{k}/c\left( U_{j}^{k}\right) $. and
the tensor product is a particular realization of $\Psi \left( U^{l}\right) $%
: 
\begin{equation*}
\Psi \left( u_{i_{1}},...u_{i_{k}}\right) =\Psi \left( u_{i_{1}}\right)
\otimes ...\otimes \Psi \left( u_{i_{k}}\right) /f\left( \left(
u_{i_{1}},...,u_{i_{k}}\right) \right)
\end{equation*}

\subsection{Subspaces, subobjects}

We generalize the notion of decomposition. To do so we detail the notion of
subobjects and maps of subobjects.

\subsubsection{Subobjects maps}

For the set:%
\begin{equation*}
U_{j}^{\left( p\right) }\subset \oplus _{k}U_{j}^{k}
\end{equation*}%
We consider a collection of subjects, that is a st of maps:%
\begin{equation*}
\left\{ U_{j}\right\} _{p}=\left\{ U_{j}^{\left( p\right) }\overset{p}{%
\hookrightarrow }\mathcal{V}\left( \oplus _{k}U_{j}^{k}\right) \right\} _{p}
\end{equation*}%
The set $\mathcal{V}\left( \oplus _{k}U_{j}^{k}\right) $ is formally a
collection of submanifold of $\oplus _{k}U_{j}^{k}$. It includes the single
point $\ast $ fr $k=0$. Maps $p$ are running over a set of indices. For a
given collection $\left( p_{l}\right) $, we define the maps by:

\begin{eqnarray*}
p_{l} &:&\left( U_{j}\right) ^{\left( p_{l}\right) }\rightarrow \mathcal{V}%
\left( \oplus _{k}U_{j}^{k}\right) \\
u^{\left( p_{l}\right) } &\rightarrow &\oplus _{k}\left\{ \left(
u_{i_{1}},...u_{i_{k}}\right) \left( u^{\left( p_{l}\right) }\right)
\right\} ,f_{k}\left( u_{i_{1}},...u_{i_{k}},u^{\left( p_{l}\right) }\right)
=0
\end{eqnarray*}%
where $\left\{ \left( u_{i_{1}},...u_{i_{k}}\right) \left( u^{\left(
p_{l}\right) }\right) \right\} $ is the subvariety defined by several
equation:%
\begin{equation*}
f_{k}\left( u_{i_{1}},...u_{i_{k}},u^{\left( p_{l}\right) }\right) =0
\end{equation*}

These maps have a translation in terms of states:%
\begin{equation*}
s_{p_{l}}:H\left( \left( U_{j}\right) ^{\left( p_{l}\right) }\right)
\hookrightarrow \oplus _{k}H^{\otimes k}\left( U_{j}\right)
\end{equation*}%
To detail $s_{p_{l}}$ we define:%
\begin{equation*}
i_{k}:%
\begin{array}{c}
\left( U_{j}\right) ^{k}\rightarrow H^{\otimes k}\left( U_{j}\right) \\ 
\left( u_{i_{1}},...,u_{i_{k}}\right) \rightarrow \left\vert
u_{i_{1}},...,u_{i_{k}}\right\rangle%
\end{array}%
\end{equation*}%
and we define $s_{p_{l}}$ as%
\begin{equation*}
s_{p_{l}}\left( u\right) =\int_{\left( U_{j}\right) ^{k}\cap p_{l}\left(
u^{\left( p_{l}\right) }\right) }g_{k,p_{l}}\left( u^{\left( p_{l}\right)
},u\right) i_{k}\left( u\right)
\end{equation*}%
The maps $g_{k,p_{l}}$ characterizes the application $p_{l}$. To shorten the
notations and since $k$ and $p_{l}$ arises explictely through $u^{\left(
p_{l}\right) }$ and $u$, we will write $g$ for $g_{k,p_{l}}$.

In coordinates:%
\begin{equation*}
s_{p_{l}}:\left\vert u^{\left( p_{l}\right) }\right\rangle \rightarrow
\sum_{k}\int_{\left( u_{i_{1}},...,u_{i_{k}}\right) \in p_{l}\left(
u^{\left( p_{l}\right) }\right) }g\left( u^{\left( p_{l}\right)
},u_{i_{1}},...,u_{i_{k}}\right) \left\vert
u_{i_{1}},...,u_{i_{k}}\right\rangle
\end{equation*}%
it includes $\left\vert 1\right\rangle $ fr $k=0$.

Then we can define $S_{p_{l}}$ that translates $p_{l}$ at states level: 
\begin{equation*}
S_{p_{l}}:\oplus H^{\otimes k_{l}}\left( \left( U_{j}\right) ^{\left(
p_{l}\right) }\right) \hookrightarrow \oplus _{k}H^{\otimes k}\left(
U_{j}\right)
\end{equation*}%
which is defined by its components $s_{p_{l}}^{\otimes k_{l}}$:%
\begin{equation*}
s_{p_{l}}^{\otimes k_{l}}:H^{\otimes k_{l}}\left( \left( U_{j}\right)
^{\left( p_{l}\right) }\right) \hookrightarrow \oplus _{k}H^{\otimes
k}\left( U_{j}\right)
\end{equation*}%
obtained by tensor products 
\begin{equation*}
s_{p_{l}}^{\otimes k_{l}}:\left\vert u_{1}^{\left( p_{l}\right)
},...,u_{k_{l}}^{\left( p_{l}\right) }\right\rangle \rightarrow
\sum_{k}\sum_{\left( \alpha _{m}\right) _{m\leqslant l}\in P\left( k\right)
}\int_{\left( \left( u_{i_{e}}\right) _{e\in \alpha _{m}}\right) \in
p_{l}\left( u_{t}^{\left( p_{l}\right) }\right) }\tprod\limits_{\alpha
_{m}}g_{\alpha _{m}}\left( u_{m}^{\left( p_{l}\right) },\left(
u_{i_{e}}\right) _{e\in \alpha _{m}}\right) \left\vert
u_{i_{1}},...,u_{i_{k}}\right\rangle
\end{equation*}

\subsubsection{Fields and functionals}

The states map allow to consider associated field, whose functional generate
the states. Defining:%
\begin{equation*}
\Psi _{J,s_{p_{l}}}^{\otimes k_{l}}\left( \left( U_{j}\right) ^{\left(
p_{l}\right) }\right) =\sum_{k}g\left( u_{1}^{\left( p_{l}\right)
},...,u_{k_{l}}^{\left( p_{l}\right) },u_{i_{1}},...,u_{i_{k}}\right) \Psi
_{J}^{\otimes k}\left( U_{j}^{k}\right)
\end{equation*}%
field $\left\{ \Psi _{J,s_{p_{l}}}^{\otimes k_{l}}\right\} _{k}$ define the
subject field define by a map $p_{l}$ of the collection. We consider this
fields as independent quantity to model the emerging agregate object that
may arise within the initial state space. As a cosequence, we consider that
the states in terms of functional are given by sums of functionals over
subjcts. The functional to consider are thus:

\begin{equation*}
\sum F_{lin}\left( \Psi _{J}^{\otimes k}\left( U_{j}^{k}\right) /f\right)
+\sum_{\left( p_{l}\right) }\sum F_{lin}^{\left( p_{l}\right) }\left( \Psi
_{J,s_{p_{l}}}^{\otimes k_{l}}\left( \left( U_{j}\right) ^{\left(
p_{l}\right) }\right) /f\right)
\end{equation*}

\subsection{Composed subobject}

More genrally, we may expect sequences of subobjects, i.e. subobjct fo
subobject and so on. We thus have to compose subobjcts maps. This is done in
two steps.

\subsubsection{Subobjects of subobjects}

We first define subobjects of subobjects in a similar manner as in previous
section. We thus consider:%
\begin{equation*}
\left\{ \left( \left( U_{j}\right) ^{\left( p\right) }\right) ^{\left(
p^{\prime }\right) }\overset{p^{\prime }}{\hookrightarrow }\mathcal{V}\oplus
_{k}\left( \left( U_{j}\right) ^{\left( p\right) }\right) ^{k}\right\}
_{p^{\prime }}
\end{equation*}%
and collections of subobjects of subobjects are given by maps:%
\begin{eqnarray*}
p_{l,u} &:&\left( \left( U_{j}\right) ^{\left( p_{u}\right) }\right)
^{\left( p_{l}\right) }\rightarrow \mathcal{V}\left( \oplus _{k}\left(
\left( U_{j}\right) ^{\left( p_{u}\right) }\right) ^{k}\right) \\
u^{\left( p_{l,u}\right) } &\rightarrow &\oplus _{k}\left( u_{i_{1}}^{\left(
u\right) },...u_{i_{k}}^{\left( u\right) }\right) \left( u^{\left(
p_{l,u}\right) }\right) ,f_{k}\left( u_{i_{1}}^{\left( u\right)
},...u_{i_{k}}^{\left( u\right) },u^{\left( p_{l,u}\right) }\right) =0
\end{eqnarray*}%
The map $s_{p_{l,u}}$, and its states translations $s_{p_{l,u}}^{\otimes
k_{l}}$ and $S_{p_{l,u}}$ are given by:%
\begin{equation*}
s_{p_{l,u}}:H\left( \left( \left( U_{j}\right) ^{\left( p_{u}\right)
}\right) ^{\left( p_{l}\right) }\right) \hookrightarrow \oplus
_{k}H^{\otimes k}\left( \left( U_{j}\right) ^{\left( p_{u}\right) }\right)
\end{equation*}%
\begin{equation*}
s_{p_{l,u}}^{\otimes k_{l}}:H^{\otimes k_{l}}\left( \left( \left(
U_{j}\right) ^{\left( p_{u}\right) }\right) ^{\left( p_{l}\right) }\right)
\hookrightarrow \oplus _{k}H^{\otimes k}\left( \left( U_{j}\right) ^{\left(
p_{u}\right) }\right)
\end{equation*}%
and:%
\begin{equation*}
S_{p_{l,u}}:\oplus H^{\otimes k_{l}}\left( \left( \left( U_{j}\right)
^{\left( p_{u}\right) }\right) ^{\left( p_{l,u}\right) }\right)
\hookrightarrow \oplus _{k}H^{\otimes k}\left( \left( U_{j}\right) ^{\left(
p_{u}\right) }\right)
\end{equation*}%
we also write:%
\begin{equation*}
\left( \left( U_{j}\right) ^{\left( p_{u},p_{lu}\right) }\right)
\end{equation*}%
for:%
\begin{equation*}
\left( \left( \left( U_{j}\right) ^{\left( p_{u}\right) }\right) ^{\left(
p_{l,u}\right) }\right)
\end{equation*}

\subsubsection{Composed Maps:}

We aim at defining the composition of subobjects maps at the level of states:%
\begin{equation*}
S_{p_{\left( l\right) ,u}}=S_{p_{u}}\circ S_{p_{l,u}}:\oplus H^{\otimes
k_{l}}\left( \left( \left( U_{j}\right) ^{\left( p_{u}\right) }\right)
^{\left( p_{l,u}\right) }\right) \hookrightarrow \oplus _{k}H^{\otimes
k}\left( U_{j}\right)
\end{equation*}%
To do so we first define the product of maps:

\paragraph{Products of maps}

We consider products of subobjcts:%
\begin{equation*}
p_{l_{1}}\times p_{l_{2}}:\left( U_{j}\right) ^{\left( p_{l_{1}}\right)
}\times \left( U_{j}\right) ^{\left( p_{l_{2}}\right) }\rightarrow \mathcal{V%
}\left( \oplus _{k}U_{j}^{k}\right)
\end{equation*}%
Define:%
\begin{equation*}
p_{l}\left( u_{p_{l}}\right) =\cup _{k}V_{k}\left( p_{l}\left(
u_{p_{l}}\right) \right)
\end{equation*}%
with the $U_{j}^{k}$ are the components of $p_{l}$.

The product is defined by:%
\begin{equation*}
p_{l_{1}}\times p_{l_{2}}\left( u_{p_{l_{1}}}\times u_{p_{l_{2}}}\right)
=\cup _{k}\left( V_{k_{1}}\left( p_{l_{1}}\left( u_{p_{l_{1}}}\right)
\right) \times V_{k-k_{1}}\left( p_{l_{2}}\left( u_{p_{l_{2}}}\right)
\right) \right)
\end{equation*}%
For example, considering:%
\begin{equation*}
U_{j_{1}}\times U_{j_{2}}/c\left( U_{j_{1}},U_{j_{2}}\right)
\end{equation*}%
the image of $p_{l_{1}}\times p_{l_{2}}$ is:%
\begin{equation*}
\left\{ p_{l_{1}}\left( u_{p_{l_{1}}}\right) \times p_{l_{2}}\left(
u_{p_{l_{2}}}\right) \right\} _{u_{p_{l_{1}}}\in \left( U_{j}\right)
^{\left( p_{l_{1}}\right) },u_{p_{l_{2}}}\in \left( U_{j}\right) ^{\left(
p_{l_{2}}\right) }}
\end{equation*}%
At level of hilbert spaces, we have the map:%
\begin{equation*}
S_{p_{l_{1}}}\otimes S_{p_{l_{2}}}:\oplus _{k_{l_{1}},k_{l_{2}}}\left(
H^{\otimes k_{l_{1}}}\left( \left( U_{j}\right) ^{\left( p_{l_{1}}\right)
}\right) \otimes H^{\otimes k_{l_{2}}}\left( \left( U_{j}\right) ^{\left(
p_{l_{2}}\right) }\right) \right) \hookrightarrow \oplus _{k}H^{\otimes
k}\left( U_{j}\right)
\end{equation*}%
written in coordinates:%
\begin{eqnarray*}
S_{p_{l}} &:&s_{p_{l_{1}}}^{\otimes k_{l_{1}}}\otimes s_{p_{l_{2}}}^{\otimes
k_{l_{2}}}:\left\vert u_{1}^{\left( p_{l_{1}}\right)
},...,u_{k_{l_{1}}}^{\left( p_{l_{1}}\right) }\right\rangle \left\vert
u_{1}^{\left( p_{l_{2}}\right) },...,u_{k_{l_{2}}}^{\left( p_{l_{2}}\right)
}\right\rangle \\
&\rightarrow &\sum_{k=k_{1}+k_{2}}\sum_{\left( \alpha _{m_{i}}\right)
_{m_{i}\leqslant l_{i}}\in P\left( k_{i}\right)
}\tprod\limits_{i=1}^{2}\int_{\left( \left( u_{i_{e}}\right) _{e\in \alpha
_{m_{i}}}\right) \in \left( p_{l_{i}}\left( u_{t}^{\left( p_{l_{i}}\right)
}\right) \right) }\tprod\limits_{\alpha _{m_{i}}}g_{\alpha _{m_{i}}}\left(
u_{m_{i}}^{\left( p_{l_{i}}\right) },\left( u_{i_{e}}\right) _{e\in \alpha
_{m_{i}}}\right) \left\vert u_{i_{1}},...,u_{i_{k}}\right\rangle
\end{eqnarray*}%
as before the maps $g_{\alpha _{m_{i}}}$ characterize $p_{l_{1}}\times
p_{l_{2}}$.

\paragraph{Compositn of mps}

We can now compose maps of subobjct to define sequences of subobjects.
Starting with:%
\begin{equation*}
p_{l}:\left( U_{j}\right) ^{\left( p_{l}\right) }\rightarrow \mathcal{V}%
\left( \oplus _{k}U_{j}^{k}\right)
\end{equation*}%
and:%
\begin{equation*}
p_{l,l^{\prime }}:\left( \left( U_{j}\right) ^{\left( p_{l}\right) }\right)
^{\left( p_{l,l^{\prime }}\right) }\rightarrow \mathcal{V}\left( \oplus
_{k}\left( \left( U_{j}\right) ^{\left( p_{l}\right) }\right) ^{k}\right)
\end{equation*}%
these maps can be composed:%
\begin{equation*}
p_{l}\circ p_{l,l^{\prime }}\rightarrow \mathcal{V}\left( \oplus
_{k}U_{j}^{k}\right)
\end{equation*}%
by considering first:%
\begin{equation*}
\left( \left( U_{j}\right) ^{\left( p_{l}\right) }\right) ^{\left(
p_{l,l^{\prime }}\right) }\rightarrow \mathcal{V}\left( \oplus
_{k_{l}}\left( \left( U_{j}\right) ^{\left( p_{l}\right) }\right)
^{k_{l}}\right) \rightrightarrows \underset{V\in \sqcup \mathcal{V}\left(
\oplus _{k}\left( \left( U_{j}\right) ^{\left( p_{l}\right) }\right)
^{k}\right) }{\sqcup }\left\{ u\in V\right\} \rightarrow \mathcal{V}\left(
\oplus _{k}\left( \left( U_{j}\right) \right) ^{k}\right)
\end{equation*}%
that transforms the parameters of the sub-subobject $u_{p_{l},p_{l,l^{\prime
}}}$ into: 
\begin{equation*}
u_{p_{l},p_{l,l^{\prime }}}\rightarrow p_{l,l^{\prime }}\left(
u_{p_{l},p_{l,l^{\prime }}}\right) \rightrightarrows \left\{ u_{p_{l}}\in
p_{l,l^{\prime }}\left( u_{p_{l},p_{l,l^{\prime }}}\right) \right\}
\rightarrow \tprod\limits_{u_{p_{l}}\in p_{l,l^{\prime }}\left(
u_{p_{l},p_{l,l^{\prime }}}\right) }p_{l}\left( u_{p_{l}}\right)
\end{equation*}%
with as before:%
\begin{equation*}
p_{l}\left( u_{p_{l}}\right) =\cup _{k}V_{k}\left( p_{l}\left(
u_{p_{l}}\right) \right)
\end{equation*}%
and:%
\begin{equation*}
p_{l}\left( u_{p_{l}}\right) \times p_{l}\left( u_{p_{l}}^{\prime }\right)
=\cup _{k}\left( V_{k_{1}}\left( p_{l}\left( u_{p_{l}}\right) \right) \times
V_{k-k_{1}}\left( p_{l}\left( u_{p_{l}}\right) \right) \right)
\end{equation*}%
For states, the composition leads to:

\begin{equation*}
S_{p_{l},p_{l,l^{\prime }}}=S_{p_{l}}\circ S_{p_{l,l^{\prime }}}:\oplus
H^{\otimes k_{l}}\left( \left( \left( U_{j}\right) ^{\left( p_{l}\right)
}\right) ^{\left( p_{l,l^{\prime }}\right) }\right) \hookrightarrow \oplus
_{k}H^{\otimes k}\left( U_{j}\right)
\end{equation*}%
On the $1$ states, it is given by $s_{p_{l,l^{\prime }}}^{\otimes
k_{l^{\prime }}}$:%
\begin{eqnarray*}
s_{p_{l,l^{\prime }}}^{\otimes k_{l^{\prime }}} &:&\left\vert u_{1}^{\left(
p_{l,l^{\prime }}\right) },...,u_{k_{l^{\prime }}}^{\left( p_{l,l^{\prime
}}\right) }\right\rangle \\
&\rightarrow &\sum_{k_{l}}\sum_{\left( \alpha _{m^{\prime }}\right)
_{m^{\prime }\leqslant k_{l^{\prime }}}\in P\left( k_{l}\right)
}\int_{\left( \left( u_{i_{e}}^{\left( p_{l}\right) }\right) _{e\in \alpha
_{m^{\prime }}}\right) \in p_{l}\left( u_{t}^{\left( p_{l,l^{\prime
}}\right) }\right) }\tprod\limits_{\alpha _{m^{\prime }}}g_{\alpha
_{m^{\prime }}}\left( u_{m}^{\left( p_{l,l^{\prime }}\right) },\left(
u_{i_{e}}^{\left( p_{l}\right) }\right) _{e\in \alpha _{m^{\prime }}}\right)
\left\vert u_{1}^{\left( p_{l}\right) },...,u_{k_{l}}^{\left( p_{l}\right)
}\right\rangle
\end{eqnarray*}%
with $s_{p_{l}}$ defined as before: 
\begin{equation*}
s_{p_{l}}:\left\vert u^{\left( p_{l}\right) }\right\rangle \rightarrow
\sum_{k}\int_{\left( u_{i_{1}},...,u_{i_{k}}\right) \in \left( p_{l}\left(
u^{\left( p_{l}\right) }\right) \right) }g_{k}\left( u^{\left( p_{l}\right)
},u_{i_{1}},...,u_{i_{k}}\right) \left\vert
u_{i_{1}},...,u_{i_{k}}\right\rangle
\end{equation*}%
Then, on the full states space, we have:%
\begin{equation*}
S_{p_{l}}=s_{p_{l}}^{\otimes k_{l}}:\left\vert u_{1}^{\left( p_{l}\right)
},...,u_{k_{l}}^{\left( p_{l}\right) }\right\rangle \rightarrow
\sum_{k}\sum_{\left( \alpha _{m}\right) _{m\leqslant l}\in P\left( k\right)
}\int_{\left( \left( u_{i_{e}}\right) _{e\in \alpha _{m}}\right) \in \left(
p_{l}\left( u_{t}^{\left( p_{l}\right) }\right) \right)
}\tprod\limits_{\alpha _{m}}g_{\alpha _{m}}\left( u_{m}^{\left( p_{l}\right)
},\left( u_{i_{e}}\right) _{e\in \alpha _{m}}\right) \left\vert
u_{i_{1}},...,u_{i_{k}}\right\rangle
\end{equation*}%
and:%
\begin{eqnarray*}
S_{p_{l},p_{l,l^{\prime }}}^{k_{l^{\prime }}} &=&S_{p_{l}}\circ
s_{p_{l,l^{\prime }}}^{\otimes k_{l^{\prime }}}:\left\vert u_{1}^{\left(
p_{l,l^{\prime }}\right) },...,u_{k_{l^{\prime }}}^{\left( p_{l,l^{\prime
}}\right) }\right\rangle \\
&\rightarrow &\sum_{k}\sum_{k_{l}}\sum_{\substack{ \left( \alpha _{m^{\prime
}}\right) _{m^{\prime }\leqslant k_{l^{\prime }}}\in P\left( k_{l}\right) 
\\ \left( \alpha _{m}\right) _{m\leqslant l}\in P\left( k\right) }}%
\int_{V\left( u_{m}^{\left( p_{l,l^{\prime }}\right) },\left(
u_{i_{e}}^{\left( p_{l}\right) }\right) _{e\in \alpha _{m^{\prime }}},\left(
u_{i_{e}}\right) _{e\in \alpha _{m}}\right) } \\
&&\times \tprod\limits_{\alpha _{m^{\prime }},\alpha _{m}}g_{\alpha
_{m^{\prime }},\alpha _{m}}\left( u_{m}^{\left( p_{l,l^{\prime }}\right)
},\left( u_{i_{e}}^{\left( p_{l}\right) }\right) _{e\in \alpha _{m^{\prime
}}},\left( u_{i_{e}}\right) _{e\in \alpha _{m}}\right) \left\vert
u_{i_{1}},...,u_{i_{k}}\right\rangle
\end{eqnarray*}%
with:%
\begin{equation*}
V\left( u_{m}^{\left( p_{l,l^{\prime }}\right) },\left( u_{i_{e}}^{\left(
p_{l}\right) }\right) _{e\in \alpha _{m^{\prime }}},\left( u_{i_{e}}\right)
_{e\in \alpha _{m}}\right) =\left\{ \left( \left( u_{i_{e}}^{\left(
p_{l}\right) }\right) _{e\in \alpha _{m^{\prime }}}\right) \in \left(
p_{l}\left( u_{t}^{\left( p_{l,l^{\prime }}\right) }\right) \right) \right\}
\times \left\{ \left( \left( u_{i_{e}}\right) _{e\in \alpha _{m}}\right) \in
\left( p_{l}\left( u_{t}^{\left( p_{l}\right) }\right) \right) \right\}
\end{equation*}%
and the functions:%
\begin{equation*}
g_{\alpha _{m^{\prime }},\alpha _{m}}\left( u_{m}^{\left( p_{l,l^{\prime
}}\right) },\left( u_{i_{e}}^{\left( p_{l}\right) }\right) _{e\in \alpha
_{m^{\prime }}},\left( u_{i_{e}}\right) _{e\in \alpha _{m}}\right)
=g_{\alpha _{m^{\prime }}}\left( u_{m}^{\left( p_{l,l^{\prime }}\right)
},\left( u_{i_{e}}^{\left( p_{l}\right) }\right) _{e\in \alpha _{m^{\prime
}}}\right) g_{\alpha _{m}}\left( u_{m}^{\left( p_{l}\right) },\left(
u_{i_{e}}\right) _{e\in \alpha _{m}}\right)
\end{equation*}%
characterizing the compositions of maps.

\section{Decomposition in subspaces}

We generalize the notion of decompositions to an arbitrary set of arbitrary
number of subobjects. We will assume that these decompositions are
compatible downward for subobjects.

\subsection{Decomposition and maps}

We consider the assumption that there is a collection:%
\begin{equation*}
p_{l}:\left( U_{j}\right) ^{\left( p_{l}\right) }\rightarrow \mathcal{V}%
\left( \oplus _{k}U_{j}^{k}\right)
\end{equation*}%
such that:%
\begin{eqnarray*}
\oplus _{k}H^{\otimes k}\left( U_{j}\right) &=&\oplus _{\left( k_{l}\right)
}\otimes _{l=1}^{m}s_{p_{l}}^{\otimes k_{l}}\left( \left( H\left( \left(
U_{j}\right) ^{\left( p_{l}\right) }\right) \right) ^{\otimes k_{l}}\right)
/\tprod\limits_{p_{1},...,p_{m}}f_{p_{1}...p_{m}} \\
&\equiv &\oplus _{\left( k_{l}\right) }\otimes _{l=1}^{m}\left( H\left(
\left( U_{j}\right) ^{\left( p_{l}\right) }\right) \right) ^{\otimes
k_{l}}/\tprod\limits_{p_{1},...,p_{m}}f_{p_{1}...p_{m}}
\end{eqnarray*}%
\begin{eqnarray*}
\left\vert u_{i_{1}},...,u_{i_{k}}\right\rangle &=&\sum_{\left( k_{l}\right)
}h_{s_{p_{l}}^{\otimes k_{l}}}\left( \left( u_{1}^{\left( p_{l}\right)
},...,u_{k_{l}}^{\left( p_{l}\right) }\right)
_{l},u_{i_{1}},...,u_{i_{k}}\right) \otimes _{l}s_{p_{l}}^{\otimes
k_{l}}\left( \left\vert u_{1}^{\left( p_{l}\right) },...,u_{k_{l}}^{\left(
p_{l}\right) }\right\rangle \right)
/\tprod\limits_{p_{1},...,p_{m}}f_{p_{1}...p_{m}} \\
&\equiv &\sum_{\left( k_{l}\right) }h_{s_{p_{l}}^{\otimes k_{l}}}\left(
\left( u_{1}^{\left( p_{l}\right) },...,u_{k_{l}}^{\left( p_{l}\right)
}\right) _{l},u_{i_{1}},...,u_{i_{k}}\right) \otimes _{l}\left( \left\vert
u_{1}^{\left( p_{l}\right) },...,u_{k_{l}}^{\left( p_{l}\right)
}\right\rangle \right) /\tprod\limits_{p_{1},...,p_{m}}f_{p_{1}...p_{m}}
\end{eqnarray*}%
The constraints $f_{p_{1}...p_{m}}$ are assumed to arise from operator
invariance:%
\begin{equation*}
K\left\vert u_{i_{1}},...,u_{i_{k}}\right\rangle =0
\end{equation*}%
that is, written in term of the previous decomposition:%
\begin{equation*}
\sum_{\left( k_{l}\right) }h_{s_{p_{l}}^{\otimes k_{l}}}\left( \left(
u_{1}^{\left( p_{l}\right) },...,u_{k_{l}}^{\left( p_{l}\right) }\right)
_{l},u_{i_{1}},...,u_{i_{k}}\right) \sum_{l}K\left( \left\vert u_{1}^{\left(
p_{l}\right) },...,u_{k_{l}}^{\left( p_{l}\right) }\right\rangle \right)
\otimes _{l^{\prime }\neq l}\left\vert u_{1}^{\left( p_{l^{\prime }}\right)
},...,u_{k_{l}}^{\left( p_{l^{\prime }}\right) }\right\rangle =0
\end{equation*}%
We may assume that implies an equation on the parameters:%
\begin{equation*}
\sum_{l}f_{l}\left( \left( u_{i}^{\left( p_{l}\right) }\right) \right) =0
\end{equation*}%
and:%
\begin{equation*}
\left\vert u_{i_{1}},...,u_{i_{k}}\right\rangle =\sum_{\left( k_{l}\right)
}h_{s_{p_{l}}^{\otimes k_{l}}}\left( \left( u_{1}^{\left( p_{l}\right)
},...,u_{k_{l}}^{\left( p_{l}\right) }\right)
_{l},u_{i_{1}},...,u_{i_{k}}\right) \tprod \delta \left( f_{p_{l}}\left(
\left( u_{i}^{\left( p_{l}\right) }\right) \right) \right) \otimes
_{l}\left\vert u_{1}^{\left( p_{l}\right) },...,u_{k_{l}}^{\left(
p_{l}\right) }\right\rangle
\end{equation*}%
We extend this relations to decompose subobject into subobject of subobjects:%
\begin{equation*}
p_{l,l^{\prime }}:\left( \left( U_{j}\right) ^{\left( p_{l}\right) }\right)
^{\left( p_{l,l^{\prime }}\right) }\rightarrow \mathcal{V}\left( \oplus
_{k_{l}}\left( \left( U_{j}\right) ^{\left( p_{l}\right) }\right)
^{k_{l}}\right)
\end{equation*}%
Using the notation:%
\begin{equation*}
\left( \left( U_{j}\right) ^{\left( p_{l}\right) }\right) ^{\left(
p_{l,l^{\prime }}\right) }\equiv \left( U_{j}\right) ^{\left(
p_{l},p_{l^{\prime }l}\right) }
\end{equation*}%
we have:%
\begin{eqnarray*}
\oplus _{k}H^{\otimes k}\left( \left( \left( U_{j}\right) ^{\left(
p_{l}\right) }\right) \right) &=&\otimes _{l^{\prime }=1}^{m}S_{p_{l^{\prime
},l}}\left( \oplus _{k_{l^{\prime }}}H^{\otimes k_{l^{\prime }}}\left(
\left( \left( U_{j}\right) ^{\left( p_{l},p_{l^{\prime }l}\right) }\right)
\right) \right) /\tprod\limits_{p_{1,l},...,p_{m,l}}f_{p_{1,l}...p_{m,l}} \\
&\equiv &\otimes _{l^{\prime }=1}^{m}\left( \oplus _{k_{l^{\prime
}}}H^{\otimes k_{l^{\prime }}}\left( \left( \left( U_{j}\right) ^{\left(
p_{l},p_{l^{\prime }l}\right) }\right) \right) \right)
/\tprod\limits_{p_{1,l},...,p_{m,l}}f_{p_{1,l}...p_{m,l}} \\
&\equiv &\oplus _{\left( k_{l^{\prime }}\right) }\otimes _{l^{\prime
}=1}^{m}\left( H^{\otimes k_{l^{\prime }}}\left( \left( \left( U_{j}\right)
^{\left( p_{l},p_{l^{\prime }l}\right) }\right) \right) \right)
/\tprod\limits_{p_{1,l},...,p_{m,l}}f_{p_{1,l}...p_{m,l}}
\end{eqnarray*}

Then at the level of entire state space, defining:%
\begin{eqnarray*}
\mathcal{H}\left( U_{j}\right) &=&\oplus _{k}H^{\otimes k}\left( U_{j}\right)
\\
\mathcal{H}\left( \left\{ \left( U_{j}\right) ^{\left( p_{l}\right)
}\right\} _{l\leqslant m}\right) &=&\oplus _{\left( k_{l}\right) }\otimes
_{l=1}^{m}s_{p_{l}}^{\otimes k_{l}}\left( \left( H\left( \left( U_{j}\right)
^{\left( p_{l}\right) }\right) \right) ^{\otimes k_{l}}\right)
/\tprod\limits_{p_{1},...,p_{m}}f_{p_{1}...p_{m}}
\end{eqnarray*}%
we have the decomposition:%
\begin{eqnarray*}
\mathcal{H}\left( \left( \left( U_{j}\right) ^{\left( p_{l}\right) }\right)
\right) &=&\oplus _{k}H^{\otimes k}\left( \left( \left( U_{j}\right)
^{\left( p_{l}\right) }\right) \right) \\
\mathcal{H}\left( \left\{ \left( U_{j}\right) ^{\left( p_{l},p_{l^{\prime
}l}\right) }\right\} _{l^{\prime }\leqslant m}\right) &=&\oplus _{\left(
k_{l^{\prime }}\right) }\otimes _{l^{\prime }=1}^{m}S_{p_{l^{\prime
},l}}\left( H^{\otimes k_{l^{\prime }}}\left( \left( \left( U_{j}\right)
^{\left( p_{l},p_{l^{\prime }l}\right) }\right) \right) \right)
/\tprod\limits_{p_{1,l},...,p_{m,l}}f_{p_{1,l}...p_{m,l}} \\
\mathcal{H}\left( \left\{ \left( U_{j}\right) ^{\left( p_{l},p_{l^{\prime
}l}\right) }\right\} _{\substack{ l\leqslant m  \\ l^{\prime }\leqslant
m^{\prime }}}\right) &=&\otimes _{l=1,l^{\prime }=1}^{m,m^{\prime
}}S_{p_{l^{\prime },l}}\left( \oplus H^{\otimes k_{l^{\prime }}}\left(
\left( \left( U_{j}\right) ^{\left( p_{l},p_{l^{\prime }l}\right) }\right)
\right) \right) /\tprod\limits_{p_{1,l},...,p_{m^{\prime
},l}}f_{p_{1,l}...p_{m^{\prime },l}}
\end{eqnarray*}

\subsection{States decomposition}

\subsubsection{General formulation: Fields decomposition}

Writing:%
\begin{eqnarray*}
\left( \left( U_{j}^{\otimes k}\right) ^{\left( p_{l}\right)
}\hookrightarrow U_{j}^{\otimes k}\right) ^{s_{p_{l}}^{\otimes k_{l}}}
&:&H\left( U_{j}^{\otimes k}\right) \rightarrow H\left( \left(
U_{j}^{\otimes k}\right) ^{\left( p_{l}\right) }\right) \\
\left\vert u_{i_{1}}\otimes ...\otimes u_{i_{k}}\right\rangle &\rightarrow
&\sum_{t}\left\vert u_{1}^{\left( p_{l}\right) }\otimes ...\otimes
u_{t}^{\left( p_{l}\right) }\right\rangle
\end{eqnarray*}%
we translate the decomposition in terms of fields. We assume that stts $%
F\left( \Psi _{J}\left( U_{j}\right) \right) $ can be decomposed along the
field decomposition:%
\begin{eqnarray*}
\Psi _{J}^{\otimes k}\left( U_{j}^{k}\right) &\rightarrow
&\tbigoplus\limits_{m}\tbigoplus\limits_{D_{j}^{p_{l},m}}h_{s_{p_{l}}^{%
\otimes k_{l}}}\left( \left( \left( \left( U_{j}\right) ^{\left(
p_{l}\right) }\right) ^{k_{l}}\right) _{l},U_{j}^{k}\right) \\
&&\times \underset{p_{l}}{\otimes }\delta \left( f_{p_{l}}\left( \left(
u_{i}^{\left( p_{l}\right) }\right) \right) \right) \Psi
_{J,s_{p_{l}}^{\otimes k_{l}}}^{\otimes k_{l}}\left( \left( \left(
U_{j}\right) ^{\left( p_{l}\right) }\right) ^{k_{l}}\right)
\end{eqnarray*}%
wth $D_{j}^{p_{l},m}$ standing for the possible decompositions:%
\begin{equation*}
\mathcal{H}\left( U_{j}\right) =\mathcal{H}\left( \left\{ \left(
U_{j}\right) ^{\left( p_{l}\right) }\right\} _{l\leqslant m}\right)
\end{equation*}%
where we define:%
\begin{equation*}
\left( \left( U_{j}\right) ^{\left( p_{l}\right) }\right) ^{k_{l}}=\left(
u_{1}^{\left( p_{l}\right) },...,u_{k_{l}}^{\left( p_{l}\right) }\right)
\end{equation*}%
and:%
\begin{equation*}
U_{j}^{k}=\left( u_{1},...,u_{k}\right) /c\left( u_{1},...,u_{k}\right)
\end{equation*}%
The $u_{i}^{\left( p_{l}\right) }$ are coordinates on $\left( U_{j}\right)
^{\left( p_{l}\right) }$, with $u_{1}$ the coordinates on $U_{j}$. We also
define the subobject fields:%
\begin{equation*}
\Psi _{J,s_{p_{l}}^{\otimes k_{l}}}^{\otimes k_{l}}\left( \left( \left(
U_{j}\right) ^{\left( p_{l}\right) }\right) ^{k_{l}}\right)
=s_{p_{l}}^{\otimes k_{l}}\left( \Psi _{J}^{\otimes k_{l}}\left( \left(
\left( U_{j}\right) ^{\left( p_{l}\right) }\right) ^{k_{l}}\right) \right)
\end{equation*}%
in a form that is similar to the states transformation:%
\begin{equation*}
s_{p_{l}}^{\otimes k_{l}}\left( \Psi _{J}^{\otimes k_{l}}\left( \left(
\left( U_{j}\right) ^{\left( p_{l}\right) }\right) ^{k_{l}}\right) \right)
=\sum_{k}\sum_{\left( \alpha _{m}\right) _{m\leqslant l}\in P\left( k\right)
}\int_{\left( \left( u_{i_{e}}\right) _{e\in \alpha _{m}}\right) \in
p_{l}\left( u_{t}^{\left( p_{l}\right) }\right) }\tprod\limits_{\alpha
_{m}}g_{\alpha _{m}}\left( u_{m}^{\left( p_{l}\right) },\left(
u_{i_{e}}\right) _{e\in \alpha _{m}}\right) \Psi _{J}^{\otimes k}\left(
u_{i_{1}},...,u_{i_{k}}\right)
\end{equation*}%
The tensor products $\underset{p_{l}}{\otimes }$ in the decomposition stands
for series expansion of products of independent copies of the $\Psi
_{J,s_{p_{l}}^{\otimes k_{l}}}^{\otimes k_{l}}\left( \left( \left(
U_{j}\right) ^{\left( p_{l}\right) }\right) ^{k_{l}}\right) $. As in the
first part, any field realization of $\Psi _{J}^{\otimes k}\left(
U_{j}^{k}\right) $ is decomposed as a series expansion of products of
realizations of the fields $\Psi _{J,s_{p_{l^{\prime },l}}^{\otimes
k_{l^{\prime }}}}^{\otimes k_{l^{\prime }}}$. The constraints are taken into
account through some Dirac functions: 
\begin{eqnarray*}
&&\underset{p_{l}}{\otimes }\delta \left( f_{p_{l}}\left( \left(
u_{i}^{\left( p_{l}\right) }\right) \right) \right) \Psi
_{J,s_{p_{l}}^{\otimes k_{l}}}^{\otimes k_{l}}\left( \left( \left(
U_{j}\right) ^{\left( p_{l}\right) }\right) ^{k_{l}}\right) \\
&\equiv &\sum_{\alpha }\delta \left( f_{p_{l}}\left( \left( u_{i}^{\left(
p_{l}\right) }\right) \right) \right) \underset{p_{l}}{\tprod }\Psi
_{J,\alpha ,s_{p_{l}}^{\otimes k_{l}}}^{\otimes k_{l}}\left( \left( \left(
U_{j}\right) ^{\left( p_{l}\right) }\right) ^{k_{l}}\right)
\end{eqnarray*}

Note that the $h_{s_{p_{l}}^{\otimes k_{l}}}\left( \left( \left( \left(
U_{j}\right) ^{\left( p_{l}\right) }\right) ^{k_{l}}\right)
_{l},U_{j}^{k}\right) $ include constraint between objects of the type:%
\begin{equation*}
\delta \left( f_{\gamma }\left( \left\{ \left( \left( U_{j}\right) ^{\left(
p_{l}\right) }\right) ^{k_{l}}\right\} _{l}\right) \right)
\end{equation*}%
where $\gamma $\ is an index for the several contsraints.

\subsubsection{Subobjects decomposition}

We also assume that subobject are decomposed similarly:

\begin{eqnarray*}
\Psi _{J,s_{p_{u}}^{\otimes k_{u}}}^{\otimes k_{u}}\left( \left( \left(
U_{j}\right) ^{\left( p_{u}\right) }\right) ^{k_{u}}\right) &\rightarrow
&\tbigoplus\limits_{m,m^{\prime }}\tbigoplus\limits_{D_{j,p_{u}}^{\left(
p_{l},p_{l^{\prime }l}\right) ,m,m^{\prime }}}h_{s_{p_{l^{\prime
},l}}^{\otimes k_{l^{\prime }}}}^{\left( p_{u}\right) }\left( \left( \left(
\left( \left( U_{j}\right) ^{\left( p_{l},p_{l^{\prime }l}\right) }\right)
\right) ^{k_{l^{\prime }}}\right) _{l},\left( \left( U_{j}\right) ^{\left(
p_{u}\right) }\right) ^{k_{u}}\right) \\
&&\times \underset{p_{l^{\prime },l}}{\otimes }\delta \left( f_{p_{l^{\prime
},l}}\left( \left( u_{i}^{\left( p_{l^{\prime },l}\right) }\right) \right)
\right) \Psi _{J,s_{p_{l^{\prime },l}}^{\otimes k_{l^{\prime }}}}^{\otimes
k_{l^{\prime }}}\left( \left( \left( \left( U_{j}\right) ^{\left(
p_{l},p_{l^{\prime }l}\right) }\right) \right) ^{k_{l^{\prime }}}\right)
\end{eqnarray*}%
where the symbols:%
\begin{equation*}
D_{j,p_{u}}^{\left( p_{l},p_{l^{\prime }l}\right) ,m,m^{\prime }}
\end{equation*}%
stand for the possible decompositions:%
\begin{equation*}
\mathcal{H}\left( \left( \left( U_{j}\right) ^{\left( p_{u}\right) }\right)
\right) =\mathcal{H}\left( \left\{ \left( U_{j}\right) ^{\left(
p_{l},p_{l^{\prime }l}\right) }\right\} _{\substack{ l\leqslant m  \\ %
l^{\prime }\leqslant m^{\prime }}}\right)
\end{equation*}%
and:%
\begin{equation*}
\left( \left( \left( U_{j}\right) ^{\left( p_{l},p_{l^{\prime }l}\right)
}\right) \right) ^{k_{l^{\prime }}}=\left( u_{1}^{\left( p_{l^{\prime
},l}\right) },...,u_{k_{l}}^{\left( p_{l^{\prime },l}\right) }\right)
^{k_{l^{\prime }}}
\end{equation*}%
where $u_{i}^{\left( p_{l^{\prime },l}\right) }$ are coordinates on $\left(
U_{j}\right) ^{\left( p_{l},p_{l^{\prime }l}\right) }$. We define the
subobject fields: 
\begin{equation*}
\Psi _{J,s_{p_{l^{\prime },l}}^{\otimes k_{l^{\prime }}}}^{\otimes
k_{l^{\prime }}}\left( \left( \left( \left( U_{j}\right) ^{\left(
p_{l},p_{l^{\prime }l}\right) }\right) \right) ^{k_{l^{\prime }}}\right)
=s_{p_{l^{\prime },l}}^{\otimes k_{l^{\prime }}}\left( \Psi _{J}^{\otimes
k_{l^{\prime }}}\left( \left( \left( \left( U_{j}\right) ^{\left(
p_{l},p_{l^{\prime }l}\right) }\right) \right) ^{k_{l^{\prime }}}\right)
\right)
\end{equation*}%
As before, the products $\underset{p_{l^{\prime },l}}{\otimes }$ stands for
series expansion of products of independent copies of the \ $\Psi
_{J,s_{p_{l^{\prime },l}}^{\otimes k_{l^{\prime }}}}^{\otimes k_{l^{\prime
}}}\left( \left( \left( \left( U_{j}\right) ^{\left( p_{l},p_{l^{\prime
}l}\right) }\right) \right) ^{k_{l^{\prime }}}\right) $. For the fields
realizations, the tensor products represent series expansions:%
\begin{eqnarray*}
&&\underset{p_{l^{\prime },l}}{\otimes }\delta \left( f_{p_{l^{\prime
},l}}\left( \left( u_{i}^{\left( p_{l^{\prime },l}\right) }\right) \right)
\right) \Psi _{J,s_{p_{l^{\prime },l}}^{\otimes k_{l^{\prime }}}}^{\otimes
k_{l^{\prime }}}\left( \left( \left( U_{j}\right) ^{\left(
p_{l},p_{l^{\prime }l}\right) }\right) ^{k_{l^{\prime }}}\right) \\
&\equiv &\sum_{\alpha }\delta \left( f_{p_{l^{\prime },l}}\left( \left(
u_{i}^{\left( p_{l^{\prime },l}\right) }\right) \right) \right) \underset{%
p_{l^{\prime },l}}{\tprod }\Psi _{J,\alpha ,s_{p_{l^{\prime },l}}^{\otimes
k_{l^{\prime }}}}^{\otimes k_{l^{\prime }}}\left( \left( \left( U_{j}\right)
^{\left( p_{l},p_{l^{\prime }l}\right) }\right) ^{k_{l^{\prime }}}\right)
\end{eqnarray*}

\paragraph{Remarks}

1. As before, the constraints:%
\begin{equation*}
h_{s_{p_{l^{\prime },l}}^{\otimes k_{l^{\prime }}}}^{\left( p_{u}\right)
}\left( \left( \left( \left( U_{j}\right) ^{\left( p_{l},p_{l^{\prime
}l}\right) }\right) ^{k_{l^{\prime }}}\right) _{l},\left( \left(
U_{j}\right) ^{\left( p_{u}\right) }\right) ^{k_{u}}\right)
\end{equation*}%
include constraints between objects:%
\begin{equation*}
\delta \left( f_{\gamma }\left( \left\{ \left( \left( U_{j}\right) ^{\left(
p_{l},p_{l^{\prime }l}\right) }\right) ^{k_{l^{\prime }}}\right\}
_{l}\right) \right)
\end{equation*}%
where $\gamma $\ is an index for the several constraints.

2. We can identify: 
\begin{equation*}
\Psi _{J,s_{p_{l^{\prime },l}}^{\otimes k_{l^{\prime }}}}^{\otimes
k_{l^{\prime }}}\left( \left( \left( U_{j}\right) ^{\left(
p_{l},p_{l^{\prime }l}\right) }\right) ^{k_{l^{\prime }}}\right)
\end{equation*}%
and:%
\begin{eqnarray*}
S_{p_{l}}\circ \left( \Psi _{J,s_{p_{l^{\prime },l}}^{\otimes k_{l^{\prime
}}}}^{\otimes k_{l^{\prime }}}\left( \left( \left( \left( U_{j}\right)
^{\left( p_{l},p_{l^{\prime }l}\right) }\right) \right) ^{k_{l^{\prime
}}}\right) \right) &\equiv &S_{p_{l^{\prime },\left( l\right) }}^{\otimes
k_{l^{\prime }}}\left( \Psi _{J}^{\otimes k_{l^{\prime }}}\left( \left(
\left( \left( U_{j}\right) ^{\left( p_{l},p_{l^{\prime }l}\right) }\right)
\right) ^{k_{l^{\prime }}}\right) \right) \\
&\equiv &\Psi _{J,S_{p_{l^{\prime },\left( l\right) }}^{\otimes k_{l^{\prime
}}}}^{\otimes k_{l^{\prime }}}\left( \left( \left( \left( U_{j}\right)
^{\left( p_{l},p_{l^{\prime }l}\right) }\right) \right) ^{k_{l^{\prime
}}}\right)
\end{eqnarray*}

\subsubsection{Fields and subobjects decomposition}

We can gather the decompositions for fields and subobjects. Writing:%
\begin{eqnarray*}
\left[ p_{u}\right] ^{k_{u}} &\equiv &\left( \left( U_{j}\right) ^{\left(
p_{u}\right) }\right) ^{k_{u}} \\
\left[ p_{l},p_{l^{\prime }l}\right] ^{k_{l^{\prime }}} &\equiv &\left(
\left( U_{j}\right) ^{\left( p_{l},p_{l^{\prime }l}\right) }\right)
^{k_{l^{\prime }}}
\end{eqnarray*}%
and expressing the constraints as:%
\begin{eqnarray*}
\delta \left( f_{p_{l}}\right) &=&\delta \left( f_{p_{l}}\left( \left(
u_{i}^{\left( p_{l}\right) }\right) \right) \right) \\
\delta \left( f_{p_{l^{\prime },l}}\right) &=&\delta \left( f_{p_{l^{\prime
},l}}\left( \left( u_{i}^{\left( p_{l^{\prime },l}\right) }\right) \right)
\right)
\end{eqnarray*}%
we can rewrite the decompsition is:%
\begin{eqnarray*}
&&\Psi _{J}^{\otimes k}\left( U_{j}^{k}\right) \tbigoplus\limits_{\left(
U_{j}\right) ^{\left( p_{u}\right) }}\Psi _{J,s_{p_{u}}^{\otimes
k_{u}}}^{\otimes k_{u}}\left( \left[ p_{u}\right] ^{k_{u}}\right) \\
&\rightarrow &\tbigoplus\limits_{m}\tbigoplus\limits_{D_{j}^{p_{l},m}}\left(
h_{s_{p_{l}}^{\otimes k_{l}}}\left( \left( \left[ p_{l}\right]
^{k_{l}}\right) _{l},U_{j}^{k}\right) \underset{p_{l}}{\otimes }\delta
\left( f_{p_{l}}\right) \Psi _{J,s_{p_{l}}^{\otimes k_{l}}}^{\otimes
k_{l}}\left( \left[ p_{l}\right] ^{k_{l}}\right) \right. \\
&&\tbigoplus\limits_{m^{\prime }}\tbigoplus\limits_{D_{j,p_{u}}^{\left(
p_{l},p_{l^{\prime }l}\right) ,m,m^{\prime }}}\left. h_{s_{p_{l^{\prime
},l}}^{\otimes k_{l^{\prime }}}}^{\left( p_{u}\right) }\left( \left( \left[
p_{l},p_{l^{\prime }l}\right] ^{k_{l^{\prime }}}\right) _{l},\left[ p_{u}%
\right] ^{k_{u}}\right) \underset{p_{l^{\prime },l}}{\otimes }\delta \left(
f_{p_{l^{\prime },l}}\right) \Psi _{J,s_{p_{l^{\prime },l}}^{\otimes
k_{l^{\prime }}}}^{\otimes k_{l^{\prime }}}\left( \left[ p_{l},p_{l^{\prime
}l}\right] ^{k_{l^{\prime }}}\right) _{l}\right)
\end{eqnarray*}%
It translates that $U_{j}^{\otimes k}$ is terminal, the relation is
"aggregated" on the $\Psi _{J}^{\otimes k}\left( U_{j}^{\otimes k}\right) $
side. However, expressing $\Psi _{J}^{\otimes k}\left( U_{j}^{\otimes
k}\right) $ as a function of $\Psi _{J,s_{p_{l}}^{\otimes k_{l}}}^{\otimes
k_{l}}\left( \left( U_{j}^{\otimes k}\right) ^{\left( p_{l}\right) }\right) $
yields intricate relations.

\subsubsection{States functional decomposition}

The decomposition of a state functional has the form:%
\begin{equation}
F_{lin}\left( \Psi _{J}^{\otimes k}\left( U_{j}^{\otimes k}\right) \right)
=\int f\left( U_{j}^{\otimes k},\hat{\Pi}_{U_{j}^{\otimes k}}\right) \Psi
_{J}^{\otimes k}\left( U_{j}^{\otimes k}\right) dU_{j}^{\otimes
k}+\sum_{\left( p_{u}\right) }F_{lin}\left( \Psi _{J,s_{p_{u}}^{\otimes
k_{u}}}^{\otimes k_{u}}\left( \left( \left( U_{j}\right) ^{\left(
p_{u}\right) }\right) ^{k_{u}}\right) \right)  \label{FSB}
\end{equation}%
where subobjects are included. Note that this can be expanded recursively:%
\begin{eqnarray*}
&&F_{lin}\left( \Psi _{J}^{\otimes k}\left( U_{j}^{\otimes k}\right) \right)
\\
&=&\int f\left( U_{j}^{\otimes k},\hat{\Pi}_{U_{j}^{\otimes k}}\right) \Psi
_{J}^{\otimes k}\left( U_{j}^{\otimes k}\right) dU_{j}^{\otimes k} \\
&&+\sum_{\left( p_{u}\right) }\int f_{1}\left( \left( \left( U_{j}\right)
^{\left( p_{u}\right) }\right) ^{k_{u}},\hat{\Pi}_{\left( \left(
U_{j}\right) ^{\left( p_{u}\right) }\right) ^{k_{u}}}\right) \Psi
_{J,s_{p_{u}}^{\otimes k_{u}}}^{\otimes k_{u}}\left( \left( \left(
U_{j}\right) ^{\left( p_{u}\right) }\right) ^{k_{u}}\right) \\
&&+\sum_{\left( p_{u^{\prime },u}\right) }\int f_{2}\left( \left( \left(
U_{j}\right) ^{\left( p_{u^{\prime },u}\right) }\right) ^{k_{u^{\prime },u}},%
\hat{\Pi}_{\left( \left( U_{j}\right) ^{\left( p_{u^{\prime },u}\right)
}\right) ^{k_{u^{\prime },u}}}\right) \Psi _{J,s_{p_{u^{\prime
},u}}^{\otimes k_{u^{\prime },u}}}^{\otimes k_{u}}\left( \left( \left(
U_{j}\right) ^{\left( p_{u^{\prime },u}\right) }\right) ^{k_{u^{\prime
},u}}\right) \\
&&+...
\end{eqnarray*}

Using the decomposition, states (\ref{FSB}) can be rewritten in the
following manner. The first term is generated by sums:%
\begin{eqnarray}
F_{lin}\left( \Psi _{J}\left( U_{j}^{\otimes k}\right) \right)
&=&F_{lin\left( i_{1}...i_{m}\right) }\left( \otimes _{l}\Psi
_{J,s_{p_{l}}^{\otimes k_{l}}}^{\otimes k_{l}}\left( \left( \left(
U_{j}\right) ^{\left( p_{l}\right) }\right) ^{k_{l}}\right) \right)
\label{fctbj} \\
&=&\underset{m}{\sum }\underset{D_{j}^{p_{l},m}}{\sum }g_{s_{p_{l}}^{\otimes
k_{l}}}\left( \left\{ \left( \left( U_{j}\right) ^{\left( p_{l}\right)
}\right) ^{k_{l}}\right\} _{l\leqslant m}\right) \underset{l\leqslant m}{%
\otimes }\Psi _{J,s_{p_{l}}^{\otimes k_{l}}}^{\otimes k_{l}}\left( \left(
\left( U_{j}\right) ^{\left( p_{l}\right) }\right) ^{k_{l}}\right)  \notag
\end{eqnarray}%
where:%
\begin{equation*}
g_{s_{p_{l}}^{\otimes k_{l}}}\left( \left\{ \left( \left( U_{j}\right)
^{\left( p_{l}\right) }\right) ^{k_{l}}\right\} _{l\leqslant m}\right) =\int
f\left( U_{j}^{\otimes k},\hat{\Pi}_{U_{j}^{\otimes k}}\right)
h_{s_{p_{l}}^{\otimes k_{l}}}\left( \left( \left( \left( U_{j}\right)
^{\left( p_{l}\right) }\right) ^{k_{l}}\right) _{l},U_{j}^{k}\right) d\left(
U_{j}^{k}\right)
\end{equation*}%
The second term describes subobjects terms\ and includes in the
decomposition:%
\begin{eqnarray}
&&F_{lin}\left( \Psi _{J,s_{p_{u}}^{\otimes k_{u}}}^{\otimes k_{u}}\left(
\left( \left( U_{j}\right) ^{\left( p_{u}\right) }\right) ^{k_{u}}\right)
\right)  \label{fctssbj} \\
&=&F_{lin\left( i_{1}...i_{m}\right) }^{\left( p_{u}\right) }\left( \otimes
_{l}\Psi _{J,s_{p_{l}}^{\otimes k_{l}}}^{\otimes k_{l}}\left( \left( \left(
U_{j}\right) ^{\left( p_{l}\right) }\right) ^{k_{l}}\right) \right)  \notag
\\
&=&\underset{m}{\sum }\underset{D_{j}^{p_{l},m}}{\sum }g_{s_{p_{l}}^{\otimes
k_{l}}}^{\left( p_{u}\right) }\left( \left\{ \left( \left( U_{j}\right)
^{\left( p_{l}\right) }\right) ^{k_{l}}\right\} _{l\leqslant m}\right) 
\underset{p_{l}}{\otimes }\delta \left( f_{p_{l}}\right)
\tprod\limits_{l\leqslant m}\Psi _{J,s_{p_{l}}^{\otimes k_{l}}}^{\otimes
k_{l}}\left( \left( \left( U_{j}\right) ^{\left( p_{l}\right) }\right)
^{k_{l}}\right)  \notag
\end{eqnarray}%
with:%
\begin{equation*}
g_{s_{p_{l}}^{\otimes k_{l}}}^{\left( p_{u}\right) }\left( \left\{ \left(
\left( U_{j}\right) ^{\left( p_{l}\right) }\right) ^{k_{l}}\right\}
_{l\leqslant m}\right) =\int f\left( U_{j}^{\otimes k},\hat{\Pi}%
_{U_{j}^{\otimes k}}\right) h_{s_{p_{l}}^{\otimes k_{l}}}^{\left(
p_{u}\right) }\left( \left( \left( \left( U_{j}\right) ^{\left( p_{l}\right)
}\right) ^{k_{l}}\right) _{l},\left( \left( U_{j}\right) ^{\left(
p_{u}\right) }\right) ^{k_{u}}\right) d\left( \left( \left( U_{j}\right)
^{\left( p_{u}\right) }\right) ^{k_{u}}\right)
\end{equation*}%
plus contributions from subobject:%
\begin{equation*}
F_{lin}\left( \Psi _{J,s_{p_{u}}^{\otimes k_{u}}}^{\otimes k_{u}}\left( 
\left[ p_{u}\right] ^{k_{u}}\right) \right)
=\sum_{m}\sum_{D_{j}^{p_{l},m}}\left( \sum_{m^{\prime
}}\sum_{D_{j,p_{u}}^{\left( p_{l},p_{l^{\prime }l}\right) ,m,m^{\prime
}}}g\left( \left( \left[ p_{l},p_{l^{\prime }l}\right] ^{k_{l^{\prime
}}}\right) _{l,l^{\prime }}\right) \underset{l^{\prime }}{\otimes }\Psi
_{J,s_{p_{l^{\prime },l}}^{\otimes k_{l^{\prime }}}}^{\otimes k_{l^{\prime
}}}\left( \left[ p_{l},p_{l^{\prime }l}\right] ^{k_{l^{\prime }}}\right)
\right)
\end{equation*}%
where we have:%
\begin{eqnarray*}
&&g\left( \left( \left[ p_{l},p_{l^{\prime }l}\right] ^{k_{l^{\prime
}}}\right) _{l,l^{\prime }}\right) \\
&=&\int f\left( \left( \left( U_{j}\right) ^{\left( p_{u}\right) }\right)
^{k_{u}},\hat{\Pi}_{\left( \left( U_{j}\right) ^{\left( p_{u}\right)
}\right) ^{k_{u}}}\right) h_{s_{p_{l^{\prime },l}}^{\otimes k_{l^{\prime
}}}}^{\left( p_{u}\right) }\left( \left( \left[ p_{l},p_{l^{\prime }l}\right]
^{k_{l^{\prime }}}\right) _{l},\left[ p_{u}\right] ^{k_{u}}\right)
\tprod\limits_{p_{l^{\prime },l}}\delta \left( f_{p_{l^{\prime },l}}\right)
d\left( \left[ p_{u}\right] ^{k_{u}}\right)
\end{eqnarray*}

Both terms can be regrouped by writing:$\ $%
\begin{equation*}
\Psi _{J,s_{p_{l}}^{\otimes k_{l}}}^{\otimes k_{l}}\left( \left[ p_{l}\right]
^{k_{l}}\right) \tbigoplus\limits_{\left( p_{l^{\prime }l}\right) }\Psi
_{J,s_{p_{l^{\prime },l}}^{\otimes k_{l^{\prime }}}}^{\otimes k_{l^{\prime
}}}\left( \left[ p_{l},p_{l^{\prime }l}\right] ^{k_{l^{\prime }}}\right)
\equiv \tbigoplus\limits_{\left( p_{l^{\prime }l}\right) }\Psi
_{J,s_{p_{l^{\prime },l}}^{\otimes k_{l^{\prime }}}}^{\otimes k_{l^{\prime
}}}\left( \left[ p_{l},p_{l^{\prime }l}\right] ^{k_{l^{\prime }}}\right)
\end{equation*}%
with:%
\begin{equation*}
\left[ p_{l},p_{l^{\prime }l}\right] ^{k_{l^{\prime }}}=\left( \left(
U_{j}\right) ^{\left( p_{l},p_{l^{\prime }l}\right) }\right) ^{k_{l^{\prime
}}}
\end{equation*}%
where the $id$ map:%
\begin{eqnarray*}
p_{l,l} &:&\left( \left( U_{j}\right) ^{\left( p_{l}\right) }\right)
\rightarrow \mathcal{V}\left( \oplus _{k}\left( \left( U_{j}\right) ^{\left(
p_{l}\right) }\right) ^{k}\right) \\
\left( U_{j}\right) ^{\left( p_{l}\right) } &\mapsto &\left( U_{j}\right)
^{\left( p_{l}\right) }
\end{eqnarray*}%
is added to the set of $\left( p_{l,l^{\prime }}\right) $. The full
functional writes:%
\begin{equation}
F_{lin}\left( \Psi _{J,s_{p_{u}}^{\otimes k_{u}}}^{\otimes k_{u}}\left( 
\left[ p_{u}\right] ^{k_{u}}\right) \right)
=\sum_{m}\sum_{D_{j}^{p_{l},m}}\left( \sum_{m^{\prime
}}\sum_{D_{j,p_{u}}^{\left( p_{l},p_{l^{\prime }l}\right) ,m,m^{\prime
}}}g\left( \left( \left[ p_{l},p_{l^{\prime }l}\right] ^{k_{l^{\prime
}}}\right) _{l,l^{\prime }}\right) \underset{l^{\prime }}{\otimes }\Psi
_{J,s_{p_{l^{\prime },l}}^{\otimes k_{l^{\prime }}}}^{\otimes k_{l^{\prime
}}}\left( \left[ p_{l},p_{l^{\prime }l}\right] ^{k_{l^{\prime }}}\right)
\right)  \label{fnctnl}
\end{equation}

\subsubsection{Partial states}

As in part one we can define partial states for one of the subobjects,
written $0$ to distinguish it from others subobjects. A partial functional
for this objects and its subobjects characterized by their fields $\Psi
_{J,s_{p_{l^{\prime },0}}^{\otimes k_{l^{\prime }}}}^{\otimes k_{l^{\prime
}}}$ is a collection:%
\begin{equation}
\left\{ v_{\left[ p_{l},p_{l^{\prime }l}\right] ^{k_{l^{\prime }}}}\left(
\Psi _{J,s_{p_{l^{\prime },0}}^{\otimes k_{l^{\prime }}}}^{\otimes
k_{l^{\prime }}}\right) \right\} _{\left[ p_{l},p_{l^{\prime }l}\right]
^{k_{l^{\prime }}}}=\left\{ v_{\left[ p_{l},p_{l^{\prime }l}\right]
^{k_{l^{\prime }}}}\left( \left[ p_{0},p_{l^{\prime }0}\right]
^{k_{l^{\prime }}}\right) \right\} _{\left[ p_{0},p_{l^{\prime }0}\right]
^{k_{l^{\prime }}},\left[ p_{l},p_{l^{\prime }l}\right] ^{k_{l^{\prime }}}}
\label{CL}
\end{equation}%
of densities that can be inserted in integrals of tensor products, so that a
partial functional is given by the collection:

\begin{equation}
\int v_{\left[ p_{l},p_{l^{\prime }l}\right] ^{k_{l^{\prime }}}}\left( \Psi
_{J,s_{p_{l^{\prime },0}}^{\otimes k_{l^{\prime }}}}^{\otimes k_{l^{\prime
}}}\right) \otimes ev_{\left[ p_{l},p_{l^{\prime }l}\right] ^{k_{l^{\prime
}}}}\left( \Psi _{J,s_{p_{l^{\prime },l}}^{\otimes k_{l^{\prime
}}}}^{\otimes k_{l^{\prime }}}\right) \delta \left( f_{\gamma }\left( \left[
p_{0},p_{l^{\prime }0}\right] ^{k_{l^{\prime }}},\left\{ \left[
p_{l},p_{l^{\prime }l}\right] ^{k_{l^{\prime }}}\right\} \right) \right)
d\left( \left[ p_{0},p_{l^{\prime }0}\right] ^{k_{l^{\prime }}},\left\{ %
\left[ p_{l},p_{l^{\prime }l}\right] ^{k_{l^{\prime }}}\right\} \right)
\label{PRTL}
\end{equation}%
where $ev_{\left[ p_{l},p_{l^{\prime }l}\right] ^{k_{l^{\prime }}}}\left(
\Psi _{J,s_{p_{l^{\prime },l}}^{\otimes k_{l^{\prime }}}}^{\otimes
k_{l^{\prime }}}\right) $ is the evaluation functional for $\Psi
_{J,s_{p_{l^{\prime },l}}^{\otimes k_{l^{\prime }}}}^{\otimes k_{l^{\prime
}}}$:%
\begin{equation*}
ev_{\left[ p_{l},p_{l^{\prime }l}\right] ^{k_{l^{\prime }}}}\left( \Psi
_{J,s_{p_{l^{\prime },l}}^{\otimes k_{l^{\prime }}}}^{\otimes k_{l^{\prime
}}}\right) =\Psi _{J,s_{p_{l^{\prime },l}}^{\otimes k_{l^{\prime
}}}}^{\otimes k_{l^{\prime }}}\left( \left[ p_{l},p_{l^{\prime }l}\right]
^{k_{l^{\prime }}}\right)
\end{equation*}

\section{Projection on one particular subobject}

\subsection{Decomposition including a given subobject}

We will consider particular decomposition:%
\begin{equation*}
D_{j}^{p_{0},p_{l},m}
\end{equation*}%
standing for:%
\begin{equation*}
\mathcal{H}\left( U_{j}\right) =\mathcal{H}\left( \left( \left( U_{j}\right)
^{\left( p_{0}\right) }\right) \right) \otimes \mathcal{H}\left( \left\{
\left( U_{j}\right) ^{\left( p_{l}\right) }\right\} _{l\leqslant m}\right)
\end{equation*}%
and project the subbjct corresponding to the $\left( p_{l}\right) $ on some
particular states. That is, we project on $\mathcal{H}\left( \left( \left(
U_{j}\right) ^{\left( p_{0}\right) }\right) \right) $.

Since we add subobjects, we have to include also decomposition:%
\begin{equation*}
D_{j,p_{u}}^{\left( p_{0},p_{l^{\prime }0},p_{l},p_{l^{\prime }l}\right)
,m,m^{\prime }}
\end{equation*}%
standing for the decomposition:%
\begin{equation*}
\mathcal{H}\left( \left( \left( U_{j}\right) ^{\left( p_{u}\right) }\right)
\right) =\mathcal{H}\left( \left( \left( U_{j}\right) ^{\left(
p_{0},p_{l^{\prime }0}\right) }\right) ^{k_{l^{\prime }}}\right) \otimes 
\mathcal{H}\left( \left\{ \left( U_{j}\right) ^{\left( p_{l},p_{l^{\prime
}l}\right) }\right\} _{\substack{ l\leqslant m  \\ l^{\prime }\leqslant
m^{\prime }}}\right)
\end{equation*}%
Considering again the $id$ map we write:%
\begin{equation*}
\Psi _{J,s_{p_{0}}^{\otimes k_{p_{0}}}}^{\otimes k_{p_{0}}}\left( \left[
p_{0}\right] ^{k_{p_{0}}}\right) \tbigoplus\limits_{\left( p_{l^{\prime
}0}\right) }\Psi _{J,s_{p_{l^{\prime },0}}^{\otimes k_{l^{\prime
}}}}^{\otimes k_{l^{\prime }}}\left( \left[ p_{0},p_{l^{\prime }0}\right]
^{k_{l^{\prime }}}\right) \equiv \tbigoplus\limits_{\left( p_{l^{\prime
}0}\right) }\Psi _{J,s_{p_{l^{\prime },0}}^{\otimes k_{l^{\prime
}}}}^{\otimes k_{l^{\prime }}}\left( \left[ p_{0},p_{l^{\prime }0}\right]
^{k_{l^{\prime }}}\right)
\end{equation*}%
and:%
\begin{equation*}
\Psi _{J,s_{p_{l^{\prime },l}}^{\otimes k_{l^{\prime }}}}^{\otimes
k_{l^{\prime }}}\left( \left[ p_{l}\right] ^{k_{l}}\right)
\tbigoplus\limits_{\left( p_{l^{\prime }l}\right) }\Psi _{J,s_{p_{l^{\prime
},l}}^{\otimes k_{l^{\prime }}}}^{\otimes k_{l^{\prime }}}\left( \left[
p_{l},p_{l^{\prime }l}\right] ^{k_{l^{\prime }}}\right) \equiv
\tbigoplus\limits_{\left( p_{l^{\prime }l}\right) }\Psi _{J,s_{p_{l^{\prime
},l}}^{\otimes k_{l^{\prime }}}}^{\otimes k_{l^{\prime }}}\left( \left[
p_{l},p_{l^{\prime }l}\right] ^{k_{l^{\prime }}}\right)
\end{equation*}

Th full functional then writes for this decomposition:%
\begin{eqnarray}
&&F_{f,lin}\left( \left\{ \Psi _{J,s_{p_{l^{\prime },0}}^{\otimes
k_{l^{\prime }}}}^{\otimes k_{l^{\prime }}}\left( \left[ p_{0},p_{l^{\prime
}0}\right] ^{k_{l^{\prime }}}\right) \right\} _{\left( p_{l^{\prime
}0}\right) }\right)  \label{fnctnlb} \\
&=&\sum_{m}\sum_{D_{j}^{p_{0},p_{l},m}}\sum_{m^{\prime
}}\sum_{D_{j,p_{u}}^{\left( p_{0},p_{l^{\prime }0},p_{l},p_{l^{\prime
}l}\right) ,m,m^{\prime }}}g\left( \left\{ \left[ p_{l},p_{l^{\prime }l}%
\right] ^{k_{l^{\prime }}}\right\} _{l,l^{\prime }},\left\{ \left( \left[
p_{0},p_{l^{\prime }0}\right] \right) \right\} _{0,l^{\prime }}\right) 
\notag \\
&&\times \left( \underset{l^{\prime }}{\otimes }\Psi _{J,s_{p_{l^{\prime
},0}}^{\otimes k_{l^{\prime }}}}^{\otimes k_{l^{\prime }}}\left( \left[
p_{0},p_{l^{\prime }0}\right] ^{k_{l^{\prime }}}\right) \right) \left( 
\underset{l^{\prime }}{\otimes }\Psi _{J,s_{p_{l^{\prime },l}}^{\otimes
k_{l^{\prime }}}}^{\otimes k_{l^{\prime }}}\left( \left[ p_{l},p_{l^{\prime
}l}\right] ^{k_{l^{\prime }}}\right) \right)  \notag
\end{eqnarray}

\subsection{Projection on $\mathcal{H}\left( \left( \left( U_{j}\right)
^{\left( p_{0}\right) }\right) \right) $}

Projection on the states of $\mathcal{H}\left( \left( \left( U_{j}\right)
^{\left( p_{0}\right) }\right) \right) $ leads to an effective functionals
integrating the degrees of freedom for the projected states of $%
\tbigoplus\limits_{\left( p_{l^{\prime }l}\right) }\Psi _{J,s_{p_{l^{\prime
},l}}^{\otimes k_{l^{\prime }}}}^{\otimes k_{l^{\prime }}}\left( \left[
p_{l},p_{l^{\prime }l}\right] ^{k_{l^{\prime }}}\right) $.

As in the first part, we will by expressing the functionals of $\Psi
_{J,s_{p_{l^{\prime },0}}^{\otimes k_{l^{\prime }}}}^{\otimes k_{l^{\prime
}}}\left( \left[ p_{0},p_{l^{\prime }0}\right] ^{k_{l^{\prime }}}\right) $
in the basis of some functionals. The remaining degrees of freedom defined
by $\Psi _{J,s_{p_{l^{\prime },l}}^{\otimes k_{l^{\prime }}}}^{\otimes
k_{l^{\prime }}}$ will then be projected on some background, depending on
the states considered.

\subsubsection{Particular basis of states $\Psi _{J,s_{p_{l^{\prime
},0}}^{\otimes k_{l^{\prime }}}}^{\otimes k_{l^{\prime }}}$}

We first describe a basis of states for $\Psi _{J,s_{p_{l^{\prime
},0}}^{\otimes k_{l^{\prime }}}}^{\otimes k_{l^{\prime }}}$. To do so, we
assume that the state space for $\Psi _{J,s_{p_{l^{\prime },0}}^{\otimes
k_{l^{\prime }}}}^{\otimes k_{l^{\prime }}}$:%
\begin{equation*}
H\left( \left\{ \Psi _{J,s_{p_{l^{\prime },0}}^{\otimes k_{l^{\prime
}}}}^{\otimes k_{l^{\prime }}}\left( \left[ p_{0},p_{l^{\prime }0}\right]
^{k_{l^{\prime }}}\right) \right\} _{\left( p_{l^{\prime }0}\right) }\right)
\end{equation*}%
decomposes as a sum of projection%
\begin{equation*}
id=\sum_{v}\tprod\limits_{v}
\end{equation*}%
where the partial states $\left\{ v\right\} $form an arbitrary basis defined
by densities, as in (\ref{CL}):%
\begin{equation*}
v\left\{ \Psi _{J}^{\otimes k_{l^{\prime }}}\left( \left[ p_{0},p_{l^{\prime
}0}\right] ^{k_{l^{\prime }}}\right) \right\} =\left\{ v_{\left[
p_{l},p_{l^{\prime }l}\right] ^{k_{l^{\prime }}}}\left( \Psi
_{J,s_{p_{l^{\prime },0}}^{\otimes k_{l^{\prime }}}}^{\otimes k_{l^{\prime
}}}\right) \right\} _{\left[ p_{l},p_{l^{\prime }l}\right] ^{k_{l^{\prime
}}}}
\end{equation*}%
The states $\left\{ v\right\} $ can be eigenstates of some operator:%
\begin{equation*}
\Phi \left( \Psi _{J}^{\otimes k_{l^{\prime }}},\frac{\delta }{\delta \Psi
_{J}^{\otimes k_{l^{\prime }}}}\right)
\end{equation*}
As before we write:%
\begin{equation*}
\left( \Psi _{J,s_{p_{l^{\prime },0}}^{\otimes k_{l^{\prime }}}}^{\otimes
k_{l^{\prime }}}\right) \left( \left[ p_{0},p_{l^{\prime }0}\right]
^{k_{l^{\prime }}}\right) \otimes \Psi _{J,s_{p_{l^{\prime },l}}^{\otimes
k_{l^{\prime }}}}^{\otimes k_{l^{\prime }}}\left( \left[ p_{l},p_{l^{\prime
}l}\right] ^{k_{l^{\prime }}}\right) =ev_{\left[ p_{0},p_{l^{\prime }0}%
\right] ^{k_{l^{\prime }}},\left[ p_{l},p_{l^{\prime }l}\right]
^{k_{l^{\prime }}}}\left( \left( \Psi _{J,s_{p_{l^{\prime },0}}^{\otimes
k_{l^{\prime }}}}^{\otimes k_{l^{\prime }}}\right) \otimes \Psi
_{J,s_{p_{l^{\prime },l}}^{\otimes k_{l^{\prime }}}}^{\otimes k_{l^{\prime
}}}\right)
\end{equation*}%
where the evaluation $ev_{\left[ p_{0},p_{l^{\prime }0}\right]
^{k_{l^{\prime }}},\left[ p_{l},p_{l^{\prime }l}\right] ^{k_{l^{\prime }}}}$
computes:%
\begin{equation*}
\left( \Psi _{J,s_{p_{l^{\prime },0}}^{\otimes k_{l^{\prime }}}}^{\otimes
k_{l^{\prime }}}\right) \otimes \Psi _{J,s_{p_{l^{\prime },l}}^{\otimes
k_{l^{\prime }}}}^{\otimes k_{l^{\prime }}}
\end{equation*}%
at a point defined by the coordinates $\left[ p_{0},p_{l^{\prime }0}\right]
^{k_{l^{\prime }}},\left[ p_{l},p_{l^{\prime }l}\right] ^{k_{l^{\prime }}}$.

The constraints $\delta \left( f_{\gamma }\left( \left[ p_{0},p_{l^{\prime
}0}\right] ^{k_{l^{\prime }}},\left[ p_{l},p_{l^{\prime }l}\right]
^{k_{l^{\prime }}}\right) \right) $ are kept implicit for the sake of
simplicity. The decomposition for functionals with respect to $v$ is
performed by multiplying the evaluation by density $v_{\left[
p_{l},p_{l^{\prime }l}\right] ^{k_{l^{\prime }}}}\left( \Psi
_{J,s_{p_{l^{\prime },0}}^{\otimes k_{l^{\prime }}}}^{\otimes k_{l^{\prime
}}}\right) $ (see \ref{PRTL}): 
\begin{equation}
\tprod\limits_{v}\rightarrow v_{\left[ p_{l},p_{l^{\prime }l}\right]
^{k_{l^{\prime }}}}\left( \Psi _{J,s_{p_{l^{\prime },0}}^{\otimes
k_{l^{\prime }}}}^{\otimes k_{l^{\prime }}}\right) \otimes ev_{\left[
p_{l},p_{l^{\prime }l}\right] ^{k_{l^{\prime }}}}\left( \Psi
_{J,s_{p_{l^{\prime },l}}^{\otimes k_{l^{\prime }}}}^{\otimes k_{l^{\prime
}}}\right) \delta \left( f_{\gamma }\left( \left[ p_{0},p_{l^{\prime }0}%
\right] ^{k_{l^{\prime }}},\left\{ \left[ p_{l},p_{l^{\prime }l}\right]
^{k_{l^{\prime }}}\right\} \right) \right)  \label{PRN}
\end{equation}%
In local coordinats, the constraints:%
\begin{equation*}
\Psi _{J,s_{p_{l^{\prime },0}}^{\otimes k_{l^{\prime }}}}^{\otimes
k_{l^{\prime }}}\left( \left[ p_{0},p_{l^{\prime }0}\right] ^{k_{l^{\prime
}}}\right)
\end{equation*}%
can be solved and $\Psi _{J,s_{p_{l^{\prime },0}}^{\otimes k_{l^{\prime
}}}}^{\otimes k_{l^{\prime }}}\left( \left[ p_{0},p_{l^{\prime }0}\right]
^{k_{l^{\prime }}}\right) $ is replaced by:%
\begin{equation*}
\Psi _{J,s_{p_{l^{\prime },0}}^{\otimes k_{l^{\prime }}}}^{\otimes
k_{l^{\prime }}}\left( \overline{\left[ p_{0},p_{l^{\prime }0}\right]
^{k_{l^{\prime }}}},\left\{ \left[ p_{l},p_{l^{\prime }l}\right]
^{k_{l^{\prime }}}\right\} \right)
\end{equation*}%
with:%
\begin{equation*}
\overline{\left[ p_{0},p_{l^{\prime }0}\right] ^{k_{l^{\prime }}}},\left\{ %
\left[ p_{l},p_{l^{\prime }l}\right] ^{k_{l^{\prime }}}\right\} \equiv \left[
p_{0},p_{l^{\prime }0}\right] ^{k_{l^{\prime
}}}/\tprod\limits_{p_{1,l},...,p_{m^{\prime },l}}f_{p_{1,l}...p_{m^{\prime
},l}},f_{p_{1,l}...p_{m^{\prime },l}}^{-1}\left\{ \left[ p_{l},p_{l^{\prime
}l}\right] ^{k_{l^{\prime }}}\right\}
\end{equation*}%
Implicitly $\Psi _{J}^{\otimes k_{l}}$ and $\Psi _{J}^{\otimes k_{p_{0}}}$
stand for $\Psi _{J,s_{p_{0}}^{\otimes k_{p_{0}}}}^{\otimes k_{p_{0}}}$ and $%
\Psi _{J,s_{p_{l}}^{\otimes k_{l}}}^{\otimes k_{l}}$.

As a consequence, (\ref{PRN}) writes:%
\begin{eqnarray*}
&&v_{\left[ p_{l},p_{l^{\prime }l}\right] ^{k_{l^{\prime }}}}\left( \Psi
_{J,s_{p_{l^{\prime },0}}^{\otimes k_{l^{\prime }}}}^{\otimes k_{l^{\prime
}}}\right) \otimes ev_{\left[ p_{l},p_{l^{\prime }l}\right] ^{k_{l^{\prime
}}}}\left( \Psi _{J,s_{p_{l^{\prime },l}}^{\otimes k_{l^{\prime
}}}}^{\otimes k_{l^{\prime }}}\right) \delta \left( f_{\gamma }\left( \left[
p_{0},p_{l^{\prime }0}\right] ^{k_{l^{\prime }}},\left\{ \left[
p_{l},p_{l^{\prime }l}\right] ^{k_{l^{\prime }}}\right\} \right) \right) \\
&=&\int v\left( \overline{\left[ p_{0},p_{l^{\prime }0}\right]
^{k_{l^{\prime }}}},\left\{ \left[ p_{l},p_{l^{\prime }l}\right]
^{k_{l^{\prime }}}\right\} \right) \\
&&\times \Psi _{J,s_{p_{l^{\prime },0}}^{\otimes k_{l^{\prime }}}}^{\otimes
k_{l^{\prime }}}\left( \overline{\left[ p_{0},p_{l^{\prime }0}\right]
^{k_{l^{\prime }}}},\left\{ \left[ p_{l},p_{l^{\prime }l}\right]
^{k_{l^{\prime }}}\right\} \right) d\left( \overline{\left[
p_{0},p_{l^{\prime }0}\right] ^{k_{l^{\prime }}}}\right) \otimes ev_{\left[
p_{l},p_{l^{\prime }l}\right] ^{k_{l^{\prime }}}}\left( \Psi
_{J,s_{p_{l^{\prime },l}}^{\otimes k_{l^{\prime }}}}^{\otimes k_{l^{\prime
}}}\right) \\
&=&v_{\left[ p_{l},p_{l^{\prime }l}\right] ^{k_{l^{\prime }}}}\left( \Psi
_{J,s_{p_{l^{\prime },0}}^{\otimes k_{l^{\prime }}}}^{\otimes k_{l^{\prime
}}}\right) \otimes ev_{\left[ p_{l},p_{l^{\prime }l}\right] ^{k_{l^{\prime
}}}}\left( \Psi _{J,s_{p_{l^{\prime },l}}^{\otimes k_{l^{\prime
}}}}^{\otimes k_{l^{\prime }}}\right)
\end{eqnarray*}%
where we have:%
\begin{eqnarray}
&&v_{\left[ p_{l},p_{l^{\prime }l}\right] ^{k_{l^{\prime }}}}\left( \Psi
_{J,s_{p_{l^{\prime },0}}^{\otimes k_{l^{\prime }}}}^{\otimes k_{l^{\prime
}}}\right)  \label{FC} \\
&=&\int v\left( \overline{\left[ p_{0},p_{l^{\prime }0}\right]
^{k_{l^{\prime }}}},\left\{ \left[ p_{l},p_{l^{\prime }l}\right]
^{k_{l^{\prime }}}\right\} \right) \times \Psi _{J,s_{p_{l^{\prime
},0}}^{\otimes k_{l^{\prime }}}}^{\otimes k_{l^{\prime }}}\left( \overline{%
\left[ p_{0},p_{l^{\prime }0}\right] ^{k_{l^{\prime }}}},\left\{ \left[
p_{l},p_{l^{\prime }l}\right] ^{k_{l^{\prime }}}\right\} \right) d\left( 
\overline{\left[ p_{0},p_{l^{\prime }0}\right] ^{k_{l^{\prime }}}}\right) 
\notag
\end{eqnarray}%
We assume below that fields are chosen eigenstates of $\Lambda $ so that the 
$\sum_{v}$ is performed for a collection

As an example, we can consider a particular case, we can choose for $v_{%
\left[ p_{l},p_{l^{\prime }l}\right] ^{k_{l^{\prime }}}}$:

\begin{equation*}
v_{\left[ p_{l},p_{l^{\prime }l}\right] ^{k_{l^{\prime }}}}\left( \Psi
_{J,s_{p_{l^{\prime },0}}^{\otimes k_{l^{\prime }}}}^{\otimes k_{l^{\prime
}}}\right) =\sum_{\overline{\left[ p_{0},p_{l^{\prime }0}\right]
^{k_{l^{\prime }}}}}v_{\left[ p_{l},p_{l^{\prime }l}\right] ^{k_{l^{\prime
}}}}^{\overline{\left[ p_{0},p_{l^{\prime }0}\right] ^{k_{l^{\prime }}}}%
}\left\{ \Psi _{J,s_{p_{l^{\prime },0}}^{\otimes k_{l^{\prime }}}}^{\otimes
k_{l^{\prime }}}\right\}
\end{equation*}%
where%
\begin{equation*}
\sum_{\overline{\left[ p_{0},p_{l^{\prime }0}\right] ^{k_{l^{\prime }}}}}v_{%
\left[ p_{l},p_{l^{\prime }l}\right] ^{k_{l^{\prime }}}}^{\overline{\left[
p_{0},p_{l^{\prime }0}\right] ^{k_{l^{\prime }}}}}\left\{ \Psi
_{J,s_{p_{l^{\prime },0}}^{\otimes k_{l^{\prime }}}}^{\otimes k_{l^{\prime
}}}\right\} =\Psi _{J}^{\otimes l}\left( \left[ p_{0},p_{l^{\prime }0}\right]
^{k_{l^{\prime }}}\right)
\end{equation*}%
This formula amounts to sum over the evaluation at points $\overline{\left[
p_{0},p_{l^{\prime }0}\right] ^{k_{l^{\prime }}}}$ and the decomposition is
performed with respect to the values of $\overline{\left[ p_{0},p_{l^{\prime
}0}\right] ^{k_{l^{\prime }}}}$. We can also consider to restrict the
projection to a subspace $V^{\subset _{\overline{\left[ p_{0},p_{l^{\prime
}0}\right] ^{k_{l^{\prime }}}}}}$ of $\overline{\left[ p_{0},p_{l^{\prime }0}%
\right] ^{k_{l^{\prime }}}}$:%
\begin{equation*}
\sum_{\overline{\left[ p_{0},p_{l^{\prime }0}\right] ^{k_{l^{\prime }}}}}v_{%
\left[ p_{l},p_{l^{\prime }l}\right] ^{k_{l^{\prime }}}}^{V^{\subset _{%
\overline{\left[ p_{0},p_{l^{\prime }0}\right] ^{k_{l^{\prime }}}}}}}\left\{
\Psi _{J,s_{p_{l^{\prime },0}}^{\otimes k_{l^{\prime }}}}^{\otimes
k_{l^{\prime }}}\right\} =\Psi _{J,s_{p_{l^{\prime },0}}^{\otimes
k_{l^{\prime }}}}^{\otimes k_{l^{\prime }}}\left( V^{\subset _{\overline{%
\left[ p_{0},p_{l^{\prime }0}\right] ^{k_{l^{\prime }}}}}},\left\{ \left[
p_{l},p_{l^{\prime }l}\right] ^{k_{l^{\prime }}}\right\} \right)
\end{equation*}%
which reduces the evaluation to $V^{\subset _{\overline{\left[
p_{0},p_{l^{\prime }0}\right] ^{k_{l^{\prime }}}}}},\left\{ \left[
p_{l},p_{l^{\prime }l}\right] ^{k_{l^{\prime }}}\right\} $.

\subsubsection{Projection along the basis}

Once the partial states $\left\{ v\right\} $ are chosen, the states
functionals of $\tbigoplus\limits_{\left( p_{l^{\prime }l}\right) }\Psi
_{J,s_{p_{l^{\prime },l}}^{\otimes k_{l^{\prime }}}}^{\otimes k_{l^{\prime
}}}\left( \left[ p_{l},p_{l^{\prime }l}\right] ^{k_{l^{\prime }}}\right) $
are projected by operator:%
\begin{equation*}
\sum_{v}\tprod\limits_{v}\otimes \tprod_{\min S\left( v\right) }
\end{equation*}%
Start with projection $\tprod\limits_{v}$. This is done by first rewriting
the functional (\ref{fnctnlb}) with the decomposition:%
\begin{eqnarray}
&&\sum_{m}\sum_{D_{j}^{p_{0},p_{l},m}}\sum_{m^{\prime
}}\sum_{D_{j,p_{u}}^{\left( p_{0},p_{l^{\prime }0},p_{l},p_{l^{\prime
}l}\right) ,m,m^{\prime }}}g\left( \left\{ \left[ p_{l},p_{l^{\prime }l}%
\right] ^{k_{l^{\prime }}}\right\} _{l,l^{\prime }},\left\{ \left( \left[
p_{0},p_{l^{\prime }0}\right] \right) \right\} _{0,l^{\prime }}\right)
\label{FCD} \\
&&\times \left( \underset{l^{\prime }}{\otimes }\Psi _{J,s_{p_{l^{\prime
},0}}^{\otimes k_{l^{\prime }}}}^{\otimes k_{l^{\prime }}}\left( \left[
p_{0},p_{l^{\prime }0}\right] ^{k_{l^{\prime }}}\right) \right) \left( 
\underset{l^{\prime }}{\otimes }\Psi _{J,s_{p_{l^{\prime },l}}^{\otimes
k_{l^{\prime }}}}^{\otimes k_{l^{\prime }}}\left( \left[ p_{l},p_{l^{\prime
}l}\right] ^{k_{l^{\prime }}}\right) \right)  \notag
\end{eqnarray}%
and express this functional in the basis of states $\left\{ v\right\} $,
each of them defined by the collection $\left\{ v_{\left[ p_{l},p_{l^{\prime
}l}\right] ^{k_{l^{\prime }}}}\left( \Psi _{J,s_{p_{l^{\prime },0}}^{\otimes
k_{l^{\prime }}}}^{\otimes k_{l^{\prime }}}\right) \right\} $ given in (\ref%
{FC}). To do so, we insert in (\ref{FCD}) the identity operator $%
\sum_{v}\tprod\limits_{v}$ so that the functionl becomes:%
\begin{eqnarray}
&&\sum_{v}\tprod\limits_{v}\sum_{m}\sum_{D_{j}^{p_{0},p_{l},m}}\sum_{m^{%
\prime }}\sum_{D_{j,p_{u}}^{\left( p_{0},p_{l^{\prime }0},p_{l},p_{l^{\prime
}l}\right) ,m,m^{\prime }}}g\left( \left\{ \left[ p_{l},p_{l^{\prime }l}%
\right] ^{k_{l^{\prime }}}\right\} _{l,l^{\prime }},\left\{ \left( \left[
p_{0},p_{l^{\prime }0}\right] \right) \right\} _{0,l^{\prime }}\right)
\label{FCCB} \\
&&\times \left( \underset{l^{\prime }}{\otimes }\Psi _{J,s_{p_{l^{\prime
},0}}^{\otimes k_{l^{\prime }}}}^{\otimes k_{l^{\prime }}}\left( \left[
p_{0},p_{l^{\prime }0}\right] ^{k_{l^{\prime }}}\right) \right) \left( 
\underset{l^{\prime }}{\otimes }\Psi _{J,s_{p_{l^{\prime },l}}^{\otimes
k_{l^{\prime }}}}^{\otimes k_{l^{\prime }}}\left( \left[ p_{l},p_{l^{\prime
}l}\right] ^{k_{l^{\prime }}}\right) \right)  \notag \\
&\rightarrow
&\sum_{v}\tprod\limits_{v}\sum_{m}\sum_{D_{j}^{p_{0},p_{l},m}}\sum_{m^{%
\prime }}\sum_{D_{j,p_{u}}^{\left( p_{0},p_{l^{\prime }0},p_{l},p_{l^{\prime
}l}\right) ,m,m^{\prime }}}g\left( \left\{ \left[ p_{l},p_{l^{\prime }l}%
\right] ^{k_{l^{\prime }}}\right\} _{l,l^{\prime }},v\right)  \notag \\
&&\otimes v_{\left[ p_{l},p_{l^{\prime }l}\right] ^{k_{l^{\prime }}}}\left(
\Psi _{J,s_{p_{l^{\prime },0}}^{\otimes k_{l^{\prime }}}}^{\otimes
k_{l^{\prime }}}\right) \left( \Psi _{J,s_{p_{l^{\prime },l}}^{\otimes
k_{l^{\prime }}}}^{\otimes k_{l^{\prime }}}\left( \left[ p_{l},p_{l^{\prime
}l}\right] ^{k_{l^{\prime }}}\right) \right)  \notag
\end{eqnarray}%
and this becomes a functional of:%
\begin{equation*}
v_{\left[ p_{l},p_{l^{\prime }l}\right] ^{k_{l^{\prime }}}}\left( \Psi
_{J,s_{p_{l^{\prime },0}}^{\otimes k_{l^{\prime }}}}^{\otimes k_{l^{\prime
}}}\right) \left( \Psi _{J,s_{p_{l^{\prime },l}}^{\otimes k_{l^{\prime
}}}}^{\otimes k_{l^{\prime }}}\left( \left[ p_{l},p_{l^{\prime }l}\right]
^{k_{l^{\prime }}}\right) \right)
\end{equation*}%
We have thus performed a change of basis in (\ref{FCCB}), by replacing the
evaluation functionals for the field $\Psi _{J,s_{p_{l^{\prime
},0}}^{\otimes k_{l^{\prime }}}}^{\otimes k_{l^{\prime }}}$ by the basis of
states $v_{\left[ p_{l},p_{l^{\prime }l}\right] ^{k_{l^{\prime }}}}\left(
\Psi _{J,s_{p_{l^{\prime },0}}^{\otimes k_{l^{\prime }}}}^{\otimes
k_{l^{\prime }}}\right) $. The functns $g\left( \left\{ \left[
p_{l},p_{l^{\prime }l}\right] ^{k_{l^{\prime }}}\right\} _{l,l^{\prime
}},v\right) $ are the coefficients of the functional in this new basis.

\subsubsection{Projection of the degrees of freedom of $\Psi
_{J,s_{p_{l^{\prime },l}}^{\otimes k_{l^{\prime }}}}^{\otimes k_{l^{\prime
}}}$}

As in the first part, we consider that the projection comes from the
minimization of a given functional of the form:%
\begin{equation*}
\exp \left( -S\left( \left\{ \underset{l^{\prime }}{\otimes }\Psi
_{J,s_{p_{l^{\prime },0}}^{\otimes k_{l^{\prime }}}}^{\otimes k_{l^{\prime
}}}\left( \left[ p_{0},p_{l^{\prime }0}\right] ^{k_{l^{\prime }}}\right) 
\underset{l^{\prime },l}{\otimes }\Psi _{J,s_{p_{l^{\prime },l}}^{\otimes
k_{l^{\prime }}}}^{\otimes k_{l^{\prime }}}\left( \left[ p_{l},p_{l^{\prime
}l}\right] ^{k_{l^{\prime }}}\right) \right\} ,v_{\left[ p_{l},p_{l^{\prime
}l}\right] ^{k_{l^{\prime }}}}\left( \Psi _{J,s_{p_{l^{\prime },0}}^{\otimes
k_{l^{\prime }}}}^{\otimes k_{l^{\prime }}}\right) \underset{l^{\prime },l}{%
\otimes }\Psi _{J,s_{p_{l^{\prime },l}}^{\otimes k_{l^{\prime }}}}^{\otimes
k_{l^{\prime }}}\left( \left[ p_{l},p_{l^{\prime }l}\right] ^{k_{l^{\prime
}}}\right) \right) \right)
\end{equation*}%
where the indices denoting the realizations are implicit.

The projection of functionals by: 
\begin{equation*}
\sum_{v}\tprod\limits_{v}\otimes \tprod_{\min S\left( v\right) }
\end{equation*}
will be: 
\begin{eqnarray*}
&&\sum_{v}\sum_{m}\sum_{D_{j}^{p_{0},p_{l},m}}\sum_{m^{\prime
}}\sum_{D_{j,p_{u}}^{\left( p_{0},p_{l^{\prime }0},p_{l},p_{l^{\prime
}l}\right) ,m,m^{\prime }}}g\left( \left\{ \left[ p_{l},p_{l^{\prime }l}%
\right] ^{k_{l^{\prime }}}\right\} _{l,l^{\prime }},v\right) \\
&&v_{\left\{ \left[ p_{l},p_{l^{\prime }l}\right] ^{k_{l^{\prime }}}\right\}
}\left( \Psi _{J,s_{p_{l^{\prime },0}}^{\otimes k_{l^{\prime }}}}^{\otimes
k_{l^{\prime }}}\right) \otimes \Psi _{J,s_{p_{l^{\prime },l}}^{\otimes
k_{l^{\prime }}}}^{\otimes k_{l^{\prime }}}\left( \left[ p_{l},p_{l^{\prime
}l}\right] ^{k_{l^{\prime }}}\right) \exp \left( -S\right) \tprod \mathcal{D}%
\left( \otimes \Psi _{J,s_{p_{l^{\prime },l}}^{\otimes k_{l^{\prime
}}}}^{\otimes k_{l^{\prime }}}\left( \left[ p_{l},p_{l^{\prime }l}\right]
^{k_{l^{\prime }}}\right) \right)
\end{eqnarray*}

Minimizing $S$ leads to $\Psi _{J,0,s_{p_{l^{\prime },l}}^{\otimes
k_{l^{\prime }}}}^{\otimes k_{l^{\prime }}}\left( \left[ p_{l},p_{l^{\prime
}l}\right] ^{k_{l^{\prime }}}\right) $ with multiplication with a background:%
\begin{equation*}
\Psi _{J,0,s_{p_{l^{\prime },l}}^{\otimes k_{l^{\prime }}}}^{\otimes
k_{l^{\prime }}}\left( \left[ p_{l},p_{l^{\prime }l}\right] ^{k_{l^{\prime
}}}\right)
\end{equation*}%
and the functional reduces to:%
\begin{eqnarray*}
&&\sum_{m}\sum_{D_{j}^{p_{0},p_{l},m}}\sum_{m^{\prime
}}\sum_{D_{j,p_{u}}^{\left( p_{0},p_{l^{\prime }0},p_{l},p_{l^{\prime
}l}\right) ,m,m^{\prime }}}g\left( \left\{ \left[ p_{l},p_{l^{\prime }l}%
\right] ^{k_{l^{\prime }}}\right\} _{l,l^{\prime }},v\right) \\
&&\otimes v_{\left\{ \left[ p_{l},p_{l^{\prime }l}\right] ^{k_{l^{\prime
}}}\right\} }\left( \Psi _{J,s_{p_{l^{\prime },0}}^{\otimes k_{l^{\prime
}}}}^{\otimes k_{l^{\prime }}}\right) \left( \Psi _{J,0,s_{p_{l^{\prime
},l}}^{\otimes k_{l^{\prime }}}}^{\otimes k_{l^{\prime }}}\left( \left[
p_{l},p_{l^{\prime }l}\right] ^{k_{l^{\prime }}}\right) \right)
\end{eqnarray*}%
The part:%
\begin{equation*}
g\left( \left\{ \left[ p_{l},p_{l^{\prime }l}\right] ^{k_{l^{\prime
}}}\right\} _{l,l^{\prime }},v\right) \times v_{\left\{ \left[
p_{l},p_{l^{\prime }l}\right] ^{k_{l^{\prime }}}\right\} }\left( \underset{%
l^{\prime }}{\otimes }\Psi _{J,s_{p_{l^{\prime },0}}^{\otimes k_{l^{\prime
}}}}^{\otimes k_{l^{\prime }}}\right) \left( \underset{l^{\prime }}{\otimes }%
\Psi _{J,0,s_{p_{l^{\prime },l}}^{\otimes k_{l^{\prime }}}}^{\otimes
k_{l^{\prime }}}\left( \left[ p_{l},p_{l^{\prime }l}\right] ^{k_{l^{\prime
}}}\right) \right)
\end{equation*}%
is a functional of:%
\begin{equation*}
v_{\left\{ \left[ p_{l},p_{l^{\prime }l}\right] ^{k_{l^{\prime }}}\right\}
}\left( \underset{l^{\prime }}{\otimes }\Psi _{J,s_{p_{l^{\prime
},0}}^{\otimes k_{l^{\prime }}}}^{\otimes k_{l^{\prime }}}\right) \left( 
\underset{l^{\prime }}{\otimes }\Psi _{J,0,s_{p_{l^{\prime },l}}^{\otimes
k_{l^{\prime }}}}^{\otimes k_{l^{\prime }}}\left( \left[ p_{l},p_{l^{\prime
}l}\right] ^{k_{l^{\prime }}}\right) \right)
\end{equation*}

\subsubsection{Saddle point solutions without degeneracy}

The background $\underset{l^{\prime }}{\otimes }\Psi _{J,0,s_{p_{l^{\prime
},l}}^{\otimes k_{l^{\prime }}}}^{\otimes k_{l^{\prime }}}\left( \left[
p_{l},p_{l^{\prime }l}\right] ^{k_{l^{\prime }}}\right) $ is obtained by
minimisation of:%
\begin{equation}
S\left( \left\{ \underset{l^{\prime }}{\otimes }\Psi _{J,s_{p_{l^{\prime
},0}}^{\otimes k_{l^{\prime }}}}^{\otimes k_{l^{\prime }}}\left( \left[
p_{0},p_{l^{\prime }0}\right] ^{k_{l^{\prime }}}\right) \underset{%
l,l^{\prime }}{\otimes }\Psi _{J,s_{p_{l^{\prime },l}}^{\otimes k_{l^{\prime
}}}}^{\otimes k_{l^{\prime }}}\left( \left[ p_{l},p_{l^{\prime }l}\right]
^{k_{l^{\prime }}}\right) \right\} ,v_{\left[ p_{l},p_{l^{\prime }l}\right]
^{k_{l^{\prime }}}}\left( \Psi _{J,s_{p_{l^{\prime },0}}^{\otimes
k_{l^{\prime }}}}^{\otimes k_{l^{\prime }}}\right) \otimes \Psi
_{J,s_{p_{l^{\prime },l}}^{\otimes k_{l^{\prime }}}}^{\otimes k_{l^{\prime
}}}\left( \left[ p_{l},p_{l^{\prime }l}\right] ^{k_{l^{\prime }}}\right)
\right)  \label{SCN}
\end{equation}%
The tensor:%
\begin{equation*}
\underset{l^{\prime }}{\otimes }\Psi _{J,s_{p_{l^{\prime },0}}^{\otimes
k_{l^{\prime }}}}^{\otimes k_{l^{\prime }}}\left( \left[ p_{0},p_{l^{\prime
}0}\right] ^{k_{l^{\prime }}}\right) \underset{l,l^{\prime }}{\otimes }\Psi
_{J,s_{p_{l^{\prime },l}}^{\otimes k_{l^{\prime }}}}^{\otimes k_{l^{\prime
}}}\left( \left[ p_{l},p_{l^{\prime }l}\right] ^{k_{l^{\prime }}}\right)
\end{equation*}%
stands for set of realizations:%
\begin{equation*}
\left\{ \underset{l^{\prime }}{\otimes }\Psi _{J,\alpha ,s_{p_{l^{\prime
},0}}^{\otimes k_{l^{\prime }}}}^{\otimes k_{l^{\prime }}}\left( \left[
p_{0},p_{l^{\prime }0}\right] ^{k_{l^{\prime }}}\right) \underset{%
l,l^{\prime }}{\otimes }\Psi _{J,\alpha ,s_{p_{l^{\prime },l}}^{\otimes
k_{l^{\prime }}}}^{\otimes k_{l^{\prime }}}\left( \left[ p_{l},p_{l^{\prime
}l}\right] ^{k_{l^{\prime }}}\right) \right\}
\end{equation*}%
and:%
\begin{equation*}
v_{\left[ p_{l},p_{l^{\prime }l}\right] ^{k_{l^{\prime }}}}\left( \Psi
_{J,s_{p_{l^{\prime },0}}^{\otimes k_{l^{\prime }}}}^{\otimes k_{l^{\prime
}}}\right) \otimes \Psi _{J,s_{p_{l^{\prime },l}}^{\otimes k_{l^{\prime
}}}}^{\otimes k_{l^{\prime }}}\left( \left[ p_{l},p_{l^{\prime }l}\right]
^{k_{l^{\prime }}}\right)
\end{equation*}%
is evaluated at the considered realization:%
\begin{equation*}
\Psi _{J,\alpha ,s_{p_{l^{\prime },0}}^{\otimes k_{l^{\prime }}}}^{\otimes
k_{l^{\prime }}}\otimes \Psi _{J,\alpha ,s_{p_{l^{\prime },l}}^{\otimes
k_{l^{\prime }}}}^{\otimes k_{l^{\prime }}}
\end{equation*}%
We will write the functional $v_{\left[ p_{l},p_{l^{\prime }l}\right]
^{k_{l^{\prime }}}}$ in components: 
\begin{eqnarray*}
&&v_{\left[ p_{l},p_{l^{\prime }l}\right] ^{k_{l^{\prime }}}}\left( \underset%
{l^{\prime }}{\otimes }\Psi _{J,s_{p_{l^{\prime },0}}^{\otimes k_{l^{\prime
}}}}^{\otimes k_{l^{\prime }}}\right) \\
&=&\int v\left( \overline{\left[ p_{0},p_{l^{\prime }0}\right]
^{k_{l^{\prime }}}},\left[ p_{l},p_{l^{\prime }l}\right] ^{k_{l^{\prime
}}}\right) \times \Psi _{J,s_{p_{l^{\prime },0}}^{\otimes k_{l^{\prime
}}}}^{\otimes k_{l^{\prime }}}\left( \overline{\left[ p_{0},p_{l^{\prime }0}%
\right] ^{k_{l^{\prime }}}},\left[ p_{l},p_{l^{\prime }l}\right]
^{k_{l^{\prime }}}\right) d\left( \overline{\left[ p_{0},p_{l^{\prime }0}%
\right] ^{k_{l^{\prime }}}}\right)
\end{eqnarray*}%
As before, including the $id$ map, we have:

\begin{equation*}
\left\{ \Psi _{J,0,s_{p_{l}}^{\otimes k_{l}}}^{\otimes k_{l}}\left( \left[
p_{l},k_{l}\right] \right) \tbigoplus\limits_{\left( p_{l^{\prime }}\right)
}\Psi _{J,0,s_{p_{l^{\prime },l}}^{\otimes k_{l^{\prime }}}}^{\otimes
k_{l^{\prime }}}\left( \left[ p_{l},p_{l^{\prime }l}\right] ^{k_{l^{\prime
}}}\right) \right\} _{\substack{ l\leqslant m  \\ l^{\prime }\leqslant
m^{\prime }}}\equiv \left\{ \tbigoplus\limits_{\left( p_{l^{\prime }}\right)
}\Psi _{J,0,s_{p_{l^{\prime },l}}^{\otimes k_{l^{\prime }}}}^{\otimes
k_{l^{\prime }}}\left( \left[ p_{l},p_{l^{\prime }l}\right] ^{k_{l^{\prime
}}}\right) \right\} _{\substack{ l\leqslant m  \\ l^{\prime }\leqslant
m^{\prime }}}
\end{equation*}%
We also write: 
\begin{equation*}
v\left\{ \Psi _{J}^{\otimes k_{l^{\prime }}}\right\} =\sum \int v\left( 
\left[ p_{0},p_{l^{\prime }0}\right] ^{k_{l^{\prime }}}\right) \Psi
_{J}^{\otimes k_{l^{\prime }}}\left( \left[ p_{0},p_{l^{\prime }0}\right]
^{k_{l^{\prime }}}\right) d\left[ p_{0},p_{l^{\prime }0}\right]
^{k_{l^{\prime }}}
\end{equation*}

As in the first part, assuming that $S$ is a series in the variables:%
\begin{equation*}
\left\{ \underset{l^{\prime }}{\otimes }\Psi _{J,s_{p_{l^{\prime
},0}}^{\otimes k_{l^{\prime }}}}^{\otimes k_{l^{\prime }}}\left( \left[
p_{0},p_{l^{\prime }0}\right] ^{k_{l^{\prime }}}\right) \underset{%
l,l^{\prime }}{\otimes }\Psi _{J,s_{p_{l^{\prime },l}}^{\otimes k_{l^{\prime
}}}}^{\otimes k_{l^{\prime }}}\left( \left[ p_{l},p_{l^{\prime }l}\right]
^{k_{l^{\prime }}}\right) \right\}
\end{equation*}%
including a quadratic term for the $\Psi _{J,s_{p_{l^{\prime },l}}^{\otimes
k_{l^{\prime }}}}^{\otimes k_{l^{\prime }}}\left( \left[ p_{l},p_{l^{\prime
}l}\right] ^{k_{l^{\prime }}}\right) $, the series expansion of the saddle
point solution is:

\begin{eqnarray}
&&\Psi _{J,0,s_{p_{l^{\prime },l}}^{\otimes k_{l^{\prime }}}}^{\otimes
k_{l^{\prime }}}\left( \left[ p_{l},p_{l^{\prime }l}\right] ^{k_{l^{\prime
}}},v\right)  \label{SL} \\
&=&\sum_{_{\substack{ \left( s,s^{\prime },\left[ p,p^{\prime },k^{\prime }%
\right] \right)  \\ \left( m,m^{\prime },\left[ p,p^{\prime },k^{\prime }%
\right] \right) _{1,2}}}}d\left( \left\{ \overline{\left[ p_{0},p_{l^{\prime
}0}\right] ^{k_{l^{\prime }}}}\right\} _{\substack{ l\leqslant s  \\ %
l^{\prime }\leqslant s^{\prime }}}\right) d\left( \left\{ \left[
p_{l_{1}},p_{l_{1}^{\prime }l_{1}}\right] ^{k_{l_{1}^{\prime }}}\right\} 
_{\substack{ l_{1}\leqslant m_{1},l_{1}\neq l  \\ l_{1}^{\prime }\leqslant
m_{1}^{\prime }}}\left\{ \left[ p_{l_{2}},p_{l_{2}^{\prime }l_{2}}\right]
^{k_{l_{2}^{\prime }}}\right\} _{\substack{ l_{2}\leqslant m_{2}  \\ %
l_{2}^{\prime }\leqslant m_{2}^{\prime }}}\right)  \notag \\
&&\times \mathcal{K}_{0}\left( \left\{ \overline{\left[ p_{0},p_{l^{\prime
}0}\right] ^{k_{l^{\prime }}}}\right\} _{_{\substack{ l\leqslant s  \\ %
l^{\prime }\leqslant s^{\prime }}}},\left\{ \left[ p_{l_{1}},p_{l_{1}^{%
\prime }l_{1}}\right] ^{k_{l_{1}^{\prime }}}\right\} _{_{\substack{ %
l_{1}\leqslant m_{1}  \\ l_{1}^{\prime }\leqslant m_{1}^{\prime }}}},\left\{ %
\left[ p_{l_{2}},p_{l_{2}^{\prime }l_{2}}\right] ^{k_{l_{2}^{\prime
}}}\right\} _{_{\substack{ l_{2}\leqslant m_{2}  \\ l_{2}^{\prime }\leqslant
m_{2}^{\prime }}}}\right)  \notag \\
&&\times \tprod_{\substack{ l\leqslant s  \\ l^{\prime }\leqslant s^{\prime
} }}\Psi _{J,s_{p_{l^{\prime },0}}^{\otimes k_{l_{i}^{\prime }}}}^{\otimes
k_{l^{\prime },i}}\left( \overline{\left[ p_{0},p_{l^{\prime }0}\right]
^{k_{l^{\prime }}}},\left\{ \left[ p_{l_{1}},p_{l_{1}^{\prime }l_{1}}\right]
^{k_{l_{1}^{\prime }}}\right\} _{_{\substack{ l_{1}\leqslant m_{1}  \\ %
l_{1}^{\prime }\leqslant m_{1}^{\prime }}}}\right) \tprod_{_{_{\substack{ %
l_{2}\leqslant m_{2}  \\ l_{2}^{\prime }\leqslant m_{2}^{\prime }}}}}v_{%
\left[ p_{l_{2}},p_{l_{2}^{\prime }l_{2}}\right] ^{k_{l_{2}^{\prime
}}}}\left\{ \Psi _{J}^{\otimes k_{l^{\prime }}}\right\}  \notag
\end{eqnarray}%
where:%
\begin{equation*}
\overline{\left[ p_{0},p_{l^{\prime }0}\right] ^{k_{l^{\prime }}}}\equiv 
\overline{\left( \left( U_{j}\right) ^{\left( p_{0},p_{l^{\prime }0}\right)
}\right) ^{k_{l^{\prime }}}}=\left( \left( U_{j}\right) ^{\left(
p_{0},p_{l^{\prime }0}\right) }\right) ^{k_{l^{\prime
}}}/\tprod\limits_{p_{1,l},...,p_{m^{\prime },l}}f_{p_{1,l}...p_{m^{\prime
},l}}
\end{equation*}%
and where the $f_{p_{l^{\prime },0},p_{1}...p_{m}}$ describe the constraint
involving $\left( U_{j}\right) ^{\left( p_{0},p_{l^{\prime }0}\right) }$ and
the $\left( U_{j}\right) ^{\left( p_{l}\right) }$ and their subobjects. Note
that:%
\begin{equation*}
\left\{ \overline{\left[ p_{0},p_{l^{\prime }0}\right] ^{k_{l^{\prime }}}}%
\right\} _{\substack{ l\leqslant s  \\ l^{\prime }\leqslant s^{\prime }}}
\end{equation*}%
are $s$ copies of:%
\begin{equation*}
\left\{ \overline{\left[ p_{0},p_{l^{\prime }0}\right] ^{k_{l^{\prime }}}}%
\right\} _{l^{\prime }\leqslant s^{\prime }}
\end{equation*}%
We have also defined:%
\begin{equation*}
\left[ p,p^{\prime },k^{\prime }\right] =\left\{ p_{0},p_{l^{\prime
}0},k_{l^{\prime }}\right\} _{\substack{ l\leqslant s  \\ l^{\prime
}\leqslant s^{\prime }}}
\end{equation*}%
and:%
\begin{equation*}
\left( m,m^{\prime },\left[ p,p^{\prime },k^{\prime }\right] \right)
_{1,2}=\left\{ \left\{ p_{l_{1}},p_{l_{1}^{\prime }l_{1}},k_{l_{1}^{\prime
}}\right\} _{_{_{\substack{ l_{1}\leqslant m_{1}  \\ l_{1}^{\prime
}\leqslant m_{1}^{\prime }}}}},\left\{ p_{l_{2}},p_{l_{2}^{\prime
}l_{2}},k_{l_{2}^{\prime }}\right\} _{_{\substack{ l_{2}\leqslant m_{2}  \\ %
l_{2}^{\prime }\leqslant m_{2}^{\prime }}}}\right\}
\end{equation*}%
The sums over indices $l^{\prime }$, $l_{1}$ and $l_{1}^{\prime }$ are
understood to run over sets:%
\begin{equation*}
\left\{ l^{\prime }\right\} ,\left\{ l_{1}\right\} ,\left\{ l_{1}^{\prime
}\right\}
\end{equation*}%
and the notation:%
\begin{equation*}
l^{\prime }\leqslant s^{\prime },l_{1}\leqslant m_{1},l_{1}^{\prime
}\leqslant m_{1}^{\prime }
\end{equation*}%
is an abreviation for:%
\begin{equation*}
\natural \left\{ l^{\prime }\right\} \leqslant s^{\prime },\natural \left\{
l_{1}\right\} \leqslant m_{1},\natural \left\{ l_{1}^{\prime }\right\}
\leqslant m_{1}^{\prime }
\end{equation*}

Gathering indices $p_{l_{1}}$, $p_{l_{2}}$,.. as $p_{l_{1}}$ and defining
the following partitions of $\left\{ \left[ p_{l_{1}},p_{l_{1}^{\prime
}l_{1}}\right] ^{k_{l_{1}^{\prime }}}\right\} _{_{\substack{ l_{1}\leqslant
m_{1}  \\ l_{1}^{\prime }\leqslant m_{1}^{\prime }}}}$:%
\begin{equation*}
\left\{ \left[ p_{l_{1}},p_{l_{1}^{\prime }l_{1}}\right] ^{k_{l_{1}^{\prime
}}}\right\} _{_{\substack{ l_{1}\leqslant m_{1}  \\ l_{1}^{\prime }\leqslant
m_{1}^{\prime }}}}=\left\{ \left[ p_{l_{1}},p_{l_{1}^{\prime }l_{1}}\right]
^{k_{l_{1}^{\prime }}}\right\} _{\mathcal{P}},\left\{ \left[
p_{l_{1}},p_{l_{1}^{\prime }l_{1}}\right] ^{k_{l_{1}^{\prime }}}\right\} _{%
\mathcal{P}^{c}}
\end{equation*}%
we find a more compact formulation for the saddle point solution without
degeneracy:%
\begin{eqnarray}
&&\Psi _{J,0,s_{p_{l^{\prime },l}}^{\otimes k_{l^{\prime }}}}^{\otimes
k_{l^{\prime }}}\left( \left[ p_{l},p_{l^{\prime }l}\right] ^{k_{l^{\prime
}}},v\right)  \label{SDL} \\
&=&\sum_{_{\substack{ \left( s,s^{\prime },\left[ p,p^{\prime },k^{\prime }%
\right] \right)  \\ \left( m,m^{\prime },\left[ p,p^{\prime },k^{\prime }%
\right] \right) _{1}}}}d\left( \left\{ \overline{\left[ p_{0},p_{l^{\prime
}0}\right] ^{k_{l^{\prime }}}}\right\} _{\substack{ l\leqslant s  \\ %
l^{\prime }\leqslant s^{\prime }}},\left\{ \left[ p_{l_{1}},p_{l_{1}^{\prime
}l_{1}}\right] ^{k_{l_{1}^{\prime }}}\right\} _{\substack{ l_{1}\leqslant
m_{1},l_{1}\neq l  \\ l_{1}^{\prime }\leqslant m_{1}^{\prime }}}\right) 
\notag \\
&&\times \mathcal{K}_{0}\left( \left\{ \overline{\left[ p_{0},p_{l^{\prime
}0}\right] ^{k_{l^{\prime }}}}\right\} _{_{\substack{ l\leqslant s  \\ %
l^{\prime }\leqslant s^{\prime }}}},\left\{ \left[ p_{l_{1}},p_{l_{1}^{%
\prime }l_{1}}\right] ^{k_{l_{1}^{\prime }}}\right\} _{_{\substack{ %
l_{1}\leqslant m_{1}  \\ l_{1}^{\prime }\leqslant m_{1}^{\prime }}}}\right) 
\notag \\
&&\times \tprod_{\substack{ l\leqslant s  \\ l^{\prime }\leqslant s^{\prime
} }}\Psi _{J,s_{p_{l^{\prime },0}}^{\otimes k_{l^{\prime }}}}^{\otimes
k_{l^{\prime }}}\left( \overline{\left[ p_{0},p_{l^{\prime }0}\right]
^{k_{l^{\prime }}}},\left\{ \left[ p_{l_{1}},p_{l_{1}^{\prime }l_{1}}\right]
^{k_{l_{1}^{\prime }}}\right\} _{_{\mathcal{P}}}\right) \tprod_{\mathcal{P}%
^{c}}v_{\left[ p_{l_{1}},p_{l_{1}^{\prime }l_{1}}\right] ^{k_{l_{1}^{\prime
}}}}\left\{ \Psi _{J}^{\otimes k_{l^{\prime }}}\right\}  \notag
\end{eqnarray}

\subsubsection{Including degeneracy: general formula}

As in part one, we conside transformations parametrized by: 
\begin{eqnarray*}
&&\left\{ \left\{ \left( \mathbf{\hat{\Lambda}}^{\left( p_{l}\right) }\left(
\Psi _{J}^{\otimes k_{p_{0}}},v\right) \right) \right\} _{p_{l}},\left\{
\left( \mathbf{\hat{\Lambda}}^{\left( p_{l},p_{l^{\prime },l},k_{l^{\prime
}}\right) }\left( \Psi _{J}^{\otimes k_{p_{0,l^{\prime }}}},v\right) \right)
\right\} _{p_{l^{\prime },l}}\right\} \\
&\rightarrow &\left\{ \left\{ \left( \mathbf{\hat{\Lambda}}^{\left(
p_{l},p_{l^{\prime },l},k_{l^{\prime }}\right) }\left( \Psi _{J}^{\otimes
k_{p_{0,l^{\prime }}}},v\right) \right) \right\} _{p_{l^{\prime },l}}\right\}
\end{eqnarray*}%
\ with in general $\left( \left( \mathbf{\hat{\Lambda}}^{\left(
p_{l},p_{l^{\prime },l},k_{l^{\prime }}\right) }\left( \Psi _{J}^{\otimes
k_{p_{0,l^{\prime }}}},v\right) \right) \right) \simeq \left( \mathbf{\hat{%
\Lambda}}^{\left( p_{l},p_{l^{\prime },l}\right) }\left( \Psi _{J}^{\otimes
k_{p_{0,l^{\prime }}}},v\right) \right) ^{k_{l^{\prime }}}$.

In the sequel, we will write the parameters in a more compact form:%
\begin{equation}
\left\{ \left\{ \left( \mathbf{\hat{\Lambda}}^{\left( p_{l},p_{l,^{\prime
}l},k_{l^{\prime }}\right) }\left( \Psi _{J}^{\otimes k_{p_{0,l^{\prime
}}}},v\right) \right) \right\} _{p_{l,l^{\prime }}}\right\} _{k_{l^{\prime
}}}\rightarrow \left[ \underset{\left[ \left\{ \Psi _{J}\left[ l^{\prime },0%
\right] \right\} ,v\right] }{\overset{\left( p_{l},p_{l,^{\prime
}l},k_{l^{\prime }}\right) }{\mathbf{\hat{\Lambda}}}}\right] \rightarrow
\left\{ \left[ 
\begin{array}{c}
\left( p_{l},p_{l,^{\prime }l},k_{l^{\prime }}\right) \\ 
\left[ \left\{ \Psi _{J}\left[ l^{\prime },0\right] \right\} ,v\right]%
\end{array}%
\right] \right\} _{\left( p_{l},p_{l,^{\prime }l},k_{l^{\prime }}\right) }
\label{PRSM}
\end{equation}

The upper indices account for the decomposition of the initial field,
indicating that the degeneracy parameters depend on the subobjects
decomposition. This dependency involves the mappings $\left(
p_{l},p_{l,^{\prime }l}\right) $ and the tensor power $k_{l^{\prime }}$ of
the subobject arising in the saddle-point solution. The lower indices
account for the dependence of symmetries on the remaining non-projected
field. The symmetry parameters are functions of the states in which this
effective system is considered\footnote{%
In fact, note that in this formula, taking account of subobject implies that
the parameters implicitely stand for:%
\begin{equation*}
\left[ \underset{\left[ \left\{ \Psi _{J}\left[ l^{\prime },0\right]
\right\} ,\left[ p_{0},p_{l^{\prime }0}\right] ^{k_{l^{\prime }}}\right] }{%
\overset{\left( p_{l_{1}},p_{l_{1}^{\prime }l_{1}},k_{l_{1}^{\prime
}}\right) }{\mathbf{\hat{\Lambda}}}}\right] =\left[ \underset{\left[ \left\{
\Psi _{J}\left[ l^{\prime },0\right] \right\} ,\Psi _{J,\alpha
,,s_{p_{l^{\prime },0}}^{\otimes k_{l^{\prime }}}}^{\otimes k_{l^{\prime
}}}\left( \left[ p_{0},p_{l^{\prime }0}\right] ^{k_{l^{\prime }}}\right) %
\right] }{\overset{\left( p_{l_{1}},p_{l_{1}^{\prime
}l_{1}},k_{l_{1}^{\prime }}\right) }{\mathbf{\hat{\Lambda}}}}\right]
\end{equation*}%
}.

In (\ref{PRSM}), the arrow indicates that we will omit $\mathbf{\hat{\Lambda}%
}$, which represented the parameters in the previous sections, to retain
only the parameters main determinants---the mappings between subobjects.

As in the first part, we define the parameters:%
\begin{equation*}
\left\{ \left[ 
\begin{array}{c}
\left( p_{l},p_{l,^{\prime }l},k_{l^{\prime }}\right) \\ 
\left[ \left\{ \Psi _{J}\left[ l^{\prime },0\right] \right\} ,v\right]%
\end{array}%
\right] \right\} =\cup _{\substack{ \left( m_{1},m_{1}^{\prime }\right) , 
\\ l\notin \left\{ l_{1}\right\} ,l^{\prime }\notin \left\{ l_{1}^{\prime
}\right\} }}\left\{ \left[ 
\begin{array}{c}
\left( p_{l},p_{l,^{\prime }l},k_{l^{\prime }}\right) ,\left\{ \left(
p_{l_{1}},p_{l_{1},^{\prime }l_{1}},k_{l_{1}^{\prime }}\right) \right\}
_{\left( \left\{ l_{1}\right\} ,\left\{ l_{1}^{\prime }\right\} \right) } \\ 
\left[ \left\{ \Psi _{J}\left[ l^{\prime },0\right] \right\} ,v\right]%
\end{array}%
\right] \right\} _{\substack{ l_{1}<m_{1}  \\ l_{1}^{\prime }<m_{1}^{\prime
} }}
\end{equation*}%
corresponding to sequences of symetry groups $G_{1}\subset ...\subset G_{k}$%
. As before the notation:%
\begin{equation*}
l^{\prime }\leqslant s^{\prime },l_{1}<m_{1},l_{1}^{\prime }<m_{1}^{\prime }
\end{equation*}%
is an abreviation for:%
\begin{equation*}
\natural \left\{ l^{\prime }\right\} \leqslant s^{\prime },\natural \left\{
l_{1}\right\} <m_{1},\natural \left\{ l_{1}^{\prime }\right\} <m_{1}^{\prime
}
\end{equation*}

This decomposition can also be written:%
\begin{equation*}
\left\{ \left[ 
\begin{array}{c}
\left( p_{l},p_{l,^{\prime }l},k_{l^{\prime }}\right) \\ 
\left[ \left\{ \Psi _{J}\left[ l^{\prime },0\right] \right\} ,v\right]%
\end{array}%
\right] \right\} =\cup _{\substack{ \left( m_{1},m_{1}^{\prime }\right) , 
\\ l\in \left\{ l_{1}\right\} ,l^{\prime }\in \left\{ l_{1}^{\prime
}\right\} }}\left\{ \left[ 
\begin{array}{c}
\left\{ \left( p_{l_{1}},p_{l_{1},^{\prime }l_{1}},k_{l_{1}^{\prime
}}\right) \right\} _{\left\{ l_{1}\right\} ,\left\{ l_{1}^{\prime }\right\} }
\\ 
\left[ \left\{ \Psi _{J}\left[ l^{\prime },0\right] \right\} ,v\right]%
\end{array}%
\right] \right\} _{_{\substack{ l_{1}\leqslant m_{1}  \\ l_{1}^{\prime
}\leqslant m_{1}^{\prime }}}}
\end{equation*}%
so that in the sequel, unless precised:%
\begin{equation*}
\left\{ \left[ 
\begin{array}{c}
\left( p_{l_{1}},p_{l_{1},^{\prime }l_{1}},k_{l_{1}^{\prime }}\right) \\ 
\left[ \left\{ \Psi _{J}\left[ l^{\prime },0\right] \right\} ,v\right]%
\end{array}%
\right] \right\} _{_{\substack{ l\subset l_{1}\leqslant m_{1}  \\ l^{\prime
}\subset l_{1}^{\prime }\leqslant m_{1}^{\prime }}}}
\end{equation*}%
will stand for:%
\begin{equation*}
\left\{ \left[ 
\begin{array}{c}
\left( p_{l},p_{l,^{\prime }l},k_{l^{\prime }}\right) ,\left\{ \left(
p_{l_{1}},p_{l_{1},^{\prime }l_{1}},k_{l_{1}^{\prime }}\right) \right\}
_{\left\{ l_{1}\right\} ,\left\{ l_{1}^{\prime }\right\} } \\ 
\left[ \left\{ \Psi _{J}\left[ l^{\prime },0\right] \right\} ,v\right]%
\end{array}%
\right] \right\} _{_{\substack{ l_{1}<m_{1}  \\ l_{1}^{\prime
}<m_{1}^{\prime }}}}
\end{equation*}

If the subobjects are included, the generators of transformations between
realizations of $\Psi _{J,0,s_{p_{l^{\prime },l}}^{\otimes k_{l^{\prime
}}}}^{\otimes k_{l^{\prime }}}$ are:%
\begin{eqnarray}
\mathbf{L}\left( \Psi _{J}^{\otimes k_{p_{0}}},v\right) &=&\sum_{ 
_{\substack{ \left( s,s^{\prime },\left[ p,p^{\prime },k^{\prime }\right]
\right)  \\ \left( m_{1},m_{1}^{\prime },\left[ p,p^{\prime },k^{\prime }%
\right] \right) }}}d\left( \left\{ \overline{\left[ p_{0},p_{l^{\prime }0}%
\right] ^{k_{l^{\prime }}}}\right\} _{\substack{ l\leqslant s  \\ l^{\prime
}\leqslant s^{\prime }}}\right) d\left( \left\{ \left[ p_{l_{1}},p_{l_{1}^{%
\prime }l_{1}}\right] ^{k_{l_{1}^{\prime }}}\right\} _{\substack{ %
l_{1}\leqslant m_{1}  \\ l_{1}^{\prime }\leqslant m^{\prime }}}\right) \\
&&\mathbf{l}\left( \mathbf{U}_{\left[ p_{l},p_{l^{\prime }l}\right]
^{k_{l^{\prime }}}},\Pi _{\left[ p_{l},p_{l^{\prime }l}\right]
^{k_{l^{\prime }}}},\overline{\left[ p_{0},p_{l^{\prime }0}\right]
^{k_{l^{\prime }}}},\left\{ \left[ p_{l_{1}},p_{l_{1}^{\prime }l_{1}}\right]
^{k_{l_{1}^{\prime }}}\right\} _{\substack{ l_{1}\leqslant m  \\ %
l_{1}^{\prime }\leqslant m^{\prime }}},v\right) v_{\left\{ \left[
p_{l_{1}},p_{l_{1}^{\prime }l_{1}}\right] ^{k_{l_{1}^{\prime }}}\right\} 
_{\substack{ l_{1}\leqslant m  \\ l_{1}^{\prime }\leqslant m^{\prime }}}%
}\left\{ \Psi _{J}^{\otimes k_{l^{\prime }}}\right\}  \notag
\end{eqnarray}%
We will write:%
\begin{equation*}
\left( m_{1},m_{1}^{\prime },\left[ p,p^{\prime },k^{\prime }\right] \right)
=\left( m,m^{\prime },\left[ p,p^{\prime },k^{\prime }\right] \right) _{1}
\end{equation*}%
and $\mathbf{l}$ has components dual to $\left[ 
\begin{array}{c}
\left( p_{l},p_{l,^{\prime }l},k_{l^{\prime }}\right) \\ 
\left[ \left\{ \Psi _{J}\left[ l^{\prime },0\right] \right\} ,v\right]%
\end{array}%
\right] $ so that we write the solution given by the group action as:%
\begin{eqnarray*}
&&\Psi _{J,0,s_{p_{l^{\prime },l}}^{\otimes k_{l^{\prime }}}}^{\otimes
k_{l^{\prime }}}\left( \left[ p_{l},p_{l^{\prime }l}\right] ^{k_{l^{\prime
}}},\left\{ \left[ 
\begin{array}{c}
\left( p_{l},p_{l,^{\prime }l},k_{l^{\prime }}\right) \\ 
\left[ \left\{ \Psi _{J}\left[ l^{\prime },0\right] \right\} ,v\right]%
\end{array}%
\right] \right\} ,v\right) \\
&=&\exp \left( i\mathbf{L}\left( \Psi _{J}^{\otimes k_{p_{0}}},\nu \right)
.\left\{ \left[ 
\begin{array}{c}
\left( p_{l},p_{l,^{\prime }l},k_{l^{\prime }}\right) \\ 
\left[ \left\{ \Psi _{J}\left[ l^{\prime },0\right] \right\} ,v\right]%
\end{array}%
\right] \right\} \right) \Psi _{J,0,s_{p_{l^{\prime },l}}^{\otimes
k_{l^{\prime }}}}^{\otimes k_{l^{\prime }}}\left( \left[ p_{l},p_{l^{\prime
}l}\right] ^{k_{l^{\prime }}},v\right)
\end{eqnarray*}%
The dependency in $\left( \Psi _{J}^{\otimes k_{p_{0}}}\right) $ is kept
implicit. Inserting this result in the saddle point equation (\ref{SDL})
yields:%
\begin{eqnarray}
&&S\left( \left\{ \underset{l^{\prime }}{\otimes }\Psi _{J,s_{p_{l^{\prime
},0}}^{\otimes k_{l^{\prime }}}}^{\otimes k_{l^{\prime }}}\left( \left[
p_{0},p_{l^{\prime }0}\right] ^{k_{l^{\prime }}}\right) \underset{%
l,l^{\prime }}{\otimes }\Psi _{J,s_{p_{l^{\prime },l}}^{\otimes k_{l^{\prime
}}}}^{\otimes k_{l^{\prime }}}\left( \left[ p_{l},p_{l^{\prime }l}\right]
^{k_{l^{\prime }}},\left[ 
\begin{array}{c}
\left( p_{l},p_{l,^{\prime }l},k_{l^{\prime }}\right) \\ 
\left[ \left\{ \Psi _{J}\left[ l^{\prime },0\right] \right\} ,v\right]%
\end{array}%
\right] \right) \right\} ,\right.  \label{Ct} \\
&&\left. v_{\left[ p_{l},p_{l^{\prime }l}\right] ^{k_{l^{\prime }}}}\left(
\Psi _{J,s_{p_{l^{\prime },0}}^{\otimes k_{l^{\prime }}}}^{\otimes
k_{l^{\prime }}}\right) \otimes \Psi _{J,s_{p_{l^{\prime },l}}^{\otimes
k_{l^{\prime }}}}^{\otimes k_{l^{\prime }}}\left( \left[ p_{l},p_{l^{\prime
}l}\right] ^{k_{l^{\prime }}},\left[ 
\begin{array}{c}
\left( p_{l},p_{l,^{\prime }l},k_{l^{\prime }}\right) \\ 
\left[ \left\{ \Psi _{J}\left[ l^{\prime },0\right] \right\} ,v\right]%
\end{array}%
\right] \right) \right)  \notag \\
&=&S\left( \left\{ \underset{l^{\prime }}{\otimes }\Psi _{J,s_{p_{l^{\prime
},0}}^{\otimes k_{l^{\prime }}}}^{\otimes k_{l^{\prime }}}\left( \left[
p_{0},p_{l^{\prime }0}\right] ^{k_{l^{\prime }}}\right) \underset{%
l,l^{\prime }}{\otimes }\Psi _{J,s_{p_{l^{\prime },l}}^{\otimes k_{l^{\prime
}}}}^{\otimes k_{l^{\prime }}}\left( \left[ p_{l},p_{l^{\prime }l}\right]
^{k_{l^{\prime }}}\right) \right\} ,v_{\left[ p_{l},p_{l^{\prime }l}\right]
^{k_{l^{\prime }}}}\left( \Psi _{J,s_{p_{l^{\prime },0}}^{\otimes
k_{l^{\prime }}}}^{\otimes k_{l^{\prime }}}\right) \otimes \Psi
_{J,s_{p_{l^{\prime },l}}^{\otimes k_{l^{\prime }}}}^{\otimes k_{l^{\prime
}}}\left( \left[ p_{l},p_{l^{\prime }l}\right] ^{k_{l^{\prime }}}\right)
\right)  \notag
\end{eqnarray}%
due to the degeneracy. The degenerate saddle point involvd in (\ref{Ct})
writes:%
\begin{eqnarray}
&&\Psi _{J,0,s_{p_{l^{\prime },l}}^{\otimes k_{l^{\prime }}}}^{\otimes
k_{l^{\prime }}}\left( \left[ p_{l},p_{l^{\prime }l}\right] ^{k_{l^{\prime
}}},\left\{ \left[ 
\begin{array}{c}
\left( p_{l},p_{l,^{\prime }l},k_{l^{\prime }}\right) \\ 
\left[ \left\{ \Psi _{J}\left[ l^{\prime },0\right] \right\} ,v\right]%
\end{array}%
\right] \right\} ,v\right)  \label{SDLPT} \\
&=&\sum_{_{\substack{ \left( s,s^{\prime },\left[ p,p^{\prime },k^{\prime }%
\right] \right)  \\ \left( m,m^{\prime },\left[ p,p^{\prime },k^{\prime }%
\right] \right) _{1}}}}d\left( \left\{ \overline{\left[ p_{0},p_{l^{\prime
}0}\right] ^{k_{l^{\prime }}}}\right\} _{\substack{ l\leqslant s  \\ %
l^{\prime }\leqslant s^{\prime }}},\left\{ \left[ p_{l_{1}},p_{l_{1}^{\prime
}l_{1}}\right] ^{k_{l_{1}^{\prime }}}\right\} _{\substack{ l_{1}\leqslant
m_{1},l_{1}\neq l  \\ l_{1}^{\prime }\leqslant m_{1}^{\prime }}}\right) 
\notag \\
&&\times \mathcal{K}_{0}\left( \left\{ \overline{\left[ p_{0},p_{l^{\prime
}0}\right] ^{k_{l^{\prime }}}}\right\} _{_{\substack{ l\leqslant s  \\ %
l^{\prime }\leqslant s^{\prime }}}},\left\{ \left[ p_{l_{1}},p_{l_{1}^{%
\prime }l_{1}}\right] ^{k_{l_{1}^{\prime }}}\right\} _{_{\substack{ %
l_{1}\leqslant m_{1}  \\ l_{1}^{\prime }\leqslant m_{1}^{\prime }}}},\left\{ %
\left[ 
\begin{array}{c}
\left( p_{l_{1}},p_{l_{1},^{\prime }l_{1}},k_{l_{1}^{\prime }}\right) \\ 
\left[ \left\{ \Psi _{J}\left[ l^{\prime },0\right] \right\} ,v\right]%
\end{array}%
\right] \right\} _{_{\substack{ l\subset l_{1}\leqslant m_{1}  \\ l^{\prime
}\subset l_{1}^{\prime }\leqslant m_{1}^{\prime }}}}\right)  \notag \\
&&\times \tprod_{\substack{ l\leqslant s  \\ l^{\prime }\leqslant s^{\prime
} }}\Psi _{J,s_{p_{l^{\prime },0}}^{\otimes k_{l^{\prime }}}}^{\otimes
k_{l^{\prime }}}\left( \overline{\left[ p_{0},p_{l^{\prime }0}\right]
^{k_{l^{\prime }}}},\left\{ \left[ p_{l_{1}},p_{l_{1}^{\prime }l_{1}}\right]
^{k_{l_{1}^{\prime }}}\right\} _{_{\mathcal{P}}}\right) \tprod_{\mathcal{P}%
^{c}}v_{\left[ p_{l_{1}},p_{l_{1}^{\prime }l_{1}}\right] ^{k_{l_{1}^{\prime
}}}}\left\{ \Psi _{J}^{\otimes k_{l^{\prime }}}\right\}  \notag
\end{eqnarray}%
with:%
\begin{eqnarray*}
&&\mathcal{K}_{0}\left( \left\{ \overline{\left[ p_{0},p_{l^{\prime }0}%
\right] ^{k_{l^{\prime }}}}\right\} _{_{\substack{ l\leqslant s  \\ %
l^{\prime }\leqslant s^{\prime }}}},\left\{ \left[ p_{l_{1}},p_{l_{1}^{%
\prime }l_{1}}\right] ^{k_{l_{1}^{\prime }}}\right\} _{\substack{ %
l_{1}\leqslant m_{1}  \\ l_{1}^{\prime }\leqslant m_{1}^{\prime }}},\left\{ %
\left[ 
\begin{array}{c}
\left( p_{l_{1}},p_{l_{1},^{\prime }l_{1}},k_{l_{1}^{\prime }}\right) \\ 
\left[ \left\{ \Psi _{J}\left[ l^{\prime },0\right] \right\} ,v\right]%
\end{array}%
\right] \right\} _{_{\substack{ l\subset l_{1}\leqslant m_{1}  \\ l^{\prime
}\subset l_{1}^{\prime }\leqslant m_{1}^{\prime }}}}\right) \\
&=&\exp \left( i\mathbf{L}\left( \Psi _{J}^{\otimes k_{p_{0}}},v\right)
.\left( \left\{ \left[ 
\begin{array}{c}
\left( p_{l_{1}},p_{l_{1},^{\prime }l_{1}},k_{l_{1}^{\prime }}\right) \\ 
\left[ \left\{ \Psi _{J}\left[ l^{\prime },0\right] \right\} ,v\right]%
\end{array}%
\right] \right\} _{_{\substack{ l\subset l_{1}\leqslant m_{1}  \\ l^{\prime
}\subset l_{1}^{\prime }\leqslant m_{1}^{\prime }}}}\right) \right) \mathcal{%
K}_{0}\left( \left\{ \overline{\left[ p_{0},p_{l^{\prime }0}\right]
^{k_{l^{\prime }}}}\right\} _{_{\substack{ l\leqslant s  \\ l^{\prime
}\leqslant s^{\prime }}}},\left\{ \left[ p_{l_{1}},p_{l_{1}^{\prime }l_{1}}%
\right] ^{k_{l_{1}^{\prime }}}\right\} _{\substack{ l_{1}\leqslant m  \\ %
l_{1}^{\prime }\leqslant m^{\prime }}}\right)
\end{eqnarray*}%
and the operator $\mathbf{L}\left( \Psi _{J}^{\otimes k_{p_{0}}}\right) $
acts on the $\overline{\left[ p_{l_{1}},p_{l_{1}^{\prime }l_{1}}\right]
^{k_{l^{\prime }}}}$ degrees of freedom. We will also write:%
\begin{equation}
\Psi \left[ l^{\prime },0\right] =\underset{l^{\prime }}{\otimes }\Psi
_{J,s_{p_{l^{\prime },0}}^{\otimes k_{l^{\prime }}}}^{\otimes k_{l^{\prime
}}}  \label{CP}
\end{equation}

\subsubsection{Projected functional: one projected subobject}

We first consider functionals involving only $\left[ p_{0},p_{l_{i}^{\prime
}0}\right] ^{k_{l^{\prime }}}$ and one of the $\left[ p_{l},p_{l^{\prime }l}%
\right] ^{k_{l^{\prime }}}$. To each state:%
\begin{equation*}
\Psi _{J,0,s_{p_{l^{\prime },l}}^{\otimes k_{l^{\prime }}}}^{\otimes
k_{l^{\prime }}}\left( \left[ p_{l},p_{l^{\prime }l}\right] ^{k_{l^{\prime
}}},\left\{ \left[ 
\begin{array}{c}
\left( p_{l},p_{l,^{\prime }l},k_{l^{\prime }}\right)  \\ 
\left[ \left\{ \Psi _{J}\left[ l^{\prime },0\right] \right\} ,v\right] 
\end{array}%
\right] \right\} ,v\right) 
\end{equation*}%
we associate coefficient $\Psi \left( \left\{ \left[ 
\begin{array}{c}
\left( p_{l},p_{l,^{\prime }l},k_{l^{\prime }}\right)  \\ 
\left[ \left\{ \Psi _{J}\left[ l^{\prime },0\right] \right\} ,v\right] 
\end{array}%
\right] \right\} ,v\right) $ to obtain a projected background state, and
this leads to define the generic projected functional for each componnt (see
appendix 7):%
\begin{eqnarray}
&&\sum_{_{\substack{ \left( s,s^{\prime },\left[ p,p^{\prime },k^{\prime }%
\right] \right)  \\ \left( m,m^{\prime },\left[ p,p^{\prime },k^{\prime }%
\right] \right) _{1}}}}\int g^{\mathcal{K}}\left( \overline{\left[
p_{0},p_{l_{i}^{\prime }0}\right] ^{k_{l^{\prime }}}},\left\{ \left[
p_{l_{1}},p_{l_{1}^{\prime }l_{1}}\right] ^{k_{l_{1}^{\prime }}}\right\} _{
_{\substack{ l_{1}\leqslant m_{1} \\ l_{1}^{\prime }\leqslant m_{1}^{\prime }
}}},v,\left\{ \left[ 
\begin{array}{c}
\left( p_{l_{1}},p_{l_{1}^{\prime }l_{1}},k_{l_{1}^{\prime }}\right)  \\ 
\left[ \left\{ \Psi _{J}\left[ l^{\prime },0\right] \right\} ,v\right] 
\end{array}%
\right] \right\} _{_{\substack{ l\subset l_{1}\leqslant m_{1} \\ l^{\prime
}\subset l_{1}^{\prime }\leqslant m_{1}^{\prime }}}}\right)   \notag \\
&&\times \tprod_{\substack{ l\leqslant s \\ l^{\prime }\leqslant s^{\prime }
}}\Psi _{J,s_{p_{l^{\prime },0}}^{\otimes k_{l_{i}^{\prime }}}}^{\otimes
k_{l_{i}^{\prime }}}\left( \overline{\left[ p_{0},p_{l_{i}^{\prime }0}\right]
^{k_{l_{i}^{\prime }}}},\left\{ \left[ p_{l_{1}},p_{l_{1}^{\prime }l_{1}}%
\right] ^{k_{l_{1}^{\prime }}}\right\} _{_{\mathcal{P}}}\right) \tprod_{%
\mathcal{P}^{c}\neq \mathcal{\emptyset }}v_{\left[ p_{l_{1}},p_{l_{1}^{%
\prime }l_{1}}\right] ^{k_{l_{1}^{\prime }}}}\left\{ \Psi
_{J,s_{p_{l_{i}^{\prime \prime },0}}^{\otimes k_{l_{i}^{\prime \prime
}}}}^{\otimes k_{l_{i}^{\prime \prime }}}\right\} \Psi \left( \left\{ \left[ 
\begin{array}{c}
\left( p_{l},p_{l,^{\prime }l},k_{l^{\prime }}\right)  \\ 
\left[ \left\{ \Psi _{J}\left[ l^{\prime },0\right] \right\} ,v\right] 
\end{array}%
\right] \right\} ,v\right)   \notag \\
&&\times d\left( \left\{ \overline{\left[ p_{0},p_{l^{\prime }0}\right]
^{k_{l^{\prime }}}}\right\} _{\substack{ l\leqslant s \\ l^{\prime
}\leqslant s^{\prime }}},\left\{ \left[ p_{l_{1}},p_{l_{1}^{\prime }l_{1}}%
\right] ^{k_{l_{1}^{\prime }}}\right\} _{\substack{ l_{1}\leqslant m_{1} \\ %
l_{1}^{\prime }\leqslant m_{1}^{\prime }}}\right) d\left( \left\{ \left[ 
\begin{array}{c}
\left( p_{l},p_{l,^{\prime }l},k_{l^{\prime }}\right)  \\ 
\left[ \left\{ \Psi _{J}\left[ l^{\prime },0\right] \right\} ,v\right] 
\end{array}%
\right] \right\} \right) dv  \label{GN}
\end{eqnarray}%
whr $g^{\mathcal{K}}$ is defined in appendix 7. The indices $l_{i}^{\prime }$%
, $l_{i}^{\prime \prime }$ are the componnts f $\left\{ l^{\prime }\right\} $%
, $\left\{ l^{\prime }\right\} $ respectivly, and: 
\begin{equation*}
\overline{\left[ p_{0},p_{l^{\prime }0}\right] ^{k_{l^{\prime }}}}=\left( 
\overline{\left\{ \left[ p_{0},p_{l_{i}^{\prime }0}\right]
^{k_{l_{i}^{\prime }}}\right\} },\overline{\left\{ \left[ p_{0},p_{l_{i}^{%
\prime \prime }0}\right] ^{k_{l_{i}^{\prime \prime }}}\right\} }\right) 
\end{equation*}%
Note that reintroducing component index $\alpha $, the equation (\ref{GN})
is a functional of the state:%
\begin{eqnarray*}
&&v_{\left\{ \left[ p_{l_{1}},p_{l_{1}^{\prime }l_{1}}\right]
^{k_{l_{1}^{\prime }}}\right\} }\Psi _{J,\alpha ,s_{p_{l^{\prime
},0}}^{\otimes k_{l_{i}^{\prime }}},s_{p_{l^{\prime },0}}^{\otimes
k_{l_{i}^{\prime \prime }}}}^{\otimes \sum k_{l_{i}^{\prime }}+\sum
k_{l_{i}^{\prime \prime }}}\left( \overline{\left[ p_{0},p_{l^{\prime }0}%
\right] ^{k_{l^{\prime }}}}\right) \Psi _{\alpha }\left( \left\{ \left[ 
\begin{array}{c}
\left( p_{l},p_{l,^{\prime }l},k_{l^{\prime }}\right)  \\ 
\left[ \left\{ \Psi _{J}\left[ l^{\prime },0\right] \right\} ,v\right] 
\end{array}%
\right] \right\} ,v\right)  \\
&=&\int \tsum\limits_{\alpha \in \left\{ \alpha _{i}\right\} \cup \left\{
\alpha _{i}^{\prime }\right\} }\tprod_{\substack{ l\leqslant s \\ l^{\prime
}\leqslant s^{\prime }}}\Psi _{J;\left\{ \alpha _{i}\right\}
,s_{p_{l^{\prime },0}}^{\otimes k_{l_{i}^{\prime }}}}^{\otimes
k_{l_{i}^{\prime }}}\left( \overline{\left[ p_{0},p_{l_{i}^{\prime }0}\right]
^{k_{l_{i}^{\prime }}}},\left\{ \left[ p_{l_{1}},p_{l_{1}^{\prime }l_{1}}%
\right] ^{k_{l_{1}^{\prime }}}\right\} _{_{\mathcal{P}}}\right)  \\
&&\times \tprod_{\mathcal{P}^{c}\neq \mathcal{\emptyset }}v_{\left[
p_{l_{1}},p_{l_{1}^{\prime }l_{1}}\right] ^{k_{l_{1}^{\prime }}}}\left\{
\Psi _{J,\left\{ \alpha _{i}^{\prime }\right\} ,s_{p_{l_{i}^{\prime \prime
},0}}^{\otimes k_{l_{i}^{\prime \prime }}}}^{\otimes k_{l_{i}^{\prime \prime
}}}\right\} \Psi _{\alpha }\left( \left\{ \left[ 
\begin{array}{c}
\left( p_{l},p_{l,^{\prime }l},k_{l^{\prime }}\right)  \\ 
\left[ \left\{ \Psi _{J}\left[ l^{\prime },0\right] \right\} ,v\right] 
\end{array}%
\right] \right\} ,v\right) 
\end{eqnarray*}

\subsubsection{State dependent effective field}

Reintroducing the realization index $\alpha $, and if the coordinates are
independent of $\alpha $, we can sum over components and the functional (\ref%
{GN}) writes:%
\begin{eqnarray}
&&\sum_{_{\substack{ \left( s,s^{\prime },\left[ p,p^{\prime },k^{\prime }%
\right] \right)  \\ \left( m,m^{\prime },\left[ p,p^{\prime },k^{\prime }%
\right] \right) _{1}}}}\int g^{\mathcal{K}}\left( \overline{\left[
p_{0},p_{l^{\prime }0}\right] ^{k_{l^{\prime }}}},\left\{ \left[
p_{l_{1}},p_{l_{1}^{\prime }l_{1}}\right] ^{k_{l_{1}^{\prime }}}\right\} _{ 
_{\substack{ l_{1}\leqslant m_{1}  \\ l_{1}^{\prime }\leqslant m_{1}^{\prime
} }}},v,\left\{ \left[ 
\begin{array}{c}
\left( p_{l_{1}},p_{l_{1}^{\prime }l_{1}},k_{l_{1}^{\prime }}\right) \\ 
\left[ \left\{ \Psi _{J}\left[ l^{\prime },0\right] \right\} ,v\right]%
\end{array}%
\right] \right\} _{_{\substack{ l\subset l_{1}\leqslant m_{1}  \\ l^{\prime
}\subset l_{1}^{\prime }\leqslant m_{1}^{\prime }}}}\right)  \label{GP} \\
&&\times v_{\left\{ \left[ p_{l_{1}},p_{l_{1}^{\prime }l_{1}}\right]
^{k_{l_{1}^{\prime }}}\right\} }\Psi _{J,s_{p_{l^{\prime },0}}^{\otimes
k_{l_{i}^{\prime }}},s_{p_{l^{\prime },0}}^{\otimes k_{l_{i}^{\prime \prime
}}}}^{\otimes \sum k_{l_{i}^{\prime }}+\sum k_{l_{i}^{\prime \prime
}}}\left( \overline{\left[ p_{0},p_{l^{\prime }0}\right] ^{k_{l^{\prime }}}}%
,\left\{ \left[ 
\begin{array}{c}
\left( p_{l},p_{l,^{\prime }l},k_{l^{\prime }}\right) \\ 
\left[ \left\{ \Psi _{J}\left[ l^{\prime },0\right] \right\} ,v\right]%
\end{array}%
\right] \right\} ,v\right)  \notag \\
&&d\left( \left\{ \overline{\left[ p_{0},p_{l^{\prime }0}\right]
^{k_{l^{\prime }}}}\right\} _{\substack{ l\leqslant s  \\ l^{\prime
}\leqslant s^{\prime }}},\left\{ \left[ p_{l_{1}},p_{l_{1}^{\prime }l_{1}}%
\right] ^{k_{l_{1}^{\prime }}}\right\} _{\substack{ l_{1}\leqslant m_{1}  \\ %
l_{1}^{\prime }\leqslant m_{1}^{\prime }}}\right) d\left( \left\{ \left[ 
\begin{array}{c}
\left( p_{l},p_{l,^{\prime }l},k_{l^{\prime }}\right) \\ 
\left[ \left\{ \Psi _{J}\left[ l^{\prime },0\right] \right\} ,v\right]%
\end{array}%
\right] \right\} \right) dv  \notag
\end{eqnarray}%
where we define the effective field:%
\begin{eqnarray}
&&v_{\left\{ \left[ p_{l_{1}},p_{l_{1}^{\prime }l_{1}}\right]
^{k_{l_{1}^{\prime }}}\right\} }\Psi _{J,s_{p_{l^{\prime },0}}^{\otimes
k_{l_{i}^{\prime }}},s_{p_{l^{\prime },0}}^{\otimes k_{l_{i}^{\prime \prime
}}}}^{\otimes \sum k_{l_{i}^{\prime }}+\sum k_{l_{i}^{\prime \prime
}}}\left( \overline{\left[ p_{0},p_{l^{\prime }0}\right] ^{k_{l^{\prime }}}}%
,\left\{ \left[ 
\begin{array}{c}
\left( p_{l},p_{l,^{\prime }l},k_{l^{\prime }}\right) \\ 
\left[ \left\{ \Psi _{J}\left[ l^{\prime },0\right] \right\} ,v\right]%
\end{array}%
\right] \right\} ,v\right)  \label{SD} \\
&=&\sum_{\alpha }v_{\left\{ \left[ p_{l_{1}},p_{l_{1}^{\prime }l_{1}}\right]
^{k_{l_{1}^{\prime }}}\right\} }\Psi _{J,\alpha ,s_{p_{l^{\prime
},0}}^{\otimes k_{l_{i}^{\prime }}},s_{p_{l^{\prime },0}}^{\otimes
k_{l_{i}^{\prime \prime }}}}^{\otimes \sum k_{l_{i}^{\prime }}+\sum
k_{l_{i}^{\prime \prime }}}\left( \overline{\left[ p_{0},p_{l_{i}^{\prime }0}%
\right] ^{k_{l^{\prime }}}}\right) \Psi _{\alpha }\left( \left\{ \left[ 
\begin{array}{c}
\left( p_{l},p_{l,^{\prime }l},k_{l^{\prime }}\right) \\ 
\left[ \left\{ \Psi _{J}\left[ l^{\prime },0\right] \right\} ,v\right]%
\end{array}%
\right] \right\} ,v\right)  \notag
\end{eqnarray}

\subsubsection{Local effective field}

In appendix 7, we rewrite the generic functional (\ref{GN}) by summing over $%
\nu $ and indices $\left( m,m^{\prime }\right) _{1}$ and by introducing
series of realizations, as in part one. It yields for the generic functional:%
\begin{eqnarray*}
&&\sum_{_{\left( s,s^{\prime },\left[ p,p^{\prime },k^{\prime }\right]
\right) }}\int \sum_{\alpha }\bar{g}^{\mathcal{K}}\left( \left\{ \left[
p_{0},p_{l^{\prime }0}\right] ^{k_{l^{\prime }}}\right\} _{\substack{ %
l\leqslant s  \\ l^{\prime }\leqslant s^{\prime }}},\left\{ \left[ 
\begin{array}{c}
\left( p_{l},p_{l,^{\prime }l},k_{l^{\prime }}\right) \\ 
\left[ \left\{ \Psi _{J}\left[ l^{\prime },0\right] \right\} ,\left[
p_{0},p_{l^{\prime }0}\right] ^{k_{l^{\prime }}},v\right]%
\end{array}%
\right] \right\} \right) \\
&&\times \tprod_{\substack{ l\leqslant s  \\ l^{\prime }\leqslant s^{\prime
} }}\Psi _{J,\alpha ,s_{p_{l^{\prime },0}}^{\otimes k_{l^{\prime
}}}}^{\otimes \sum k_{l^{\prime }}}\left( \left[ p_{0},p_{l^{\prime }0}%
\right] ^{k_{l^{\prime }}}\right) \Psi _{\alpha }\left( \left\{ \left[ 
\begin{array}{c}
\left( p_{l},p_{l,^{\prime }l},k_{l^{\prime }}\right) \\ 
\left[ \left\{ \Psi _{J}\left[ l^{\prime },0\right] \right\} ,\left[
p_{0},p_{l^{\prime }0}\right] ^{k_{l^{\prime }}},v\right]%
\end{array}%
\right] \right\} ,v\right) \\
&&d\left( \left\{ \left[ p_{0},p_{l^{\prime }0}\right] ^{k_{l^{\prime
}}}\right\} _{\substack{ l\leqslant s  \\ l^{\prime }\leqslant s^{\prime }}}%
\right) d\left( \left\{ \left[ 
\begin{array}{c}
\left( p_{l},p_{l,^{\prime }l},k_{l^{\prime }}\right) \\ 
\left[ \left\{ \Psi _{J}\left[ l^{\prime },0\right] \right\} ,v\right]%
\end{array}%
\right] \right\} \right)
\end{eqnarray*}

If the set $\left\{ \left[ 
\begin{array}{c}
\left( p_{l},p_{l,^{\prime }l},k_{l^{\prime }}\right) \\ 
\left[ \left\{ \Psi _{J}\left[ l^{\prime },0\right] \right\} ,\left[
p_{0},p_{l^{\prime }0}\right] ^{k_{l^{\prime }}},v\right]%
\end{array}%
\right] \right\} $ is independent of the copies $\Psi _{J,\alpha
,s_{p_{l^{\prime },0}}^{\otimes k_{l^{\prime }}}}^{\otimes \sum k_{l^{\prime
}}}$, this becomes:%
\begin{eqnarray}
&&\sum_{_{\substack{ \left( s,s^{\prime },\left[ p,p^{\prime },k^{\prime }%
\right] \right)  \\ \left( \left[ p,p^{\prime },k^{\prime }\right] \right)
_{1}}}}\int \bar{g}^{\mathcal{K}}\left( \left\{ \left[ p_{0},p_{l^{\prime }0}%
\right] ^{k_{l^{\prime }}}\right\} _{\substack{ l\leqslant s  \\ l^{\prime
}\leqslant s^{\prime }}},\left\{ \left[ 
\begin{array}{c}
\left( p_{l},p_{l,^{\prime }l},k_{l^{\prime }}\right) \\ 
\left[ \left\{ \Psi _{J}\left[ l^{\prime },0\right] \right\} ,\left[
p_{0},p_{l^{\prime }0}\right] ^{k_{l^{\prime }}},v\right]%
\end{array}%
\right] \right\} \right)  \label{FL} \\
&&\times \Psi _{J,\tprod_{l^{\prime }}s_{p_{l^{\prime },0}}^{\otimes
k_{l^{\prime }}}}^{\otimes \sum_{l^{\prime }}k_{l^{\prime }}}\left( \left\{ 
\left[ p_{0},p_{l^{\prime }0}\right] ^{k_{l^{\prime }}}\right\} _{ 
_{\substack{ l\leqslant s  \\ l^{\prime }\leqslant s^{\prime }}}},\left\{ %
\left[ 
\begin{array}{c}
\left( p_{l},p_{l,^{\prime }l},k_{l^{\prime }}\right) \\ 
\left[ \left\{ \Psi _{J}\left[ l^{\prime },0\right] \right\} ,\left[
p_{0},p_{l^{\prime }0}\right] ^{k_{l^{\prime }}},v\right]%
\end{array}%
\right] \right\} ,v\right)  \notag \\
&&d\left( \left\{ \left[ p_{0},p_{l^{\prime }0}\right] ^{k_{l^{\prime
}}}\right\} _{\substack{ l\leqslant s  \\ l^{\prime }\leqslant s^{\prime }}}%
\right) d\left( \left\{ \left[ 
\begin{array}{c}
\left( p_{l},p_{l,^{\prime }l},k_{l^{\prime }}\right) \\ 
\left[ \left\{ \Psi _{J}\left[ l^{\prime },0\right] \right\} ,v\right]%
\end{array}%
\right] \right\} \right)  \notag \\
&\equiv &F_{f,lin}\left( \left\{ \Psi _{J,\tprod_{l^{\prime
}}s_{p_{l^{\prime },0}}^{\otimes k_{l^{\prime }}}}^{\otimes \sum_{l^{\prime
}}k_{l^{\prime }}}\left( \left\{ \left[ p_{0},p_{l^{\prime }0}\right]
^{k_{l^{\prime }}}\right\} _{_{\substack{ l\leqslant s  \\ l^{\prime
}\leqslant s^{\prime }}}},\left\{ \left[ 
\begin{array}{c}
\left( p_{l},p_{l,^{\prime }l},k_{l^{\prime }}\right) \\ 
\left[ \left\{ \Psi _{J}\left[ l^{\prime },0\right] \right\} ,\left[
p_{0},p_{l^{\prime }0}\right] ^{k_{l^{\prime }}},v\right]%
\end{array}%
\right] \right\} ,v\right) \right\} _{l^{\prime }}\right)  \notag
\end{eqnarray}%
where the local field is defined by:%
\begin{eqnarray}
&&\Psi _{J,\tprod_{l^{\prime }}s_{p_{l^{\prime },0}}^{\otimes k_{l^{\prime
}}}}^{\otimes k_{l^{\prime }}}\left( \left\{ \left[ p_{0},p_{l^{\prime }0}%
\right] ^{k_{l^{\prime }}}\right\} _{_{\substack{ l\leqslant s  \\ l^{\prime
}\leqslant s^{\prime }}}},\left\{ \left[ 
\begin{array}{c}
\left( p_{l},p_{l,^{\prime }l},k_{l^{\prime }}\right) \\ 
\left[ \left\{ \Psi _{J}\left[ l^{\prime },0\right] \right\} ,\left[
p_{0},p_{l^{\prime }0}\right] ^{k_{l^{\prime }}},v\right]%
\end{array}%
\right] \right\} ,v\right)  \label{LC} \\
&=&\sum_{\alpha }\tprod_{\substack{ l\leqslant s  \\ l^{\prime }\leqslant
s^{\prime }}}\Psi _{J,\alpha ,s_{p_{l^{\prime },0}}^{\otimes k_{l^{\prime
}}}}^{\otimes \sum k_{l^{\prime }}}\left( \left[ p_{0},p_{l^{\prime }0}%
\right] ^{k_{l^{\prime }}}\right) \Psi _{\alpha }\left( \left\{ \left[ 
\begin{array}{c}
\left( p_{l},p_{l,^{\prime }l},k_{l^{\prime }}\right) \\ 
\left[ \left\{ \Psi _{J}\left[ l^{\prime },0\right] \right\} ,\left[
p_{0},p_{l^{\prime }0}\right] ^{k_{l^{\prime }}},v\right]%
\end{array}%
\right] \right\} ,v\right)  \notag
\end{eqnarray}%
If the $\left\{ \left[ 
\begin{array}{c}
\left( p_{l},p_{l,^{\prime }l},k_{l^{\prime }}\right) \\ 
\left[ \left\{ \Psi _{J}\left[ l^{\prime },0\right] \right\} ,\left[
p_{0},p_{l^{\prime }0}\right] ^{k_{l^{\prime }}},v\right]%
\end{array}%
\right] \right\} $ are not independent of the copies $\Psi _{J,\alpha
}^{\otimes k_{p_{0}}}$, identification is local.

\subsubsection{Projected functional: general case}

When several objects $\left[ p_{l_{i,}},p_{l_{i}^{\prime }l_{i}}\right]
^{k_{l_{i}^{\prime }}}$ are involved in the functional, the generalization
of (\ref{GP}) is straightforward: 
\begin{eqnarray}
&&\sum_{_{\substack{ \left( s,s^{\prime },\left[ p,p^{\prime },k^{\prime }%
\right] \right)  \\ \left( m,m^{\prime },\left[ p,p^{\prime },k^{\prime }%
\right] \right) _{i,1}}}}\int g^{\mathcal{K}}\left( \overline{\left[
p_{0},p_{l^{\prime }0}\right] ^{k_{l^{\prime }}}},\left\{ \left[
p_{l_{i,1}},p_{l_{i,1}^{\prime }l_{i,1}}\right] ^{k_{l_{i,1}^{\prime
}}}\right\} _{_{\substack{ l_{i,1}\leqslant m_{i,1}  \\ l_{i,1}^{\prime
}\leqslant m_{i,1}^{\prime }}}},v,\left\{ \left[ 
\begin{array}{c}
\left( p_{l_{i,1}},p_{l_{i,1}^{\prime }l_{i,1}},k_{l_{i,1}^{\prime }}\right)
\\ 
\left[ \left\{ \Psi _{J}\left[ l^{\prime },0\right] \right\} ,v\right]%
\end{array}%
\right] \right\} _{_{\substack{ l_{i}\subset l_{i,1}\leqslant m_{i,1}  \\ %
l_{i}^{\prime }\subset l_{i,1}^{\prime }\leqslant m_{i,1}^{\prime }}}}\right)
\label{Gp} \\
&&\times v_{\left\{ \left[ p_{l_{i,1}},p_{l_{i,1}^{\prime }l_{i,1}}\right]
^{k_{l_{i,1}^{\prime }}}\right\} }\Psi _{J,s_{p_{l^{\prime },0}}^{\otimes
k_{l_{i}^{\prime }}},s_{p_{l^{\prime },0}}^{\otimes k_{l_{i}^{\prime \prime
}}}}^{\otimes \sum k_{l_{i}^{\prime }}+\sum k_{l_{i}^{\prime \prime
}}}\left( \overline{\left[ p_{0},p_{l^{\prime }0}\right] ^{k_{l^{\prime }}}}%
,\left\{ \left[ 
\begin{array}{c}
\left( p_{l_{i}},p_{l_{i},^{\prime }l_{i}},k_{l_{i}^{\prime }}\right) \\ 
\left[ \left\{ \Psi _{J}\left[ l^{\prime },0\right] \right\} ,v\right]%
\end{array}%
\right] \right\} _{i},v\right)  \notag \\
&&d\left( \left\{ \overline{\left[ p_{0},p_{l^{\prime }0}\right]
^{k_{l^{\prime }}}}\right\} _{\substack{ l\leqslant s  \\ l^{\prime
}\leqslant s^{\prime }}},\left\{ \left[ p_{l_{i,1}},p_{l_{i,1}^{\prime
}l_{i,1}}\right] ^{k_{l_{i,1}^{\prime }}}\right\} _{\substack{ %
l_{i,1}\leqslant m_{i,1}  \\ l_{i,1}^{\prime }\leqslant m_{i,1}^{\prime }}}%
\right) d\left( \left\{ \left[ 
\begin{array}{c}
\left( p_{l_{i}},p_{l_{i},^{\prime }l_{i}},k_{l_{i}^{\prime }}\right) \\ 
\left[ \left\{ \Psi _{J}\left[ l^{\prime },0\right] \right\} ,v\right]%
\end{array}%
\right] \right\} _{i}\right) dv  \notag
\end{eqnarray}%
with:%
\begin{eqnarray}
&&v_{\left\{ \left[ p_{l_{i,1}},p_{l_{i,1}^{\prime }l_{i,1}}\right]
^{k_{l_{i,1}^{\prime }}}\right\} }\Psi _{J,s_{p_{l^{\prime },0}}^{\otimes
k_{l_{i}^{\prime }}},s_{p_{l^{\prime },0}}^{\otimes k_{l_{i}^{\prime \prime
}}}}^{\otimes \sum k_{l_{i}^{\prime }}+\sum k_{l_{i}^{\prime \prime
}}}\left( \overline{\left[ p_{0},p_{l^{\prime }0}\right] ^{k_{l^{\prime }}}}%
,\left\{ \left[ 
\begin{array}{c}
\left( p_{l_{i}},p_{l_{i},^{\prime }l_{i}},k_{l_{i}^{\prime }}\right) \\ 
\left[ \left\{ \Psi _{J}\left[ l^{\prime },0\right] \right\} ,v\right]%
\end{array}%
\right] \right\} _{i},v\right)  \label{Sd} \\
&=&\sum_{\alpha }v_{\left\{ \left[ p_{l_{1}},p_{l_{1}^{\prime }l_{1}}\right]
^{k_{l_{1}^{\prime }}}\right\} }\Psi _{J,\alpha ,s_{p_{l^{\prime
},0}}^{\otimes k_{l_{i}^{\prime }}},s_{p_{l^{\prime },0}}^{\otimes
k_{l_{i}^{\prime \prime }}}}^{\otimes \sum k_{l_{i}^{\prime }}+\sum
k_{l_{i}^{\prime \prime }}}\left( \overline{\left[ p_{0},p_{l_{i}^{\prime }0}%
\right] ^{k_{l^{\prime }}}}\right) \Psi _{\alpha }\left( \left\{ \left[ 
\begin{array}{c}
\left( p_{l_{i}},p_{l_{i},^{\prime }l_{i}},k_{l_{i}^{\prime }}\right) \\ 
\left[ \left\{ \Psi _{J}\left[ l^{\prime },0\right] \right\} ,v\right]%
\end{array}%
\right] \right\} _{i},v\right)  \notag
\end{eqnarray}%
while (\ref{FL}) becomes:%
\begin{eqnarray}
&&\sum_{_{\left( s,s^{\prime },\left[ p,p^{\prime },k^{\prime }\right]
\right) }}\int \bar{g}^{\mathcal{K}}\left( \left\{ \left[ p_{0},p_{l^{\prime
}0}\right] ^{k_{l^{\prime }}}\right\} _{\substack{ l\leqslant s  \\ %
l^{\prime }\leqslant s^{\prime }}},\left\{ \left[ 
\begin{array}{c}
\left( p_{l_{i}},p_{l_{i},^{\prime }l_{i}},k_{l_{i}^{\prime }}\right) \\ 
\left[ \left\{ \Psi _{J}\left[ l^{\prime },0\right] \right\} ,\left[
p_{0},p_{l^{\prime }0}\right] ^{k_{l^{\prime }}},v\right]%
\end{array}%
\right] \right\} _{i}\right)  \label{Fl} \\
&&\times \Psi _{J,\tprod_{l^{\prime }}s_{p_{l^{\prime },0}}^{\otimes
k_{l^{\prime }}}}^{\otimes \sum_{l^{\prime }}k_{l^{\prime }}}\left( \left\{ 
\left[ p_{0},p_{l^{\prime }0}\right] ^{k_{l^{\prime }}}\right\} _{ 
_{\substack{ l\leqslant s  \\ l^{\prime }\leqslant s^{\prime }}}},\left\{ %
\left[ 
\begin{array}{c}
\left( p_{l_{i}},p_{l_{i},^{\prime }l_{i}},k_{l_{i}^{\prime }}\right) \\ 
\left[ \left\{ \Psi _{J}\left[ l^{\prime },0\right] \right\} ,\left[
p_{0},p_{l^{\prime }0}\right] ^{k_{l^{\prime }}},v\right]%
\end{array}%
\right] \right\} _{i},v\right)  \notag \\
&&\times d\left( \left\{ \left[ 
\begin{array}{c}
\left( p_{l_{i}},p_{l_{i},^{\prime }l_{i}},k_{l_{i}^{\prime }}\right) \\ 
\left[ \left\{ \Psi _{J}\left[ l^{\prime },0\right] \right\} ,v\right]%
\end{array}%
\right] \right\} _{i}\right) d\left( \left\{ \left[ p_{0},p_{l^{\prime }0}%
\right] ^{k_{l^{\prime }}}\right\} _{\substack{ l\leqslant s  \\ l^{\prime
}\leqslant s^{\prime }}}\right)  \notag
\end{eqnarray}

\section{Variations}

\subsection{Local invariance}

The principle is similar to part one. Starting with the effective field
after projection:%
\begin{eqnarray}
&&\Psi _{J,\tprod_{l^{\prime }}s_{p_{l^{\prime },0}}^{\otimes k_{l^{\prime
}}}}^{\otimes \sum_{l^{\prime }}k_{l^{\prime }}}\left( \left\{ \left[
p_{0},p_{l^{\prime }0}\right] ^{k_{l^{\prime }}}\right\} _{_{\substack{ %
l\leqslant s  \\ l^{\prime }\leqslant s^{\prime }}}},\left\{ \left[ 
\begin{array}{c}
\left( p_{l},p_{l,^{\prime }l},k_{l^{\prime }}\right) \\ 
\left[ \left\{ \Psi _{J}\left[ l^{\prime },0\right] \right\} ,v\right]%
\end{array}%
\right] \right\} ,v\right)  \label{L} \\
&=&\sum_{\alpha }\Psi _{J,\tprod_{l^{\prime }}s_{p_{l^{\prime },0}}^{\otimes
k_{l^{\prime }}}}^{\otimes \sum_{l^{\prime }}k_{l^{\prime }}}\left( \left\{ 
\left[ p_{0},p_{l^{\prime }0}\right] ^{k_{l^{\prime }}}\right\} _{ 
_{\substack{ l\leqslant s  \\ l^{\prime }\leqslant s^{\prime }}}},\left\{ %
\left[ 
\begin{array}{c}
\left( p_{l},p_{l,^{\prime }l},k_{l^{\prime }}\right) \\ 
\left[ \left\{ \Psi _{J}\left[ l^{\prime },0\right] \right\} ,v\right]%
\end{array}%
\right] \right\} ,v,\alpha \right)  \notag \\
&=&\sum_{\alpha }\tprod_{\substack{ l\leqslant s  \\ l^{\prime }\leqslant
s^{\prime }}}\Psi _{J,s_{p_{l^{\prime },0}}^{\otimes k_{l^{\prime
}}}}^{\otimes \sum k_{l^{\prime }}}\left( \left[ p_{0},p_{l^{\prime }0}%
\right] ^{k_{l^{\prime }}},\alpha \right) \Psi _{\alpha }\left( \left\{ %
\left[ 
\begin{array}{c}
\left( p_{l},p_{l,^{\prime }l},k_{l^{\prime }}\right) \\ 
\left[ \left\{ \Psi _{J}\left[ l^{\prime },0\right] \right\} ,v\right]%
\end{array}%
\right] \right\} ,v\right)  \notag
\end{eqnarray}%
We request that variations of fields states\ does not affect the parameter
space. More generally we will consider the state dependent field (\ref{SD}).
Using (\ref{L}), it can be rewritten more compactly as: 
\begin{equation*}
\int v\left( \left\{ \left[ p_{0},p_{l^{\prime }0}\right] ^{k_{l^{\prime
}}}\right\} _{_{\substack{ l\leqslant s  \\ l^{\prime }\leqslant s^{\prime } 
}}}\right) \Psi _{J,\tprod_{l^{\prime }}s_{p_{l^{\prime },0}}^{\otimes
k_{l^{\prime }}}}^{\otimes \sum_{l^{\prime }}k_{l^{\prime }}}\left( \left\{ %
\left[ p_{0},p_{l^{\prime }0}\right] ^{k_{l^{\prime }}}\right\} _{ 
_{\substack{ l\leqslant s  \\ l^{\prime }\leqslant s^{\prime }}}},\left\{ %
\left[ 
\begin{array}{c}
\left( p_{l},p_{l,^{\prime }l},k_{l^{\prime }}\right) \\ 
\left[ \left\{ \Psi _{J}\left[ l^{\prime },0\right] \right\} ,v\right]%
\end{array}%
\right] \right\} ,v\right) d\left\{ \left[ p_{0},p_{l^{\prime }0}\right]
^{k_{l^{\prime }}}\right\} _{_{\substack{ l\leqslant s  \\ l^{\prime
}\leqslant s^{\prime }}}}
\end{equation*}%
where:%
\begin{equation*}
v\left( \left\{ \left[ p_{0},p_{l^{\prime }0}\right] ^{k_{l^{\prime
}}}\right\} _{_{\substack{ l\leqslant s  \\ l^{\prime }\leqslant s^{\prime } 
}}}\right)
\end{equation*}%
includes some delta functions to account for the partition $\left\{ \left[
p_{l_{1}},p_{l_{1}^{\prime }l_{1}}\right] ^{k_{l_{1}^{\prime }}}\right\} $, $%
\overline{\left[ p_{0},p_{l_{i}^{\prime }0}\right] ^{k_{l^{\prime }}}}$ in (%
\ref{SD}). The integral element $d\left\{ \left[ p_{0},p_{l^{\prime }0}%
\right] ^{k_{l^{\prime }}}\right\} _{_{\substack{ l\leqslant s  \\ l^{\prime
}\leqslant s^{\prime }}}}$ will be omitted.

The variation of $\Psi _{J,s_{p_{l^{\prime },0}}^{\otimes k_{l^{\prime
}}}}^{\otimes k_{l^{\prime }}}\left( \left[ \bar{p}_{0},\bar{p}_{l^{\prime
}0}\right] ^{k_{l^{\prime }}},\beta \right) $ in the previous expression
leads to:%
\begin{eqnarray*}
&&\int v\left( \left\{ \left[ p_{0},p_{l^{\prime }0}\right] ^{k_{l^{\prime
}}}\right\} _{_{\substack{ l\leqslant s  \\ l^{\prime }\leqslant s^{\prime } 
}}}\right) \frac{\delta \Psi _{J,\tprod_{l^{\prime }}s_{p_{l^{\prime
},0}}^{\otimes k_{l^{\prime }}}}^{\otimes \sum_{l^{\prime }}k_{l^{\prime
}}}\left( \left\{ \left[ p_{0},p_{l^{\prime }0}\right] ^{k_{l^{\prime
}}}\right\} _{_{\substack{ l\leqslant s  \\ l^{\prime }\leqslant s^{\prime } 
}}},\left\{ \left[ 
\begin{array}{c}
\left( p_{l_{1}},p_{l_{1}^{\prime }l_{1}},k_{l_{1}^{\prime }}\right) \\ 
\left[ \left\{ \Psi _{J}\left[ l^{\prime },0\right] \right\} ,v\right]%
\end{array}%
\right] \right\} ,v\right) }{\delta \Psi _{J,s_{p_{l^{\prime },0}}^{\otimes
k_{l^{\prime }}}}^{\otimes k_{l^{\prime }}}\left( \left[ \bar{p}_{0},\bar{p}%
_{l^{\prime }0}\right] ^{k_{l^{\prime }}},\beta \right) } \\
&=&\int v\left( \left\{ \left[ p_{0},p_{l^{\prime }0}\right] ^{k_{l^{\prime
}}}\right\} _{_{\substack{ l\leqslant s  \\ l^{\prime }\leqslant s^{\prime } 
}}}\right) \frac{\delta ^{\prime }\Psi _{J,\tprod_{l^{\prime
}}s_{p_{l^{\prime },0}}^{\otimes k_{l^{\prime }}}}^{\otimes \sum_{l^{\prime
}}k_{l^{\prime }}}\left( \left\{ \left[ p_{0},p_{l^{\prime }0}\right]
^{k_{l^{\prime }}}\right\} _{_{\substack{ l\leqslant s  \\ l^{\prime
}\leqslant s^{\prime }}}},\left\{ \left[ 
\begin{array}{c}
\left( p_{l_{1}},p_{l_{1}^{\prime }l_{1}},k_{l_{1}^{\prime }}\right) \\ 
\left[ \left\{ \Psi _{J}\left[ l^{\prime },0\right] \right\} ,v\right]%
\end{array}%
\right] \right\} ,v,\alpha \right) }{\delta ^{\prime }\Psi
_{J,s_{p_{l^{\prime },0}}^{\otimes k_{l^{\prime }}}}^{\otimes k_{l^{\prime
}}}\left( \left[ \bar{p}_{0},\bar{p}_{l^{\prime }0}\right] ^{k_{l^{\prime
}}},\beta \right) } \\
&&+\int v\left( \left\{ \left[ p_{0},p_{l^{\prime }0}\right] ^{k_{l^{\prime
}}}\right\} _{_{\substack{ l\leqslant s  \\ l^{\prime }\leqslant s^{\prime } 
}}}\right) \frac{\delta \left\{ \left[ 
\begin{array}{c}
\left( p_{l_{1}},p_{l_{1}^{\prime }l_{1}},k_{l_{1}^{\prime }}\right) \\ 
\left[ \left\{ \Psi _{J}\left[ l^{\prime },0\right] \right\} ,v\right]%
\end{array}%
\right] \right\} }{\delta \Psi _{J,s_{p_{l^{\prime },0}}^{\otimes
k_{l^{\prime }}}}^{\otimes k_{l^{\prime }}}\left( \left[ \bar{p}_{0},\bar{p}%
_{l^{\prime }0}\right] ^{k_{l^{\prime }}},\beta \right) } \\
&&\times \nabla _{\left\{ \left[ 
\begin{array}{c}
\left( p_{l_{1}},p_{l_{1}^{\prime }l_{1}},k_{l_{1}^{\prime }}\right) \\ 
\left[ \left\{ \Psi _{J}\left[ l^{\prime },0\right] \right\} ,v\right]%
\end{array}%
\right] \right\} }\Psi _{J,\tprod_{l^{\prime }}s_{p_{l^{\prime
},0}}^{\otimes k_{l^{\prime }}}}^{\otimes \sum_{l^{\prime }}k_{l^{\prime
}}}\left( \left\{ \left[ p_{0},p_{l^{\prime }0}\right] ^{k_{l^{\prime
}}}\right\} _{_{\substack{ l\leqslant s  \\ l^{\prime }\leqslant s^{\prime } 
}}},\left\{ \left[ 
\begin{array}{c}
\left( p_{l_{1}},p_{l_{1}^{\prime }l_{1}},k_{l_{1}^{\prime }}\right) \\ 
\left[ \left\{ \Psi _{J}\left[ l^{\prime },0\right] \right\} ,v\right]%
\end{array}%
\right] \right\} ,v,\alpha \right)
\end{eqnarray*}%
where $\delta ^{\prime }$ is the variation with constant parametrs. The
first order independence of $\left[ 
\begin{array}{c}
\left( p_{l_{1}},p_{l_{1}^{\prime }l_{1}},k_{l_{1}^{\prime }}\right) \\ 
\left[ \left\{ \Psi _{J}\left[ l^{\prime },0\right] \right\} ,v\right]%
\end{array}%
\right] $ writes:

\begin{eqnarray}
0 &=&\int v\left( \left\{ \left[ p_{0},p_{l^{\prime }0}\right]
^{k_{l^{\prime }}}\right\} _{_{\substack{ l\leqslant s \\ l^{\prime
}\leqslant s^{\prime }}}}\right) \frac{\delta \left\{ \left[ 
\begin{array}{c}
\left( p_{l_{1}},p_{l_{1}^{\prime }l_{1}},k_{l_{1}^{\prime }}\right)  \\ 
\left[ \left\{ \Psi _{J}\left[ l^{\prime },0\right] \right\} ,v\right] 
\end{array}%
\right] \right\} }{\delta \Psi _{J,s_{p_{l^{\prime },0}}^{\otimes
k_{l^{\prime }}}}^{\otimes k_{l^{\prime }}}\left( \left[ \bar{p}_{0},\bar{p}%
_{l^{\prime }0}\right] ^{k_{l^{\prime }}},\beta \right) }  \label{Cdn} \\
&&\times \nabla _{\left\{ \left[ 
\begin{array}{c}
\left( p_{l_{1}},p_{l_{1}^{\prime }l_{1}},k_{l_{1}^{\prime }}\right)  \\ 
\left[ \left\{ \Psi _{J}\left[ l^{\prime },0\right] \right\} ,v\right] 
\end{array}%
\right] \right\} }\Psi _{J,\tprod_{l^{\prime }}s_{p_{l^{\prime
},0}}^{\otimes k_{l^{\prime }}}}^{\otimes \sum_{l^{\prime }}k_{l^{\prime
}}}\left( \left\{ \left[ p_{0},p_{l^{\prime }0}\right] ^{k_{l^{\prime
}}}\right\} _{_{\substack{ l\leqslant s \\ l^{\prime }\leqslant s^{\prime }}}%
},\left\{ \left[ 
\begin{array}{c}
\left( p_{l_{1}},p_{l_{1}^{\prime }l_{1}},k_{l_{1}^{\prime }}\right)  \\ 
\left[ \left\{ \Psi _{J}\left[ l^{\prime },0\right] \right\} ,v\right] 
\end{array}%
\right] \right\} ,v,\alpha \right)   \notag
\end{eqnarray}

This independence can also be written in local coordinates, as in the first
part:%
\begin{eqnarray}
&&\int \sum_{\alpha }\left[ \Xi ^{\left( k_{i}\right) }\left( \left\{
\left\{ \Psi _{J}\left[ l^{\prime },0\right] \right\} ,\mathbf{\hat{\Lambda}}%
^{\left( k_{l_{1}^{\prime }}\right) ,\alpha }\right\} \right) \right]
_{\beta ,\left[ \bar{p}_{0},\bar{p}_{l^{\prime }0}\right] ^{k_{l^{\prime
}}}}^{\alpha ,v}\nabla _{\mathbf{\hat{\Lambda}}^{\left( k_{i}\right) }}\hat{%
\Psi}_{J,\alpha }^{\otimes \sum_{i}l_{i}}\left( \left\{ \left\{ \left[
p_{0},p_{l^{\prime }0}\right] ^{k_{l^{\prime }}}\right\} _{_{\substack{ %
l\leqslant s \\ l^{\prime }\leqslant s^{\prime }}}},\mathbf{\hat{\Lambda}}%
^{\left( k_{l_{1}^{\prime }}\right) ,\alpha }\right\} _{i},v,\alpha \right) 
\notag \\
&=&\frac{\delta }{\delta \Psi _{J,\alpha ,s_{p_{l^{\prime },0}}^{\otimes
k_{l^{\prime }}}}^{\otimes k_{l^{\prime }}}\left( \left[ \bar{p}_{0},\bar{p}%
_{l^{\prime }0}\right] ^{k_{l^{\prime }}},\beta \right) }V\left( \left\{ 
\hat{\Psi}_{J,\alpha }^{\otimes \sum_{i}l_{i}}\left( \left\{ \left\{ \left[
p_{0},p_{l^{\prime }0}\right] ^{k_{l^{\prime }}}\right\} _{_{\substack{ %
l\leqslant s \\ l^{\prime }\leqslant s^{\prime }}}},\mathbf{\hat{\Lambda}}%
^{\left( k_{l_{1}^{\prime }}\right) ,\alpha }\right\} _{i},v,\alpha \right)
\right\} \right)   \label{Cdt}
\end{eqnarray}

where:%
\begin{eqnarray*}
&&\left[ \Xi ^{\left( k_{i}\right) }\left( \left\{ \left\{ \Psi _{J}\left[
l^{\prime },0\right] \right\} ,\mathbf{\hat{\Lambda}}^{\left(
k_{l_{1}^{\prime }}\right) ,\alpha }\right\} \right) \right] _{\beta ,\left[ 
\bar{p}_{0},\bar{p}_{l^{\prime }0}\right] ^{k_{l^{\prime }}}}^{\alpha ,v} \\
&=&\int v\left( \left\{ \left[ p_{0},p_{l^{\prime }0}\right] ^{k_{l^{\prime
}}}\right\} _{_{\substack{ l\leqslant s  \\ l^{\prime }\leqslant s^{\prime } 
}}}\right) \frac{\delta \left\{ \left[ 
\begin{array}{c}
\left( p_{l_{1}},p_{l_{1}^{\prime }l_{1}},k_{l_{1}^{\prime }}\right) \\ 
\left[ \left\{ \Psi _{J}\left[ l^{\prime },0\right] \right\} ,v\right]%
\end{array}%
\right] \right\} }{\delta \Psi _{J,s_{p_{l^{\prime },0}}^{\otimes
k_{l^{\prime }}}}^{\otimes k_{l^{\prime }}}\left( \left[ \bar{p}_{0},\bar{p}%
_{l^{\prime }0}\right] ^{k_{l^{\prime }}},\beta \right) }\left( \frac{\delta %
\left[ 
\begin{array}{c}
\left( p_{l_{1}},p_{l_{1}^{\prime }l_{1}},k_{l_{1}^{\prime }}\right) \\ 
\left[ \left\{ \Psi _{J}\left[ l^{\prime },0\right] \right\} ,v\right]%
\end{array}%
\right] }{\delta \mathbf{\hat{\Lambda}}^{\left( k_{l_{1}^{\prime }}\right)
,\alpha }}\right) ^{-1}
\end{eqnarray*}%
As in part 1, the derivative: 
\begin{equation*}
\frac{\delta \left[ 
\begin{array}{c}
\left( p_{l_{1}},p_{l_{1}^{\prime }l_{1}},k_{l_{1}^{\prime }}\right) \\ 
\left[ \left\{ \Psi _{J}\left[ l^{\prime },0\right] \right\} ,v\right]%
\end{array}%
\right] }{\delta \mathbf{\hat{\Lambda}}^{\left( k_{l_{1}^{\prime }}\right)
,\alpha }}
\end{equation*}%
measures the dependency of the dependent coordinates in some local reference
independent coordinates.

\subsection{Averaged field and global invariance}

As in the first part, we can rather require a global invariance for an
averaged field. This one is defined as a local field, defined in one point,
with cloud of parameters that are considered in an integrated manner.

\subsubsection{Averaged field}

To explain the forementioned claim, we assume that for a sequence of
included group of symmetry:%
\begin{equation*}
\left\{ \left[ 
\begin{array}{c}
\left( p_{l_{1}},p_{l_{1}^{\prime }l_{1}},k_{l_{1}^{\prime }}\right) \\ 
\left[ \left\{ \Psi _{J}\left[ l^{\prime },0\right] \right\} ,v\right]%
\end{array}%
\right] \right\} =\left\{ \left[ 
\begin{array}{c}
\left[ p_{l_{i}},p_{l_{i}^{\prime }l_{i}},k_{l_{i}^{\prime }}\right] \\ 
\left[ \left\{ \Psi _{J}\left[ l^{\prime },0\right] \right\} ,v\right]%
\end{array}%
\right] \right\} _{\infty }
\end{equation*}%
with:%
\begin{equation*}
\left\{ \left[ 
\begin{array}{c}
\left[ p_{l_{i}},p_{l_{i}^{\prime }l_{i}},k_{l_{i}^{\prime }}\right] \\ 
\left[ \left\{ \Psi _{J}\left[ l^{\prime },0\right] \right\} ,v\right]%
\end{array}%
\right] \right\} _{\infty }=\left( \left\{ \left[ 
\begin{array}{c}
\left[ p_{l_{i}},p_{l_{i}^{\prime }l_{i}},k_{l_{i}^{\prime }}\right]
_{i\leqslant n} \\ 
\left[ \left\{ \Psi _{J}\left[ l^{\prime },0\right] \right\} ,v\right]%
\end{array}%
\right] \right\} \right) _{n}
\end{equation*}%
and:%
\begin{equation*}
\left\{ \left[ 
\begin{array}{c}
\left[ p_{l_{i}},p_{l_{i}^{\prime }l_{i}},k_{l_{i}^{\prime }}\right]
_{i\leqslant n} \\ 
\left[ \left\{ \Psi _{J}\left[ l^{\prime },0\right] \right\} ,v\right]%
\end{array}%
\right] \right\} _{\infty }=\left( \left[ 
\begin{array}{c}
\left( p_{l_{1}},p_{l_{1}^{\prime }l_{1}},k_{l_{1}^{\prime }}\right) \\ 
\left[ \left\{ \Psi _{J}\left[ l^{\prime },0\right] \right\} ,v\right]%
\end{array}%
\right] ,\left\{ \left[ 
\begin{array}{c}
\left[ p_{l_{i}},p_{l_{i}^{\prime }l_{i}},k_{l_{i}^{\prime }}\right]
_{i\leqslant n} \\ 
\left[ \left\{ \Psi _{J}\left[ l^{\prime },0\right] \right\} ,v\right]%
\end{array}%
\right] \right\} \right) _{n}
\end{equation*}%
with $\left( p_{l_{1}},p_{l_{1}^{\prime }l_{1}},k_{l_{1}^{\prime }}\right) $
an initial st f prmtrs. That is the parameter spaces is a set of infinite
flag manifold starting with one set of parameters:%
\begin{equation*}
\left[ 
\begin{array}{c}
\left( p_{l_{1}},p_{l_{1}^{\prime }l_{1}},k_{l_{1}^{\prime }}\right) \\ 
\left[ \left\{ \Psi _{J}\left[ l^{\prime },0\right] \right\} ,v\right]%
\end{array}%
\right]
\end{equation*}%
and the field can be described by:%
\begin{equation*}
\Psi _{J,\tprod_{l^{\prime }}s_{p_{l^{\prime },0}}^{\otimes k_{l^{\prime
}}}}^{\otimes \sum_{l^{\prime }}k_{l^{\prime }}}\left( \left\{ \left[
p_{0},p_{l^{\prime }0}\right] ^{k_{l^{\prime }}}\right\} _{_{\substack{ %
l\leqslant s  \\ l^{\prime }\leqslant s^{\prime }}}},\left( \left[ 
\begin{array}{c}
\left( p_{l_{1}},p_{l_{1}^{\prime }l_{1}},k_{l_{1}^{\prime }}\right) \\ 
\left[ \left\{ \Psi _{J}\left[ l^{\prime },0\right] \right\} ,v\right]%
\end{array}%
\right] ,\left\{ \left[ 
\begin{array}{c}
\left[ p_{l_{i}},p_{l_{i}^{\prime }l_{i}},k_{l_{i}^{\prime }}\right]
_{i\leqslant n} \\ 
\left[ \left\{ \Psi _{J}\left[ l^{\prime },0\right] \right\} ,v\right]%
\end{array}%
\right] \right\} \right) _{n},v\right)
\end{equation*}%
which accounts both the integrated presence of a cloud, that is the points
of a space, with distinguished points $\left[ 
\begin{array}{c}
\left( p_{l_{1}},p_{l_{1}^{\prime }l_{1}},k_{l_{1}^{\prime }}\right) \\ 
\left[ \left\{ \Psi _{J}\left[ l^{\prime },0\right] \right\} ,v\right]%
\end{array}%
\right] $. \ These points bear the physical quantities dependency.

\subsubsection{Series expansion of the field in cloud variables}

As in part one, we can expand the field in series of the cloud variables.
Given that:%
\begin{equation*}
\left( \left\{ \left[ 
\begin{array}{c}
\left[ p_{l_{i}},p_{l_{i}^{\prime }l_{i}},k_{l_{i}^{\prime }}\right]
_{i\leqslant n} \\ 
\left[ \left\{ \Psi _{J}\left[ l^{\prime },0\right] \right\} ,v\right]%
\end{array}%
\right] \right\} \right) _{n}
\end{equation*}%
is a set of sequences:%
\begin{equation*}
\left\{ \left[ 
\begin{array}{c}
\left( p_{l_{1}},p_{l_{1}^{\prime }l_{1}},k_{l_{1}^{\prime }}\right) \\ 
\left[ \left\{ \Psi _{J}\left[ l^{\prime },0\right] \right\} ,v\right]%
\end{array}%
\right] \right\} ,\left\{ \left[ 
\begin{array}{c}
\left( p_{l_{1}},p_{l_{1}^{\prime }l_{1}},k_{l_{1}^{\prime }}\right) ,\left(
p_{l_{2}},p_{l_{2}^{\prime }l_{2}},k_{l_{2}^{\prime }}\right) \\ 
\left[ \left\{ \Psi _{J}\left[ l^{\prime },0\right] \right\} ,v\right]%
\end{array}%
\right] \right\} ....
\end{equation*}%
and given the hypothesis of group inclusions, we can decompose the
parameters depending on several indices as:%
\begin{equation*}
\left[ 
\begin{array}{c}
\left( p_{l_{1}},p_{l_{1}^{\prime }l_{1}},k_{l_{1}^{\prime }}\right) ,\left(
p_{l_{2}},p_{l_{2}^{\prime }l_{2}},k_{l_{2}^{\prime }}\right) \\ 
\left[ \left\{ \Psi _{J}\left[ l^{\prime },0\right] \right\} ,v\right]%
\end{array}%
\right] =\left( \left[ 
\begin{array}{c}
\left( p_{l_{1}},p_{l_{1}^{\prime }l_{1}},k_{l_{1}^{\prime }}\right) \\ 
\left[ \left\{ \Psi _{J}\left[ l^{\prime },0\right] \right\} ,v\right]%
\end{array}%
\right] ,\left[ 
\begin{array}{c}
\left( p_{l_{2}},p_{l_{2}^{\prime }l_{2}},k_{l_{2}^{\prime }}\right) \\ 
\left[ \left\{ \Psi _{J}\left[ l^{\prime },0\right] \right\} ,v\right]%
\end{array}%
\right] \right)
\end{equation*}%
and, as a consequenc, we can expand $\Psi _{J,\tprod_{l^{\prime
}}s_{p_{l^{\prime },0}}^{\otimes k_{l^{\prime }}}}^{\otimes \sum_{l^{\prime
}}k_{l^{\prime }}}$ as series of one point, two points..:%
\begin{eqnarray*}
&&\Psi _{J,\tprod_{l^{\prime }}s_{p_{l^{\prime },0}}^{\otimes k_{l^{\prime
}}}}^{\otimes \sum_{l^{\prime }}k_{l^{\prime }}}\left( \left\{ \left[
p_{0},p_{l^{\prime }0}\right] ^{k_{l^{\prime }}}\right\} _{_{\substack{ %
l\leqslant s  \\ l^{\prime }\leqslant s^{\prime }}}},\left( \left[ 
\begin{array}{c}
\left( p_{l_{1}},p_{l_{1}^{\prime }l_{1}},k_{l_{1}^{\prime }}\right) \\ 
\left[ \left\{ \Psi _{J}\left[ l^{\prime },0\right] \right\} ,v\right]%
\end{array}%
\right] ,\left\{ \left[ 
\begin{array}{c}
\left[ p_{l_{i}},p_{l_{i}^{\prime }l_{i}},k_{l_{i}^{\prime }}\right]
_{i\leqslant n} \\ 
\left[ \left\{ \Psi _{J}\left[ l^{\prime },0\right] \right\} ,v\right]%
\end{array}%
\right] \right\} \right) _{n},v\right) \\
&=&\sum \Psi _{J,\tprod_{l^{\prime }}s_{p_{l^{\prime },0}}^{\otimes
k_{l^{\prime }}}}^{\otimes \sum_{l^{\prime }}k_{l^{\prime }}}\left( \left\{ 
\left[ p_{0},p_{l^{\prime }0}\right] ^{k_{l^{\prime }}}\right\} _{ 
_{\substack{ l\leqslant s  \\ l^{\prime }\leqslant s^{\prime }}}},\left[ 
\begin{array}{c}
\left( p_{l_{1}},p_{l_{1}^{\prime }l_{1}},k_{l_{1}^{\prime }}\right) \\ 
\left[ \left\{ \Psi _{J}\left[ l^{\prime },0\right] \right\} ,v\right]%
\end{array}%
\right] ,\left\{ \left[ 
\begin{array}{c}
\left( p_{l_{i}},p_{l_{i}^{\prime }l_{i}},k_{l_{i}^{\prime }}\right) \\ 
\left[ \left\{ \Psi _{J}\left[ l^{\prime },0\right] \right\} ,v\right]%
\end{array}%
\right] \right\} ,v\right) \\
&&+\sum \Psi _{J,\tprod_{l^{\prime }}s_{p_{l^{\prime },0}}^{\otimes
k_{l^{\prime }}}}^{\otimes \sum_{l^{\prime }}k_{l^{\prime }}}\left( \left\{ 
\left[ p_{0},p_{l^{\prime }0}\right] ^{k_{l^{\prime }}}\right\} _{ 
_{\substack{ l\leqslant s  \\ l^{\prime }\leqslant s^{\prime }}}},\left[ 
\begin{array}{c}
\left( p_{l_{1}},p_{l_{1}^{\prime }l_{1}},k_{l_{1}^{\prime }}\right) \\ 
\left[ \left\{ \Psi _{J}\left[ l^{\prime },0\right] \right\} ,v\right]%
\end{array}%
\right] ,\left\{ \left[ 
\begin{array}{c}
\left( p_{l_{i_{1}}},p_{l_{i_{1}}^{\prime }l_{i_{1}}},k_{l_{i_{1}}^{\prime
}}\right) \\ 
\left[ \left\{ \Psi _{J}\left[ l^{\prime },0\right] \right\} ,v\right]%
\end{array}%
\right] ,\left[ 
\begin{array}{c}
\left( p_{l_{i_{2}}},p_{l_{i_{2}}^{\prime }l_{i_{2}}},k_{l_{i_{2}}^{\prime
}}\right) \\ 
\left[ \left\{ \Psi _{J}\left[ l^{\prime },0\right] \right\} ,v\right]%
\end{array}%
\right] \right\} ,v\right) \\
&&+...
\end{eqnarray*}%
In this sum, the set:%
\begin{equation*}
\left[ 
\begin{array}{c}
\left( p_{l_{i}},p_{l_{i}^{\prime }l_{i}},k_{l_{i}^{\prime }}\right) \\ 
\left[ \left\{ \Psi _{J}\left[ l^{\prime },0\right] \right\} ,v\right]%
\end{array}%
\right]
\end{equation*}%
appears through an infinite number of representants, since there are an
infinite number of flags arising in the sum. As a consequence, the field is
thus a series:%
\begin{eqnarray*}
&&\Psi _{J,\tprod_{l^{\prime }}s_{p_{l^{\prime },0}}^{\otimes k_{l^{\prime
}}}}^{\otimes \sum_{l^{\prime }}k_{l^{\prime }}}\left( \left\{ \left[
p_{0},p_{l^{\prime }0}\right] ^{k_{l^{\prime }}}\right\} _{_{\substack{ %
l\leqslant s  \\ l^{\prime }\leqslant s^{\prime }}}},\left( \left[ 
\begin{array}{c}
\left( p_{l_{1}},p_{l_{1}^{\prime }l_{1}},k_{l_{1}^{\prime }}\right) \\ 
\left[ \left\{ \Psi _{J}\left[ l^{\prime },0\right] \right\} ,v\right]%
\end{array}%
\right] ,\left\{ \left[ 
\begin{array}{c}
\left[ \left( p_{l_{i}},p_{l_{i}^{\prime }l_{i}},k_{l_{i}^{\prime }}\right) %
\right] _{i\leqslant n} \\ 
\left[ \left\{ \Psi _{J}\left[ l^{\prime },0\right] \right\} ,v\right]%
\end{array}%
\right] \right\} \right) _{n},v\right) \\
&=&\sum \Psi _{J,\tprod_{l^{\prime }}s_{p_{l^{\prime },0}}^{\otimes
k_{l^{\prime }}}}^{\otimes \sum_{l^{\prime }}k_{l^{\prime }}}\left( \left\{ 
\left[ p_{0},p_{l^{\prime }0}\right] ^{k_{l^{\prime }}}\right\} _{ 
_{\substack{ l\leqslant s  \\ l^{\prime }\leqslant s^{\prime }}}},\left[ 
\begin{array}{c}
\left( p_{l_{1}},p_{l_{1}^{\prime }l_{1}},k_{l_{1}^{\prime }}\right) \\ 
\left[ \left\{ \Psi _{J}\left[ l^{\prime },0\right] \right\} ,v\right]%
\end{array}%
\right] ,\left[ 
\begin{array}{c}
\left( p_{l_{i}},p_{l_{i}^{\prime }l_{i}},k_{l_{i}^{\prime }}\right) \\ 
\left[ \left\{ \Psi _{J}\left[ l^{\prime },0\right] \right\} ,v\right]%
\end{array}%
\right] ,v\right) \\
&&+\sum \Psi _{J,\tprod_{l^{\prime }}s_{p_{l^{\prime },0}}^{\otimes
k_{l^{\prime }}}}^{\otimes \sum_{l^{\prime }}k_{l^{\prime }}}\left( \left\{ 
\left[ p_{0},p_{l^{\prime }0}\right] ^{k_{l^{\prime }}}\right\} _{ 
_{\substack{ l\leqslant s  \\ l^{\prime }\leqslant s^{\prime }}}},\left[ 
\begin{array}{c}
\left( p_{l_{1}},p_{l_{1}^{\prime }l_{1}},k_{l_{1}^{\prime }}\right) \\ 
\left[ \left\{ \Psi _{J}\left[ l^{\prime },0\right] \right\} ,v\right]%
\end{array}%
\right] ,\left\{ \left[ 
\begin{array}{c}
\left( p_{l_{i_{1}}},p_{l_{i_{1}}^{\prime }l_{i_{1}}},k_{l_{i_{1}}^{\prime
}}\right) \\ 
\left[ \left\{ \Psi _{J}\left[ l^{\prime },0\right] \right\} ,v\right]%
\end{array}%
\right] ,\left[ 
\begin{array}{c}
\left( p_{l_{i_{2}}},p_{l_{i_{2}}^{\prime }l_{i_{2}}},k_{l_{i_{2}}^{\prime
}}\right) \\ 
\left[ \left\{ \Psi _{J}\left[ l^{\prime },0\right] \right\} ,v\right]%
\end{array}%
\right] \right\} ,v\right) \\
&&+...
\end{eqnarray*}%
and this series enables to rewrite $\Psi _{J,\tprod_{l^{\prime
}}s_{p_{l^{\prime },0}}^{\otimes k_{l^{\prime }}}}^{\otimes \sum_{l^{\prime
}}k_{l^{\prime }}}$ as the following functional:%
\begin{eqnarray}
&&\Psi _{J,\tprod_{l^{\prime }}s_{p_{l^{\prime },0}}^{\otimes k_{l^{\prime
}}}}^{\otimes \sum_{l^{\prime }}k_{l^{\prime }}}\left( \left\{ \left[
p_{0},p_{l^{\prime }0}\right] ^{k_{l^{\prime }}}\right\} _{_{\substack{ %
l\leqslant s  \\ l^{\prime }\leqslant s^{\prime }}}},\left[ 
\begin{array}{c}
\left( p_{l_{1}},p_{l_{1}^{\prime }l_{1}},k_{l_{1}^{\prime }}\right) \\ 
\left[ \left\{ \Psi _{J}\left[ l^{\prime },0\right] \right\} ,v\right]%
\end{array}%
\right] ,V_{\left( p_{l_{i}},p_{l_{i}^{\prime }l_{i}},k_{l_{i}^{\prime
}}\right) },v\right)  \label{SV} \\
&&+\Psi _{J,\tprod_{l^{\prime }}s_{p_{l^{\prime },0}}^{\otimes k_{l^{\prime
}}}}^{\otimes \sum_{l^{\prime }}k_{l^{\prime }}}\left( \left\{ \left[
p_{0},p_{l^{\prime }0}\right] ^{k_{l^{\prime }}}\right\} _{_{\substack{ %
l\leqslant s  \\ l^{\prime }\leqslant s^{\prime }}}},\left[ 
\begin{array}{c}
\left( p_{l_{1}},p_{l_{1}^{\prime }l_{1}},k_{l_{1}^{\prime }}\right) \\ 
\left[ \left\{ \Psi _{J}\left[ l^{\prime },0\right] \right\} ,v\right]%
\end{array}%
\right] ,V_{\left( p_{l_{i}},p_{l_{i}^{\prime }l_{i}},k_{l_{i}^{\prime
}}\right) }^{2},v\right) +...  \notag
\end{eqnarray}%
for the set of points:%
\begin{equation*}
V_{\left( p_{l_{i}},p_{l_{i}^{\prime }l_{i}},k_{l_{i}^{\prime }}\right)
}=\left\{ \left[ 
\begin{array}{c}
\left( p_{l_{i}},p_{l_{i}^{\prime }l_{i}},k_{l_{i}^{\prime }}\right) \\ 
\left[ \left\{ \Psi _{J}\left[ l^{\prime },0\right] \right\} ,v\right]%
\end{array}%
\right] \right\} _{\left( p_{l_{i}},p_{l_{i}^{\prime
}l_{i}},k_{l_{i}^{\prime }}\right) }
\end{equation*}

\subsubsection{Averaging over cloud variables}

As in the first part, this can be rewritten as an average over cloud
variables:%
\begin{eqnarray}
&&\Psi _{J,\tprod_{l^{\prime }}s_{p_{l^{\prime },0}}^{\otimes k_{l^{\prime
}}}}^{\otimes \sum_{l^{\prime }}k_{l^{\prime }}}\left( \left\{ \left[
p_{0},p_{l^{\prime }0}\right] ^{k_{l^{\prime }}}\right\} _{_{\substack{ %
l\leqslant s  \\ l^{\prime }\leqslant s^{\prime }}}},\left( \left[ 
\begin{array}{c}
\left( p_{l_{1}},p_{l_{1}^{\prime }l_{1}},k_{l_{1}^{\prime }}\right) \\ 
\left[ \left\{ \Psi _{J}\left[ l^{\prime },0\right] \right\} ,v\right]%
\end{array}%
\right] ,\left\{ \left[ 
\begin{array}{c}
\left[ \left( p_{l_{i}},p_{l_{i}^{\prime }l_{i}},k_{l_{i}^{\prime }}\right) %
\right] _{i\leqslant n} \\ 
\left[ \left\{ \Psi _{J}\left[ l^{\prime },0\right] \right\} ,v\right]%
\end{array}%
\right] \right\} \right) _{n},v\right)  \label{FGN} \\
&=&\sum_{r}\int_{V_{\left( p_{l_{i}},p_{l_{i}^{\prime
}l_{i}},k_{l_{i}^{\prime }}\right) }^{r}}\Psi _{J,\tprod_{l^{\prime
}}s_{p_{l^{\prime },0}}^{\otimes k_{l^{\prime }}}}^{\otimes \sum_{l^{\prime
}}k_{l^{\prime }}}\left( \left\{ \left[ p_{0},p_{l^{\prime }0}\right]
^{k_{l^{\prime }}}\right\} _{_{\substack{ l\leqslant s  \\ l^{\prime
}\leqslant s^{\prime }}}},\left[ 
\begin{array}{c}
\left( p_{l_{1}},p_{l_{1}^{\prime }l_{1}},k_{l_{1}^{\prime }}\right) \\ 
\left[ \left\{ \Psi _{J}\left[ l^{\prime },0\right] \right\} ,v\right]%
\end{array}%
\right] ,\left( \Lambda _{\left( p_{l_{i}},p_{l_{i}^{\prime
}l_{i}},k_{l_{i}^{\prime }}\right) }\right) ^{r},v\right) d\left( \Lambda
_{\left( p_{l_{i}},p_{l_{i}^{\prime }l_{i}},k_{l_{i}^{\prime }}\right)
}\right) ^{r}  \notag
\end{eqnarray}%
where the $\Lambda _{\left( p_{l_{i}},p_{l_{i}^{\prime
}l_{i}},k_{l_{i}^{\prime }}\right) }$ are local coordinates for $V_{\left(
p_{l_{i}},p_{l_{i}^{\prime }l_{i}},k_{l_{i}^{\prime }}\right) }$.

\subsubsection{Invariance for averaged field}

For the averaged field, the local invariance of the parameter space with
respect to the fluctuations of $\Psi _{J,\tprod_{l^{\prime }}s_{p_{l^{\prime
},0}}^{\otimes k_{l^{\prime }}}}^{\otimes \sum_{l^{\prime }}k_{l^{\prime }}}$
may be diregarded if we consider that only the entire cloud should not be
affected. This corresponds to impose that the variations at the border are
cancelled. Starting with the equation (\ref{SV}) and imposing the invariance
(we only write the first term of (\ref{FGN}), the computation is similar for
the higher order terms):%
\begin{eqnarray}
0 &=&\int v\left( \left\{ \left[ p_{0},p_{l^{\prime }0}\right]
^{k_{l^{\prime }}}\right\} _{_{\substack{ l\leqslant s  \\ l^{\prime
}\leqslant s^{\prime }}}}\right) \frac{\delta \left\{ \left[ 
\begin{array}{c}
\left( p_{l_{1}},p_{l_{1}^{\prime }l_{1}},k_{l_{1}^{\prime }}\right) \\ 
\left[ \left\{ \Psi _{J}\left[ l^{\prime },0\right] \right\} ,v\right]%
\end{array}%
\right] \right\} }{\delta \Psi _{J,s_{p_{l^{\prime },0}}^{\otimes
k_{l^{\prime }}}}^{\otimes k_{l^{\prime }}}\left( \left[ \bar{p}_{0},\bar{p}%
_{l^{\prime }0}\right] ^{k_{l^{\prime }}},\beta \right) }  \label{Cdv} \\
&&\times \nabla _{\left\{ \left[ 
\begin{array}{c}
\left( p_{l_{1}},p_{l_{1}^{\prime }l_{1}},k_{l_{1}^{\prime }}\right) \\ 
\left[ \left\{ \Psi _{J}\left[ l^{\prime },0\right] \right\} ,v\right]%
\end{array}%
\right] \right\} }\Psi _{J,\tprod_{l^{\prime }}s_{p_{l^{\prime
},0}}^{\otimes k_{l^{\prime }}}}^{\otimes \sum_{l^{\prime }}k_{l^{\prime
}}}\left( \left\{ \left[ p_{0},p_{l^{\prime }0}\right] ^{k_{l^{\prime
}}}\right\} _{_{\substack{ l\leqslant s  \\ l^{\prime }\leqslant s^{\prime } 
}}},\left[ 
\begin{array}{c}
\left( p_{l_{1}},p_{l_{1}^{\prime }l_{1}},k_{l_{1}^{\prime }}\right) \\ 
\left[ \left\{ \Psi _{J}\left[ l^{\prime },0\right] \right\} ,v\right]%
\end{array}%
\right] ,V_{\left( p_{l_{i}},p_{l_{i}^{\prime }l_{i}},k_{l_{i}^{\prime
}}\right) },v\right)  \notag \\
&&+\int v\left( \left\{ \left[ p_{0},p_{l^{\prime }0}\right] ^{k_{l^{\prime
}}}\right\} _{_{\substack{ l\leqslant s  \\ l^{\prime }\leqslant s^{\prime } 
}}}\right) \frac{\delta V_{\left( p_{l_{i}},p_{l_{i}^{\prime
}l_{i}},k_{l_{i}^{\prime }}\right) }}{\delta \Psi _{J,s_{p_{l^{\prime
},0}}^{\otimes k_{l^{\prime }}}}^{\otimes k_{l^{\prime }}}\left( \left[ \bar{%
p}_{0},\bar{p}_{l^{\prime }0}\right] ^{k_{l^{\prime }}},\beta \right) } 
\notag \\
&&\times \nabla _{V_{\left( p_{l_{i}},p_{l_{i}^{\prime
}l_{i}},k_{l_{i}^{\prime }}\right) }}\Psi _{J,\tprod_{l^{\prime
}}s_{p_{l^{\prime },0}}^{\otimes k_{l^{\prime }}}}^{\otimes \sum_{l^{\prime
}}k_{l^{\prime }}}\left( \left\{ \left[ p_{0},p_{l^{\prime }0}\right]
^{k_{l^{\prime }}}\right\} _{_{\substack{ l\leqslant s  \\ l^{\prime
}\leqslant s^{\prime }}}},\left[ 
\begin{array}{c}
\left( p_{l_{1}},p_{l_{1}^{\prime }l_{1}},k_{l_{1}^{\prime }}\right) \\ 
\left[ \left\{ \Psi _{J}\left[ l^{\prime },0\right] \right\} ,v\right]%
\end{array}%
\right] ,V_{\left( p_{l_{i}},p_{l_{i}^{\prime }l_{i}},k_{l_{i}^{\prime
}}\right) },v\right)  \notag
\end{eqnarray}%
\bigskip A derivation similar to (\ref{BNF}) leads to the following relation:%
\begin{eqnarray*}
0 &=&\int v\left( \left\{ \left[ p_{0},p_{l^{\prime }0}\right]
^{k_{l^{\prime }}}\right\} _{_{\substack{ l\leqslant s  \\ l^{\prime
}\leqslant s^{\prime }}}}\right) \frac{\delta \left\{ \left[ 
\begin{array}{c}
\left( p_{l_{1}},p_{l_{1}^{\prime }l_{1}},k_{l_{1}^{\prime }}\right) \\ 
\left[ \left\{ \Psi _{J}\left[ l^{\prime },0\right] \right\} ,v\right]%
\end{array}%
\right] \right\} }{\delta \Psi _{J,s_{p_{l^{\prime },0}}^{\otimes
k_{l^{\prime }}}}^{\otimes k_{l^{\prime }}}\left( \left[ \bar{p}_{0},\bar{p}%
_{l^{\prime }0}\right] ^{k_{l^{\prime }}},\beta \right) } \\
&&\times \nabla _{\left\{ \left[ 
\begin{array}{c}
\left( p_{l_{1}},p_{l_{1}^{\prime }l_{1}},k_{l_{1}^{\prime }}\right) \\ 
\left[ \left\{ \Psi _{J}\left[ l^{\prime },0\right] \right\} ,v\right]%
\end{array}%
\right] \right\} }\Psi _{J,\tprod_{l^{\prime }}s_{p_{l^{\prime
},0}}^{\otimes k_{l^{\prime }}}}^{\otimes \sum_{l^{\prime }}k_{l^{\prime
}}}\left( \left\{ \left[ p_{0},p_{l^{\prime }0}\right] ^{k_{l^{\prime
}}}\right\} _{_{\substack{ l\leqslant s  \\ l^{\prime }\leqslant s^{\prime } 
}}},\left[ 
\begin{array}{c}
\left( p_{l_{1}},p_{l_{1}^{\prime }l_{1}},k_{l_{1}^{\prime }}\right) \\ 
\left[ \left\{ \Psi _{J}\left[ l^{\prime },0\right] \right\} ,v\right]%
\end{array}%
\right] ,V_{\left( p_{l_{i}},p_{l_{i}^{\prime }l_{i}},k_{l_{i}^{\prime
}}\right) },v\right) \\
&&+\int v\left( \left\{ \left[ p_{0},p_{l^{\prime }0}\right] ^{k_{l^{\prime
}}}\right\} _{_{\substack{ l\leqslant s  \\ l^{\prime }\leqslant s^{\prime } 
}}}\right) \left( \epsilon \frac{\delta V_{\left( p_{l_{i}},p_{l_{i}^{\prime
}l_{i}},k_{l_{i}^{\prime }}\right) }}{\delta \Psi _{J,s_{p_{l^{\prime
},0}}^{\otimes k_{l^{\prime }}}}^{\otimes k_{l^{\prime }}}\left( \left[ \bar{%
p}_{0},\bar{p}_{l^{\prime }0}\right] ^{k_{l^{\prime }}},\beta \right) }%
-h\left( V_{\left( p_{l_{i}},p_{l_{i}^{\prime }l_{i}},k_{l_{i}^{\prime
}}\right) },\Psi _{J,\tprod_{l^{\prime }}s_{p_{l^{\prime },0}}^{\otimes
k_{l^{\prime }}}}^{\otimes \sum_{l^{\prime }}k_{l^{\prime }}}\left( \left\{ %
\left[ p_{0},p_{l^{\prime }0}\right] ^{k_{l^{\prime }}}\right\} _{ 
_{\substack{ l\leqslant s  \\ l^{\prime }\leqslant s^{\prime }}}}\right)
\right) \right) \\
&&\times \Psi _{J,\tprod_{l^{\prime }}s_{p_{l^{\prime },0}}^{\otimes
k_{l^{\prime }}}}^{\otimes \sum_{l^{\prime }}k_{l^{\prime }}}\left( \left\{ 
\left[ p_{0},p_{l^{\prime }0}\right] ^{k_{l^{\prime }}}\right\} _{ 
_{\substack{ l\leqslant s  \\ l^{\prime }\leqslant s^{\prime }}}},\left[ 
\underset{\left[ \left\{ \Psi _{J}\left[ l^{\prime },0\right] \right\} ,v%
\right] }{\overset{\left( p_{l_{1}},p_{l_{1}^{\prime
}l_{1}},k_{l_{1}^{\prime }}\right) }{\mathbf{\hat{\Lambda}}}}\right]
,V_{\left( p_{l_{i}},p_{l_{i}^{\prime }l_{i}},k_{l_{i}^{\prime }}\right)
},v\right)
\end{eqnarray*}%
where:%
\begin{equation*}
h\left( V_{\left( p_{l_{i}},p_{l_{i}^{\prime }l_{i}},k_{l_{i}^{\prime
}}\right) },\Psi _{J,\tprod_{l^{\prime }}s_{p_{l^{\prime },0}}^{\otimes
k_{l^{\prime }}}}^{\otimes \sum_{l^{\prime }}k_{l^{\prime }}}\left( \left\{ 
\left[ p_{0},p_{l^{\prime }0}\right] ^{k_{l^{\prime }}}\right\} _{ 
_{\substack{ l\leqslant s  \\ l^{\prime }\leqslant s^{\prime }}}}\right)
\right)
\end{equation*}%
is the average of:%
\begin{equation*}
h\left( \left( \Lambda _{\left( p_{l_{i}},p_{l_{i}^{\prime
}l_{i}},k_{l_{i}^{\prime }}\right) }\right) ^{r},\Psi _{J,\tprod_{l^{\prime
}}s_{p_{l^{\prime },0}}^{\otimes k_{l^{\prime }}}}^{\otimes \sum_{l^{\prime
}}k_{l^{\prime }}}\left( \left\{ \left[ p_{0},p_{l^{\prime }0}\right]
^{k_{l^{\prime }}}\right\} _{_{\substack{ l\leqslant s  \\ l^{\prime
}\leqslant s^{\prime }}}}\right) \right) =\frac{\delta \left( \Lambda
_{\left( p_{l_{i}},p_{l_{i}^{\prime }l_{i}},k_{l_{i}^{\prime }}\right)
}\right) ^{r}}{\delta \Psi _{J,\tprod_{l^{\prime }}s_{p_{l^{\prime
},0}}^{\otimes k_{l^{\prime }}}}^{\otimes \sum_{l^{\prime }}k_{l^{\prime
}}}\left( \left\{ \left[ p_{0},p_{l^{\prime }0}\right] ^{k_{l^{\prime
}}}\right\} _{_{\substack{ l\leqslant s  \\ l^{\prime }\leqslant s^{\prime } 
}}}\right) }
\end{equation*}%
over $V_{\left( p_{l_{i}},p_{l_{i}^{\prime }l_{i}},k_{l_{i}^{\prime
}}\right) }$.

\section{Constraints}

\subsection{General dependency}

So far we left aside the form of the constraints. The treatment is very
similar to the particular case of the first part, so that we only generalize
the results. The constraint describes relations on the smtr operators for $%
n\in 
\mathbb{N}
$:%
\begin{equation}
h_{k_{n}}\left( \left\{ \mathbf{L}_{\left\{ \alpha _{i}\right\} }\left( \Psi
_{J}^{\otimes l}\right) \right\} _{i\leqslant n},\left\{ U_{i}^{k}\right\}
_{i},h_{p}\left( \left( \Psi _{J}\right) ,U_{j}^{l},\nu \right) \right) =0
\label{GRC}
\end{equation}%
where the $\left\{ U_{i}^{k}\right\} _{i}$ are the initial parameters for
the projected fields, the $\mathbf{L}_{\left\{ \alpha _{i}\right\} }\left(
\Psi _{J}^{\otimes l}\right) $ act on the set of projected fields and where
the functionals $h_{l}\left( \left( \Psi _{J}\right) ,U_{j}^{l},\nu \right) $
have the form: 
\begin{equation*}
h_{p}\left( \left( \Psi _{J}\right) ,\left\{ U_{j}^{l_{i}}\right\} ,\nu
\right) =\left[ \int h_{p}\left( \left( U_{j}^{l_{i}^{\prime }}\right) ,\nu
\right) \left( \tprod\limits_{i}\Psi _{J}^{\otimes l_{i}^{\prime }}\left(
U_{j}^{l_{i}^{\prime }}\right) \right) d\left( U_{j}^{l_{i}^{\prime
}}\right) \right] \tprod\limits_{i}\Psi _{J}^{\otimes l_{i}}\left(
U_{j}^{l_{i}}\right)
\end{equation*}

Evaluated on the projectd states, and assuming the independence of
parameters from the realizations, this becomes a relation:%
\begin{eqnarray}
0 &=&\int h_{k_{n}}\left( \left\{ \left[ 
\begin{array}{c}
\left( p_{l_{i}},p_{l_{i}^{\prime }l_{i}},k_{l_{i}^{\prime }}\right) \\ 
\left[ \left\{ \Psi _{J}\left[ l^{\prime },0\right] \right\} ,v\right]%
\end{array}%
\right] \right\} ,\left\{ U_{i}^{k}\right\} _{i},h_{p}\left( \left( \Psi
_{J}\right) ,U_{j}^{l},\nu \right) \right) \\
&&\tprod\limits_{k_{l^{\prime }}}\left\vert \Psi _{J,0,s_{p_{l^{\prime
},l}}^{\otimes k_{l^{\prime }}}}^{\otimes k_{l^{\prime }}}\left( \left[
p_{l},p_{l^{\prime }l}\right] ^{k_{l^{\prime }}},\left\{ \left[ 
\begin{array}{c}
\left( p_{l_{i}},p_{l_{i}^{\prime }l_{i}},k_{l_{i}^{\prime }}\right) \\ 
\left[ \left\{ \Psi _{J}\left[ l^{\prime },0\right] \right\} ,v\right]%
\end{array}%
\right] \right\} ,v\right) \right\vert ^{2}d\left\{ U_{i}^{k}\right\} _{i} 
\notag
\end{eqnarray}%
where $\Psi _{J,0,s_{p_{l^{\prime },l}}^{\otimes k_{l^{\prime }}}}^{\otimes
k_{l^{\prime }}}$ has been defined in (\ref{SDLPT}). A series expansion \ of
the constrnt in the $\left\{ U_{i}^{k}\right\} _{i}$ shows that the
constraint become functional relations between the parameters, and:%
\begin{equation*}
\left\{ \Psi \left[ l^{\prime },0\right] \right\} ,\nu ,\left\{ \Psi \left[
l^{\prime },0\right] \right\} ,\bar{H}_{m}\left( \Psi _{J,0,s_{p_{l^{\prime
},l}}^{\otimes k_{l^{\prime }}}}^{\otimes k_{l^{\prime }}}\right)
\end{equation*}%
where the fld $\Psi \left[ l^{\prime },0\right] $ defined in (\ref{CP}) and:%
\begin{equation*}
\bar{H}_{m}=\int \bar{h}_{m}\left( \left\{ U_{i}^{k}\right\} _{i}\right)
\tprod\limits_{k_{l^{\prime }}}\left\vert \Psi _{J,0,s_{p_{l^{\prime
},l}}^{\otimes k_{l^{\prime }}}}^{\otimes k_{l^{\prime }}}\left( \left[
p_{l},p_{l^{\prime }l}\right] ^{k_{l^{\prime }}},\left\{ \left[ 
\begin{array}{c}
\left( p_{l_{i}},p_{l_{i}^{\prime }l_{i}},k_{l_{i}^{\prime }}\right) \\ 
\left[ \left\{ \Psi _{J}\left[ l^{\prime },0\right] \right\} ,v\right]%
\end{array}%
\right] \right\} ,v\right) \right\vert ^{2}d\left\{ U_{i}^{k}\right\} _{i}
\end{equation*}%
for some functions $\bar{h}_{m}\left( \left\{ U_{i}^{k}\right\} _{i}\right) $%
.

Alternatively, the constraint depends on the kernel defind in (\ref{SDLPT}):%
\begin{equation*}
\left\vert \mathcal{K}_{0}\left( \left\{ \overline{\left[ p_{0},p_{l^{\prime
}0}\right] ^{k_{l^{\prime }}}}\right\} _{_{\substack{ l\leqslant s  \\ %
l^{\prime }\leqslant s^{\prime }}}},\left\{ \left[ p_{l_{1}},p_{l_{1}^{%
\prime }l_{1}}\right] ^{k_{l_{1}^{\prime }}}\right\} _{_{\substack{ %
l_{1}\leqslant m_{1}  \\ l_{1}^{\prime }\leqslant m_{1}^{\prime }}}},\left\{ %
\left[ 
\begin{array}{c}
\left( p_{l_{i}},p_{l_{i}^{\prime }l_{i}},k_{l_{i}^{\prime }}\right) \\ 
\left[ \left\{ \Psi _{J}\left[ l^{\prime },0\right] \right\} ,v\right]%
\end{array}%
\right] \right\} _{_{\substack{ l\subset l_{1}\leqslant m_{1}  \\ l^{\prime
}\subset l_{1}^{\prime }\leqslant m_{1}^{\prime }}}}\right) \right\vert ^{2}
\end{equation*}%
that keeps track of the projected states in the definition of the constraint.

\subsection{Lowest order expansion and Metrics on projected states}

A second order expansion of the previous constraint enables to define a
metric on the parameter space. It depends both on the \ field $\left\{ \Psi %
\left[ l^{\prime },0\right] \right\} ,$ and the backgrnd $\Psi
_{J,0,s_{p_{l^{\prime },l}}^{\otimes k_{l^{\prime }}}}^{\otimes k_{l^{\prime
}}}$ through $\bar{H}_{m}\left( \Psi _{J,0,s_{p_{l^{\prime },l}}^{\otimes
k_{l^{\prime }}}}^{\otimes k_{l^{\prime }}}\right) $.

For a local functional at the lowest order we find:%
\begin{eqnarray}
0 &=&\gamma \left( \left\{ \Psi \left[ l^{\prime },0\right] \right\} ,\nu
\right) +\left\{ \left[ 
\begin{array}{c}
\left[ p_{l_{i}},p_{l_{i}^{\prime }l_{i}},k_{l_{i}^{\prime }}\right] \\ 
\left[ \left\{ \Psi _{J}\left[ l^{\prime },0\right] \right\} ,v\right]%
\end{array}%
\right] _{i\leqslant n_{1}}\right\}  \label{FM} \\
&&\times M\left( \left\{ \left[ p_{l_{i}},p_{l_{i}^{\prime
}l_{i}},k_{l_{i}^{\prime }}\right] ,\left[ p_{l_{j}},p_{l_{j}^{\prime
}l_{j}},k_{l_{j}^{\prime }}\right] \right\} ,\nu ,\left\{ \left\{ \Psi \left[
l^{\prime },0\right] \right\} ,\bar{H}_{m}\right\} \right) \left\{ \left[ 
\begin{array}{c}
\left[ p_{l_{j}},p_{l_{j}^{\prime }l_{j}},k_{l_{j}^{\prime }}\right] \\ 
\left[ \left\{ \Psi _{J}\left[ l^{\prime },0\right] \right\} ,v\right]%
\end{array}%
\right] _{j\leqslant n_{2}}\right\}  \notag
\end{eqnarray}

The scalar functional is a series expansion of integrals for the fields $%
\left\{ \Psi \left[ l^{\prime },0\right] \right\} $.

Formula (\ref{FM}) implies that in the projected space, the metric is
dynamicaly an object defined by the projected field:%
\begin{equation*}
\Psi _{J,0,s_{p_{l^{\prime },l}}^{\otimes k_{l^{\prime }}}}^{\otimes
k_{l^{\prime }}}\left( \left[ p_{l},p_{l^{\prime }l}\right] ^{k_{l^{\prime
}}},\left\{ \left[ 
\begin{array}{c}
\left[ p_{l_{i}},p_{l_{i}^{\prime }l_{i}},k_{l_{i}^{\prime }}\right] \\ 
\left[ \left\{ \Psi _{J}\left[ l^{\prime },0\right] \right\} ,v\right]%
\end{array}%
\right] \right\} ,v\right)
\end{equation*}%
and that the metric tensor for one point is a functional of:%
\begin{equation*}
g\left( \left\{ \left( p_{l_{i}},p_{l_{i}^{\prime }l_{i}},k_{l_{i}^{\prime
}}\right) ,\left( p_{l_{j}},p_{l_{j}^{\prime }l_{j}},k_{l_{j}^{\prime
}}\right) \right\} ,\nu ,\left\{ \Psi \left[ l^{\prime },0\right] \mathcal{K}%
_{0}\right\} \right)
\end{equation*}%
Mr prcsly, cnstrn (\ref{FM}) cn b dignlzd s:%
\begin{eqnarray}
0 &=&\gamma \left( \left\{ \Psi \left[ l^{\prime },0\right] \right\} ,\nu
\right) +\left( \left\{ \left[ 
\begin{array}{c}
\left[ p_{l_{i}},p_{l_{i}^{\prime }l_{i}},k_{l_{i}^{\prime }}\right] \\ 
\left[ \left\{ \Psi _{J}\left[ l^{\prime },0\right] \right\} ,v\right]%
\end{array}%
\right] _{i\leqslant n_{1}}\right\} ^{\prime }\right) ^{t}\left\{ \left[ 
\begin{array}{c}
\left[ p_{l_{j}},p_{l_{j}^{\prime }l_{j}},k_{l_{j}^{\prime }}\right] \\ 
\left[ \left\{ \Psi _{J}\left[ l^{\prime },0\right] \right\} ,v\right]%
\end{array}%
\right] _{j\leqslant n_{2}}\right\} ^{\prime } \\
&&\times M\left( \left\{ \left[ p_{l_{i}},p_{l_{i}^{\prime
}l_{i}},k_{l_{i}^{\prime }}\right] ,\left[ p_{l_{j}},p_{l_{j}^{\prime
}l_{j}},k_{l_{j}^{\prime }}\right] \right\} ,\nu ,\left\{ \left\{ \Psi \left[
l^{\prime },0\right] \right\} ,\bar{H}_{m}\right\} \right)  \notag
\end{eqnarray}%
\begin{equation*}
\left\{ \left[ 
\begin{array}{c}
\left[ p_{l_{j}},p_{l_{j}^{\prime }l_{j}},k_{l_{j}^{\prime }}\right] \\ 
\left[ \left\{ \Psi _{J}\left[ l^{\prime },0\right] \right\} ,v\right]%
\end{array}%
\right] _{j\leqslant n_{2}}\right\} ^{\prime }=\sqrt{D}P^{-1}
\end{equation*}%
whr:%
\begin{equation*}
M=PDP^{-1}
\end{equation*}%
nd th mtrc s btnd b ntrnsc crdnts $\widetilde{\left[ 
\begin{array}{c}
\left[ p_{l_{j}},p_{l_{j}^{\prime }l_{j}},k_{l_{j}^{\prime }}\right] \\ 
\left[ \left\{ \Psi _{J}\left[ l^{\prime },0\right] \right\} ,v\right]%
\end{array}%
\right] }_{j\leqslant n_{2}}$:%
\begin{equation*}
\frac{\partial \left\{ \left[ 
\begin{array}{c}
\left[ p_{l_{j}},p_{l_{j}^{\prime }l_{j}},k_{l_{j}^{\prime }}\right] \\ 
\left[ \left\{ \Psi _{J}\left[ l^{\prime },0\right] \right\} ,v\right]%
\end{array}%
\right] _{j\leqslant n_{2}}\right\} ^{\prime }}{\partial \widetilde{\left[ 
\begin{array}{c}
\left[ p_{l_{j}},p_{l_{j}^{\prime }l_{j}},k_{l_{j}^{\prime }}\right] \\ 
\left[ \left\{ \Psi _{J}\left[ l^{\prime },0\right] \right\} ,v\right]%
\end{array}%
\right] }_{j\leqslant n_{2}}}.\frac{\partial \left\{ \left[ 
\begin{array}{c}
\left[ p_{l_{j}},p_{l_{j}^{\prime }l_{j}},k_{l_{j}^{\prime }}\right] \\ 
\left[ \left\{ \Psi _{J}\left[ l^{\prime },0\right] \right\} ,v\right]%
\end{array}%
\right] _{j\leqslant n_{2}}\right\} ^{\prime }}{\partial \widetilde{\left[ 
\begin{array}{c}
\left[ p_{l_{j}},p_{l_{j}^{\prime }l_{j}},k_{l_{j}^{\prime }}\right] \\ 
\left[ \left\{ \Psi _{J}\left[ l^{\prime },0\right] \right\} ,v\right]%
\end{array}%
\right] }_{j\leqslant n_{2}}}
\end{equation*}%
whr $\mathbf{u}$ s dfnd b $\left\vert \gamma \left( \left\{ \Psi \left[
l^{\prime },0\right] \right\} ,\nu \right) \right\vert $ mltpld b sqnc f $%
\cos $ $\cosh $.... dpndng n th sgn f gnvls f $M$.

See dicussion in part one.

\subsection{General constraint and intrinsc coordnts}

Th cn b sn s mbdd cnstrnt fr th:%
\begin{equation*}
\bar{H}_{m}=\int \bar{h}_{m}\left( \left\{ U_{i}^{k}\right\} _{i}\right)
\tprod\limits_{k_{l^{\prime }}}\left\vert \Psi _{J,0,s_{p_{l^{\prime
},l}}^{\otimes k_{l^{\prime }}}}^{\otimes k_{l^{\prime }}}\left( \left[
p_{l},p_{l^{\prime }l}\right] ^{k_{l^{\prime }}},\left\{ \left[ 
\begin{array}{c}
\left( p_{l_{i}},p_{l_{i}^{\prime }l_{i}},k_{l_{i}^{\prime }}\right) \\ 
\left[ \left\{ \Psi _{J}\left[ l^{\prime },0\right] \right\} ,v\right]%
\end{array}%
\right] \right\} ,v\right) \right\vert ^{2}d\left\{ U_{i}^{k}\right\} _{i}
\end{equation*}

r, s ntrnsc crdnts $\widetilde{\left[ 
\begin{array}{c}
\left( p_{l_{i}},p_{l_{i}^{\prime }l_{i}},k_{l_{i}^{\prime }}\right) \\ 
\left[ \left\{ \Psi _{J}\left[ l^{\prime },0\right] \right\} ,v\right]%
\end{array}%
\right] }$ lng wth sm mtrcs n grp f pnts. s n prt n, snc th cld f pnt gnrt
sm flg mnfld, w cn cnsdr tht th mbdd crdnts fllw th sm knd f nclsn rltns nd
th mtrc cn b wrtt s srs:%
\begin{equation*}
\sum_{n}d\widetilde{\left[ 
\begin{array}{c}
\left( p_{l_{i}},p_{l_{i}^{\prime }l_{i}},k_{l_{i}^{\prime }}\right) \\ 
\left[ \left\{ \Psi _{J}\left[ l^{\prime },0\right] \right\} ,v\right]%
\end{array}%
\right] }_{n}G_{n}\left( \widetilde{\left[ 
\begin{array}{c}
\left( p_{l_{i}},p_{l_{i}^{\prime }l_{i}},k_{l_{i}^{\prime }}\right) \\ 
\left[ \left\{ \Psi _{J}\left[ l^{\prime },0\right] \right\} ,v\right]%
\end{array}%
\right] }\right) d\widetilde{\left[ 
\begin{array}{c}
\left( p_{l_{i}},p_{l_{i}^{\prime }l_{i}},k_{l_{i}^{\prime }}\right) \\ 
\left[ \left\{ \Psi _{J}\left[ l^{\prime },0\right] \right\} ,v\right]%
\end{array}%
\right] }_{n}
\end{equation*}%
whr th ndcs rfr t th strts f th flg.

\section{Constraints, and reparametrization}

\subsection{Constraints defining \textbf{the parameters}}

Assuming that the dependency in $\Psi _{J}^{\otimes k_{p}}$ can be described
by the constraints: 
\begin{equation}
H\left( \left\{ \left[ 
\begin{array}{c}
\left( p_{l_{i}},p_{l_{i}^{\prime }l_{i}},k_{l_{i}^{\prime }}\right) \\ 
\left[ \left\{ \Psi _{J}\left[ l^{\prime }\eta ,\eta \right] \right\} ,\left[
v\right] \right]%
\end{array}%
\right] \right\} ,\left\{ \Psi _{J}^{\otimes k_{p}},\left[
p_{0},p_{l^{\prime }0}\right] \right\} \right) =0  \label{Ctn}
\end{equation}%
whr $H$ hs cmpnn $\bar{H}_{m}$.

We implement this condition in functionals by including $\delta \left(
H\left( \left\{ \left[ 
\begin{array}{c}
\left( p_{l_{i}},p_{l_{i}^{\prime }l_{i}},k_{l_{i}^{\prime }}\right) \\ 
\left[ \left\{ \Psi _{J}\left[ l^{\prime }\eta ,\eta \right] \right\} ,\left[
v\right] \right]%
\end{array}%
\right] \right\} ,\Psi _{J}^{\otimes k_{p}}\right) \right) $:%
\begin{eqnarray*}
&&\int \bar{g}\left( \left\{ \left[ p_{0},p_{l^{\prime }0}\right] \right\}
_{l^{\prime }},\left\{ \left\{ \left[ \left( p_{l_{i}},p_{l_{i}^{\prime
}l_{i}},k_{l_{i}^{\prime }}\right) \right] \right\} \right\} \right) \\
&&\Psi _{J,\underset{i}{\tprod }s_{p_{l^{\prime },0}}^{\otimes k_{l^{\prime
}}}}^{\otimes \sum_{i}k_{l^{\prime }}}\left( \left\{ \left[
p_{0},p_{l^{\prime }0}\right] ^{k_{l^{\prime },i}},\left[ \left(
p_{l_{i}},p_{l_{i}^{\prime }l_{i}},k_{l_{i}^{\prime }}\right) \right]
\right\} \right) \delta \left( H\left( \left\{ \left[ \left(
p_{l_{i}},p_{l_{i}^{\prime }l_{i}},k_{l_{i}^{\prime }}\right) \right]
\right\} ,\Psi _{J}^{\otimes k_{p}}\right) \right)
\end{eqnarray*}

where the $\left( \left( p_{l_{i}},p_{l_{i}^{\prime }l_{i}},k_{l_{i}^{\prime
}}\right) \right) $ are local coordinates. The relation $H\left( \left(
p_{l_{i}},p_{l_{i}^{\prime }l_{i}},k_{l_{i}^{\prime }}\right) ,\Psi
_{J}^{\otimes k_{p}}\right) $ is global. In coordinates, we\ assume a
relation involving the linear function:%
\begin{equation*}
\mathbf{h.}\left( \left( p_{l_{i}},p_{l_{i}^{\prime }l_{i}},k_{l_{i}^{\prime
}}\right) \right) =h_{i}^{j}\left( p_{l_{i}},p_{l_{i}^{\prime
}l_{i}},k_{l_{i}^{\prime }}\right)
\end{equation*}%
and relation:

\begin{equation*}
0=H\left( \mathbf{h.}\left( \left( p_{l_{i}},p_{l_{i}^{\prime
}l_{i}},k_{l_{i}^{\prime }}\right) \right) ,\Psi _{J}^{\otimes k_{p}}\left(
\left\{ \left( p_{l_{i}},p_{l_{i}^{\prime }l_{i}},k_{l_{i}^{\prime }}\right)
\right\} \right) \right)
\end{equation*}%
where $\Psi _{J}^{\otimes k_{p}}\left( \left\{ \left(
p_{l_{i}},p_{l_{i}^{\prime }l_{i}},k_{l_{i}^{\prime }}\right) \right\}
\right) $ is an averaged field over independent degrees of freedom:%
\begin{eqnarray*}
&&\Psi _{J}^{\otimes k_{p}}\left( \left\{ \left( p_{l_{i}},p_{l_{i}^{\prime
}l_{i}},k_{l_{i}^{\prime }}\right) \right\} \right) \\
&=&\int v\left( \overline{\left[ p_{0},p_{l^{\prime }0}\right]
^{k_{l^{\prime }}}}\right) \Psi _{J}^{\otimes k_{l^{\prime }}}\left( 
\overline{\left[ p_{0},p_{l^{\prime }0}\right] ^{k_{l^{\prime }}}},\left\{
\left( p_{l_{i}},p_{l_{i}^{\prime }l_{i}},k_{l_{i}^{\prime }}\right)
\right\} \right) d\left( \overline{\left[ p_{0},p_{l^{\prime }0}\right]
^{k_{l^{\prime }}}}\right)
\end{eqnarray*}

for some function $v\left( \overline{\left( \left( U_{j}\right) ^{\left(
p_{0},p_{l^{\prime }0}\right) }\right) ^{k_{l^{\prime }}}}\right) $. The
constraint expresses parameters of $\left( \mathbf{L}\left( \Psi
_{J}^{\otimes k_{p}}\right) \right) $ as functions of $\Psi _{J}^{\otimes
k_{p}}$. The $\left\{ \left( p_{l_{i}},p_{l_{i}^{\prime
}l_{i}},k_{l_{i}^{\prime }}\right) \right\} $ become metric spaces, with
metrics, functional of $\Psi _{J}^{\otimes k_{p}}$.

\subsection{Reparametrization and field transformation}

We do not include subobject, for the sake of simplicity.

\subsubsection{Changes of variables in the parameters}

We assume that in the sets: 
\begin{equation*}
V_{\left( p_{l_{i}},p_{l_{i}^{\prime }l_{i}},k_{l_{i}^{\prime }}\right)
}=\left\{ \left[ 
\begin{array}{c}
\left( p_{l_{i}},p_{l_{i}^{\prime }l_{i}},k_{l_{i}^{\prime }}\right) \\ 
\left[ \left\{ \Psi _{J}\left[ l^{\prime },0\right] \right\} ,v\right]%
\end{array}%
\right] \right\}
\end{equation*}

the variables $\left[ 
\begin{array}{c}
\left( p_{l_{i}},p_{l_{i}^{\prime }l_{i}},k_{l_{i}^{\prime }}\right) \\ 
\left[ \left\{ \Psi _{J}\left[ l^{\prime },0\right] \right\} ,v\right]%
\end{array}%
\right] $ transform as the initial parameters $U_{i}$ in a change of
variables.

Under these assumptions, we consider some transformations described by some
group elements $\mathbf{g}$ acting on the field $\Psi _{J,s_{p_{l^{\prime
},0}}^{\otimes k_{l_{i}^{\prime }}}}^{\otimes k_{l^{\prime },i}}$. We
consider the tranformation:%
\begin{eqnarray*}
\Psi _{J,s_{p_{l^{\prime },0}}^{\otimes k_{l_{i}^{\prime }}}}^{\otimes
k_{l^{\prime },i}}\left( \left[ p_{0},p_{l^{\prime }0}\right] ^{k_{l^{\prime
},i}}\right) &\rightarrow &\mathbf{g.}\Psi _{J,s_{p_{l^{\prime
},0}}^{\otimes k_{l_{i}^{\prime }}}}^{\otimes k_{l^{\prime },i}}\left( \left[
p_{0},p_{l^{\prime }0}\right] ^{k_{l^{\prime },i}}\right) \\
&=&\Psi _{J,s_{p_{l^{\prime },0}}^{\otimes k_{l_{i}^{\prime }}}}^{\otimes
k_{l^{\prime },i}}\left( \mathbf{g}^{-1}\mathbf{.}\left[ p_{0},p_{l^{\prime
}0}\right] ^{k_{l^{\prime },i}}\right)
\end{eqnarray*}%
We also assume that the kernel $\mathcal{K}_{0}$ arising in the saddle point
equation:%
\begin{eqnarray*}
&&\mathcal{K}_{0}\left( \left\{ \left[ p_{0},p_{l^{\prime }0}\right]
^{k_{l^{\prime },i}}\right\} _{l^{\prime },i},\left\{ \overline{\left[
p_{l_{i}},p_{l_{i}^{\prime }l_{i}}\right] ^{k_{l_{i}^{\prime }}}},\left[ 
\begin{array}{c}
\left( p_{l_{i}},p_{l_{i}^{\prime }l_{i}},k_{l_{i}^{\prime }}\right) \\ 
\left[ \left\{ \Psi _{J}\left[ l^{\prime },0\right] \right\} ,v\right]%
\end{array}%
\right] \right\} _{i}\right) \\
&=&\exp \left( i\mathbf{L}\left( \Psi _{J}^{\otimes k_{p_{0}}}\right)
.\left\{ \left[ 
\begin{array}{c}
\left( p_{l_{i}},p_{l_{i}^{\prime }l_{i}},k_{l_{i}^{\prime }}\right) \\ 
\left[ \left\{ \Psi _{J}\left[ l^{\prime },0\right] \right\} ,v\right]%
\end{array}%
\right] \right\} \right) \mathcal{K}_{0}\left( \left\{ \left[
p_{0},p_{l^{\prime }0}\right] ^{k_{l^{\prime },i}}\right\} _{l^{\prime
},i},\left\{ \overline{\left[ p_{l_{i}},p_{l_{i}^{\prime }l_{i}}\right]
^{k_{l_{i}^{\prime }}}}\right\} _{i}\right)
\end{eqnarray*}%
transforms in the following way under $\mathbf{g}$:%
\begin{eqnarray*}
&&\mathbf{g.}\mathcal{K}_{0}\left( \left\{ \left[ p_{0},p_{l^{\prime }0}%
\right] ^{k_{l^{\prime },i}}\right\} _{l^{\prime },i},\left\{ \overline{%
\left[ p_{l_{i}},p_{l_{i}^{\prime }l_{i}}\right] ^{k_{l_{i}^{\prime }}}},%
\left[ 
\begin{array}{c}
\left( p_{l_{i}},p_{l_{i}^{\prime }l_{i}},k_{l_{i}^{\prime }}\right) \\ 
\left[ \left\{ \Psi _{J}\left[ l^{\prime },0\right] \right\} ,v\right]%
\end{array}%
\right] \right\} _{i}\right) \\
&\rightarrow &\mathcal{K}_{0}\left( \left\{ \mathbf{g}^{-1}\mathbf{.}\left[
p_{0},p_{l^{\prime }0}\right] ^{k_{l^{\prime },i}}\right\} _{l^{\prime
},i},\left\{ \overline{\left[ p_{l_{i}},p_{l_{i}^{\prime }l_{i}}\right]
^{k_{l_{i}^{\prime }}}},\left[ 
\begin{array}{c}
\left( p_{l_{i}},p_{l_{i}^{\prime }l_{i}},k_{l_{i}^{\prime }}\right) \\ 
\left[ \left\{ \mathbf{g.}\Psi _{J}\left[ l^{\prime },0\right] \right\} ,v%
\right]%
\end{array}%
\right] \right\} _{i}\right)
\end{eqnarray*}%
These kernel arise in integrals while computing the saddle points. This
involves products of variables $\left\{ \left[ p_{0},p_{l^{\prime }0}\right]
^{k_{l^{\prime },i}}\right\} _{l^{\prime },i}$ and $\overline{\left[
p_{l_{i}},p_{l_{i}^{\prime }l_{i}}\right] ^{k_{l_{i}^{\prime }}}}$ in the
kernel. We can assume a translational invariance, so that by change of
variable::%
\begin{eqnarray}
&&\mathbf{g.}\mathcal{K}_{0}\left( \left\{ \left[ p_{0},p_{l^{\prime }0}%
\right] ^{k_{l^{\prime },i}}\right\} _{l^{\prime },i},\left\{ \overline{%
\left[ p_{l_{i}},p_{l_{i}^{\prime }l_{i}}\right] ^{k_{l_{i}^{\prime }}}},%
\left[ 
\begin{array}{c}
\left( p_{l_{i}},p_{l_{i}^{\prime }l_{i}},k_{l_{i}^{\prime }}\right) \\ 
\left[ \left\{ \Psi _{J}\left[ l^{\prime },0\right] \right\} ,v\right]%
\end{array}%
\right] \right\} _{i}\right)  \label{TKR} \\
&\rightarrow &\mathcal{K}_{0}\left( \left\{ \left[ p_{0},p_{l^{\prime }0}%
\right] ^{k_{l^{\prime },i}}\right\} _{l^{\prime },i},\left\{ \mathbf{g.}%
\overline{\left[ p_{l_{i}},p_{l_{i}^{\prime }l_{i}}\right]
^{k_{l_{i}^{\prime }}}},\left[ 
\begin{array}{c}
\left( p_{l_{i}},p_{l_{i}^{\prime }l_{i}},k_{l_{i}^{\prime }}\right) \\ 
\left[ \left\{ \mathbf{g.}\Psi _{J}\left[ l^{\prime },0\right] \right\} ,v%
\right]%
\end{array}%
\right] \right\} _{i}\right)  \notag
\end{eqnarray}

\subsubsection{Transformations for degeneracy generators}

We rewrite (\ref{TKR}) by considering the transformation properties of the
degeneracies operators. We define the operators:%
\begin{equation*}
\mathbf{U}_{\left[ p_{l},p_{l^{\prime }l}\right] ^{k_{l^{\prime }}}},\Pi _{%
\left[ p_{l},p_{l^{\prime }l}\right] ^{k_{l^{\prime }}}}
\end{equation*}%
of multiplication and derivation by coordinates of $\left( \left(
U_{j}\right) ^{\left( p_{l},p_{l^{\prime }l}\right) }\right) ^{k_{l^{\prime
}}}$.

To study the invariance properties of the states, we assume the general form
for the generators $\mathbf{L}\left( \Psi _{J}^{\otimes k_{p_{0}}}\right) $: 
\begin{eqnarray}
\mathbf{L}\left( \Psi _{J}^{\otimes k_{p_{0}}}\right) &=&\sum_{i,k_{p_{i}}}%
\mathbf{l}\left( \mathbf{U}_{\left[ p_{l},p_{l^{\prime }l}\right]
^{k_{l^{\prime }}}},\Pi _{\left[ p_{l},p_{l^{\prime }l}\right]
^{k_{l^{\prime }}}},\overline{\left[ p_{0},p_{l^{\prime }0}\right]
^{k_{p_{0}}^{\left( i\right) }}},\left\{ \left[ p_{l},p_{l^{\prime }l}\right]
^{k_{l^{\prime }}}\right\} _{\substack{ l\leqslant m  \\ l^{\prime
}\leqslant m^{\prime }}}\right)  \label{GR} \\
&&\times \left\{ \tprod \int d\overline{\left[ p_{0},p_{l^{\prime }0}\right]
^{k_{p_{0}}^{\left( i\right) }}}d\left\{ \left[ p_{l},p_{l^{\prime }l}\right]
^{k_{l^{\prime }}}\right\} _{\substack{ l\leqslant m  \\ l^{\prime
}\leqslant m^{\prime }}}\right.  \notag \\
&&\times v\left( \left( \overline{\left[ p_{0},p_{l^{\prime }0}\right]
^{k_{p_{0}}^{\left( i\right) }}},\left\{ \left[ p_{l},p_{l^{\prime }l}\right]
^{k_{l^{\prime }}}\right\} _{\substack{ l\leqslant m  \\ l^{\prime
}\leqslant m^{\prime }}}\right) \right) \left. \Psi _{J}^{\otimes
k_{p_{i}}}\left( \overline{\left[ p_{0},p_{l^{\prime }0}\right]
^{k_{p_{0}}^{\left( i\right) }}},\left\{ \left[ p_{l},p_{l^{\prime }l}\right]
^{k_{l^{\prime }}}\right\} _{\substack{ l\leqslant m  \\ l^{\prime
}\leqslant m^{\prime }}}\right) \right\}  \notag
\end{eqnarray}%
where the $\mathbf{l}$ are defined in (\ref{gn1}):%
\begin{equation*}
\mathbf{l}\left( \mathbf{U}_{\left[ p_{l},p_{l^{\prime }l}\right]
^{k_{l^{\prime }}}},\Pi _{\left[ p_{l},p_{l^{\prime }l}\right]
^{k_{l^{\prime }}}},\overline{\left[ p_{0},p_{l^{\prime }0}\right]
^{k_{p_{0}}^{\left( i\right) }}},\left\{ \left[ p_{l},p_{l^{\prime }l}\right]
^{k_{l^{\prime }}}\right\} _{\substack{ l\leqslant m  \\ l^{\prime
}\leqslant m^{\prime }}}\right)
\end{equation*}%
If these operators satisfy also some linear transformations properties:%
\begin{eqnarray*}
&&\mathbf{g}.\mathbf{l}\left( \mathbf{U}_{\left[ p_{l},p_{l^{\prime }l}%
\right] ^{k_{l^{\prime }}}},\Pi _{\left[ p_{l},p_{l^{\prime }l}\right]
^{k_{l^{\prime }}}},\overline{\left[ p_{0},p_{l^{\prime }0}\right]
^{k_{l^{\prime }}^{\left( i\right) }}},\mathbf{g.}\left\{ \left[
p_{l},p_{l^{\prime }l}\right] ^{k_{l^{\prime }}}\right\} _{\substack{ %
l\leqslant m  \\ l^{\prime }\leqslant m^{\prime }}}\right) \\
&=&\mathbf{l}\left( \mathbf{g}.\mathbf{U}_{\left[ p_{l},p_{l^{\prime }l}%
\right] ^{k_{l^{\prime }}}},\Pi _{\mathbf{g}.\left[ p_{l},p_{l^{\prime }l}%
\right] ^{k_{l^{\prime }}}},\overline{\left[ p_{0},p_{l^{\prime }0}\right]
^{k_{l^{\prime }}^{\left( i\right) }}},\left\{ \left[ p_{l},p_{l^{\prime }l}%
\right] ^{k_{l^{\prime }}}\right\} _{\substack{ l\leqslant m  \\ l^{\prime
}\leqslant m^{\prime }}}\right)
\end{eqnarray*}%
and if this action of $\mathbf{g}$ is given by the action of some operators: 
\begin{equation}
M_{\mathbf{g}}\left( \mathbf{U}_{\overline{\left[ p_{l_{i}},p_{l_{i}^{\prime
}l_{i}}\right] ^{k_{l_{i}^{\prime }}}}},\hat{\Pi}_{\overline{\left[
p_{l_{i}},p_{l_{i}^{\prime }l_{i}}\right] ^{k_{l_{i}^{\prime }}}}}\right)
\label{MG}
\end{equation}

\subsubsection{Transformations for the kernel}

Using (\ref{MG}), the transformation (\ref{TKR}) for the kernel is thus:%
\begin{eqnarray*}
&&\mathbf{g.}\mathcal{K}_{0}\left( \left\{ \left[ p_{0},p_{l^{\prime }0}%
\right] ^{k_{l^{\prime },i}}\right\} _{l^{\prime },i},\left\{ \overline{%
\left[ p_{l_{i}},p_{l_{i}^{\prime }l_{i}}\right] ^{k_{l_{i}^{\prime }}}},%
\left[ 
\begin{array}{c}
\left( p_{l_{i}},p_{l_{i}^{\prime }l_{i}},k_{l_{i}^{\prime }}\right) \\ 
\left[ \left\{ \Psi _{J}\left[ l^{\prime },0\right] \right\} ,v\right]%
\end{array}%
\right] \right\} _{i}\right) \\
&=&\mathcal{K}_{0}\left( \left\{ \left[ p_{0},p_{l^{\prime }0}\right]
^{k_{l^{\prime },i}}\right\} _{l^{\prime },i},\left\{ \mathbf{g.}\overline{%
\left[ p_{l_{i}},p_{l_{i}^{\prime }l_{i}}\right] ^{k_{l_{i}^{\prime }}}},%
\left[ 
\begin{array}{c}
\left( p_{l_{i}},p_{l_{i}^{\prime }l_{i}},k_{l_{i}^{\prime }}\right) \\ 
\left[ \left\{ \mathbf{g.}\Psi _{J}\left[ l^{\prime },0\right] \right\} ,v%
\right]%
\end{array}%
\right] \right\} _{i}\right) \\
&=&\exp \left( iM_{\mathbf{g}}\left( \mathbf{U}_{\overline{\left[
p_{l_{i}},p_{l_{i}^{\prime }l_{i}}\right] ^{k_{l_{i}^{\prime }}}}},\hat{\Pi}%
_{\overline{\left[ p_{l_{i}},p_{l_{i}^{\prime }l_{i}}\right]
^{k_{l_{i}^{\prime }}}}}\right) \right) \mathcal{K}_{0}\left( \left\{ \left[
p_{0},p_{l^{\prime }0}\right] ^{k_{l^{\prime },i}}\right\} _{l^{\prime
},i},\left\{ \overline{\left[ p_{l_{i}},p_{l_{i}^{\prime }l_{i}}\right]
^{k_{l_{i}^{\prime }}}},%
\begin{array}{c}
\left( p_{l_{i}},p_{l_{i}^{\prime }l_{i}},k_{l_{i}^{\prime }}\right) \\ 
\left[ \left\{ \mathbf{g.}\Psi _{J}\left[ l^{\prime },0\right] \right\} ,v%
\right]%
\end{array}%
\right\} _{i}\right)
\end{eqnarray*}%
The linear transformation on the $\left[ p_{l_{i}},p_{l_{i}^{\prime }l_{i}}%
\right] ^{k_{l_{i}^{\prime }}}$ translates on the operator generating
degeneracies:%
\begin{eqnarray}
&&\exp \left( iM_{\mathbf{g}}\left( \mathbf{U}_{\overline{\left[
p_{l_{i}},p_{l_{i}^{\prime }l_{i}}\right] ^{k_{l_{i}^{\prime }}}}},\hat{\Pi}%
_{\overline{\left[ p_{l_{i}},p_{l_{i}^{\prime }l_{i}}\right]
^{k_{l_{i}^{\prime }}}}}\right) \right)  \label{CMT} \\
&&\times \exp \left( i\mathbf{L}\left( \mathbf{g.}\Psi _{J}^{\otimes
k_{l}}\right) .\left( \left[ 
\begin{array}{c}
\left( p_{l_{i}},p_{l_{i}^{\prime }l_{i}},k_{l_{i}^{\prime }}\right) \\ 
\left[ \left\{ \mathbf{g.}\Psi _{J}\left[ l^{\prime },0\right] \right\} ,v%
\right]%
\end{array}%
\right] \right) \right) \exp \left( -iM_{\mathbf{g}}\left( \mathbf{U}_{%
\overline{\left[ p_{l_{i}},p_{l_{i}^{\prime }l_{i}}\right]
^{k_{l_{i}^{\prime }}}}},\hat{\Pi}_{\overline{\left[ p_{l_{i}},p_{l_{i}^{%
\prime }l_{i}}\right] ^{k_{l_{i}^{\prime }}}}}\right) \right)  \notag \\
&=&\exp \left( i\mathbf{L}^{\mathbf{g}^{-1}}\left( \mathbf{g.}\Psi
_{J}^{\otimes k_{l}}\right) .\left( \left[ 
\begin{array}{c}
\left( p_{l_{i}},p_{l_{i}^{\prime }l_{i}},k_{l_{i}^{\prime }}\right) \\ 
\left[ \left\{ \mathbf{g.}\Psi _{J}\left[ l^{\prime },0\right] \right\} ,v%
\right]%
\end{array}%
\right] \right) \right)  \notag
\end{eqnarray}%
where $\mathbf{L}^{\mathbf{g}}\left( \mathbf{g.}\Psi _{J}^{\otimes
k_{l}}\right) $ is the operator with generator:%
\begin{equation*}
\mathbf{l}\left( \mathbf{U}_{\mathbf{g.}\left[ p_{l},p_{l^{\prime }l}\right]
^{k_{l^{\prime }}}},\Pi _{\mathbf{g.}\left[ p_{l},p_{l^{\prime }l}\right]
^{k_{l^{\prime }}}},\overline{\left[ p_{0},p_{l^{\prime }0}\right]
^{k_{p_{0}}^{\left( i\right) }}},\left\{ \mathbf{g.}\left[
p_{l},p_{l^{\prime }l}\right] ^{k_{l^{\prime }}}\right\} _{\substack{ %
l\leqslant m  \\ l^{\prime }\leqslant m^{\prime }}}\right)
\end{equation*}%
Given that we assumed dual linear transformation for the parameters $\left( %
\left[ 
\begin{array}{c}
\left( p_{l_{i}},p_{l_{i}^{\prime }l_{i}},k_{l_{i}^{\prime }}\right) \\ 
\left[ \left\{ \mathbf{g.}\Psi _{J}\left[ l^{\prime },0\right] \right\} ,v%
\right]%
\end{array}%
\right] \right) $, we find:%
\begin{eqnarray}
&&\exp \left( i\mathbf{L}^{\mathbf{g}}\left( \mathbf{g.}\Psi _{J}^{\otimes
k_{l}}\right) .\left( \left[ 
\begin{array}{c}
\left( p_{l_{i}},p_{l_{i}^{\prime }l_{i}},k_{l_{i}^{\prime }}\right) \\ 
\left[ \left\{ \mathbf{g.}\Psi _{J}\left[ l^{\prime },0\right] \right\} ,v%
\right]%
\end{array}%
\right] \right) \right)  \label{DGR} \\
&\equiv &\exp \left( i\mathbf{L}\left( \mathbf{g.}\Psi _{J}^{\otimes
k_{l}}\right) .\left( \mathbf{g.}\left[ 
\begin{array}{c}
\left( p_{l_{i}},p_{l_{i}^{\prime }l_{i}},k_{l_{i}^{\prime }}\right) \\ 
\left[ \left\{ \mathbf{g.}\Psi _{J}\left[ l^{\prime },0\right] \right\} ,v%
\right]%
\end{array}%
\right] \right) \right)  \notag
\end{eqnarray}%
Then, using (\ref{CMT}) nd (\ref{DGR}), we can deduce the transformation
property for the kernel: 
\begin{eqnarray*}
&&\mathcal{K}_{0}\left( \left\{ \left[ p_{0},p_{l^{\prime }0}\right]
^{k_{l^{\prime },i}}\right\} _{l^{\prime },i},\left\{ \mathbf{g.}\overline{%
\left[ p_{l_{i}},p_{l_{i}^{\prime }l_{i}}\right] ^{k_{l_{i}^{\prime }}}},%
\left[ 
\begin{array}{c}
\left( p_{l_{i}},p_{l_{i}^{\prime }l_{i}},k_{l_{i}^{\prime }}\right) \\ 
\left[ \left\{ \mathbf{g.}\Psi _{J}\left[ l^{\prime },0\right] \right\} ,v%
\right]%
\end{array}%
\right] \right\} _{i}\right) \\
&=&\exp \left( iM_{\mathbf{g}}\left( \mathbf{U}_{\overline{\left[
p_{l_{i}},p_{l_{i}^{\prime }l_{i}}\right] ^{k_{l_{i}^{\prime }}}}},\hat{\Pi}%
_{\overline{\left[ p_{l_{i}},p_{l_{i}^{\prime }l_{i}}\right]
^{k_{l_{i}^{\prime }}}}}\right) \right) \\
&&\exp \left( i\mathbf{L}\left( \mathbf{g.}\Psi _{J}^{\otimes k_{l}}\right)
.\left( \left[ 
\begin{array}{c}
\left( p_{l_{i}},p_{l_{i}^{\prime }l_{i}},k_{l_{i}^{\prime }}\right) \\ 
\left[ \left\{ \mathbf{g.}\Psi _{J}\left[ l^{\prime },0\right] \right\} ,v%
\right]%
\end{array}%
\right] \right) \right) \exp \left( -iM_{\mathbf{g}}\left( \mathbf{U}_{%
\overline{\left[ p_{l_{i}},p_{l_{i}^{\prime }l_{i}}\right]
^{k_{l_{i}^{\prime }}}}},\hat{\Pi}_{\overline{\left[ p_{l_{i}},p_{l_{i}^{%
\prime }l_{i}}\right] ^{k_{l_{i}^{\prime }}}}}\right) \right) \\
&&\times \exp \left( iM_{\mathbf{g}}\left( \mathbf{U}_{\overline{\left[
p_{l_{i}},p_{l_{i}^{\prime }l_{i}}\right] ^{k_{l_{i}^{\prime }}}}},\hat{\Pi}%
_{\overline{\left[ p_{l_{i}},p_{l_{i}^{\prime }l_{i}}\right]
^{k_{l_{i}^{\prime }}}}}\right) \right) \mathcal{K}_{0}\left( \left\{ \left[
p_{0},p_{l^{\prime }0}\right] ^{k_{l^{\prime },i}}\right\} _{l^{\prime
},i},\left\{ \overline{\left[ p_{l_{i}},p_{l_{i}^{\prime }l_{i}}\right]
^{k_{l_{i}^{\prime }}}}\right\} _{i}\right) \\
&=&\exp \left( i\mathbf{L}\left( \mathbf{g.}\Psi _{J}^{\otimes k_{l}}\right)
.\left( \mathbf{g.}\left[ 
\begin{array}{c}
\left( p_{l_{i}},p_{l_{i}^{\prime }l_{i}},k_{l_{i}^{\prime }}\right) \\ 
\left[ \left\{ \mathbf{g.}\Psi _{J}\left[ l^{\prime },0\right] \right\} ,v%
\right]%
\end{array}%
\right] \right) \right) \mathcal{K}_{0}\left( \left\{ \left[
p_{0},p_{l^{\prime }0}\right] ^{k_{l^{\prime },i}}\right\} _{l^{\prime },i},%
\mathbf{g.}\left\{ \overline{\left[ p_{l_{i}},p_{l_{i}^{\prime }l_{i}}\right]
^{k_{l_{i}^{\prime }}}}\right\} _{i}\right)
\end{eqnarray*}

\subsubsection{Transformation for functionals and effectiv fld}

Inserting previous formula in (\ref{GN}), the function characterizing the
generic formula is transformed as:%
\begin{equation*}
\int g^{\mathcal{K}}\left( \mathbf{g.}\overline{\left[ p_{0},p_{l_{i}^{%
\prime }0}\right] ^{k_{l^{\prime }}}},\mathbf{g}\left\{ \left[
p_{l_{1}},p_{l_{1}^{\prime }l_{1}}\right] ^{k_{l_{1}^{\prime }}}\right\} _{ 
_{\substack{ l_{1}\leqslant m_{1}  \\ l_{1}^{\prime }\leqslant m_{1}^{\prime
} }}},v,\left\{ \left[ 
\begin{array}{c}
\left( p_{l_{1}},p_{l_{1}^{\prime }l_{1}},k_{l_{1}^{\prime }}\right) \\ 
\left[ \left\{ \Psi _{J}\left[ l^{\prime },0\right] \right\} ,v\right]%
\end{array}%
\right] \right\} _{_{\substack{ l\subset l_{1}\leqslant m_{1}  \\ l^{\prime
}\subset l_{1}^{\prime }\leqslant m_{1}^{\prime }}}}\right)
\end{equation*}%
where the actions $\mathbf{g.}\overline{\left[ p_{0},p_{l_{i}^{\prime }0}%
\right] ^{k_{l^{\prime }}}}$,\ $\mathbf{g}\left\{ \left[
p_{l_{1}},p_{l_{1}^{\prime }l_{1}}\right] ^{k_{l_{1}^{\prime }}}\right\} $
take into account the initial change of variables. Since we have assumed
some invariance, we can assume that the integral of $g^{\mathcal{K}}$ over $%
\left\{ \left[ p_{l_{1}},p_{l_{1}^{\prime }l_{1}}\right] ^{k_{l_{1}^{\prime
}}}\right\} $ is invariant, so that we can replace:%
\begin{equation*}
\mathbf{g}\left\{ \left[ p_{l_{1}},p_{l_{1}^{\prime }l_{1}}\right]
^{k_{l_{1}^{\prime }}}\right\} \rightarrow \left\{ \left[
p_{l_{1}},p_{l_{1}^{\prime }l_{1}}\right] ^{k_{l_{1}^{\prime }}}\right\}
\end{equation*}%
A second change of variable:%
\begin{equation*}
\mathbf{g.}\overline{\left[ p_{0},p_{l_{i}^{\prime }0}\right] ^{k_{l^{\prime
}}}}\rightarrow \mathbf{.}\overline{\left[ p_{0},p_{l_{i}^{\prime }0}\right]
^{k_{l^{\prime }}}}
\end{equation*}%
As a consequence, in (\ref{FL}), $\bar{g}^{\mathcal{K}}$ is invariant, and
the effective field (\ref{LC}) is replaced by:%
\begin{equation*}
\Psi _{J,\underset{i}{\tprod }s_{p_{l^{\prime },0}}^{\otimes
k_{l_{i}^{\prime }}}}^{\otimes \sum_{i}k_{l^{\prime },i}}\left( \left\{ 
\mathbf{g}^{-1}.\left[ p_{0},p_{l^{\prime }0}\right] ^{k_{l^{\prime },i}},%
\mathbf{g}.\left[ 
\begin{array}{c}
\left( p_{l_{i}},p_{l_{i}^{\prime }l_{i}},k_{l_{i}^{\prime }}\right) \\ 
\left[ \left\{ \mathbf{g.}\Psi _{J}\left[ l^{\prime },0\right] \right\} ,v%
\right]%
\end{array}%
\right] \right\} _{i}\right)
\end{equation*}

and this implies the transformation property of the effective field:%
\begin{equation*}
\mathbf{g.}\Psi _{J,\underset{i}{\tprod }s_{p_{l^{\prime },0}}^{\otimes
k_{l_{i}^{\prime }}}}^{\otimes \sum_{i}k_{l^{\prime },i}}\left( \left\{ %
\left[ p_{0},p_{l^{\prime }0}\right] ^{k_{l^{\prime },i}},\left[ 
\begin{array}{c}
\left( p_{l_{i}},p_{l_{i}^{\prime }l_{i}},k_{l_{i}^{\prime }}\right) \\ 
\left[ \left\{ \mathbf{g.}\Psi _{J}\left[ l^{\prime },0\right] \right\} ,v%
\right]%
\end{array}%
\right] \right\} _{i}\right) \equiv \Psi _{J,\underset{i}{\tprod }%
s_{p_{l^{\prime },0}}^{\otimes k_{l_{i}^{\prime }}}}^{\otimes
\sum_{i}k_{l^{\prime },i}}\left( \left\{ \mathbf{g}^{-1}\mathbf{.}\left[
p_{0},p_{l^{\prime }0}\right] ^{k_{l^{\prime },i}},\mathbf{g.}\left[ 
\begin{array}{c}
\left( p_{l_{i}},p_{l_{i}^{\prime }l_{i}},k_{l_{i}^{\prime }}\right) \\ 
\left[ \left\{ \mathbf{g.}\Psi _{J}\left[ l^{\prime },0\right] \right\} ,v%
\right]%
\end{array}%
\right] \right\} _{i}\right)
\end{equation*}

\subsection{Constraints and transformation}

If the parameters $\left[ 
\begin{array}{c}
\left( p_{l_{i}},p_{l_{i}^{\prime }l_{i}},k_{l_{i}^{\prime }}\right) \\ 
\left[ \left\{ \Psi _{J}\left[ l^{\prime },0\right] \right\} ,v\right]%
\end{array}%
\right] $ satisfy equations of the form:%
\begin{equation}
0=f\left( \left[ 
\begin{array}{c}
\left( p_{l_{i}},p_{l_{i}^{\prime }l_{i}},k_{l_{i}^{\prime }}\right) \\ 
\left[ \left\{ \Psi _{J}\left[ l^{\prime },0\right] \right\} ,v\right]%
\end{array}%
\right] ,\left\{ h_{p}\left( \left( \Psi _{J}\right) ,\left[
p_{0},p_{l^{\prime }0}\right] \right) \right\} \right)  \label{C}
\end{equation}%
where the vector functions $h_{p}\left( \left( \Psi _{J}\right) \right) $
have the following functional form: 
\begin{equation*}
h_{p}\left( \left( \Psi _{J}\right) ,\left[ p_{0},p_{l^{\prime }0}\right]
\right) =\left[ \int h_{p}\left( \left[ \bar{p}_{0},\bar{p}_{l^{\prime }0}%
\right] ^{k_{l^{\prime }}}\right) \Psi _{J,s_{\bar{p}_{l^{\prime
},0}}^{\otimes k_{l_{i}^{\prime }}}}^{\otimes k_{l^{\prime },i}}\left( \left[
\bar{p}_{0},\bar{p}_{l^{\prime }0}\right] ^{k_{l^{\prime }}}\right) d\left[ 
\bar{p}_{0},\bar{p}_{l^{\prime }0}\right] ^{k_{l^{\prime }}}\right] \Psi
_{J,s_{p_{l^{\prime },0}}^{\otimes k_{l_{i}^{\prime }}}}^{\otimes
k_{l^{\prime },i}}\left( \left[ p_{0},p_{l^{\prime }0}\right] ^{k_{l^{\prime
},i}}\right)
\end{equation*}%
We can write the transformation property for these functionals:%
\begin{eqnarray*}
\mathbf{g.}h_{p}\left( \left( \Psi _{J}\right) \right) &=&h_{p}\left( 
\mathbf{g.}\left( \Psi _{J}\right) \right) \\
&=&\left[ \int h_{p}\left( \mathbf{g.}\left[ \bar{p}_{0},\bar{p}_{l^{\prime
}0}\right] ^{k_{l^{\prime }}}\right) \mathbf{g.}\Psi _{J,s_{\bar{p}%
_{l^{\prime },0}}^{\otimes k_{l_{i}^{\prime }}}}^{\otimes k_{l^{\prime
},i}}\left( \left[ \bar{p}_{0},\bar{p}_{l^{\prime }0}\right] ^{k_{l^{\prime
}}}\right) d\left[ \bar{p}_{0},\bar{p}_{l^{\prime }0}\right] ^{k_{l^{\prime
}}}\right] \mathbf{g.}\Psi _{J,s_{p_{l^{\prime },0}}^{\otimes
k_{l_{i}^{\prime }}}}^{\otimes k_{l^{\prime },i}}\left( \left[
p_{0},p_{l^{\prime }0}\right] ^{k_{l^{\prime },i}}\right)
\end{eqnarray*}%
Consider first a set of functions: 
\begin{equation*}
\left\{ h_{p}\left( \mathbf{g.}\left[ p_{0},p_{l^{\prime }0}\right]
^{k_{l^{\prime },i}}\right) \mathbf{g.}\Psi _{J,s_{p_{l^{\prime
},0}}^{\otimes k_{l_{i}^{\prime }}}}^{\otimes k_{l^{\prime },i}}\left( \left[
p_{0},p_{l^{\prime }0}\right] ^{k_{l^{\prime },i}}\right) \right\}
\end{equation*}%
belonging to some representation $\mathcal{R}\left( \mathbf{g}\right) $ of $%
\mathbf{g}$, and transforming as:%
\begin{equation*}
\left\{ h_{p}\left( \mathbf{g.}\left[ p_{0},p_{l^{\prime }0}\right]
^{k_{l^{\prime },i}}\right) \mathbf{g.}\Psi _{J,s_{p_{l^{\prime
},0}}^{\otimes k_{l_{i}^{\prime }}}}^{\otimes k_{l^{\prime },i}}\left( \left[
p_{0},p_{l^{\prime }0}\right] ^{k_{l^{\prime },i}}\right) \right\} =\mathcal{%
R}\left( \mathbf{g}\right) \left\{ h_{p}\left( \left[ p_{0},p_{l^{\prime }0}%
\right] ^{k_{l^{\prime },i}}\right) \Psi _{J,s_{p_{l^{\prime },0}}^{\otimes
k_{l_{i}^{\prime }}}}^{\otimes k_{l^{\prime },i}}\left( \left[
p_{0},p_{l^{\prime }0}\right] ^{k_{l^{\prime },i}}\right) \right\}
\end{equation*}%
so that:%
\begin{equation*}
\mathbf{g.}h_{p}\left( \left( \Psi _{J}\right) \right) =\mathcal{R}\left( 
\mathbf{g}\right) h_{p}\left( \left( \Psi _{J}\right) \right)
\end{equation*}%
Consider also that at the lowest order approxmtn, the function $f$ involves
some scalar products:%
\begin{equation*}
f\left( \left[ 
\begin{array}{c}
\left( p_{l_{i}},p_{l_{i}^{\prime }l_{i}},k_{l_{i}^{\prime }}\right) \\ 
\left[ \left\{ \Psi _{J}\left[ l^{\prime },0\right] \right\} ,v\right]%
\end{array}%
\right] ,\left\{ h_{p}\left( \left( \Psi _{J}\right) \right) \right\}
\right) =\bar{f}\left( 
\begin{array}{c}
\left( p_{l_{i}},p_{l_{i}^{\prime }l_{i}},k_{l_{i}^{\prime }}\right) \\ 
\left[ \left\{ \Psi _{J}\left[ l^{\prime },0\right] \right\} ,v\right]%
\end{array}%
.\left\{ h_{p}\left( \left( \Psi _{J}\right) \right) \right\} \right)
\end{equation*}

Under these assumptions, the constraints (\ref{C}) satisfy some invariance
property:%
\begin{equation*}
f\left( \left[ 
\begin{array}{c}
\left( p_{l_{i}},p_{l_{i}^{\prime }l_{i}},k_{l_{i}^{\prime }}\right) \\ 
\left[ \left\{ \Psi _{J}\left[ l^{\prime },0\right] \right\} ,v\right]%
\end{array}%
\right] ,\mathcal{R}\left( \mathbf{g}\right) \left\{ h_{p}\left( \left( \Psi
_{J}\right) \right) \right\} \right) =f\left( \mathcal{R}\left( \mathbf{g}%
^{-1}\right) \left[ 
\begin{array}{c}
\left( p_{l_{i}},p_{l_{i}^{\prime }l_{i}},k_{l_{i}^{\prime }}\right) \\ 
\left[ \left\{ \Psi _{J}\left[ l^{\prime },0\right] \right\} ,v\right]%
\end{array}%
\right] ,\left\{ h_{p}\left( \left( \Psi _{J}\right) \right) \right\} \right)
\end{equation*}%
This relation is satisfied if we consider the lowest order constraint:

\begin{equation}
f\left( \sum \left( \left[ 
\begin{array}{c}
\left( p_{l_{i}},p_{l_{i}^{\prime }l_{i}},k_{l_{i}^{\prime }}\right) \\ 
\left[ \left\{ \Psi _{J}\left[ l^{\prime },0\right] \right\} ,v\right]%
\end{array}%
\right] G\left( \left[ 
\begin{array}{c}
\left( p_{l_{i}},p_{l_{i}^{\prime }l_{i}},k_{l_{i}^{\prime }}\right) \\ 
\left[ \left\{ \Psi _{J}\left[ l^{\prime },0\right] \right\} ,v\right]%
\end{array}%
\right] ,\Psi _{J}\right) \left[ 
\begin{array}{c}
\left( p_{l_{i}},p_{l_{i}^{\prime }l_{i}},k_{l_{i}^{\prime }}\right) \\ 
\left[ \left\{ \Psi _{J}\left[ l^{\prime },0\right] \right\} ,v\right]%
\end{array}%
\right] -\gamma \left( \Psi _{J}\right) \right) ^{2}\right)  \label{Ctp}
\end{equation}%
and if we also assume that $\gamma \left( U^{\left( j\right) }\right) $
arising in the constraints are invariant under transfrmation of the
parameters.%
\begin{equation*}
\gamma \left( \mathbf{g.}\Psi _{J}\right) =\gamma \left( \Psi _{J}\right)
\end{equation*}%
As a consequence, function $G\left( \left[ 
\begin{array}{c}
\left( p_{l_{i}},p_{l_{i}^{\prime }l_{i}},k_{l_{i}^{\prime }}\right) \\ 
\left[ \left\{ \Psi _{J}\left[ l^{\prime },0\right] \right\} ,v\right]%
\end{array}%
\right] ,\Psi _{J}\right) $ transforms as: 
\begin{equation*}
G\left( \mathcal{R}_{0}\left( \mathbf{g}\right) \left[ 
\begin{array}{c}
\left( p_{l_{i}},p_{l_{i}^{\prime }l_{i}},k_{l_{i}^{\prime }}\right) \\ 
\left[ \left\{ \Psi _{J}\left[ l^{\prime },0\right] \right\} ,v\right]%
\end{array}%
\right] ,\mathbf{g.}\Psi _{J}\right) =\left( \mathcal{R}_{0}^{t}\left( 
\mathbf{g}\right) \right) ^{-1}G\left( \left[ 
\begin{array}{c}
\left( p_{l_{i}},p_{l_{i}^{\prime }l_{i}},k_{l_{i}^{\prime }}\right) \\ 
\left[ \left\{ \Psi _{J}\left[ l^{\prime },0\right] \right\} ,v\right]%
\end{array}%
\right] ,\Psi _{J}\right) \mathcal{R}_{0}^{-1}\left( \mathbf{g}\right)
\end{equation*}%
or written differently:%
\begin{equation*}
\mathcal{R}_{0}\left( \mathbf{g}\right) .G\left( \left[ 
\begin{array}{c}
\left( p_{l_{i}},p_{l_{i}^{\prime }l_{i}},k_{l_{i}^{\prime }}\right) \\ 
\left[ \left\{ \Psi _{J}\left[ l^{\prime },0\right] \right\} ,v\right]%
\end{array}%
\right] ,\Psi _{J}\right) =\left( \mathcal{R}_{0}^{t}\left( \mathbf{g}%
\right) \right) ^{-1}G\left( \mathcal{R}_{0}^{-1}\left( \mathbf{g}\right) %
\left[ 
\begin{array}{c}
\left( p_{l_{i}},p_{l_{i}^{\prime }l_{i}},k_{l_{i}^{\prime }}\right) \\ 
\left[ \left\{ \Psi _{J}\left[ l^{\prime },0\right] \right\} ,v\right]%
\end{array}%
\right] ,\mathbf{g}^{-1}\mathbf{.}\Psi _{J}\right) \mathcal{R}%
_{0}^{-1}\left( \mathbf{g}\right)
\end{equation*}

\section{Several components}

\subsection{General set up}

The same procedure applies if there are several fields corresponding to
sevral types of non projected sbbjcts. We replac:%
\begin{equation*}
U_{j}^{\left( p_{0}\right) }\overset{p_{0}}{\hookrightarrow }\mathcal{V}%
\left( \oplus _{k}U_{j}^{k}\right)
\end{equation*}%
with the collection:%
\begin{equation*}
\left\{ U_{j}^{\left( p_{\eta }\right) }\overset{p_{\eta }}{\hookrightarrow }%
\mathcal{V}\left( \oplus _{k}U_{j}^{k}\right) \right\} _{p_{\eta }}
\end{equation*}%
The corresponding field for subobject $\eta $\ writes as a vector component:%
\begin{equation*}
\Psi _{J,\alpha ,s_{p_{\eta ^{\prime }\eta }}^{\otimes k_{\eta ^{\prime
}\eta }}}^{\otimes k_{\eta ^{\prime }\eta }}\left( \left[ p_{\eta },p_{\eta
^{\prime }\eta }\right] ^{k_{\eta ^{\prime }\eta }}\right) =\left[ \Psi
_{J,\alpha ,s_{p_{l^{\prime },\eta }}^{\otimes k_{l^{\prime }\eta
}}}^{\otimes k_{\eta ^{\prime }\eta }}\left( \left[ p_{\eta },p_{\eta
^{\prime }\eta }\right] ^{k_{\eta ^{\prime }\eta }}\right) \right] _{\eta }
\end{equation*}%
The whole set of remaining subobject is thus:%
\begin{equation*}
\left[ \Psi _{J,\alpha ,s_{p_{\eta ^{\prime }\eta }}^{\otimes k_{\eta
^{\prime }\eta }}}^{\otimes k_{\eta ^{\prime }\eta }}\left( \left[ p_{\eta
},p_{\eta ^{\prime }\eta }\right] ^{k_{\eta ^{\prime }\eta }}\right) \right]
\end{equation*}%
Similarly, we define the vectors: 
\begin{equation*}
\left[ \left[ p_{\eta },p_{\eta ^{\prime }\eta }\right] ^{k_{\eta ^{\prime
}\eta }}\right] \text{ and }\left[ \Psi _{J,\alpha ,s_{p_{\eta ^{\prime
}\eta }}^{\otimes k_{\eta ^{\prime }\eta }}}^{\otimes k_{\eta ^{\prime }\eta
}}\left( \overline{\left[ p_{\eta },p_{\eta ^{\prime }\eta }\right]
^{k_{\eta ^{\prime }\eta }}},\left\{ \left[ p_{l},p_{l^{\prime }l}\right]
^{k_{l^{\prime }}}\right\} \right) \right]
\end{equation*}
with components:%
\begin{equation*}
\left[ \left[ p_{\eta },p_{\eta ^{\prime }\eta }\right] ^{k_{\eta ^{\prime
}\eta }}\right] _{\eta }=\left[ p_{\eta },p_{\eta ^{\prime }\eta }\right]
^{k_{\eta ^{\prime }\eta }}
\end{equation*}%
and:%
\begin{equation*}
\left[ \Psi _{J,\alpha ,s_{p_{\eta ^{\prime }\eta }}^{\otimes k_{\eta
^{\prime }\eta }}}^{\otimes k_{\eta ^{\prime }\eta }}\left( \overline{\left[
p_{\eta },p_{\eta ^{\prime }\eta }\right] ^{k_{\eta ^{\prime }\eta }}}%
,\left\{ \left[ p_{l},p_{l^{\prime }l}\right] ^{k_{l^{\prime }}}\right\}
\right) \right] _{\eta }=\Psi _{J,\alpha ,s_{p_{\eta ^{\prime }\eta
}}^{\otimes k_{\eta ^{\prime }\eta }}}^{\otimes k_{\eta ^{\prime }\eta
}}\left( \overline{\left[ p_{\eta },p_{\eta ^{\prime }\eta }\right]
^{k_{\eta ^{\prime }\eta }}},\left\{ \left[ p_{l},p_{l^{\prime }l}\right]
^{k_{l^{\prime }}}\right\} \right)
\end{equation*}%
The sets $\overline{\left[ p_{\eta },p_{\eta ^{\prime }\eta }\right]
^{k_{\eta ^{\prime }\eta }}}$ are $\left[ p_{\eta },p_{\eta ^{\prime }\eta }%
\right] ^{k_{\eta ^{\prime }\eta }}$ quotiented by constraints imposed by $%
\left\{ \left[ p_{l},p_{l^{\prime }l}\right] ^{k_{l^{\prime }}}\right\} $.

We will also consider the product of remaining fields:%
\begin{equation*}
\tprod_{\eta }\Psi _{J,\alpha ,s_{p_{\eta ^{\prime }\eta }}^{\otimes k_{\eta
^{\prime }\eta }}}^{\otimes k_{\eta ^{\prime }\eta }}\left( \left[ p_{\eta
},p_{\eta ^{\prime }\eta }\right] ^{k_{\eta ^{\prime }\eta }}\right)
\end{equation*}%
and for their subobjects:%
\begin{equation*}
\tprod_{\eta }\Psi _{J,\alpha ,s_{p_{\eta ^{\prime }\eta }}^{\otimes k_{\eta
^{\prime }\eta }}}^{\otimes k_{\eta ^{\prime }\eta }}\left( \overline{\left[
p_{\eta },p_{\eta ^{\prime }\eta }\right] ^{k_{\eta ^{\prime }\eta }}}%
,\left\{ \left[ p_{l},p_{l^{\prime }l}\right] ^{k_{l^{\prime }}}\right\}
\right)
\end{equation*}

\subsection{Projected states}

The states are decomposed as:%
\begin{equation*}
1=\sum_{\left\{ v_{\eta }\right\} }\tprod\limits_{\left\{ v_{\eta }\right\}
}=\sum_{\left[ v\right] }\tprod\limits_{\left[ v\right] }
\end{equation*}%
which accounts for a\ decomposition relative to multiple states.

Including the projection on background states for some subobjcts yields the
projection:%
\begin{equation*}
\sum_{\left[ v\right] }\tprod\limits_{\left[ v\right] }\otimes \tprod_{\min
S\left( \left[ v\right] \right) }
\end{equation*}%
We start with the projection $\tprod\limits_{v}$. The basis of states is: 
\begin{equation*}
v\left( \left[ \Psi _{J,\alpha ,s_{p_{\eta ^{\prime }\eta }}^{\otimes
k_{\eta ^{\prime }\eta }}}^{\otimes k_{\eta ^{\prime }\eta }}\right] \right)
=\tprod_{\eta }v_{\eta }\left( \Psi _{J,\alpha ,s_{p_{\eta ^{\prime }\eta
}}^{\otimes k_{\eta ^{\prime }\eta }}}^{\otimes k_{\eta ^{\prime }\eta
}}\left( \left[ p_{\eta },p_{\eta ^{\prime }\eta }\right] ^{k_{\eta ^{\prime
}\eta }}\right) \right)
\end{equation*}%
and including the evaluation functional on other fields yields as before:%
\begin{eqnarray*}
&&\left[ v\left( \Psi _{J,\alpha ,s_{p_{\eta ^{\prime }\eta }}^{\otimes
k_{\eta ^{\prime }\eta }}}^{\otimes k_{\eta ^{\prime }\eta }}\right) \right]
\otimes ev_{\left[ p_{l},p_{l^{\prime }l}\right] ^{k_{l^{\prime }}}}\left(
\Psi _{J,s_{p_{l^{\prime },l}}^{\otimes k_{l^{\prime }}}}^{\otimes
k_{l^{\prime }}}\right) \delta \left( f_{\gamma }\left( \left[ \left[
p_{\eta },p_{\eta ^{\prime }\eta }\right] ^{k_{\eta ^{\prime }\eta }}\right]
,\left\{ \left[ p_{l},p_{l^{\prime }l}\right] ^{k_{l^{\prime }}}\right\}
\right) \right) \\
&=&\int \tprod_{\eta }v_{\eta }\left( \overline{\left[ p_{\eta },p_{\eta
^{\prime }\eta }\right] ^{k_{\eta ^{\prime }\eta }}},\left\{ \left[
p_{l},p_{l^{\prime }l}\right] ^{k_{l^{\prime }}}\right\} \right) \\
&&\times \left[ \Psi _{J,\alpha ,s_{p_{l^{\prime },\eta }}^{\otimes
k_{l^{\prime }\eta }}}^{\otimes k_{l^{\prime }\eta }}\left( \overline{\left[
p_{\eta },p_{\eta ^{\prime }\eta }\right] ^{k_{\eta ^{\prime }\eta }}}%
,\left\{ \left[ p_{l},p_{l^{\prime }l}\right] ^{k_{l^{\prime }}}\right\}
\right) \right] d\left( \overline{\left[ p_{\eta },p_{\eta ^{\prime }\eta }%
\right] ^{k_{\eta ^{\prime }\eta }}}\right) \otimes ev_{\left[
p_{l},p_{l^{\prime }l}\right] ^{k_{l^{\prime }}}}\left( \Psi
_{J,s_{p_{l^{\prime },l}}^{\otimes k_{l^{\prime }}}}^{\otimes k_{l^{\prime
}}}\right) \\
&=&\left[ v_{\left[ p_{l},p_{l^{\prime }l}\right] ^{k_{l^{\prime }}}}\left(
\Psi _{J,\alpha ,s_{p_{\eta ^{\prime }\eta }}^{\otimes k_{\eta ^{\prime
}\eta }}}^{\otimes k_{\eta ^{\prime }\eta }}\right) \right] \otimes ev_{%
\left[ p_{l},p_{l^{\prime }l}\right] ^{k_{l^{\prime }}}}\left( \Psi
_{J,s_{p_{l^{\prime },l}}^{\otimes k_{l^{\prime }}}}^{\otimes k_{l^{\prime
}}}\right)
\end{eqnarray*}%
with the implicit notation:%
\begin{equation*}
\left[ \overline{\left[ p_{\eta },p_{\eta ^{\prime }\eta }\right] ^{k_{\eta
^{\prime }\eta }}}\right] ,\left\{ \left[ p_{l},p_{l^{\prime }l}\right]
^{k_{l^{\prime }}}\right\} \equiv \left[ \overline{\left[ p_{\eta },p_{\eta
^{\prime }\eta }\right] ^{k_{\eta ^{\prime }\eta }}}\right]
/\tprod\limits_{p_{1,l},...,p_{m^{\prime },l}}f_{p_{1,l}...p_{m^{\prime
},l}},f_{p_{1,l}...p_{m^{\prime },l}}^{-1}\left\{ \left[ p_{l},p_{l^{\prime
}l}\right] ^{k_{l^{\prime }}}\right\}
\end{equation*}%
As a consequence, projection on the states $\left[ v\right] $, amounts to
include the following contributions in the functional : 
\begin{equation*}
\tprod\limits_{\left[ v\right] }\rightarrow \left[ v_{\left[
p_{l},p_{l^{\prime }l}\right] ^{k_{l^{\prime }}}}\left( \Psi _{J,\alpha
,s_{p_{\eta ^{\prime }\eta }}^{\otimes k_{\eta ^{\prime }\eta }}}^{\otimes
k_{\eta ^{\prime }\eta }}\right) \right] \otimes ev_{\left[
p_{l},p_{l^{\prime }l}\right] ^{k_{l^{\prime }}}}\left( \Psi
_{J,s_{p_{l^{\prime },l}}^{\otimes k_{l^{\prime }}}}^{\otimes k_{l^{\prime
}}}\right) \tprod\limits_{\eta }\delta \left( f_{\eta }\left( \overline{%
\left[ p_{\eta },p_{\eta ^{\prime }\eta }\right] ^{k_{\eta ^{\prime }\eta }}}%
,\left\{ \left[ p_{l},p_{l^{\prime }l}\right] ^{k_{l^{\prime }}}\right\}
\right) \right)
\end{equation*}%
where:%
\begin{eqnarray*}
&&\left[ v_{\left[ p_{l},p_{l^{\prime }l}\right] ^{k_{l^{\prime }}}}\left(
\Psi _{J,\alpha ,s_{p_{\eta ^{\prime }\eta }}^{\otimes k_{\eta ^{\prime
}\eta }}}^{\otimes k_{\eta ^{\prime }\eta }}\right) \right] \\
&=&\int \tprod_{\eta }v_{\eta }\left( \overline{\left[ p_{\eta },p_{\eta
^{\prime }\eta }\right] ^{k_{\eta ^{\prime }\eta }}},\left\{ \left[
p_{l},p_{l^{\prime }l}\right] ^{k_{l^{\prime }}}\right\} \right) \left[ \Psi
_{J,\alpha ,s_{p_{l^{\prime },0}}^{\otimes k_{l^{\prime }}}}^{\otimes
k_{l^{\prime }}}\left( \overline{\left[ p_{\eta },p_{\eta ^{\prime }\eta }%
\right] ^{k_{\eta ^{\prime }\eta }}},\left\{ \left[ p_{l},p_{l^{\prime }l}%
\right] ^{k_{l^{\prime }}}\right\} \right) \right] d\left(
\tprod\limits_{\eta }\overline{\left[ p_{\eta },p_{\eta ^{\prime }\eta }%
\right] ^{k_{\eta ^{\prime }\eta }}}\right)
\end{eqnarray*}%
Once the decomposition with respect to the states $\left[ v\right] $ is
performed, we consider the projection coming from:%
\begin{eqnarray*}
&&\exp \left( -S\left( \left\{ \otimes \left[ \Psi _{J,\alpha ,s_{p_{\eta
^{\prime }\eta }}^{\otimes k_{\eta ^{\prime }\eta }}}^{\otimes k_{\eta
^{\prime }\eta }}\left( \left[ p_{\eta },p_{\eta ^{\prime }\eta }\right]
^{k_{\eta ^{\prime }\eta }}\right) \right] \underset{l^{\prime }}{\otimes }%
\Psi _{J,s_{p_{l^{\prime },l}}^{\otimes k_{l^{\prime }}}}^{\otimes
k_{l^{\prime }}}\left( \left[ p_{l},p_{l^{\prime }l}\right] ^{k_{l^{\prime
}}}\right) \right\} ,\right. \right. \\
&&\left. \left. \left[ v_{\left[ p_{l},p_{l^{\prime }l}\right]
^{k_{l^{\prime }}}}\left( \Psi _{J,\alpha ,s_{p_{\eta ^{\prime }\eta
}}^{\otimes k_{\eta ^{\prime }\eta }}}^{\otimes k_{\eta ^{\prime }\eta
}}\right) \right] \left( \Psi _{J,s_{p_{l^{\prime },l}}^{\otimes
k_{l^{\prime }}}}^{\otimes k_{l^{\prime }}}\left( \left[ p_{l},p_{l^{\prime
}l}\right] ^{k_{l^{\prime }}}\right) \right) \right) \right)
\end{eqnarray*}

\subsection{Projected functionals and effective fields}

Computations similar to the previous sections leads to the projected
functionals (for one realization $\alpha $). Assume a collection $\left\{
\eta _{i}\right\} $ among the remaining subobjects. Choosing the valuation
for $v_{\eta _{i}}$:%
\begin{equation*}
v_{\eta _{i}}\rightarrow \left[ p_{\eta _{i}},p_{\eta _{i^{\prime }}^{\prime
}\eta _{i}}\right]
\end{equation*}%
the projected functional generlzing (\ref{Fl}) is:%
\begin{eqnarray}
&&\sum_{_{\left\{ \left( s,s^{\prime },\left[ p_{\eta },p_{\eta ^{\prime
}\eta }\right] _{i,i^{\prime }}\right) \right\} }}\int \bar{g}^{\mathcal{K}%
}\left( \left[ \left\{ \left[ p_{\eta _{i}},p_{\eta _{i^{\prime }}^{\prime
}\eta _{i}}\right] ^{k_{\eta _{i^{\prime }}^{\prime }\eta _{i}}}\right\} 
_{\substack{ i\leqslant s  \\ i^{\prime }\leqslant s^{\prime }}}\right]
,\left\{ \left[ 
\begin{array}{c}
\left( p_{l_{i}},p_{l_{i}^{\prime }l_{i}},k_{l_{i}^{\prime }}\right) \\ 
\left[ \left\{ \Psi _{J}\left[ p_{\eta },p_{\eta ^{\prime }\eta }\right]
\right\} ,v\right]%
\end{array}%
\right] \right\} _{i},\left[ v\right] \right)  \label{Fls} \\
&&\times \left( \tprod_{_{_{_{\substack{ i\leqslant s  \\ i^{\prime
}\leqslant s^{\prime }}}}}}\left[ \Psi _{J,\alpha _{i},s_{p_{\eta
_{i^{\prime }}^{\prime }\eta _{i}}}^{\otimes k_{\eta _{i^{\prime }}^{\prime
}\eta _{i}}}}^{\otimes k_{\eta _{i^{\prime }}^{\prime }\eta _{i}}}\left( %
\left[ p_{\eta _{i}},p_{\eta _{i^{\prime }}^{\prime }\eta _{i}}\right]
^{k_{\eta _{i^{\prime }}^{\prime }\eta _{i}}}\right) \right] \right) \Psi
\left( \left\{ \left[ 
\begin{array}{c}
\left( p_{l_{i}},p_{l_{i}^{\prime }l_{i}},k_{l_{i}^{\prime }}\right) \\ 
\left[ \left\{ \Psi _{J}\left[ p_{\eta },p_{\eta ^{\prime }\eta }\right]
\right\} ,v\right]%
\end{array}%
\right] \right\} _{i},\left[ v\right] \right)  \notag \\
&&\times d\left( \left\{ \left[ 
\begin{array}{c}
\left( p_{l_{i}},p_{l_{i},^{\prime }l_{i}},k_{l_{i}^{\prime }}\right) \\ 
\left[ \left\{ \Psi _{J}\left[ l^{\prime },0\right] \right\} ,v\right]%
\end{array}%
\right] \right\} _{i}\right) d\left( \left\{ \left[ p_{\eta _{i}},p_{\eta
_{i^{\prime }}^{\prime }\eta _{i}}\right] ^{k_{\eta _{i^{\prime }}^{\prime
}\eta _{i}}}\right\} _{\substack{ l\leqslant s  \\ l^{\prime }\leqslant
s^{\prime }}}\right)  \notag
\end{eqnarray}
with:%
\begin{equation*}
\left( \left[ p,p^{\prime },k^{\prime }\right] \right) _{i}=\left\{ p_{\eta
_{i}},p_{\eta _{i}^{\prime }\eta _{i}},k_{\eta _{i}^{\prime }\eta
_{i}}\right\} _{_{\substack{ i\leqslant s  \\ i\leqslant s^{\prime }}}}
\end{equation*}%
and:%
\begin{equation*}
\left( m,m^{\prime },\left[ p,p^{\prime },k^{\prime }\right] \right)
=\left\{ p_{l},p_{l^{\prime }l},k_{l^{\prime }}\right\} _{_{_{\substack{ %
l\leqslant m  \\ l^{\prime }\leqslant m^{\prime }}}}}
\end{equation*}%
and:%
\begin{equation*}
\left[ v\right] =\left\{ v_{\eta _{i}}\right\}
\end{equation*}

Constraints between components coming from $\overline{\left[ \left[ p_{\eta
_{i}},p_{\eta _{i^{\prime }}^{\prime }\eta _{i}}\right] ^{k_{\eta
_{i^{\prime }}^{\prime }\eta _{i}}}\right] },\left\{ \left[
p_{l},p_{l^{\prime }l}\right] ^{k_{l^{\prime }}}\right\} $ and the variables 
$\mathbf{\hat{\Lambda}}^{\left( p_{l}\right) }\left( \Psi _{J}^{\otimes
k_{p_{0}}},\left[ v\right] \right) $ are implicit. Considering general
states:%
\begin{equation*}
v\left( \left\{ \left[ p_{\eta _{i}},p_{\eta _{i^{\prime }}^{\prime }\eta
_{i}}\right] ^{k_{\eta _{i^{\prime }}^{\prime }\eta _{i}}}\right\} _{ 
_{\substack{ i\leqslant s  \\ i^{\prime }\leqslant s^{\prime }}}}\right)
=\tprod\limits_{_{\substack{ i\leqslant s  \\ i^{\prime }\leqslant s^{\prime
} }}}v_{\eta _{i}}\left( \left[ p_{\eta _{i}},p_{\eta _{i^{\prime }}^{\prime
}\eta _{i}}\right] ^{k_{\eta _{i^{\prime }}^{\prime }\eta _{i}}}\right)
\end{equation*}%
the effective fields have the form:%
\begin{eqnarray}
&&\int d\left\{ \left[ p_{\eta _{i}},p_{\eta _{i^{\prime }}^{\prime }\eta
_{i}}\right] ^{k_{\eta _{i^{\prime }}^{\prime }\eta _{i}}}\right\} _{ 
_{\substack{ i\leqslant s  \\ i^{\prime }\leqslant s^{\prime }}}}v\left(
\left\{ \left[ p_{\eta _{i}},p_{\eta _{i^{\prime }}^{\prime }\eta _{i}}%
\right] ^{k_{\eta _{i^{\prime }}^{\prime }\eta _{i}}}\right\} _{_{\substack{ %
i\leqslant s  \\ i^{\prime }\leqslant s^{\prime }}}}\right)  \label{F} \\
&&\times \Psi _{J,\alpha _{i},\tprod\limits_{i^{\prime }}s_{p_{\eta
_{i^{\prime }}^{\prime }\eta _{i}}}^{\otimes k_{\eta _{i^{\prime }}^{\prime
}\eta _{i}}}}^{\otimes \sum_{i^{\prime }}k_{\eta _{i^{\prime }}^{\prime
}\eta _{i}}}\left( \left\{ \left[ p_{\eta _{i}},p_{\eta _{i^{\prime
}}^{\prime }\eta _{i}}\right] ^{k_{\eta _{i^{\prime }}^{\prime }\eta
_{i}}}\right\} _{_{_{_{\substack{ i\leqslant s  \\ i^{\prime }\leqslant
s^{\prime }}}}}},\left\{ \left[ 
\begin{array}{c}
\left( p_{l_{i}},p_{l_{i}^{\prime }l_{i}},k_{l_{i}^{\prime }}\right) \\ 
\left[ \left\{ \Psi _{J,\alpha _{i}}\left[ p_{\eta },p_{\eta ^{\prime }\eta }%
\right] \right\} ,v\right]%
\end{array}%
\right] \right\} _{i},\left[ v\right] \right)  \notag
\end{eqnarray}%
We can define a global field by summing over realizations if the variables $%
\left[ 
\begin{array}{c}
\left( p_{l_{i}},p_{l_{i}^{\prime }l_{i}},k_{l_{i}^{\prime }}\right) \\ 
\left[ \left\{ \Psi _{J,\alpha _{i}}\left[ p_{\eta },p_{\eta ^{\prime }\eta }%
\right] \right\} ,v\right]%
\end{array}%
\right] $ are independent from realizations. In this case, we find a single
composed field: \ 
\begin{eqnarray*}
&&\Psi _{J,\tprod\limits_{i^{\prime }}s_{p_{\eta _{i^{\prime }}^{\prime
}\eta _{i}}}^{\otimes k_{\eta _{i^{\prime }}^{\prime }\eta _{i}}}}^{\otimes
\sum_{i^{\prime }}k_{\eta _{i^{\prime }}^{\prime }\eta _{i}}}\left( \left\{ 
\left[ p_{\eta _{i}},p_{\eta _{i^{\prime }}^{\prime }\eta _{i}}\right]
^{k_{\eta _{i^{\prime }}^{\prime }\eta _{i}}}\right\} _{_{_{_{\substack{ %
i\leqslant s  \\ i^{\prime }\leqslant s^{\prime }}}}}},\left\{ \left[ 
\begin{array}{c}
\left( p_{l_{i}},p_{l_{i}^{\prime }l_{1}},k_{l_{1}^{\prime }}\right) \\ 
\left[ \left\{ \Psi _{J}\left[ p_{\eta },p_{\eta ^{\prime }\eta }\right]
\right\} ,v\right]%
\end{array}%
\right] \right\} _{i},\left[ v\right] \right) \\
&=&\sum_{\alpha _{i}}\left( \tprod_{i,i^{\prime }}\left[ \Psi _{J,\alpha
_{i},s_{p_{\eta _{i^{\prime }}^{\prime }\eta _{i}}}^{\otimes k_{\eta
_{i^{\prime }}^{\prime }\eta _{i}}}}^{\otimes k_{\eta _{i^{\prime }}^{\prime
}\eta _{i}}}\left( \left[ p_{\eta _{i}},p_{\eta _{i^{\prime }}^{\prime }\eta
_{i}}\right] ^{k_{\eta _{i^{\prime }}^{\prime }\eta _{i}}}\right) \right]
\right) \Psi \left( \left\{ \left[ 
\begin{array}{c}
\left( p_{l_{i}},p_{l_{i}^{\prime }l_{i}},k_{l_{i}^{\prime }}\right) \\ 
\left[ \left\{ \Psi _{J,\alpha _{i}}\left[ p_{\eta },p_{\eta ^{\prime }\eta }%
\right] \right\} ,v\right]%
\end{array}%
\right] \right\} _{i},\left[ v\right] ,\alpha _{i}\right)
\end{eqnarray*}%
The projection has led to a single composed field. However, when the
transformation group are, at least in first approximation, independent and
can be written $\mathbf{L}_{\eta }\left( \Psi _{J,\eta }^{\otimes
k_{p_{0}}}\right) $ with a formula similar to (\ref{GR}), we can assume that
the parameters $\left[ 
\begin{array}{c}
\left( p_{l_{i}},p_{l_{i}^{\prime }l_{i}},k_{l_{i}^{\prime }}\right) \\ 
\left[ \left\{ \Psi _{J,\alpha _{i}}\left[ p_{\eta },p_{\eta ^{\prime }\eta }%
\right] \right\} ,v\right]%
\end{array}%
\right] $\ are independent, and that, the effective field becomes:%
\begin{eqnarray}
&&\tprod_{i}\sum_{\alpha _{i}}\left( \left[ \Psi _{J,\alpha _{i},s_{p_{\eta
_{i^{\prime }}^{\prime }\eta _{i}}}^{\otimes k_{\eta _{i^{\prime }}^{\prime
}\eta _{i}}}}^{\otimes k_{\eta _{i^{\prime }}^{\prime }\eta _{i}}}\left( %
\left[ p_{\eta _{i}},p_{\eta _{i^{\prime }}^{\prime }\eta _{i}}\right]
^{k_{\eta _{i^{\prime }}^{\prime }\eta _{i}}}\right) \right] \right) \Psi
_{\alpha _{i}}\left( \left\{ \left[ 
\begin{array}{c}
\left( p_{l_{i}},p_{l_{i}^{\prime }l_{i}},k_{l_{i}^{\prime }}\right) \\ 
\left[ \left\{ \Psi _{J,\alpha }\left[ p_{\eta },p_{\eta ^{\prime }\eta }%
\right] \right\} ,v\right]%
\end{array}%
\right] \right\} ,\left[ v\right] \right)  \label{FS} \\
&=&\tprod_{i}\Psi _{J,\alpha _{i},s_{p_{\eta _{i^{\prime }}^{\prime }\eta
_{i}}}^{\otimes k_{\eta _{i^{\prime }}^{\prime }\eta _{i}}}}^{\otimes
k_{\eta _{i^{\prime }}^{\prime }\eta _{i}}}\left( \left[ p_{\eta
_{i}},p_{\eta _{i^{\prime }}^{\prime }\eta _{i}}\right] ^{k_{\eta
_{i^{\prime }}^{\prime }\eta _{i}}},\left\{ \left[ 
\begin{array}{c}
\left( p_{l_{i}},p_{l_{i}^{\prime }l_{i}},k_{l_{i}^{\prime }}\right) \\ 
\left[ \left\{ \Psi _{J,\alpha }\left[ p_{\eta },p_{\eta ^{\prime }\eta }%
\right] \right\} ,v\right]%
\end{array}%
\right] \right\} ,v_{\eta }\right)  \notag
\end{eqnarray}%
and the system is described by several field $\Psi _{J,s_{p_{l^{\prime
},\eta }}^{\otimes k_{l^{\prime }\eta }}}^{\otimes k_{l^{\prime }\eta }}$.

In ths case the functionl (\ref{Fls}) wrts:%
\begin{eqnarray}
&&\sum_{_{\left\{ \left( s,s^{\prime },\left[ p_{\eta },p_{\eta ^{\prime
}\eta }\right] _{i,i^{\prime }}\right) \right\} }}\int \bar{g}^{\mathcal{K}%
}\left( \left[ \left\{ \left[ p_{\eta _{i}},p_{\eta _{i^{\prime }}^{\prime
}\eta _{i}}\right] ^{k_{\eta _{i^{\prime }}^{\prime }\eta _{i}}}\right\} 
_{\substack{ i\leqslant s  \\ i^{\prime }\leqslant s^{\prime }}}\right]
,\left\{ \left[ 
\begin{array}{c}
\left( p_{l_{i}},p_{l_{i}^{\prime }l_{i}},k_{l_{i}^{\prime }}\right) \\ 
\left[ \left\{ \Psi _{J}\left[ p_{\eta },p_{\eta ^{\prime }\eta }\right]
\right\} ,v\right]%
\end{array}%
\right] \right\} _{i},\left[ v\right] \right)  \label{Fs} \\
&&\times \Psi _{J,\tprod\limits_{i^{\prime }}s_{p_{\eta _{i^{\prime
}}^{\prime }\eta _{i}}}^{\otimes k_{\eta _{i^{\prime }}^{\prime }\eta
_{i}}}}^{\otimes \sum_{i^{\prime }}k_{\eta _{i^{\prime }}^{\prime }\eta
_{i}}}\left( \left\{ \left[ p_{\eta _{i}},p_{\eta _{i^{\prime }}^{\prime
}\eta _{i}}\right] ^{k_{\eta _{i^{\prime }}^{\prime }\eta _{i}}}\right\}
_{_{_{_{\substack{ i\leqslant s  \\ i^{\prime }\leqslant s^{\prime }}}%
}}},\left\{ \left[ 
\begin{array}{c}
\left( p_{l_{i}},p_{l_{i}^{\prime }l_{1}},k_{l_{1}^{\prime }}\right) \\ 
\left[ \left\{ \Psi _{J}\left[ p_{\eta },p_{\eta ^{\prime }\eta }\right]
\right\} ,v\right]%
\end{array}%
\right] \right\} _{i},\left[ v\right] \right)  \notag \\
&&\times d\left( \left\{ \left[ 
\begin{array}{c}
\left( p_{l_{i}},p_{l_{i},^{\prime }l_{i}},k_{l_{i}^{\prime }}\right) \\ 
\left[ \left\{ \Psi _{J}\left[ l^{\prime },0\right] \right\} ,v\right]%
\end{array}%
\right] \right\} _{i}\right) d\left( \left\{ \left[ p_{\eta _{i}},p_{\eta
_{i^{\prime }}^{\prime }\eta _{i}}\right] ^{k_{\eta _{i^{\prime }}^{\prime
}\eta _{i}}}\right\} _{\substack{ l\leqslant s  \\ l^{\prime }\leqslant
s^{\prime }}}\right)  \notag
\end{eqnarray}

In general the decomposition (\ref{FS}) is not ensured. We may assume that
the constraints defining the generators $\mathbf{L}_{\eta }\left( \Psi
_{J,\eta }^{\otimes k_{p_{0}}}\right) $ and thus the parameters depend on
all the $\Psi _{J,\eta }^{\otimes k_{p_{0}}}$. Expanding the series in
constraints and assuming the independency in first approximation, we can
assume the constraints to have the frm:%
\begin{eqnarray*}
0 &=&\sum_{\eta }h_{k_{n}}^{\left( \eta \right) }\left( \left\{ \mathbf{L}%
_{\eta ,\left\{ \alpha _{i}\right\} }\left( \Psi _{J,\eta }^{\otimes
k_{p_{0}}}\right) \right\} _{i\leqslant n},\left\{ U_{i}^{k}\right\}
_{i},h_{p}\left( \left( \Psi _{J,\eta }^{\otimes k_{p_{0}}}\right)
,U_{j,\eta }^{l},\nu _{\eta }\right) \right) \\
&&+\sum_{\left[ \eta \right] }h_{k_{n}}^{\left[ \eta \right] }\left( \left\{
\left\{ \mathbf{L}_{\eta ,\left\{ \alpha _{i}\right\} }\left( \Psi _{J,\eta
}^{\otimes k_{p_{0}}}\right) \right\} _{i\leqslant n}\right\} _{\eta
},\left\{ U_{i}^{k}\right\} _{i},\left\{ h_{p}\left( \left( \Psi _{J,\eta
}^{\otimes k_{p_{0}}}\right) ,U_{j,\eta }^{l},\nu _{\eta }\right) \right\}
_{\eta }\right)
\end{eqnarray*}%
where the sum is over any arbitrary set $\left[ \eta \right] $ of nn prjctd
stts. If the $h_{k_{n}}^{\left( \eta \right) }$ are positive, we may expect
that the parameters are defined such that:%
\begin{equation*}
h_{k_{n}}^{\left( \eta \right) }\left( \left\{ \mathbf{L}_{\eta ,\left\{
\alpha _{i}\right\} }\left( \Psi _{J,\eta }^{\otimes k_{p_{0}}}\right)
\right\} _{i\leqslant n},\left\{ U_{i}^{k}\right\} _{i},h_{p}\left( \left(
\Psi _{J,\eta }^{\otimes k_{p_{0}}}\right) ,U_{j,\eta }^{l},\nu _{\eta
}\right) \right) =0
\end{equation*}%
so that we recover flds $\Psi _{J,s_{p_{l^{\prime },\eta }}^{\otimes
k_{l^{\prime }\eta }}}^{\otimes k_{l^{\prime }\eta }}\left( \left[ p_{\eta
},p_{\eta ^{\prime }\eta }\right] ^{k_{\eta ^{\prime }\eta }},\left\{ \left[ 
\begin{array}{c}
\left( p_{l_{i}},p_{l_{i}^{\prime }l_{i}},k_{l_{i}^{\prime }}\right) \\ 
\left[ \left\{ \Psi _{J,\alpha }\left[ p_{\eta },p_{\eta ^{\prime }\eta }%
\right] \right\} ,v\right]%
\end{array}%
\right] \right\} ,v_{\eta }\right) $. The remaining relations: 
\begin{equation*}
h_{k_{n}}^{\left[ \eta \right] }\left( \left\{ \left\{ \mathbf{L}_{\eta
,\left\{ \alpha _{i}\right\} }\left( \Psi _{J,\eta }^{\otimes
k_{p_{0}}}\right) \right\} _{i\leqslant n}\right\} _{\eta },\left\{
U_{i}^{k}\right\} _{i},\left\{ h_{p}\left( \left( \Psi _{J,\eta }^{\otimes
k_{p_{0}}}\right) ,U_{j,\eta }^{l},\nu _{\eta }\right) \right\} _{\eta
}\right) =0
\end{equation*}%
are constraints between some parametrs. Changing of variables leads to
reinterpret this constraints as $\delta $ functions in the functionls that
becomes series involvng products:%
\begin{equation*}
\tprod_{i}\Psi _{J,s_{p_{\eta _{i^{\prime }}^{\prime }\eta _{i}}}^{\otimes
k_{\eta _{i^{\prime }}^{\prime }\eta _{i}}}}^{\otimes k_{\eta _{i^{\prime
}}^{\prime }\eta _{i}}}\left( \left[ p_{\eta _{i}},p_{\eta _{i^{\prime
}}^{\prime }\eta _{i}}\right] ^{k_{\eta _{i^{\prime }}^{\prime }\eta
_{i}}},\left\{ \left[ 
\begin{array}{c}
\left( p_{l_{i}},p_{l_{i}^{\prime }l_{i}},k_{l_{i}^{\prime }}\right) \\ 
\left[ \left\{ \Psi _{J,\alpha }\left[ p_{\eta },p_{\eta ^{\prime }\eta }%
\right] \right\} ,v\right]%
\end{array}%
\right] \right\} ,v_{\eta _{i}}\right) \tprod\limits_{\eta }\delta \left(
f\left( \left\{ \left[ 
\begin{array}{c}
\left( p_{l_{i}},p_{l_{i}^{\prime }l_{i}},k_{l_{i}^{\prime }}\right) \\ 
\left[ \left\{ \Psi _{J,\alpha }\left[ p_{\eta },p_{\eta ^{\prime }\eta }%
\right] \right\} ,v\right]%
\end{array}%
\right] \right\} _{\eta }\right) \right)
\end{equation*}%
Ths constraints imply that some localty in this functionl has to be
introduced.

\subsection{General sequencs of subobjcts}

Until now we have considered sequences of two mps of subobjcts; $%
p_{l_{i}},p_{l_{i}^{\prime }l_{i}}$. Globlly, we should consider sequences:%
\begin{equation*}
\left[ p_{l_{i}},p_{l_{i}^{\prime }l_{i}},...p_{...l_{i}^{\prime \prime
}l_{i}^{\prime }l_{i}},..\right] 
\end{equation*}%
that is:%
\begin{equation}
\left[ p_{\left( \mathbf{l}_{i}\right) _{1}},p_{\left( \mathbf{l}_{i}\right)
_{2}},...p_{\left( \mathbf{l}_{i}\right) _{n}},...\right]   \label{MS}
\end{equation}%
where the $\left( \mathbf{l}_{i}\right) _{n}$ are the $n$ first terms of any
infinit seqnc. We show in appendix 7 that we can wrt (\ref{MS}) s:%
\begin{equation*}
\left[ \mathbf{p}_{\left( \mathbf{l}_{i}\right) _{\infty }}\right] 
\end{equation*}%
representing an infinite sequence of subjcts. This sequenc can b truncted at
order $k$ $\left[ \mathbf{p}_{\left( \mathbf{l}_{i}\right) _{\infty }<k}%
\right] $. In appendix 7 we show tht it leads to define fields:%
\begin{equation}
\tprod_{i}\Psi _{J,,s_{\left[ \mathbf{p}_{\left( \eta _{i},\mathbf{l}%
_{i}\right) _{<k}}\right] }^{\otimes \mathbf{k}_{\mathbf{p}_{\left( \eta
_{i},\mathbf{l}_{i}\right) _{<k}}}}}^{\otimes \mathbf{k}_{\mathbf{p}_{\left(
\eta _{i},\mathbf{l}_{i}\right) _{<k}}}}\left( \left[ \mathbf{p}_{\left(
\eta _{i},\mathbf{l}_{i}\right) _{<k}}\right] ^{\mathbf{k}_{\mathbf{p}%
_{\left( \eta _{i},\mathbf{l}_{i}\right) _{<k}}}},\left\{ \left[ 
\begin{array}{c}
\left( \left[ \mathbf{p}_{\left( \eta _{i},\mathbf{l}_{i}\right) _{<k}}%
\right] ,\mathbf{k}_{\mathbf{p}_{\left( \eta _{i},\mathbf{l}_{i}\right)
_{<k}}}\right)  \\ 
\left[ \left\{ \Psi _{J,\alpha }\left[ \mathbf{p}_{\left( \eta _{i},\mathbf{l%
}_{i}\right) _{<k}}\right] \right\} ,v\right] 
\end{array}%
\right] \right\} _{i},v_{\eta }\right) 
\end{equation}%
with functional:%
\begin{eqnarray*}
&&\sum_{_{\left\{ \left( s,s^{\prime },\left[ \mathbf{p}_{\left( \eta _{i},%
\mathbf{l}_{i}\right) _{<k}}\right] \right) \right\} }}\int \bar{g}^{%
\mathcal{K}}\left( \left[ \left\{ \left[ \mathbf{p}_{\left( \eta _{i},%
\mathbf{l}_{i}\right) _{<k}}\right] ^{\mathbf{k}_{\mathbf{p}_{\left( \eta
_{i},\mathbf{l}_{i}\right) _{<k}}}}\right\} _{\substack{ i\leqslant s \\ %
i^{\prime }\leqslant s^{\prime }}}\right] ,\left\{ \left[ 
\begin{array}{c}
\left( \left[ \mathbf{p}_{\left( \eta _{i},\mathbf{l}_{i}\right) _{<k}}%
\right] ,\mathbf{k}_{\mathbf{p}_{\left( \eta _{i},\mathbf{l}_{i}\right)
_{<k}}}\right)  \\ 
\left[ \left\{ \Psi _{J,\alpha }\left[ \mathbf{p}_{\left( \eta _{i},\mathbf{l%
}_{i}\right) _{<k}}\right] \right\} ,v\right] 
\end{array}%
\right] \right\} _{i},\left[ v\right] \right)  \\
&&\times \Psi _{J,\tprod\limits_{i^{\prime }}s_{p_{\eta _{i^{\prime
}}^{\prime }\eta _{i}}}^{\otimes k_{\eta _{i^{\prime }}^{\prime }\eta
_{i}}}}^{\otimes \sum_{i^{\prime }}k_{\eta _{i^{\prime }}^{\prime }\eta
_{i}}}\left( \left[ \left\{ \left[ \mathbf{p}_{\left( \eta _{i},\mathbf{l}%
_{i}\right) _{<k}}\right] ^{\mathbf{k}_{\mathbf{p}_{\left( \eta _{i},\mathbf{%
l}_{i}\right) _{<k}}}}\right\} _{\substack{ i\leqslant s \\ i^{\prime
}\leqslant s^{\prime }}}\right] ,\left\{ \left[ 
\begin{array}{c}
\left( \left[ \mathbf{p}_{\left( \eta _{i},\mathbf{l}_{i}\right) _{<k}}%
\right] ,\mathbf{k}_{\mathbf{p}_{\left( \eta _{i},\mathbf{l}_{i}\right)
_{<k}}}\right)  \\ 
\left[ \left\{ \Psi _{J,\alpha }\left[ \mathbf{p}_{\left( \eta _{i},\mathbf{l%
}_{i}\right) _{<k}}\right] \right\} ,v\right] 
\end{array}%
\right] \right\} _{i},\left[ v\right] \right)  \\
&&\times d\left( \left\{ \left[ 
\begin{array}{c}
\left( \left[ \mathbf{p}_{\left( \eta _{i},\mathbf{l}_{i}\right) _{<k}}%
\right] ,\mathbf{k}_{\mathbf{p}_{\left( \eta _{i},\mathbf{l}_{i}\right)
_{<k}}}\right)  \\ 
\left[ \left\{ \Psi _{J,\alpha }\left[ \mathbf{p}_{\left( \eta _{i},\mathbf{l%
}_{i}\right) _{<k}}\right] \right\} ,v\right] 
\end{array}%
\right] \right\} _{ii}\right) d\left( \left\{ \left[ p_{\eta _{i}},p_{\eta
_{i^{\prime }}^{\prime }\eta _{i}}\right] ^{k_{\eta _{i^{\prime }}^{\prime
}\eta _{i}}}\right\} _{\substack{ l\leqslant s \\ l^{\prime }\leqslant
s^{\prime }}}\right) 
\end{eqnarray*}

We also define in appendix 7, rank $l$ subobject and their functional.

\subsection{Variation and averaged field}

Writing the variation of the effective fld:%
\begin{equation*}
0=\int v\left( \left\{ \left[ p_{\eta _{i}},p_{\eta _{i^{\prime }}^{\prime
}\eta _{i}}\right] ^{k_{\eta _{i^{\prime }}^{\prime }\eta _{i}}}\right\} _{ 
_{\substack{ i\leqslant s  \\ i^{\prime }\leqslant s^{\prime }}}}\right) 
\frac{\delta \Psi _{J,\tprod\limits_{i^{\prime }}s_{p_{\eta _{i^{\prime
}}^{\prime }\eta _{i}}}^{\otimes k_{\eta _{i^{\prime }}^{\prime }\eta
_{i}}}}^{\otimes \sum_{i^{\prime }}k_{\eta _{i^{\prime }}^{\prime }\eta
_{i}}}\left( \left\{ \left[ p_{\eta _{i}},p_{\eta _{i^{\prime }}^{\prime
}\eta _{i}}\right] ^{k_{\eta _{i^{\prime }}^{\prime }\eta _{i}}}\right\} _{ 
_{\substack{ i\leqslant s  \\ i^{\prime }\leqslant s^{\prime }}}},\left\{ %
\left[ 
\begin{array}{c}
\left( p_{l_{i}},p_{l_{i}^{\prime }l_{i}},k_{l_{i}^{\prime }}\right) \\ 
\left[ \left\{ \Psi _{J}\left[ p_{\eta },p_{\eta ^{\prime }\eta }\right]
\right\} ,v\right]%
\end{array}%
\right] \right\} ,\left[ v\right] \right) }{\delta \Psi _{J,s_{p_{\eta
_{^{\prime }}^{\prime }\eta }}^{\otimes k\eta _{^{\prime }}^{\prime }\eta
}}^{\otimes \sum_{i^{\prime }}k_{\eta _{^{\prime }}^{\prime }\eta }}\left( %
\left[ \bar{p}_{\eta },\bar{p}_{\eta _{^{\prime }}^{\prime }\eta }\right]
^{k_{\eta _{^{\prime }}^{\prime }\eta }}\right) }
\end{equation*}%
Imposing that the variables $\left\{ \left[ 
\begin{array}{c}
\left( p_{l_{i}},p_{l_{i}^{\prime }l_{i}},k_{l_{i}^{\prime }}\right) \\ 
\left[ \left\{ \Psi _{J,\alpha }\left[ p_{\eta },p_{\eta ^{\prime }\eta }%
\right] \right\} ,v\right]%
\end{array}%
\right] \right\} $ are independent at the first order from the field
variations, we are led to the field equation: 
\begin{eqnarray}
&&0=\int v\left( \left\{ \left[ p_{\eta _{i}},p_{\eta _{i^{\prime }}^{\prime
}\eta _{i}}\right] ^{k_{\eta _{i^{\prime }}^{\prime }\eta _{i}}}\right\} _{ 
_{\substack{ i\leqslant s  \\ i^{\prime }\leqslant s^{\prime }}}}\right) 
\frac{\delta \left\{ \left[ 
\begin{array}{c}
\left( p_{l_{i}},p_{l_{i}^{\prime }l_{i}},k_{l_{i}^{\prime }}\right) \\ 
\left[ \left\{ \Psi _{J}\left[ p_{\eta },p_{\eta ^{\prime }\eta }\right]
\right\} ,v\right]%
\end{array}%
\right] \right\} }{\delta \Psi _{J,s_{p_{\eta _{^{\prime }}^{\prime }\eta
}}^{\otimes k\eta _{^{\prime }}^{\prime }\eta }}^{\otimes \sum_{i^{\prime
}}k_{\eta _{^{\prime }}^{\prime }\eta }}\left( \left[ \bar{p}_{\eta },\bar{p}%
_{\eta _{^{\prime }}^{\prime }\eta }\right] ^{k_{\eta _{^{\prime }}^{\prime
}\eta }}\right) }  \label{Fn} \\
&&\times \nabla _{\left\{ \left[ 
\begin{array}{c}
\left( p_{l_{i}},p_{l_{i}^{\prime }l_{i}},k_{l_{i}^{\prime }}\right) \\ 
\left[ \left\{ \Psi _{J,\alpha }\left[ p_{\eta },p_{\eta ^{\prime }\eta }%
\right] \right\} ,v\right]%
\end{array}%
\right] \right\} }\Psi _{J,\tprod\limits_{i^{\prime }}s_{p_{\eta _{i^{\prime
}}^{\prime }\eta _{i}}}^{\otimes k_{\eta _{i^{\prime }}^{\prime }\eta
_{i}}}}^{\otimes \sum_{i^{\prime }}k_{\eta _{i^{\prime }}^{\prime }\eta
_{i}}}\left( \left\{ \left[ p_{\eta _{i}},p_{\eta _{i^{\prime }}^{\prime
}\eta _{i}}\right] ^{k_{\eta _{i^{\prime }}^{\prime }\eta _{i}}}\right\} _{ 
_{\substack{ i\leqslant s  \\ i^{\prime }\leqslant s^{\prime }}}},\left\{ %
\left[ 
\begin{array}{c}
\left( p_{l_{i}},p_{l_{i}^{\prime }l_{i}},k_{l_{i}^{\prime }}\right) \\ 
\left[ \left\{ \Psi _{J}\left[ p_{\eta },p_{\eta ^{\prime }\eta }\right]
\right\} ,v\right]%
\end{array}%
\right] \right\} ,v\right)  \notag
\end{eqnarray}%
which becomes in local coordinates $\left[ \left( p_{l_{i}},p_{l_{i}^{\prime
}l_{i}},k_{l_{i}^{\prime }}\right) ,\eta \right] $:%
\begin{eqnarray}
&&0=\int \left[ \Xi ^{\left( k_{i}\right) }\left( \left\{ \left\{ \Psi _{J}%
\left[ p_{\eta },p_{\eta ^{\prime }\eta }\right] \right\} ,\left[ \left(
p_{l_{i}},p_{l_{i}^{\prime }l_{i}},k_{l_{i}^{\prime }}\right) ,\eta \right]
\right\} \right) \right] _{\left[ \bar{p}_{\eta },\bar{p}_{\eta _{^{\prime
}}^{\prime }\eta }\right] ^{k_{\eta _{^{\prime }}^{\prime }\eta }}}^{v}
\label{Fc} \\
&&\times \nabla _{\left\{ \left[ \left( p_{l_{i}},p_{l_{i}^{\prime
}l_{i}},k_{l_{i}^{\prime }}\right) ,\eta \right] \right\} }\Psi
_{J,\tprod\limits_{i^{\prime }}s_{p_{\eta _{i^{\prime }}^{\prime }\eta
_{i}}}^{\otimes k_{\eta _{i^{\prime }}^{\prime }\eta _{i}}}}^{\otimes
\sum_{i^{\prime }}k_{\eta _{i^{\prime }}^{\prime }\eta _{i}}}\left( \left\{ %
\left[ p_{\eta _{i}},p_{\eta _{i^{\prime }}^{\prime }\eta _{i}}\right]
^{k_{\eta _{i^{\prime }}^{\prime }\eta _{i}}}\right\} _{_{\substack{ %
i\leqslant s  \\ i^{\prime }\leqslant s^{\prime }}}},\left\{ \left[ \left(
p_{l_{i}},p_{l_{i}^{\prime }l_{i}},k_{l_{i}^{\prime }}\right) ,\eta \right]
\right\} ,v\right)  \notag
\end{eqnarray}%
where:%
\begin{eqnarray*}
&&\left[ \Xi ^{\left( k_{i}\right) }\left( \left\{ \left\{ \Psi _{J}\left[
l^{\prime },0\right] \right\} ,\left[ \left( p_{l_{i}},p_{l_{i}^{\prime
}l_{i}},k_{l_{i}^{\prime }}\right) ,\eta \right] \right\} \right) \right] _{%
\left[ \bar{p}_{\eta },\bar{p}_{\eta _{^{\prime }}^{\prime }\eta }\right]
^{k_{\eta _{^{\prime }}^{\prime }\eta }}}^{v} \\
&=&\int v\left( \left\{ \left[ p_{\eta },p_{l^{\prime }\eta }\right]
^{k_{l^{\prime }}}\right\} _{_{\substack{ l\leqslant s  \\ l^{\prime
}\leqslant s^{\prime }}}}\right) \frac{\delta \left\{ \left[ 
\begin{array}{c}
\left( p_{l_{i}},p_{l_{i}^{\prime }l_{i}},k_{l_{i}^{\prime }}\right) \\ 
\left[ \left\{ \Psi _{J,\alpha }\left[ p_{\eta },p_{\eta ^{\prime }\eta }%
\right] \right\} ,v\right]%
\end{array}%
\right] \right\} }{\delta \Psi _{J,s_{p_{l^{\prime },\eta }}^{\otimes
k_{l^{\prime }\eta }}}^{\otimes k_{l^{\prime }\eta }}\left( \left[ \bar{p}%
_{\eta },\bar{p}_{l^{\prime }\eta }\right] ^{k_{l_{\eta }^{\prime }}}\right) 
}\left( \frac{\delta \left[ 
\begin{array}{c}
\left( p_{l_{i}},p_{l_{i}^{\prime }l_{i}},k_{l_{i}^{\prime }}\right) \\ 
\left[ \left\{ \Psi _{J,\alpha }\left[ p_{\eta },p_{\eta ^{\prime }\eta }%
\right] \right\} ,v\right]%
\end{array}%
\right] }{\delta \left[ \left( p_{l_{i}},p_{l_{i}^{\prime
}l_{i}},k_{l_{i}^{\prime }}\right) ,\eta \right] }\right) ^{-1}
\end{eqnarray*}%
with identifications:%
\begin{equation*}
v\rightarrow \left[ p_{\eta },p_{\eta ^{\prime }\eta }\right] ^{k_{\eta
^{\prime }\eta }}
\end{equation*}%
and this equation mixes components due to the constraints.

As before, the derivatice:%
\begin{equation*}
\frac{\delta \left[ 
\begin{array}{c}
\left( p_{l_{i}},p_{l_{i}^{\prime }l_{i}},k_{l_{i}^{\prime }}\right) \\ 
\left[ \left\{ \Psi _{J,\alpha }\left[ p_{\eta },p_{\eta ^{\prime }\eta }%
\right] \right\} ,v\right]%
\end{array}%
\right] }{\delta \left[ \left( p_{l_{i}},p_{l_{i}^{\prime
}l_{i}},k_{l_{i}^{\prime }}\right) ,\eta \right] }
\end{equation*}%
measures the variation of the dependent coordinates $\left[ 
\begin{array}{c}
\left( p_{l_{i}},p_{l_{i}^{\prime }l_{i}},k_{l_{i}^{\prime }}\right) \\ 
\left[ \left\{ \Psi _{J,\alpha }\left[ p_{\eta },p_{\eta ^{\prime }\eta }%
\right] \right\} ,v\right]%
\end{array}%
\right] $ on some local independent reference coordntes.

\subsection{Averaged field variation}

As before considered successive inclusions of grps of transfrmations, the
effective fld is a series:%
\begin{eqnarray}
&&\Psi _{J,\tprod\limits_{i^{\prime }}s_{p_{\eta _{i^{\prime }}^{\prime
}\eta _{i}}}^{\otimes k_{\eta _{i^{\prime }}^{\prime }\eta _{i}}}}^{\otimes
\sum_{i^{\prime }}k_{\eta _{i^{\prime }}^{\prime }\eta _{i}}}\left( \left\{ 
\left[ p_{\eta _{i}},p_{\eta _{i^{\prime }}^{\prime }\eta _{i}}\right]
^{k_{\eta _{i^{\prime }}^{\prime }\eta _{i}}}\right\} _{_{\substack{ %
i\leqslant s  \\ i^{\prime }\leqslant s^{\prime }}}},\left[ 
\begin{array}{c}
\left( p_{l_{i}},p_{l_{i}^{\prime }l_{i}},k_{l_{i}^{\prime }}\right) \\ 
\left[ \left\{ \Psi _{J}\left[ p_{\eta },p_{\eta ^{\prime }\eta }\right]
\right\} ,v\right]%
\end{array}%
\right] ,V_{\left( p_{l_{i}},p_{l_{i}^{\prime }l_{i}},k_{l_{i}^{\prime
}}\right) },v\right) \\
&&+\Psi _{J,\tprod\limits_{i^{\prime }}s_{p_{\eta _{i^{\prime }}^{\prime
}\eta _{i}}}^{\otimes k_{\eta _{i^{\prime }}^{\prime }\eta _{i}}}}^{\otimes
\sum_{i^{\prime }}k_{\eta _{i^{\prime }}^{\prime }\eta _{i}}}\left( \left\{ 
\left[ p_{\eta _{i}},p_{\eta _{i^{\prime }}^{\prime }\eta _{i}}\right]
^{k_{\eta _{i^{\prime }}^{\prime }\eta _{i}}}\right\} _{_{\substack{ %
i\leqslant s  \\ i^{\prime }\leqslant s^{\prime }}}},\left[ 
\begin{array}{c}
\left( p_{l_{i}},p_{l_{i}^{\prime }l_{i}},k_{l_{i}^{\prime }}\right) \\ 
\left[ \left\{ \Psi _{J}\left[ p_{\eta },p_{\eta ^{\prime }\eta }\right]
\right\} ,v\right]%
\end{array}%
\right] ,V_{\left( p_{l_{i}},p_{l_{i}^{\prime }l_{i}},k_{l_{i}^{\prime
}}\right) }^{2},v\right) +...  \notag
\end{eqnarray}%
for:%
\begin{equation*}
V_{\left( p_{l_{i}},p_{l_{i}^{\prime }l_{i}},k_{l_{i}^{\prime }}\right)
}=\left\{ \left[ 
\begin{array}{c}
\left( p_{l_{i}},p_{l_{i}^{\prime }l_{i}},k_{l_{i}^{\prime }}\right) \\ 
\left[ \left\{ \Psi _{J}\left[ p_{\eta },p_{\eta ^{\prime }\eta }\right]
\right\} ,v\right]%
\end{array}%
\right] \right\}
\end{equation*}%
and the variation equation for this averaged field is:%
\begin{eqnarray*}
0 &=&\int d\left\{ \left[ p_{\eta _{i}},p_{\eta _{i^{\prime }}^{\prime }\eta
_{i}}\right] ^{k_{\eta _{i^{\prime }}^{\prime }\eta _{i}}}\right\} _{ 
_{\substack{ i\leqslant s  \\ i^{\prime }\leqslant s^{\prime }}}}v\left(
\left\{ \left[ p_{\eta _{i}},p_{\eta _{i^{\prime }}^{\prime }\eta _{i}}%
\right] ^{k_{\eta _{i^{\prime }}^{\prime }\eta _{i}}}\right\} _{_{\substack{ %
i\leqslant s  \\ i^{\prime }\leqslant s^{\prime }}}}\right) \frac{\delta
\left\{ \left[ 
\begin{array}{c}
\left( p_{l_{i}},p_{l_{i}^{\prime }l_{i}},k_{l_{i}^{\prime }}\right) \\ 
\left[ \left\{ \Psi _{J}\left[ p_{\eta },p_{\eta ^{\prime }\eta }\right]
\right\} ,v\right]%
\end{array}%
\right] \right\} }{\delta \Psi _{J,s_{p_{l^{\prime },\eta }}^{\otimes
k_{l^{\prime }\eta }}}^{\otimes k_{l^{\prime }\eta }}\left( \left[ \bar{p}%
_{\eta },\bar{p}_{l^{\prime }\eta }\right] ^{k_{l_{\eta }^{\prime }}}\right) 
} \\
&&\times \nabla _{\left\{ \left[ 
\begin{array}{c}
\left( p_{l_{i}},p_{l_{i}^{\prime }l_{i}},k_{l_{i}^{\prime }}\right) \\ 
\left[ \left\{ \Psi _{J}\left[ p_{\eta },p_{\eta ^{\prime }\eta }\right]
\right\} ,v\right]%
\end{array}%
\right] \right\} }\Psi _{J,\tprod\limits_{i^{\prime }}s_{p_{\eta _{i^{\prime
}}^{\prime }\eta _{i}}}^{\otimes k_{\eta _{i^{\prime }}^{\prime }\eta
_{i}}}}^{\otimes \sum_{i^{\prime }}k_{\eta _{i^{\prime }}^{\prime }\eta
_{i}}}\left( \left\{ \left[ p_{\eta _{i}},p_{\eta _{i^{\prime }}^{\prime
}\eta _{i}}\right] ^{k_{\eta _{i^{\prime }}^{\prime }\eta _{i}}}\right\} _{ 
_{\substack{ i\leqslant s  \\ i^{\prime }\leqslant s^{\prime }}}},\left[ 
\begin{array}{c}
\left( p_{l_{i}},p_{l_{i}^{\prime }l_{i}},k_{l_{i}^{\prime }}\right) \\ 
\left[ \left\{ \Psi _{J}\left[ p_{\eta },p_{\eta ^{\prime }\eta }\right]
\right\} ,v\right]%
\end{array}%
\right] ,V_{\left( p_{l_{i}},p_{l_{i}^{\prime }l_{i}},k_{l_{i}^{\prime
}}\right) },v\right) \\
&&+\int d\left\{ \left[ p_{\eta _{i}},p_{\eta _{i^{\prime }}^{\prime }\eta
_{i}}\right] ^{k_{\eta _{i^{\prime }}^{\prime }\eta _{i}}}\right\} _{ 
_{\substack{ i\leqslant s  \\ i^{\prime }\leqslant s^{\prime }}}}v\left(
\left\{ \left[ p_{\eta _{i}},p_{\eta _{i^{\prime }}^{\prime }\eta _{i}}%
\right] ^{k_{\eta _{i^{\prime }}^{\prime }\eta _{i}}}\right\} _{_{\substack{ %
i\leqslant s  \\ i^{\prime }\leqslant s^{\prime }}}}\right) \\
&&\times \left( \epsilon \frac{\delta V_{\left( p_{l_{i}},p_{l_{i}^{\prime
}l_{i}},k_{l_{i}^{\prime }}\right) }}{\delta \Psi _{J,s_{p_{l^{\prime },\eta
}}^{\otimes k_{l^{\prime }\eta }}}^{\otimes k_{l^{\prime }\eta }}\left( %
\left[ \bar{p}_{\eta },\bar{p}_{l^{\prime }\eta }\right] ^{k_{l_{\eta
}^{\prime }}}\right) }-h\left( V_{\left( p_{l_{i}},p_{l_{i}^{\prime
}l_{i}},k_{l_{i}^{\prime }}\right) },\Psi _{J,\tprod_{l^{\prime
}}s_{p_{l^{\prime },0}}^{\otimes k_{l^{\prime }}}}^{\otimes \sum_{l^{\prime
}}k_{l^{\prime }}}\left( \left\{ \left[ p_{0},p_{l^{\prime }0}\right]
^{k_{l^{\prime }}}\right\} _{_{\substack{ l\leqslant s  \\ l^{\prime
}\leqslant s^{\prime }}}}\right) \right) \right) \\
&&\times \Psi _{J,\tprod\limits_{i^{\prime }}s_{p_{\eta _{i^{\prime
}}^{\prime }\eta _{i}}}^{\otimes k_{\eta _{i^{\prime }}^{\prime }\eta
_{i}}}}^{\otimes \sum_{i^{\prime }}k_{\eta _{i^{\prime }}^{\prime }\eta
_{i}}}\left( \left\{ \left[ p_{0},p_{l^{\prime }0}\right] ^{k_{l^{\prime
}}}\right\} _{_{\substack{ l\leqslant s  \\ l^{\prime }\leqslant s^{\prime } 
}}},\left[ 
\begin{array}{c}
\left( p_{l_{1}},p_{l_{1}^{\prime }l_{1}},k_{l_{1}^{\prime }}\right) \\ 
\left[ \left\{ \Psi _{J}\left[ p_{\eta },p_{\eta ^{\prime }\eta }\right]
\right\} ,v\right]%
\end{array}%
\right] ,V_{\left( p_{l_{i}},p_{l_{i}^{\prime }l_{i}},k_{l_{i}^{\prime
}}\right) },v\right)
\end{eqnarray*}

\section{Average values on eigenstates of operators}

This is an equivalent description already presntd in the first part. We
start with the functionals without path integration:

\begin{eqnarray*}
&&F_{f,lin}\left( \left\{ \Psi _{J,s_{p_{l^{\prime },0}}^{\otimes
k_{l^{\prime }}}}^{\otimes k_{l^{\prime }}}\left[ p_{0},p_{l^{\prime }0}%
\right] ^{k_{l^{\prime }}}\right\} _{l^{\prime }}\right) \\
&=&\sum_{m}\sum_{D_{j}^{p_{0},p_{l},m}}\sum_{m^{\prime
}}\sum_{D_{j,p_{u}}^{\left( p_{0},p_{l^{\prime }0},p_{l},p_{l^{\prime
}l}\right) ,m,m^{\prime }}}g\left( \left\{ \left[ p_{l},p_{l^{\prime }l}%
\right] \right\} _{l,l^{\prime }},\left\{ \left[ p_{0},p_{l^{\prime }0}%
\right] \right\} _{l^{\prime }}\right) \\
&&\tprod\limits_{l^{\prime }}\Psi _{J,s_{p_{l^{\prime },0}}^{\otimes
k_{l^{\prime }}}}^{\otimes k_{l^{\prime }}}\left( \left[ p_{0},p_{l^{\prime
}0}\right] ^{k_{l^{\prime }}}\right) \tprod\limits_{l^{\prime }}\Psi
_{J,s_{p_{l^{\prime },l}}^{\otimes k_{l^{\prime }}}}^{\otimes k_{l^{\prime
}}}\left( \left[ p_{l},p_{l^{\prime }l}\right] ^{k_{l^{\prime }}}\right)
\end{eqnarray*}

and replace:%
\begin{equation*}
\Psi _{J,s_{p_{l^{\prime },0}}^{\otimes k_{l^{\prime }}}}^{\otimes
k_{l^{\prime }}}\left( \left[ p_{0},p_{l^{\prime }0}\right] ^{k_{l^{\prime
}}}\right)
\end{equation*}%
by:%
\begin{equation*}
\Psi _{J,s_{p_{l^{\prime },0}}^{\otimes k_{l^{\prime }}}}^{\otimes
k_{l^{\prime }}}\left( \left[ p_{0},p_{l^{\prime }0}\right] ^{k_{l^{\prime
}}}/\tprod\limits_{p_{1,l},...,p_{m^{\prime },l}}f_{p_{1,l}...p_{m^{\prime
},l}},f_{p_{1,l}...p_{m^{\prime },l}}^{-1}\left\{ \left[ p_{l},p_{l^{\prime
}l}\right] ^{k_{l^{\prime }}}\right\} _{\substack{ l\leqslant m  \\ %
l^{\prime }\leqslant m^{\prime }}}\right)
\end{equation*}%
Implicitly $\Psi _{J}^{\otimes k_{l}}$ and $\Psi _{J}^{\otimes k_{p_{0}}}$
stand for $\Psi _{J,s_{p_{0}}^{\otimes k_{p_{0}}}}^{\otimes k_{p_{0}}}$ and $%
\Psi _{J,s_{p_{l}}^{\otimes k_{l}}}^{\otimes k_{l}}$.

We will consider the average of the field:%
\begin{equation*}
\tprod\limits_{l^{\prime }}\Psi _{J,s_{p_{l^{\prime },l}}^{\otimes
k_{l^{\prime }}}}^{\otimes k_{l^{\prime }}}\left( \left[ p_{l},p_{l^{\prime
}l}\right] ^{k_{l^{\prime }}}\right)
\end{equation*}%
on minimal eigenstates of an operator:%
\begin{equation*}
\mathcal{H}\left( \left\{ \Psi _{J,s_{p_{l^{\prime },l}}^{\otimes
k_{l^{\prime }}}}^{\otimes k_{l^{\prime }}}\left( \left[ p_{l},p_{l^{\prime
}l}\right] ^{k_{l^{\prime }}}\right) ,\Pi _{\Psi _{J,s_{p_{l^{\prime
},l}}^{\otimes k_{l^{\prime }}}}^{\otimes k_{l^{\prime }}}\left( \left[
p_{l},p_{l^{\prime }l}\right] ^{k_{l^{\prime }}}\right) }\right\}
_{l,l^{\prime }},\left\{ \Psi _{J,s_{p_{l^{\prime },0}}^{\otimes
k_{l^{\prime }}}}^{\otimes k_{l^{\prime }}}\left( \left[ p_{0},p_{l^{\prime
}0}\right] ^{k_{l^{\prime }}}\right) \right\} _{l^{\prime }}\right)
\end{equation*}%
with:%
\begin{equation*}
\Pi _{\Psi _{J,s_{p_{l^{\prime },l}}^{\otimes k_{l^{\prime }}}}^{\otimes
k_{l^{\prime }}}\left( \left[ p_{l},p_{l^{\prime }l}\right] ^{k_{l^{\prime
}}}\right) }=\frac{\delta }{\delta \Psi _{J,s_{p_{l^{\prime },l}}^{\otimes
k_{l^{\prime }}}}^{\otimes k_{l^{\prime }}}\left( \left[ p_{l},p_{l^{\prime
}l}\right] ^{k_{l^{\prime }}}\right) }
\end{equation*}

\subsection{Average over one eigenspas, one remaining subobject}

If there is only one eigenstate, the projctd functional has the follwing
form (we assume one remaining subobject, the general case will be studied
belw):%
\begin{eqnarray*}
&&F_{0}\left[ \left[ \Psi _{J}^{\otimes k_{p_{0}}}\right] ,\left\{ \Psi
_{J,s_{p_{l^{\prime },l}}^{\otimes k_{l^{\prime }}}}^{\otimes k_{l^{\prime
}}}\right\} \right] \\
&\equiv &\int F_{0}\left[ \left[ p_{l},p_{l^{\prime }l}\right]
^{k_{l^{\prime }}},\left[ \Psi _{J}^{\otimes k_{p_{0}}}\right] \right]
\otimes \Psi _{J,s_{p_{l^{\prime },l}}^{\otimes k_{l^{\prime }}}}^{\otimes
k_{l^{\prime }}}\left( \left[ p_{l},p_{l^{\prime }l}\right] ^{k_{l^{\prime
}}}\right) d\left( \left[ p_{l},p_{l^{\prime }l}\right] ^{k_{l^{\prime
}}}\right) \\
&=&\int F_{0}\left[ \left[ p_{l},p_{l^{\prime }l}\right] ^{k_{l^{\prime }}},%
\left[ \Psi _{J}^{\otimes k_{p_{0}}}\right] \right] \otimes \Psi
_{J,s_{p_{l^{\prime },l}}^{\otimes k_{l^{\prime }}}}^{\otimes k_{l^{\prime
}}}\left( \left[ p_{l},p_{l^{\prime }l}\right] ^{k_{l^{\prime }}}\right)
d\left( \left[ p_{l},p_{l^{\prime }l}\right] ^{k_{l^{\prime }}}\right)
\end{eqnarray*}%
The dependence $\left[ \Psi _{J}^{\otimes k_{p_{0}}}\right] $ is through a
functional of the type $v\left( \Psi _{J}^{\otimes k_{p_{0}}}\right) $, as
in the previous sections. As before, the tensor $\otimes $ stands for series
of products of identical copies.

ncldng symtris parmtrized by $\left[ \underset{\left[ \left\{ \Psi _{J}\left[
l^{\prime },0\right] \right\} ,v\right] }{\overset{\left(
p_{l_{1}},p_{l_{1}^{\prime }l_{1}},k_{l_{1}^{\prime }}\right) }{\mathbf{\hat{%
\Lambda}}}}\right] $, th functnls bcms:%
\begin{equation*}
F_{0}\left[ \left[ \Psi _{J}^{\otimes k_{p_{0}}}\right] ,\left\{ \Psi
_{J,s_{p_{l^{\prime },l}}^{\otimes k_{l^{\prime }}}}^{\otimes k_{l^{\prime
}}}\right\} ,\left\{ \left[ 
\begin{array}{c}
\left[ p_{l_{i}},p_{l_{i}^{\prime }l_{i}},k_{l_{i}^{\prime }}\right] \\ 
\left[ \left\{ \Psi _{J}\left[ l^{\prime },0\right] \right\} ,v\right]%
\end{array}%
\right] \right\} \right]
\end{equation*}%
average values in this state:%
\begin{equation*}
\left\langle \tprod\limits_{l^{\prime }}\Psi _{J,s_{p_{l^{\prime
},l}}^{\otimes k_{l^{\prime }}}}^{\otimes k_{l^{\prime }}}\left( \left[
p_{l},p_{l^{\prime }l}\right] ^{k_{l^{\prime }}}\right) \right\rangle
=\tprod\limits_{l^{\prime }}\Psi _{J,s_{p_{l^{\prime },l}}^{\otimes
k_{l^{\prime }}}}^{\otimes k_{l^{\prime }}}\left( \left[ p_{l},p_{l^{\prime
}l}\right] ^{k_{l^{\prime }}},\left[ \Psi _{J}^{\otimes k_{p_{0}}}\right]
,\left\{ \left[ 
\begin{array}{c}
\left[ p_{l_{i}},p_{l_{i}^{\prime }l_{i}},k_{l_{i}^{\prime }}\right] \\ 
\left[ \left\{ \Psi _{J}\left[ l^{\prime },0\right] \right\} ,v\right]%
\end{array}%
\right] \right\} \right)
\end{equation*}%
where:%
\begin{eqnarray*}
&&\tprod\limits_{l^{\prime }}\Psi _{J,s_{p_{l^{\prime },l}}^{\otimes
k_{l^{\prime }}}}^{\otimes k_{l^{\prime }}}\left( \left[ p_{l},p_{l^{\prime
}l}\right] ^{k_{l^{\prime }}},\left[ \Psi _{J}^{\otimes k_{p_{0}}}\right]
,\left\{ \left[ 
\begin{array}{c}
\left[ p_{l_{i}},p_{l_{i}^{\prime }l_{i}},k_{l_{i}^{\prime }}\right] \\ 
\left[ \left\{ \Psi _{J}\left[ l^{\prime },0\right] \right\} ,v\right]%
\end{array}%
\right] \right\} \right) \\
&=&\int \tprod\limits_{l^{\prime }}\Psi _{J,s_{p_{l^{\prime },l}}^{\otimes
k_{l^{\prime }}}}^{\otimes k_{l^{\prime }}}\left( \left[ p_{l},p_{l^{\prime
}l}\right] ^{k_{l^{\prime }}}\right) \left\vert F_{0}\left[ \left[ \Psi
_{J}^{\otimes k_{p_{0}}}\right] ,\left\{ \Psi _{J,s_{p_{l^{\prime
},l}}^{\otimes k_{l^{\prime }}}}^{\otimes k_{l^{\prime }}}\right\} ,\left\{ %
\left[ 
\begin{array}{c}
\left[ p_{l_{i}},p_{l_{i}^{\prime }l_{i}},k_{l_{i}^{\prime }}\right] \\ 
\left[ \left\{ \Psi _{J}\left[ l^{\prime },0\right] \right\} ,v\right]%
\end{array}%
\right] \right\} \right] \right\vert ^{2}\tprod \mathcal{D}\left\{ \Psi
_{J,s_{p_{l^{\prime },l}}^{\otimes k_{l^{\prime }}}}^{\otimes k_{l^{\prime
}}}\right\}
\end{eqnarray*}%
nd th measure is invariant for this transformation. Then: 
\begin{eqnarray*}
&&\tprod\limits_{l^{\prime }}\Psi _{J,s_{p_{l^{\prime },l}}^{\otimes
k_{l^{\prime }}}}^{\otimes k_{l^{\prime }}}\left( \left[ p_{l},p_{l^{\prime
}l}\right] ^{k_{l^{\prime }}},\left[ \Psi _{J}^{\otimes k_{p_{0}}}\right]
,\left\{ \left[ 
\begin{array}{c}
\left[ p_{l_{i}},p_{l_{i}^{\prime }l_{i}},k_{l_{i}^{\prime }}\right] \\ 
\left[ \left\{ \Psi _{J}\left[ l^{\prime },0\right] \right\} ,v\right]%
\end{array}%
\right] \right\} \right) \\
&=&\int \tprod\limits_{l^{\prime }}\Psi _{J,s_{p_{l^{\prime },l}}^{\otimes
k_{l^{\prime }}}}^{\otimes k_{l^{\prime }}}\left( \left[ p_{l},p_{l^{\prime
}l}\right] ^{k_{l^{\prime }}}\right) \left\vert F_{0}\left[ \left[ \Psi
_{J}^{\otimes k_{p_{0}}}\right] ,\left\{ \Psi _{J,s_{p_{l^{\prime
},l}}^{\otimes k_{l^{\prime }}}}^{\otimes k_{l^{\prime }}}\right\} ,\left\{ %
\left[ 
\begin{array}{c}
\left[ p_{l_{i}},p_{l_{i}^{\prime }l_{i}},k_{l_{i}^{\prime }}\right] \\ 
\left[ \left\{ \Psi _{J}\left[ l^{\prime },0\right] \right\} ,v\right]%
\end{array}%
\right] \right\} \right] \right\vert ^{2}\tprod \mathcal{D}\left\{ \Psi
_{J,s_{p_{l^{\prime },l}}^{\otimes k_{l^{\prime }}}}^{\otimes k_{l^{\prime
}}}\right\}
\end{eqnarray*}%
As a consequence, replacing:%
\begin{equation*}
\tprod\limits_{l^{\prime }}\Psi _{J,s_{p_{l^{\prime },l}}^{\otimes
k_{l^{\prime }}}}^{\otimes k_{l^{\prime }}}\left( \left[ p_{l},p_{l^{\prime
}l}\right] ^{k_{l^{\prime }}}\right)
\end{equation*}%
by its average over the eigenstate considered in the functional, and
considering linear combinations over these states:%
\begin{equation*}
\int \Psi \left( \mathbf{\hat{O}}_{i}^{\left( p_{l}\right) }\left( \Psi
_{J}^{\otimes k_{p_{0}}}\right) \right) \tprod\limits_{l^{\prime }}\Psi
_{J,s_{p_{l^{\prime },l}}^{\otimes k_{l^{\prime }}}}^{\otimes k_{l^{\prime
}}}\left( \left[ p_{l},p_{l^{\prime }l}\right] ^{k_{l^{\prime }}},\left[
\Psi _{J}^{\otimes k_{p_{0}}}\right] ,\left\{ \left[ 
\begin{array}{c}
\left[ p_{l_{i}},p_{l_{i}^{\prime }l_{i}},k_{l_{i}^{\prime }}\right] \\ 
\left[ \left\{ \Psi _{J}\left[ l^{\prime },0\right] \right\} ,v\right]%
\end{array}%
\right] \right\} \right)
\end{equation*}%
and:%
\begin{equation*}
\Psi _{J,s_{p_{l^{\prime },l}}^{\otimes k_{l^{\prime }}}}^{\otimes
k_{l^{\prime }}}\left( \left[ p_{l},p_{l^{\prime }l}\right] ^{k_{l^{\prime
}}},\left[ \Psi _{J}^{\otimes k_{p_{0}}}\right] ,\left\{ \left[ 
\begin{array}{c}
\left[ p_{l_{i}},p_{l_{i}^{\prime }l_{i}},k_{l_{i}^{\prime }}\right] \\ 
\left[ \left\{ \Psi _{J}\left[ l^{\prime },0\right] \right\} ,v\right]%
\end{array}%
\right] \right\} \right)
\end{equation*}%
has an expansion:%
\begin{eqnarray*}
&&\Psi _{J,s_{p_{l^{\prime },l}}^{\otimes k_{l^{\prime }}}}^{\otimes
k_{l^{\prime }}}\left( \left[ p_{l},p_{l^{\prime }l}\right] ^{k_{l^{\prime
}}},v,\left\{ \left[ 
\begin{array}{c}
\left[ p_{l_{i}},p_{l_{i}^{\prime }l_{i}},k_{l_{i}^{\prime }}\right] \\ 
\left[ \left\{ \Psi _{J}\left[ l^{\prime },0\right] \right\} ,v\right]%
\end{array}%
\right] \right\} \right) \\
&=&\sum_{_{\substack{ \left( s,s^{\prime },\left[ p,p^{\prime },k^{\prime }%
\right] \right)  \\ \left( m,m^{\prime },\left[ p,p^{\prime },k^{\prime }%
\right] \right) _{1}}}}d\left( \left\{ \overline{\left[ p_{0},p_{l^{\prime
}0}\right] ^{k_{l^{\prime }}}}\right\} _{\substack{ l\leqslant s  \\ %
l^{\prime }\leqslant s^{\prime }}},\left\{ \left[ p_{l_{1}},p_{l_{1}^{\prime
}l_{1}}\right] ^{k_{l_{1}^{\prime }}}\right\} _{\substack{ l_{1}\leqslant
m_{1},l_{1}\neq l  \\ l_{1}^{\prime }\leqslant m_{1}^{\prime }}}\right) \\
&&\times \mathcal{V}_{0}\left( \left\{ \overline{\left[ p_{0},p_{l^{\prime
}0}\right] ^{k_{l^{\prime }}}}\right\} _{_{\substack{ l\leqslant s  \\ %
l^{\prime }\leqslant s^{\prime }}}},\left\{ \left[ p_{l_{1}},p_{l_{1}^{%
\prime }l_{1}}\right] ^{k_{l_{1}^{\prime }}}\right\} _{_{\substack{ %
l_{1}\leqslant m_{1}  \\ l_{1}^{\prime }\leqslant m_{1}^{\prime }}}%
}^{k_{l^{\prime }}},\left\{ \left[ 
\begin{array}{c}
\left[ p_{l_{i}},p_{l_{i}^{\prime }l_{i}},k_{l_{i}^{\prime }}\right] \\ 
\left[ \left\{ \Psi _{J}\left[ l^{\prime },0\right] \right\} ,v\right]%
\end{array}%
\right] \right\} \right) \\
&&\times \tprod_{\substack{ l\leqslant s  \\ l^{\prime }\leqslant s^{\prime
} }}\Psi _{J,s_{p_{l^{\prime },0}}^{\otimes k_{l^{\prime }}}}^{\otimes
k_{l^{\prime }}}\left( \overline{\left[ p_{0},p_{l^{\prime }0}\right]
^{k_{l^{\prime }}}},\left\{ \left[ p_{l_{1}},p_{l_{1}^{\prime }l_{1}}\right]
^{k_{l_{1}^{\prime }}}\right\} _{_{\mathcal{P}}}\right) \tprod_{\mathcal{P}%
^{c}}v_{\left[ p_{l_{1}},p_{l_{1}^{\prime }l_{1}}\right] ^{k_{l_{1}^{\prime
}}}}\left\{ \Psi _{J}^{\otimes k_{l^{\prime }}}\right\}
\end{eqnarray*}

and the functionals decompoe for one realization:

\begin{eqnarray}
&&F_{f,lin}\left( \left\{ \Psi _{J,s_{p_{l^{\prime },0}}^{\otimes
k_{l^{\prime }}}}^{\otimes k_{l^{\prime }}}\left( \left[ p_{0},p_{l^{\prime
}0}\right] ^{k_{l^{\prime }}}\right) \right\} _{\left( p_{l^{\prime
}0}\right) }\right)  \label{FPT} \\
&=&\sum_{m}\sum_{m^{\prime }}\int \sum_{_{\substack{ \left( s,s^{\prime },%
\left[ p,p^{\prime },k^{\prime }\right] \right)  \\ \left( m,m^{\prime },%
\left[ p,p^{\prime },k^{\prime }\right] \right) _{1}}}}d\left( \left\{ 
\overline{\left[ p_{0},p_{l^{\prime }0}\right] ^{k_{l^{\prime }}}}\right\} 
_{\substack{ l\leqslant s  \\ l^{\prime }\leqslant s^{\prime }}},\left\{ %
\left[ p_{l_{1}},p_{l_{1}^{\prime }l_{1}}\right] ^{k_{l_{1}^{\prime
}}}\right\} _{\substack{ l_{1}\leqslant m_{1}  \\ l_{1}^{\prime }\leqslant
m_{1}^{\prime }}}\right) d\left\{ \left[ 
\begin{array}{c}
\left[ p_{l_{i}},p_{l_{i}^{\prime }l_{i}},k_{l_{i}^{\prime }}\right] \\ 
\left[ \left\{ \Psi _{J}\left[ l^{\prime },0\right] \right\} ,v\right]%
\end{array}%
\right] \right\}  \notag \\
&&\times g\left( \left\{ \left[ p_{l},p_{l^{\prime }l}\right] \right\}
_{l,l^{\prime }},\left\{ \overline{\left[ p_{0},p_{l^{\prime }0}\right] }%
\right\} _{l^{\prime }}\right) \mathcal{V}_{0}\left( \left\{ \overline{\left[
p_{0},p_{l^{\prime }0}\right] ^{k_{l^{\prime }}}}\right\} _{_{\substack{ %
l\leqslant s  \\ l^{\prime }\leqslant s^{\prime }}}},\left\{ \left[
p_{l_{1}},p_{l_{1}^{\prime }l_{1}}\right] ^{k_{l_{1}^{\prime }}}\right\} _{ 
_{\substack{ l_{1}\leqslant m_{1}  \\ l_{1}^{\prime }\leqslant m_{1}^{\prime
} }}}^{k_{l^{\prime }}},\left\{ \left[ 
\begin{array}{c}
\left[ p_{l_{i}},p_{l_{i}^{\prime }l_{i}},k_{l_{i}^{\prime }}\right] \\ 
\left[ \left\{ \Psi _{J}\left[ l^{\prime },0\right] \right\} ,v\right]%
\end{array}%
\right] \right\} \right)  \notag \\
&&\times \tprod_{\substack{ l\leqslant s  \\ l^{\prime }\leqslant s^{\prime
} }}\Psi _{J,s_{p_{l^{\prime },0}}^{\otimes k_{l^{\prime }}}}^{\otimes
k_{l^{\prime }}}\left( \overline{\left[ p_{0},p_{l^{\prime }0}\right]
^{k_{l^{\prime }}}},\left\{ \left[ p_{l_{1}},p_{l_{1}^{\prime }l_{1}}\right]
^{k_{l_{1}^{\prime }}}\right\} _{_{\mathcal{P}}}\right) \tprod_{\mathcal{P}%
^{c}}v_{\left[ p_{l_{1}},p_{l_{1}^{\prime }l_{1}}\right] ^{k_{l_{1}^{\prime
}}}}\left\{ \Psi _{J}^{\otimes k_{l^{\prime }}}\right\} \Psi \left( \left\{ %
\left[ 
\begin{array}{c}
\left[ p_{l_{i}},p_{l_{i}^{\prime }l_{i}},k_{l_{i}^{\prime }}\right] \\ 
\left[ \left\{ \Psi _{J}\left[ l^{\prime },0\right] \right\} ,v\right]%
\end{array}%
\right] \right\} \right)  \notag
\end{eqnarray}%
where $D_{j}^{p_{0},p_{l},m}$ stand for decomposition%
\begin{equation*}
\mathcal{H}\left( U_{j}\right) =\mathcal{H}\left( \left( \left( U_{j}\right)
^{\left( p_{0}\right) }\right) \right) \otimes \mathcal{H}\left( \left\{
\left( U_{j}\right) ^{\left( p_{l}\right) }\right\} _{l\leqslant m}\right)
\end{equation*}%
and $D_{j,p_{u}}^{\left( p_{0},p_{l^{\prime }0},p_{l},p_{l^{\prime
}l}\right) ,m,m^{\prime }}$ for decomposition: 
\begin{equation*}
\mathcal{H}\left( \left( \left( U_{j}\right) ^{\left( p_{u}\right) }\right)
\right) =\mathcal{H}\left( \left( \left( U_{j}\right) ^{\left(
p_{0},p_{l^{\prime }0}\right) }\right) ^{k_{l^{\prime }}}\right) \otimes 
\mathcal{H}\left( \left\{ \left( U_{j}\right) ^{\left( p_{l},p_{l^{\prime
}l}\right) }\right\} _{\substack{ l\leqslant m  \\ l^{\prime }\leqslant
m^{\prime }}}\right)
\end{equation*}%
By changing variables in (\ref{FPT}), we show in appendix 7 that:%
\begin{eqnarray*}
&&F_{f,lin}\left( \left\{ \Psi _{J,s_{p_{l^{\prime },0}}^{\otimes
k_{l^{\prime }}}}^{\otimes k_{l^{\prime }}}\left( \left[ p_{0},p_{l^{\prime
}0}\right] ^{k_{l^{\prime }}}\right) \right\} _{\left( p_{l^{\prime
}0}\right) }\right) \\
&=&\sum_{m}\sum_{m^{\prime }}\int \sum_{_{\substack{ \left( s,s^{\prime },%
\left[ p,p^{\prime },k^{\prime }\right] \right)  \\ \left( m,m^{\prime },%
\left[ p,p^{\prime },k^{\prime }\right] \right) _{1}}}}d\left\{ \left[
p_{0},p_{l^{\prime }0}\right] ^{k_{l^{\prime }}}\right\} _{\substack{ %
l\leqslant s  \\ l^{\prime }\leqslant s^{\prime }}}d\left\{ \left[ 
\begin{array}{c}
\left[ p_{l_{i}},p_{l_{i}^{\prime }l_{i}},k_{l_{i}^{\prime }}\right] \\ 
\left[ \left\{ \Psi _{J}\left[ l^{\prime },0\right] \right\} ,v\right]%
\end{array}%
\right] \right\} \\
&&\times \bar{g}\left( \left\{ \left[ p_{0},p_{l^{\prime }0}\right]
^{k_{l^{\prime }}}\right\} _{_{\substack{ l\leqslant s  \\ l^{\prime
}\leqslant s^{\prime }}}},\left\{ \left[ 
\begin{array}{c}
\left[ p_{l_{i}},p_{l_{i}^{\prime }l_{i}},k_{l_{i}^{\prime }}\right] \\ 
\left[ \left\{ \Psi _{J}\left[ l^{\prime },0\right] \right\} ,v\right]%
\end{array}%
\right] \right\} \right) \Psi _{J,\tprod\limits_{l^{\prime }}s_{p_{l^{\prime
},0}}^{\otimes k_{l^{\prime }}}}^{\otimes \sum_{l^{\prime }}k_{l^{\prime
}}}\left( \tprod\limits_{l^{\prime }}\left[ p_{0},p_{l^{\prime }0}\right]
^{k_{l^{\prime },i}},\left\{ \left[ 
\begin{array}{c}
\left[ p_{l_{i}},p_{l_{i}^{\prime }l_{i}},k_{l_{i}^{\prime }}\right] \\ 
\left[ \left\{ \Psi _{J}\left[ l^{\prime },0\right] \right\} ,v\right]%
\end{array}%
\right] \right\} \right)
\end{eqnarray*}%
Introducing explicitely the eigenvalue on which the projection occurs, the
effective field is similar to the one component case:%
\begin{eqnarray*}
&&\Psi _{J,\tprod\limits_{l^{\prime }}s_{p_{l^{\prime },0}}^{\otimes
k_{l^{\prime }}}}^{\otimes \sum_{l^{\prime }}k_{l^{\prime }}}\left(
\tprod\limits_{l^{\prime }}\left[ p_{0},p_{l^{\prime }0}\right]
^{k_{l^{\prime },i}},\left\{ \left[ 
\begin{array}{c}
\left[ p_{l_{i}},p_{l_{i}^{\prime }l_{i}},k_{l_{i}^{\prime }}\right] \\ 
\left[ \left\{ \Psi _{J,\alpha }\left[ l^{\prime },0\right] \right\} ,v%
\right]%
\end{array}%
\right] \right\} ,\lambda \left( \left[ \left\{ \Psi _{J}\left[ l^{\prime },0%
\right] \right\} ,v\right] \right) \right) \\
&=&\sum_{\mathcal{P},\alpha }\tprod_{\substack{ l\leqslant s  \\ l^{\prime
}\leqslant s^{\prime }}}\Psi _{J,s_{p_{l^{\prime },0}}^{\otimes k_{l^{\prime
}}}}^{\otimes k_{l^{\prime }}}\left( \overline{\left[ p_{0},p_{l^{\prime }0}%
\right] ^{k_{l^{\prime }}}},\left\{ \left[ p_{l_{1}},p_{l_{1}^{\prime }l_{1}}%
\right] ^{k_{l_{1}^{\prime }}}\right\} _{_{_{\mathcal{P}}}},\left\{ \left[ 
\begin{array}{c}
\left[ p_{l_{i}},p_{l_{i}^{\prime }l_{i}},k_{l_{i}^{\prime }}\right] \\ 
\left[ \left\{ \Psi _{J,\alpha }\left[ l^{\prime },0\right] \right\} ,v%
\right]%
\end{array}%
\right] \right\} \right) \\
&&\times \tprod_{\mathcal{P}^{c}}v_{\left[ p_{l_{1}},p_{l_{1}^{\prime }l_{1}}%
\right] ^{k_{l_{1}^{\prime }}}}\left\{ \Psi _{J}^{\otimes k_{l^{\prime
}}}\right\} \Psi _{\alpha }\left( \left\{ \left[ 
\begin{array}{c}
\left[ p_{l_{i}},p_{l_{i}^{\prime }l_{i}},k_{l_{i}^{\prime }}\right] \\ 
\left[ \left\{ \Psi _{J,\alpha }\left[ l^{\prime },0\right] \right\} ,v%
\right]%
\end{array}%
\right] \right\} \right) d\left\{ \left[ 
\begin{array}{c}
\left[ p_{l_{i}},p_{l_{i}^{\prime }l_{i}},k_{l_{i}^{\prime }}\right] \\ 
\left[ \left\{ \Psi _{J,\alpha }\left[ l^{\prime },0\right] \right\} ,v%
\right]%
\end{array}%
\right] \right\}
\end{eqnarray*}

\subsection{Average over eigenspaces, several remaining subobjects}

This case is similar to the case one remaining subobject. If the remaining
subobjects cannot be separated, the projection yields the composed effective
field similar to (\ref{F}):%
\begin{eqnarray*}
&&\int d\left\{ \left[ p_{\eta _{i}},p_{\eta _{i^{\prime }}^{\prime }\eta
_{i}}\right] ^{k_{\eta _{i^{\prime }}^{\prime }\eta _{i}}}\right\} _{ 
_{\substack{ i\leqslant s  \\ i^{\prime }\leqslant s^{\prime }}}}v\left(
\left\{ \left[ p_{\eta _{i}},p_{\eta _{i^{\prime }}^{\prime }\eta _{i}}%
\right] ^{k_{\eta _{i^{\prime }}^{\prime }\eta _{i}}}\right\} _{_{\substack{ %
i\leqslant s  \\ i^{\prime }\leqslant s^{\prime }}}}\right) \\
&&\times \Psi _{J,\tprod\limits_{i^{\prime }}s_{p_{\eta _{i^{\prime
}}^{\prime }\eta _{i}}}^{\otimes k_{\eta _{i^{\prime }}^{\prime }\eta
_{i}}}}^{\otimes \sum_{i^{\prime }}k_{\eta _{i^{\prime }}^{\prime }\eta
_{i}}}\left( \left\{ \left[ p_{\eta _{i}},p_{\eta _{i^{\prime }}^{\prime
}\eta _{i}}\right] ^{k_{\eta _{i^{\prime }}^{\prime }\eta _{i}}}\right\}
,\left\{ \left[ 
\begin{array}{c}
\left( p_{l_{1}},p_{l_{1}^{\prime }l_{1}},k_{l_{1}^{\prime }}\right) \\ 
\left[ \left\{ \Psi _{J}\left[ p_{\eta },p_{\eta ^{\prime }\eta }\right]
\right\} ,v\right]%
\end{array}%
\right] \right\} ,\left[ v\right] ,\lambda \left( \left[ \left\{ \Psi _{J}%
\left[ p_{\eta },p_{\eta ^{\prime }\eta }\right] \right\} ,v\right] \right)
\right)
\end{eqnarray*}

If, on the contrary several components can be isolated (see (\ref{FL})), the
system is described by the fields, indexed by $\eta $, the remaining objects:%
\begin{equation}
\Psi _{J,s_{p_{\eta _{i^{\prime }}^{\prime }\eta _{i}}}^{\otimes k_{\eta
_{i^{\prime }}^{\prime }\eta _{i}}}}^{\otimes k_{\eta _{i^{\prime }}^{\prime
}\eta _{i}}}\left( \left[ p_{\eta _{i}},p_{\eta _{i^{\prime }}^{\prime }\eta
_{i}}\right] ^{k_{\eta _{i^{\prime }}^{\prime }\eta _{i}}},\left\{ \left[ 
\begin{array}{c}
\left( p_{l_{1}},p_{l_{1}^{\prime }l_{1}},k_{l_{1}^{\prime }}\right) \\ 
\left[ \left\{ \Psi _{J}\left[ p_{\eta },p_{\eta ^{\prime }\eta }\right]
\right\} ,v\right]%
\end{array}%
\right] \right\} ,\left[ v\right] ,\lambda \left( \left[ \left\{ \Psi _{J}%
\left[ p_{\eta },p_{\eta ^{\prime }\eta }\right] \right\} ,v\right] \right)
\right)  \label{SFLD}
\end{equation}

\subsection{Transitions}

As in part one, we can redefine the fields (\ref{SFLD}) if the remaining
subobject can be disentangled:%
\begin{equation*}
\Psi _{J,s_{p_{\eta ^{\prime }\eta }}^{\otimes k_{\eta ^{\prime }\eta
}}}^{\otimes k_{\eta ^{\prime }\eta }}\left( \left[ p_{\eta },p_{\eta
^{\prime }\eta }\right] ^{k_{\eta ^{\prime }\eta }}/\left[ \left[ p_{\eta
},p_{\eta ^{\prime }\eta }\right] ^{k_{\eta ^{\prime }\eta }}\right]
,\left\{ \left[ 
\begin{array}{c}
\left( p_{l_{i}},p_{l_{i}^{\prime }l_{i}},k_{l_{1}^{\prime }}\right) \\ 
\left[ \left\{ \Psi _{J}\left[ l^{\prime }\eta ,\eta \right] \right\} ,v%
\right]%
\end{array}%
\right] \right\} ,\left[ v\right] ,\lambda \right)
\end{equation*}%
where $\left[ p_{\eta },p_{\eta ^{\prime }\eta }\right] ^{k_{\eta ^{\prime
}\eta }}/\left[ \left[ p_{\eta },p_{\eta ^{\prime }\eta }\right] ^{k_{\eta
^{\prime }\eta }}\right] $ describes the remaining free parameters of $\left[
p_{\eta },p_{\eta ^{\prime }\eta }\right] ^{k_{\eta ^{\prime }\eta }}$ when
the constraints with the parameters $\left\{ \left[ 
\begin{array}{c}
\left( p_{l_{i}},p_{l_{i}^{\prime }l_{i}},k_{l_{1}^{\prime }}\right) \\ 
\left[ \left\{ \Psi _{J}\left[ l^{\prime }\eta ,\eta \right] \right\} ,v%
\right]%
\end{array}%
\right] \right\} $ are solved. Equivalently, effective field write:%
\begin{equation*}
v\left( \left[ p_{\eta },p_{\eta ^{\prime }\eta }\right] ^{k_{\eta ^{\prime
}\eta }}/\left[ \left[ p_{\eta },p_{\eta ^{\prime }\eta }\right] ^{k_{\eta
^{\prime }\eta }}\right] \right) \Psi _{J,s_{p_{\eta ^{\prime }\eta
}}^{\otimes k_{\eta ^{\prime }\eta }}}^{\otimes k_{\eta ^{\prime }\eta
}}\left( \left[ p_{\eta },p_{\eta ^{\prime }\eta }\right] ^{k_{\eta ^{\prime
}\eta }}/\left[ \left[ p_{\eta },p_{\eta ^{\prime }\eta }\right] ^{k_{\eta
^{\prime }\eta }}\right] ,\left\{ \left[ 
\begin{array}{c}
\left( p_{l_{i}},p_{l_{i}^{\prime }l_{i}},k_{l_{1}^{\prime }}\right) \\ 
\left[ \left\{ \Psi _{J}\left[ l^{\prime }\eta ,\eta \right] \right\} ,v%
\right]%
\end{array}%
\right] \right\} ,\left[ v\right] ,\lambda \right)
\end{equation*}

If the subobjects remain composite, an effective field for a collection of
subobjects $\eta _{i},\eta _{i^{\prime }}^{\prime }$ rewrites%
\begin{equation}
\int v\left( \left\{ \overline{\left[ p_{\eta _{i}},p_{\eta _{i^{\prime
}}^{\prime }\eta _{i}}\right] ^{k_{\eta _{i^{\prime }}^{\prime }\eta _{i}}}}%
\right\} _{_{\substack{ i\leqslant s  \\ i^{\prime }\leqslant s^{\prime }}}%
}\right) \Psi _{J,\tprod\limits_{i^{\prime }}s_{p_{\eta _{i^{\prime
}}^{\prime }\eta _{i}}}^{\otimes k_{\eta _{i^{\prime }}^{\prime }\eta
_{i}}}}^{\otimes \sum_{i^{\prime }}k_{\eta _{i^{\prime }}^{\prime }\eta
_{i}}}\left( \left\{ \overline{\left[ p_{\eta _{i}},p_{\eta _{i^{\prime
}}^{\prime }\eta _{i}}\right] ^{k_{\eta _{i^{\prime }}^{\prime }\eta _{i}}}}%
\right\} _{_{\substack{ i\leqslant s  \\ i^{\prime }\leqslant s^{\prime }}}%
},\left\{ \left[ 
\begin{array}{c}
\left( p_{l_{i}},p_{l_{i}^{\prime }l_{i}},k_{l_{i}^{\prime }}\right) \\ 
\left[ \left\{ \Psi _{J}\left[ p_{\eta },p_{\eta ^{\prime }\eta }\right]
\right\} ,v\right]%
\end{array}%
\right] \right\} ,\left[ v\right] \right)  \label{SML}
\end{equation}%
with:%
\begin{equation*}
\overline{\left[ p_{\eta _{i}},p_{\eta _{i^{\prime }}^{\prime }\eta _{i}}%
\right] ^{k_{\eta _{i^{\prime }}^{\prime }\eta _{i}}}}=\left[ p_{\eta
_{i}},p_{\eta _{i^{\prime }}^{\prime }\eta _{i}}\right] ^{k_{\eta
_{i^{\prime }}^{\prime }\eta _{i}}}/\left\{ \left[ p_{\eta _{i}},p_{\eta
_{i^{\prime }}^{\prime }\eta _{i}}\right] ^{k_{\eta _{i^{\prime }}^{\prime
}\eta _{i}}}\right\} _{_{_{_{\substack{ i\leqslant s  \\ i^{\prime
}\leqslant s^{\prime }}}}}}
\end{equation*}

The transitions for the fields (\ref{SFLD}) (similar formula are obtained
for (\ref{SML})) are generated by a functional:%
\begin{equation*}
S\left( \Psi _{J,s_{p_{\eta ^{\prime }\eta }}^{\otimes k_{\eta ^{\prime
}\eta }}}^{\otimes k_{\eta ^{\prime }\eta }},\frac{\delta }{\delta \Psi
_{J,s_{p_{\eta ^{\prime }\eta }}^{\otimes k_{\eta ^{\prime }\eta
}}}^{\otimes k_{\eta ^{\prime }\eta }}},\underline{\nabla }_{\left[ 
\begin{array}{c}
\left( p_{l_{i}},p_{l_{i}^{\prime }l_{i}},k_{l_{1}^{\prime }}\right) \\ 
\left[ \left\{ \Psi _{J}\left[ p_{\eta },p_{\eta ^{\prime }\eta }\right]
\right\} ,v\right]%
\end{array}%
\right] }\underline{\nabla }_{_{\lambda }}\right)
\end{equation*}%
with $\Psi _{J,s_{p_{l^{\prime },\eta }}^{\otimes k_{l^{\prime }\eta
}}}^{\otimes k_{l^{\prime }\eta }}$ defined in (\ref{SFLD}), and the
covariant derivatives are:

\begin{eqnarray*}
&&\underline{\nabla }_{\left\{ \underline{\mathbf{\hat{\Lambda}}}^{\left[
k_{i}\right] }\right\} }\Psi _{J,s_{p_{\eta ^{\prime }\eta }}^{\otimes
k_{\eta ^{\prime }\eta }}}^{\otimes k_{\eta ^{\prime }\eta }}\left( \left[
p_{\eta },p_{\eta ^{\prime }\eta }\right] ^{k_{\eta ^{\prime }\eta }}/\left[ %
\left[ p_{\eta },p_{\eta ^{\prime }\eta }\right] ^{k_{\eta ^{\prime }\eta }}%
\right] ,\left\{ \left[ 
\begin{array}{c}
\left( p_{l_{i}},p_{l_{i}^{\prime }l_{i}},k_{l_{1}^{\prime }}\right) \\ 
\left[ \left\{ \Psi _{J}\left[ p_{\eta },p_{\eta ^{\prime }\eta }\right]
\right\} ,v\right]%
\end{array}%
\right] \right\} ,\left[ v\right] ,\lambda \right) \\
&=&\nabla _{\left\{ \underline{\mathbf{\hat{\Lambda}}}^{\left[ k_{i}\right]
}\right\} }\Psi _{J,s_{p_{\eta ^{\prime }\eta }}^{\otimes k_{\eta ^{\prime
}\eta }}}^{\otimes k_{\eta ^{\prime }\eta }}\left( \left[ p_{\eta },p_{\eta
^{\prime }\eta }\right] ^{k_{\eta ^{\prime }\eta }}/\left[ \left[ p_{\eta
},p_{\eta ^{\prime }\eta }\right] ^{k_{\eta ^{\prime }\eta }}\right]
,\left\{ \left[ 
\begin{array}{c}
\left( p_{l_{i}},p_{l_{i}^{\prime }l_{i}},k_{l_{1}^{\prime }}\right) \\ 
\left[ \left\{ \Psi _{J}\left[ p_{\eta },p_{\eta ^{\prime }\eta }\right]
\right\} ,v\right]%
\end{array}%
\right] \right\} ,\left[ v\right] ,\lambda \right) \\
&&+\left( \left( \mathbf{A}\right) _{k}^{k^{\prime }}\Psi _{J,s_{p_{\eta
^{\prime }\eta }}^{\otimes k_{\eta ^{\prime }\eta }}}^{\otimes k_{\eta
^{\prime }\eta }}\left( \left( \left[ p_{\eta },p_{\eta ^{\prime }\eta }%
\right] ^{k_{\eta ^{\prime }\eta }}/\left[ \left[ p_{\eta },p_{\eta ^{\prime
}\eta }\right] ^{k_{\eta ^{\prime }\eta }}\right] \right) _{k^{\prime
}},\left\{ \left[ 
\begin{array}{c}
\left( p_{l_{i}},p_{l_{i}^{\prime }l_{i}},k_{l_{1}^{\prime }}\right) \\ 
\left[ \left\{ \Psi _{J}\left[ p_{\eta },p_{\eta ^{\prime }\eta }\right]
\right\} ,v\right]%
\end{array}%
\right] \right\} ,\left[ v\right] ,\lambda \right) \right)
\end{eqnarray*}%
with:%
\begin{equation*}
\mathbf{A}=A_{\left\{ \left[ 
\begin{array}{c}
\left( p_{l_{i}},p_{l_{i}^{\prime }l_{i}},k_{l_{1}^{\prime }}\right) \\ 
\left[ \left\{ \Psi _{J}\left[ p_{\eta },p_{\eta ^{\prime }\eta }\right]
\right\} ,v\right]%
\end{array}%
\right] \right\} }
\end{equation*}%
and:%
\begin{eqnarray*}
&&\delta \lambda \underline{\nabla }_{\lambda }\Psi _{J,s_{p_{\eta ^{\prime
}\eta }}^{\otimes k_{\eta ^{\prime }\eta }}}^{\otimes k_{\eta ^{\prime }\eta
}}\left( \left[ p_{\eta },p_{\eta ^{\prime }\eta }\right] ^{k_{\eta ^{\prime
}\eta }}/\left[ \left[ p_{\eta },p_{\eta ^{\prime }\eta }\right] ^{k_{\eta
^{\prime }\eta }}\right] ,\left\{ \left[ 
\begin{array}{c}
\left( p_{l_{i}},p_{l_{i}^{\prime }l_{i}},k_{l_{1}^{\prime }}\right) \\ 
\left[ \left\{ \Psi _{J}\left[ p_{\eta },p_{\eta ^{\prime }\eta }\right]
\right\} ,v\right]%
\end{array}%
\right] \right\} ,\left[ v\right] ,\lambda \left( \left[ \left\{ \Psi _{J}%
\left[ p_{\eta },p_{\eta ^{\prime }\eta }\right] \right\} ,v\right] \right)
\right) \\
&=&\frac{\partial }{\partial \lambda }\Psi _{J,s_{p_{\eta ^{\prime }\eta
}}^{\otimes k_{\eta ^{\prime }\eta }}}^{\otimes k_{\eta ^{\prime }\eta
}}\left( \left[ p_{\eta },p_{\eta ^{\prime }\eta }\right] ^{k_{\eta ^{\prime
}\eta }}/\left[ \left[ p_{\eta },p_{\eta ^{\prime }\eta }\right] ^{k_{\eta
^{\prime }\eta }}\right] ,\left\{ \left[ 
\begin{array}{c}
\left( p_{l_{i}},p_{l_{i}^{\prime }l_{i}},k_{l_{1}^{\prime }}\right) \\ 
\left[ \left\{ \Psi _{J}\left[ l^{\prime }\eta ,\eta \right] \right\} ,v%
\right]%
\end{array}%
\right] \right\} ,\left[ v\right] ,\lambda \right) \\
&&+\left( \left( A_{\lambda }\right) _{k}^{k^{\prime }}\Psi _{J,s_{p_{\eta
^{\prime }\eta }}^{\otimes k_{\eta ^{\prime }\eta }}}^{\otimes k_{\eta
^{\prime }\eta }}\left( \left( \left[ p_{\eta },p_{\eta ^{\prime }\eta }%
\right] ^{k_{\eta ^{\prime }\eta }}/\left[ \left[ p_{\eta },p_{\eta ^{\prime
}\eta }\right] ^{k_{\eta ^{\prime }\eta }}\right] \right) _{k^{\prime
}},\left\{ \left[ 
\begin{array}{c}
\left( p_{l_{i}},p_{l_{i}^{\prime }l_{i}},k_{l_{1}^{\prime }}\right) \\ 
\left[ \left\{ \Psi _{J}\left[ p_{\eta },p_{\eta ^{\prime }\eta }\right]
\right\} ,v\right]%
\end{array}%
\right] \right\} ,\left[ v\right] ,\lambda \right) \right)
\end{eqnarray*}%
are some covariant derivatives. Indices $k$ are some local coordinates in
the space of maps:%
\begin{equation*}
\left( \left[ p_{\eta },p_{\eta ^{\prime }\eta }\right] ^{k_{\eta ^{\prime
}\eta }}/\left[ \left[ p_{\eta },p_{\eta ^{\prime }\eta }\right] ^{k_{\eta
^{\prime }\eta }}\right] \right)
\end{equation*}%
so that the connections are matrices.

The transitions generalizes (\ref{TRM}). For two states $A$ and $A^{\prime }$%
, it writes: 
\begin{eqnarray}
&&\left\langle A^{\prime },\lambda ^{\prime }\right\vert T_{\lambda \lambda
^{\prime }}\left\vert \Psi _{J}\left( U_{j}/\left[ U_{j}\right] ,\left\{ 
\underline{\mathbf{\hat{\Lambda}}}^{\left[ k_{i}\right] }\right\} ,\lambda
\right) \right\rangle \\
&=&\left\langle \Psi _{J}\left( U_{j}/\left[ U_{j}\right] ,\left\{ 
\underline{\mathbf{\hat{\Lambda}}}^{\left[ k_{i}\right] }\right\} ,\lambda
+\delta \lambda \right) \right\vert  \notag \\
&&\times N\left( \left( \left\{ \left[ p_{\eta },p_{\eta ^{\prime }\eta }%
\right] ^{k_{\eta ^{\prime }\eta }}\right\} \right) _{\lambda
}\hookrightarrow \left( \left\{ \left[ p_{\eta },p_{\eta ^{\prime }\eta }%
\right] ^{k_{\eta ^{\prime }\eta }}\right\} \right) _{\lambda ^{\prime
}}\right)  \notag \\
&&\times \exp \left( i\int \delta \lambda S\left( \Psi _{J,s_{p_{\eta
^{\prime }\eta }}^{\otimes k_{\eta ^{\prime }\eta }}}^{\otimes k_{\eta
^{\prime }\eta }},\frac{\delta }{\delta \Psi _{J,s_{p_{\eta ^{\prime }\eta
}}^{\otimes k_{\eta ^{\prime }\eta }}}^{\otimes k_{\eta ^{\prime }\eta }}},%
\underline{\nabla }_{\left[ 
\begin{array}{c}
\left( p_{l_{i}},p_{l_{i}^{\prime }l_{i}},k_{l_{1}^{\prime }}\right) \\ 
\left[ \left\{ \Psi _{J}\left[ p_{\eta },p_{\eta ^{\prime }\eta }\right]
\right\} ,v\right]%
\end{array}%
\right] }\underline{\nabla }_{_{\lambda }}\right) \right) \left\vert
A,\lambda \right\rangle  \notag
\end{eqnarray}%
with $N\left( \left( \left\{ \left[ p_{\eta },p_{\eta ^{\prime }\eta }\right]
^{k_{\eta ^{\prime }\eta }}\right\} \right) _{\lambda }\hookrightarrow
\left( \left\{ \left[ p_{\eta },p_{\eta ^{\prime }\eta }\right] ^{k_{\eta
^{\prime }\eta }}\right\} \right) _{\lambda ^{\prime }}\right) $ the number,
or the volume of set of maps between the two sets of maps: 
\begin{equation*}
\left( \left\{ \left[ p_{\eta },p_{\eta ^{\prime }\eta }\right] ^{k_{\eta
^{\prime }\eta }}\right\} \right) _{\lambda }\hookrightarrow \left( \left\{ %
\left[ p_{\eta },p_{\eta ^{\prime }\eta }\right] ^{k_{\eta ^{\prime }\eta
}}\right\} \right) _{\lambda ^{\prime }}
\end{equation*}%
Note these maps do not always factor as:%
\begin{equation*}
\tprod \left( \left( \left[ p_{\eta },p_{\eta ^{\prime }\eta }\right]
^{k_{\eta ^{\prime }\eta }}\right) _{\lambda }\hookrightarrow \left( \left[
p_{\eta },p_{\eta ^{\prime }\eta }\right] ^{k_{\eta ^{\prime }\eta }}\right)
_{\lambda ^{\prime }}\right)
\end{equation*}%
Depending on the values of $\lambda $ some fields may be independent or
glued as one global object (see comments after (\ref{F})).

\part*{Part III States and operators approach}

We introduce an alternative approach, which is less general than the field
formalism described previously but retains the key aspect of decomposing
state space into entangled spaces. Beginning with a direct description of
states and operators, we retrieve the essential elements of the field
formalism. Notably, spaces of parameters dependent on the states emerge as a
characteristic feature.

\section{\textbf{States formulation}}

We begin by delineating states resulting from the decomposition of a
parameter space and the associated state space into entangled subspaces.
Projecting onto one of these subspaces allows us to describe states as
dependent on certain seemingly exogenous parameters. In fact, the
entanglement of the initial states is still present, by the way of
constraints conditioning hese parameters. This description retains the main
features introduced in the preceding sections of this work.

\subsection{States and \textbf{projection on partial states}}

We start with the decomposition of a state space into two subspaces
entangled through some constraints. These constraints arise as eigenvalues
of operators defined on the entire initial states space.

\subsubsection{Set up}

Consider $U$ divided in two different collections (plus identification
constraints) $\left( U^{\left( i\right) }\right) $ and $\left( U^{\left(
j\right) }\right) $. The $U^{\left( i\right) }$ nd $U^{\left( j\right) }$
stand for the previous $\cup U_{i}^{k}$, \ $\cup U_{j}^{l}$ in the field
approach. In term of states they encompass tensor products.

Disregarding first the identifications, one is left with a collection of
parameters: 
\begin{equation*}
\left\{ \left( U^{\left( i\right) }\mid U^{\left( j\right) }\right) \right\}
\end{equation*}%
with associated states:

\begin{equation}
\left( \hat{U}^{\left( i\right) }\mid \hat{U}^{\left( j\right) }\right)
\rightarrow \left\{ 
\begin{array}{c}
\left\vert \hat{U}^{\left( i\right) }\mid \hat{U}^{\left( j\right)
}\right\rangle \\ 
\left\langle \hat{U}^{\left( i\right) }\mid \hat{U}^{\left( j\right)
}\right\vert%
\end{array}%
\right.  \label{STTS}
\end{equation}%
which is a short cut for some combinations, up to constraints, of basis
states:%
\begin{equation*}
\sum_{k,l}a\left( u_{1}^{\left( i\right) }...u_{k}^{\left( i\right)
},u_{1}^{\left( j\right) }...u_{l}^{\left( j\right) }\right) \left\vert
u_{1}^{\left( i\right) }...u_{k}^{\left( i\right) }\mid u_{1}^{\left(
j\right) }...u_{l}^{\left( j\right) }\right\rangle
\end{equation*}%
The states (\ref{STTS}) form a basis of an internal spce.

The whole space being set of linear combinations:%
\begin{equation*}
\sum a\left( \hat{U}^{\left( i\right) },\hat{U}^{\left( j\right) }\right)
\left\vert \hat{U}^{\left( i\right) }\mid \hat{U}^{\left( j\right)
}\right\rangle
\end{equation*}%
We assume that the stts $\left\vert \hat{U}^{\left( i\right) }\mid \hat{U}%
^{\left( j\right) }\right\rangle $ can be divided in sectors through some
operator $M$ eigenvalues (or equivalently, they are defined as eigenstts of
this operator). This implies%
\begin{equation*}
\tprod\limits_{M=m}\left\vert \hat{U}^{\left( i\right) }\mid \hat{U}^{\left(
j\right) }\right\rangle
\end{equation*}

Assume also that $M$ is a combination of two operators $M=f\left(
M_{i},M_{j}\right) $ acting on the $\hat{U}^{\left( i\right) }$ and $\hat{U}%
^{\left( j\right) }$ part. Thus:%
\begin{equation*}
\sum_{m}\tprod\limits_{M=m}\left\vert \hat{U}^{\left( i\right) }\mid \hat{U}%
^{\left( j\right) }\right\rangle =\sum_{m}\tprod\limits_{\delta \left(
f\left( M_{i},U^{\left( j\right) }\right) -m\right) }\left\vert \hat{U}%
^{\left( i\right) }\right\rangle \left\vert \hat{U}^{\left( j\right)
}\right\rangle
\end{equation*}%
For a basis $\left\vert \hat{U}^{\left( j\right) }\right\rangle $ of
eigenstates of $M_{j}$, the constraint becomes $\delta \left( f\left( M_{i},%
\hat{U}^{\left( j\right) }\right) -m\right) $ and:%
\begin{equation*}
\sum_{m}\tprod\limits_{\delta \left( f\left( M_{i},U^{\left( j\right)
}\right) -m\right) }\equiv \tprod\limits_{\hat{U}^{\left( j\right)
}}^{\left( i\right) }
\end{equation*}

\subsubsection{Partial diagonalization of projection operator}

We will project the states space on the eigenstates of some operator. This
projection accounts for constraints between the two states spaces, so that
it follows several steps. An example base on some basic model is given in
appendix 8.

We define:%
\begin{equation*}
\hat{H}_{i,j}=\sum_{l}\hat{H}_{j}^{\left( l\right) }\hat{H}_{i}^{\left(
l\right) }
\end{equation*}%
where $\hat{H}_{j}^{\left( l\right) }$ and $\hat{H}_{i}^{\left( l\right) }$
act on the $\left\vert \hat{U}^{\left( i\right) }\right\rangle $ and $%
\left\vert \hat{U}^{\left( j\right) }\right\rangle $ respectvl. We assume $%
\hat{H}_{j}^{\left( l\right) }$ commute with $M_{j}$ and have common
eigenstates $\left\vert U^{\left( j\right) }\right\rangle $:%
\begin{eqnarray*}
\tprod\limits_{\hat{U}^{\left( j\right) }}^{\left( i\right)
}\sum_{l}\left\vert U^{\left( j\right) }\right\rangle \left\langle U^{\left(
j\right) }\right\vert \hat{H}_{j}^{\left( l\right) }\left\vert U^{\left(
j\right) }\right\rangle \left\langle U^{\left( j\right) }\right\vert \hat{H}%
_{i}^{\left( l\right) }\tprod\limits_{\hat{U}^{\left( j\right) }}^{\left(
i\right) } &=&\left\vert U^{\left( j\right) }\right\rangle
\tprod\limits_{U^{\left( j\right) }}^{\left( i\right) }\sum_{l}\left\langle
U^{\left( j\right) }\right\vert \hat{H}_{j}^{\left( l\right) }\left\vert
U^{\left( j\right) }\right\rangle \hat{H}_{i}^{\left( l\right)
}\tprod\limits_{U^{\left( j\right) }}^{\left( i\right) }\left\langle
U^{\left( j\right) }\right\vert \\
&\equiv &\left\vert U^{\left( j\right) }\right\rangle
\tprod\limits_{U^{\left( j\right) }}^{\left( i\right) }\hat{H}_{i}\left(
U^{\left( j\right) }\right) \tprod\limits_{U^{\left( j\right) }}^{\left(
i\right) }\left\langle U^{\left( j\right) }\right\vert
\end{eqnarray*}%
We consider the operators $\tprod\limits_{U^{\left( j\right) }}^{\left(
i\right) }\hat{H}_{i}\left( U^{\left( j\right) }\right)
\tprod\limits_{U^{\left( j\right) }}^{\left( i\right) }$ and their
eigenstates:%
\begin{equation}
\tprod\limits_{U^{\left( j\right) }}^{\left( i\right) }\hat{H}_{i}\left(
U^{\left( j\right) }\right) \tprod\limits_{U^{\left( j\right) }}^{\left(
i\right) }=\tprod\limits_{U^{\left( j\right) }}^{\left( i\right)
}\sum_{l}\left\langle U^{\left( j\right) }\right\vert \hat{H}_{j}^{\left(
l\right) }\left\vert U^{\left( j\right) }\right\rangle \hat{H}_{i}^{\left(
l\right) }\tprod\limits_{U^{\left( j\right) }}^{\left( i\right) }\rightarrow
\left\vert \lambda \left( U^{\left( j\right) }\right) ,\left\{ \mathbf{%
\Lambda }_{i}\left( U^{\left( j\right) }\right) \right\} _{i}\right\rangle
\label{P}
\end{equation}%
The $\mathbf{\Lambda }_{i}\left( U^{\left( j\right) }\right) $ eigenvalues
of set of operators $\left\{ \mathbf{\hat{\Lambda}}_{i}\left( U^{\left(
j\right) }\right) \right\} $. Operators$\left\{ \mathbf{\hat{\Lambda}}%
_{i}\left( U^{\left( j\right) }\right) \right\} $ describe some degeneracy.
As an example we can consider: 
\begin{equation*}
\left\{ \mathbf{\Lambda }_{i}\left( U^{\left( j\right) }\right) \right\}
_{i}=\cup _{n}\left\{ \mathbf{\Lambda }_{i}^{n}\left( U^{\left( j\right)
}\right) \right\}
\end{equation*}%
for some fixed set of parameters $\mathbf{\Lambda }_{i}\left( U^{\left(
j\right) }\right) $. This corresponds to a constraint 
\begin{equation}
f\left( H,\left\{ \mathbf{\hat{\Lambda}}_{i}\left( U^{\left( j\right)
}\right) \right\} \right)  \label{FHL}
\end{equation}%
and the state space decomposing accordingly to:%
\begin{equation*}
\oplus \left\vert \left( \mathbf{\Lambda }_{i}\left( U^{\left( j\right)
}\right) \right) _{1}...\left( \mathbf{\Lambda }_{i}\left( U^{\left(
j\right) }\right) \right) _{n}\right\rangle
\end{equation*}%
with:%
\begin{equation*}
\lambda \left( U^{\left( j\right) }\right) =\hat{f}\left( \left( \mathbf{%
\Lambda }_{i}\left( U^{\left( j\right) }\right) \right) _{1}...\left( 
\mathbf{\Lambda }_{i}\left( U^{\left( j\right) }\right) \right) _{n}\right)
\end{equation*}%
solving (\ref{FHL}) on state:%
\begin{equation*}
\left\vert \left( \mathbf{\Lambda }_{i}\left( U^{\left( j\right) }\right)
\right) _{1}...\left( \mathbf{\Lambda }_{i}\left( U^{\left( j\right)
}\right) \right) _{n}\right\rangle
\end{equation*}

We can write the states arising in (\ref{P}) as a linear combination:%
\begin{equation}
\left\vert \lambda \left( U^{\left( j\right) }\right) ,\left\{ \mathbf{%
\Lambda }_{i}\left( U^{\left( j\right) }\right) \right\} _{i}\right\rangle
=\sum \psi \left( \mathbf{\Lambda }^{i}\left( U^{\left( j\right) }\right)
\right) \left\vert \lambda \left( U^{\left( j\right) }\right) ,\mathbf{%
\Lambda }^{i}\left( U^{\left( j\right) }\right) \right\rangle  \label{ps}
\end{equation}%
As a consequence:%
\begin{equation*}
\left\vert \lambda \left( U^{\left( j\right) }\right) ,\left\{ \mathbf{%
\Lambda }_{i}\left( U^{\left( j\right) }\right) \right\} _{i}\right\rangle
=\left\vert \lambda \left( U^{\left( j\right) }\right) ,\mathbf{S}\left( 
\mathbf{\Lambda }\left( U^{\left( j\right) }\right) \right) \right\rangle
\end{equation*}%
where $\mathbf{S}\left( \mathbf{\Lambda }\left( U^{\left( j\right) }\right)
\right) $ is the entire space defined by the $\mathbf{\Lambda }\left(
U^{\left( j\right) }\right) $. \ The states arising in (\ref{ps}) are the of
the same type as the ones defined in the first part. They represent states
that are series depending on increasing sequences of parameters, forming a
vld of pnts.

In terms of functionals, this translates in states:%
\begin{equation*}
\psi \left( \lambda \left( U^{\left( j\right) }\right) ,\mathbf{S}\left( 
\mathbf{\Lambda }\left( U^{\left( j\right) }\right) \right) \right)
\end{equation*}%
a functional defined on the space $\mathbf{S}\left( \mathbf{\Lambda }\left(
U^{\left( j\right) }\right) \right) $.

The $\mathbf{\Lambda }_{i}\left( U^{\left( j\right) }\right) $ are defined
on projected space $\tprod\limits_{U^{\left( j\right) }}^{\left( i\right) }$%
. The $\mathbf{\Lambda }_{i}\left( U^{\left( j\right) }\right) $ are
function of the constraint:%
\begin{equation*}
\mathbf{\Lambda }_{i}\left( U^{\left( j\right) }\right) \equiv \mathbf{%
\Lambda }_{i}\left( \left[ U^{\left( j\right) }\right] \right)
\end{equation*}%
wth $\left[ U^{\left( j\right) }\right] $ denotes the components of $%
U^{\left( j\right) }$ that define $\tprod\limits_{U^{\left( j\right)
}}^{\left( i\right) }$.

As an exemple, we can consider in first approximation that the constrait at
the quadratic rdr:%
\begin{eqnarray*}
&&H\left( \left( \left( \left\{ \mathbf{\hat{\Lambda}}_{i}\left( U^{\left(
j\right) }\right) \right\} .\left\{ \mathbf{\hat{\Lambda}}_{i}\left(
U^{\left( j\right) }\right) \right\} -\lambda ^{2}\left( U^{\left( j\right)
}\right) -\alpha \left( U^{\left( j\right) }\right) \right) ^{2}\right)
^{\otimes n}\right) \\
&\rightarrow &\left( \left\{ \mathbf{\Lambda }_{i}\left( U^{\left( j\right)
}\right) \right\} .\left\{ \mathbf{\Lambda }_{i}\left( U^{\left( j\right)
}\right) \right\} -\lambda ^{2}\left( U^{\left( j\right) }\right) -\alpha
\left( U^{\left( j\right) }\right) \right) ^{2}=0
\end{eqnarray*}%
where the cnstraint has one realization on each state $\left\vert \left( 
\mathbf{\Lambda }_{i}\left( U^{\left( j\right) }\right) \right)
_{1}...\left( \mathbf{\Lambda }_{i}\left( U^{\left( j\right) }\right)
\right) _{n}\right\rangle $:%
\begin{equation*}
\left( \left( \left( \mathbf{\Lambda }_{i}\left( U^{\left( j\right) }\right)
\right) _{1},...,\left( \mathbf{\Lambda }_{i}\left( U^{\left( j\right)
}\right) \right) _{n}\right) ^{2}-\lambda ^{2}\left( U^{\left( j\right)
}\right) -\alpha \left( U^{\left( j\right) }\right) \right) ^{2}=0
\end{equation*}

\subsubsection{Projected states}

We assume that the projectn $\tprod\limits_{U^{\left( j\right) }}^{\left(
i\right) }$ are defined as some eigenstates of an operator $M_{i}$:%
\begin{equation*}
\tprod\limits_{U^{\left( j\right) }}^{\left( i\right)
}=\tprod\limits_{M_{i}=m\left( U^{\left( j\right) }\right) }^{\left(
i\right) }=\tprod\limits_{M_{i}=m\left( \left[ U^{\left( j\right) }\right]
\right) }^{\left( i\right) }
\end{equation*}%
The $\mathbf{\hat{\Lambda}}_{i}\left( U^{\left( j\right) }\right) $ commute
with $M_{i}$ and thus there is a relation:%
\begin{equation*}
f\left( \left\{ \mathbf{\Lambda }_{i}\left( \left[ U^{\left( j\right) }%
\right] \right) \right\} \right) =m\left( \left[ U^{\left( j\right) }\right]
\right)
\end{equation*}%
we assume that this relation may be inverted, which implies that: 
\begin{equation*}
vect\left\{ \left\vert U^{\left( j\right) }\right\rangle \left\vert \lambda
\left( U^{\left( j\right) }\right) ,\left\{ \mathbf{\Lambda }_{i}\left( %
\left[ U^{\left( j\right) }\right] \right) \right\} _{i}\right\rangle
\right\}
\end{equation*}%
can also be written:%
\begin{equation*}
vect\left\{ \left\vert U^{\left( j\right) }/\left[ U^{\left( j\right) }%
\right] ,\left[ U^{\left( j\right) }\left( \left\{ \mathbf{\Lambda }%
_{i}\right\} \right) \right] \right\rangle \left\vert \lambda \left(
U^{\left( j\right) }\right) ,\left\{ \mathbf{\Lambda }_{i}\right\}
\right\rangle \right\} =vect\left\{ \left\vert U^{\left( j\right) /\left\{ 
\mathbf{\Lambda }_{i}\right\} },\left[ U^{\left( j\right) }\left( \left\{ 
\mathbf{\Lambda }_{i}\right\} \right) \right] \right\rangle \left\vert
\lambda \left( U^{\left( j\right) }\right) ,\left\{ \mathbf{\Lambda }%
_{i}\right\} \right\rangle \right\}
\end{equation*}%
for any given set of $\left\{ \mathbf{\Lambda }_{i}\right\} $, whereas as in
part one $\left[ U^{\left( j\right) }\left( \left\{ \mathbf{\Lambda }%
_{i}\right\} \right) \right] $ is the subspace of parameters that is
expressed as a function of the $\left\{ \mathbf{\Lambda }_{i}\right\} $. The
remaining independent parameters $U^{\left( j\right) }/\left[ U^{\left(
j\right) }\right] $ wil be also written $U^{\left( j\right) /\left\{ \mathbf{%
\Lambda }_{i}\right\} }$.

We gather the $\left\vert U^{\left( j\right) /\left\{ \mathbf{\Lambda }%
_{i}\right\} },U^{\left( j\right) }\left( \left\{ \mathbf{\Lambda }%
_{i}\right\} \right) \right\rangle \left\vert \lambda \left( U^{\left(
j\right) }\right) ,\mathbf{\Lambda }_{i}\right\rangle $ with a given $%
\lambda \left( U^{\left( j\right) }\right) $ and consider:%
\begin{equation*}
vect\left\{ \left\vert U^{\left( j\right) /\left\{ \mathbf{\Lambda }%
_{i}\right\} },U^{\left( j\right) }\left( \left\{ \mathbf{\Lambda }%
_{i}\right\} \right) \right\rangle \left\vert \lambda \left( U^{\left(
j\right) }\right) ,\left\{ \mathbf{\Lambda }_{i}\right\} \right\rangle
\right\} _{\lambda \left( U^{\left( j\right) }\right) =\lambda }
\end{equation*}

Thus any states in this space writes:%
\begin{eqnarray*}
&&\int_{\lambda \left( U^{\left( j\right) }\right) =\lambda }\psi \left(
\lambda ,U^{\left( j\right) /\left\{ \mathbf{\Lambda }_{i}\right\} },\left[
U^{\left( j\right) }\left( \left\{ \mathbf{\Lambda }_{i}\right\} \right) %
\right] \right) \left\vert U^{\left( j\right) /\left\{ \mathbf{\Lambda }%
_{i}\right\} },U^{\left( j\right) }\left( \left\{ \mathbf{\Lambda }%
_{i}\right\} \right) \right\rangle \left\vert \lambda ,\mathbf{\Lambda }%
_{i}\right\rangle d\mathbf{\Lambda }_{i} \\
&=&\sum \int_{\lambda \left( U^{\left( j\right) }\right) =\lambda }\psi
\left( \lambda ,U^{\left( j\right) /\mathbf{\Lambda }^{i}},\left[ U^{\left(
j\right) }\left( \left\{ \mathbf{\Lambda }_{i}\right\} \right) \right]
\right) \left\vert U^{\left( j\right) /\mathbf{\Lambda }^{i}},\left[
U^{\left( j\right) }\left( \left\{ \mathbf{\Lambda }_{i}\right\} \right) %
\right] \right\rangle \left\vert \lambda ,\mathbf{\Lambda }^{i}\right\rangle
d\mathbf{\Lambda }_{i}
\end{eqnarray*}%
which leads to describe such state by the functional:%
\begin{equation*}
\psi \left( \lambda ,U^{\left( j\right) /\mathbf{\Lambda }^{i}},\left[
U^{\left( j\right) }\left( \mathbf{\Lambda }\right) \right] ,\mathbf{S}%
\left( \mathbf{\Lambda }\right) \right)
\end{equation*}%
a functional of the function $U^{\left( j\right) /\mathbf{\Lambda }%
^{i}}\left( \mathbf{\Lambda }\right) $ over $\mathbf{S}\left( \mathbf{%
\Lambda }\right) $.

\subsubsection{Remark: inverting the relation}

If we assume the $\left\vert \lambda \left( U^{\left( j\right) }\right)
,\left\{ \mathbf{\Lambda }_{i}\right\} \right\rangle $ can be written in
terms of a common basis $\left\vert \mathbf{\bar{\Lambda}}_{i}\right\rangle $
independent of $U^{\left( j\right) }$:

\begin{equation}
\left\vert \lambda \left( U^{\left( j\right) }\right) ,\mathbf{\Lambda }%
_{i}\left( U^{\left( j\right) }\right) \right\rangle =\int h\left( \lambda
\left( U^{\left( j\right) }\right) ,\mathbf{\Lambda }_{i}\left( U^{\left(
j\right) }\right) ,\mathbf{\bar{\Lambda}}_{i}\right) \left\vert \mathbf{\bar{%
\Lambda}}_{i}\right\rangle d\mathbf{\bar{\Lambda}}_{i}  \label{CHNG}
\end{equation}%
\begin{equation*}
\left\vert U^{\left( j\right) /\left\{ \mathbf{\hat{\Lambda}}_{i}\right\} },%
\left[ U^{\left( j\right) }\left( \left\{ \mathbf{\Lambda }_{i}\right\}
\right) \right] \right\rangle \left\vert \lambda \left( U^{\left( j\right)
}\right) ,\left\{ \mathbf{\Lambda }_{i}\right\} \right\rangle =\sum_{\mathbf{%
\Lambda }_{i}}\left\vert U^{\left( j\right) /\mathbf{\hat{\Lambda}}_{i}},%
\left[ U^{\left( j\right) }\left( \left\{ \mathbf{\Lambda }_{i}\right\}
\right) \right] \right\rangle h\left( \lambda \left( U^{\left( j\right)
}\right) ,\mathbf{\Lambda }_{i}\left( U^{\left( j\right) }\right) ,\mathbf{%
\bar{\Lambda}}_{i}\right) \left\vert \mathbf{\bar{\Lambda}}_{i}\right\rangle
\end{equation*}%
where as before $\left[ U^{\left( j\right) }\left( \left\{ \mathbf{\bar{%
\Lambda}}_{i}\right\} \right) \right] $ represents the degrees of freedom of 
$U^{\left( j\right) }$ that can be expressed as function of $\left\{ \mathbf{%
\bar{\Lambda}}_{i}\right\} $.

The $\mathbf{\Lambda }_{i}$ are of the form $\mathbf{\Lambda }^{\otimes i}$
and $\left[ U^{\left( j\right) }\right] $ denotes the components of $%
U^{\left( j\right) }$ that define $\tprod\limits_{U^{\left( j\right)
}}^{\left( i\right) }$. If $h\left( \lambda \left( U^{\left( j\right)
}\right) ,\mathbf{\hat{\Lambda}}_{i}\left( U^{\left( j\right) }\right) ,%
\mathbf{\bar{\Lambda}}_{i}\right) $ is invertible, we have: 
\begin{eqnarray*}
&&\sum_{\mathbf{\Lambda }_{i}\left( U^{\left( j\right) }\right) }h^{\dag
}\left( \lambda \left( U^{\left( j\right) }\right) ,\left\{ \mathbf{\Lambda }%
_{i}\left( U^{\left( j\right) }\right) \right\} ,\left\{ \mathbf{\bar{\Lambda%
}}_{i}\right\} \right) \left\vert U^{\left( j\right) /\mathbf{\Lambda }_{i}},%
\left[ U^{\left( j\right) }\left( \left\{ \mathbf{\bar{\Lambda}}_{i}\right\}
\right) \right] \right\rangle \left\vert \lambda \left( U^{\left( j\right)
}\right) ,\left\{ \mathbf{\bar{\Lambda}}_{i}\right\} \right\rangle \\
&=&\sum_{\mathbf{\Lambda }_{i}\left( U^{\left( j\right) }\right) }\sum_{%
\mathbf{\bar{\Lambda}}_{i}^{\prime }}\left\vert U^{\left( j\right) /\mathbf{%
\bar{\Lambda}}_{i}},\mathbf{\bar{\Lambda}}_{i}\right\rangle h^{\dag }\left(
\lambda \left( U^{\left( j\right) }\right) ,\mathbf{\hat{\Lambda}}_{i}\left(
U^{\left( j\right) }\right) ,\mathbf{\bar{\Lambda}}_{i}\right) h\left(
\lambda \left( U^{\left( j\right) }\right) ,\mathbf{\hat{\Lambda}}_{i}\left(
U^{\left( j\right) }\right) ,\left\{ \mathbf{\bar{\Lambda}}_{i}^{\prime
}\right\} \right) \left\vert \left\{ \mathbf{\bar{\Lambda}}_{i}^{\prime
}\right\} \right\rangle \\
&=&\left\vert U^{\left( j\right) /\mathbf{\Lambda }_{i}},\left[ U^{\left(
j\right) }\left( \left\{ \mathbf{\bar{\Lambda}}_{i}\right\} \right) \right]
\right\rangle \left\vert \lambda \left( U^{\left( j\right) }\right) ,\left\{ 
\mathbf{\bar{\Lambda}}_{i}\right\} \right\rangle
\end{eqnarray*}%
However, if $h\left( \lambda \left( U^{\left( j\right) }\right) ,\mathbf{%
\Lambda }_{i}\left( U^{\left( j\right) }\right) ,\mathbf{\bar{\Lambda}}%
_{i}\right) $\ is not invertible, we can write only on some subspac:%
\begin{eqnarray*}
&&\sum_{\mathbf{\Lambda }_{i}\left( U^{\left( j\right) }\right) }h^{\dag
}\left( \lambda \left( U^{\left( j\right) }\right) ,\mathbf{\Lambda }%
_{i}\left( U^{\left( j\right) }\right) ,o_{i}\right) \left\vert U^{\left(
j\right) /\mathbf{\Lambda }_{i}},\mathbf{\bar{\Lambda}}_{i}\right\rangle
\left\vert \lambda \left( U^{\left( j\right) }\right) ,\mathbf{\hat{\Lambda}}%
_{i}\right\rangle \\
&=&\sum_{\mathbf{\Lambda }_{i}\left( U^{\left( j\right) }\right) }\sum_{%
\mathbf{\Lambda }_{i}^{\prime }\left( U^{\left( j\right) }\right)
}\left\vert U^{\left( j\right) /\mathbf{\Lambda }_{i}},\mathbf{\bar{\Lambda}}%
_{i}\right\rangle h^{\dag }\left( \lambda \left( U^{\left( j\right) }\right)
,\mathbf{\Lambda }_{i}\left( U^{\left( j\right) }\right) ,\mathbf{\bar{%
\Lambda}}_{i}\right) h\left( \lambda \left( U^{\left( j\right) }\right) ,%
\mathbf{\Lambda }_{i}\left( U^{\left( j\right) }\right) ,\mathbf{\bar{\Lambda%
}}_{i}^{\prime }\right) \left\vert \mathbf{\bar{\Lambda}}_{i}^{\prime
}\right\rangle \\
&=&\sum_{\mathbf{\bar{\Lambda}}_{i}\left( U^{\left( j\right) }\right)
}\left\vert U^{\left( j\right) /\mathbf{\Lambda }_{i}},\left( \mathbf{%
\Lambda }_{i}\right) \right\rangle \left\vert \lambda \left( U^{\left(
j\right) }\right) ,\mathbf{\bar{\Lambda}}_{i}\right\rangle \left\vert 
\mathbf{\bar{\Lambda}}_{i}\right\rangle
\end{eqnarray*}

\subsection{States \textbf{Covariantly}}

The degeneracy for the projected prtr wrts:%
\begin{equation*}
\left[ \left[ \left( U^{\left( i\right) }\right) ^{\prime }\right] \left(
U^{\left( j\right) }\right) H_{i}^{\left( l\right) }\left( U^{\left(
j\right) }\right) _{\times \bar{V}_{U^{\left( i\right) },\left( U^{\left(
i\right) }\right) ^{\prime },U^{\left( j\right) }}^{l}}\left[ U^{\left(
i\right) }\right] ,\mathbf{\Lambda }_{i}\left( U^{\left( j\right) }\right) %
\right] =0
\end{equation*}%
for some of the $\left\{ \mathbf{\Lambda }_{i}\left( U^{\left( j\right)
}\right) \right\} $ arises for the more compact form (\ref{frma})%
\begin{equation*}
H_{i}\left( U^{\left( j\right) }\right) =H\left[ U^{\left( j\right) },%
\mathbf{\Lambda }_{i}\left( U^{\left( j\right) }\right) \right] 
\end{equation*}%
We rewrite this relation as:%
\begin{equation*}
G\left( H_{i}\left( U^{\left( j\right) }\right) ,\left\{ \mathbf{\Lambda }%
_{i}\left( U^{\left( j\right) }\right) \right\} \right) =0
\end{equation*}%
and consider that this arises from more general constraint on operators:%
\begin{equation}
G\left( \left\{ \mathbf{\Lambda }_{\alpha }\left( U^{\left( j\right)
}\right) \right\} \right) =0  \label{TN}
\end{equation}%
where operators $\mathbf{\Lambda }_{\alpha }\left( U^{\left( j\right)
}\right) $ are given by:%
\begin{equation*}
\mathbf{\Lambda }_{\alpha }\left( U^{\left( j\right) }\right) =\mathbf{%
\Lambda }_{0}\left( U^{\left( j\right) }\right) ,\mathbf{\Lambda }_{i}\left(
U^{\left( j\right) }\right) 
\end{equation*}%
Assuming that there are transformations $M^{\left( i\right) }\left\{ \mathbf{%
\Lambda }_{\alpha }\left( U^{\left( j\right) }\right) \right\} $ preserving:%
\begin{equation*}
G\left( M^{\left( i\right) }\left\{ \mathbf{\Lambda }_{\alpha }\left(
U^{\left( j\right) }\right) \right\} \right) =0
\end{equation*}%
These transformations may arise from:%
\begin{equation*}
M^{\left( i\right) }\left\{ \mathbf{\Lambda }_{\alpha }\left( U^{\left(
j\right) }\right) \right\} =\left\{ \mathbf{\Lambda }_{\alpha }\left( \left(
M^{\left( i\right) }\right) ^{-1}U^{\left( j\right) }\right) \right\} 
\end{equation*}%
We assume that the transformations translate on states: 
\begin{equation*}
\left\vert M^{\left( i\right) }\left\{ \mathbf{\Lambda }_{\alpha }\left(
U^{\left( j\right) }\right) \right\} \right\rangle =R\left( M^{\left(
i\right) }\right) \left\vert \left\{ \mathbf{\Lambda }_{\alpha }\left(
U^{\left( j\right) }\right) \right\} \right\rangle 
\end{equation*}%
so that we have:%
\begin{equation*}
\left\vert \left\{ \mathbf{\Lambda }_{\alpha }\left( U^{\left( j\right)
}\right) \right\} \right\rangle =R\left( M^{\left( i\right) }\right)
\left\vert \left( M^{\left( i\right) }\right) ^{-1}\left\{ \mathbf{\Lambda }%
_{\alpha }\left( U^{\left( j\right) }\right) \right\} \right\rangle 
\end{equation*}%
This means that the states $\left\vert \left\{ \mathbf{\Lambda }_{\alpha
}\left( U^{\left( j\right) }\right) \right\} \right\rangle $ depends on the
entire space (\ref{TN}): 
\begin{equation*}
\left\vert \left\{ \mathbf{\Lambda }_{\alpha }\left( U^{\left( j\right)
}\right) \right\} \right\rangle =\left\vert \Sigma \left( U^{\left( j\right)
}\right) \right\rangle 
\end{equation*}%
\begin{equation*}
\left\{ \mathbf{\Lambda }_{\alpha }\left( U^{\left( j\right) }\right)
\right\} =\cup \left\{ \left( \Lambda _{\alpha }\left( U^{\left( j\right)
}\right) \right) ^{k}\right\} 
\end{equation*}%
\begin{equation*}
\left\vert \left\{ \mathbf{\Lambda }_{\alpha }\left( U^{\left( j\right)
}\right) \right\} \right\rangle =\sum \int_{G\left( \left\{ \mathbf{\Lambda }%
_{\alpha }\left( U^{\left( j\right) }\right) \right\} \right) =0}g\left(
\Lambda _{\alpha ,1}\left( U^{\left( j\right) }\right) ,...\Lambda _{\alpha
,k}\left( U^{\left( j\right) }\right) \right) \left\vert \left( \Lambda
_{\alpha }\left( U^{\left( j\right) }\right) \right) ^{k}\right\rangle 
\end{equation*}%
where $\left\vert \left( \Lambda _{\alpha }\left( U^{\left( j\right)
}\right) \right) ^{k}\right\rangle $ is defined for given values of the
variables. The invariant state is thus an averaged state.

Considering rather representation $R\left( M^{\left( i\right) }\right) $, we
are led to consider components:

\begin{equation*}
\left\vert \Sigma \left( U^{\left( j\right) }\right) \right\rangle _{a}=\sum
\int_{G\left( \left\{ \mathbf{\Lambda }_{\alpha }\left( U^{\left( j\right)
}\right) \right\} \right) =0}g_{a}\left( \Lambda _{\alpha ,1}\left(
U^{\left( j\right) }\right) ,...\Lambda _{\alpha ,k}\left( U^{\left(
j\right) }\right) \right) \left\vert \left\{ \mathbf{\Lambda }_{\alpha
}\left( U^{\left( j\right) }\right) \right\} \right\rangle
\end{equation*}%
where $R\left( M^{\left( i\right) }\right) $ acts on these components:%
\begin{eqnarray*}
&&\left( R\left( M^{\left( i\right) }\right) \left\vert \Sigma \left(
U^{\left( j\right) }\right) \right\rangle \right) _{a} \\
&=&R\left( M^{\left( i\right) }\right) _{a}^{b}\sum \int_{G\left( \left\{ 
\mathbf{\Lambda }_{\alpha }\left( U^{\left( j\right) }\right) \right\}
\right) =0}g_{b}\left( \Lambda _{\alpha ,1}\left( U^{\left( j\right)
}\right) ,...\Lambda _{\alpha ,k}\left( U^{\left( j\right) }\right) \right)
\left\vert \left\{ \mathbf{\Lambda }_{\alpha }\left( U^{\left( j\right)
}\right) \right\} \right\rangle
\end{eqnarray*}%
assuming a transformation:%
\begin{equation*}
\left( M^{\left( i\right) }\right) ^{-1}\left\{ \mathbf{\Lambda }_{\alpha
}^{\prime }\left( U^{\left( j\right) }\right) \right\} =\mathbf{\lambda }%
\left( U^{\left( j\right) }\right) ,\left\{ \mathbf{\Lambda }_{i}\left(
U^{\left( j\right) }\right) \right\}
\end{equation*}%
\begin{equation*}
\left\vert \Sigma \left( U^{\left( j\right) }\right) \right\rangle
=\left\vert \left\{ \mathbf{\Lambda }_{\alpha }^{\prime }\left( U^{\left(
j\right) }\right) \right\} \right\rangle =\left\vert \left( M^{\left(
i\right) }\right) ^{-1}\left\{ \mathbf{\Lambda }_{\alpha }^{\prime }\left(
U^{\left( j\right) }\right) \right\} \right\rangle =\left\vert \mathbf{%
\lambda }\left( U^{\left( j\right) }\right) ,\left\{ \mathbf{\Lambda }%
_{i}\left( U^{\left( j\right) }\right) \right\} \right\rangle
\end{equation*}%
\begin{equation*}
\left\vert \left\{ \mathbf{\Lambda }_{\alpha }\left( U^{\left( j\right)
}\right) \right\} \right\rangle =\sum \int g\left( \mathbf{\lambda }\left(
U^{\left( j\right) }\right) ,\Lambda _{i,1}\left( U^{\left( j\right)
}\right) ...\Lambda _{i,k}\left( U^{\left( j\right) }\right) \right)
\left\vert \mathbf{\lambda }\left( U^{\left( j\right) }\right) ,\left\{ 
\mathbf{\Lambda }_{i}\left( U^{\left( j\right) }\right) \right\}
\right\rangle =\left\vert \mathbf{\lambda }\left( U^{\left( j\right)
}\right) ,\left\{ \mathbf{\Lambda }_{i}\left( U^{\left( j\right) }\right)
\right\} \right\rangle
\end{equation*}%
and we recover our previous formulation by choice of coordinates.

\subsection{Operators induced by states}

The states $\left\vert U^{\left( j\right) /\mathbf{\Lambda }_{i}},\left( 
\mathbf{\Lambda }_{i}\right) \right\rangle \left\vert \lambda \left(
U^{\left( j\right) }\right) ,\mathbf{\Lambda }_{i}\right\rangle $ and spaces:%
\begin{equation*}
vect\left\{ \left\vert U^{\left( j\right) /\mathbf{\Lambda }_{i}},\left\{ 
\mathbf{\Lambda }_{i}\right\} \right\rangle \left\vert \lambda \left(
U^{\left( j\right) }\right) ,\left\{ \mathbf{\Lambda }_{i}\right\}
\right\rangle \right\} _{\lambda \left( U^{\left( j\right) }\right) =\lambda
}
\end{equation*}%
can be derived from some field theory, if we consider the field operator:%
\begin{equation*}
\left\vert U^{\left( i^{\prime }\right) }\right\rangle \left\vert \hat{U}%
^{\left( j^{\prime }\right) }\right\rangle \Psi \left( \hat{U}^{\left(
j\right) },\hat{U}^{\left( j^{\prime }\right) },U^{\left( i\right)
},U^{\left( i^{\prime }\right) }\right) \left\langle \hat{U}^{\left(
j\right) }\right\vert \left\langle U^{\left( i\right) }\right\vert
\end{equation*}%
that can be projected through operator $\tprod\limits_{U^{\left( j\right)
},\lambda \left( U^{\left( j\right) }\right) }$. This projection yields:%
\begin{equation*}
\tprod\limits_{U^{\left( j\right) },\lambda \left( U^{\left( j\right)
}\right) }\rightarrow \Psi \left( U^{\left( j\right) /\mathbf{\Lambda }%
_{i}}\left( \mathbf{\Lambda }_{i}\right) \right) \left\vert \lambda ,\mathbf{%
\Lambda }_{i}\right\rangle \rightarrow \left\vert \lambda ,\mathbf{\Lambda }%
_{i}\right\rangle \Psi \left( U^{\left( j\right) /\mathbf{\Lambda }%
_{i}}\left( \mathbf{\Lambda }_{i}\right) ,\lambda \right) \left\langle
\lambda ,\mathbf{\Lambda }_{i}\right\vert
\end{equation*}%
The spcc $\left( \hat{o}_{i}\right) $ ndwd wt $G\left( \lambda \left(
U^{\left( j\right) }\right) ,\mathbf{\hat{\Lambda}}_{i}\left( U^{\left(
j\right) }\right) \right) $ depends on $\Psi \left( U^{\left( j\right) /%
\mathbf{\Lambda }_{i}}\left( \mathbf{\Lambda }_{i}\right) \right) $:%
\begin{equation*}
G\left( \lambda \left( U^{\left( j\right) }\right) ,\mathbf{\hat{\Lambda}}%
_{i}\left( U^{\left( j\right) }\right) \right) \rightarrow G\left( \Psi
\left( U^{\left( j\right) /\mathbf{\Lambda }_{i}}\left( \mathbf{\Lambda }%
_{i}\right) \right) \right)
\end{equation*}

\section{Operator formalism}

In this section, we rather start from an operator formalism and describe
directly the constraints and projections in terms of operators.

\subsection{Operators and constraints}

Keeping the decomposition (\ref{STTS}) presented in the previous section, we
write an operator under the following frm:%
\begin{equation}
\left[ \left( \hat{U}^{\left( i\right) }\right) ^{\prime }\mid \left( \hat{U}%
^{\left( j\right) }\right) ^{\prime }\right] \Psi \left[ \hat{U}^{\left(
i\right) }\mid \hat{U}^{\left( j\right) }\right] \equiv \int \left\vert
\left( \hat{U}^{\left( j\right) }\right) ^{\prime }\right\rangle \left\vert
\left( \hat{U}^{\left( i\right) }\right) ^{\prime }\right\rangle \Psi \left(
\left( \hat{U}^{\left( i\right) }\right) ^{\prime },\left( \hat{U}^{\left(
j\right) }\right) ^{\prime },\hat{U}^{\left( i\right) },\hat{U}^{\left(
j\right) }\right) \left\langle \hat{U}^{\left( i\right) }\right\vert
\left\langle \hat{U}^{\left( j\right) }\right\vert  \label{PRST}
\end{equation}%
where the $\left\vert \left( \hat{U}^{\left( j\right) }\right) ^{\prime
}\right\rangle \left\vert \left( \hat{U}^{\left( i\right) }\right) ^{\prime
}\right\rangle $ represent a basis (\ref{STTS}). Operator (\ref{PRST}) can
be decomposed as a sum:%
\begin{equation*}
\left[ \hat{U}^{\prime }\right] \Psi \left[ \hat{U}\right] \rightarrow
\sum_{\alpha }\int \left[ \left( \hat{U}^{\left( i\right) }\right) ^{\prime
}\mid \left( \hat{U}^{\left( j\right) }\right) ^{\prime }\right] \left( \Psi
_{i}^{\left( V\right) }\right) _{\times V_{\hat{U}^{\left( i\right) },\left( 
\hat{U}^{\left( i\right) }\right) ^{\prime },\hat{U}^{\left( j\right)
},\left( \hat{U}^{\left( j\right) }\right) ^{\prime }}^{\alpha }}\left( \Psi
_{j}^{\left( V\right) }\right) \left[ \hat{U}^{\left( i\right) }\mid \hat{U}%
^{\left( j\right) }\right] dV_{\hat{U}^{\left( i\right) },\left( \hat{U}%
^{\left( i\right) }\right) ^{\prime },\hat{U}^{\left( j\right) },\left( \hat{%
U}^{\left( j\right) }\right) ^{\prime }}^{\alpha }
\end{equation*}%
where the $V_{\hat{U}^{\left( i\right) },\left( \hat{U}^{\left( i\right)
}\right) ^{\prime },\hat{U}^{\left( j\right) },\left( \hat{U}^{\left(
j\right) }\right) ^{\prime }}^{\alpha }$ model some constraints and are
defined by equations:%
\begin{equation*}
V_{\hat{U}^{\left( i\right) },\left( \hat{U}^{\left( i\right) }\right)
^{\prime },\hat{U}^{\left( j\right) },\left( \hat{U}^{\left( j\right)
}\right) ^{\prime }}^{\alpha }:f^{\alpha }\left( \hat{U}^{\left( i\right)
},\left( \hat{U}^{\left( i\right) }\right) ^{\prime },\hat{U}^{\left(
j\right) },\left( \hat{U}^{\left( j\right) }\right) ^{\prime }\right) =0
\end{equation*}%
The sum is over the various constraints, or functions $f^{\alpha }$ ($\alpha 
$ describing any set of constraints).

For example assuming constraints factoring as:%
\begin{equation*}
g_{i}\left( \hat{U}^{\left( i\right) },\left( \hat{U}^{\left( i\right)
}\right) ^{\prime }\right) =g_{j}\left( \hat{U}^{\left( j\right) },\left( 
\hat{U}^{\left( j\right) }\right) ^{\prime }\right)
\end{equation*}%
where $g_{i}$ and $g_{j}$ may be multicomponents and take values $v\in V$,
one has:%
\begin{equation*}
\left[ \hat{U}^{\prime }\right] \Psi \left[ \hat{U}\right] \rightarrow
\int_{v\in V}\left[ \left( \hat{U}^{\left( i\right) }\right) ^{\prime }\mid
\left( \hat{U}^{\left( j\right) }\right) ^{\prime }\right] \underset{%
_{g_{i}\left( \hat{U}^{\left( i\right) },\left( \hat{U}^{\left( i\right)
}\right) ^{\prime }\right) =g_{j}\left( \hat{U}^{\left( j\right) },\left( 
\hat{U}^{\left( j\right) }\right) ^{\prime }\right) =v}}{\left( \Psi
_{i}^{\left( V\right) }\right) \times \left( \Psi _{j}^{\left( V\right)
}\right) }\left[ \hat{U}^{\left( i\right) }\mid \hat{U}^{\left( j\right) }%
\right] dv
\end{equation*}%
Usually we can think the constraints as some conservation relations. If $%
\hat{U}^{\left( i\right) }$ and $\hat{U}^{\left( j\right) }$ depend on some
variables $k_{i}$ and $k_{j}$, the constraints are of the form: $%
k_{i}-k_{i}^{\prime }=k_{j}-k_{j}^{\prime }$.

The constraints may be written using $\delta $ functions:%
\begin{eqnarray*}
\left[ \hat{U}^{\prime }\right] \Psi \left[ \hat{U}\right] &=&\sum_{\alpha
}\int \left[ \left( \hat{U}^{\left( i\right) }\right) ^{\prime }\mid \left( 
\hat{U}^{\left( j\right) }\right) ^{\prime }\right] \left( \Psi _{i}^{\left(
V\right) }\right) \delta \left( V_{\hat{U}^{\left( i\right) },\left( \hat{U}%
^{\left( i\right) }\right) ^{\prime },\hat{U}^{\left( j\right) },\left( \hat{%
U}^{\left( j\right) }\right) ^{\prime }}^{\alpha }\right) \left( \Psi
_{j}^{\left( V\right) }\right) \left[ \hat{U}^{\left( i\right) }\mid \hat{U}%
^{\left( j\right) }\right] \\
&&\times d\left( \hat{U}^{\left( i\right) },\left( \hat{U}^{\left( i\right)
}\right) ^{\prime },\hat{U}^{\left( j\right) },\left( \hat{U}^{\left(
j\right) }\right) ^{\prime }\right)
\end{eqnarray*}

More is said about the constraints in appendix 9. We describe these
constraints as equations satisfied by operators, and recover the states
resulting from these constraints.

\subsection{Projection operators}

\subsubsection{Global projection operator}

The operator defining the projection are defined following the previous
decomposition: $\Psi _{i}^{\left( V\right) }\rightarrow \hat{H}_{i}^{\left(
V\right) }$, $\Psi _{j}^{\left( V\right) }\rightarrow \hat{H}_{j}^{\left(
V\right) }$:%
\begin{equation}
\sum_{V}\left[ \left( \hat{U}^{\left( i\right) }\right) ^{\prime }\mid
\left( \hat{U}^{\left( j\right) }\right) ^{\prime }\right] \left( \hat{H}%
_{i}^{\left( V\right) }\right) \underset{V_{\hat{U}^{\left( i\right)
},\left( \hat{U}^{\left( i\right) }\right) ^{\prime },\hat{U}^{\left(
j\right) },\left( \hat{U}^{\left( j\right) }\right) ^{\prime }}}{\times }%
\left( \hat{H}_{j}^{\left( V\right) }\right) \left[ \hat{U}^{\left( i\right)
}\mid \hat{U}^{\left( j\right) }\right]   \label{PRPRT}
\end{equation}%
We assume that for all $\hat{U}^{\left( i\right) },\left( \hat{U}^{\left(
i\right) }\right) ^{\prime }$, operators $\delta \left( V_{\hat{U}^{\left(
i\right) },\left( \hat{U}^{\left( i\right) }\right) ^{\prime },\hat{U}%
^{\left( j\right) },\left( \hat{U}^{\left( j\right) }\right) ^{\prime
}}^{\alpha }\right) \hat{H}_{j}^{\left( V\right) }$ have common eigenstates $%
\left\vert U^{\left( j\right) }\right\rangle $ with eigenvalues: 
\begin{equation*}
h_{\alpha }\left( U^{\left( j\right) },\hat{U}^{\left( i\right) },\left( 
\hat{U}^{\left( i\right) }\right) ^{\prime }\right) \delta \left( \bar{V}_{%
\hat{U}^{\left( i\right) },\left( \hat{U}^{\left( i\right) }\right) ^{\prime
},U^{\left( j\right) }}^{\alpha }\right) 
\end{equation*}%
where $\bar{V}_{\hat{U}^{\left( i\right) },\left( \hat{U}^{\left( i\right)
}\right) ^{\prime }}$ are possible remaining constraints such as $\left\{
g_{i}\left( \hat{U}^{\left( i\right) },\left( \hat{U}^{\left( i\right)
}\right) ^{\prime }\right) =v\in \bar{V}_{\hat{U}^{\left( i\right) },\left( 
\hat{U}^{\left( i\right) }\right) ^{\prime },U^{\left( j\right) }}^{\alpha
}\right\} $. Disregarding $\bar{V}_{\hat{U}^{\left( i\right) },\left( \hat{U}%
^{\left( i\right) }\right) ^{\prime },U^{\left( j\right) }}^{\alpha }$ we
also assume that the eigenvalue $h_{\alpha }\left( U^{\left( j\right) },\hat{%
U}^{\left( i\right) },\left( \hat{U}^{\left( i\right) }\right) ^{\prime
}\right) $ is build from operators $\mathbf{\Lambda }_{i}\left( U^{\left(
j\right) }\right) $ commuting with $\hat{H}_{j}^{\left( V\right) }$ written
as $\hat{H}_{j}^{\left( V\right) }\left( \mathbf{\Lambda }_{i}\left(
U^{\left( j\right) }\right) \right) $ (similarly to a band hamiltonian)
including some degeneracy. Operator (\ref{PRPRT}) writes:%
\begin{eqnarray*}
&&\int \left\vert U^{\left( j\right) }\right\rangle \left\langle U^{\left(
j\right) }\right\vert \left[ \left( \hat{U}^{\left( i\right) }\right)
^{\prime }\mid \left( \hat{U}^{\left( j\right) }\right) ^{\prime }\right]
\left( \hat{H}_{i}^{\left( V\right) }\right) \left( H_{j}^{\left( V\right)
}\right) \left[ \hat{U}^{\left( i\right) }\mid \hat{U}^{\left( j\right) }%
\right] \left\vert U^{\left( j\right) }\right\rangle \left\langle U^{\left(
j\right) }\right\vert  \\
&&\times \delta \left( V_{\hat{U}^{\left( i\right) },\left( \hat{U}^{\left(
i\right) }\right) ^{\prime },\hat{U}^{\left( j\right) },\left( \hat{U}%
^{\left( j\right) }\right) ^{\prime }}^{\alpha }\right) d\left( \hat{U}%
^{\left( j\right) },\left( \hat{U}^{\left( j\right) }\right) ^{\prime
}\right)  \\
&=&\int \left\vert U^{\left( j\right) }\right\rangle \left[ \left( \hat{U}%
^{\left( i\right) }\right) ^{\prime }\right] \left( \hat{H}_{i}^{\left(
V\right) }\right) \left[ \hat{U}^{\left( i\right) }\right] h_{\alpha
}^{V}\left( U^{\left( j\right) },\hat{U}^{\left( i\right) },\left( \hat{U}%
^{\left( i\right) }\right) ^{\prime }\right) \left\langle U^{\left( j\right)
}\right\vert  \\
&\equiv &\int \left\vert U^{\left( j\right) }\right\rangle \left[ \left( 
\hat{U}^{\left( i\right) }\right) ^{\prime }\right] \left( \hat{H}_{i}\left(
U^{\left( j\right) }\right) \right) \left[ \hat{U}^{\left( i\right) }\right]
\left\langle U^{\left( j\right) }\right\vert 
\end{eqnarray*}%
The operator $\left[ \left( \hat{U}^{\left( i\right) }\right) ^{\prime }%
\right] \left( \hat{H}_{i}\left( U^{\left( j\right) }\right) \right) \left[ 
\hat{U}^{\left( i\right) }\right] $ may include constraints $\bar{V}_{\hat{U}%
^{\left( i\right) },\left( \hat{U}^{\left( i\right) }\right) ^{\prime
},U^{\left( j\right) }}^{\alpha }$ so that (\ref{PRPRT} writes:%
\begin{equation}
\equiv \sum_{\alpha }\left[ \left( \hat{U}^{\left( i\right) }\right)
^{\prime }\mid U^{\left( j\right) }\right] \left( \hat{H}_{i}^{\left(
V\right) }\right) \underset{\bar{V}_{\hat{U}^{\left( i\right) },\left( \hat{U%
}^{\left( i\right) }\right) ^{\prime },U^{\left( j\right) }}^{\alpha }}{%
\times }\left( H_{j}^{\left( V\right) }\right) \left[ \hat{U}^{\left(
i\right) }\mid U^{\left( j\right) }\right]   \label{PRPT}
\end{equation}%
where the $h_{\alpha }\left( U^{\left( j\right) },\hat{U}^{\left( i\right)
},\left( \hat{U}^{\left( i\right) }\right) ^{\prime }\right) $ are
eigenvalues of the $H_{j}^{\left( V\right) }$ in states $\left\vert
U^{\left( j\right) }\right\rangle $.

\subsubsection{Generalization: local diagonalization}

The previous set up can be generalized. If $\delta \left( V_{\hat{U}^{\left(
i\right) },\left( \hat{U}^{\left( i\right) }\right) ^{\prime },\hat{U}%
^{\left( j\right) },\left( \hat{U}^{\left( j\right) }\right) ^{\prime
}}^{\alpha }\right) \hat{H}_{j}^{\left( V\right) }$ have common eigenstates: 
\begin{equation*}
\left\vert U^{\left( j\right) }\right\rangle _{\left( \hat{U}^{\left(
i\right) },\left( \hat{U}^{\left( i\right) }\right) ^{\prime }\right)
}\equiv \left\vert U^{\left( j\right) }\right\rangle _{i}
\end{equation*}%
depending on $\left( \hat{U}^{\left( i\right) },\left( \hat{U}^{\left(
i\right) }\right) ^{\prime }\right) $ with eigenvalues: 
\begin{equation*}
h_{\alpha }\left( U^{\left( j\right) },\hat{U}^{\left( i\right) },\left( 
\hat{U}^{\left( i\right) }\right) ^{\prime }\right) \delta \left( \bar{V}_{%
\hat{U}^{\left( i\right) },\left( \hat{U}^{\left( i\right) }\right) ^{\prime
},U^{\left( j\right) }}^{\alpha }\right) 
\end{equation*}%
where $\bar{V}_{\hat{U}^{\left( i\right) },\left( \hat{U}^{\left( i\right)
}\right) ^{\prime },U^{\left( j\right) }}^{\alpha }$ are the possible
remaining constraints such as: 
\begin{equation*}
\left\{ g_{i}\left( \hat{U}^{\left( i\right) },\left( \hat{U}^{\left(
i\right) }\right) ^{\prime }\right) =v\in \bar{V}_{\hat{U}^{\left( i\right)
},\left( \hat{U}^{\left( i\right) }\right) ^{\prime },U^{\left( j\right)
}}^{\alpha }\right\} 
\end{equation*}%
and the decomposition (\ref{PRPRT}) becomes:%
\begin{eqnarray}
&&\sum_{\alpha }\int \left\vert U^{\left( j\right) }\right\rangle
_{i}\left\langle U^{\left( j\right) }\right\vert _{i}\left[ \left( \hat{U}%
^{\left( i\right) }\right) ^{\prime }\mid \left( \hat{U}^{\left( j\right)
}\right) ^{\prime }\right] \left( \hat{H}_{i}^{\left( V\right) }\right) 
\underset{V_{\hat{U}^{\left( i\right) },\left( \hat{U}^{\left( i\right)
}\right) ^{\prime },\hat{U}^{\left( j\right) },\left( \hat{U}^{\left(
j\right) }\right) ^{\prime }}^{\alpha }}{\times }\left( H_{j}^{\left(
V\right) }\right) \left[ \hat{U}^{\left( i\right) }\mid \hat{U}^{\left(
j\right) }\right] \left\vert U^{\left( j\right) }\right\rangle
_{i}\left\langle U^{\left( j\right) }\right\vert _{i}  \notag \\
&=&\sum_{\alpha }\int \left\vert U^{\left( j\right) }\right\rangle _{i}\left[
\left( \hat{U}^{\left( i\right) }\right) ^{\prime }\right] \left( \hat{H}%
_{i}^{\left( V\right) }\right) \left[ \hat{U}^{\left( i\right) }\right]
h_{\alpha }\left( U^{\left( j\right) },\hat{U}^{\left( i\right) },\left( 
\hat{U}^{\left( i\right) }\right) ^{\prime }\right) \delta \left( \bar{V}_{%
\hat{U}^{\left( i\right) },\left( \hat{U}^{\left( i\right) }\right) ^{\prime
},U^{\left( j\right) }}^{\alpha }\right) \left\langle U^{\left( j\right)
}\right\vert _{i}d\left( \hat{U}^{\left( i\right) },\left( \hat{U}^{\left(
i\right) }\right) ^{\prime }\right)   \notag \\
&\equiv &\sum_{\alpha }\left[ \left( \hat{U}^{\left( i\right) }\right)
^{\prime }\mid U_{\left( \hat{U}^{\left( i\right) },\left( \hat{U}^{\left(
i\right) }\right) ^{\prime }\right) }^{\left( j\right) }\right] \left( \hat{H%
}_{i}^{\left( V\right) }\right) \underset{\bar{V}_{\hat{U}^{\left( i\right)
},\left( \hat{U}^{\left( i\right) }\right) ^{\prime },U^{\left( j\right)
}}^{\alpha }}{\times }\left( H_{j}^{\left( V\right) }\right) \left[ \hat{U}%
^{\left( i\right) }\mid U_{\left( \hat{U}^{\left( i\right) },\left( \hat{U}%
^{\left( i\right) }\right) ^{\prime }\right) }^{\left( j\right) }\right] 
\label{DBS}
\end{eqnarray}%
Writing $F_{\hat{U}^{\left( i\right) },\left( \hat{U}^{\left( i\right)
}\right) ^{\prime }}^{\alpha }$ the fiber of $V_{\hat{U}^{\left( i\right)
},\left( \hat{U}^{\left( i\right) }\right) ^{\prime },\hat{U}^{\left(
j\right) },\left( \hat{U}^{\left( j\right) }\right) ^{\prime }}^{\alpha }$
over a given $\hat{U}^{\left( i\right) },\left( \hat{U}^{\left( i\right)
}\right) ^{\prime }$, we can decompose: 
\begin{equation*}
\left\{ \hat{U}^{\left( i\right) },\left( \hat{U}^{\left( i\right) }\right)
^{\prime }\right\} =\cup _{F^{\alpha }}\left\{ \hat{U}^{\left( i\right)
},\left( \hat{U}^{\left( i\right) }\right) ^{\prime }\right\} _{F^{\alpha }}
\end{equation*}%
with: 
\begin{equation*}
\left\{ \hat{U}^{\left( i\right) },\left( \hat{U}^{\left( i\right) }\right)
^{\prime }\right\} _{F^{\alpha }}=\left\{ \hat{U}^{\left( i\right) },\left( 
\hat{U}^{\left( i\right) }\right) ^{\prime }/F_{\hat{U}^{\left( i\right)
},\left( \hat{U}^{\left( i\right) }\right) ^{\prime }}^{\alpha }=F^{\alpha
}\right\} 
\end{equation*}

Along a basis $\left\{ \hat{U}^{\left( i\right) },\left( \hat{U}^{\left(
i\right) }\right) ^{\prime }\right\} _{F^{\alpha }}$, the operator: 
\begin{equation*}
\delta \left( V_{\hat{U}^{\left( i\right) },\left( \hat{U}^{\left( i\right)
}\right) ^{\prime },\hat{U}^{\left( j\right) },\left( \hat{U}^{\left(
j\right) }\right) ^{\prime }}^{\alpha }\right) \hat{H}_{j}^{\left( V\right) }
\end{equation*}%
writes: 
\begin{equation*}
\delta \left( \left( \hat{U}^{\left( j\right) },\left( \hat{U}^{\left(
j\right) }\right) ^{\prime }\right) \subset F^{\alpha }\right) \hat{H}%
_{j}^{\left( V\right) }
\end{equation*}%
We can assume that the $\delta \left( \left( \hat{U}^{\left( j\right)
},\left( \hat{U}^{\left( j\right) }\right) ^{\prime }\right) \subset
F^{\alpha }\right) \hat{H}_{j}^{\left( V\right) }$ have common eigenstates $%
\left\vert U^{\left( j\right) }\right\rangle _{F^{\alpha }}$.Decomposing the
measure of integration along the fiber:%
\begin{equation*}
d\left( \hat{U}^{\left( i\right) },\left( \hat{U}^{\left( i\right) }\right)
^{\prime }\right) =d\left\{ \hat{U}^{\left( i\right) },\left( \hat{U}%
^{\left( i\right) }\right) ^{\prime }\right\} _{F^{\alpha }}dF^{\alpha }
\end{equation*}%
formula (\ref{DBS}) becomes ultimately:

\begin{eqnarray*}
&&\sum_{\alpha }\int \left\vert U^{\left( j\right) }\right\rangle
_{F^{\alpha }}\left[ \left( \hat{U}^{\left( i\right) }\right) ^{\prime }%
\right] \left( \hat{H}_{i}^{\left( V\right) }\right) \left[ \hat{U}^{\left(
i\right) }\right] h_{\alpha }\left( U^{\left( j\right) },\hat{U}^{\left(
i\right) },\left( \hat{U}^{\left( i\right) }\right) ^{\prime }\right)
d\left\{ \hat{U}^{\left( i\right) },\left( \hat{U}^{\left( i\right) }\right)
^{\prime }\right\} _{F^{\alpha }}\left\langle U^{\left( j\right)
}\right\vert _{F^{\alpha }}dF^{\alpha } \\
&\equiv &\sum_{\alpha }\int \left[ \left( \hat{U}^{\left( i\right) }\right)
^{\prime }\mid U_{F^{\alpha }}^{\left( j\right) }\right] \left( \hat{H}%
_{iF^{\alpha }}^{\left( V\right) }\right) \left( H_{j}^{\left( V\right)
}\right) \left[ \hat{U}^{\left( i\right) }\mid U_{F^{\alpha }}^{\left(
j\right) }\right] dF^{\alpha }
\end{eqnarray*}%
with:%
\begin{equation*}
\hat{H}_{iF^{\alpha }}^{\left( V\right) }=\hat{H}_{i}^{\left( V\right)
}\delta \left( \hat{U}^{\left( i\right) },\left( \hat{U}^{\left( i\right)
}\right) ^{\prime }\subset F^{\alpha }\right)
\end{equation*}

\subsection{Projection on partial states: Basis of eigenstates depending on $%
U^{\left( j\right) }$}

Consider the sum of projections along the eigenstates $U^{\left( j\right) }$:%
\begin{equation*}
\sum \dprod_{j}\oplus \dprod_{j}=1
\end{equation*}

and decompose the operator (\ref{PRPT} along this decomposition. For each
eigenstate, we obtain an operator acting on the $\left( U^{\left( i\right)
}\right) $ degrees of freedom of the states space:%
\begin{eqnarray*}
&&\sum_{\alpha }\left[ \left( \hat{U}^{\left( i\right) }\right) ^{\prime
}\mid U^{\left( j\right) }\right] \left( \hat{H}_{i}^{\left( V\right)
}\right) \underset{\bar{V}_{\hat{U}^{\left( i\right) },\left( \hat{U}%
^{\left( i\right) }\right) ^{\prime },U^{\left( j\right) }}^{\alpha }}{%
\times }\left( H_{j}^{\left( V\right) }\right) \left[ \hat{U}^{\left(
i\right) }\mid U^{\left( j\right) }\right] \\
&&\overset{\dprod_{j}\oplus \dprod_{j}}{{\Huge \rightarrow }}\sum_{\alpha }%
\left[ \left( \hat{U}^{\left( i\right) }\right) ^{\prime }\right] \left(
U_{j}\right) \left( \hat{H}_{i}^{\left( l\right) }\right) \underset{\bar{V}_{%
\hat{U}^{\left( i\right) },\left( \hat{U}^{\left( i\right) }\right) ^{\prime
},U^{\left( j\right) }}^{\alpha }}{\times }\left( U_{j}\right) \left[ \hat{U}%
^{\left( i\right) }\right]
\end{eqnarray*}

As before, we assume that: 
\begin{equation*}
\left[ \left( \hat{U}^{\left( i\right) }\right) ^{\prime }\right] \left(
U_{j}\right) \left( \hat{H}_{i}^{\left( l\right) }\right) \underset{\bar{V}_{%
\hat{U}^{\left( i\right) },\left( \hat{U}^{\left( i\right) }\right) ^{\prime
},U^{\left( j\right) }}^{\alpha }}{\times }\left( U_{j}\right) \left[ \hat{U}%
^{\left( i\right) }\right] 
\end{equation*}%
commute with a family of operators $\mathbf{\hat{\Lambda}}_{i}\left(
U^{\left( j\right) }\right) $ depending on $U^{\left( j\right) }$ and that
the relation:%
\begin{equation*}
f\left( H\left( U^{\left( j\right) }\right) ,\mathbf{\Lambda }_{i}\left(
U^{\left( j\right) }\right) ,\alpha \left( U^{\left( j\right) }\right)
\right) =0
\end{equation*}%
stands for each of the $U_{j}$. As a consequence, the diagonalization of:%
\begin{equation*}
\sum_{\alpha }\left[ \left( \hat{U}^{\left( i\right) }\right) ^{\prime }%
\right] \left( U_{j}\right) \left( \hat{H}_{i}^{\left( l\right) }\right) 
\underset{\bar{V}_{\hat{U}^{\left( i\right) },\left( \hat{U}^{\left(
i\right) }\right) ^{\prime },U^{\left( j\right) }}^{\alpha }}{\times }\left(
U_{j}\right) \left[ \hat{U}^{\left( i\right) }\right] 
\end{equation*}%
yields degenerated spaces: 
\begin{equation*}
\sum_{\alpha }\left[ \left( \hat{U}^{\left( i\right) }\right) ^{\prime }%
\right] \left( U_{j}\right) \left( \hat{H}_{i}^{\left( l\right) }\right) 
\underset{\bar{V}_{\hat{U}^{\left( i\right) },\left( \hat{U}^{\left(
i\right) }\right) ^{\prime },U^{\left( j\right) }}^{\alpha }}{\times }\left(
U_{j}\right) \left[ \hat{U}^{\left( i\right) }\right] \rightarrow \left[
\left( \hat{U}^{\left( i\right) }\right) ^{\prime }\right] \left(
U_{j}\right) H_{i}\left( U_{j}\right) \left[ \hat{U}^{\left( i\right) }%
\right] \rightarrow \left\vert \lambda \left( U_{j}\right) ,\mathbf{\Lambda }%
_{i}\left( U^{\left( j\right) }\right) \right\rangle 
\end{equation*}%
The constraints $\bar{V}_{\hat{U}^{\left( i\right) },\left( \hat{U}^{\left(
i\right) }\right) ^{\prime },U^{\left( j\right) }}^{\alpha }$ imply that the
states $\left\vert \lambda \left( U_{j}\right) ,\mathbf{\Lambda }_{i}\left(
U^{\left( j\right) }\right) \right\rangle $ do not generate the entire
initial states space. The relation%
\begin{equation}
f\left( H\left( U^{\left( j\right) }\right) ,\mathbf{\hat{\Lambda}}%
_{i}\left( U^{\left( j\right) }\right) ,\alpha \left( U^{\left( j\right)
}\right) \right) =0  \label{CTF}
\end{equation}%
allows thus to replace some parameters $\left[ U_{j}\right] $ as functional
of the eigenvalues $\left\{ \mathbf{\Lambda }_{i}\left( U^{\left( j\right)
}\right) ,\lambda \left( U_{j}\right) \right\} $ and states:%
\begin{equation*}
\left\vert \left( U_{j}\right) \right\rangle \left\vert \lambda \left(
U_{j}\right) ,\mathbf{\Lambda }_{i}\left( U^{\left( j\right) }\right)
\right\rangle 
\end{equation*}%
are linear combinations of states of the form:%
\begin{equation}
\left\vert \left( U_{j}/\left[ U_{j}\right] ,\left[ U_{j}\right] \left(
\left\{ \mathbf{\Lambda }_{i}\left( U^{\left( j\right) }\right) \right\}
\right) ,\lambda \left( U_{j}\right) \right) \right\rangle \left\vert
\lambda \left( U_{j}\right) ,\mathbf{\Lambda }_{i}\left( U^{\left( j\right)
}\right) \right\rangle   \label{CFR}
\end{equation}

The parameters $U_{j}/\left[ U_{j}\right] $ represent the remaining
parameters after using the constraint (\ref{CTF}) has been imposed and $%
\left[ U_{j}\right] \left( \left\{ \mathbf{\Lambda }_{i}\left( U^{\left(
j\right) }\right) \right\} \right) $ describes the part of $U_{j}$ written
as a function of $\left\{ \mathbf{\Lambda }_{i}\left( U^{\left( j\right)
}\right) \right\} $. Note that this replacement depends on (\ref{CTF}) and
thus on $\lambda \left( U_{j}\right) ,\mathbf{\Lambda }_{i}\left( U^{\left(
j\right) }\right) $. This change of variable is local: the parameters $%
U^{\left( j\right) }$ describing some fundamental state, determines the
eigenstates defined by $\left( \lambda \left( U_{j}\right) ,\mathbf{\Lambda }%
_{i}\left( U^{\left( j\right) }\right) \right) $ and thus the decomposition.
The entanglement between the $U^{\left( i\right) }$ and $U^{\left( j\right)
} $ degrees of freedom translate in this interdependence between the
parameters and the apparent system's degrees of freedom $U_{j}/\left[ U_{j}%
\right] $.

Remark also that the states (\ref{CFR}) are similar to the states described
in the first and second part of this work. There is local relation between
the apparent degrees of freedom and the effective parameter space described
by $\left( \lambda \left( U_{j}\right) ,\mathbf{\Lambda }_{i}\left(
U^{\left( j\right) }\right) \right) $.

\subsection{Projected operators for fixed value of $\protect\lambda $}

As in the first part, we can consider the linear span of the states such
that $\lambda \left( U_{j}\right) =\lambda $ and project operators along the
basis (\ref{CFR}). Starting with an operator:%
\begin{equation*}
\left[ \left( \hat{U}^{\left( i\right) }\right) ^{\prime }\mid \left( \hat{U}%
^{\left( j\right) }\right) ^{\prime }\right] \Psi \left[ \hat{U}^{\left(
i\right) }\mid \hat{U}^{\left( j\right) }\right]
\end{equation*}%
it can be expressed in a basis corresponding to the projection operator.
Introducing the change of basis:%
\begin{eqnarray*}
&&\left\vert \left( U_{j}/\left[ U_{j}\right] ,\left[ U_{j}\right] \left(
\left\{ \mathbf{\Lambda }_{i}\left( U^{\left( j\right) }\right) \right\}
\right) ,\lambda \left( U_{j}\right) \right) \right\rangle \left\vert
\lambda \left( U_{j}\right) ,\mathbf{\Lambda }_{i}\left( U^{\left( j\right)
}\right) \right\rangle \\
&=&\int g\left( U_{j},\lambda \left( U_{j}\right) ,\mathbf{\Lambda }%
_{i}\left( U^{\left( j\right) }\right) ,\hat{U}^{\left( i\right) },\hat{U}%
^{\left( j\right) }\right) \left\vert \hat{U}^{\left( i\right) }\mid \hat{U}%
^{\left( j\right) }\right\rangle d\hat{U}^{\left( i\right) }d\hat{U}^{\left(
i\right) }
\end{eqnarray*}%
we can write:%
\begin{eqnarray}
&&\left[ \left( \hat{U}^{\left( i\right) }\right) ^{\prime }\mid \left( \hat{%
U}^{\left( j\right) }\right) ^{\prime }\right] \Psi \left[ \hat{U}^{\left(
i\right) }\mid \hat{U}^{\left( j\right) }\right]  \label{FMR} \\
&=&\int g\left( U_{j},\lambda ,\mathbf{\Lambda }_{i}\left( U^{\left(
j\right) }\right) ,\left( \hat{U}^{\left( i\right) }\right) ^{\prime
},\left( \hat{U}^{\left( j\right) }\right) ^{\prime }\right)  \notag \\
&&\times \left[ \left\{ \mathbf{\Lambda }_{i}\left( U^{\left( j^{\prime
}\right) }\right) \right\} ,\lambda \mid U_{j^{\prime }}/\left[ U_{j^{\prime
}}\right] ,\left[ U_{j^{\prime }}\left( \left\{ \mathbf{\Lambda }_{i}\left(
U^{\left( j^{\prime }\right) }\right) \right\} \right) \right] \right] \Psi %
\left[ \left\{ \mathbf{\Lambda }_{i}\left( U^{\left( j\right) }\right)
\right\} ,\lambda \mid U_{j}/\left[ U_{j}\right] ,\left[ U_{j}\left( \left\{ 
\mathbf{\Lambda }_{i}\left( U^{\left( j\right) }\right) \right\} \right) %
\right] \right]  \notag \\
&&\times \int g^{\dag }\left( U_{j^{\prime }},\lambda \left( U_{j^{\prime
}}\right) ,\mathbf{\Lambda }_{i}\left( U^{\left( j\right) }\right) ,\hat{U}%
^{\left( i\right) },\hat{U}^{\left( j\right) }\right) d\left( \hat{U}%
^{\left( i\right) }\right) ^{\prime }d\left( \hat{U}^{\left( j\right)
}\right) ^{\prime }d\hat{U}^{\left( i\right) }d\hat{U}^{\left( j\right)
}d\lambda  \notag
\end{eqnarray}%
Formula (\ref{FMR}) defines an operator:%
\begin{equation}
\left[ \left\{ \mathbf{\Lambda }_{i}\left( U^{\left( j^{\prime }\right)
}\right) \right\} ,\lambda \mid U_{i^{\prime }}/\left[ U_{i^{\prime }}\right]
,\left[ U_{i^{\prime }}\left( \left\{ \mathbf{\Lambda }_{i}\left( U^{\left(
j^{\prime }\right) }\right) \right\} \right) \right] \right] \Psi \left[
\left\{ \mathbf{\Lambda }_{i}\left( U^{\left( j\right) }\right) \right\}
,\lambda \mid U_{i}/\left[ U_{i}\right] ,\left[ U_{i}\left( \left\{ \mathbf{%
\Lambda }_{i}\left( U^{\left( j\right) }\right) \right\} \right) \right] %
\right]  \label{PRTS}
\end{equation}

acting on states such that $\lambda \left( U_{j}\right) =\lambda $.

In the sequel, we will consider the notation:%
\begin{equation*}
U_{j}/\left[ U_{j}\right] ,\left[ U_{j}\left( \left\{ \mathbf{\Lambda }%
_{i}\left( U^{\left( j\right) }\right) \right\} \right) \right] =U_{j}/\left[
U_{j}\right] ,\left[ U_{j}\right] 
\end{equation*}%
to depict the decomposition of $U_{j}$ in which $\left[ U_{j}\left( \left\{ 
\mathbf{\Lambda }_{i}\left( U^{\left( j\right) }\right) \right\} \right) %
\right] $ is the part of $U_{j}$ rewritten using the constraints as
functions of $\left\{ \mathbf{\Lambda }_{i}\left( U^{\left( j\right)
}\right) \right\} $ and where $U_{j}/\left[ U_{j}\right] $ are the remaining
free parameters. As a consequence, the operator rewrites: 
\begin{equation*}
\left[ \left\{ \mathbf{\Lambda }_{i}\left( U^{\left( j^{\prime }\right)
}\right) \right\} ,\lambda \mid U_{j^{\prime }}/\left[ U_{j^{\prime }}\right]
,\left[ U_{j^{\prime }}\right] \right] \Psi \left[ \left\{ \mathbf{\Lambda }%
_{i}\left( U^{\left( j\right) }\right) \right\} ,\lambda \mid U_{j}/\left[
U_{j}\right] ,\left[ U_{j}\right] \right] 
\end{equation*}

Consider that the eigenstates of operators $\mathbf{\hat{\Lambda}}_{i}\left(
U^{\left( j\right) }\right) $ generate:%
\begin{equation*}
vect\left\{ \left\vert \lambda \left( U_{j}\right) ,\mathbf{\Lambda }%
_{i}\left( U^{\left( j\right) }\right) \right\rangle \right\} 
\end{equation*}%
so that a change of variables:%
\begin{eqnarray*}
\left\{ \mathbf{\Lambda }_{i}\left( U^{\left( j^{\prime }\right) }\right)
\right\}  &\rightarrow &\left\{ \mathbf{\Lambda }_{i}^{\prime }\right\}  \\
\left\{ \mathbf{\Lambda }_{i}\left( U^{\left( j\right) }\right) \right\} 
&\rightarrow &\left\{ \mathbf{\Lambda }_{i}\right\} 
\end{eqnarray*}%
allows to consider the parameters $\left\{ \mathbf{\Lambda }_{i}^{\prime
}\right\} $ and $\left\{ \mathbf{\Lambda }_{i}\right\} $ as exogeneous
variables. The change of variable is performed by integrals:%
\begin{eqnarray}
&&\left[ \left\{ \mathbf{\Lambda }_{i}^{\prime }\right\} ,\lambda \mid
U_{j^{\prime }}/\left[ U_{j^{\prime }}\right] ,\left[ U_{j^{\prime }}\right] %
\right] \Psi \left[ \left\{ \mathbf{\Lambda }_{i}\right\} ,\lambda \mid
U_{j}/\left[ U_{j}\right] ,\left[ U_{j}\right] \right]   \label{PRL} \\
&=&\int h^{\dag }\left( \lambda \left( U^{\left( j^{\prime }\right) }\right)
,\mathbf{\Lambda }_{i}\left( U^{\left( j^{\prime }\right) }\right) ,\mathbf{%
\Lambda }_{i}^{\prime }\right)   \notag \\
&&\times \left[ \left\{ \mathbf{\Lambda }_{i}\left( U^{\left( j^{\prime
}\right) }\right) \right\} ,\lambda \mid U_{j^{\prime }}/\left[ U_{j^{\prime
}}\right] ,\left[ U_{j^{\prime }}\right] \right] \Psi \left[ \left\{ \mathbf{%
\Lambda }_{i}\left( U^{\left( j\right) }\right) \right\} ,\lambda \mid U_{j}/%
\left[ U_{j}\right] ,\left[ U_{j}\right] \right]   \notag \\
&&\times h\left( \lambda \left( U^{\left( j\right) }\right) ,\mathbf{\Lambda 
}_{i}\left( U^{\left( j\right) }\right) ,\mathbf{\Lambda }_{i}\right)  
\notag
\end{eqnarray}%
where $h\left( \lambda \left( U^{\left( j\right) }\right) ,\mathbf{\Lambda }%
_{i}\left( U^{\left( j\right) }\right) ,\mathbf{\Lambda }_{i}\right) $
performs the change of basis, see (\ref{CHNG}).

\subsection{\textbf{Remark:}}

Similarly to part one, the previous approach can be generalized to a
multiple decomposition of the initial parametres space: $\left[ U^{\left(
i\right) }\mid U^{\left( j\right) }\mid U^{\left( k\right) }...\right] $.
Somes stats are prjctd, and some constraints $V_{lk}$ between $U^{\left(
l\right) }$, $U^{\left( k\right) }$ translates on the projected states
between the remaining degrees of freedom and the parameters arising in the
projection.

\section{Transformation $\left\vert U_{0}^{\left( j\right) }\right\rangle
\rightarrow \left\vert U^{\left( j\right) }\right\rangle $ and amplitudes}

\subsection{States description}

We start with the states where part of the $U^{\left( i\right) }$ are
written as functions of $\left\{ \mathbf{\bar{\Lambda}}_{i}\right\} $: 
\begin{equation}
\left\vert \lambda \left( U^{\left( j\right) }\right) ,\left\{ \mathbf{\bar{%
\Lambda}}_{i}\right\} \right\rangle \left\vert U^{\left( j\right) /\left\{ 
\mathbf{\hat{\Lambda}}_{i}\right\} },\left[ U^{\left( j\right) }\left(
\left\{ \mathbf{\Lambda }_{i}\right\} \right) \right] \right\rangle
=\left\vert \lambda \left( U^{\left( j\right) }\right) ,\left\{ \mathbf{\bar{%
\Lambda}}_{i}\right\} \right\rangle \left\vert h\left( \left\{ \mathbf{\bar{%
\Lambda}}_{i}\right\} \right) \right\rangle _{U^{\left( i\right) }}
\label{PRM}
\end{equation}%
where as before $\left[ U^{\left( j\right) }\left( \left\{ \mathbf{\bar{%
\Lambda}}_{i}\right\} \right) \right] $ represents the degrees of freedom of 
$U^{\left( j\right) }$ that can be expressed as function of $\left\{ \mathbf{%
\bar{\Lambda}}_{i}\right\} $. The subscript $U^{\left( i\right) }$ reminds
that 
\begin{equation*}
\left\vert h\left( \left\{ \mathbf{\bar{\Lambda}}_{i}\right\} \right)
\right\rangle _{U^{\left( i\right) }}=\left\vert U^{\left( j\right) /\left\{ 
\mathbf{\hat{\Lambda}}_{i}\right\} },\left[ U^{\left( j\right) }\left(
\left\{ \mathbf{\Lambda }_{i}\right\} \right) \right] \right\rangle 
\end{equation*}%
depends on $U^{\left( j\right) }$ through the constraints and the free
parameters $U^{\left( j\right) /\left\{ \mathbf{\hat{\Lambda}}_{i}\right\} }$%
.

Note that the identification is only local in general, due to the
constraints sets $\left\{ \mathbf{\bar{\Lambda}}_{i}\right\} $ and states $%
h\left( \left\{ \mathbf{\bar{\Lambda}}_{i}\right\} \right) $ are not
independent. This impacts the amplitudes between two states with different
parameters. We will use this remark below.

Gathering states such that $\lambda \left( U^{\left( j\right) }\right)
=\lambda $ the state writes:%
\begin{equation*}
\mathcal{H}_{\left( \lambda ,\left\{ \left\{ \mathbf{\bar{\Lambda}}%
_{i}\right\} \left( U_{\lambda }^{\left( j\right) }\right) ,U_{\lambda
}^{\left( j\right) }\right\} \right) }=\left\{ \left\vert \lambda ,\left\{ 
\mathbf{\bar{\Lambda}}_{i}\right\} \right\rangle \left\vert h\left( \left\{ 
\mathbf{\bar{\Lambda}}_{i}\right\} \right) \right\rangle _{U^{\left(
i\right) }}\right\} _{\lambda =\lambda \left( U^{\left( j\right) }\right) }
\end{equation*}

For later purpose, we consider a decomposition:%
\begin{equation*}
U^{\left( j\right) }\rightarrow \left( U^{\left( j\right) /p},U^{\left(
j\right) p}\right) 
\end{equation*}%
which is preserved for the eigenstates parameters:%
\begin{equation*}
\left\{ \mathbf{\bar{\Lambda}}_{i}\right\} =\left( \left\{ \mathbf{\bar{%
\Lambda}}_{i,p}\right\} ,\left\{ \mathbf{\bar{\Lambda}}_{i}\right\}
_{/p}\right) 
\end{equation*}%
so that the state $\left\vert h\left( \left\{ \mathbf{\bar{\Lambda}}%
_{i}\right\} \right) \right\rangle _{U^{\left( j\right) }}$ of this space
decomposes as:%
\begin{equation}
\left\vert \lambda ,\left\{ \mathbf{\bar{\Lambda}}_{i}\right\} \right\rangle
\left\vert h\left( \left\{ \mathbf{\bar{\Lambda}}_{i}\right\} \right)
\right\rangle _{U^{\left( j\right) }}=\left\vert \lambda ,\left\{ \mathbf{%
\bar{\Lambda}}_{i}\right\} _{/p}\right\rangle \left\vert h_{/p}\left(
\left\{ \mathbf{\bar{\Lambda}}_{i}\right\} _{/p}\right) \right\rangle
_{U^{\left( i\right) }}\left\vert \left\{ \mathbf{\bar{\Lambda}}%
_{i,p}\right\} \right\rangle \left\vert h_{p}\left( \left\{ \mathbf{\bar{%
\Lambda}}_{i,p}\right\} \right) \right\rangle _{U^{\left( i\right) }}
\label{TD}
\end{equation}%
Subspace spanned by states (\ref{TD}) are series expansion:%
\begin{equation}
\int g\left( \lambda ,\left\{ \mathbf{\bar{\Lambda}}_{i}\right\} _{/p},%
\mathbf{\bar{\Lambda}}_{i,p},h\left( \left\{ \mathbf{\bar{\Lambda}}%
_{i}\right\} _{/p}\right) ,h\left( \left\{ \mathbf{\bar{\Lambda}}%
_{i,p}\right\} \right) \right) \left\vert \lambda ,\left\{ \mathbf{\bar{%
\Lambda}}_{i}\right\} _{/p}\right\rangle \left\vert h\left( \left\{ \mathbf{%
\bar{\Lambda}}_{i}\right\} _{/p}\right) \right\rangle _{U^{\left( i\right)
}}\left\vert \left\{ \mathbf{\bar{\Lambda}}_{i,p}\right\} \right\rangle
\left\vert h\left( \left\{ \mathbf{\bar{\Lambda}}_{i,p}\right\} \right)
\right\rangle _{U^{\left( i\right) }}  \label{PM}
\end{equation}

States in (\ref{TD}) decompose the system between few degrees of freedom and
background.

\subsection{Transformation $\left\vert U_{0}^{\left( j\right) }\right\rangle
\rightarrow \left\vert U^{\left( j\right) }\right\rangle $}

As in first and second part, we aim at consider transitions between two stts
wth different values of $\lambda $. Our assumption is that

for $U_{0}^{\left( j\right) }$ and $U^{\left( j\right) }$ there exists: 
\begin{equation*}
T_{\lambda _{0}\lambda }:\mathcal{H}_{\left( \lambda _{0},\left\{ \left\{ 
\mathbf{\bar{\Lambda}}_{i}\right\} \left( U_{\lambda _{0}}^{\left( j\right)
}\right) ,U_{\lambda _{0}}^{\left( j\right) }\right\} \right) }\mathcal{%
\simeq H}_{\left( \lambda ,\left\{ \left\{ \mathbf{\bar{\Lambda}}%
_{i}\right\} \left( U_{\lambda }^{\left( j\right) }\right) ,U_{\lambda
}^{\left( j\right) }\right\} \right) }
\end{equation*}%
that is, the transfrormation $T_{\lambda _{0}\lambda }$:%
\begin{equation*}
T_{\lambda _{0}\lambda }:\left\{ \left\vert \lambda \left( U_{0}^{\left(
j\right) }\right) ,\mathbf{\bar{\Lambda}}_{i}\right\rangle \left\vert
h\left( \left\{ \mathbf{\bar{\Lambda}}_{i}\right\} \right) \right\rangle
_{U_{0}^{\left( j\right) }}\right\} _{\lambda _{0}}\rightarrow \left\{
\left\vert \lambda \left( U^{\left( j\right) }\right) ,\mathbf{\bar{\Lambda}}%
_{i}\right\rangle \left\vert h\left( \left\{ \mathbf{\bar{\Lambda}}%
_{i}\right\} \right) \right\rangle _{U^{\left( j\right) }}\right\} _{\lambda
}
\end{equation*}

\subsection{Amplitudes of transitions}

Given that the states for which we compute the transitions are analog to
those in the first part, the derivation follows the same stp.

We consider the infinitesimal transition $\delta T_{\lambda ^{\prime
}\lambda }$.

Given that $\delta T_{\lambda ^{\prime }\lambda }$ transforms states from
one spaces to another space, the transition include some transport operator.
In addition, the transition depends on the operator acting on the set wth
gnvl $\lambda $. As a consequence, the transition depends on two parts.

The transport, involves "derivatives" $\left\{ \mathbf{\hat{\Lambda}}%
_{i}\right\} _{\lambda ^{\prime }}-\left\{ \mathbf{\hat{\Lambda}}%
_{i}\right\} _{\lambda }+A\left( \left\{ \mathbf{\hat{\Lambda}}_{i}\right\}
_{\lambda }\right) $, the connection $A\left( \left\{ \mathbf{\hat{\Lambda}}%
_{i}\right\} _{\lambda }\right) $ correcting the fact that $\left\{ \mathbf{%
\hat{\Lambda}}_{i}\right\} _{\lambda ^{\prime }}$ nd $\left\{ \mathbf{\hat{%
\Lambda}}_{i}\right\} _{\lambda }$ \ do not act on the same space.

The second part of the transition depends on operators $\left\{ \mathbf{\hat{%
\Lambda}}_{i}\right\} $ and $\left\{ \Pi _{\mathbf{\hat{\Lambda}}%
_{i}}\right\} $ acting with multiplication by $\mathbf{\bar{\Lambda}}_{i}$
and differentiation by $\mathbf{\bar{\Lambda}}_{i}$:%
\begin{eqnarray*}
\mathbf{\hat{\Lambda}}_{i}\left( \lambda \right) \left\vert \lambda ,\left\{ 
\mathbf{\bar{\Lambda}}_{i}\right\} \right\rangle \left\vert h\left( \left\{ 
\mathbf{\bar{\Lambda}}_{i}\right\} \right) \right\rangle _{U^{\left(
j\right) }} &=&\mathbf{\bar{\Lambda}}_{i}\left\vert \lambda ,\left\{ \mathbf{%
\bar{\Lambda}}_{i}\right\} \right\rangle \left\vert h\left( \left\{ \mathbf{%
\bar{\Lambda}}_{i}\right\} \right) \right\rangle _{U^{\left( j\right) }} \\
\Pi _{\mathbf{\hat{\Lambda}}_{i}}\left( \lambda \right) \left\vert \lambda
,\left\{ \mathbf{\bar{\Lambda}}_{i}\right\} \right\rangle \left\vert h\left(
\left\{ \mathbf{\bar{\Lambda}}_{i}\right\} \right) \right\rangle _{U^{\left(
j\right) }} &=&\nabla _{\mathbf{\bar{\Lambda}}_{i}}\left\vert \lambda
,\left\{ \mathbf{\bar{\Lambda}}_{i}\right\} \right\rangle \left\vert h\left(
\left\{ \mathbf{\bar{\Lambda}}_{i}\right\} \right) \right\rangle _{U^{\left(
j\right) }}
\end{eqnarray*}%
However, the second action also has to be corrected due to the constraints.
Modifying $\left\{ \mathbf{\bar{\Lambda}}_{i}\right\} \rightarrow \left\{ 
\mathbf{\bar{\Lambda}}_{i}\right\} +\delta \left\{ \mathbf{\bar{\Lambda}}%
_{i}\right\} $ through $\nabla _{\mathbf{\bar{\Lambda}}_{i}}$ modifies the
state $h\left( \left\{ \mathbf{\bar{\Lambda}}_{i}\right\} \right) $ and
should be modified including a covariant derivative.

We present the technical derivation of the transition in the next
paragraphs, starting with the case where the parameters are global, and then
consider the implications of only local identification.

\subsubsection{Global identification}

We assume in first approximation that the parameters $\mathbf{\bar{\Lambda}}%
_{i}$ are global. The amplitudes%
\begin{equation}
\left\langle \lambda +\delta \lambda ,\left\{ \mathbf{\bar{\Lambda}}%
_{i}\right\} ^{\prime }\right\vert _{U^{\left( j\right) }}\left\langle
h^{\prime }\left( \left\{ \mathbf{\bar{\Lambda}}_{i}\right\} ^{\prime
}\right) \right\vert \delta T_{\lambda \lambda +\delta \lambda }\left\vert
\lambda ,\left\{ \mathbf{\bar{\Lambda}}_{i}\right\} \right\rangle \left\vert
h\left( \left\{ \mathbf{\bar{\Lambda}}_{i}\right\} \right) \right\rangle
_{U^{\left( j\right) }}  \label{PLTS}
\end{equation}%
are computed by pulling back the field state:%
\begin{equation*}
\left\langle \lambda +\delta \lambda ,\left\{ \mathbf{\bar{\Lambda}}%
_{i}\right\} ^{\prime }\right\vert _{U^{\left( j\right) }}\left\langle
h^{\prime }\left( \left\{ \mathbf{\bar{\Lambda}}_{i}\right\} ^{\prime
}\right) \right\vert \rightarrow \left\langle \lambda ,\left\{ \mathbf{\bar{%
\Lambda}}_{i}\right\} ^{\prime }\right\vert _{U^{\left( j\right)
}}\left\langle h\left( \left\{ \mathbf{\bar{\Lambda}}_{i}\right\} ^{\prime
}\right) \right\vert
\end{equation*}%
from $\mathcal{H}_{\left( \lambda +\delta \lambda ,\left\{ \left\{ \mathbf{%
\bar{\Lambda}}_{i}\right\} ^{\prime }\left( U_{\lambda +\delta \lambda
}^{\left( j\right) }\right) ,U_{\lambda +\delta \lambda }^{\left( j\right)
}\right\} \right) }$ to $\mathcal{H}_{\left( \lambda ,\left\{ \mathbf{\bar{%
\Lambda}}_{i}\left( U_{\lambda }^{\left( j\right) }\right) ,U_{\lambda
}^{\left( j\right) }\right\} \right) }$ by some parallel transport operator $%
P_{\lambda \lambda +\delta \lambda }$ and then computing the matrix elements
of:%
\begin{equation*}
P_{\lambda \lambda +\delta \lambda }\delta T_{\lambda \lambda +\delta
\lambda }
\end{equation*}

Assuming that the parameters $\left\{ \mathbf{\bar{\Lambda}}_{i}\right\} $
are global, the transport is trivial: we can identify the various spaces,
the transport involves only derivatives.

Formally, it transports $h\left( \left\{ \mathbf{\bar{\Lambda}}_{i}\right\}
\left( \lambda +\delta \lambda \right) \right) $ to $h\left( \left\{ \mathbf{%
\bar{\Lambda}}_{i}\right\} \left( \lambda \right) \right) $ and the parallel
transport is generated by the operator: 
\begin{equation*}
\left( \left\{ \mathbf{\bar{\Lambda}}_{i}\left( \lambda +\delta \lambda
\right) \right\} -\left\{ \mathbf{\bar{\Lambda}}_{i}\left( \lambda \right)
\right\} \right) \frac{\delta }{\delta \left\{ \mathbf{\bar{\Lambda}}%
_{i}\left( \lambda \right) \right\} }=\delta \lambda \frac{\partial \left\{ 
\mathbf{\bar{\Lambda}}_{i}\left( \lambda \right) \right\} }{\partial \lambda 
}\frac{\delta }{\delta \left\{ \mathbf{\bar{\Lambda}}_{i}\left( \lambda
\right) \right\} }
\end{equation*}%
so that exponentiation, the transport is:%
\begin{equation*}
P_{\lambda \lambda +\delta \lambda }=\exp \left( \int i\delta \lambda 
\underline{\nabla }_{\lambda }\left\{ \mathbf{\bar{\Lambda}}_{i}\left(
\lambda \right) \right\} \frac{\delta }{\delta \left\{ \mathbf{\bar{\Lambda}}%
_{i}\left( \lambda \right) \right\} }\mathcal{D}\left\{ \mathbf{\bar{\Lambda}%
}_{i}\left( \lambda \right) \right\} \right) 
\end{equation*}%
Inserting this opertor in the matrix elements of an arbitrary expression $%
\exp \left( iF\left( \Pi _{\left\{ \mathbf{\bar{\Lambda}}_{i}\left( \lambda
\right) \right\} }\right) \right) $, the saddle point computation allows to
replace:%
\begin{equation*}
\Pi _{\left\{ \mathbf{\bar{\Lambda}}_{i}\left( \lambda \right) \right\}
}\rightarrow \frac{\partial \left\{ \mathbf{\bar{\Lambda}}_{i}\left( \lambda
\right) \right\} }{\partial \lambda }
\end{equation*}%
As a consequence the transition operator involves the variables $\mathbf{%
\bar{\Lambda}}_{i}\left( \lambda \right) $ nd $\frac{\partial \left\{ 
\mathbf{\bar{\Lambda}}_{i}\left( \lambda \right) \right\} }{\partial \lambda 
}$. By exponentiation of infinitesimal matrix elements leads to transitions:%
\begin{eqnarray}
&&\left\langle \lambda ,\left\{ \mathbf{\bar{\Lambda}}_{i}\right\} ^{\prime
}\right\vert _{U^{\left( j\right) }}\left\langle h\left( \left\{ \mathbf{%
\bar{\Lambda}}_{i}\right\} ^{\prime }\right) \right\vert \exp \left(
i\int_{\lambda \left( \left( U^{\left( j\right) }\right) _{0}\right)
}^{\lambda \left( U^{\left( j\right) }\right) }L\left( \left\{ \mathbf{\bar{%
\Lambda}}_{i}\left( \lambda \right) \right\} ,\left\{ \frac{\partial \left\{ 
\mathbf{\bar{\Lambda}}_{i}\left( \lambda \right) \right\} }{\partial \lambda 
}\right\} \right) d\lambda \right) \left\vert \lambda _{0},\left\{ \mathbf{%
\bar{\Lambda}}_{i}\right\} \right\rangle \left\vert h_{0}\left( \left\{ 
\mathbf{\bar{\Lambda}}_{i}\right\} \right) \right\rangle _{U_{0}^{\left(
j\right) }}  \notag \\
&=&\int_{U^{\left( j\right) }}\left\langle h\left( \left\{ \mathbf{\bar{%
\Lambda}}_{i}\right\} ^{\prime }\right) \right\vert \int \exp \left(
i\int_{\left\{ \mathbf{\bar{\Lambda}}_{i}\left( \lambda _{0}\right) \right\}
=\left\{ \mathbf{\bar{\Lambda}}_{i}\right\} }^{\left\{ \mathbf{\bar{\Lambda}}%
_{i}\left( \lambda \right) \right\} =\left\{ \mathbf{\bar{\Lambda}}%
_{i}\right\} ^{\prime }}L\left( \left\{ \mathbf{\bar{\Lambda}}_{i}\left(
\lambda \right) \right\} ,\left\{ \frac{\partial \left\{ \mathbf{\bar{\Lambda%
}}_{i}\left( \lambda \right) \right\} }{\partial \lambda }\right\} \right)
d\lambda \right)   \notag \\
&&\times \mathcal{D}\mathbf{\bar{\Lambda}}_{i}\left( \lambda \right)
\left\vert h_{0}\left( \left\{ \mathbf{\bar{\Lambda}}_{i}\right\} \right)
\right\rangle _{U_{0}^{\left( j\right) }}d\left\{ \mathbf{\bar{\Lambda}}%
_{i}\right\} ^{\prime }d\left\{ \mathbf{\bar{\Lambda}}_{i}\right\} 
\label{GLT}
\end{eqnarray}

\subsubsection{Local identification}

Again the amplitudes:%
\begin{equation}
\left\langle \lambda +\delta \lambda ,\left\{ \mathbf{\bar{\Lambda}}%
_{i}\right\} ^{\prime }\right\vert _{U^{\left( j\right) }}\left\langle
h^{\prime }\left( \left\{ \mathbf{\bar{\Lambda}}_{i}\right\} ^{\prime
}\right) \right\vert \delta T_{\lambda \lambda +\delta \lambda }\left\vert
\lambda ,\left\{ \mathbf{\bar{\Lambda}}_{i}\right\} \right\rangle \left\vert
h\left( \left\{ \mathbf{\bar{\Lambda}}_{i}\right\} \right) \right\rangle
_{U^{\left( j\right) }}  \label{PL}
\end{equation}%
are computed by composing some parrallel transport operator $P_{\lambda
\lambda +\delta \lambda }$ and then computing the matrix elements of:%
\begin{equation*}
P_{\lambda \lambda +\delta \lambda }\delta T_{\lambda \lambda +\delta
\lambda }
\end{equation*}%
As in the global case, the transport $P_{\lambda \lambda +\delta \lambda }$
is performed through generator of translation operator:%
\begin{equation*}
\left( \left\{ \mathbf{\bar{\Lambda}}_{i}\left( \lambda +\delta \lambda
\right) \right\} -\left\{ \mathbf{\bar{\Lambda}}_{i}\left( \lambda \right)
\right\} \right) \frac{\delta }{\delta \left\{ \mathbf{\bar{\Lambda}}%
_{i}\left( \lambda \right) \right\} }
\end{equation*}

However, since $\mathbf{\bar{\Lambda}}_{i}^{\prime }\left( \lambda +\delta
\lambda \right) $ and $\mathbf{\bar{\Lambda}}_{i}\left( \lambda \right) $ do
not act on the same space, their difference is not the derivative of a set
of variable $\mathbf{\bar{\Lambda}}_{i}\left( \lambda \right) $ that be
defined for every $\lambda $. This derivative has to be corrected to account
for the change in spaces the modification of $\lambda $ accounts for.

Given the constraints:%
\begin{eqnarray*}
&&\left( \left\{ \mathbf{\bar{\Lambda}}_{i}^{\prime }\left( \lambda +\delta
\lambda \right) \right\} -\left\{ \mathbf{\bar{\Lambda}}_{i}\left( \lambda
\right) \right\} \right) \\
&=&\delta \lambda \left( \frac{\partial \left\{ \mathbf{\bar{\Lambda}}%
_{i}\left( \lambda \right) \right\} }{\partial \lambda }+\left( M\left\{ 
\mathbf{\bar{\Lambda}}_{i}\left( \lambda \right) \right\} \right) \right)
=\delta \lambda \underline{\nabla }_{\lambda }\left\{ \mathbf{\bar{\Lambda}}%
_{i}\left( \lambda \right) \right\}
\end{eqnarray*}%
\bigskip and the transport writes:%
\begin{equation*}
P_{\lambda \lambda +\delta \lambda }=\exp \left( \int i\delta \lambda 
\underline{\nabla }_{\lambda }\left\{ \mathbf{\bar{\Lambda}}_{i}\left(
\lambda \right) \right\} \frac{\delta }{\delta \left\{ \mathbf{\bar{\Lambda}}%
_{i}\left( \lambda \right) \right\} }\mathcal{D}\left\{ \mathbf{\bar{\Lambda}%
}_{i}\left( \lambda \right) \right\} \right)
\end{equation*}%
The matrix elements of $\frac{\delta }{\delta \left\{ \mathbf{\bar{\Lambda}}%
_{i}\left( \lambda \right) \right\} }$ are computed using dual basis,
involving amplitudes of the form:%
\begin{equation*}
\exp \left( i\left( \left\{ \mathbf{\bar{\Lambda}}_{i}\left( \lambda \right)
\right\} ^{\prime }-\left\{ \mathbf{\bar{\Lambda}}_{i}\left( \lambda \right)
\right\} \right) \Pi _{\left\{ \mathbf{\bar{\Lambda}}_{i}\left( \lambda
\right) \right\} }\right)
\end{equation*}%
where $\Pi _{J}\left( U_{j}/\left[ U_{j}\right] ,\left\{ \underline{\mathbf{%
\hat{\Lambda}}}^{\left[ k_{i}\right] }\right\} ,\lambda \right) $ are
elements of dual basis. By parallel transport back to $\lambda +\delta
\lambda $ we can write:%
\begin{equation*}
\left\{ \mathbf{\bar{\Lambda}}_{i}\left( \lambda \right) \right\} ^{\prime
}\rightarrow \left\{ \mathbf{\bar{\Lambda}}_{i}\left( \lambda +\delta
\lambda \right) \right\} ^{\prime }
\end{equation*}%
so that matrices elements involving $\frac{\delta }{\delta \mathbf{\bar{%
\Lambda}}_{i}\left( \lambda \right) }$ write:%
\begin{equation*}
\exp \left( i\delta \lambda \underline{\nabla }_{\lambda }\left\{ \mathbf{%
\bar{\Lambda}}_{i}\left( \lambda \right) \right\} \Pi _{\left\{ \mathbf{\bar{%
\Lambda}}_{i}\left( \lambda \right) \right\} }\right)
\end{equation*}%
The matrix contribution of any operator depending on $\Pi _{\left\{ \mathbf{%
\bar{\Lambda}}_{i}\left( \lambda \right) \right\} }$ is:%
\begin{equation*}
\exp \left( i\delta \lambda \underline{\nabla }_{\lambda }\left\{ \mathbf{%
\bar{\Lambda}}_{i}\left( \lambda \right) \right\} \Pi _{\left\{ \mathbf{\bar{%
\Lambda}}_{i}\left( \lambda \right) \right\} }\right) \exp \left( iF\left(
\Pi _{\left\{ \mathbf{\bar{\Lambda}}_{i}\left( \lambda \right) \right\}
}\right) \right)
\end{equation*}%
obtained by saddle point equation:%
\begin{equation}
\Pi _{\left\{ \mathbf{\bar{\Lambda}}_{i}\left( \lambda \right) \right\} }=%
\underline{\nabla }_{\lambda }\left\{ \mathbf{\bar{\Lambda}}_{i}\left(
\lambda \right) \right\} +\text{something including the field}  \label{DCN}
\end{equation}%
Moreover, matrix elements of $\mathbf{\hat{\Lambda}}_{i}\left( \lambda
\right) $ involve terms:%
\begin{equation*}
\left\{ \mathbf{\bar{\Lambda}}_{i}+\delta \mathbf{\bar{\Lambda}}_{i}\right\}
-\left\{ \mathbf{\bar{\Lambda}}_{i}\right\}
\end{equation*}%
since in amplitudes $\mathbf{\hat{\Lambda}}_{i}\left( \lambda \right) $ acts
on different values of $\mathbf{\hat{\Lambda}}_{i}\left( \lambda \right) $,
see (\ref{PL}). Due to constraints this variation in parameters $\left\{ 
\mathbf{\bar{\Lambda}}_{i}\right\} $ involves a change in the state, so that
the matrix element $\mathbf{\hat{\Lambda}}_{i}\left( \lambda \right) $
between two states evaluated at $\left\{ \mathbf{\bar{\Lambda}}_{i}+\delta 
\mathbf{\bar{\Lambda}}_{i}\right\} $ and $\left\{ \mathbf{\bar{\Lambda}}%
_{i}\right\} $ induces a contribution:

\begin{eqnarray}
h\left( \left\{ \mathbf{\bar{\Lambda}}_{i}+\delta \mathbf{\bar{\Lambda}}%
_{i}\right\} ,\lambda \right) -h\left( \left\{ \mathbf{\bar{\Lambda}}%
_{i}\right\} ,\lambda \right) &=&\delta \mathbf{\bar{\Lambda}}_{i}\left( 
\frac{\delta }{\delta \left\{ \mathbf{\bar{\Lambda}}_{i}\left( \lambda
\right) \right\} }h\left( \left\{ \mathbf{\bar{\Lambda}}_{i}\right\}
,\lambda \right) +A_{\mathbf{\bar{\Lambda}}_{i}\left( \lambda \right)
}h\left( \left\{ \mathbf{\bar{\Lambda}}_{i}\right\} ,\lambda \right) \right)
\label{DCT} \\
&=&\delta \mathbf{\bar{\Lambda}}_{i}\underline{\nabla }_{\left\{ \mathbf{%
\bar{\Lambda}}_{i}\left( \lambda \right) \right\} }h\left( \left\{ \mathbf{%
\bar{\Lambda}}_{i}\right\} ,\lambda \right)  \notag
\end{eqnarray}%
Gathering (\ref{DCN}) and (\ref{DCT}), we find that the infinitesimal
transitions have the form:%
\begin{equation*}
L\left( \left\{ \mathbf{\bar{\Lambda}}_{i}\left( \lambda \right) \right\}
,\left\{ \underline{\nabla }_{\lambda }\left\{ \mathbf{\bar{\Lambda}}%
_{i}\left( \lambda \right) \right\} ,\underline{\nabla }_{\left\{ \mathbf{%
\bar{\Lambda}}_{i}\left( \lambda \right) \right\} }\right\} \right)
\end{equation*}%
where the $\left\{ \mathbf{\bar{\Lambda}}_{i}\left( \lambda \right) \right\} 
$ are independent sts of variables $\left\{ \mathbf{\bar{\Lambda}}%
_{i}\right\} $, one for each $\lambda $ and $\left( \underline{\nabla }%
_{\lambda }\left\{ \mathbf{\bar{\Lambda}}_{i}\left( \lambda \right) \right\}
,\underline{\nabla }_{\left\{ \mathbf{\bar{\Lambda}}_{i}\left( \lambda
\right) \right\} }\right) $ are defined in (\ref{DCN}) and (\ref{DCT}).
Exponentiating and composing the infinitesimal transformations yields the
amplitudes:%
\begin{eqnarray}
&&\left\langle \lambda ,\left\{ \mathbf{\bar{\Lambda}}_{i}\right\} ^{\prime
}\right\vert _{U^{\left( j\right) }}\left\langle h\left( \left\{ \mathbf{%
\bar{\Lambda}}_{i}\right\} ^{\prime }\right) \right\vert \exp \left(
i\int_{\lambda \left( \left( U^{\left( j\right) }\right) _{0}\right)
}^{\lambda \left( U^{\left( j\right) }\right) }L\left( \left\{ \mathbf{\bar{%
\Lambda}}_{i}\left( \lambda \right) \right\} ,\left\{ \underline{\nabla }%
_{\lambda }\left\{ \mathbf{\bar{\Lambda}}_{i}\left( \lambda \right) \right\}
,\underline{\nabla }_{\left\{ \mathbf{\bar{\Lambda}}_{i}\left( \lambda
\right) \right\} }\right\} \right) d\lambda \right)  \label{pltd} \\
&&\left\vert \lambda _{0},\left\{ \mathbf{\bar{\Lambda}}_{i}\right\}
\right\rangle \left\vert h_{0}\left( \left\{ \mathbf{\bar{\Lambda}}%
_{i}\right\} \right) \right\rangle _{U_{0}^{\left( j\right) }}  \notag \\
&=&\int_{U^{\left( j\right) }}\left\langle h\left( \left\{ \mathbf{\bar{%
\Lambda}}_{i}\right\} ^{\prime }\right) \right\vert \int \exp \left(
i\int_{\left\{ \mathbf{\bar{\Lambda}}_{i}\left( \lambda _{0}\right) \right\}
=\left\{ \mathbf{\bar{\Lambda}}_{i}\right\} }^{\left\{ \mathbf{\bar{\Lambda}}%
_{i}\left( \lambda \right) \right\} =\left\{ \mathbf{\bar{\Lambda}}%
_{i}\right\} ^{\prime }}L\left( \left\{ \mathbf{\bar{\Lambda}}_{i}\left(
\lambda \right) \right\} ,\left\{ \underline{\nabla }_{\lambda }\left\{ 
\mathbf{\bar{\Lambda}}_{i}\left( \lambda \right) \right\} ,\underline{\nabla 
}_{\left\{ \mathbf{\bar{\Lambda}}_{i}\left( \lambda \right) \right\}
}\right\} \right) d\lambda \right)  \notag \\
&&\times \mathcal{D}\mathbf{\bar{\Lambda}}_{i}\left( \lambda \right)
\left\vert h_{0}\left( \left\{ \mathbf{\bar{\Lambda}}_{i}\right\} \right)
\right\rangle _{U_{0}^{\left( j\right) }}d\left\{ \mathbf{\bar{\Lambda}}%
_{i}\right\} ^{\prime }d\left\{ \mathbf{\bar{\Lambda}}_{i}\right\}  \notag
\end{eqnarray}

For states (\ref{TD}) the system decomposes as:%
\begin{eqnarray}
&&\left\langle \lambda ,\left\{ \mathbf{\bar{\Lambda}}_{i}\right\}
_{/p}^{\prime }\right\vert _{U^{\left( j\right) }}\left\langle h\left(
\left\{ \mathbf{\bar{\Lambda}}_{i,p}\right\} ^{\prime }\right) \right\vert
\exp \left( i\int_{\lambda \left( \left( U^{\left( j\right) }\right)
_{0}\right) }^{\lambda \left( U^{\left( j\right) }\right) }L\left( \left\{ 
\mathbf{\bar{\Lambda}}_{i}\left( \lambda \right) \right\} ,\left\{ 
\underline{\nabla }_{\lambda }\left\{ \mathbf{\bar{\Lambda}}_{i}\left(
\lambda \right) \right\} ,\underline{\nabla }_{\left\{ \mathbf{\bar{\Lambda}}%
_{i}\left( \lambda \right) \right\} }\right\} \right) d\lambda \right) \\
&&\left\vert \lambda _{0},\left\{ \mathbf{\bar{\Lambda}}_{i}\right\}
_{/p}\right\rangle \left\vert h_{0}\left( \left\{ \mathbf{\bar{\Lambda}}%
_{i,p}\right\} \right) \right\rangle _{U_{0}^{\left( j\right) }}  \notag \\
&=&\int_{U^{\left( j\right) }}\left\langle \lambda ,\left\{ \mathbf{\bar{%
\Lambda}}_{i}\right\} _{/p}^{\prime }\right\vert _{U^{\left( j\right) }}\int
\exp \left( i\int_{\left\{ \mathbf{\bar{\Lambda}}_{i}\left( \lambda
_{0}\right) \right\} =\left\{ \mathbf{\bar{\Lambda}}_{i}\right\} }^{\left\{ 
\mathbf{\bar{\Lambda}}_{i}\left( \lambda \right) \right\} =\left\{ \mathbf{%
\bar{\Lambda}}_{i}\right\} ^{\prime }}L\left( \left\{ \mathbf{\bar{\Lambda}}%
_{i}\left( \lambda \right) \right\} ,\left\{ \underline{\nabla }_{\lambda
}\left\{ \mathbf{\bar{\Lambda}}_{i}\left( \lambda \right) \right\} ,%
\underline{\nabla }_{\left\{ \mathbf{\bar{\Lambda}}_{i}\left( \lambda
\right) \right\} }\right\} \right) d\lambda \right) \left\vert \lambda
_{0},\left\{ \mathbf{\bar{\Lambda}}_{i}\right\} _{/p}\right\rangle  \notag \\
&&\times \mathcal{D}\mathbf{\bar{\Lambda}}_{i}\left( \lambda \right)
d\left\{ \mathbf{\bar{\Lambda}}_{i}\right\} _{/p}^{\prime }d\left\{ \mathbf{%
\bar{\Lambda}}_{i,p}\right\} ^{\prime }d\left\{ \mathbf{\bar{\Lambda}}%
_{i,p}\right\} d\left\{ \mathbf{\bar{\Lambda}}_{i,p}\right\}
\end{eqnarray}%
and:%
\begin{eqnarray*}
&&L\left( \left\{ \mathbf{\bar{\Lambda}}_{i}\left( \lambda \right) \right\}
,\left\{ \underline{\nabla }_{\lambda }\left\{ \mathbf{\bar{\Lambda}}%
_{i}\left( \lambda \right) \right\} ,\underline{\nabla }_{\left\{ \mathbf{%
\bar{\Lambda}}_{i}\left( \lambda \right) \right\} }\right\} \right) \\
&=&L_{/p}\left( \left\{ \mathbf{\bar{\Lambda}}_{i}\right\} _{/p},\left\{ 
\underline{\nabla }_{\lambda }\left\{ \mathbf{\bar{\Lambda}}_{i}\left(
\lambda \right) \right\} ,\underline{\nabla }_{\left\{ \mathbf{\bar{\Lambda}}%
_{i}\right\} _{/p}}\right\} \right) +L_{p}\left( \left\{ \mathbf{\bar{\Lambda%
}}_{i}\right\} _{p},\left\{ \underline{\nabla }_{\lambda }\left\{ \mathbf{%
\bar{\Lambda}}_{i,p}\left( \lambda \right) \right\} ,\underline{\nabla }%
_{\left\{ \mathbf{\bar{\Lambda}}_{i}\right\} _{p}}\right\} \right) \\
&&+L_{/p,p}\left( \left\{ \mathbf{\bar{\Lambda}}_{i}\right\} _{/p},\left\{ 
\underline{\nabla }_{\lambda }\left\{ \mathbf{\bar{\Lambda}}_{i/p}\left(
\lambda \right) \right\} ,\underline{\nabla }_{\left\{ \mathbf{\bar{\Lambda}}%
_{i}\right\} _{/p}}\right\} ,\left\{ \underline{\nabla }_{\lambda }\left\{ 
\mathbf{\bar{\Lambda}}_{i,p}\left( \lambda \right) \right\} ,\underline{%
\nabla }_{\left\{ \mathbf{\bar{\Lambda}}_{i}\right\} _{p}}\right\} \right)
\end{eqnarray*}

The first term describes the main part of the system acting as a background,
the second one, the subsystem partly isolated from the background, while the
third one represents the interaction between them. Note that the entire
system determines the covariant derivatives.

\subsubsection{Field formulation of the transition}

In the previous paragraphs we have considered a "first quantized" formalism,
where the number of states is preserved. More generally we can consider the
transitions to depend on the field from which states derive:%
\begin{equation*}
\Psi _{J}\left( U^{\left( j\right) /\left\{ \mathbf{\hat{\Lambda}}%
_{i}\right\} },\left[ U^{\left( j\right) }\left( \left\{ \mathbf{\Lambda }%
_{i}\right\} \right) \right] ,\lambda \right)
\end{equation*}%
A derivation similar to that presented in part 1, shows that the transitions
are generated by:%
\begin{equation*}
\left\langle \lambda ,\left\{ \mathbf{\bar{\Lambda}}_{i}\right\}
_{/p}^{\prime }\right\vert _{U^{\left( j\right) }}\left\langle h\left(
\left\{ \mathbf{\bar{\Lambda}}_{i,p}\right\} ^{\prime }\right) \right\vert
\exp \left( i\int_{\lambda \left( \left( U^{\left( j\right) }\right)
_{0}\right) }^{\lambda \left( U^{\left( j\right) }\right) }L\left( \Psi
_{J}\right) d\lambda \right) \left\vert \lambda _{0},\left\{ \mathbf{\bar{%
\Lambda}}_{i}\right\} _{/p}\right\rangle \left\vert h_{0}\left( \left\{ 
\mathbf{\bar{\Lambda}}_{i,p}\right\} \right) \right\rangle _{U_{0}^{\left(
j\right) }}
\end{equation*}%
where:

\begin{eqnarray*}
&&L\left( \Psi _{J}\right) \\
&=&S\left( \left( \underline{\nabla }_{\lambda },\underline{\nabla }%
_{\left\{ \mathbf{\bar{\Lambda}}_{i}\left( \lambda \right) \right\} }\right)
\Psi _{J}\left( U^{\left( j\right) /\left\{ \mathbf{\hat{\Lambda}}%
_{i}\right\} },\left[ U^{\left( j\right) }\left( \left\{ \mathbf{\Lambda }%
_{i}\right\} \right) \right] ,\lambda \right) ,\Psi _{J}\left( U^{\left(
j\right) /\left\{ \mathbf{\hat{\Lambda}}_{i}\right\} },\left[ U^{\left(
j\right) }\left( \left\{ \mathbf{\Lambda }_{i}\right\} \right) \right]
,\lambda \right) \right)
\end{eqnarray*}

\subsection{\textbf{Transitions for the operators}:}

\subsubsection{Operatorial form of transition}

Under the apprximation of global identification the identification $\left\{ 
\mathbf{\bar{\Lambda}}_{i}\right\} $, the spaces $\mathcal{H}_{\left(
\lambda _{0},\left\{ \left\{ \mathbf{\bar{\Lambda}}_{i}\right\} \left(
U_{\lambda _{0}}^{\left( j\right) }\right) ,U_{\lambda _{0}}^{\left(
j\right) }\right\} \right) }$ and $\mathcal{H}_{\left( \lambda ,\left\{
\left\{ \mathbf{\bar{\Lambda}}_{i}\right\} \left( U_{\lambda }^{\left(
j\right) }\right) ,U_{\lambda }^{\left( j\right) }\right\} \right) }$ can be
identified. so that the transitions can be seen as an operator of the global
parameters $\mathbf{\bar{\Lambda}}_{i}$, $\Pi _{\left\{ \mathbf{\bar{\Lambda}%
}_{i}\right\} }$:%
\begin{equation}
T_{\lambda _{0}\lambda }=\exp \left( i\left( \lambda -\lambda _{0}\right)
V\left( \left\{ \mathbf{\hat{\Lambda}}_{i}\right\} ,\Pi _{\left\{ \mathbf{%
\hat{\Lambda}}_{i}\right\} }\right) \right)  \label{TP}
\end{equation}%
such that the transitions: 
\begin{equation}
\left\langle \lambda ,\left\{ \mathbf{\bar{\Lambda}}_{i}\right\} ^{\prime
}\right\vert _{U^{\left( j\right) }}\left\langle h\left( \left\{ \mathbf{%
\bar{\Lambda}}_{i}\right\} ^{\prime }\right) \right\vert \exp \left( i\left(
\lambda -\lambda _{0}\right) V\left( \left\{ \mathbf{\hat{\Lambda}}%
_{i}\right\} ,\Pi _{\left\{ \mathbf{\hat{\Lambda}}_{i}\right\} }\right)
\right) \left\vert \lambda _{0},\left\{ \mathbf{\bar{\Lambda}}_{i}\right\}
\right\rangle \left\vert h_{0}\left( \left\{ \mathbf{\bar{\Lambda}}%
_{i}\right\} \right) \right\rangle _{U_{0}^{\left( j\right) }}  \label{GLP}
\end{equation}%
are equal to (\ref{GLT}). This is achieved by the usual change of basis: at
the infinitesimal level, the matrix elements of a term $\exp \left( i\delta
\lambda F\left( \Pi _{\left\{ \mathbf{\hat{\Lambda}}_{i}\right\} }\right)
\right) $ is computed as:%
\begin{equation*}
\int \exp \left( i\delta \lambda \left( \left\{ \mathbf{\bar{\Lambda}}%
_{i}\right\} ^{\prime }-\left\{ \mathbf{\bar{\Lambda}}_{i}\right\} \right)
\pi _{\left\{ \mathbf{\bar{\Lambda}}_{i}\right\} }\right) \exp \left(
i\delta \lambda F\left( \pi _{\left\{ \mathbf{\bar{\Lambda}}_{i}\right\}
}\right) \right) d\pi _{\left\{ \mathbf{\bar{\Lambda}}_{i}\right\} }
\end{equation*}%
where $\pi _{\left\{ \mathbf{\bar{\Lambda}}_{i}\right\} }$ are eigenstates
of $\Pi _{\left\{ \mathbf{\bar{\Lambda}}_{i}\left( \lambda \right) \right\}
} $. These integral lead to replace $\pi _{\left\{ \mathbf{\bar{\Lambda}}%
_{i}\right\} }\rightarrow \frac{\partial \left\{ \mathbf{\bar{\Lambda}}%
_{i}\left( \lambda \right) \right\} }{\partial \lambda }$ and transitions of
the form (\ref{GLT}).

\subsubsection{Transition for operators in the eignstates basis}

We can use (\ref{TP}) to rewrite operators as a function from of their value
at a given $\lambda _{0}$.

Using the form (\ref{PRM}) for states, formula (\ref{PRL}) for operators:%
\begin{equation*}
\left[ \left\{ \mathbf{\Lambda }_{i}^{\prime }\right\} ,\lambda \mid h\left(
\left\{ \mathbf{\bar{\Lambda}}_{i}^{\prime }\right\} \right) _{_{U^{\left(
j^{\prime }\right) }}}\right] \Psi \left[ \left\{ \mathbf{\Lambda }%
_{i}\right\} ,\lambda \mid U_{j}/\left[ U_{j}\right] ,\left[ U_{j}\right] %
\right] =\left[ \left\{ \mathbf{\Lambda }_{i}^{\prime }\right\} ,\lambda
\mid U_{j^{\prime }}/\left[ U_{j^{\prime }}\right] ,\left[ U_{j^{\prime }}%
\right] \right] \Psi \left[ \left\{ \mathbf{\Lambda }_{i}\right\} ,\lambda
\mid h\left( \left\{ \mathbf{\bar{\Lambda}}_{i}\right\} \right)
_{_{U^{\left( j\right) }}}\right]
\end{equation*}%
along with (\ref{GLP}), the transition for operators is:%
\begin{eqnarray*}
&&\left[ \left\{ \mathbf{\Lambda }_{i}^{\prime }\right\} ,\lambda \mid
U_{j^{\prime }}/\left[ U_{j^{\prime }}\right] ,\left[ U_{j^{\prime }}\right] %
\right] \Psi \left[ \left\{ \mathbf{\Lambda }_{i}\right\} ,\lambda \mid
U_{j}/\left[ U_{j}\right] ,\left[ U_{j}\right] \right] \\
&=&\exp \left( -i\left( \lambda -\lambda _{0}\right) V\left( \left\{ \mathbf{%
\hat{\Lambda}}_{i}\right\} ,\Pi _{\left\{ \mathbf{\hat{\Lambda}}_{i}\right\}
}\right) \right) \\
&&\times \left[ \left\{ \mathbf{\Lambda }_{i}^{\prime }\right\} ,\lambda
_{0}\mid h\left( \left\{ \mathbf{\bar{\Lambda}}_{i}^{\prime }\right\}
\right) _{_{U^{\left( j^{\prime }\right) }}}\right] \Psi \left[ \left\{ 
\mathbf{\Lambda }_{i}\right\} ,\lambda _{0}\mid h\left( \left\{ \mathbf{\bar{%
\Lambda}}_{i}\right\} \right) _{_{U^{\left( j\right) }}}\right] \\
&&\times \exp \left( i\left( \lambda -\lambda _{0}\right) V\left( \left\{ 
\mathbf{\hat{\Lambda}}_{i}\right\} ,\Pi _{\left\{ \mathbf{\hat{\Lambda}}%
_{i}\right\} }\right) \right)
\end{eqnarray*}%
\bigskip If we use the decomposition (\ref{TD}), the projection of this
operator on the states: 
\begin{equation*}
\left\vert \lambda ,\left\{ \mathbf{\bar{\Lambda}}_{i}\right\}
_{/p}\right\rangle \left\vert h_{/p}\left( \left\{ \mathbf{\bar{\Lambda}}%
_{i}\right\} _{/p}\right) \right\rangle _{U^{\left( j\right) }}\left\vert
\left\{ \mathbf{\bar{\Lambda}}_{i,p}\right\} \right\rangle \left\vert
h_{p}\left( \left\{ \mathbf{\bar{\Lambda}}_{i,p}\right\} \right)
\right\rangle _{U^{\left( j\right) }}
\end{equation*}%
we find an operator:%
\begin{equation*}
\left[ \lambda ,\left\{ \mathbf{\bar{\Lambda}}_{i}^{\prime }\right\}
_{/p},\left\{ \mathbf{\bar{\Lambda}}_{i,p}^{\prime }\right\} ,h_{/p}\left(
\left\{ \mathbf{\bar{\Lambda}}_{i}^{\prime }\right\} _{/p}\right)
,h_{p}\left( \left\{ \mathbf{\bar{\Lambda}}_{i,p}\right\} \right) \right]
\Psi \left[ \lambda ,\left\{ \mathbf{\bar{\Lambda}}_{i}\right\}
_{/p},h_{/p}\left( \left\{ \mathbf{\bar{\Lambda}}_{i}\right\} _{/p}\right)
,h_{p}\left( \left\{ \mathbf{\bar{\Lambda}}_{i,p}\right\} \right) \right]
\end{equation*}

If we consider an operator depending only on $h\left( \left\{ \mathbf{\bar{%
\Lambda}}_{i,p}^{\prime }\right\} \right) $, and which is identity over the
other components, we obtain a decomposition:%
\begin{equation*}
\left\vert \lambda ,\left\{ \mathbf{\bar{\Lambda}}_{i}\right\}
_{/p},h_{/p}\left( \left\{ \mathbf{\bar{\Lambda}}_{i}^{\prime }\right\}
_{/p}\right) \right\rangle \left\vert \left\{ \mathbf{\bar{\Lambda}}%
_{i,p}\right\} \right\rangle \left[ h_{p}\left( \left\{ \mathbf{\bar{\Lambda}%
}_{i,p}^{\prime }\right\} ,\left( U^{\left( i^{\prime }\right) /p}\right)
\right) \right] \Psi \left[ h_{p}\left( \left\{ \mathbf{\bar{\Lambda}}%
_{i,p}\right\} \right) ,\left( U^{\left( i\right) /p}\right) \right]
\left\langle \left\{ \mathbf{\bar{\Lambda}}_{i,p}\right\} \right\vert
\left\langle \lambda ,\left\{ \mathbf{\bar{\Lambda}}_{i}\right\}
_{/p},h_{p}\left( \left\{ \mathbf{\bar{\Lambda}}_{i,p}\right\} \right)
\right\vert 
\end{equation*}%
Moreover, if the transition operator respects the decomposition:%
\begin{eqnarray*}
&&\exp \left( i\left( \lambda -\lambda _{0}\right) V\left( \left\{ \mathbf{%
\hat{\Lambda}}_{i}\right\} ,\Pi _{\left\{ \mathbf{\hat{\Lambda}}_{i}\right\}
}\right) \right)  \\
&=&\exp \left( i\left( \lambda -\lambda _{0}\right) \left( V_{1}\left(
\left\{ \mathbf{\bar{\Lambda}}_{i}\right\} _{/p},\Pi _{\left\{ \mathbf{\bar{%
\Lambda}}_{i}\right\} _{/p}}\right) +V_{2}\left( \left\{ \mathbf{\bar{\Lambda%
}}_{i}\right\} _{p},\Pi _{\left\{ \mathbf{\bar{\Lambda}}_{i}\right\}
_{p}}\right) \right) \right) 
\end{eqnarray*}%
and that globally $\left\vert \lambda ,\left\{ \mathbf{\bar{\Lambda}}%
_{i}\right\} _{/p},h_{/p}\left( \left\{ \mathbf{\bar{\Lambda}}_{i}^{\prime
}\right\} _{/p}\right) \right\rangle \left\vert \left\{ \mathbf{\bar{\Lambda}%
}_{i,p}\right\} \right\rangle $ is invariant by transition, we are left with:%
\begin{eqnarray*}
&&\left\vert \lambda ,\left\{ \mathbf{\bar{\Lambda}}_{i}\right\}
_{/p},h_{/p}\left( \left\{ \mathbf{\bar{\Lambda}}_{i}^{\prime }\right\}
_{/p}\right) \right\rangle \left\vert \left\{ \mathbf{\bar{\Lambda}}%
_{i,p}\right\} \right\rangle \left[ h_{p}\left( \left\{ \mathbf{\bar{\Lambda}%
}_{i,p}^{\prime }\right\} \right) \right] \Psi \left[ h_{p}\left( \left\{ 
\mathbf{\bar{\Lambda}}_{i,p}\right\} \right) \right] \left\langle \left\{ 
\mathbf{\bar{\Lambda}}_{i,p}\right\} \right\vert \left\langle \lambda
,\left\{ \mathbf{\bar{\Lambda}}_{i}\right\} _{/p},h_{p}\left( \left\{ 
\mathbf{\bar{\Lambda}}_{i,p}\right\} \right) \right\vert  \\
&=&\left\vert \lambda ,\left\{ \mathbf{\bar{\Lambda}}_{i}\right\}
_{/p},h_{/p}\left( \left\{ \mathbf{\bar{\Lambda}}_{i}^{\prime }\right\}
_{/p}\right) \right\rangle \left\vert \left\{ \mathbf{\bar{\Lambda}}%
_{i,p}\right\} \right\rangle \exp \left( -i\left( \lambda -\lambda
_{0}\right) V_{2}\left( \left\{ \mathbf{\bar{\Lambda}}_{i}\right\} _{p},\Pi
_{\left\{ \mathbf{\bar{\Lambda}}_{i}\right\} _{p}}\right) \right)  \\
&&\times \left[ h_{p}\left( \left\{ \mathbf{\bar{\Lambda}}_{i,p}^{\prime
}\right\} \right) \right] \Psi \left[ h_{p}\left( \left\{ \mathbf{\bar{%
\Lambda}}_{i,p}\right\} \right) \right]  \\
&&\times \exp \left( i\left( \lambda -\lambda _{0}\right) V_{2}\left(
\left\{ \mathbf{\bar{\Lambda}}_{i}\right\} _{p},\Pi _{\left\{ \mathbf{\bar{%
\Lambda}}_{i}\right\} _{p}}\right) \right) \left\langle \left\{ \mathbf{\bar{%
\Lambda}}_{i,p}\right\} \right\vert \left\langle \lambda ,\left\{ \mathbf{%
\bar{\Lambda}}_{i}\right\} _{/p},h_{p}\left( \left\{ \mathbf{\bar{\Lambda}}%
_{i,p}\right\} \right) \right\vert 
\end{eqnarray*}%
which corresponds to the usual definition of field evolution, where the
space is considered as an inert surrounding as a first approximation.

\subsubsection{Transformation of operators in the initial basis}

Considering only the subsystem wth evolution:%
\begin{equation*}
\exp \left( -i\left( \lambda -\lambda _{0}\right) V_{2}\right) \left\vert
\left\{ \mathbf{\bar{\Lambda}}_{i,p}^{\prime }\right\} \right\rangle \left[
h_{p}\left( \left\{ \mathbf{\bar{\Lambda}}_{i,p}^{\prime }\right\} \right) %
\right] \Psi \left[ h_{p}\left( \left\{ \mathbf{\bar{\Lambda}}_{i,p}\right\}
\right) \right] \left\langle \left\{ \mathbf{\bar{\Lambda}}_{i,p}\right\}
\right\vert \exp \left( i\left( \lambda -\lambda _{0}\right) V_{2}\right) 
\end{equation*}%
and rewrite this operator as a function of the initial basis of operator:%
\begin{equation*}
\left[ \left( \hat{U}^{\left( i\right) }\right) ^{\prime }\mid \left( \hat{U}%
^{\left( j\right) }\right) ^{\prime }\right] \Psi \left[ \hat{U}^{\left(
i\right) }\mid \hat{U}^{\left( j\right) }\right] 
\end{equation*}%
We can decompose this basis according to our decomposition:%
\begin{equation*}
U^{\left( j\right) }\rightarrow \left( U^{\left( j\right) /p},U^{\left(
j\right) p}\right) 
\end{equation*}%
and consider an operator depending only on $U^{\left( j\right) p}$, so that
it acts as the identity on other coordinates:%
\begin{equation*}
\left[ \left( U^{\left( j\right) p}\right) ^{\prime }\right] \Psi \left[
U^{\left( j\right) p}\right] 
\end{equation*}%
The change of basis similar to the inverse transform of (\ref{FMR}) writes:%
\begin{eqnarray*}
&&\left[ h_{p}\left( \left\{ \mathbf{\bar{\Lambda}}_{i,p}^{\prime }\right\}
\right) \right] \Psi \left[ h_{p}\left( \left\{ \mathbf{\bar{\Lambda}}%
_{i,p}\right\} \right) \right]  \\
&=&\int g^{\dag }\left( \left( U^{\left( j\right) p}\right) ^{\prime
},h_{p}\left( \left\{ \mathbf{\bar{\Lambda}}_{i,p}^{\prime }\right\} \right)
\right) \left[ \left( U^{\left( j\right) p}\right) ^{\prime }\right] \Psi %
\left[ U^{\left( j\right) p}\right] g\left( \left( U^{\left( j\right)
p}\right) ^{\prime },h_{p}\left( \left\{ \mathbf{\bar{\Lambda}}%
_{i,p}\right\} \right) \right) d\left( U^{\left( j\right) p}\right) ^{\prime
}dU^{\left( j\right) p}
\end{eqnarray*}%
and using the transitions:%
\begin{eqnarray*}
&&\left\vert \left\{ \mathbf{\bar{\Lambda}}_{i,p}^{\prime }\right\}
\right\rangle \int g^{\dag }\left( \left( U^{\left( j\right) p}\right)
^{\prime },h_{p}\left( \left\{ \mathbf{\bar{\Lambda}}_{i,p}^{\prime
}\right\} \right) \right) \left[ \lambda ,\left( U^{\left( j\right)
p}\right) ^{\prime }\right] \Psi \left[ \lambda ,U^{\left( j\right) p}\right]
g\left( \left( U^{\left( j\right) p}\right) ^{\prime },h_{p}\left( \left\{ 
\mathbf{\bar{\Lambda}}_{i,p}\right\} \right) \right)  \\
&&\times d\left( U^{\left( j\right) p}\right) ^{\prime }dU^{\left( j\right)
p}\left\langle \left\{ \mathbf{\bar{\Lambda}}_{i,p}\right\} \right\vert  \\
&=&\left\vert \left\{ \mathbf{\bar{\Lambda}}_{i,p}^{\prime }\right\}
\right\rangle \exp \left( -i\left( \lambda -\lambda _{0}\right) V_{2}\right) 
\\
&&\times \int g^{\dag }\left( \left( U^{\left( j\right) p}\right) ^{\prime
},h_{p}\left( \left\{ \mathbf{\bar{\Lambda}}_{i,p}^{\prime }\right\} \right)
\right) \left[ \left( U^{\left( j\right) p}\right) ^{\prime }\right] \Psi %
\left[ U^{\left( j\right) p}\right] g\left( \left( U^{\left( j\right)
p}\right) ^{\prime },h_{p}\left( \left\{ \mathbf{\bar{\Lambda}}%
_{i,p}\right\} \right) \right) d\left( U^{\left( j\right) p}\right) ^{\prime
}dU^{\left( j\right) p} \\
&&\times \exp \left( i\left( \lambda -\lambda _{0}\right) V_{2}\right)
\left\langle \left\{ \mathbf{\bar{\Lambda}}_{i,p}\right\} \right\vert 
\end{eqnarray*}%
where we identified:%
\begin{equation*}
\left[ \left( U^{\left( j\right) p}\right) ^{\prime }\right] \Psi \left[
U^{\left( j\right) p}\right] =\left[ \lambda _{0},\left( U^{\left( j\right)
p}\right) ^{\prime }\right] \Psi \left[ \lambda _{0},U^{\left( j\right) p}%
\right] 
\end{equation*}%
The usual transformation corresponds to the case where $V$ is diagonal in
the $U^{\left( j\right) p}$:%
\begin{eqnarray*}
&&\exp \left( -i\left( \lambda \left( U^{\left( j\right) }\right) -\lambda
_{0}\right) V\right) \left[ \left( U^{\left( j\right) p}\right) ^{\prime }%
\right] \Psi _{0}\left[ U^{\left( j\right) p}\right] \exp \left( i\left(
\lambda \left( U^{\left( j\right) }\right) -\lambda _{0}\right) V\right)  \\
&=&\exp \left( -i\left( \lambda \left( U^{\left( j\right) }\right) -\lambda
_{0}\right) \left( V\left( \left( U^{\left( j\right) p}\right) ^{\prime
}\right) -V\left( U^{\left( j\right) p}\right) \right) \right) \left[ \left(
U^{\left( j\right) p}\right) ^{\prime }\right] \Psi _{0}\left[ U^{\left(
j\right) p}\right] 
\end{eqnarray*}%
and:%
\begin{eqnarray*}
&&\left\vert \left\{ \mathbf{\bar{\Lambda}}_{i,p}^{\prime }\right\}
\right\rangle \int g^{\dag }\left( \left( U^{\left( j\right) p}\right)
^{\prime },h_{p}\left( \left\{ \mathbf{\bar{\Lambda}}_{i,p}^{\prime
}\right\} \right) \right) \left[ \lambda ,\left( U^{\left( j\right)
p}\right) ^{\prime }\right] \Psi \left[ \lambda ,U^{\left( j\right) p}\right]
\\
&&\times g\left( \left( U^{\left( j\right) p}\right) ^{\prime },h_{p}\left(
\left\{ \mathbf{\bar{\Lambda}}_{i,p}\right\} \right) \right) d\left(
U^{\left( j\right) p}\right) ^{\prime }dU^{\left( j\right) p}\left\langle
\left\{ \mathbf{\bar{\Lambda}}_{i,p}\right\} \right\vert  \\
&=&\left\vert \left\{ \mathbf{\bar{\Lambda}}_{i,p}^{\prime }\right\}
\right\rangle \int \exp \left( -i\left( \lambda \left( U^{\left( j\right)
}\right) -\lambda _{0}\right) \left( V\left( \left( U^{\left( j\right)
p}\right) ^{\prime }\right) -V\left( U^{\left( j\right) p}\right) \right)
\right) \left[ \left( U^{\left( j\right) p}\right) ^{\prime }\right]  \\
&&\times g^{\dag }\left( \left( U^{\left( j\right) p}\right) ^{\prime
},h_{p}\left( \left\{ \mathbf{\bar{\Lambda}}_{i,p}^{\prime }\right\} \right)
\right) \left[ \left( U^{\left( j\right) p}\right) ^{\prime }\right] \Psi %
\left[ U^{\left( j\right) p}\right] g\left( \left( U^{\left( j\right)
p}\right) ^{\prime },h_{p}\left( \left\{ \mathbf{\bar{\Lambda}}%
_{i,p}\right\} \right) \right) d\left( U^{\left( j\right) p}\right) ^{\prime
}dU^{\left( j\right) p}\left\langle \left\{ \mathbf{\bar{\Lambda}}%
_{i,p}\right\} \right\vert 
\end{eqnarray*}

\paragraph{Exemple\textbf{\ }}

For operators:%
\begin{equation*}
\left[ \left( U^{\left( i\right) p}\right) ^{\prime }\right] \Psi \left[
U^{\left( i\right) p}\right] =\left\vert U^{\left( j\right)
p+k}\right\rangle \left\langle U^{\left( j\right) p}\right\vert
\end{equation*}%
the conjugate representation writes:%
\begin{equation*}
\left\vert \left\{ \mathbf{\bar{\Lambda}}_{i,p+k}\right\} \right\rangle \int
g^{\dag }\left( U^{\left( j\right) p+k},h_{p+k}\left( \left\{ \mathbf{\bar{%
\Lambda}}_{i,p+k}\right\} \right) \right) \left\vert U^{\left( j\right)
p+k}\right\rangle \left\langle U^{\left( j\right) p}\right\vert g\left(
U^{\left( j\right) p},h_{p}\left( \left\{ \mathbf{\bar{\Lambda}}%
_{i,p}^{\prime }\right\} \right) \right) dU^{\left( j\right) p}d\left(
U^{\left( j\right) p}\right) ^{\prime }\left\langle \left\{ \mathbf{\bar{%
\Lambda}}_{i,p}\right\} \right\vert
\end{equation*}%
In the usual set up, where translation invariance is considered: 
\begin{equation*}
g^{\dag }\left( U^{\left( j\right) p+k},h_{p+k}\left( \left\{ \mathbf{\bar{%
\Lambda}}_{i,p+k}\right\} \right) \right) g\left( U^{\left( j\right)
p},h_{p}\left( \left\{ \mathbf{\bar{\Lambda}}_{i,p}\right\} \right) \right)
=\exp \left( iU^{\left( j\right) p+k/p}.\left\{ \mathbf{\bar{\Lambda}}%
_{i,p+k/p}\right\} \right)
\end{equation*}%
where $\left\{ \mathbf{\bar{\Lambda}}_{i,p+k/p}\right\} $\ corresponds to
the the degrees of freedom of $\left\{ \mathbf{\bar{\Lambda}}%
_{i,p+k}\right\} $ except those of $\mathbf{\bar{\Lambda}}_{i,p}$.

Consequently, the operator becomes:%
\begin{equation*}
\left[ \lambda ^{\left( i\right) },\left( U^{\left( i\right) p}\right)
^{\prime }\right] \Psi \left[ \lambda ,U^{\left( i\right) p}\right]
=\left\vert \left\{ \mathbf{\bar{\Lambda}}_{i,p+k}\right\} \right\rangle
\int \exp \left( iU^{\left( j\right) p+k/p}.\left\{ \mathbf{\bar{\Lambda}}%
_{i,p+k/p}\right\} \right) \left\vert U^{\left( j\right) p+k}\right\rangle
\left\langle U^{\left( j\right) p}\right\vert dU^{\left( j\right) p}d\left(
U^{\left( j\right) p}\right) ^{\prime }\left\langle \left\{ \mathbf{\bar{%
\Lambda}}_{i,p}\right\} \right\vert
\end{equation*}%
similar to a usual field in terms of creation and annihilation operators.

\part*{Part IV\ Further developments}

This fourth part serves as a conclusion, presenting various potential
developments arising from the present work. We connect our formalism to more
standard descriptions based on fields defined on some partly exogeneous
space. Such descriptions appear as apparent dynamical system, the dynamic
aspects deriving from modifications in the parameter space $\left\{ 
\underline{\mathbf{\hat{\Lambda}}}^{\left[ k_{i}\right] }\right\} $ we
defined above.

We first focus on the role of the constraints relating the degeneracy
parameters $\left\{ \underline{\mathbf{\hat{\Lambda}}}^{\left[ k_{i}\right]
}\right\} $ and the state $v$.in the structure of the effective formalism.
We posit that this formalism should be locally described by fields on a
fibered space. Subsequently, our attention turns to the modifications of the
constraints induced by modifications of the apparent state. resulting from
changes in the apparent state $v$. From this, we derive linear equations for
the modified states similar to some dynamical equations and establish
commutation relations between generators of modifications. Thus, local field
and states arise from modifications both of the state $v$ and the constraint
defind by this stat. Then, we consider effective action functionals derived
from the projected states and fields defined in part II. These effective
actions can be locally considered as action functionals for fields defined
on the parameter space. Ultimately, applying our description in terms of
slices in the parameter spaces (see part III), such fields and actions can
be considered as dynamical objects, the dynamics being given by slices in
the parameter space. These slices are considered to be connected by arrows
translatng some transitions, these transitions are not reversible, since
their amplitude depend on the parameter defining the slices. The higher
amplitudes correspond to an increases in the parameter. We thus recover
general standard features of field formalism. We argue that this description
can only be local. Consdering the full prmt, the action functional should
include a term for the parameters space. These global term should influence
the apparent dynamic fields and be part of the whole dynamic for field but
also for the constraint and the corrsponding metric on the parameter space.

\section{Collection of several states and parameters spaces}

Given that the space of parameters varies with the state $\nu $, there is no
field defined globally on the set:%
\begin{equation*}
\left\{ \left[ 
\begin{array}{c}
\left( p_{l_{i}},p_{l_{i}^{\prime }l_{i}},k_{l_{i}^{\prime }}\right)  \\ 
\left[ \left\{ \Psi _{J}\left[ p_{\eta },p_{\eta ^{\prime }\eta }\right]
\right\} ,v\right] 
\end{array}%
\right] \right\} 
\end{equation*}%
In fact, we should rather consider collections of objects:%
\begin{equation*}
\left\{ \left[ 
\begin{array}{c}
\left\{ p_{l_{i}},p_{l_{i}^{\prime }l_{i}},k_{l_{i}^{\prime }}\right\}
,\left\{ \underline{\mathbf{\hat{\Lambda}}}^{\left[ k_{i}\right] }\right\} 
\\ 
\left\{ p_{\eta },p_{\eta ^{\prime }\eta }\right\} ,v\left( \left\{ p_{\eta
},p_{\eta ^{\prime }\eta }\right\} \right) 
\end{array}%
\right] \right\} 
\end{equation*}%
defined by sets of clouds of parameters $\left\{ \underline{\mathbf{\hat{%
\Lambda}}}^{\left[ k_{i}\right] }\right\} $. These parameters are involved
in the constraints through symetry operators $\exp \left( i\left\{ 
\underline{\mathbf{\hat{\Lambda}}}^{\left[ k_{i}\right] }\right\} .\mathbf{l}%
\right) $ (see parts I and II). The $\left\{ \underline{\mathbf{\hat{\Lambda}%
}}^{\left[ k_{i}\right] }\right\} $ describe themselves the eigenstates of
some families of operatrs $\left\{ \mathbf{L}^{\left[ k_{i}\right] }\right\} 
$, written $\mathbf{L}$ for shrt, dual to $\mathbf{l}$ satisfying the
constraints:%
\begin{equation}
H\left( \left\{ \underline{\mathbf{\hat{\Lambda}}}^{\left[ k_{i}\right]
}\right\} ,\mathcal{K}_{0}\left( \left\{ p_{l_{i}},p_{l_{i}^{\prime
}l_{i}},k_{l_{i}^{\prime }}\right\} ,\left\{ v^{\left( \eta \right) }\left(
\left\{ p_{\eta },p_{\eta ^{\prime }\eta }\right\} \right) \right\} \right)
\right) =0  \label{PRS}
\end{equation}%
where $H$ is a vector with infinite number of components, gathering the
components equations of (\ref{GRC}). $\left\{ v^{\left( \eta \right) }\left(
\left\{ p_{\eta },p_{\eta ^{\prime }\eta }\right\} \right) \right\} $ is a
vector of functionals for several remaining subobjects. The kernel $\mathcal{%
K}_{0}\left( \left\{ p_{l_{i}},p_{l_{i}^{\prime }l_{i}},k_{l_{i}^{\prime
}}\right\} ,\left\{ v^{\left( \eta \right) }\left( \left\{ p_{\eta },p_{\eta
^{\prime }\eta }\right\} \right) \right\} \right) $ is defined by the
background states for some degrees of freedom.

The set of constraints (\ref{PRS}) defines one or several manifolds $\left\{
V\left( v^{\left( \eta \right) }\right) \right\} $ with sets of points $%
\left\{ \underline{\mathbf{\hat{\Lambda}}}^{\left[ k_{i}\right] }\right\} $
and depending on states $v^{\left( \eta \right) }$. These manifolds do
depend on the symmetries conditionned by the states $\left\vert \left\{
v^{\left( \eta \right) }\left( \left\{ p_{\eta },p_{\eta ^{\prime }\eta
}\right\} \right) \right\} \right\rangle $, so do their dimensions.$\left\{
V_{\alpha }\left( v^{\left( \eta \right) }\right) \right\} $

When local variations $v^{\left( \eta \right) }$ deform continuously $%
V\left( v^{\left( \eta \right) }\right) $, and $V\left( v^{\left( \eta
\right) }\right) \simeq V\left( v^{\left( \eta \right) }+\delta v^{\left(
\eta \right) }\right) $ we can consider that the $V\left( v^{\left( \eta
\right) }\right) $ form a collection of spaces $\left\{ V_{\alpha }\right\} $
fibered on the subspaces of the state space defined by the $v^{\left( \eta
\right) }$. Locally the $V_{\alpha }$ are defined by coordinates $\left(
\left\{ \underline{\mathbf{\hat{\Lambda}}}^{\left[ k_{i}\right] }\right\}
,v^{\left( \eta \right) }\right) $ solving (\ref{PRS}).

Defining fields over a set $V_{\alpha }$ corresponds to consider state
spaces $H\left( V_{\alpha }\right) $ described locally by the fields $\Psi
_{v^{\left( \eta \right) }}\left( \underline{\mathbf{\hat{\Lambda}}}^{\left[
k_{i}\right] }\right) $ or $\Psi \left( v^{\left( \eta \right) },\underline{%
\mathbf{\hat{\Lambda}}}^{\left[ k_{i}\right] }\right) $.

Maps of inclusion between such spaces:%
\begin{equation*}
V_{\beta }\rightarrow V_{\alpha }
\end{equation*}%
correspond to consider degeneracy subspaces, with inclusion of subsets of
degeneracy generators $\left\{ \mathbf{L}\right\} _{\beta }\subset \left\{ 
\mathbf{L}\right\} _{\alpha }$. Such maps should translate into maps:%
\begin{equation*}
H\left( V_{\alpha }\right) \rightarrow H\left( V_{\beta }\right)
\end{equation*}%
and products of maps:%
\begin{equation*}
V_{\beta }\underset{V_{\beta ,\beta ^{\prime }}}{\times }V_{\beta ^{\prime
}}\rightarrow V_{\alpha }
\end{equation*}%
where $V_{\beta ,\beta ^{\prime }}$ represents the constraints between both
parameters spaces $V_{\beta }$ and $V_{\beta ^{\prime }}$ should translate
into the decomposition of fields:%
\begin{equation*}
\Psi _{v^{\left( \eta \right) }}^{k,\alpha }\left( \underline{\mathbf{\hat{%
\Lambda}}}_{\alpha }^{\left[ k_{i}\right] }\right) \rightarrow \sum \Psi
_{v^{\left( \eta \right) }}^{k,\beta }\left( \underline{\mathbf{\hat{\Lambda}%
}}_{\beta }^{\left[ k_{i}\right] }\right) \otimes \Psi _{v^{\left( \eta
\right) }}^{k,\beta ^{\prime }}\left( \underline{\mathbf{\hat{\Lambda}}}%
_{\beta ^{\prime }}^{\left[ k_{i}\right] }\right)
\end{equation*}%
where the decomposition is for realizations, as described in the first and
second part.

The full system should thus be given by the sets $\left\{ V_{\alpha
}\right\} $ together with maps between these sets and decomposition of
fields corresponding to these maps.

In this context, the state $\left\vert \left\{ v^{\left( \eta \right)
}\left( \left\{ p_{\eta },p_{\eta ^{\prime }\eta }\right\} \right) \right\}
\right\rangle $ conditioning the constraints (\ref{PRS}) can be interpreted
as a type of global state of the system. The set of parameters changes along
these states. However, considering small deviations $\left\vert \left\{
\delta v^{\left( \eta \right) }\left( \left\{ p_{\eta },p_{\eta ^{\prime
}\eta }\right\} \right) \right\} \right\rangle $ from the state $\left\vert
\left\{ v^{\left( \eta \right) }\left( \left\{ p_{\eta },p_{\eta ^{\prime
}\eta }\right\} \right) \right\} \right\rangle $ may maintain the global
state by keeping the constraints invariant. In this case, $\delta v$
represents a small variation of the state, akin to a microstate. In the next
section, we examine the impact of the constraints on this variation.

\section{Continuous variation of state and constraint modification.}

\subsection{First order variation}

To express the variation of the constraints resulting from a change in
state, we refer to (\ref{GRC}), disregarding the realization index and the
dependence on $\left( \Psi _{J}^{\otimes l}\right) $. Additionally, we
implicitly consider the dependencies $\left\{ U_{i}^{k}\right\} _{i}$ and $%
\left( \Psi _{J}\right) ,U_{j}^{l}$ to formulate the set of constrained
variables in a manner analog to (\ref{PRS}): 
\begin{equation}
H\left( \left\{ \mathbf{L}\right\} ,\nu \right) =0  \label{PRTV}
\end{equation}%
or, if we have several types of states as in part II:%
\begin{equation*}
H\left( \left\{ \mathbf{L}\right\} ,\left\{ \nu ^{\left( \eta \right)
}\right\} \right) =0
\end{equation*}%
The constraint has been rewritten in terms of generators of parameters
states.

Modifying states $\left\vert \left\{ v^{\left( \eta \right) }\right\}
\right\rangle $ to $\left\vert \left\{ v^{\left( \eta \right) }+\delta
v^{\left( \eta \right) }\right\} \right\rangle $ modifies the symetry
parameters $\mathbf{L}$. This may both consist in a modification $\mathbf{%
L\rightarrow L+AL+}\Delta \mathbf{L}$ where $\mathbf{A}$ is a continuous
infinitesimal transformation, and $\Delta \mathbf{L}$ is a discrete
deformation changing the number of parameters. For instance, we can suppose
a deformation $\Delta \mathbf{L}$ generated by some $\Delta \mathbf{L}%
^{+}+\Delta \mathbf{L}^{-}$where $\Delta \mathbf{L}^{+}$ adds a symmetry
generator, and $\Delta \mathbf{L}$ removes one generator. The modified set
operator writes:%
\begin{equation*}
\mathbf{L\rightarrow }\left( \mathbf{L,}\Delta \mathbf{L}\right)
\end{equation*}%
The $\Delta \mathbf{L}$ have also to be considered as acting on an
additional set of eigenstates. This could correspond to the creation or
destruction of a parameter. We write the full deformation of parameters $%
\mathbf{\delta L=AL+}\left( \mathbf{1+A}\right) \Delta \mathbf{L}$. This
modifies the constraint (\ref{PRTV}) as:%
\begin{equation}
H\left( \left\{ \mathbf{L+\delta L}\right\} ,\left\{ v^{\left( \eta \right)
}+\delta v^{\left( \eta \right) }\right\} \right) =0  \label{PRTZ}
\end{equation}%
Considering the transformation $\left( 1+\mathbf{A}\right) $, this becomes,
applied on states:%
\begin{equation}
H\left( \left\{ \mathbf{L+\delta L}\right\} ,\left\{ v^{\left( \eta \right)
}+\delta v^{\left( \eta \right) }\right\} \right) \left\vert \left\{ \left(
1+\mathbf{A}\right) \underline{\mathbf{\hat{\Lambda}}}^{\left[ k_{i}\right]
}\right\} \right\rangle \left\vert \underline{\mathbf{\hat{\Lambda}}}^{\left[
\Delta _{1}\right] }\right\rangle =0  \label{CNPS}
\end{equation}%
Then, using that $\mathbf{L+\delta L=}\left( 1+\mathbf{A}\right) \left( 
\mathbf{L+\Delta L}\right) $ and applying a global transformation $\left( 1-%
\mathbf{A}\right) $ yields:%
\begin{equation*}
\left( 1-\mathbf{A}\right) H\left( \left\{ \mathbf{L+\delta L}\right\}
,\left\{ v^{\left( \eta \right) }+\delta v^{\left( \eta \right) }\right\}
\right) \left( 1+\mathbf{A}\right) \left( 1-\mathbf{A}\right) \left\vert
\left\{ \left( 1+\mathbf{A}\right) \underline{\mathbf{\hat{\Lambda}}}^{\left[
k_{i}\right] }\right\} \right\rangle \left\vert \underline{\mathbf{\hat{%
\Lambda}}}^{\left[ \Delta _{1}\right] }\right\rangle =0
\end{equation*}%
Using that:%
\begin{equation*}
\left( 1+\mathbf{A}\right) \left( 1-\mathbf{A}\right) =1
\end{equation*}%
at the lowest order and that:%
\begin{equation*}
\left( 1-\mathbf{A}\right) \left\{ \mathbf{L+\delta L}\right\} \rightarrow 
\mathbf{L+\Delta L}
\end{equation*}%
we find:%
\begin{equation*}
H\left( \left\{ \mathbf{L+\Delta L}\right\} ,\left\{ v^{\left( \eta \right)
}+\delta v^{\left( \eta \right) }\right\} \right) \left\vert \left\{ 
\underline{\mathbf{\hat{\Lambda}}}^{\left[ k_{i}\right] }\right\}
\right\rangle \left\vert \left( 1-\mathbf{A}\right) \underline{\mathbf{\hat{%
\Lambda}}}^{\left[ \Delta _{1}\right] }\right\rangle =0
\end{equation*}%
This last formula is also written:%
\begin{equation}
\left( 1-\mathbf{A}\right) _{\underline{\mathbf{\hat{\Lambda}}}^{\left[
\Delta _{1}\right] }}H\left( \left\{ \left\{ \underline{\mathbf{\hat{\Lambda}%
}}^{\left[ k_{i}\right] }\right\} ,\underline{\mathbf{\hat{\Lambda}}}^{\left[
\Delta _{1}\right] }\right\} ,\left\{ v^{\left( \eta \right) }+\delta
v^{\left( \eta \right) }\right\} \right) \left\vert \left\{ \underline{%
\mathbf{\hat{\Lambda}}}^{\left[ k_{i}\right] }\right\} \right\rangle
\left\vert \underline{\mathbf{\hat{\Lambda}}}^{\left[ \Delta _{1}\right]
}\right\rangle =0  \label{PRZ}
\end{equation}%
which is a developd form of (\ref{PRTZ}. We consider a first order expansion
of (\ref{PRZ}) of the form:%
\begin{eqnarray}
0 &=&\left( 1-\mathbf{A}\right) _{\underline{\mathbf{\hat{\Lambda}}}^{\left[
\Delta _{1}\right] }}H\left( \left\{ \underline{\mathbf{\hat{\Lambda}}}^{%
\left[ k_{i}\right] }\right\} ,\left\{ v^{\left( \eta \right) }+\delta
v^{\left( \eta \right) }\right\} \right) \left\vert \left\{ \underline{%
\mathbf{\hat{\Lambda}}}^{\left[ k_{i}\right] }\right\} \right\rangle
\left\vert \underline{\mathbf{\hat{\Lambda}}}^{\left[ \Delta _{1}\right]
}\right\rangle  \label{PNSV} \\
&&+\left( 1-\mathbf{A}\right) _{\underline{\mathbf{\hat{\Lambda}}}^{\left[
\Delta _{1}\right] }}H_{\underline{\mathbf{\hat{\Lambda}}}^{\left[ \Delta
_{1}\right] }\delta v^{\left( \eta \right) }}\left( \left\{ \underline{%
\mathbf{\hat{\Lambda}}}^{\left[ k_{i}\right] }\right\} ,\left\{ v^{\left(
\eta \right) }\right\} \right) \delta v^{\left( \eta \right) }H\left( 
\underline{\mathbf{\hat{\Lambda}}}^{\left[ \Delta _{1}\right] }\right)
\left\vert \left\{ \underline{\mathbf{\hat{\Lambda}}}^{\left[ k_{i}\right]
}\right\} \right\rangle \left\vert \underline{\mathbf{\hat{\Lambda}}}^{\left[
\Delta _{1}\right] }\right\rangle  \notag
\end{eqnarray}%
where $H\left( \underline{\mathbf{\hat{\Lambda}}}^{\left[ \Delta _{1}\right]
}\right) $ encompasses the action of the modified parameters $\underline{%
\mathbf{\hat{\Lambda}}}^{\left[ \Delta _{1}\right] }$. For the sake of
simplicity, we have assumed a separable form for this action.

We then write the first order expansion of $H\left( \left\{ \underline{%
\mathbf{\hat{\Lambda}}}^{\left[ k_{i}\right] }\right\} ,\left\{ v^{\left(
\eta \right) }+\delta v^{\left( \eta \right) }\right\} \right) $ in $\delta
v^{\left( \eta \right) }$\ as:%
\begin{equation*}
H\left( \left\{ \underline{\mathbf{\hat{\Lambda}}}^{\left[ k_{i}\right]
}\right\} ,\left\{ v^{\left( \eta \right) }+\delta v^{\left( \eta \right)
}\right\} \right) =H\left( \left\{ \underline{\mathbf{\hat{\Lambda}}}^{\left[
k_{i}\right] }\right\} ,\left\{ v^{\left( \eta \right) }\right\} \right)
+H_{\delta v^{\left( \eta \right) }}\left( \left\{ \underline{\mathbf{\hat{%
\Lambda}}}^{\left[ k_{i}\right] }\right\} ,\left\{ v^{\left( \eta \right)
}\right\} \right) 
\end{equation*}%
Using that the following constraint is satisfied at the zeroth order:%
\begin{equation*}
\left( 1-\mathbf{A}\right) _{\underline{\mathbf{\hat{\Lambda}}}^{\left[
\Delta _{1}\right] }}H\left( \left\{ \underline{\mathbf{\hat{\Lambda}}}^{%
\left[ k_{i}\right] }\right\} ,\left\{ v^{\left( \eta \right) }\right\}
\right) \left\vert \left\{ \underline{\mathbf{\hat{\Lambda}}}^{\left[ k_{i}%
\right] }\right\} \right\rangle \left\vert \underline{\mathbf{\hat{\Lambda}}}%
^{\left[ \Delta _{1}\right] }\right\rangle =0
\end{equation*}%
formula (\ref{PNSV}) becomes:%
\begin{eqnarray}
0 &=&\left( 1-\mathbf{A}\right) _{\underline{\mathbf{\hat{\Lambda}}}^{\left[
\Delta _{1}\right] }}H_{\delta v^{\left( \eta \right) }}\left( \left\{ 
\underline{\mathbf{\hat{\Lambda}}}^{\left[ k_{i}\right] }\right\} ,\left\{
v^{\left( \eta \right) }\right\} \right) \delta v^{\left( \eta \right)
}\left\vert \left\{ \underline{\mathbf{\hat{\Lambda}}}^{\left[ k_{i}\right]
}\right\} \right\rangle \left\vert \underline{\mathbf{\hat{\Lambda}}}^{\left[
\Delta _{1}\right] }\right\rangle   \label{PCM} \\
&&+\left( 1-\mathbf{A}\right) _{\underline{\mathbf{\hat{\Lambda}}}^{\left[
\Delta _{1}\right] }}H_{\underline{\mathbf{\hat{\Lambda}}}^{\left[ \Delta
_{1}\right] }\delta v^{\left( \eta \right) }}\left( \left\{ \underline{%
\mathbf{\hat{\Lambda}}}^{\left[ k_{i}\right] }\right\} ,\left\{ v^{\left(
\eta \right) }\right\} \right) \delta v^{\left( \eta \right) }H\left( 
\underline{\mathbf{\hat{\Lambda}}}^{\left[ \Delta _{1}\right] }\right)
\left\vert \left\{ \underline{\mathbf{\hat{\Lambda}}}^{\left[ k_{i}\right]
}\right\} \right\rangle \left\vert \underline{\mathbf{\hat{\Lambda}}}^{\left[
\Delta _{1}\right] }\right\rangle   \notag
\end{eqnarray}%
Equation (\ref{PCM}) can be written more compactly:%
\begin{equation*}
H_{\delta v^{\left( \eta \right) }}\left( 1-\mathbf{A}\right) _{\underline{%
\mathbf{\hat{\Lambda}}}^{\left[ \Delta _{1}\right] }}\delta v^{\left( \eta
\right) }\left\vert \underline{\mathbf{\hat{\Lambda}}}^{\left[ \Delta _{1}%
\right] }\right\rangle +\left( 1-\mathbf{A}\right) _{\underline{\mathbf{\hat{%
\Lambda}}}^{\left[ \Delta _{1}\right] }}H_{\underline{\mathbf{\hat{\Lambda}}}%
^{\left[ \Delta _{1}\right] }\delta v^{\left( \eta \right) }}H\left( 
\underline{\mathbf{\hat{\Lambda}}}^{\left[ \Delta _{1}\right] }\right)
\delta v^{\left( \eta \right) }\left\vert \underline{\mathbf{\hat{\Lambda}}}%
^{\left[ \Delta _{1}\right] }\right\rangle =0
\end{equation*}%
or equivalently as an equation for a state:%
\begin{equation*}
\left\vert \delta v^{\left( \eta \right) },\underline{\mathbf{\hat{\Lambda}}}%
^{\left[ \Delta _{1}\right] }\right\rangle =\delta v^{\left( \eta \right)
}\left\vert \underline{\mathbf{\hat{\Lambda}}}^{\left[ \Delta _{1}\right]
}\right\rangle 
\end{equation*}%
as:%
\begin{equation}
\left( H_{\delta v^{\left( \eta \right) }}+H_{\underline{\mathbf{\hat{\Lambda%
}}}^{\left[ \Delta _{1}\right] }\delta v^{\left( \eta \right) }}H\left( 
\underline{\mathbf{\hat{\Lambda}}}^{\left[ \Delta _{1}\right] }\right) -%
\mathbf{A}_{\underline{\mathbf{\hat{\Lambda}}}^{\left[ \Delta _{1}\right]
}}\left( H_{\delta v^{\left( \eta \right) }}+H_{\underline{\mathbf{\hat{%
\Lambda}}}^{\left[ \Delta _{1}\right] }\delta v^{\left( \eta \right)
}}H\left( \underline{\mathbf{\hat{\Lambda}}}^{\left[ \Delta _{1}\right]
}\right) \right) \right) \left\vert \delta v^{\left( \eta \right) },%
\underline{\mathbf{\hat{\Lambda}}}^{\left[ \Delta _{1}\right] }\right\rangle
=0  \label{TDF}
\end{equation}%
Equation (\ref{TDF}) for state $\left\vert \delta v^{\left( \eta \right) },%
\underline{\mathbf{\hat{\Lambda}}}^{\left[ \Delta _{1}\right] }\right\rangle 
$ describes the tangent deformation of the constraint by a state
characterized by internal degrees of freedom $\delta v^{\left( \eta \right) }
$, and apparently a-priori parameters $\underline{\mathbf{\hat{\Lambda}}}^{%
\left[ \Delta _{1}\right] }$. In fact, this parameter keeps track of the
projected background, as described in Part I and II.

\subsection{Composition of modifications}

We can compose two infinitesimal variations preserving the constraints. As
before, see (\ref{CNPS}), we can write for the first modification:%
\begin{equation}
H\left( \left\{ \mathbf{L+\delta L}\right\} ,\left\{ v^{\left( \eta \right)
}+\delta v^{\left( \eta \right) }\right\} \right) \left\vert \left\{ \left(
1+\mathbf{A}_{1}\right) \underline{\mathbf{\hat{\Lambda}}}^{\left[ k_{i}%
\right] }\right\} \right\rangle \left\vert \underline{\mathbf{\hat{\Lambda}}}%
^{\left[ \Delta _{1}\right] }\right\rangle =0
\end{equation}%
Inserting the operator $\left( 1-\mathbf{A}_{1}\right) $, this becomes:%
\begin{equation*}
0=\left( 1-\mathbf{A}_{1}\right) H\left( \left\{ \mathbf{L+\delta }_{1}%
\mathbf{L}\right\} ,\left\{ v^{\left( \eta \right) }+\delta v_{1}^{\left(
\eta \right) }\right\} \right) \left( 1+\mathbf{A}_{1}\right) \left( 1-%
\mathbf{A}_{1}\right) \left\vert \left\{ \left( 1+\mathbf{A}_{1}\right) 
\underline{\mathbf{\hat{\Lambda}}}^{\left[ k_{i}\right] }\right\}
\right\rangle \left\vert \underline{\mathbf{\hat{\Lambda}}}^{\left[ \Delta
_{1}\right] }\right\rangle
\end{equation*}%
that can be simplified as:%
\begin{equation*}
0=H\left( \left\{ \mathbf{L+\Delta }_{1}\mathbf{L}\right\} ,\left\{
v^{\left( \eta \right) }+\delta v_{1}^{\left( \eta \right) }\right\} \right)
\left\vert \left\{ \underline{\mathbf{\hat{\Lambda}}}^{\left[ k_{i}\right]
}\right\} \right\rangle \left\vert \left( 1-\mathbf{A}_{1}\right) \underline{%
\mathbf{\hat{\Lambda}}}^{\left[ \Delta _{1}\right] }\right\rangle
\end{equation*}%
This transformation can be composed with a second modification $\delta
v_{2}^{\left( \eta \right) }$ and we are led to:%
\begin{equation*}
0=\left( 1-\mathbf{A}_{21}\right) H\left( \left\{ \mathbf{L+\Delta }_{1}%
\mathbf{L+\delta }_{21}\mathbf{L}\right\} ,\left\{ v^{\left( \eta \right)
}+\delta v_{1}^{\left( \eta \right) }+\delta v_{2}^{\left( \eta \right)
}\right\} \right) \left\vert \left\{ \underline{\mathbf{\hat{\Lambda}}}^{%
\left[ k_{i}\right] }\right\} \right\rangle \left\vert \left( 1-\mathbf{A}%
_{1}\right) \underline{\mathbf{\hat{\Lambda}}}^{\left[ \Delta _{1}\right]
}\right\rangle
\end{equation*}%
The second transformation $\mathbf{A}_{21}$ models that transformation acts
on states already transformed by $\mathbf{A}_{1}$ and should a priori depend
locally on this transformation. The invariance of the constraints with
respect to the composition of transformations becomes:

\begin{equation*}
0=H\left( \left\{ \mathbf{L+\Delta }_{1}\mathbf{L+\Delta }_{21}\mathbf{L}%
\right\} ,\left\{ v^{\left( \eta \right) }+\delta v_{1}^{\left( \eta \right)
}+\delta v_{2}^{\left( \eta \right) }\right\} \right) \left\vert \left\{ 
\underline{\mathbf{\hat{\Lambda}}}^{\left[ k_{i}\right] }\right\}
\right\rangle \left\vert \left( 1-\mathbf{A}_{21}\right) \left( 1-\mathbf{A}%
_{1}\right) \underline{\mathbf{\hat{\Lambda}}}^{\left[ \Delta _{1}\right]
}\right\rangle
\end{equation*}

At least, even if the constraints are modified by both transformations, we
can derive conditions for these two deformations to commute:%
\begin{eqnarray*}
0 &=&H_{\delta v^{\left( \eta \right) }}\left( \left( 1-\mathbf{A}%
_{21}\right) \left( 1-\mathbf{A}_{1}\right) -\left( 1-\mathbf{A}_{12}\right)
\left( 1-\mathbf{A}_{2}\right) \right) \delta v^{\left( \eta \right)
}\left\vert \underline{\mathbf{\hat{\Lambda}}}^{\left[ \Delta _{1}\right]
}\right\rangle  \\
&&+\left( 1-\mathbf{A}_{21}\right) \left( 1-\mathbf{A}_{1}\right) H_{%
\underline{\mathbf{\hat{\Lambda}}}^{\left[ \Delta _{1}\right] }\delta
v^{\left( \eta \right) }}H_{21}\left( \underline{\mathbf{\hat{\Lambda}}}^{%
\left[ \Delta _{1}\right] }\right) -\left( 1-\mathbf{A}_{12}\right) \left( 1-%
\mathbf{A}_{2}\right) H_{\underline{\mathbf{\hat{\Lambda}}}^{\left[ \Delta
_{1}\right] }\delta v^{\left( \eta \right) }}H_{12}\left( \underline{\mathbf{%
\hat{\Lambda}}}^{\left[ \Delta _{1}\right] }\right) \delta v^{\left( \eta
\right) }
\end{eqnarray*}%
This expression can be factored in the following manner:%
\begin{eqnarray}
0 &=&\left( \left( 1-\mathbf{A}_{21}\right) \left( 1-\mathbf{A}_{1}\right)
-\left( 1-\mathbf{A}_{12}\right) \left( 1-\mathbf{A}_{2}\right) \right)  
\notag \\
&&\times \left( H_{\delta v^{\left( \eta \right) }}+\frac{1}{2}\left\{ H_{%
\underline{\mathbf{\hat{\Lambda}}}^{\left[ \Delta _{1}\right] }\delta
v^{\left( \eta \right) }}H_{21}\left( \underline{\mathbf{\hat{\Lambda}}}^{%
\left[ \Delta _{1}\right] }\right) +H_{\underline{\mathbf{\hat{\Lambda}}}^{%
\left[ \Delta _{1}\right] }\delta v^{\left( \eta \right) }}H_{12}\left( 
\underline{\mathbf{\hat{\Lambda}}}^{\left[ \Delta _{1}\right] }\right)
\right\} \right) \delta v^{\left( \eta \right) }  \notag \\
&&+\left( \left( 1-\mathbf{A}_{21}\right) \left( 1-\mathbf{A}_{1}\right)
+\left( 1-\mathbf{A}_{12}\right) \left( 1-\mathbf{A}_{2}\right) \right) 
\frac{1}{2}\left\{ H_{\underline{\mathbf{\hat{\Lambda}}}^{\left[ \Delta _{1}%
\right] }\delta v^{\left( \eta \right) }}H_{21}\left( \underline{\mathbf{%
\hat{\Lambda}}}^{\left[ \Delta _{1}\right] }\right) -H_{\underline{\mathbf{%
\hat{\Lambda}}}^{\left[ \Delta _{1}\right] }\delta v^{\left( \eta \right)
}}H_{12}\left( \underline{\mathbf{\hat{\Lambda}}}^{\left[ \Delta _{1}\right]
}\right) \right\} \delta v^{\left( \eta \right) }  \label{CMTNS}
\end{eqnarray}

We may assume that the coordinate transformations write, at least in first
approximation:%
\begin{equation*}
\mathbf{A}_{21}=\mathbf{A}_{2}\delta v_{2}^{\left( \eta \right) }+\mathbf{A}%
^{\prime }\left( \delta v_{2}^{\left( \eta \right) }-\delta v_{1}^{\left(
\eta \right) }\right) 
\end{equation*}%
and symmetrically:%
\begin{equation*}
\mathbf{A}_{12}=\mathbf{A}_{1}\delta v_{1}^{\left( \eta \right) }+\mathbf{A}%
^{\prime }\left( \delta v_{1}^{\left( \eta \right) }-\delta v_{2}^{\left(
\eta \right) }\right) 
\end{equation*}%
since for $\left( \delta v_{2}^{\left( \eta \right) }-\delta v_{1}^{\left(
\eta \right) }\right) =0$, the two modifications are identical and commute.

Similarly, for the modifications of generators we assume:%
\begin{equation*}
H_{21}\left( \underline{\mathbf{\hat{\Lambda}}}^{\left[ \Delta _{1}\right]
}\right) =H^{\prime }\left( \underline{\mathbf{\hat{\Lambda}}}^{\left[
\Delta _{1}\right] }\right) \left( \delta v_{2}^{\left( \eta \right)
}-\delta v_{1}^{\left( \eta \right) }\right) 
\end{equation*}%
and:%
\begin{equation*}
H_{12}\left( \underline{\mathbf{\hat{\Lambda}}}^{\left[ \Delta _{1}\right]
}\right) =H^{\prime }\left( \underline{\mathbf{\hat{\Lambda}}}^{\left[
\Delta _{1}\right] }\right) \left( \delta v_{1}^{\left( \eta \right)
}-\delta v_{2}^{\left( \eta \right) }\right) 
\end{equation*}%
We can compute the quantities involved in (\ref{CMTNS}) with these formula: 
\begin{eqnarray*}
&&\left( \left( 1-\mathbf{A}_{21}\right) \left( 1-\mathbf{A}_{1}\right)
-\left( 1-\mathbf{A}_{12}\right) \left( 1-\mathbf{A}_{2}\right) \right)  \\
&=&\left( \left( 1-\mathbf{A}_{2}\right) \left( 1-\mathbf{A}_{1}\right)
-\left( 1-\mathbf{A}_{1}\right) \left( 1-\mathbf{A}_{2}\right) \right) -%
\mathbf{A}^{\prime }\left( \delta v_{2}^{\left( \eta \right) }-\delta
v_{1}^{\left( \eta \right) }\right) \left( \left( 1-\mathbf{A}_{1}\right)
+\left( 1-\mathbf{A}_{2}\right) \right) 
\end{eqnarray*}%
and:%
\begin{equation*}
H_{\underline{\mathbf{\hat{\Lambda}}}^{\left[ \Delta _{1}\right] }\delta
v^{\left( \eta \right) }}H_{21}\left( \underline{\mathbf{\hat{\Lambda}}}^{%
\left[ \Delta _{1}\right] }\right) -H_{\underline{\mathbf{\hat{\Lambda}}}^{%
\left[ \Delta _{1}\right] }\delta v^{\left( \eta \right) }}H_{12}\left( 
\underline{\mathbf{\hat{\Lambda}}}^{\left[ \Delta _{1}\right] }\right) =H_{%
\underline{\mathbf{\hat{\Lambda}}}^{\left[ \Delta _{1}\right] }\delta
v^{\left( \eta \right) }}\left( \delta v_{2}^{\left( \eta \right) }-\delta
v_{1}^{\left( \eta \right) }\right) \left( H^{\prime }\left( \underline{%
\mathbf{\hat{\Lambda}}}^{\left[ \Delta _{1}\right] }\right) +H^{\prime
}\left( \underline{\mathbf{\hat{\Lambda}}}^{\left[ \Delta _{1}\right]
}\right) \right) 
\end{equation*}%
and:%
\begin{equation*}
H_{\underline{\mathbf{\hat{\Lambda}}}^{\left[ \Delta _{1}\right] }\delta
v^{\left( \eta \right) }}H_{21}\left( \underline{\mathbf{\hat{\Lambda}}}^{%
\left[ \Delta _{1}\right] }\right) +H_{\underline{\mathbf{\hat{\Lambda}}}^{%
\left[ \Delta _{1}\right] }\delta v^{\left( \eta \right) }}H_{12}\left( 
\underline{\mathbf{\hat{\Lambda}}}^{\left[ \Delta _{1}\right] }\right) =H_{%
\underline{\mathbf{\hat{\Lambda}}}^{\left[ \Delta _{1}\right] }\delta
v^{\left( \eta \right) }}\left( \delta v_{2}^{\left( \eta \right) }-\delta
v_{1}^{\left( \eta \right) }\right) \left( H^{\prime }\left( \underline{%
\mathbf{\hat{\Lambda}}}^{\left[ \Delta _{1}\right] }\right) -H^{\prime
}\left( \underline{\mathbf{\hat{\Lambda}}}^{\left[ \Delta _{1}\right]
}\right) \right) 
\end{equation*}%
The commutation relations should be independent from $\left( \delta
v_{2}^{\left( \eta \right) }-\delta v_{1}^{\left( \eta \right) }\right) $,
so that any decomposition $\delta v^{\left( \eta \right) }=\delta
v_{2}^{\left( \eta \right) }+\delta v_{1}^{\left( \eta \right) }$ should
lead ultimately to the same result. As a consequence, assuming:%
\begin{equation*}
\left( 1-\mathbf{A}_{2}\right) \left( 1-\mathbf{A}_{1}\right) -\left( 1-%
\mathbf{A}_{1}\right) \left( 1-\mathbf{A}_{2}\right) =0
\end{equation*}%
the commutation relations imply $H^{\prime }\left( \underline{\mathbf{\hat{%
\Lambda}}}^{\left[ \Delta _{1}\right] }\right) =0$ and $\mathbf{A}^{\prime
}=0$ .That is both "global" part of the transformations commute, and the
local modifications cancel. This means that the transformation is global.

These conditions are restrictive, so that we can expect that some holonomy
should appear while composing transformations. This is studied in the next
section.

\subsection{Second order expansion}

We consider the second order expansion of a transformation preserving the
constraint. This enables to study the consequence of this invariance for a
product of states. Define first:%
\begin{equation*}
\delta v^{\left( \eta \right) }=\delta v_{a}^{\left( \eta \right) }\delta
v_{b}^{\left( \eta \right) }
\end{equation*}%
for two states $\delta v_{a}^{\left( \eta \right) }$ and $\delta
v_{b}^{\left( \eta \right) }$, and:%
\begin{equation*}
\underline{\mathbf{\hat{\Lambda}}}^{\left[ \Delta _{1}\right] }=\left\vert 
\underline{\mathbf{\hat{\Lambda}}}_{a}^{\left[ \Delta _{1}\right]
}\right\rangle \left\vert \underline{\mathbf{\hat{\Lambda}}}_{b}^{\left[
\Delta _{1}\right] }\right\rangle
\end{equation*}%
Assuming that the constraints are invariant at the first order in
modifications, the second order variation of the constraints with respect to 
$\delta v_{a}^{\left( \eta \right) }$ and $\delta v_{b}^{\left( \eta \right)
}$ writes with these notations:%
\begin{eqnarray}
0 &=&\left( 1-\mathbf{A}\right) _{\underline{\mathbf{\hat{\Lambda}}}^{\left[
\Delta _{1}\right] }}\delta v_{a}^{\left( \eta \right) }H_{\delta
v_{a}^{\left( \eta \right) }\delta v_{b}^{\left( \eta \right) }}\delta
v_{b}^{\left( \eta \right) }\left\vert \underline{\mathbf{\hat{\Lambda}}}^{%
\left[ \Delta _{1}\right] }\right\rangle  \label{CLF} \\
&&+\left( 1-\mathbf{A}\right) _{\underline{\mathbf{\hat{\Lambda}}}^{\left[
\Delta _{1}\right] }}\delta v_{a}^{\left( \eta \right) }H_{\underline{%
\mathbf{\hat{\Lambda}}}^{\left[ \Delta _{1}\right] }\delta v_{a}^{\left(
\eta \right) }\delta v_{b}^{\left( \eta \right) }}\delta v_{b}^{\left( \eta
\right) }H\left( \underline{\mathbf{\hat{\Lambda}}}^{\left[ \Delta _{1}%
\right] }\right) \left\vert \underline{\mathbf{\hat{\Lambda}}}^{\left[
\Delta _{1}\right] }\right\rangle  \notag
\end{eqnarray}%
Assume that at least in first approximation the terms $H_{\delta v^{\left(
\eta \right) }\delta v^{\left( \eta ^{\prime }\right) }}$ and $H_{\underline{%
\mathbf{\hat{\Lambda}}}^{\left[ \Delta _{1}\right] }\delta v^{\left( \eta
\right) }\delta v^{\left( \eta ^{\prime }\right) }}$ can be factored:%
\begin{equation*}
H_{\delta v^{\left( \eta \right) }\delta v^{\left( \eta ^{\prime }\right)
}}=H_{a,b}=cH_{a}H_{b}
\end{equation*}%
and:%
\begin{equation*}
H_{\underline{\mathbf{\hat{\Lambda}}}^{\left[ \Delta _{1}\right] }\delta
v^{\left( \eta \right) }\delta v^{\left( \eta ^{\prime }\right) }}H\left( 
\underline{\mathbf{\hat{\Lambda}}}^{\left[ \Delta _{1}\right] }\right)
=c^{\prime }H_{\underline{\mathbf{\hat{\Lambda}}}^{\left[ \Delta _{1}\right]
}a}H_{\underline{\mathbf{\hat{\Lambda}}}^{\left[ \Delta _{1}\right]
}b}H\left( \underline{\mathbf{\hat{\Lambda}}}_{a}^{\left[ \Delta _{1}\right]
}\right) H\left( \underline{\mathbf{\hat{\Lambda}}}_{b}^{\left[ \Delta _{1}%
\right] }\right)
\end{equation*}%
corresponding to a constraints which is a sum of tensor products of
operators.

In this case, averaging equation (\ref{CLF}) over $\delta v_{a}^{\left( \eta
^{\prime }\right) }\delta v_{b}^{\left( \eta ^{\prime }\right) }\left\vert 
\underline{\mathbf{\hat{\Lambda}}}_{a}^{\left[ \Delta _{1}\right]
}\right\rangle \left\vert \underline{\mathbf{\hat{\Lambda}}}_{b}^{\left[
\Delta _{1}\right] }\right\rangle $: 
\begin{eqnarray*}
0 &=&\left\langle \underline{\mathbf{\hat{\Lambda}}}_{a}^{\left[ \Delta _{1}%
\right] }\right\vert \left\langle \underline{\mathbf{\hat{\Lambda}}}_{b}^{%
\left[ \Delta _{1}\right] }\right\vert \left( 1-\mathbf{A}\right) _{%
\underline{\mathbf{\hat{\Lambda}}}_{a}^{\left[ \Delta _{1}\right] }}\left( 1-%
\mathbf{A}\right) _{\underline{\mathbf{\hat{\Lambda}}}_{b}^{\left[ \Delta
_{1}\right] }}\left\vert \underline{\mathbf{\hat{\Lambda}}}_{a}^{\left[
\Delta _{1}\right] }\right\rangle \left\vert \underline{\mathbf{\hat{\Lambda}%
}}_{b}^{\left[ \Delta _{1}\right] }\right\rangle \\
&&\times c\delta v_{a}^{\left( \eta ^{\prime }\right) }\delta v_{b}^{\left(
\eta ^{\prime }\right) }H_{a}\delta v_{a}^{\left( \eta \right) }H_{b}\delta
v_{b}^{\left( \eta \right) } \\
&&+\left\langle \underline{\mathbf{\hat{\Lambda}}}_{a}^{\left[ \Delta _{1}%
\right] }\right\vert \left\langle \underline{\mathbf{\hat{\Lambda}}}_{b}^{%
\left[ \Delta _{1}\right] }\right\vert \left( 1-\mathbf{A}\right) _{%
\underline{\mathbf{\hat{\Lambda}}}_{a}^{\left[ \Delta _{1}\right] }}\left( 1-%
\mathbf{A}\right) _{\underline{\mathbf{\hat{\Lambda}}}_{b}^{\left[ \Delta
_{1}\right] }}H\left( \underline{\mathbf{\hat{\Lambda}}}_{a}^{\left[ \Delta
_{1}\right] }\right) H\left( \underline{\mathbf{\hat{\Lambda}}}_{b}^{\left[
\Delta _{1}\right] }\right) \left\vert \underline{\mathbf{\hat{\Lambda}}}%
_{a}^{\left[ \Delta _{1}\right] }\right\rangle \left\vert \underline{\mathbf{%
\hat{\Lambda}}}_{b}^{\left[ \Delta _{1}\right] }\right\rangle \\
&&\times c^{\prime }\delta v_{a}^{\left( \eta ^{\prime }\right) }\delta
v_{b}^{\left( \eta ^{\prime }\right) }H_{\underline{\mathbf{\hat{\Lambda}}}^{%
\left[ \Delta _{1}\right] }a}H_{\underline{\mathbf{\hat{\Lambda}}}^{\left[
\Delta _{1}\right] }b}\delta v_{a}^{\left( \eta \right) }\delta
v_{b}^{\left( \eta \right) }
\end{eqnarray*}%
can be expanded as:%
\begin{eqnarray*}
0 &=&c\left\langle \underline{\mathbf{\hat{\Lambda}}}_{a}^{\left[ \Delta _{1}%
\right] },\delta v_{a}^{\left( \eta ^{\prime }\right) }\right\vert \left( 1-%
\mathbf{A}\right) _{\underline{\mathbf{\hat{\Lambda}}}_{a}^{\left[ \Delta
_{1}\right] }}H_{a}\left\vert \underline{\mathbf{\hat{\Lambda}}}_{a}^{\left[
\Delta _{1}\right] },\delta v_{a}^{\left( \eta \right) }\right\rangle
\left\langle \underline{\mathbf{\hat{\Lambda}}}_{b}^{\left[ \Delta _{1}%
\right] }\delta v_{b}^{\left( \eta ^{\prime }\right) }\right\vert \left( 1-%
\mathbf{A}\right) _{\underline{\mathbf{\hat{\Lambda}}}_{b}^{\left[ \Delta
_{1}\right] }}H_{b}\left\vert \underline{\mathbf{\hat{\Lambda}}}_{b}^{\left[
\Delta _{1}\right] },\delta v_{b}^{\left( \eta ^{\prime }\right)
}\right\rangle \\
&&+c^{\prime }\left\langle \underline{\mathbf{\hat{\Lambda}}}_{a}^{\left[
\Delta _{1}\right] },\delta v_{a}^{\left( \eta ^{\prime }\right)
}\right\vert \left( 1-\mathbf{A}\right) _{\underline{\mathbf{\hat{\Lambda}}}%
_{a}^{\left[ \Delta _{1}\right] }}H_{\underline{\mathbf{\hat{\Lambda}}}^{%
\left[ \Delta _{1}\right] }a}H\left( \underline{\mathbf{\hat{\Lambda}}}_{a}^{%
\left[ \Delta _{1}\right] }\right) \left\vert \underline{\mathbf{\hat{\Lambda%
}}}_{a}^{\left[ \Delta _{1}\right] },\delta v_{a}^{\left( \eta \right)
}\right\rangle \\
&&\times \left\langle \underline{\mathbf{\hat{\Lambda}}}_{b}^{\left[ \Delta
_{1}\right] }\delta v_{b}^{\left( \eta ^{\prime }\right) }\right\vert \left(
1-\mathbf{A}\right) _{\underline{\mathbf{\hat{\Lambda}}}_{b}^{\left[ \Delta
_{1}\right] }}H_{\underline{\mathbf{\hat{\Lambda}}}^{\left[ \Delta _{1}%
\right] }b}H\left( \underline{\mathbf{\hat{\Lambda}}}_{b}^{\left[ \Delta _{1}%
\right] }\right) \left\vert \underline{\mathbf{\hat{\Lambda}}}_{b}^{\left[
\Delta _{1}\right] },\delta v_{b}^{\left( \eta ^{\prime }\right)
}\right\rangle
\end{eqnarray*}

In the case $\left\vert \underline{\mathbf{\hat{\Lambda}}}_{b}^{\left[
\Delta _{1}\right] },\delta v_{b}^{\left( \eta ^{\prime }\right)
}\right\rangle =\left\vert \underline{\mathbf{\hat{\Lambda}}}_{a}^{\left[
\Delta _{1}\right] },\delta v_{a}^{\left( \eta \right) }\right\rangle $ that
is, when two identical modifications are performed, we have:%
\begin{eqnarray*}
0 &=&c\left\langle \underline{\mathbf{\hat{\Lambda}}}_{a}^{\left[ \Delta _{1}%
\right] },\delta v_{a}^{\left( \eta ^{\prime }\right) }\right\vert \left( 1-%
\mathbf{A}\right) _{\underline{\mathbf{\hat{\Lambda}}}_{a}^{\left[ \Delta
_{1}\right] }}H_{a}\left\vert \underline{\mathbf{\hat{\Lambda}}}_{a}^{\left[
\Delta _{1}\right] },\delta v_{a}^{\left( \eta \right) }\right\rangle ^{2} \\
&&+c^{\prime }\left\langle \underline{\mathbf{\hat{\Lambda}}}_{b}^{\left[
\Delta _{1}\right] }\delta v_{b}^{\left( \eta ^{\prime }\right) }\right\vert
\left( 1-\mathbf{A}\right) _{\underline{\mathbf{\hat{\Lambda}}}_{a}^{\left[
\Delta _{1}\right] }}H_{\underline{\mathbf{\hat{\Lambda}}}^{\left[ \Delta
_{1}\right] }a}H\left( \underline{\mathbf{\hat{\Lambda}}}_{a}^{\left[ \Delta
_{1}\right] }\right) \left\vert \underline{\mathbf{\hat{\Lambda}}}_{a}^{%
\left[ \Delta _{1}\right] },\delta v_{a}^{\left( \eta \right) }\right\rangle
^{2}
\end{eqnarray*}%
and depending on the value of $c$ and $c^{\prime }$, states $\left\vert 
\underline{\mathbf{\hat{\Lambda}}}_{a}^{\left[ \Delta _{1}\right] },\delta
v_{a}^{\left( \eta \right) }\right\rangle \left\vert \underline{\mathbf{\hat{%
\Lambda}}}_{a}^{\left[ \Delta _{1}\right] },\delta v_{a}^{\left( \eta
\right) }\right\rangle $ may exist or not. This constraint on the number of
states depends on the sign of the modificatn in the number of genertrs. It
should als arise in the context of commutations relations, which is the
topic of the next section.

\section{Discrete transformations}

Up until now, our focus has been on continuous deformations of the
constraints, and we have explored the implications of state modifications
that preserve these constraints. We now delve into the possibilities of
discrete deformations that entail a change in the number of parameters.

We will consider two transformations of state $v^{\left( \eta \right) }$
that modify the constraints by changing the number of generators. In other
words, these transformations modify the space $V\left( v^{\left( \eta
\right) }\right) $. Moreover, we do not assume that these transformations
induce the same modification of the constraints but rather impose some
conditions on the difference between these modifications. These conditions
translates into commutation relations between the generators of
transformations. To illustrate these ideas we will only consider a
simplified example.

Suppose the states space to which the $v^{\left( \eta \right) }$ belong, is
composed of sums of tensor products, contingent on the parameters $\left\{ 
\underline{\mathbf{\hat{\Lambda}}}^{\left[ k_{i}\right] }\right\} $ through
the constraints for the degeneracy operators $\exp \left( i\left\{ 
\underline{\mathbf{\hat{\Lambda}}}^{\left[ k_{i}\right] }\right\} .\mathbf{l}%
\right) $. These degeneracy operators depend themselves on the considered
state $v^{\left( \eta \right) }$ and a modification of this state may imply
a change of the set of generators, and thus a modification of the parameters 
$\left\{ \underline{\mathbf{\hat{\Lambda}}}^{\left[ k_{i}\right] }\right\} $.

We consider in the present example that modifying $v^{\left( \eta \right) }$
by tensoring or destroying one state results in transforming the set $%
\left\{ \underline{\mathbf{\hat{\Lambda}}}^{\left[ k_{i}\right] }\right\} $
by adding, or removing one or several parameters, with multiplicity. In part
I and II this transformation arose from the projection onto some background
state (see section 4.3). Tensoring a given state $v^{\left( \eta \right) }$
by some state $\delta v^{\left( \eta \right) }$ involves more realizations
of the non-projected field in the series expansion of the background state.
This increase in the number of realizations induces more symmetries
generatrs that mix the realizations of the fields.

Actually, starting with the symmetry generators $\mathbf{L}\left( \Psi
_{J}^{\otimes k_{p_{0}}},v\otimes \delta \nu \right) $:

\begin{eqnarray}
\mathbf{L}\left( \Psi _{J}^{\otimes k_{p_{0}}},v\otimes \delta \nu \right)
&=&\sum_{_{\substack{ \left( s,s^{\prime },\left[ p,p^{\prime },k^{\prime }%
\right] \right)  \\ \left( m_{1},m_{1}^{\prime },\left[ p,p^{\prime
},k^{\prime }\right] \right) }}}d\left( \left\{ \overline{\left[
p_{0},p_{l^{\prime }0}\right] ^{k_{l^{\prime }}}}\right\} _{\substack{ %
l\leqslant s  \\ l^{\prime }\leqslant s^{\prime }}}\right) d\left( \left\{ %
\left[ p_{l_{1}},p_{l_{1}^{\prime }l_{1}}\right] ^{k_{l_{1}^{\prime
}}}\right\} _{\substack{ l_{1}\leqslant m_{1}  \\ l_{1}^{\prime }\leqslant
m^{\prime }}}\right) \\
&&\mathbf{l}\left( \mathbf{U}_{\left[ p_{l},p_{l^{\prime }l}\right]
^{k_{l^{\prime }}}},\Pi _{\left[ p_{l},p_{l^{\prime }l}\right]
^{k_{l^{\prime }}}},\overline{\left[ p_{0},p_{l^{\prime }0}\right]
^{k_{l^{\prime }}}},\left\{ \left[ p_{l_{1}},p_{l_{1}^{\prime }l_{1}}\right]
^{k_{l_{1}^{\prime }}}\right\} _{\substack{ l_{1}\leqslant m  \\ %
l_{1}^{\prime }\leqslant m^{\prime }}},\left[ p_{l_{\delta }},p_{l_{\delta
}^{\prime }l_{\delta }}\right] ^{k_{l_{\delta }^{\prime }}},v\otimes \delta
\nu \right)  \notag \\
&&\times v_{\left\{ \left[ p_{l_{1}},p_{l_{1}^{\prime }l_{1}}\right]
^{k_{l_{1}^{\prime }}}\right\} _{\substack{ l_{1}\leqslant m  \\ %
l_{1}^{\prime }\leqslant m^{\prime }}}}\left\{ \Psi _{J}^{\otimes
k_{l^{\prime }}}\right\} \delta \nu _{\left[ p_{l_{\delta }},p_{l_{\delta
}^{\prime }l_{\delta }}\right] ^{k_{l_{\delta }^{\prime }}}}\left\{ \Psi
_{J}^{\otimes k_{l^{\prime }}}\right\}  \notag
\end{eqnarray}%
and assuming that operators:%
\begin{equation*}
\mathbf{l}\left( \mathbf{U}_{\left[ p_{l},p_{l^{\prime }l}\right]
^{k_{l^{\prime }}}},\Pi _{\left[ p_{l},p_{l^{\prime }l}\right]
^{k_{l^{\prime }}}},\overline{\left[ p_{0},p_{l^{\prime }0}\right]
^{k_{l^{\prime }}}},\left\{ \left[ p_{l_{1}},p_{l_{1}^{\prime }l_{1}}\right]
^{k_{l_{1}^{\prime }}}\right\} _{\substack{ l_{1}\leqslant m  \\ %
l_{1}^{\prime }\leqslant m^{\prime }}},\left[ p_{l_{\delta }},p_{l_{\delta
}^{\prime }l_{\delta }}\right] ^{k_{l_{\delta }^{\prime }}},v\otimes \delta
\nu \right)
\end{equation*}%
decompose as two set of generators, one "large" initial set and a set
corresponding to the modification:%
\begin{equation*}
\mathbf{l}\left( \mathbf{U}_{\left[ p_{l},p_{l^{\prime }l}\right]
^{k_{l^{\prime }}}},\Pi _{\left[ p_{l},p_{l^{\prime }l}\right]
^{k_{l^{\prime }}}},\overline{\left[ p_{0},p_{l^{\prime }0}\right]
^{k_{l^{\prime }}}},\left\{ \left[ p_{l_{1}},p_{l_{1}^{\prime }l_{1}}\right]
^{k_{l_{1}^{\prime }}}\right\} _{\substack{ l_{1}\leqslant m  \\ %
l_{1}^{\prime }\leqslant m^{\prime }}},v\right) ,\delta \mathbf{l}\left( 
\mathbf{U}_{\left[ p_{l},p_{l^{\prime }l}\right] ^{k_{l^{\prime }}}},\Pi _{%
\left[ p_{l},p_{l^{\prime }l}\right] ^{k_{l^{\prime }}}},\left[ p_{l_{\delta
}},p_{l_{\delta }^{\prime }l_{\delta }}\right] ^{k_{l_{\delta }^{\prime
}}},\delta \nu \right)
\end{equation*}%
the modification of the symetry transformations can be written: 
\begin{eqnarray*}
&&\exp \left( i\mathbf{L}\left( \Psi _{J}^{\otimes k_{p_{0}}},v\right)
.\left\{ \left[ 
\begin{array}{c}
\left( p_{l_{i}},p_{l_{i}^{\prime }l_{i}},k_{l_{i}^{\prime }}\right) \\ 
\left[ \left\{ \Psi _{J}\left[ l^{\prime },0\right] \right\} ,v\right]%
\end{array}%
\right] \right\} \right) \\
&\rightarrow &\exp \left( i\mathbf{L}\left( \Psi _{J}^{\otimes
k_{p_{0}}},v\right) +\delta \mathbf{l}\left( \Psi _{J}^{\otimes
k_{p_{0}}},\delta \nu \right) .\left\{ \left[ 
\begin{array}{c}
\left( p_{l_{i}},p_{l_{i}^{\prime }l_{i}},k_{l_{i}^{\prime }}\right) +\left(
p_{l_{\delta }},p_{l_{\delta }^{\prime }l_{\delta }}k_{l_{\delta }^{\prime
}}\right) \\ 
\left[ \left\{ \Psi _{J}\left[ l^{\prime },0\right] \right\} ,v,\delta \nu %
\right]%
\end{array}%
\right] \right\} \right)
\end{eqnarray*}%
where the parameters decompose as a large set of parameters, plus the
parameters corresponding to the modification of generators:%
\begin{equation*}
\left\{ \left[ 
\begin{array}{c}
\left( p_{l_{i}},p_{l_{i}^{\prime }l_{i}},k_{l_{i}^{\prime }}\right) +\left(
p_{l_{\delta }},p_{l_{\delta }^{\prime }l_{\delta }}k_{l_{\delta }^{\prime
}}\right) \\ 
\left[ \left\{ \Psi _{J}\left[ l^{\prime },0\right] \right\} ,v,\delta \nu %
\right]%
\end{array}%
\right] \right\} =\left\{ \left[ 
\begin{array}{c}
\left( p_{l_{i}},p_{l_{i}^{\prime }l_{i}},k_{l_{i}^{\prime }}\right) \\ 
\left[ \left\{ \Psi _{J}\left[ l^{\prime },0\right] \right\} ,v,\delta \nu %
\right]%
\end{array}%
\right] \right\} +\left\{ \left[ 
\begin{array}{c}
\left( p_{l_{\delta }},p_{l_{\delta }^{\prime }l_{\delta }}k_{l_{\delta
}^{\prime }}\right) \\ 
\left[ \left\{ \Psi _{J}\left[ l^{\prime },0\right] \right\} ,v,\delta \nu %
\right]%
\end{array}%
\right] \right\}
\end{equation*}%
This decomposition implies that the set of parameters is modified by the
adjunction of some new variables:%
\begin{equation*}
\left\{ \underline{\mathbf{\hat{\Lambda}}}^{\left[ k_{i}\right] }\right\}
\rightarrow \left\{ \underline{\mathbf{\hat{\Lambda}}}^{\left[ k_{i}\right]
},\underline{\mathbf{\hat{\Lambda}}}^{\left[ \Delta \right] }\right\}
\end{equation*}

When considering two successive modifications creating and destroying a
state:%
\begin{equation*}
v^{\left( \eta \right) }\rightarrow A_{\left( \delta v^{\left( \eta \right)
}\right) ^{\prime }}^{-}A_{\delta v^{\left( \eta \right) }}^{+}v^{\left(
\eta \right) }
\end{equation*}%
or destroying and creating a state:%
\begin{equation*}
v^{\left( \eta \right) }\rightarrow A_{\delta v^{\left( \eta \right)
}}^{+}A_{\left( \delta v^{\left( \eta \right) }\right) ^{\prime
}}^{-}v^{\left( \eta \right) }
\end{equation*}%
the first modification increases the number of sets of parameters from $%
\left\{ \underline{\mathbf{\hat{\Lambda}}}^{\left[ k_{i}\right] }\right\} $
to $\left\{ \left\{ \underline{\mathbf{\hat{\Lambda}}}^{\left[ k_{i}\right]
}\right\} ,\underline{\mathbf{\hat{\Lambda}}}^{\left[ \Delta _{1}\right]
}\right\} $ and then reduces this set to $\left\{ \left\{ \underline{\mathbf{%
\hat{\Lambda}}}^{\left[ k_{i}\right] }\right\} ,\underline{\mathbf{\hat{%
\Lambda}}}^{\left[ \Delta _{1}\right] }\right\} /\underline{\mathbf{\hat{%
\Lambda}}}^{\left[ \Delta _{2}\right] }$ while the second one modifies the
set $\left\{ \left\{ \underline{\mathbf{\hat{\Lambda}}}^{\left[ k_{i}\right]
}\right\} /\underline{\mathbf{\hat{\Lambda}}}^{\left[ \Delta _{2}\right] },%
\underline{\mathbf{\hat{\Lambda}}}^{\left[ \Delta _{1}\right] }\right\} $. \
Here, we define $\left\{ \underline{\mathbf{\hat{\Lambda}}}^{\left[ k_{i}%
\right] }\right\} /\underline{\mathbf{\hat{\Lambda}}}^{\left[ \Delta _{2}%
\right] }$ as:%
\begin{equation*}
\left\{ \underline{\mathbf{\hat{\Lambda}}}^{\left[ k_{i}\right] }\right\} /%
\underline{\mathbf{\hat{\Lambda}}}^{\left[ \Delta _{2}\right] }=\underset{%
S\cup \underline{\mathbf{\hat{\Lambda}}}^{\left[ \Delta _{2}\right]
}=\left\{ \underline{\mathbf{\hat{\Lambda}}}^{\left[ k_{i}\right] }\right\} }%
{\overset{\circ }{\cup }}S
\end{equation*}%
where $\overset{\circ }{\cup }$ is the disjoint union. Note that in the
definition of $\left\{ \underline{\mathbf{\hat{\Lambda}}}^{\left[ k_{i}%
\right] }\right\} $, the value of any parameter $\underline{\mathbf{\hat{%
\Lambda}}}^{\left[ k_{i}\right] }$,\ including $\underline{\mathbf{\hat{%
\Lambda}}}^{\left[ \Delta _{2}\right] }$, can arise with multiplicity.
Removing $\underline{\mathbf{\hat{\Lambda}}}^{\left[ \Delta _{2}\right] }$
corresponds to reduce this multiplicity by $1$.

The two sets $\left\{ \left\{ \underline{\mathbf{\hat{\Lambda}}}^{\left[
k_{i}\right] }\right\} ,\underline{\mathbf{\hat{\Lambda}}}^{\left[ \Delta
_{1}\right] }\right\} /\underline{\mathbf{\hat{\Lambda}}}^{\left[ \Delta _{2}%
\right] }$ and $\left\{ \left\{ \underline{\mathbf{\hat{\Lambda}}}^{\left[
k_{i}\right] }\right\} /\underline{\mathbf{\hat{\Lambda}}}^{\left[ \Delta
_{2}\right] },\underline{\mathbf{\hat{\Lambda}}}^{\left[ \Delta _{1}\right]
}\right\} $\ are identical if $\underline{\mathbf{\hat{\Lambda}}}^{\left[
\Delta _{2}\right] }\neq \underline{\mathbf{\hat{\Lambda}}}^{\left[ \Delta
_{1}\right] }$ whereas: 
\begin{equation*}
\left\{ \left\{ \underline{\mathbf{\hat{\Lambda}}}^{\left[ k_{i}\right]
}\right\} /\underline{\mathbf{\hat{\Lambda}}}^{\left[ \Delta _{1}\right] },%
\underline{\mathbf{\hat{\Lambda}}}^{\left[ \Delta _{1}\right] }\right\} 
\overset{\circ }{\cup }\left\{ \underline{\mathbf{\hat{\Lambda}}}^{\left[
k_{i}\right] }\right\} =\left\{ \left\{ \underline{\mathbf{\hat{\Lambda}}}^{%
\left[ k_{i}\right] }\right\} ,\underline{\mathbf{\hat{\Lambda}}}^{\left[
\Delta _{1}\right] }\right\} /\underline{\mathbf{\hat{\Lambda}}}^{\left[
\Delta _{1}\right] }
\end{equation*}

In terms of modification operators, we can write:%
\begin{equation*}
\left\{ \left\{ \underline{\mathbf{\hat{\Lambda}}}^{\left[ k_{i}\right]
}\right\} /\underline{\mathbf{\hat{\Lambda}}}^{\left[ \Delta _{2}\right] },%
\underline{\mathbf{\hat{\Lambda}}}^{\left[ \Delta _{1}\right] }\right\} =A_{%
\underline{\mathbf{\hat{\Lambda}}}^{\left[ \Delta _{1}\right] }}^{+}A_{%
\underline{\mathbf{\hat{\Lambda}}}^{\left[ \Delta _{2}\right] }}^{-}\left\{ 
\underline{\mathbf{\hat{\Lambda}}}^{\left[ k_{i}\right] }\right\} 
\end{equation*}%
and:%
\begin{equation*}
\left\{ \left\{ \underline{\mathbf{\hat{\Lambda}}}^{\left[ k_{i}\right]
}\right\} ,\underline{\mathbf{\hat{\Lambda}}}^{\left[ \Delta _{1}\right]
}\right\} /\underline{\mathbf{\hat{\Lambda}}}^{\left[ \Delta _{1}\right]
}=A_{\underline{\mathbf{\hat{\Lambda}}}^{\left[ \Delta _{2}\right] }}^{-}A_{%
\underline{\mathbf{\hat{\Lambda}}}^{\left[ \Delta _{1}\right] }}^{+}\left\{ 
\underline{\mathbf{\hat{\Lambda}}}^{\left[ k_{i}\right] }\right\} 
\end{equation*}%
where $A_{\underline{\mathbf{\hat{\Lambda}}}^{\left[ \Delta _{1}\right]
}}^{+}$ increases the multiplicity of a value $\underline{\mathbf{\hat{%
\Lambda}}}^{\left[ \Delta _{1}\right] }$ of the parameter by $1$, while $A_{%
\underline{\mathbf{\hat{\Lambda}}}^{\left[ \Delta _{2}\right] }}^{-}$
reduces it by $1$. These sets gather eigenstates of the set of operators $%
\left\{ A_{L^{\left[ \Delta _{2}\right] }}^{-}A_{L^{\left[ \Delta _{1}\right]
}}^{+}\mathbf{L}^{\left[ k_{i}\right] }\right\} $ and $\left\{ A_{L^{\left[
\Delta _{1}\right] }}^{+}A_{L^{\left[ \Delta _{2}\right] }}^{-}\mathbf{L}^{%
\left[ k_{i}\right] }\right\} $ respectively, where $A_{L^{\left[ \Delta _{1}%
\right] }}^{+}$ adds a generator $L^{\left[ \Delta _{1}\right] }$ and $A_{L^{%
\left[ \Delta _{2}\right] }}^{-}$ removes a generator $L^{\left[ \Delta _{2}%
\right] }$.

In the sequel, we will neglect the continuous transformations $\left( 1-%
\mathbf{A}_{i}\right) $ on the parameters. As a consequence, given our
assumptions, the first modification of the constraints writes in terms f
generators:%
\begin{equation}
0=H\left( \left\{ A_{\delta v_{2}^{\left( \eta \right) }}^{-}A_{\delta
v_{1}^{\left( \eta \right) }}^{+}v^{\left( \eta \right) }\right\} ,\left\{
A_{L^{\left[ \Delta _{2}\right] }}^{-}A_{L^{\left[ \Delta _{1}\right] }}^{+}%
\mathbf{L}^{\left[ k_{i}\right] }\right\} \right)   \label{Cnf}
\end{equation}%
and the second one:%
\begin{equation}
0=H\left( \left\{ A_{\delta v_{1}^{\left( \eta \right) }}^{+}A_{\delta
v_{2}^{\left( \eta \right) }}^{-}v^{\left( \eta \right) }\right\} ,\left\{
A_{L^{\left[ \Delta _{1}\right] }}^{+}A_{L^{\left[ \Delta _{2}\right] }}^{-}%
\mathbf{L}^{\left[ k_{i}\right] }\right\} \right)   \label{Cns}
\end{equation}%
Since the two constraints involve a different number of generator, we may
assume a discontinuity in the modification, so that the expression of these
two constraints (\ref{Cnf}) nd (\ref{Cns}) does not arise necessarily from
the same expansion of the initial constraint:%
\begin{equation}
0=H\left( \left\{ v^{\left( \eta \right) }\right\} ,\left\{ \mathbf{L}^{%
\left[ k_{i}\right] }\right\} \right)   \label{rtd}
\end{equation}%
In the sequel we assume the expansion:%
\begin{equation}
H\left( \left\{ v^{\left( \eta \right) }\right\} ,\left\{ \mathbf{L}^{\left[
k_{i}\right] }\right\} \right) =\sum_{s}H^{\left( s\right) }\left( \left\{
v^{\left( \eta \right) }\right\} \right) J^{\left( s\right) }\left( \left\{ 
\mathbf{L}^{\left[ k_{i}\right] }\right\} \right)   \label{pns}
\end{equation}%
Gven this assumption, we derive:%
\begin{equation*}
H\left( \left\{ A_{\delta v_{2}^{\left( \eta \right) }}^{-}A_{\delta
v_{1}^{\left( \eta \right) }}^{+}v^{\left( \eta \right) }\right\} ,A_{L^{
\left[ \Delta _{2}\right] }}^{-}A_{L^{\left[ \Delta _{1}\right]
}}^{+}\left\{ \mathbf{L}^{\left[ k_{i}\right] }\right\} \right) 
\end{equation*}%
by considerng the constraints for the group operator:%
\begin{equation*}
\exp \left( \left\{ \mathbf{A}_{\mathbf{L}^{\left[ k_{i}\right]
}}^{+}\right\} \mathbf{.\alpha }\right) 
\end{equation*}%
nd $v^{\left( \eta \right) }$ being some coherent stats:%
\begin{equation*}
\exp \left( A_{v^{\left( \eta \right) }}^{+}\right) \left\vert
Vac\right\rangle 
\end{equation*}%
and consider the finite transformation:%
\begin{equation*}
T_{1}=\exp \left( A_{\delta v_{2}^{\left( \eta \right) }}^{-}A_{\delta
v_{1}^{\left( \eta \right) }}^{+}\right) \exp \left( A_{L^{\left[ \Delta _{2}%
\right] }}^{-}A_{L^{\left[ \Delta _{1}\right] }}^{+}\right) 
\end{equation*}%
The constraint for the state transformd by $T_{1}$ is written:%
\begin{equation*}
H_{1}\left( \left\{ \exp \left( A_{\delta v_{2}^{\left( \eta \right)
}}^{-}A_{\delta v_{1}^{\left( \eta \right) }}^{+}\right) v^{\left( \eta
\right) }\right\} ,\exp \left( A_{L^{\left[ \Delta _{2}\right] }}^{-}A_{L^{%
\left[ \Delta _{1}\right] }}^{+}\right) \left\{ \mathbf{L}^{\left[ k_{i}%
\right] }\right\} \right) 
\end{equation*}%
and we assume that this constraint can be expanded in series:%
\begin{eqnarray*}
&&H_{1}\left( \left\{ \exp \left( A_{\delta v_{2}^{\left( \eta \right)
}}^{-}A_{\delta v_{1}^{\left( \eta \right) }}^{+}\right) v^{\left( \eta
\right) }\right\} ,\exp \left( A_{L^{\left[ \Delta _{2}\right] }}^{-}A_{L^{%
\left[ \Delta _{1}\right] }}^{+}\right) \left\{ \mathbf{L}^{\left[ k_{i}%
\right] }\right\} \right)  \\
&=&\sum_{s}H_{1}^{\left( s\right) }\left( \left\{ \exp \left( A_{\delta
v_{2}^{\left( \eta \right) }}^{-}A_{\delta v_{1}^{\left( \eta \right)
}}^{+}\right) v^{\left( \eta \right) }\right\} \right) J_{1}^{\left(
s\right) }\left( \exp \left( A_{L^{\left[ \Delta _{2}\right] }}^{-}A_{L^{%
\left[ \Delta _{1}\right] }}^{+}\right) \left\{ \mathbf{L}^{\left[ k_{i}%
\right] }\right\} \right) 
\end{eqnarray*}

Assuming multiplicative modification for group elements, the constraint for
first modification is:%
\begin{equation*}
\sum_{s}H_{1}^{-+s}\left( \exp \left( A_{\delta v_{2}^{\left( \eta \right)
}}^{-}A_{\delta v_{1}^{\left( \eta \right) }}^{+}\right) \right)
H_{1}^{\left( s\right) }\left( \left\{ v^{\left( \eta \right) }\right\}
\right) J_{1}^{-+s}\left( \exp \left( A_{L^{\left[ \Delta _{2}\right]
}}^{-}A_{L^{\left[ \Delta _{1}\right] }}^{+}\right) \right) J_{1}^{\left(
s\right) }\left( \left\{ \mathbf{L}^{\left[ k_{i}\right] }\right\} \right) 
\end{equation*}%
We then consider that the modification operators do not depend in $s$: 
\begin{eqnarray*}
H_{1}^{-+s}\left( \exp \left( A_{\delta v_{2}^{\left( \eta \right)
}}^{-}A_{\delta v_{1}^{\left( \eta \right) }}^{+}\right) \right) 
&=&H_{1}^{-+}\left( \exp \left( A_{\delta v_{2}^{\left( \eta \right)
}}^{-}A_{\delta v_{1}^{\left( \eta \right) }}^{+}\right) \right)  \\
J_{1}^{-+s}\left( \exp \left( A_{\underline{\mathbf{\hat{\Lambda}}}^{\left[
\Delta _{2}\right] }}^{-}A_{\underline{\mathbf{\hat{\Lambda}}}^{\left[
\Delta _{1}\right] }}^{+}\right) \right)  &=&J_{1}^{-+}\left( \exp \left(
A_{L^{\left[ \Delta _{2}\right] }}^{-}A_{L^{\left[ \Delta _{1}\right]
}}^{+}\right) \right) 
\end{eqnarray*}%
and: 
\begin{eqnarray*}
H_{1}^{-+}\left( \exp \left( A_{\delta v_{2}^{\left( \eta \right)
}}^{-}A_{\delta v_{1}^{\left( \eta \right) }}^{+}\right) \right)  &\equiv
&H_{1}^{-+}\left( \exp \left( A_{\delta v_{2}^{\left( \eta \right)
}}^{-}A_{\delta v_{1}^{\left( \eta \right) }}^{+}\right) ,v^{\left( \eta
\right) }\right)  \\
J_{1}^{-+}\left( \exp \left( A_{\underline{\mathbf{\hat{\Lambda}}}^{\left[
\Delta _{2}\right] }}^{-}A_{\underline{\mathbf{\hat{\Lambda}}}^{\left[
\Delta _{1}\right] }}^{+}\right) \right)  &\equiv &J_{1}^{-+}\left( \exp
\left( A_{L^{\left[ \Delta _{2}\right] }}^{-}A_{L^{\left[ \Delta _{1}\right]
}}^{+}\right) ,v^{\left( \eta \right) },\left\{ \mathbf{L}^{\left[ k_{i}%
\right] }\right\} \right) 
\end{eqnarray*}%
where the dependency in $v^{\left( \eta \right) }$ and $\left\{ \mathbf{L}^{%
\left[ k_{i}\right] }\right\} $ are kept implicit.

We also assume that the modification of the constraint is included in $%
H_{1}^{-+}$ and $J_{1}^{-+}$, so that:%
\begin{eqnarray*}
H_{1}^{\left( s\right) }\left( \left\{ v^{\left( \eta \right) }\right\}
\right) &=&H^{\left( s\right) }\left( \left\{ v^{\left( \eta \right)
}\right\} \right) \\
J_{1}^{\left( s\right) }\left( \left\{ \mathbf{L}^{\left[ k_{i}\right]
}\right\} \right) &=&J^{\left( s\right) }\left( \left\{ \mathbf{L}^{\left[
k_{i}\right] }\right\} \right)
\end{eqnarray*}%
and using (\ref{pns}) the constraint writes:%
\begin{eqnarray*}
&&\sum_{s}H_{1}^{-+}\left( \exp \left( A_{\delta v_{2}^{\left( \eta \right)
}}^{-}A_{\delta v_{1}^{\left( \eta \right) }}^{+}\right) \right) H^{\left(
s\right) }\left( \left\{ v^{\left( \eta \right) }\right\} \right)
J_{1}^{-+}\left( \exp \left( A_{L^{\left[ \Delta _{2}\right] }}^{-}A_{L^{%
\left[ \Delta _{1}\right] }}^{+}\right) \right) J^{\left( s\right) }\left(
\left\{ \mathbf{L}^{\left[ k_{i}\right] }\right\} \right) \\
&&H_{1}^{-+}\left( \exp \left( A_{\delta v_{2}^{\left( \eta \right)
}}^{-}A_{\delta v_{1}^{\left( \eta \right) }}^{+}\right) \right)
J_{1}^{-+}\left( \exp \left( A_{L^{\left[ \Delta _{2}\right] }}^{-}A_{L^{%
\left[ \Delta _{1}\right] }}^{+}\right) \right) H\left( \left\{ v^{\left(
\eta \right) }\right\} ,\left\{ \mathbf{L}^{\left[ k_{i}\right] }\right\}
\right)
\end{eqnarray*}%
We have also to take into account that parameters depending on $v$ and
include the modification:

\begin{equation*}
\partial _{\left\{ \underline{\mathbf{\hat{\Lambda}}}^{\left[ k_{i}\right]
}\right\} }H\left( \left\{ v^{\left( \eta \right) }\right\} ,\left\{ \mathbf{%
L}^{\left[ k_{i}\right] }\right\} \right) \frac{\partial \left\{ \mathbf{L}^{%
\left[ k_{i}\right] }\right\} }{\partial v^{\left( \eta \right) }}\left[
A_{\delta v_{2}^{\left( \eta \right) }}^{-}A_{\delta v_{1}^{\left( \eta
\right) }}^{+}v^{\left( \eta \right) }-v^{\left( \eta \right) }\right]
\end{equation*}%
using (\ref{rtd}) the constraint becomes at the first orders:%
\begin{eqnarray}
0 &=&\partial _{\left\{ \underline{\mathbf{\hat{\Lambda}}}^{\left[ k_{i}%
\right] }\right\} }H\left( \left\{ v^{\left( \eta \right) }\right\} ,\left\{ 
\mathbf{L}^{\left[ k_{i}\right] }\right\} \right) \frac{\partial \left\{ 
\mathbf{L}^{\left[ k_{i}\right] }\right\} }{\partial v^{\left( \eta \right) }%
}\left[ A_{\delta v_{2}^{\left( \eta \right) }}^{-}A_{\delta v_{1}^{\left(
\eta \right) }}^{+}v^{\left( \eta \right) }-v^{\left( \eta \right) }\right]
\label{CTn} \\
&&+\left( h_{1}^{-+}A_{\delta v_{2}^{\left( \eta \right) }}^{-}A_{\delta
v_{1}^{\left( \eta \right) }}^{+}+j_{1}^{-+}A_{L^{\left[ \Delta _{2}\right]
}}^{-}A_{L^{\left[ \Delta _{1}\right] }}^{+}\right) H\left( \left\{
v^{\left( \eta \right) }\right\} ,\left\{ \mathbf{L}^{\left[ k_{i}\right]
}\right\} \right)  \notag \\
&&+h_{1}^{-+}j_{1}^{-+}A_{\delta v_{2}^{\left( \eta \right) }}^{-}A_{\delta
v_{1}^{\left( \eta \right) }}^{+}A_{L^{\left[ \Delta _{2}\right] }}^{-}A_{L^{%
\left[ \Delta _{1}\right] }}^{+}H\left( \left\{ v^{\left( \eta \right)
}\right\} ,\left\{ \mathbf{L}^{\left[ k_{i}\right] }\right\} \right)  \notag
\end{eqnarray}%
and the coefficients $h_{1}^{-+}$, $j_{1}^{-+}$ are the derivatives of \ $%
H_{1}^{-+}$ with respect to $\exp \left( A_{\delta v_{2}^{\left( \eta
\right) }}^{-}A_{\delta v_{1}^{\left( \eta \right) }}^{+}\right) $, and of $%
J_{1}^{-+}\exp \left( A_{L^{\left[ \Delta _{2}\right] }}^{-}A_{L^{\left[
\Delta _{1}\right] }}^{+}\right) $ evaluated at $1$. The dependency of $%
h_{1}^{-+}$ and $j_{1}^{-+}$ in $v^{\left( \eta \right) }$ and $v^{\left(
\eta \right) },\left\{ \mathbf{L}^{\left[ k_{i}\right] }\right\} $
respectively are kept implict.

Similarly, the second modification writes:%
\begin{eqnarray}
0 &=&\partial _{\left\{ \underline{\mathbf{\hat{\Lambda}}}^{\left[ k_{i}%
\right] }\right\} }H\left( \left\{ v^{\left( \eta \right) }\right\} ,\left\{ 
\mathbf{L}^{\left[ k_{i}\right] }\right\} \right) \frac{\partial \left\{ 
\mathbf{L}^{\left[ k_{i}\right] }\right\} }{\partial v^{\left( \eta \right) }%
}\left[ A_{\delta v_{1}^{\left( \eta \right) }}^{+}A_{\delta v_{2}^{\left(
\eta \right) }}^{-}v^{\left( \eta \right) }-v^{\left( \eta \right) }\right]
\label{CTd} \\
&&+\left( h_{2}^{+-}A_{\delta v_{2}^{\left( \eta \right) }}^{+}A_{\delta
v_{1}^{\left( \eta \right) }}^{-}+j_{2}^{+-}A_{L^{\left[ \Delta _{1}\right]
}}^{+}A_{L^{\left[ \Delta _{2}\right] }}^{-}\right) H\left( \left\{
v^{\left( \eta \right) }\right\} ,\left\{ \mathbf{L}^{\left[ k_{i}\right]
}\right\} \right)  \notag \\
&&+h_{2}^{+-}j_{2}^{+-}A_{\delta v_{2}^{\left( \eta \right) }}^{+}A_{\delta
v_{1}^{\left( \eta \right) }}^{-}A_{L^{\left[ \Delta _{1}\right] }}^{+}A_{L^{%
\left[ \Delta _{2}\right] }}^{-}H\left( \left\{ v^{\left( \eta \right)
}\right\} ,\left\{ \mathbf{L}^{\left[ k_{i}\right] }\right\} \right)  \notag
\end{eqnarray}%
Removing the indices $1$ and $2$ for coefficients $h_{i}$, $j_{i}$, the
difference of (\ref{CTn}) and (\ref{CTd}) this writes:%
\begin{eqnarray*}
&&-\partial _{\left\{ \underline{\mathbf{\hat{\Lambda}}}^{\left[ k_{i}\right]
}\right\} }H\left( \left\{ v^{\left( \eta \right) }\right\} ,\left\{ \mathbf{%
L}^{\left[ k_{i}\right] }\right\} \right) \frac{\partial \left\{ \mathbf{L}^{%
\left[ k_{i}\right] }\right\} }{\partial v^{\left( \eta \right) }}\delta
\left( \delta v_{2}^{\left( \eta \right) }-\delta v_{1}^{\left( \eta \right)
}\right) \\
&=&\left( \left\{ h^{-+}A_{\delta v_{2}^{\left( \eta \right) }}^{-}A_{\delta
v_{1}^{\left( \eta \right) }}^{+}-h^{+-}A_{\delta v_{1}^{\left( \eta \right)
}}^{+}A_{\delta v_{2}^{\left( \eta \right) }}^{-}\right\} +\left\{
j^{-+}A_{L^{\left[ \Delta _{2}\right] }}^{-}A_{L^{\left[ \Delta _{1}\right]
}}^{+}-j^{+-}A_{L^{\left[ \Delta _{1}\right] }}^{+}A_{L^{\left[ \Delta _{2}%
\right] }}^{-}\right\} \right) H\left( \left\{ v^{\left( \eta \right)
}\right\} ,\left\{ \mathbf{L}^{\left[ k_{i}\right] }\right\} \right) \\
&&+\left\{ h^{+-}j^{+-}A_{\delta v_{2}^{\left( \eta \right) }}^{-}A_{%
\underline{\mathbf{\hat{\Lambda}}}^{\left[ \Delta _{2}\right]
}}^{-}A_{\delta v_{1}^{\left( \eta \right) }}^{+}A_{\underline{\mathbf{\hat{%
\Lambda}}}^{\left[ \Delta _{1}\right] }}^{+}-h^{-+}j^{-+}A_{\delta
v_{1}^{\left( \eta \right) }}^{+}A_{\underline{\mathbf{\hat{\Lambda}}}^{%
\left[ \Delta _{1}\right] }}^{+}A_{\delta v_{2}^{\left( \eta \right)
}}^{-}A_{\underline{\mathbf{\hat{\Lambda}}}^{\left[ \Delta _{2}\right]
}}^{-}\right\} H\left( \left\{ v^{\left( \eta \right) }\right\} ,\left\{ 
\mathbf{L}^{\left[ k_{i}\right] }\right\} \right)
\end{eqnarray*}

We assume that $h_{1}^{-+}h_{1}^{+}=h_{2}^{+-}h_{2}^{-}=h$, so that the
deformation is in the modification of generator and if we evaluate the
commutator between $\left\langle \underline{\mathbf{\hat{\Lambda}}}^{\left[
k_{i}\right] },\underline{\mathbf{\hat{\Lambda}}}^{\left[ \Delta _{2}\right]
}\right\vert $ and $\left\vert \underline{\mathbf{\hat{\Lambda}}}^{\left[
k_{i}\right] },\underline{\mathbf{\hat{\Lambda}}}^{\left[ \Delta _{1}\right]
}\right\rangle $,\ we find:

\begin{eqnarray}
&&-\partial _{\left\{ \underline{\mathbf{\hat{\Lambda}}}^{\left[ k_{i}\right]
}\right\} }H\left( \left\{ v^{\left( \eta \right) }\right\} ,\left\{ 
\underline{\mathbf{\hat{\Lambda}}}^{\left[ k_{i}\right] }\right\} \right) 
\frac{\partial \left\{ \underline{\mathbf{\hat{\Lambda}}}^{\left[ k_{i}%
\right] }\right\} }{\partial v^{\left( \eta \right) }}\delta \left( \delta
v_{1}^{\left( \eta \right) }-\delta v_{2}^{\left( \eta \right) }\right)
\delta \left( \underline{\mathbf{\hat{\Lambda}}}^{\left[ \Delta _{2}\right]
}-\underline{\mathbf{\hat{\Lambda}}}^{\left[ \Delta _{1}\right] }\right)
\label{CM} \\
&\rightarrow &\left( h\left( v^{\left( \eta \right) }\right) \left\{
A_{\delta v_{2}^{\left( \eta \right) }}^{-}A_{\delta v_{1}^{\left( \eta
\right) }}^{+}-A_{\delta v_{1}^{\left( \eta \right) }}^{+}A_{\delta
v_{2}^{\left( \eta \right) }}^{-}\right\} \delta \left( \underline{\mathbf{%
\hat{\Lambda}}}^{\left[ \Delta _{2}\right] }-\underline{\mathbf{\hat{\Lambda}%
}}^{\left[ \Delta _{1}\right] }\right) \right.  \notag \\
&&\left. +\left\{ j^{-+}\left( \left\{ v^{\left( \eta \right) },\underline{%
\mathbf{\hat{\Lambda}}}^{\left[ k_{i}\right] }\right\} \right) A_{\underline{%
\mathbf{\hat{\Lambda}}}^{\left[ \Delta _{2}\right] }}^{-}A_{\underline{%
\mathbf{\hat{\Lambda}}}^{\left[ \Delta _{1}\right] }}^{+}-j^{+-}\left(
\left\{ v^{\left( \eta \right) },\underline{\mathbf{\hat{\Lambda}}}^{\left[
k_{i}\right] }\right\} \right) A_{\underline{\mathbf{\hat{\Lambda}}}^{\left[
\Delta _{1}\right] }}^{+}A_{\underline{\mathbf{\hat{\Lambda}}}^{\left[
\Delta _{2}\right] }}^{-}\right\} \right) H\left( \left\{ v^{\left( \eta
\right) }\right\} ,\left\{ \underline{\mathbf{\hat{\Lambda}}}^{\left[ k_{i}%
\right] }\right\} \right)  \notag \\
&&+h\left( v^{\left( \eta \right) }\right) \left\{ j^{+-}\left( \left\{
v^{\left( \eta \right) },\underline{\mathbf{\hat{\Lambda}}}^{\left[ k_{i}%
\right] }\right\} \right) A_{\delta v_{2}^{\left( \eta \right) }}^{-}A_{%
\underline{\mathbf{\hat{\Lambda}}}^{\left[ \Delta _{2}\right]
}}^{-}A_{\delta v_{1}^{\left( \eta \right) }}^{+}A_{\underline{\mathbf{\hat{%
\Lambda}}}^{\left[ \Delta _{1}\right] }}^{+}\right.  \notag \\
&&\left. -j^{-+}\left( \left\{ v^{\left( \eta \right) },\underline{\mathbf{%
\hat{\Lambda}}}^{\left[ k_{i}\right] }\right\} \right) A_{\delta
v_{1}^{\left( \eta \right) }}^{+}A_{\underline{\mathbf{\hat{\Lambda}}}^{%
\left[ \Delta _{1}\right] }}^{+}A_{\delta v_{2}^{\left( \eta \right)
}}^{-}A_{\underline{\mathbf{\hat{\Lambda}}}^{\left[ \Delta _{2}\right]
}}^{-}\right\} H\left( \left\{ v^{\left( \eta \right) }\right\} ,\left\{ 
\underline{\mathbf{\hat{\Lambda}}}^{\left[ k_{i}\right] }\right\} \right) 
\notag
\end{eqnarray}%
where the dependency of $h$, $j^{-+}$ and $j^{+-}$ in $v^{\left( \eta
\right) }$ and $v^{\left( \eta \right) },\left\{ \underline{\mathbf{\hat{%
\Lambda}}}^{\left[ k_{i}\right] }\right\} $ have been introduced.

These operators have been evaluated between $\left\langle \underline{\mathbf{%
\hat{\Lambda}}}^{\left[ \Delta _{1}\right] },\delta v_{1}^{\left( \eta
\right) }\right\vert \left\langle \underline{\mathbf{\hat{\Lambda}}}^{\left[
\Delta _{2}\right] },\delta v_{2}^{\left( \eta \right) }\right\vert $ and $%
\left\vert \underline{\mathbf{\hat{\Lambda}}}^{\left[ \Delta _{1}\right]
},\delta v_{1}^{\left( \eta \right) }\right\rangle \left\vert \underline{%
\mathbf{\hat{\Lambda}}}^{\left[ \Delta _{2}\right] },\delta v_{2}^{\left(
\eta \right) }\right\rangle $. Since the modification are discrete, nothing
guarantees the equality:%
\begin{equation*}
j^{-+}\left( \left\{ v^{\left( \eta \right) },\underline{\mathbf{\hat{\Lambda%
}}}^{\left[ k_{i}\right] }\right\} \right) =j^{+-}\left( \left\{ v^{\left(
\eta \right) },\underline{\mathbf{\hat{\Lambda}}}^{\left[ k_{i}\right]
}\right\} \right)
\end{equation*}%
Assume to simplify that we can normalize $\left\vert j^{-+}\right\vert
=\left\vert j^{+-}\right\vert =h\left( v^{\left( \eta \right) }\right) =1$
this implies that:%
\begin{eqnarray}
A_{\delta v_{1}^{\left( \eta \right) }}^{+}A_{\delta v_{2}^{\left( \eta
\right) }}^{-}\pm A_{\delta v_{2}^{\left( \eta \right) }}^{-}A_{\delta
v_{1}^{\left( \eta \right) }}^{+} &=&\alpha _{1}\delta \left( \delta
v_{2}^{\left( \eta \right) }-\delta v_{1}^{\left( \eta \right) }\right)
\label{cr} \\
A_{\underline{\mathbf{\hat{\Lambda}}}^{\left[ \Delta _{1}\right] }}^{+}A_{%
\underline{\mathbf{\hat{\Lambda}}}^{\left[ \Delta _{2}\right] }}^{-}\pm A_{%
\underline{\mathbf{\hat{\Lambda}}}^{\left[ \Delta _{2}\right] }}^{-}A_{%
\underline{\mathbf{\hat{\Lambda}}}^{\left[ \Delta _{1}\right] }}^{+}
&=&\alpha _{2}\delta \left( \underline{\mathbf{\hat{\Lambda}}}^{\left[
\Delta _{2}\right] }-\underline{\mathbf{\hat{\Lambda}}}^{\left[ \Delta _{1}%
\right] }\right) \delta \left( \delta v_{2}^{\left( \eta \right) }-\delta
v_{1}^{\left( \eta \right) }\right)  \notag \\
A_{\delta v_{1}^{\left( \eta \right) },\underline{\mathbf{\hat{\Lambda}}}^{%
\left[ \Delta _{1}\right] }}^{+}A_{\delta v_{2}^{\left( \eta \right) },%
\underline{\mathbf{\hat{\Lambda}}}^{\left[ \Delta _{2}\right] }}^{-}\pm
A_{\delta v_{2}^{\left( \eta \right) },\underline{\mathbf{\hat{\Lambda}}}^{%
\left[ \Delta _{2}\right] }}^{-}A_{\delta v_{1}^{\left( \eta \right) },%
\underline{\mathbf{\hat{\Lambda}}}^{\left[ \Delta _{1}\right] }}^{+}
&=&\beta \delta \left( \underline{\mathbf{\hat{\Lambda}}}^{\left[ \Delta _{2}%
\right] }-\underline{\mathbf{\hat{\Lambda}}}^{\left[ \Delta _{1}\right]
}\right) \delta \left( \delta v_{2}^{\left( \eta \right) }-\delta
v_{1}^{\left( \eta \right) }\right)  \notag
\end{eqnarray}%
where the sum of coefficients $\alpha _{i}$ and $\beta $ yields the left
hand side of (\ref{CM}). These relations are leading to some commutation or
anticommutation relations between operators involved in the modifications of
the constraints.

\subsubsection{Product of modifications and product of fields}

We start by assuming the following modifications $\left\{ \underline{\mathbf{%
\hat{\Lambda}}}^{\left[ \Delta _{i}\right] }\right\} _{i}$:%
\begin{equation*}
v^{\left( \eta \right) }\otimes \delta v_{1}^{\left( \eta \right) }\otimes
\delta v_{2}^{\left( \eta \right) }\rightarrow v^{\left( \eta \right)
}\otimes \delta v_{1}^{\left( \eta \right) }\otimes \delta v_{2}^{\left(
\eta \right) }\otimes \underline{\mathbf{\hat{\Lambda}}}^{\left[ \Delta _{1}%
\right] }\otimes \underline{\mathbf{\hat{\Lambda}}}^{\left[ \Delta _{2}%
\right] }
\end{equation*}%
If the symmetries generators are differents for $\delta v_{1}^{\left( \eta
\right) }$ and $\delta v_{2}^{\left( \eta \right) }$, the states can be
associated to:%
\begin{equation*}
\delta v_{1}^{\left( \eta \right) }\otimes \underline{\mathbf{\hat{\Lambda}}}%
^{\left[ \Delta _{1}\right] }
\end{equation*}%
and:%
\begin{equation*}
\delta v_{2}^{\left( \eta \right) }\otimes \underline{\mathbf{\hat{\Lambda}}}%
^{\left[ \Delta _{2}\right] }
\end{equation*}

\section{Projected action functional and constraints}

We have described the effect of small variations in the states $v^{\left(
\eta \right) }$ in terms of infinitesmal states creations or destructions.
Now, we consider an effective action point of view, for the system as whole
given a state $\nu $, and then for elementary variations $\delta \nu $.

Given such state $\nu $, we first inspect the expansion in the projected
field of the action functional and focus on the specific part of the action
describing the cloud of points. Then we will consider the additional
effective action corrsponding to the variation $\delta \nu $.

\subsection{Action for the whole system and the cloud of points:}

\subsubsection{2 objects decomposition, no subobject}

Coming back to the full action of a state, we consider the expansion of:%
\begin{equation*}
S\left( \left\{ \Psi _{J,\alpha }^{\otimes l}\left( U_{j}^{l}\right) 
\underset{l,k}{\otimes }\Psi _{I,\alpha }^{\otimes k}\left( U_{i}^{k}\right)
\right\} _{\alpha },\left\{ v_{\left\{ U_{i}^{k}\right\} }\left( \Psi
_{J,\alpha }^{\otimes k}\right) \otimes \Psi _{I,\alpha }^{\otimes
k}\right\} _{\left\{ U_{i}^{k}\right\} }\right) 
\end{equation*}%
where the saddle point was derived previously:%
\begin{eqnarray}
&&\Psi _{I,\alpha ,0}^{\otimes k}\left( U_{i}^{k},\mathbf{\hat{\Lambda}}%
_{\infty }^{\left[ k\right] }\left[ \Psi _{J},\nu \right] ,\left\{ \Psi
_{J}^{\otimes l}\right\} _{l}\right)  \\
&=&\sum_{s}\sum_{s,l_{1},...,l_{s},\left\{ \alpha _{i}^{\prime }\right\}
_{_{i\leqslant s}}}d\left( \left\{ U_{j}^{l_{i}}\right\} _{l_{i}}\right)
d\left( \left\{ U_{i}^{k_{i}}/f_{k_{i}l_{i}}\right\} _{k_{i},l_{i}}\right)  
\notag \\
&&\times \tprod_{i}\Psi _{J,\alpha _{i}^{\prime }}^{\otimes l_{i}}\left(
U_{j}^{l_{i}}\right) \mathcal{K}_{0}^{v}\left( U_{i}^{k},\left\{
U_{j}^{l_{i}}\right\} _{l_{i}},\left\{ U_{i}^{k_{i}}/f_{k_{i}l_{i}}\right\}
_{i},\mathbf{\hat{\Lambda}}_{\infty }^{\left[ k\right] }\left[ \Psi _{J},\nu %
\right] \right)   \notag
\end{eqnarray}%
with:%
\begin{eqnarray*}
&&\mathcal{K}_{0}^{v}\left( U_{i}^{k},\left\{ U_{j}^{l_{i}}\right\} _{l_{i}},%
\mathbf{\hat{\Lambda}}_{\infty }^{\left[ k\right] }\left[ \Psi _{J},\nu %
\right] \right)  \\
&=&\mathcal{K}_{0}\left( U_{i}^{k},\left\{ U_{j}^{l_{i}}\right\} _{l_{i}},%
\mathbf{\hat{\Lambda}}_{\infty }^{\left[ k\right] }\left[ \Psi _{J},\nu %
\right] \right) \tprod_{i}v\left(
U_{i}^{k_{i}},U_{j}^{l_{i}}/f_{k_{i}l_{i}}\right) 
\end{eqnarray*}%
The kernel includes possible derivatives and the field involved in the
action writes:%
\begin{eqnarray*}
&&\Psi _{J,\alpha }^{\otimes l}\left( U_{j}^{l}\right) \underset{l,k}{%
\otimes }\Psi _{I,\alpha }^{\otimes k}\left( U_{i}^{k}\right)  \\
&\rightarrow &\Psi _{J,\alpha }^{\otimes l}\left( U_{j}^{l}\right) \underset{%
l,k}{\otimes }\sum_{s}\sum_{s,l_{1},...,l_{s},\left\{ \alpha _{i}^{\prime
}\right\} _{_{i\leqslant s}}}d\left( \left\{ U_{j}^{l_{i}}\right\}
_{l_{i}}\right) d\left( \left\{ U_{i}^{k_{i}}/f_{k_{i}l_{i}}\right\}
_{k_{i},l_{i}}\right)  \\
&&\times \tprod_{i}\Psi _{J,\alpha _{i}^{\prime }}^{\otimes l_{i}}\left(
U_{j}^{l_{i}}\right) \mathcal{K}_{0}^{v}\left( U_{i}^{k},\left\{
U_{j}^{l_{i}}\right\} _{l_{i}},\left\{ U_{i}^{k_{i}}/f_{k_{i}l_{i}}\right\}
_{i},\mathbf{\hat{\Lambda}}_{\infty }^{\left[ k\right] }\left[ \Psi _{J},\nu %
\right] \right) 
\end{eqnarray*}%
As a consequence, including in the action combinations of this fields with
coefficients:%
\begin{equation*}
\Psi \left( \mathbf{\hat{\Lambda}}_{\infty }^{\left[ k\right] }\left[ \Psi
_{J},\nu \right] ,v\right) 
\end{equation*}%
the action is a sum of terms (see (\ref{PFT}) and (\ref{Sr})):%
\begin{equation}
\int \sum_{k}\bar{s}_{k}\left( \left\{ U_{j}^{l_{i}}\right\} _{i},\mathbf{%
\hat{\Lambda}}_{\infty }^{\left[ k\right] }\left[ \Psi _{J},\nu ,\left\{
U_{j}^{l_{i}}\right\} \right] \right) \Psi _{J}^{\otimes \left(
\sum_{i}l_{i}\right) }\left( \left\{ U_{j}^{l_{i}}\right\} _{i},\mathbf{\hat{%
\Lambda}}_{\infty }^{\left[ k\right] }\left[ \Psi _{J},\nu ,\left\{
U_{j}^{l_{i}}\right\} \right] ,v\right) 
\end{equation}%
with the field given by a sum over realizations:%
\begin{eqnarray}
&&\Psi _{J}^{\otimes \left( \sum_{i}l_{i}\right) }\left( \left\{
U_{j}^{l_{i}}\right\} _{i},\mathbf{\hat{\Lambda}}_{\infty }^{\left[ k\right]
}\left[ \Psi _{J},\nu ,\left\{ U_{j}^{l_{i}}\right\} \right] ,v\right)  \\
&=&\sum_{\alpha }\Psi _{J,\alpha }^{\otimes \left( \sum_{i}l_{i}\right)
}\left( \left\{ U_{j}^{l_{i}}\right\} _{i},\mathbf{\hat{\Lambda}}_{\infty }^{%
\left[ k\right] }\left[ \Psi _{J},\nu ,\left\{ U_{j}^{l_{i}}\right\} \right]
,v\right)   \notag
\end{eqnarray}%
and the action should include the cloud variables, since the lowest order
term has the form:%
\begin{equation*}
S_{0}=\int \sum_{k}\bar{s}_{k}\left( \mathbf{\hat{\Lambda}}_{\infty }^{\left[
k\right] }\left[ \Psi _{J},\nu ,\left\{ U_{j}^{l_{i}}\right\} \right]
\right) \Psi _{J}\left( \mathbf{\hat{\Lambda}}_{\infty }^{\left[ k\right] }%
\left[ \Psi _{J},\nu ,\left\{ U_{j}^{l_{i}}\right\} \right] ,v\right) 
\end{equation*}%
when the coordinates $\mathbf{\hat{\Lambda}}_{\infty }^{\left[ k\right] }$
can be decomposed:%
\begin{equation*}
\mathbf{\hat{\Lambda}}_{\infty }^{\left[ k\right] }\left[ \Psi _{J},\nu
,\left\{ U_{j}^{l_{i}}\right\} \right] =\left( \mathbf{\hat{\Lambda}}%
_{\infty }\left[ \Psi _{J},\nu ,\left\{ U_{j}^{l_{i}}\right\} \right]
\right) ^{k}
\end{equation*}%
The action can be written using powers of $\mathbf{\hat{\Lambda}}_{\infty }%
\left[ \Psi _{J},\nu ,\left\{ U_{j}^{l_{i}}\right\} \right] $:%
\begin{equation}
\int \sum_{k}\bar{s}_{k}\left( \left\{ U_{j}^{l_{i}}\right\} _{i},\left\{
\left( \mathbf{\hat{\Lambda}}_{\infty }\left[ \Psi _{J},\nu ,\left\{
U_{j}^{l_{i}}\right\} \right] \right) _{i}\right\} _{i\leqslant k}\right)
\Psi _{J}^{\otimes \left( \sum_{i}l_{i}\right) }\left( \left\{
U_{j}^{l_{i}}\right\} _{i},\left\{ \left( \mathbf{\hat{\Lambda}}_{\infty }%
\left[ \Psi _{J},\nu ,\left\{ U_{j}^{l_{i}}\right\} \right] \right)
_{i}\right\} _{i\leqslant k},v\right) 
\end{equation}%
where the $\left( \mathbf{\hat{\Lambda}}_{\infty }\left[ \Psi _{J},\nu
,\left\{ U_{j}^{l_{i}}\right\} \right] \right) _{i}$ are the components of $%
\mathbf{\hat{\Lambda}}_{\infty }\left[ \Psi _{J},\nu ,\left\{
U_{j}^{l_{i}}\right\} \right] $.

These terms are not local and we could consider infinite expansion and some $%
\mathbf{\hat{\Lambda}}_{\infty }\left[ \Psi _{J},\nu ,\left\{
U_{j}^{l_{i}}\right\} \right] $:%
\begin{eqnarray}
&&\sum_{k}\int \hat{s}_{k}\left( \left\{ U_{j}^{l_{i}}\right\} _{i},\mathbf{%
\hat{\Lambda}}_{\infty }\left[ \Psi _{J},\nu ,\left\{ U_{j}^{l_{i}}\right\} %
\right] \right) \Psi _{J}^{\otimes \left( \sum_{i}l_{i}\right) }\left(
\left\{ U_{j}^{l_{i}}\right\} _{i},\mathbf{\hat{\Lambda}}_{\infty }\left[
\Psi _{J},\nu ,\left\{ U_{j}^{l_{i}}\right\} \right] ,v\right)   \label{Sn}
\\
&&+\sum_{k}\sum_{p}\int \hat{s}_{k}^{\left( p\right) }\left( \left\{
U_{j}^{l_{i}}\right\} _{i},\mathbf{\hat{\Lambda}}_{\infty }\left[ \Psi
_{J},\nu ,\left\{ U_{j}^{l_{i}}\right\} \right] \right) \nabla _{\mathbf{%
\hat{\Lambda}}_{\infty }}^{p}\Psi _{J}^{\otimes \left( \sum_{i}l_{i}\right)
}\left( \left\{ U_{j}^{l_{i}}\right\} _{i},\mathbf{\hat{\Lambda}}_{\infty }%
\left[ \Psi _{J},\nu ,\left\{ U_{j}^{l_{i}}\right\} \right] ,v\right)  
\notag
\end{eqnarray}%
and the action for cloud of parameters:%
\begin{eqnarray}
&&\sum_{k}\int \hat{s}_{k}\left( \mathbf{\hat{\Lambda}}_{\infty }\left[ \Psi
_{J},\nu ,\left\{ U_{j}^{l_{i}}\right\} \right] \right) \Psi _{J}\left( 
\mathbf{\hat{\Lambda}}_{\infty }\left[ \Psi _{J},\nu ,\left\{
U_{j}^{l_{i}}\right\} \right] ,v\right)  \\
&&+\sum_{k}\sum_{p}\int \hat{s}_{k}^{\left( p\right) }\left( \mathbf{\hat{%
\Lambda}}_{\infty }\left[ \Psi _{J},\nu ,\left\{ U_{j}^{l_{i}}\right\} %
\right] \right) \nabla _{\mathbf{\hat{\Lambda}}_{\infty }}^{p}\Psi
_{J}\left( \mathbf{\hat{\Lambda}}_{\infty }\left[ \Psi _{J},\nu ,\left\{
U_{j}^{l_{i}}\right\} \right] ,v\right)   \notag
\end{eqnarray}%
The constraint implies that the first contribution should include
derivatives so that in first approximtion, equation (\ref{MLN}):%
\begin{equation}
0=\sum_{U_{j}^{l}}\hat{\Gamma}\left( \left( U_{j}^{l_{i}}\right) ^{\prime
},U_{j}^{l}\right) \nabla _{\mathbf{\hat{\Lambda}}_{\infty }^{\left[ k\right]
}}\hat{\Psi}_{J}^{\otimes \sum_{i}l_{i}}\left( \left\{ U_{j}^{l_{i}}\right\}
_{i},\mathbf{\hat{\Lambda}}_{\infty }^{\left[ k\right] },v\right) 
\end{equation}%
is satisfied. This implies that in first approximation, the series (\ref{Sn}%
) includes terms of the form:%
\begin{equation*}
\int \hat{\Psi}_{J}^{\otimes \sum_{i}l_{i}}\left( \left\{ \left(
U_{j}^{l_{i^{\prime }}}\right) ^{\prime }\right\} _{i},\mathbf{\hat{\Lambda}}%
_{\infty }^{\left[ k\right] },v\right) \sum_{U_{j}^{l}}\hat{\Gamma}\left(
\left( U_{j}^{l_{i^{\prime }}}\right) ^{\prime },U_{j}^{l_{i}}\right) \nabla
_{\mathbf{\hat{\Lambda}}_{\infty }^{\left[ k\right] }}\hat{\Psi}%
_{J}^{\otimes \sum_{i}l_{i}}\left( \left\{ U_{j}^{l_{i}}\right\} _{i},%
\mathbf{\hat{\Lambda}}_{\infty }^{\left[ k\right] },v\right) d\mathbf{\hat{%
\Lambda}}_{\infty }^{\left[ k\right] }
\end{equation*}%
or terms including massive terms if we consider average invariance.

In the local approximation:%
\begin{equation*}
\int \hat{\Psi}_{J}^{\otimes \sum_{i}l_{i}}\left( \left\{ \left(
U_{j}^{l_{i^{\prime }}}\right) ^{\prime }\right\} _{i},\mathbf{\hat{\Lambda}}%
_{\infty },v\right) \sum_{U_{j}^{l}}\hat{\Gamma}\left( \left(
U_{j}^{l_{i^{\prime }}}\right) ^{\prime },U_{j}^{l_{i}}\right) \nabla _{%
\mathbf{\hat{\Lambda}}_{\infty }}\hat{\Psi}_{J}^{\otimes
\sum_{i}l_{i}}\left( \left\{ U_{j}^{l_{i}}\right\} _{i},\mathbf{\hat{\Lambda}%
}_{\infty },v\right) d\mathbf{\hat{\Lambda}}_{\infty }^{\left[ k\right] }
\end{equation*}

\subsubsection{Remark 1}

This action is similar to some spinor field action. This description also
includes vectors, or $U\left( 1\right) $ or $SU\left( 2\right) $ boson
fields in their spinorial form. Note the differences with the usual spinor
description. Here, no conjugate field $\left( \hat{\Psi}_{J}^{\otimes
\sum_{i}l_{i}}\right) ^{\dag }$ appears since the conjugate is itself
included in the definition of the field. For instance, if we were to write
the sole Dirac action in our notation, we would have to rewrite functionals:%
\begin{equation}
S_{1}=\int \left( \hat{\Psi}_{J}^{\otimes \sum_{i}l_{i}}\left( \left\{
\left( s_{j}^{l_{i^{\prime }}}\right) ^{\prime }\right\} ,\mathbf{\hat{%
\Lambda}}_{\infty }^{\left[ k\right] }\right) \right) ^{\dag }\hat{\Gamma}%
\left( \left\{ \left( s_{j}^{l_{i^{\prime }}}\right) ^{\prime }\right\}
,\left\{ s_{j}^{l_{i}}\right\} _{i}\right) .\nabla \hat{\Psi}_{J}^{\otimes
\sum_{i}l_{i}}\left( \left\{ s_{j}^{l_{i}}\right\} _{i},\mathbf{\hat{\Lambda}%
}_{\infty }^{\left[ k\right] }\right)   \label{SP}
\end{equation}%
by first introducting field encompassing both the spinors and thir
conjugates:%
\begin{eqnarray*}
&&\left( \hat{\Psi}_{J}^{\otimes \sum_{i}l_{i}}\left( \left\{ \left(
s_{j}^{l_{i^{\prime }}}\right) ^{\prime }\right\} ,\mathbf{\hat{\Lambda}}%
_{\infty }^{\left[ k\right] }\right) \right) ^{\dag },\hat{\Psi}%
_{J}^{\otimes \sum_{i}l_{i}}\left( \left\{ s_{j}^{l_{i}}\right\} _{i},%
\mathbf{\hat{\Lambda}}_{\infty }^{\left[ k\right] }\right)  \\
&\rightarrow &\hat{\Psi}_{J}^{\otimes \sum_{i}l_{i}}\left( \left\{ \left\{
\left( s_{j}^{l_{i}}\right) _{b}\right\} \right\} _{b},\mathbf{\hat{\Lambda}}%
_{\infty }^{\left[ k\right] }\right) 
\end{eqnarray*}%
with $b=1,2$ and:%
\begin{equation*}
S_{1}\rightarrow \frac{1}{2}\int \left( \hat{\Psi}_{J}^{\otimes
\sum_{i}l_{i}}\left( \left\{ \left\{ \left( s_{j}^{l_{i^{\prime }}}\right)
_{b^{\prime }}^{\prime }\right\} \right\} _{b^{\prime }},\mathbf{\hat{\Lambda%
}}_{\infty }^{\left[ k\right] }\right) \right) \left[ \hat{\Gamma}%
_{b^{\prime }b}\right] \left( \left\{ \left\{ \left( s_{j}^{l_{i^{\prime
}}}\right) _{b^{\prime }}^{\prime }\right\} \right\} _{b^{\prime }},\left\{
\left\{ \left( s_{j}^{l_{i}}\right) _{b}\right\} \right\} _{b}\right)
.\nabla \hat{\Psi}_{J}^{\otimes \sum_{i}l_{i}}\left( \left\{ \left\{ \left(
s_{j}^{l_{i}}\right) _{b}\right\} \right\} _{b},\mathbf{\hat{\Lambda}}%
_{\infty }^{\left[ k\right] }\right) 
\end{equation*}%
To recover (\ref{SP}), we set:%
\begin{eqnarray*}
\left[ \hat{\Gamma}_{11}\right]  &=&\left[ \hat{\Gamma}_{22}\right] =0 \\
\left[ \hat{\Gamma}_{12}\left( \left\{ \left\{ \left( s_{j}^{l_{i^{\prime
}}}\right) _{b^{\prime }}^{\prime }\right\} \right\} _{b^{\prime }},\left\{
\left\{ \left( s_{j}^{l_{i}}\right) _{b}\right\} \right\} _{b}\right) \right]
&=&\hat{\Gamma}\left( \left\{ \left( s_{j}^{l_{i^{\prime }}}\right) ^{\prime
}\right\} ,\left\{ s_{j}^{l_{i}}\right\} _{i}\right)  \\
\left[ \hat{\Gamma}_{21}\left( \left\{ \left\{ \left( s_{j}^{l_{i^{\prime
}}}\right) _{b^{\prime }}^{\prime }\right\} \right\} _{b^{\prime }},\left\{
\left\{ \left( s_{j}^{l_{i}}\right) _{b}\right\} \right\} _{b}\right) \right]
&=&-\hat{\Gamma}\left( \left\{ s_{j}^{l_{i}}\right\} _{i},\left\{ \left(
s_{j}^{l_{i^{\prime }}}\right) ^{\prime }\right\} \right) 
\end{eqnarray*}%
wher we assume that intgration by part is possible without boundary trms.
Otherwise some mass terms can appear, as explained in part one.

\subsubsection{Remark 2}

To recover the standard formalism of action functional involving products of
fields, we can expand:%
\begin{eqnarray*}
&&\Psi _{J}^{\otimes \left( \sum_{i}l_{i}\right) }\left( \left\{
U_{j}^{l_{i}}\right\} _{i},\mathbf{\hat{\Lambda}}_{\infty }\left[ \Psi
_{J},\nu ,\left\{ U_{j}^{l_{i}}\right\} \right] ,v\right) \\
&\rightarrow &\sum_{\left\{ a_{i}\right\} _{i}}\tprod\limits_{i}\Psi
_{J,a_{i}}^{l_{i}}\left( U_{j}^{l_{i}},\left( \mathbf{\hat{\Lambda}}_{\infty
}\left[ \Psi _{J},\nu ,\left\{ U_{j}^{l_{i}}\right\} \right] \right)
_{i},v\right) \\
&\rightarrow &\sum_{\left\{ a_{i}\right\} _{i}}\tprod\limits_{i}\Psi
_{J}^{l_{i}}\left( U_{j}^{l_{i}},a_{i},\left( \mathbf{\hat{\Lambda}}_{\infty
}\left[ \Psi _{J},\nu ,\left\{ U_{j}^{l_{i}}\right\} \right] \right)
_{i},v\right)
\end{eqnarray*}%
\begin{eqnarray}
&\rightarrow &\sum_{k}\sum_{\left\{ a_{i}\right\} _{i}}\int \hat{s}%
_{k}\left( \left\{ U_{j}^{l_{i}}\right\} _{i},\left\{ \left( \mathbf{\hat{%
\Lambda}}_{\infty }\left[ \Psi _{J},\nu ,\left\{ U_{j}^{l_{i}}\right\} %
\right] \right) _{i}\right\} _{i\leqslant k}\right) \tprod\limits_{i}\Psi
_{J}^{l_{i}}\left( \left\{ U_{j}^{l_{i}}\right\} _{i},a_{i},\left( \mathbf{%
\hat{\Lambda}}_{\infty }\left[ \Psi _{J},\nu ,\left\{ U_{j}^{l_{i}}\right\} %
\right] \right) _{i},v\right) \\
&&+\sum_{k}\sum_{\left\{ p_{i}\right\} _{i=2,k}}\sum_{\left\{ a_{i}\right\}
_{i}}\int \hat{s}_{k}^{\left( \left\{ p_{i}\right\} _{i=2,k}\right) }\left(
\left\{ U_{j}^{l_{i}}\right\} _{i},\mathbf{\hat{\Lambda}}_{\infty }\left[
\Psi _{J},\nu ,\left\{ U_{j}^{l_{i}}\right\} \right] ,\left\{ a_{i}\right\}
_{i}\right) \tprod\limits_{i}\nabla _{\mathbf{\hat{\Lambda}}_{\infty
}}^{p_{i}}\Psi _{J}^{l_{i}}\left( \left\{ U_{j}^{l_{i}}\right\}
_{i},a_{i},\left( \mathbf{\hat{\Lambda}}_{\infty }\left[ \Psi _{J},\nu
,\left\{ U_{j}^{l_{i}}\right\} \right] \right) _{i},v\right)  \notag
\end{eqnarray}

Taking into account that the sum over the $\left\{ U_{j}^{l_{i}}\right\}
_{i} $ is implicit, we can set:%
\begin{equation*}
\left( U_{j}^{l_{i}},a_{i}\right) =\bar{U}_{j}^{l_{i}}
\end{equation*}%
and write the sum over $\bar{U}_{j}^{l_{i}}$ implicitly: 
\begin{eqnarray}
&\rightarrow &\sum_{k}\int \hat{s}_{k}\left( \left\{ \bar{U}%
_{j}^{l_{i}}\right\} _{i},\left\{ \left( \mathbf{\hat{\Lambda}}_{\infty }%
\left[ \Psi _{J},\nu ,\left\{ U_{j}^{l_{i}}\right\} \right] \right)
_{i}\right\} _{i\leqslant k}\right) \tprod\limits_{i}\Psi _{J}^{l_{i}}\left( 
\bar{U}_{j}^{l_{i}},\left( \mathbf{\hat{\Lambda}}_{\infty }\left[ \Psi
_{J},\nu ,\left\{ U_{j}^{l_{i}}\right\} \right] \right) _{i},v\right)
\label{Sp} \\
&&+\sum_{k}\sum_{\left( \left\{ p_{i}\right\} _{i=2,k}\right) }\int \hat{s}%
_{k}^{\left( \left\{ p_{i}\right\} _{i=2,k}\right) }\left( \left\{ \bar{U}%
_{j}^{l_{i}}\right\} _{i},\mathbf{\hat{\Lambda}}_{\infty }\left[ \Psi
_{J},\nu ,\left\{ U_{j}^{l_{i}}\right\} \right] \right)
\tprod\limits_{i}\nabla _{\mathbf{\hat{\Lambda}}_{\infty }}^{p_{i}}\Psi
_{J}^{l_{i}}\left( \bar{U}_{j}^{l_{i}},\left( \mathbf{\hat{\Lambda}}_{\infty
}\left[ \Psi _{J},\nu ,\left\{ U_{j}^{l_{i}}\right\} \right] \right)
_{i},v\right)  \notag
\end{eqnarray}

The cloud action writes:%
\begin{eqnarray*}
&&\sum_{k}\int \hat{s}_{0,k}\left( \left\{ \left( \mathbf{\hat{\Lambda}}%
_{\infty }\left[ \Psi _{J},\nu ,\left\{ U_{j}^{l_{i}}\right\} \right]
\right) _{i}\right\} _{i\leqslant k}\right) \tprod\limits_{i}\Psi _{J}\left(
\left( \mathbf{\hat{\Lambda}}_{\infty }\left[ \Psi _{J},\nu ,\left\{
U_{j}^{l_{i}}\right\} \right] \right) _{i},v\right) \\
&&+\sum_{k}\sum_{\left( \left\{ p_{i}\right\} _{i=2,k}\right) }\int \hat{s}%
_{k}^{\left( \left\{ p_{i}\right\} _{i=2,k}\right) }\left( \mathbf{\hat{%
\Lambda}}_{\infty }\left[ \Psi _{J},\nu ,\left\{ U_{j}^{l_{i}}\right\} %
\right] \right) \tprod\limits_{i}\nabla _{\mathbf{\hat{\Lambda}}_{\infty
}}^{p_{i}}\Psi _{J}\left( \left( \mathbf{\hat{\Lambda}}_{\infty }\left[ \Psi
_{J},\nu ,\left\{ U_{j}^{l_{i}}\right\} \right] \right) _{i},v\right)
\end{eqnarray*}%
\bigskip

\subsection{General case: including subobjects}

\subsubsection{Projection for several field decomposition}

More generally, considering the states: 
\begin{equation*}
v_{\left[ p_{l_{1}},p_{l_{1}^{\prime }l_{1}}\right] ^{k_{l_{1}^{\prime
}}}}\left\{ \Psi _{J}^{\otimes k_{l^{\prime }}}\right\} =\Psi _{J}^{\otimes
k_{l^{\prime }}}
\end{equation*}%
the constraint for fields is given by (\ref{Cdn}), (\ref{Cdt}) or (\ref{Cdv}%
) if we consider average invariance. The action series expansion should
encompass this constraint at least at first order.

Using (\ref{SCN}) and (\ref{SDLPT}), the minimum of the action:

\begin{eqnarray*}
&&\left( -S\left( \left\{ \otimes \left[ \Psi _{J,\alpha ,s_{p_{\eta
^{\prime }\eta }}^{\otimes k_{\eta ^{\prime }\eta }}}^{\otimes k_{\eta
^{\prime }\eta }}\left( \left[ p_{\eta },p_{\eta ^{\prime }\eta }\right]
^{k_{\eta ^{\prime }\eta }}\right) \right] \underset{l^{\prime }}{\otimes }%
\Psi _{J,s_{p_{l^{\prime },l}}^{\otimes k_{l^{\prime }}}}^{\otimes
k_{l^{\prime }}}\left( \left[ p_{l},p_{l^{\prime }l}\right] ^{k_{l^{\prime
}}}\right) \right\} ,\right. \right. \\
&&\left. \left[ v_{\left[ p_{l},p_{l^{\prime }l}\right] ^{k_{l^{\prime
}}}}\left( \Psi _{J,\alpha ,s_{p_{\eta ^{\prime }\eta }}^{\otimes k_{\eta
^{\prime }\eta }}}^{\otimes k_{\eta ^{\prime }\eta }}\right) \right] \left(
\Psi _{J,s_{p_{l^{\prime },l}}^{\otimes k_{l^{\prime }}}}^{\otimes
k_{l^{\prime }}}\left( \left[ p_{l},p_{l^{\prime }l}\right] ^{k_{l^{\prime
}}}\right) \right) \right)
\end{eqnarray*}%
is given by realizations:

\begin{equation*}
\underset{l^{\prime },\eta ^{\prime }\eta }{\otimes }\Psi _{J,\alpha
,s_{p_{\eta ^{\prime }\eta }}^{\otimes k_{\eta ^{\prime }\eta }}}^{\otimes
k_{\eta ^{\prime }\eta }}\left( \left[ p_{\eta },p_{\eta ^{\prime }\eta }%
\right] ^{k_{\eta ^{\prime }\eta }}\right) \underset{l^{\prime },l}{\otimes }%
\Psi _{J,s_{p_{l^{\prime },l}}^{\otimes k_{l^{\prime }}}}^{\otimes
k_{l^{\prime }}}\left( \left[ p_{l},p_{l^{\prime }l}\right] ^{k_{l^{\prime
}}}\right) 
\end{equation*}%
and the complete description is given by the set of these realizations:%
\begin{equation*}
\rightarrow \left\{ \underset{l^{\prime },\eta ^{\prime }\eta }{\otimes }%
\Psi _{J,\alpha ,s_{p_{\eta ^{\prime }\eta }}^{\otimes k_{\eta ^{\prime
}\eta }}}^{\otimes k_{\eta ^{\prime }\eta }}\left( \left[ p_{\eta },p_{\eta
^{\prime }\eta }\right] ^{k_{\eta ^{\prime }\eta }}\right) \underset{%
l^{\prime },l}{\otimes }\Psi _{J,s_{p_{l^{\prime },l}}^{\otimes k_{l^{\prime
}}}}^{\otimes k_{l^{\prime }}}\left( \left[ p_{l},p_{l^{\prime }l}\right]
^{k_{l^{\prime }}}\right) \right\} 
\end{equation*}%
For such saddle point solutions, including as before combinations with
coefficients:%
\begin{equation*}
\Psi \left( \left\{ \left[ 
\begin{array}{c}
\left( p_{l},p_{l,^{\prime }l},k_{l^{\prime }}\right)  \\ 
\left[ \left\{ \Psi _{J}\left[ l^{\prime },0\right] \right\} ,\left[
p_{0},p_{l^{\prime }0}\right] ^{k_{l^{\prime }}},v\right] 
\end{array}%
\right] \right\} _{i},v\right) 
\end{equation*}%
the action writes (the several symbols of integrations are implicit) as
series expansion summing terms of the form (\ref{FL}) or its general form (%
\ref{Fl}), but with at least quadratic terms in field:

\begin{eqnarray}
&&\sum_{_{\left( s,s^{\prime },\left[ p,p^{\prime },k^{\prime }\right]
\right) }}\sum_{l}\int \bar{S}^{\mathcal{K}}\left( \left[ \left\{ \left[
p_{\eta _{i}},p_{\eta _{i^{\prime }}^{\prime }\eta _{i}}\right] ^{k_{\eta
_{i^{\prime }}^{\prime }\eta _{i}}}\right\} _{\substack{ i\leqslant s  \\ %
i^{\prime }\leqslant s^{\prime }}}\right] ,\left\{ \left[ 
\begin{array}{c}
\left( p_{l_{i}},p_{l_{i},^{\prime }l_{i}},k_{l_{i}^{\prime }}\right) \\ 
\left[ \left\{ \Psi _{J}\left[ l^{\prime },0\right] \right\} ,\left[
p_{0},p_{l^{\prime }0}\right] ^{k_{l^{\prime }}},v\right]%
\end{array}%
\right] \right\} _{i}\right)  \label{Rs} \\
&&\times \tprod\limits_{\left( l_{i},^{\prime }l_{i}\right) }\Psi
_{J,\tprod_{i}s_{p_{\eta _{i}},p_{\eta _{i^{\prime }}^{\prime }\eta
_{i}}}^{\otimes k_{\eta _{i^{\prime }}^{\prime }\eta _{i}}}}^{\otimes
\sum_{\left( \eta _{i^{\prime }}^{\prime }\eta _{i}\right) }k_{\eta
_{i^{\prime }}^{\prime }\eta _{i}}}\left( \left[ \left\{ \left[ p_{\eta
_{i}},p_{\eta _{i^{\prime }}^{\prime }\eta _{i}}\right] ^{k_{\eta
_{i^{\prime }}^{\prime }\eta _{i}}}\right\} _{\substack{ i\leqslant s  \\ %
i^{\prime }\leqslant s^{\prime }}}\right] ^{\mathcal{P}_{i}},\left\{ \left[ 
\begin{array}{c}
\left( p_{l_{i}},p_{l_{i},^{\prime }l_{i}},k_{l_{i}^{\prime }}\right) \\ 
\left[ \left\{ \Psi _{J}\left[ l^{\prime },0\right] \right\} ,\left[
p_{0},p_{l^{\prime }0}\right] ^{k_{l^{\prime }}},v\right]%
\end{array}%
\right] \right\} _{i\in \mathcal{P}_{l}},v\right)  \notag \\
&&\times \tprod\limits_{i}\left( d\left\{ \left[ 
\begin{array}{c}
\left( p_{l_{i}},p_{l_{i},^{\prime }l_{i}},k_{l_{i}^{\prime }}\right) \\ 
\left[ \left\{ \Psi _{J}\left[ l^{\prime },0\right] \right\} ,\left[
p_{0},p_{l^{\prime }0}\right] ^{k_{l^{\prime }}},v\right]%
\end{array}%
\right] \right\} _{i}\right) d\left( \left\{ \left[ p_{\eta _{i}},p_{\eta
_{i^{\prime }}^{\prime }\eta _{i}}\right] ^{k_{\eta _{i^{\prime }}^{\prime
}\eta _{i}}}\right\} _{\substack{ i\leqslant s  \\ i^{\prime }\leqslant
s^{\prime }}}\right)  \notag
\end{eqnarray}%
where $\mathcal{P}_{l}$ partitions the set: 
\begin{equation*}
\left\{ \left[ p_{\eta _{i}},p_{\eta _{i^{\prime }}^{\prime }\eta _{i}}%
\right] ^{k_{\eta _{i^{\prime }}^{\prime }\eta _{i}}}\right\} _{\substack{ %
i\leqslant s  \\ i^{\prime }\leqslant s^{\prime }}},\left\{ \left[ 
\begin{array}{c}
\left( p_{l_{i}},p_{l_{i},^{\prime }l_{i}},k_{l_{i}^{\prime }}\right) \\ 
\left[ \left\{ \Psi _{J}\left[ l^{\prime },0\right] \right\} ,\left[
p_{0},p_{l^{\prime }0}\right] ^{k_{l^{\prime }}},v\right]%
\end{array}%
\right] \right\}
\end{equation*}%
so that:%
\begin{eqnarray*}
\left\{ \left[ p_{\eta _{i}},p_{\eta _{i^{\prime }}^{\prime }\eta _{i}}%
\right] ^{k_{\eta _{i^{\prime }}^{\prime }\eta _{i}}}\right\} _{\substack{ %
i\leqslant s  \\ i^{\prime }\leqslant s^{\prime }}} &=&\tprod\limits_{l}%
\left\{ \left[ p_{\eta _{i}},p_{\eta _{i^{\prime }}^{\prime }\eta _{i}}%
\right] ^{k_{\eta _{i^{\prime }}^{\prime }\eta _{i}}}\right\} _{\substack{ %
i\leqslant s  \\ i^{\prime }\leqslant s^{\prime }}}^{\mathcal{P}_{l}} \\
\left\{ \left[ 
\begin{array}{c}
\left( p_{l_{i}},p_{l_{i},^{\prime }l_{i}},k_{l_{i}^{\prime }}\right) \\ 
\left[ \left\{ \Psi _{J}\left[ l^{\prime },0\right] \right\} ,\left[
p_{0},p_{l^{\prime }0}\right] ^{k_{l^{\prime }}},v\right]%
\end{array}%
\right] \right\} _{i} &=&\tprod\limits_{l}\left\{ \left[ 
\begin{array}{c}
\left( p_{l_{i}},p_{l_{i},^{\prime }l_{i}},k_{l_{i}^{\prime }}\right) \\ 
\left[ \left\{ \Psi _{J}\left[ l^{\prime },0\right] \right\} ,\left[
p_{0},p_{l^{\prime }0}\right] ^{k_{l^{\prime }}},v\right]%
\end{array}%
\right] \right\} _{i\in \mathcal{P}_{l}}
\end{eqnarray*}%
The first term in the action is:

\begin{eqnarray}
&&S_{0}=\sum_{_{\left( s,s^{\prime },\left[ p,p^{\prime },k^{\prime }\right]
\right) }}\int \bar{S}_{0}^{\mathcal{K}}\left( \left\{ \left[ 
\begin{array}{c}
\left( p_{l_{i}},p_{l_{i},^{\prime }l_{i}},k_{l_{i}^{\prime }}\right) \\ 
\left[ \left\{ \Psi _{J}\left[ l^{\prime },0\right] \right\} ,\left[
p_{0},p_{l^{\prime }0}\right] ^{k_{l^{\prime }}},v\right]%
\end{array}%
\right] \right\} _{i}\right)  \label{Pr} \\
&&\times \Psi _{J}\left( \left\{ \left[ 
\begin{array}{c}
\left( p_{l_{i}},p_{l_{i},^{\prime }l_{i}},k_{l_{i}^{\prime }}\right) \\ 
\left[ \left\{ \Psi _{J}\left[ l^{\prime },0\right] \right\} ,\left[
p_{0},p_{l^{\prime }0}\right] ^{k_{l^{\prime }}},v\right]%
\end{array}%
\right] \right\} _{i\in \mathcal{P}_{l}},v\right) \tprod\limits_{i}\left(
d\left\{ \left[ 
\begin{array}{c}
\left( p_{l},p_{l,^{\prime }l},k_{l^{\prime }}\right) \\ 
\left[ \left\{ \Psi _{J}\left[ l^{\prime },0\right] \right\} ,\left[
p_{0},p_{l^{\prime }0}\right] ^{k_{l^{\prime }}},v\right]%
\end{array}%
\right] \right\} _{i}\right)  \notag
\end{eqnarray}

\subsubsection{Local expansion and constraint}

We can write the local expansion for this action (\ref{Pr}):

\begin{eqnarray}
S_{0} &=&\sum_{m}\sum_{l,\mathcal{P}_{l}}\int d\left( \left\{ \left[ 
\begin{array}{c}
\left( p_{l},p_{l,^{\prime }l},k_{l^{\prime }}\right) \\ 
\left[ \left\{ \Psi _{J}\left[ l^{\prime },0\right] \right\} ,v\right]%
\end{array}%
\right] \right\} \right) \Psi _{J}\left( \left\{ \left[ 
\begin{array}{c}
\left( p_{l},p_{l,^{\prime }l},k_{l^{\prime }}\right) \\ 
\left[ \left\{ \Psi _{J}\left[ l^{\prime },0\right] \right\} ,v\right]%
\end{array}%
\right] \right\} ,v\right)  \label{Ps} \\
&&\times \bar{S}_{0,\left\{ m_{_{\mathcal{P}_{l}}}\right\} }^{\mathcal{K}%
}\left( \left\{ \left[ 
\begin{array}{c}
\left( p_{l},p_{l,^{\prime }l},k_{l^{\prime }}\right) \\ 
\left[ \left\{ \Psi _{J}\left[ l^{\prime },0\right] \right\} ,v\right]%
\end{array}%
\right] \right\} \right) \tprod\limits_{l>1}\frac{\nabla ^{m_{_{\mathcal{P}%
_{l}}}}\Psi _{J}\left( \left\{ \left[ 
\begin{array}{c}
\left( p_{l},p_{l,^{\prime }l},k_{l^{\prime }}\right) \\ 
\left[ \left\{ \Psi _{J}\left[ l^{\prime },0\right] \right\} ,v\right]%
\end{array}%
\right] \right\} ,v\right) }{\nabla \left\{ \left[ 
\begin{array}{c}
\left( p_{l},p_{l,^{\prime }l},k_{l^{\prime }}\right) \\ 
\left[ \left\{ \Psi _{J}\left[ l^{\prime },0\right] \right\} ,v\right]%
\end{array}%
\right] \right\} ^{m}}  \notag
\end{eqnarray}%
This term $S_{0}$ is the specific action for the space of parameters.

Similarly, the local series for the action is:%
\begin{eqnarray}
&&\sum_{\left( m_{\mathcal{P}_{l}},l>1\right) }\sum_{_{\left( s,s^{\prime },%
\left[ p,p^{\prime },k^{\prime }\right] \right) }}\int \bar{S}_{\left( m_{%
\mathcal{P}_{l}}\right) }^{\mathcal{K}}\left( \left[ \left\{ \left[ p_{\eta
_{i}},p_{\eta _{i^{\prime }}^{\prime }\eta _{i}}\right] ^{k_{\eta
_{i^{\prime }}^{\prime }\eta _{i}}}\right\} _{\substack{ i\leqslant s \\ %
i^{\prime }\leqslant s^{\prime }}}\right] ,\left\{ \left[ 
\begin{array}{c}
\left( p_{l},p_{l,^{\prime }l},k_{l^{\prime }}\right)  \\ 
\left[ \left\{ \Psi _{J}\left[ l^{\prime },0\right] \right\} ,\left[
p_{0},p_{l^{\prime }0}\right] ^{k_{l^{\prime }}},v\right] 
\end{array}%
\right] \right\} \right)   \notag \\
&&\times \Psi _{J,\tprod_{i}s_{p_{\eta _{i}},p_{\eta _{i^{\prime }}^{\prime
}\eta _{i}}}^{\otimes k_{\eta _{i^{\prime }}^{\prime }\eta _{i}}}}^{\otimes
\sum_{\left( \eta _{i^{\prime }}^{\prime }\eta _{i}\right) }k_{\eta
_{i^{\prime }}^{\prime }\eta _{i}}}\left( \left[ \left\{ \left[ p_{\eta
_{i}},p_{\eta _{i^{\prime }}^{\prime }\eta _{i}}\right] ^{k_{\eta
_{i^{\prime }}^{\prime }\eta _{i}}}\right\} _{\substack{ i\leqslant s \\ %
i^{\prime }\leqslant s^{\prime }}}\right] ^{\mathcal{P}_{1}},\left\{ \left[ 
\begin{array}{c}
\left( p_{l},p_{l,^{\prime }l},k_{l^{\prime }}\right)  \\ 
\left[ \left\{ \Psi _{J}\left[ l^{\prime },0\right] \right\} ,\left[
p_{0},p_{l^{\prime }0}\right] ^{k_{l^{\prime }}},v\right] 
\end{array}%
\right] \right\} ,v\right)   \notag \\
&&\times \tprod\limits_{l>1}\nabla ^{m_{\mathcal{P}_{l}}}\Psi
_{J,\tprod_{i}s_{p_{\eta _{i}},p_{\eta _{i^{\prime }}^{\prime }\eta
_{i}}}^{\otimes k_{\eta _{i^{\prime }}^{\prime }\eta _{i}}}}^{\otimes
\sum_{\left( \eta _{i^{\prime }}^{\prime }\eta _{i}\right) }k_{\eta
_{i^{\prime }}^{\prime }\eta _{i}}}\left( \left[ \left\{ \left[ p_{\eta
_{i}},p_{\eta _{i^{\prime }}^{\prime }\eta _{i}}\right] ^{k_{\eta
_{i^{\prime }}^{\prime }\eta _{i}}}\right\} _{\substack{ i\leqslant s \\ %
i^{\prime }\leqslant s^{\prime }}}\right] ^{\mathcal{P}_{l}},\left\{ \left[ 
\begin{array}{c}
\left( p_{l},p_{l,^{\prime }l},k_{l^{\prime }}\right)  \\ 
\left[ \left\{ \Psi _{J}\left[ l^{\prime },0\right] \right\} ,\left[
p_{0},p_{l^{\prime }0}\right] ^{k_{l^{\prime }}},v\right] 
\end{array}%
\right] \right\} ,v\right)   \notag \\
&&\times d\left( \left\{ \left[ 
\begin{array}{c}
\left( p_{l},p_{l,^{\prime }l},k_{l^{\prime }}\right)  \\ 
\left[ \left\{ \Psi _{J}\left[ l^{\prime },0\right] \right\} ,\left[
p_{0},p_{l^{\prime }0}\right] ^{k_{l^{\prime }}},v\right] 
\end{array}%
\right] \right\} \right) d\left( \left[ \left\{ \left[ p_{\eta _{i}},p_{\eta
_{i^{\prime }}^{\prime }\eta _{i}}\right] ^{k_{\eta _{i^{\prime }}^{\prime
}\eta _{i}}}\right\} _{\substack{ i\leqslant s \\ i^{\prime }\leqslant
s^{\prime }}}\right] \right)   \label{Rn}
\end{eqnarray}%
where we define:%
\begin{equation*}
\nabla ^{m_{\mathcal{P}_{l}}}=\nabla _{\left\{ \left[ 
\begin{array}{c}
\left( p_{l},p_{l,^{\prime }l},k_{l^{\prime }}\right)  \\ 
\left[ \left\{ \Psi _{J}\left[ l^{\prime },0\right] \right\} ,v\right] 
\end{array}%
\right] \right\} }^{m_{\mathcal{P}_{l}}}
\end{equation*}%
Due to non locality, this form allows to impose (\ref{Cdn}) or (\ref{Cdt}).
The expansion should include the contribution of the type:%
\begin{eqnarray*}
&&\bar{S}_{\left( 1\right) }^{\mathcal{K}}\left( \left[ \left\{ \left[
p_{\eta _{i}},p_{\eta _{i^{\prime }}^{\prime }\eta _{i}}\right] ^{k_{\eta
_{i^{\prime }}^{\prime }\eta _{i}}}\right\} _{\substack{ i\leqslant s \\ %
i^{\prime }\leqslant s^{\prime }}}\right] ,\left\{ \left[ 
\begin{array}{c}
\left( p_{l},p_{l,^{\prime }l},k_{l^{\prime }}\right)  \\ 
\left[ \left\{ \Psi _{J}\left[ l^{\prime },0\right] \right\} ,\left[
p_{0},p_{l^{\prime }0}\right] ^{k_{l^{\prime }}},v\right] 
\end{array}%
\right] \right\} \right)  \\
&=&\left[ \Xi ^{\left( k_{i}\right) }\left( \left\{ \Psi _{J}\left[ \left\{ %
\left[ p_{\eta _{i}},p_{\eta _{i^{\prime }}^{\prime }\eta _{i}}\right]
^{k_{\eta _{i^{\prime }}^{\prime }\eta _{i}}}\right\} _{\substack{ %
i\leqslant s \\ i^{\prime }\leqslant s^{\prime }}}\right] ^{\mathcal{P}},%
\left[ \left\{ \left[ p_{\eta _{i}},p_{\eta _{i^{\prime }}^{\prime }\eta
_{i}}\right] ^{k_{\eta _{i^{\prime }}^{\prime }\eta _{i}}}\right\} 
_{\substack{ i\leqslant s \\ i^{\prime }\leqslant s^{\prime }}}\right] ^{%
\mathcal{P}^{\prime }}\left[ \left( p_{l},p_{l,^{\prime }l},k_{l^{\prime
}}\right) ,\eta \right] \right\} \right) \right] 
\end{eqnarray*}%
where:%
\begin{eqnarray*}
&&\left[ \Xi ^{\left( k_{i}\right) }\left( \left\{ \Psi _{J}\left[ \left\{ %
\left[ p_{\eta _{i}},p_{\eta _{i^{\prime }}^{\prime }\eta _{i}}\right]
^{k_{\eta _{i^{\prime }}^{\prime }\eta _{i}}}\right\} _{\substack{ %
i\leqslant s \\ i^{\prime }\leqslant s^{\prime }}}\right] ^{\mathcal{P}},%
\left[ \left\{ \left[ p_{\eta _{i}},p_{\eta _{i^{\prime }}^{\prime }\eta
_{i}}\right] ^{k_{\eta _{i^{\prime }}^{\prime }\eta _{i}}}\right\} 
_{\substack{ i\leqslant s \\ i^{\prime }\leqslant s^{\prime }}}\right] ^{%
\mathcal{P}^{\prime }},\left[ \left( p_{l},p_{l,^{\prime }l},k_{l^{\prime
}}\right) ,\eta \right] \right\} \right) \right] ^{v} \\
&=&\int v\left( \left\{ \left[ p_{\eta },p_{l^{\prime }\eta }\right]
^{k_{l^{\prime }}}\right\} _{_{\substack{ l\leqslant s \\ l^{\prime
}\leqslant s^{\prime }}}}\right) \frac{\delta \left\{ \left[ 
\begin{array}{c}
\left( p_{l},p_{l,^{\prime }l},k_{l^{\prime }}\right)  \\ 
\left[ \left\{ \Psi _{J,\alpha }\left[ p_{\eta },p_{\eta ^{\prime }\eta }%
\right] \right\} ,v\right] 
\end{array}%
\right] \right\} }{\delta \Psi _{J,s_{p_{l^{\prime },\eta }}^{\otimes
k_{l^{\prime }\eta }}}^{\otimes k_{l^{\prime }\eta }}\left( \left\{ \left[
p_{\eta _{i}},p_{\eta _{i^{\prime }}^{\prime }\eta _{i}}\right] ^{k_{\eta
_{i^{\prime }}^{\prime }\eta _{i}}}\right\} _{\substack{ i\leqslant s \\ %
i^{\prime }\leqslant s^{\prime }}}\right) }\left( \frac{\delta \left[ 
\begin{array}{c}
\left( p_{l_{i}},p_{l_{i}^{\prime }l_{i}},k_{l_{i}^{\prime }}\right)  \\ 
\left[ \left\{ \Psi _{J,\alpha }\left[ p_{\eta },p_{\eta ^{\prime }\eta }%
\right] \right\} ,v\right] 
\end{array}%
\right] }{\delta \left[ \left( p_{l},p_{l,^{\prime }l},k_{l^{\prime
}}\right) ,\eta \right] }\right) ^{-1}
\end{eqnarray*}%
The $\left[ \left( p_{l},p_{l,^{\prime }l},k_{l^{\prime }}\right) ,\eta %
\right] $ represents some local coordinates of reference.

\subsection{Constraint and effective action in local parametrs}

The local expansion of (\ref{Rs}) is similar to (\ref{Rn}) and should be
expanded in apparent independent parameters:%
\begin{equation}
\left[ 
\begin{array}{c}
\left( p_{l_{1}},p_{l_{1}^{\prime }l_{1}},k_{l_{1}^{\prime }}\right)  \\ 
\left[ \left\{ \Psi _{J}\left[ l^{\prime },0\right] \right\} ,v\right] 
\end{array}%
\right] \rightarrow \left[ \mathbf{\hat{\Lambda}}_{\infty }^{\left[ k\right]
},\lambda \right]   \label{Lcn}
\end{equation}%
The volume of space: 
\begin{equation*}
Vol\left( \left\{ \left[ 
\begin{array}{c}
\left( p_{l_{1}},p_{l_{1}^{\prime }l_{1}},k_{l_{1}^{\prime }}\right)  \\ 
\left[ \left\{ \Psi _{J}\left[ l^{\prime },0\right] \right\} ,v\right] 
\end{array}%
\right] \right\} \right) 
\end{equation*}%
and the metric $g\left( \left[ 
\begin{array}{c}
\left( p_{l_{1}},p_{l_{1}^{\prime }l_{1}},k_{l_{1}^{\prime }}\right)  \\ 
\left[ \left\{ \Psi _{J}\left[ l^{\prime },0\right] \right\} ,v\right] 
\end{array}%
\right] \right) $ should be taken into account by including some volume
element in the approximation in the local parameters:%
\begin{equation*}
\sqrt{g\left[ \mathbf{\hat{\Lambda}}_{\infty }^{\left[ k\right] },\lambda %
\right] }d\left[ \mathbf{\hat{\Lambda}}_{\infty }^{\left[ k\right] },\lambda %
\right] 
\end{equation*}%
and the local expansion writes:%
\begin{eqnarray}
&&\sum_{\left( m_{\mathcal{P}_{l}},l>1\right) }\sum_{_{\left( s,s^{\prime },%
\left[ p,p^{\prime },k^{\prime }\right] \right) }}\int \sqrt{g}d\left( \left[
\mathbf{\hat{\Lambda}}_{\infty }^{\left[ k\right] },\lambda \right] \right)
d\left( \left[ \left\{ \left[ p_{\eta _{i}},p_{\eta _{i^{\prime }}^{\prime
}\eta _{i}}\right] ^{k_{\eta _{i^{\prime }}^{\prime }\eta _{i}}}\right\} 
_{\substack{ i\leqslant s \\ i^{\prime }\leqslant s^{\prime }}}\right]
\right)  \\
&&\times \bar{S}_{\left( m_{\mathcal{P}_{l}}\right) }^{\mathcal{K}}\left( %
\left[ \left\{ \left[ p_{\eta _{i}},p_{\eta _{i^{\prime }}^{\prime }\eta
_{i}}\right] ^{k_{\eta _{i^{\prime }}^{\prime }\eta _{i}}}\right\} 
_{\substack{ i\leqslant s \\ i^{\prime }\leqslant s^{\prime }}}\right] ,%
\left[ \mathbf{\hat{\Lambda}}_{\infty }^{\left[ k\right] },\lambda \right]
\right) \Psi _{J,\tprod_{i}s_{p_{\eta _{i}},p_{\eta _{i^{\prime }}^{\prime
}\eta _{i}}}^{\otimes k_{\eta _{i^{\prime }}^{\prime }\eta _{i}}}}^{\otimes
\sum_{\left( \eta _{i^{\prime }}^{\prime }\eta _{i}\right) }k_{\eta
_{i^{\prime }}^{\prime }\eta _{i}}}\left( \left[ \left\{ \left[ p_{\eta
_{i}},p_{\eta _{i^{\prime }}^{\prime }\eta _{i}}\right] ^{k_{\eta
_{i^{\prime }}^{\prime }\eta _{i}}}\right\} _{\substack{ i\leqslant s \\ %
i^{\prime }\leqslant s^{\prime }}}\right] ^{\mathcal{P}_{1}},\left[ \mathbf{%
\hat{\Lambda}}_{\infty }^{\left[ k\right] },\lambda \right] ,v\right)  
\notag \\
&&\times \tprod\limits_{l>1}\nabla ^{m_{\mathcal{P}_{l}}}\Psi
_{J,\tprod_{i}s_{p_{\eta _{i}},p_{\eta _{i^{\prime }}^{\prime }\eta
_{i}}}^{\otimes k_{\eta _{i^{\prime }}^{\prime }\eta _{i}}}}^{\otimes
\sum_{\left( \eta _{i^{\prime }}^{\prime }\eta _{i}\right) }k_{\eta
_{i^{\prime }}^{\prime }\eta _{i}}}\left( \left[ \left\{ \left[ p_{\eta
_{i}},p_{\eta _{i^{\prime }}^{\prime }\eta _{i}}\right] ^{k_{\eta
_{i^{\prime }}^{\prime }\eta _{i}}}\right\} _{\substack{ i\leqslant s \\ %
i^{\prime }\leqslant s^{\prime }}}\right] ^{\mathcal{P}_{l}},\left[ \mathbf{%
\hat{\Lambda}}_{\infty }^{\left[ k\right] },\lambda \right] ,v\right)  
\notag
\end{eqnarray}%
nd:%
\begin{eqnarray}
S_{0} &=&\sum_{m}\sum_{l,\mathcal{P}_{l}}\int \sqrt{g}d\left( \left\{ \left[ 
\mathbf{\hat{\Lambda}}_{\infty }^{\left[ k\right] },\lambda \right] \right\}
\right) \Psi _{J}\left( \left\{ \left[ \mathbf{\hat{\Lambda}}_{\infty }^{%
\left[ k\right] },\lambda \right] \right\} ,v\right)  \\
&&\times \bar{S}_{0,\left\{ m_{_{\mathcal{P}_{l}}}\right\} }^{\mathcal{K}%
}\left( \left\{ \left[ \mathbf{\hat{\Lambda}}_{\infty }^{\left[ k\right]
},\lambda \right] \right\} \right) \tprod\limits_{l>1}\frac{\nabla ^{m_{_{%
\mathcal{P}_{l}}}}\Psi _{J}\left( \left\{ \left[ \mathbf{\hat{\Lambda}}%
_{\infty }^{\left[ k\right] },\lambda \right] \right\} ,v\right) }{\nabla
\left\{ \left[ \mathbf{\hat{\Lambda}}_{\infty }^{\left[ k\right] },\lambda %
\right] \right\} ^{m}}  \notag
\end{eqnarray}%
However, since the coordinates are globaly constrained, and this constraint
is expressed in the metric $g$. The constraint can ths be imposed by
including an action for the metric, and then we should consider:%
\begin{equation*}
S_{g}=\int s_{g}\sqrt{g}d\left[ \mathbf{\hat{\Lambda}}_{\infty }^{\left[ k%
\right] },\lambda \right] 
\end{equation*}%
The action:%
\begin{equation*}
S_{g}+S
\end{equation*}%
corresponds to the sole space plus mattr actions. The choice for $S_{g}$
could be impsd by usual arguments to obtain scalar action, involving the
curvature of space, while the form for $S$ can be chosen to impose the
invariance or average invariance developed before.

This implies that in first apprximation, the matter action should be written
in some spinor form using (\ref{Fn}) and (\ref{Fc}) in local coordinates:%
\begin{eqnarray}
&&S=\int d\left[ \mathbf{\hat{\Lambda}}_{\infty }^{\left[ k\right] },\lambda %
\right] _{i}\Psi _{J,\tprod\limits_{i^{\prime }}s_{p_{\eta _{i^{\prime
}}^{\prime }\eta _{i}}}^{\otimes k_{\eta _{i^{\prime }}^{\prime }\eta
_{i}}}}^{\otimes \sum_{i^{\prime }}k_{\eta _{i^{\prime }}^{\prime }\eta
_{i}}}\left( \left\{ \left[ p_{\eta _{i}},p_{\eta _{i^{\prime }}^{\prime
}\eta _{i}}\right] ^{k_{\eta _{i^{\prime }}^{\prime }\eta _{i}}}\right\} _{
_{\substack{ i\leqslant s \\ i^{\prime }\leqslant s^{\prime }}}},\left[ 
\mathbf{\hat{\Lambda}}_{\infty }^{\left[ k\right] },\lambda \right]
_{i},v\right)   \label{Ctns} \\
&&\times \left[ \Xi ^{\left( k_{i}\right) }\left( \left\{ \left\{ \Psi _{J}%
\left[ p_{\eta },p_{\eta ^{\prime }\eta }\right] \right\} ,\left[ \mathbf{%
\hat{\Lambda}}_{\infty }^{\left[ k\right] },\lambda \right] _{i}\right\}
\right) \right] _{\left[ \bar{p}_{\eta },\bar{p}_{\eta _{^{\prime }}^{\prime
}\eta }\right] ^{k_{\eta _{^{\prime }}^{\prime }\eta }}}^{v}\nabla _{\left[ 
\mathbf{\hat{\Lambda}}_{\infty }^{\left[ k\right] },\lambda \right]
_{i}}\Psi _{J,\tprod\limits_{i^{\prime }}s_{p_{\eta _{i^{\prime }}^{\prime
}\eta _{i}}}^{\otimes k_{\eta _{i^{\prime }}^{\prime }\eta _{i}}}}^{\otimes
\sum_{i^{\prime }}k_{\eta _{i^{\prime }}^{\prime }\eta _{i}}}\left( \left\{ %
\left[ p_{\eta _{i}},p_{\eta _{i^{\prime }}^{\prime }\eta _{i}}\right]
^{k_{\eta _{i^{\prime }}^{\prime }\eta _{i}}}\right\} _{_{\substack{ %
i\leqslant s \\ i^{\prime }\leqslant s^{\prime }}}},\left[ \mathbf{\hat{%
\Lambda}}_{\infty }^{\left[ k\right] },\lambda \right] _{i},v\right)   \notag
\end{eqnarray}%
where:%
\begin{eqnarray*}
&&\left[ \Xi ^{\left( k_{i}\right) }\left( \left\{ \left\{ \Psi _{J}\left[
l^{\prime },0\right] \right\} ,\left[ \mathbf{\hat{\Lambda}}_{\infty }^{%
\left[ k\right] },\lambda \right] _{i}\right\} \right) \right] _{\left[ \bar{%
p}_{\eta },\bar{p}_{\eta _{^{\prime }}^{\prime }\eta }\right] ^{k_{\eta
_{^{\prime }}^{\prime }\eta }}}^{v} \\
&=&\int v\left( \left\{ \left[ p_{\eta },p_{l^{\prime }\eta }\right]
^{k_{l^{\prime }}}\right\} _{_{\substack{ l\leqslant s \\ l^{\prime
}\leqslant s^{\prime }}}}\right) \frac{\delta \left[ \mathbf{\hat{\Lambda}}%
_{\infty }^{\left[ k\right] },\lambda \right] _{i}}{\delta \Psi
_{J,s_{p_{\eta _{^{\prime }}^{\prime }\eta }}^{\otimes k\eta _{^{\prime
}}^{\prime }\eta }}^{\otimes \sum_{i^{\prime }}k_{\eta _{^{\prime }}^{\prime
}\eta }}\left( \left[ \bar{p}_{\eta },\bar{p}_{\eta _{^{\prime }}^{\prime
}\eta }\right] ^{k_{\eta _{^{\prime }}^{\prime }\eta }}\right) }\left( \frac{%
\delta \left[ 
\begin{array}{c}
\left( p_{l_{i}},p_{l_{i}^{\prime }l_{i}},k_{l_{i}^{\prime }}\right)  \\ 
\left[ \left\{ \Psi _{J,\alpha }\left[ p_{\eta },p_{\eta ^{\prime }\eta }%
\right] \right\} ,v\right] 
\end{array}%
\right] }{\delta \left[ \mathbf{\hat{\Lambda}}_{\infty }^{\left[ k\right]
},\lambda \right] _{i}}\right) ^{-1}
\end{eqnarray*}%
and identifications:%
\begin{equation*}
v\rightarrow \left[ p_{\eta },p_{\eta ^{\prime }\eta }\right] ^{k_{\eta
^{\prime }\eta }}
\end{equation*}%
Some massive terms or potentials should be included if we consider the
average invariance. The derivative:%
\begin{equation*}
\frac{\delta \left[ 
\begin{array}{c}
\left( p_{l_{i}},p_{l_{i}^{\prime }l_{i}},k_{l_{i}^{\prime }}\right)  \\ 
\left[ \left\{ \Psi _{J,\alpha }\left[ p_{\eta },p_{\eta ^{\prime }\eta }%
\right] \right\} ,v\right] 
\end{array}%
\right] }{\delta \left[ \mathbf{\hat{\Lambda}}_{\infty }^{\left[ k\right]
},\lambda \right] _{i}}
\end{equation*}%
reminds that the "true" parameters $\left[ 
\begin{array}{c}
\left( p_{l_{i}},p_{l_{i}^{\prime }l_{i}},k_{l_{i}^{\prime }}\right)  \\ 
\left[ \left\{ \Psi _{J,\alpha }\left[ p_{\eta },p_{\eta ^{\prime }\eta }%
\right] \right\} ,v\right] 
\end{array}%
\right] $ are state $v$ \ and field $\Psi _{J,\alpha }\left[ p_{\eta
},p_{\eta ^{\prime }\eta }\right] $ dependent, so that the dependncy in $%
\left[ \mathbf{\hat{\Lambda}}_{\infty }^{\left[ k\right] },\lambda \right]
_{i}$ is functional.

Note that the action $S_{0}$ for parameters should also be included n (\ref%
{Ctns}):%
\begin{equation*}
S_{0}+S_{g}+S
\end{equation*}%
In local coordinates, the cloud part of the action writes:%
\begin{equation}
S_{0}=\sum_{m}\int d\left( \left[ \mathbf{\hat{\Lambda}}_{\infty }^{\left[ k%
\right] },\lambda \right] \right) \Psi \left( \left[ \mathbf{\hat{\Lambda}}%
_{\infty }^{\left[ k\right] },\lambda \right] ,v\right) \bar{S}_{0/m}^{%
\mathcal{K}}\left( \left[ \mathbf{\hat{\Lambda}}_{\infty }^{\left[ k\right]
},\lambda \right] \right) \frac{\nabla ^{m}\Psi \left( \left[ \mathbf{\hat{%
\Lambda}}_{\infty }^{\left[ k\right] },\lambda \right] ,v\right) }{\nabla %
\left[ \mathbf{\hat{\Lambda}}_{\infty }^{\left[ k\right] },\lambda \right]
^{m}}  \label{CLh}
\end{equation}%
This action should modify the apparent equation for metric $g$, that is
modify the energy momentum tensor. This term is specific of the projected
state and inherent to the nature of parameters.\ The cloud part of the
action should thus matter and take place within the apparent dynamics of the
states. We develop this point in the next section.

\subsection{Effective local field and action}

We have described the effective and its the expansion of the action for the
saddle point solution given a state $\nu $ We consder now the effective
action for a given state variation.

\subsubsection{Local effective action}

In an effective point of view, the previous description is equivalent to
consider that the fields describing the system:%
\begin{equation*}
\Psi \left( \left\{ v^{\left( \eta \right) },\underline{\mathbf{\hat{\Lambda}%
}}^{\left[ k_{i}\right] }\right\} ,\delta v^{\left( \eta \right) },%
\underline{\mathbf{\hat{\Lambda}}}^{\left[ \Delta \right] }\right)
\end{equation*}%
can be replaced by a descrption based on:%
\begin{equation*}
\psi \left( \delta v^{\left( \eta \right) },\underline{\mathbf{\hat{\Lambda}}%
}^{\left[ \Delta \right] }\right)
\end{equation*}%
where the clouds $\left\{ \underline{\mathbf{\hat{\Lambda}}}^{\left[ k_{i}%
\right] }\right\} $, the background state $v^{\left( \eta \right) }$ the
field and $\Psi \left( \left\{ v^{\left( \eta \right) },\underline{\mathbf{%
\hat{\Lambda}}}^{\left[ k_{i}\right] }\right\} \right) $\ are considered as
a prori given. Given the relation (\ref{cr}), the effective fields $\psi
\left( \underline{\mathbf{\hat{\Lambda}}}^{\left[ \Delta \right] }\right) $
and their adjoints $\psi \dag \left( \underline{\mathbf{\hat{\Lambda}}}^{%
\left[ \Delta \right] }\right) $ satisfy commutation or anticommutation
relations.

In terms of the action, and assuming separable contributions we have:

\begin{equation*}
S\left( \Psi \left( \left\{ v^{\left( \eta \right) },\underline{\mathbf{\hat{%
\Lambda}}}^{\left[ k_{i}\right] },\underline{\mathbf{\hat{\Lambda}}}^{\left[
\Delta \right] }\right\} \right) \right) =S\left( \Psi \left( \left\{
v^{\left( \eta \right) },\underline{\mathbf{\hat{\Lambda}}}^{\left[ k_{i}%
\right] }\right\} \right) \right) +S\left( \Psi \left( \left\{ v^{\left(
\eta \right) },\underline{\mathbf{\hat{\Lambda}}}^{\left[ k_{i}\right]
}\right\} \right) ,\Psi \left( \underline{\mathbf{\hat{\Lambda}}}^{\left[
\Delta \right] }\right) \right)
\end{equation*}%
The second term includes $\Psi \left( \left\{ v^{\left( \eta \right) },%
\underline{\mathbf{\hat{\Lambda}}}^{\left[ k_{i}\right] }\right\} \right) $,
shaping the action functional of $\Psi \left( \underline{\mathbf{\hat{\Lambda%
}}}^{\left[ \Delta \right] }\right) $.

\subsubsection{Product of modification and product of fields}

We start with the local identification (\ref{Lcn}) and by assuming the
following modifications $\left\{ \underline{\mathbf{\hat{\Lambda}}}^{\left[
\Delta _{i}\right] }\right\} _{i}$:%
\begin{equation*}
v^{\left( \eta \right) }\otimes \delta v_{1}^{\left( \eta \right) }\otimes
\delta v_{2}^{\left( \eta \right) }\rightarrow v^{\left( \eta \right)
}\otimes \delta v_{1}^{\left( \eta \right) }\otimes \delta v_{2}^{\left(
\eta \right) }\otimes \underline{\mathbf{\hat{\Lambda}}}^{\left[ \Delta _{1}%
\right] }\otimes \underline{\mathbf{\hat{\Lambda}}}^{\left[ \Delta _{2}%
\right] }
\end{equation*}%
If the symmetries generators are differents for $\delta v_{1}^{\left( \eta
\right) }$ and $\delta v_{2}^{\left( \eta \right) }$, the states can be
associated to:%
\begin{equation*}
\delta v_{1}^{\left( \eta \right) }\otimes \underline{\mathbf{\hat{\Lambda}}}%
^{\left[ \Delta _{1}\right] }
\end{equation*}%
and:%
\begin{equation*}
\delta v_{2}^{\left( \eta \right) }\otimes \underline{\mathbf{\hat{\Lambda}}}%
^{\left[ \Delta _{2}\right] }
\end{equation*}%
If the generators are identical:%
\begin{equation*}
\left( \delta v_{1}^{\left( \eta \right) }\otimes \underline{\mathbf{\hat{%
\Lambda}}}^{\left[ \Delta _{1}\right] }\right) \otimes \left( \delta
v_{2}^{\left( \eta \right) }\otimes \underline{\mathbf{\hat{\Lambda}}}^{%
\left[ \Delta _{2}\right] }\right) \pm \left( \delta v_{1}^{\left( \eta
\right) }\otimes \underline{\mathbf{\hat{\Lambda}}}^{\left[ \Delta _{2}%
\right] }\right) \otimes \left( \delta v_{2}^{\left( \eta \right) }\otimes 
\underline{\mathbf{\hat{\Lambda}}}^{\left[ \Delta _{1}\right] }\right) 
\end{equation*}%
and this relation implies that some terms in $S\left( \Psi \left( \left\{
v^{\left( \eta \right) },\underline{\mathbf{\hat{\Lambda}}}^{\left[ k_{i}%
\right] }\right\} \right) ,\Psi \left( \underline{\mathbf{\hat{\Lambda}}}^{%
\left[ \Delta \right] }\right) \right) $ have the form:%
\begin{equation}
\Psi \left( \delta v_{1}^{\left( \eta \right) },\underline{\mathbf{\hat{%
\Lambda}}}^{\left[ \Delta _{1}\right] }\right) \Psi \left( \delta
v_{2}^{\left( \eta \right) },\underline{\mathbf{\hat{\Lambda}}}^{\left[
\Delta _{2}\right] }\right)   \label{Pt}
\end{equation}%
where $\Psi \left( \underline{\mathbf{\hat{\Lambda}}}^{\left[ \Delta \right]
}\right) $ represents commuting or anticommuting fields. These commutation
relation depends on the constraint. Actually, coming back to the field
dependent parameters, the constraints (\ref{Ctn}):%
\begin{equation}
H\left( \left\{ \left[ 
\begin{array}{c}
\left( p_{l_{i}},p_{l_{i}^{\prime }l_{i}},k_{l_{i}^{\prime }}\right)  \\ 
\left[ \left\{ \Psi _{J}\left[ l^{\prime }\eta ,\eta \right] \right\} ,\left[
v\right] \right] 
\end{array}%
\right] \right\} ,\left\{ \Psi _{J}^{\otimes k_{p}},\left[
p_{0},p_{l^{\prime }0}\right] \right\} \right) =0
\end{equation}%
or (\ref{Ctp}):

\begin{equation}
f\left( \sum \left( \left[ 
\begin{array}{c}
\left( p_{l_{i}},p_{l_{i}^{\prime }l_{i}},k_{l_{i}^{\prime }}\right) \\ 
\left[ \left\{ \Psi _{J}\left[ l^{\prime },0\right] \right\} ,v\right]%
\end{array}%
\right] G\left( \left[ 
\begin{array}{c}
\left( p_{l_{i}},p_{l_{i}^{\prime }l_{i}},k_{l_{i}^{\prime }}\right) \\ 
\left[ \left\{ \Psi _{J}\left[ l^{\prime },0\right] \right\} ,v\right]%
\end{array}%
\right] ,\Psi _{J}\right) \left[ 
\begin{array}{c}
\left( p_{l_{i}},p_{l_{i}^{\prime }l_{i}},k_{l_{i}^{\prime }}\right) \\ 
\left[ \left\{ \Psi _{J}\left[ l^{\prime },0\right] \right\} ,v\right]%
\end{array}%
\right] -\gamma \left( \Psi _{J}\right) \right) ^{2}\right)
\end{equation}%
write in frst approximation:%
\begin{eqnarray}
&&0=\left\{ \left[ 
\begin{array}{c}
\left( p_{l_{i}},p_{l_{i}^{\prime }l_{i}},k_{l_{i}^{\prime }}\right) \\ 
\left[ \left\{ \Psi _{J}\left[ l^{\prime },0\right] \right\} ,v\right]%
\end{array}%
\right] \right\} G\left\{ \left[ 
\begin{array}{c}
\left( p_{l_{i}},p_{l_{i}^{\prime }l_{i}},k_{l_{i}^{\prime }}\right) \\ 
\left[ \left\{ \Psi _{J}\left[ l^{\prime },0\right] \right\} ,v,\right]%
\end{array}%
\right] \right\}  \label{Tc} \\
&&\pm \left\{ \left[ 
\begin{array}{c}
\left( p_{l_{\delta }},p_{l_{\delta }^{\prime }l_{\delta }}k_{l_{\delta
}^{\prime }}\right) \\ 
\left[ \left\{ \Psi _{J}\left[ l^{\prime },0\right] \right\} ,v,\delta \nu %
\right]%
\end{array}%
\right] \right\} \delta G\left\{ \left[ 
\begin{array}{c}
\left( p_{l_{\delta }},p_{l_{\delta }^{\prime }l_{\delta }}k_{l_{\delta
}^{\prime }}\right) \\ 
\left[ \left\{ \Psi _{J}\left[ l^{\prime },0\right] \right\} ,v,\delta \nu %
\right]%
\end{array}%
\right] \right\}  \notag
\end{eqnarray}%
We have seen in the previous paragraph that, depending on the sign of the
second term in (\ref{Tc}) for each modification, the fields will be
commuting or anticommuting. Actually this sign corresponds to the derivative
of the constraint at each modifcation encompassed in the coefficients $%
h_{i}^{-+}$, $j_{i}^{-+}$, $h_{i}^{+-}$, $j_{i}^{+-}$ defined previously.

Note that the product (\ref{Pt}), is a priori non local. Locality of the
product implies some conditions on modification. Considering some
modification:%
\begin{equation*}
\left( \delta v_{1}^{\left( \eta \right) }\otimes \underline{\mathbf{\hat{%
\Lambda}}}^{\left[ \Delta _{1}\right] }\right) \otimes \left( \delta
v_{2}^{\left( \eta \right) }\otimes \underline{\mathbf{\hat{\Lambda}}}^{%
\left[ \Delta _{2}\right] }\right)
\end{equation*}%
and consider states of the form:%
\begin{equation*}
v^{\left( \eta \right) }\otimes \delta v_{1}^{\left( \eta \right) }\otimes
\delta v_{2}^{\left( \eta \right) }
\end{equation*}%
the generators of modification are:%
\begin{equation*}
\left\{ \underline{\mathbf{\hat{\Lambda}}}^{\left[ k_{i}\right] },\underline{%
\mathbf{\hat{\Lambda}}}^{\left[ \Delta _{1,2}\right] }\right\}
\end{equation*}%
As a consequence, products of fields in the effective action are local and
write:%
\begin{equation*}
\Psi \left( \delta v_{1}^{\left( \eta \right) },\underline{\mathbf{\hat{%
\Lambda}}}^{\left[ \Delta _{1}\right] }\right) \Psi \left( \delta
v_{2}^{\left( \eta \right) },\underline{\mathbf{\hat{\Lambda}}}^{\left[
\Delta _{1}\right] }\right)
\end{equation*}%
only if $\underline{\mathbf{\hat{\Lambda}}}^{\left[ \Delta _{1,2}\right] }=%
\underline{\mathbf{\hat{\Lambda}}}^{\left[ \Delta _{1}\right] }$.

\subsubsection{Action for additonal states}

We start with the formulas of part 1. We consider equations (\ref{MLN}) or (%
\ref{CT}) and the local identification:%
\begin{equation}
0=\sum_{U_{j}^{l}}\hat{\Gamma}\left( \left( U_{j}^{l_{i}}\right) ^{\prime
},U_{j}^{l}\right) \nabla _{\mathbf{\hat{\Lambda}}_{\infty }^{\left[ k\right]
}}\hat{\Psi}_{J}^{\otimes \sum_{i}l_{i}}\left( \left\{ U_{j}^{l_{i}}\right\}
_{i},\mathbf{\hat{\Lambda}}_{\infty }^{\left[ k\right] },v,\alpha \right) 
\label{MLn}
\end{equation}%
\begin{eqnarray}
0 &=&\int v\left( U_{i}^{l}\right) \Gamma \left( \left[ \Psi _{J},\nu
,U_{j}^{l}\right] ,\left( U_{j}^{l_{i}}\right) ^{\prime }\right) \nabla _{%
\mathbf{\hat{\Lambda}}_{\alpha }^{\left[ \left\{ k,k_{i\leqslant
n-1}\right\} \right] }}\hat{\Psi}_{J}^{\otimes \sum_{i}l_{i}}\left( \left\{
U_{j}^{l_{i}}\right\} _{i},\left( \left\{ \mathbf{\hat{\Lambda}}^{\left[
\left\{ k,k_{i\leqslant n-1}\right\} \right] }\right\} \right) _{n},v,\alpha
\right)   \notag \\
&&+C\left( \Psi _{J}^{\otimes \left( \sum_{i}l_{i}\right) }\left( \left(
U_{j}^{l_{i}}\right) ^{\prime },\mathbf{\hat{\Lambda}}^{\left[ k\right]
},v\right) \right) \hat{\Psi}_{J}^{\otimes \sum_{i}l_{i}}\left( \left\{
U_{j}^{l_{i}}\right\} _{i},\left( \left\{ \mathbf{\hat{\Lambda}}^{\left[
\left\{ k,k_{i\leqslant n-1}\right\} \right] }\right\} \right) _{n},v,\alpha
\right)   \label{CTP}
\end{eqnarray}%
The form of the action is conditioned by these constraints and have the form:%
\begin{equation*}
\sum_{U_{j}^{l},\left( U_{j}^{l_{i}}\right) ^{\prime }}\int \hat{\Psi}%
_{J}^{\otimes \sum_{i}l_{i}}\left( \left\{ \left( U_{j}^{l_{i}}\right)
^{\prime }\right\} _{i},\mathbf{\hat{\Lambda}}_{\infty }^{\left[ k\right]
},v,\alpha \right) \hat{\Gamma}\left( \left( U_{j}^{l_{i}}\right) ^{\prime
},U_{j}^{l}\right) \nabla _{\mathbf{\hat{\Lambda}}_{\infty }^{\left[ k\right]
}}\hat{\Psi}_{J}^{\otimes \sum_{i}l_{i}}\left( \left\{ U_{j}^{l_{i}}\right\}
_{i},\mathbf{\hat{\Lambda}}_{\infty }^{\left[ k\right] },v,\alpha \right) 
\end{equation*}%
or for average invariance:%
\begin{eqnarray}
&&S\left( v\right) =\sum_{U_{j}^{l},\left( U_{j}^{l_{i}}\right) ^{\prime
}}\int \hat{\Psi}_{J}^{\otimes \sum_{i}l_{i}}\left( \left\{ \left(
U_{j}^{l_{i}}\right) ^{\prime }\right\} _{i},\left( \left\{ \mathbf{\hat{%
\Lambda}}^{\left[ \left\{ k,k_{i\leqslant n-1}\right\} \right] }\right\}
\right) _{n},v,\alpha \right)  \\
&&\times \left[ v\left( U_{i}^{l}\right) \Gamma \left( \left[ \Psi _{J},\nu
,U_{j}^{l}\right] ,\left( U_{j}^{l_{i}}\right) ^{\prime }\right) \nabla _{%
\mathbf{\hat{\Lambda}}_{\alpha }^{\left[ \left\{ k,k_{i\leqslant
n-1}\right\} \right] }}\hat{\Psi}_{J}^{\otimes \sum_{i}l_{i}}\left( \left\{
U_{j}^{l_{i}}\right\} _{i},\left( \left\{ \mathbf{\hat{\Lambda}}^{\left[
\left\{ k,k_{i\leqslant n-1}\right\} \right] }\right\} \right) _{n},v,\alpha
\right) \right.   \notag \\
&&\left. +C\left( \Psi _{J}^{\otimes \left( \sum_{i}l_{i}\right) }\left(
\left( U_{j}^{l_{i}}\right) ^{\prime },\mathbf{\hat{\Lambda}}^{\left[ k%
\right] },v\right) \right) \hat{\Psi}_{J}^{\otimes \sum_{i}l_{i}}\left(
\left\{ U_{j}^{l_{i}}\right\} _{i},\left( \left\{ \mathbf{\hat{\Lambda}}^{%
\left[ \left\{ k,k_{i\leqslant n-1}\right\} \right] }\right\} \right)
_{n},v,\alpha \right) \right]   \notag
\end{eqnarray}%
so that the minimization leads to the equation (\ref{MLn}) or (\ref{CTP}).

In the case of additive action, including a modification implies that:%
\begin{eqnarray*}
&&S\left( v+\delta \nu \right) \\
&=&S\left( v\right) +\sum_{U_{j}^{l},\left( U_{j}^{l_{i}}\right) ^{\prime
}}\int \hat{\Psi}_{J}^{\otimes \sum_{i}l_{i}}\left( \left\{ \left(
U_{j}^{l_{i}}\right) ^{\prime }\right\} _{i},\underline{\mathbf{\hat{\Lambda}%
}}^{\left[ \Delta \right] },\delta v,\alpha \right) \hat{\Gamma}\left(
\left( U_{j}^{l_{i}}\right) ^{\prime },U_{j}^{l}\right) \nabla _{\underline{%
\mathbf{\hat{\Lambda}}}^{\left[ \Delta \right] }}\hat{\Psi}_{J}^{\otimes
\sum_{i}l_{i}}\left( \left\{ U_{j}^{l_{i}}\right\} _{i},\underline{\mathbf{%
\hat{\Lambda}}}^{\left[ \Delta \right] },\delta v,\alpha \right)
\end{eqnarray*}%
or for the average field:%
\begin{eqnarray}
&&\sum_{U_{j}^{l},\left( U_{j}^{l_{i}}\right) ^{\prime }}\int \hat{\Psi}%
_{J}^{\otimes \sum_{i}l_{i}}\left( \left\{ \left( U_{j}^{l_{i}}\right)
^{\prime }\right\} _{i},\underline{\mathbf{\hat{\Lambda}}}^{\left[ \Delta %
\right] },v,\alpha \right) \\
&&\times \left[ v\left( U_{i}^{l}\right) \Gamma \left( \left[ \Psi _{J},\nu
,U_{j}^{l}\right] ,\left( U_{j}^{l_{i}}\right) ^{\prime }\right) \nabla _{%
\underline{\mathbf{\hat{\Lambda}}}^{\left[ \Delta \right] }}\hat{\Psi}%
_{J}^{\otimes \sum_{i}l_{i}}\left( \left\{ U_{j}^{l_{i}}\right\} _{i},%
\underline{\mathbf{\hat{\Lambda}}}^{\left[ \Delta \right] },\delta v,\alpha
\right) \right.  \notag \\
&&\left. +C\left( \Psi _{J}^{\otimes \left( \sum_{i}l_{i}\right) }\left(
\left( U_{j}^{l_{i}}\right) ^{\prime },\left\{ \mathbf{\hat{\Lambda}}%
_{\infty }^{\left[ k\right] }\right\} ,\underline{\mathbf{\hat{\Lambda}}}^{%
\left[ \Delta \right] },v\right) \right) \hat{\Psi}_{J}^{\otimes
\sum_{i}l_{i}}\left( \left\{ U_{j}^{l_{i}}\right\} _{i},\left( \underline{%
\mathbf{\hat{\Lambda}}}^{\left[ \Delta \right] }\right) ,\delta v,\alpha
\right) \right]  \notag
\end{eqnarray}%
that are usual action functionals for local fields in spinor representation.

In first approximation this can be written:%
\begin{eqnarray}
&&\sum_{U_{j}^{l},\left( U_{j}^{l_{i}}\right) ^{\prime }}\int \hat{\Psi}%
_{J}^{\otimes \sum_{i}l_{i}}\left( \left\{ \left( U_{j}^{l_{i}}\right)
^{\prime }\right\} _{i},\underline{\mathbf{\hat{\Lambda}}}^{\left[ \Delta %
\right] }\right) \\
&&\times \left[ \Gamma \left( \left( U_{j}^{l}\right) ,\left(
U_{j}^{l_{i}}\right) ^{\prime }\right) \nabla _{\underline{\mathbf{\hat{%
\Lambda}}}^{\left[ \Delta \right] }}+C\left( \Psi _{J}^{\otimes \left(
\sum_{i}l_{i}\right) }\left( \left( U_{j}^{l_{i}}\right) ^{\prime },%
\underline{\mathbf{\hat{\Lambda}}}^{\left[ \Delta \right] },v\right) \right) %
\right] \hat{\Psi}_{J}^{\otimes \sum_{i}l_{i}}\left( \left\{
U_{j}^{l_{i}}\right\} _{i},\left( \underline{\mathbf{\hat{\Lambda}}}^{\left[
\Delta \right] }\right) \right)  \notag
\end{eqnarray}%
To include the constraint for these states, the gradient $\nabla _{\mathbf{%
\hat{\Lambda}}_{\alpha }^{\left[ \left\{ k,k_{i\leqslant n-1}\right\} \right]
}}$ should be replaced by some covariant gradient:%
\begin{equation*}
\nabla _{\mathbf{\hat{\Lambda}}_{\alpha }^{\left[ \left\{ k,k_{i\leqslant
n-1}\right\} \right] }}+A\left( \hat{\Psi}_{J}^{\otimes \sum_{i}l_{i}}\left(
\left\{ U_{j}^{l_{i}}\right\} _{i},\mathbf{\hat{\Lambda}}_{\infty }^{\left[ k%
\right] },v\right) \right)
\end{equation*}%
the connection $A$ depending on the whole state described by $\hat{\Psi}%
_{J}^{\otimes \sum_{i}l_{i}}\left( \left\{ U_{j}^{l_{i}}\right\} _{i},%
\mathbf{\hat{\Lambda}}_{\infty }^{\left[ k\right] },v\right) $.

\section{Landscape of states and induced parameter spaces}

So far, we have studied the impact on the projected state space while $v$
varies. Any variation $\delta \nu $ comes along with continuous
transformations, with compatibility conditions, and discrete variations that
translates in creations of supplementary points in first approximation or
more generally of cloud of points.

The continuous transformation can be identified with transformation with
apparent parameter (\ref{TP}) and (\ref{GLP}) in part III. We also studied,
for given state $v$, the general form for the effective action functionals
in the local identificatn (\ref{Lcn}) for the fields $\hat{\Psi}%
_{J}^{\otimes \sum_{i}l_{i}}\left( \left\{ U_{j}^{l_{i}}\right\} _{i},%
\mathbf{\hat{\Lambda}}_{\infty }^{\left[ k\right] },v\right) $ including
could of points action functional. We can gather these results to provide a
landscape of states variation and its interpretation in terms of parameters
variations.

In the previous paragraph, the action functionl depend on the state $v$
defining the projection and thus the parameter space. To obtain a more
global perspective, we consider small variations of $\nu $. Gathering the
states $\left\vert \nu \right\rangle $:%
\begin{equation*}
\left\{ v_{\left[ p_{l_{1}},p_{l_{1}^{\prime }l_{1}}\right]
^{k_{l_{1}^{\prime }}}}\left\{ \Psi _{J}^{\otimes k_{l^{\prime }}}\right\}
\right\} 
\end{equation*}%
we can consider above these states, the parameter spaces $\mathbf{\hat{%
\Lambda}}_{\infty }^{\left[ k\right] }$, along with the metric being derived
from the constraint (see (\ref{Tc})):%
\begin{equation*}
g\left\{ \left[ 
\begin{array}{c}
\left( p_{l_{i}},p_{l_{i}^{\prime }l_{i}},k_{l_{i}^{\prime }}\right)  \\ 
\left[ \left\{ \Psi _{J}\left[ l^{\prime },0\right] \right\} ,v\right] 
\end{array}%
\right] \right\} =\left\{ g\left\{ \left[ 
\begin{array}{c}
\left( p_{l_{i}},p_{l_{i}^{\prime }l_{i}},k_{l_{i}^{\prime }}\right)  \\ 
\left[ \left\{ \Psi _{J}\left[ l^{\prime },0\right] \right\} ,v\right] 
\end{array}%
\right] \right\} _{n}\right\} 
\end{equation*}%
The sequence\ is induced by the flag manifold property (as in part 1 and 2),
and depends on the fields:%
\begin{equation*}
\hat{\Psi}_{J}^{\otimes \sum_{i}l_{i}}\left( \left\{ U_{j}^{l_{i}}\right\}
_{i},\mathbf{\hat{\Lambda}}_{\infty }^{\left[ k\right] },v\right) 
\end{equation*}%
and the parameter $\lambda $. This leads to the graph:%
\begin{equation*}
\left\{ 
\begin{array}{c}
\left[ g,\mathbf{\hat{\Lambda}}_{\infty }^{\left[ k\right] },\hat{\Psi}%
_{J}^{\otimes \sum_{i}l_{i}},\lambda \right] _{\nu } \\ 
\downarrow  \\ 
v_{\left[ p_{l_{1}},p_{l_{1}^{\prime }l_{1}}\right] ^{k_{l_{1}^{\prime
}}}}\left\{ \Psi _{J}^{\otimes k_{l^{\prime }}}\right\} 
\end{array}%
\right\} 
\end{equation*}%
that should correspond to some fibration:%
\begin{equation*}
\begin{array}{c}
\left\{ \left[ g,\mathbf{\hat{\Lambda}}_{\infty }^{\left[ k\right] },\hat{%
\Psi}_{J}^{\otimes \sum_{i}l_{i}},\lambda \right] _{\nu }\right\}  \\ 
\downarrow  \\ 
\left\{ v_{\left[ p_{l_{1}},p_{l_{1}^{\prime }l_{1}}\right]
^{k_{l_{1}^{\prime }}}}\left\{ \Psi _{J}^{\otimes k_{l^{\prime }}}\right\}
\right\} 
\end{array}%
\end{equation*}%
if we can define some parallel transport:%
\begin{equation*}
T_{\left\{ v,v+\delta \nu \right\} }:\left[ g,\mathbf{\hat{\Lambda}}_{\infty
}^{\left[ k\right] },\hat{\Psi}_{J}^{\otimes \sum_{i}l_{i}},\lambda \right]
_{\nu }\rightarrow \left[ g,\mathbf{\hat{\Lambda}}_{\infty }^{\left[ k\right]
},\hat{\Psi}_{J}^{\otimes \sum_{i}l_{i}},\lambda \right] _{\nu +\delta \nu }
\end{equation*}%
As seen in the previous paragraphs, this parallel transport should contain a
continuous and a discret part. We saw that the states are described by their
transitions:%
\begin{equation*}
\left[ g,\mathbf{\hat{\Lambda}}_{\infty }^{\left[ k\right] },\hat{\Psi}%
_{J}^{\otimes \sum_{i}l_{i}},\lambda \right] _{\nu }\rightarrow \left[
g+\delta g,\mathbf{\hat{\Lambda}}_{\infty }^{\left[ k\right] },\delta
\Lambda ,\hat{\Psi}_{J}^{\otimes \sum_{i}l_{i}},\delta \hat{\Psi},\lambda
+\delta \lambda \right] _{\nu +\delta \nu }
\end{equation*}%
Locally in first approximation for additive action functional, we can
consider that $\nu $, $\mathbf{\hat{\Lambda}}_{\infty }^{\left[ k\right] }$,
are kept constant, so that $\delta \lambda $ can be considerd as exogeneous
parameter and that:%
\begin{eqnarray*}
&&\left[ g+\delta g,\mathbf{\hat{\Lambda}}_{\infty }^{\left[ k\right]
},\delta \Lambda ,\hat{\Psi}_{J}^{\otimes \sum_{i}l_{i}},\delta \hat{\Psi}%
,\lambda +\delta \lambda \right] _{\nu +\delta \nu } \\
&=&\left[ g,\mathbf{\hat{\Lambda}}_{\infty }^{\left[ k\right] },\hat{\Psi}%
_{J}^{\otimes \sum_{i}l_{i}},\lambda +\delta \lambda \right] _{\nu }\otimes
\psi _{\delta g}\left( \delta \Lambda \right) 
\end{eqnarray*}%
where the tensor product with the field $\psi _{\delta g}\left( \delta
\Lambda \right) $ represnts the discrete part of the parallel transport. The
field $\psi _{\delta g}\left( \delta \Lambda \right) $ depends on the
constraint $g$ and variations $\delta g$ and $\delta \Lambda $. In this
formula:%
\begin{equation*}
\left[ g,\mathbf{\hat{\Lambda}}_{\infty }^{\left[ k\right] },\hat{\Psi}%
_{J}^{\otimes \sum_{i}l_{i}},\lambda +\delta \lambda \right] _{\nu }
\end{equation*}%
represents the continuous variation obtained by the infinitesimal version of
parallel transport (\ref{Tp}), written $T_{\left\{ \lambda ,\lambda +\delta
\lambda \right\} }$ with apparent parametr $\delta \lambda $ of:%
\begin{equation*}
\left[ g,\mathbf{\hat{\Lambda}}_{\infty }^{\left[ k\right] },\hat{\Psi}%
_{J}^{\otimes \sum_{i}l_{i}},\lambda \right] _{\nu }
\end{equation*}%
with amplitude (\ref{GLP}). Practically this corresponds to describe the
constraint $g$ and $\hat{\Psi}_{J}^{\otimes \sum_{i}l_{i}}$ can be
considered locally as function of $\lambda $. Considering the set of
variations $\delta \nu $, represented by an effective parameter $\lambda $
we have an infinite number of possible variations. This yields the local
landscape:%
\begin{equation*}
\begin{array}{ccc}
\begin{array}{c}
\left[ g,\mathbf{\hat{\Lambda}}_{\infty }^{\left[ k\right] },\hat{\Psi}%
_{J}^{\otimes \sum_{i}l_{i}},\lambda +\delta \lambda _{3}\right] _{\nu } \\ 
{\small \otimes }\left( \psi _{\delta g}\left( \delta \Lambda \right)
\right) _{3}%
\end{array}
& 
\begin{array}{c}
\left[ g,\mathbf{\hat{\Lambda}}_{\infty }^{\left[ k\right] },\hat{\Psi}%
_{J}^{\otimes \sum_{i}l_{i}},\lambda +\delta \lambda _{3}^{\prime }\right]
_{\nu } \\ 
{\small \otimes }\left( \psi _{\delta g}\left( \delta \Lambda \right)
\right) _{3}^{\prime }%
\end{array}
& ... \\ 
& \overset{\delta \lambda _{3}}{\nwarrow }{\small \qquad \qquad \qquad
\qquad \qquad \overset{\delta \lambda _{3}^{\prime }}{\uparrow }\qquad
\qquad \nearrow \qquad \qquad \qquad \!\!\!\!\!\!\!\!\!\!\!\!\!\!\!\!\!\!\!}
& \uparrow  \\ 
\begin{array}{c}
\left[ g,\mathbf{\hat{\Lambda}}_{\infty }^{\left[ k\right] },\hat{\Psi}%
_{J}^{\otimes \sum_{i}l_{i}},\lambda +\delta \lambda _{2}\right] _{\nu } \\ 
{\small \otimes }\left( \psi _{\delta g}\left( \delta \Lambda \right)
\right) _{2}%
\end{array}
& \overset{\delta \lambda _{2}}{{\large \leftarrow }\qquad }\left[ g,\mathbf{%
\hat{\Lambda}}_{\infty }^{\left[ k\right] },\hat{\Psi}_{J}^{\otimes
\sum_{i}l_{i}},\lambda \right] _{\nu }{\small \qquad }\overset{\delta
\lambda _{1}}{{\large \rightarrow }}{\small \!\!} & 
\begin{array}{c}
\left[ g,\mathbf{\hat{\Lambda}}_{\infty }^{\left[ k\right] },\hat{\Psi}%
_{J}^{\otimes \sum_{i}l_{i}},\lambda +\delta \lambda _{1}\right] _{\nu } \\ 
{\small \otimes }\left( \psi _{\delta g}\left( \delta \Lambda \right)
\right) _{1}%
\end{array}
\\ 
& {\small \qquad \qquad \qquad \downarrow \delta \lambda }_{4}{\small \qquad
\qquad \qquad \qquad \qquad }\overset{\delta \lambda _{5}}{{\small %
\!\!\!\!\!\!\!\!\!\!\!\!\!\!}\searrow } & {\small \downarrow \delta \lambda }%
_{5}{\small -\delta \lambda }_{1} \\ 
& 
\begin{array}{c}
\left[ g,\mathbf{\hat{\Lambda}}_{\infty }^{\left[ k\right] },\hat{\Psi}%
_{J}^{\otimes \sum_{i}l_{i}},\lambda +\delta \lambda _{4}\right] _{\nu } \\ 
{\small \otimes }\left( \psi _{\delta g}\left( \delta \Lambda \right)
\right) _{4}%
\end{array}
& 
\begin{array}{c}
\left[ g,\mathbf{\hat{\Lambda}}_{\infty }^{\left[ k\right] },\hat{\Psi}%
_{J}^{\otimes \sum_{i}l_{i}},\lambda +\delta \lambda _{5}\right] _{\nu } \\ 
{\small \otimes }\left( \psi _{\delta g}\left( \delta \Lambda \right)
\right) _{1}%
\end{array}%
\end{array}%
\end{equation*}

which describes a "vacuum" or reference state $\left[ g,\mathbf{\hat{\Lambda}%
}_{\infty }^{\left[ k\right] },\hat{\Psi}_{J}^{\otimes
\sum_{i}l_{i}},\lambda \right] _{\nu }$ with space and metrics constrained
by the state $\nu $ and the felds. The surrounding landscap corresponds to
modify this state while $\lambda $ is modified. As described above the
arrows are overweighted for the positive variations $\delta \lambda $. Thus
we can consider statistical combinations of the various possible
modifications for increasing $\lambda $. The statistical variations are
described by fields:%
\begin{equation*}
\psi _{\delta g}\left( \delta \Lambda ,\delta \lambda \right) 
\end{equation*}%
that involve the space variables $\left( \mathbf{\hat{\Lambda}}_{\infty }^{%
\left[ k\right] },\lambda \right) $ arising from the discrete part of the
transformation $\nu \rightarrow \nu +\delta \nu $. Using intrinsic
coordinates, so tht $\mathbf{\hat{\Lambda}}_{\infty }^{\left[ k\right] }$
refers to: $\widetilde{\left[ 
\begin{array}{c}
\left( p_{l_{i}},p_{l_{i}^{\prime }l_{i}},k_{l_{i}^{\prime }}\right)  \\ 
\left[ \left\{ \Psi _{J}\left[ l^{\prime },0\right] \right\} ,v,\delta \nu %
\right] 
\end{array}%
\right] }$ along with some metrics implied by the constraint on group of
points:%
\begin{equation}
\sum_{n}d\widetilde{\left[ 
\begin{array}{c}
\left( p_{l_{i}},p_{l_{i}^{\prime }l_{i}},k_{l_{i}^{\prime }}\right)  \\ 
\left[ \left\{ \Psi _{J}\left[ l^{\prime },0\right] \right\} ,\delta \nu %
\right] 
\end{array}%
\right] }_{n}\delta g_{n}\left( \widetilde{\left[ 
\begin{array}{c}
\left( p_{l_{i}},p_{l_{i}^{\prime }l_{i}},k_{l_{i}^{\prime }}\right)  \\ 
\left[ \left\{ \Psi _{J}\left[ l^{\prime },0\right] \right\} ,v,\delta \nu %
\right] 
\end{array}%
\right] }\right) d\widetilde{\left[ 
\begin{array}{c}
\left( p_{l_{i}},p_{l_{i}^{\prime }l_{i}},k_{l_{i}^{\prime }}\right)  \\ 
\left[ \left\{ \Psi _{J}\left[ l^{\prime },0\right] \right\} ,\delta \nu %
\right] 
\end{array}%
\right] }_{n}  \label{Mtr}
\end{equation}%
This term represents the metric for any group of points arising in the
perturbation $\delta \nu $. The sum is over the various components of sm
flag manifold, the lowest component of such manifold defining what
corresponds to som usual physical space. For small $\delta \nu $, the metric 
$\delta g$ corresponds to usual metrics for points, or several states of
points. Here $\delta g$ is not a small variation of the constraint $g$, but
rather additional constraint for the supplementary points, that is
supplementary generator of degeneracy. We can consider that $\delta g$ is
rather the componnts of the constraint $g$ restricted to the new points.

The fact that the constraint is modified along arrows, or statistical
combinations of arrows, justifies that, considering $\left( \delta \Lambda
,\delta \lambda \right) $ as exogeneous parameters in a first approximation,
the apparent dynamics can be described by some field action functional for
fields and metric, the metric satisfying field dependent dynamics.

However, globally, the state cannot be considered as unchanged, and:%
\begin{equation*}
\left[ g,\mathbf{\hat{\Lambda}}_{\infty }^{\left[ k\right] },\hat{\Psi}%
_{J}^{\otimes \sum_{i}l_{i}},\lambda \right] _{\nu }\rightarrow \left[ g,%
\mathbf{\hat{\Lambda}}_{\infty }^{\left[ k\right] },\hat{\Psi}_{J}^{\otimes
\sum_{i}l_{i}},\lambda ^{\prime }\right] _{\nu ^{\prime }}
\end{equation*}%
cannot be considered as the evolution of systems:%
\begin{equation*}
\left\{ \psi _{\delta g}\left( \delta \Lambda \right) \right\} \rightarrow
\left\{ \psi _{\delta g}\left( \delta \Lambda \right) \right\} ^{\prime }
\end{equation*}%
since the cloud space, that is the parameter space, itself evolves with $\nu 
$ and invalidates globally the identification:%
\begin{equation*}
\left\{ \left[ 
\begin{array}{c}
\left( p_{l_{1}},p_{l_{1}^{\prime }l_{1}},k_{l_{1}^{\prime }}\right)  \\ 
\left[ \left\{ \Psi _{J}\left[ l^{\prime },0\right] \right\} ,v\right] 
\end{array}%
\right] \right\} \rightarrow \mathbf{\hat{\Lambda}}_{\infty }^{\left[ k%
\right] }
\end{equation*}

Thus, the change in cloud space:%
\begin{equation*}
\left[ \left\{ \left[ 
\begin{array}{c}
\left( p_{l_{1}},p_{l_{1}^{\prime }l_{1}},k_{l_{1}^{\prime }}\right)  \\ 
\left[ \left\{ \Psi _{J}\left[ l^{\prime },0\right] \right\} ,v\right] 
\end{array}%
\right] \right\} ,\lambda \right] _{\nu }\rightarrow \left[ \left\{ \left[ 
\begin{array}{c}
\left( p_{l_{1}},p_{l_{1}^{\prime }l_{1}},k_{l_{1}^{\prime }}\right)  \\ 
\left[ \left\{ \Psi _{J}\left[ l^{\prime },0\right] \right\} ,v^{\prime }%
\right] 
\end{array}%
\right] \right\} ,\lambda ^{\prime }\right] _{\nu ^{\prime }}
\end{equation*}%
should matter, even at the effective level, in the apparent transition in
which $\lambda $ is modified to $\lambda ^{\prime }$.

The cloud part of the action (\ref{Pr}) or its local expansion (\ref{Ps})
should take place within the apparent dynamic. The cloud should be
considered as a dynamic object when considered from variation of $\lambda $.
We saw that transitions:%
\begin{equation*}
\lambda \rightarrow \lambda ^{\prime }
\end{equation*}%
favours an increase in the number of states. The transition are most likely
for an increase in: 
\begin{equation*}
Vol\left( \left\{ \left[ 
\begin{array}{c}
\left( p_{l_{1}},p_{l_{1}^{\prime }l_{1}},k_{l_{1}^{\prime }}\right) \\ 
\left[ \left\{ \Psi _{J}\left[ l^{\prime },0\right] \right\} ,v\right]%
\end{array}%
\right] \right\} \right)
\end{equation*}%
that is for an increased $\lambda $. This means that the action $S_{0}$
defind n (\ref{CLh}) for the cloud should not be hermitian when considerd
from the point of view of apparent dynamics.

To connect this approach with the usual framework, we can still work with
the coordinates $\left[ \mathbf{\hat{\Lambda}}_{\infty }^{\left[ k\right]
},\lambda \right] $. The impact of the cloud could be evaluated by writing
the equation for the cloud field $\Psi \left( \left[ \mathbf{\hat{\Lambda}}%
_{\infty }^{\left[ k\right] },\lambda \right] ,v\right) $, and then account
for this dynamics as a modification for the constraint:%
\begin{equation}
0=\frac{\delta S_{g}\left( g\left( \left[ \mathbf{\hat{\Lambda}}_{\infty }^{%
\left[ k\right] },\lambda \right] \right) \right) }{\delta g\left( \left[ 
\mathbf{\hat{\Lambda}}_{\infty }^{\left[ k\right] },\lambda \right] \right) }%
+\frac{\delta S\left( \Psi \left( \left[ \mathbf{\hat{\Lambda}}_{\infty }^{%
\left[ k\right] },\lambda \right] ,v\right) \right) }{\delta g\left( \left[ 
\mathbf{\hat{\Lambda}}_{\infty }^{\left[ k\right] },\lambda \right] \right) }%
+\frac{\delta S_{0}\left( \Psi \left( \left[ \mathbf{\hat{\Lambda}}_{\infty
}^{\left[ k\right] },\lambda \right] ,v\right) \right) }{\delta g\left( %
\left[ \mathbf{\hat{\Lambda}}_{\infty }^{\left[ k\right] },\lambda \right]
\right) }  \label{Qtg}
\end{equation}%
The cloud field should itself satisfies some saddle point equation:%
\begin{equation}
0=\frac{\delta S_{0}\left( \Psi \left( \left[ \mathbf{\hat{\Lambda}}_{\infty
}^{\left[ k\right] },\lambda \right] ,v\right) \right) }{\delta \Psi \left( %
\left[ \mathbf{\hat{\Lambda}}_{\infty }^{\left[ k\right] },\lambda \right]
,v\right) }+\frac{\delta S\left( \hat{\Psi}_{J}^{\otimes
\sum_{i}l_{i}}\left( \left\{ U_{j}^{l_{i}}\right\} _{i},\mathbf{\hat{\Lambda}%
}_{\infty }^{\left[ k\right] },v\right) \right) }{\delta \Psi \left( \left[ 
\mathbf{\hat{\Lambda}}_{\infty }^{\left[ k\right] },\lambda \right]
,v\right) }  \label{Qtk}
\end{equation}%
since $\hat{\Psi}_{J}^{\otimes \sum_{i}l_{i}}\left( \left\{
U_{j}^{l_{i}}\right\} _{i},\mathbf{\hat{\Lambda}}_{\infty }^{\left[ k\right]
},v\right) $ and $\Psi \left( \left[ \mathbf{\hat{\Lambda}}_{\infty }^{\left[
k\right] },\lambda \right] ,v\right) $ may interact through (\ref{Ctns}).
The last equation (\ref{Qtk}) allows to replace $\Psi \left( \left[ \mathbf{%
\hat{\Lambda}}_{\infty }^{\left[ k\right] },\lambda \right] ,v\right) $ by
some function of $\hat{\Psi}_{J}^{\otimes \sum_{i}l_{i}}\left( \left\{
U_{j}^{l_{i}}\right\} _{i},\mathbf{\hat{\Lambda}}_{\infty }^{\left[ k\right]
},v\right) $, where, necessarily, the indices $\left\{ U_{j}^{l_{i}}\right\}
_{i}$ will be summed. This leads to consider that the equation (\ref{Qtg})
for $g\left( \left[ \mathbf{\hat{\Lambda}}_{\infty }^{\left[ k\right]
},\lambda \right] \right) $ should include some supplementary terms
translating the impact of the cloud effective action in terms of \ "standard
fields" $\hat{\Psi}_{J}^{\otimes \sum_{i}l_{i}}\left( \left\{
U_{j}^{l_{i}}\right\} _{i},\mathbf{\hat{\Lambda}}_{\infty }^{\left[ k\right]
},v\right) $.

\section{References}

\begin{enumerate}
\item Y. Nambu (1950). "Force Potentials in Quantum Field Theory". Progress
of Theoretical Physics. 5 (4): 614

\item Gell-Mann, M., \& Low, F. "Bound States in Quantum Field Theory."
Physical Review, vol. 84, no. 2, 1951, pp. 350-355

\item Salpeter, E. E. "Mass Corrections to the Fine Structure of
Hydrogen-Like Atoms." Physical Review, vol. 87, no. 2, 1952, pp. 328-343.{}

\item Nakanishi, N. "A General Survey of the Theory of the Bethe-Salpeter
Equation." Progress of Theoretical Physics Supplement, vol. 43, 1969, pp.
1-81.{}

\item Lucha, W., Sch\"{o}berl, F. F., and Gromes, D. "Bound States of
Quarks." Physics Reports, vol. 200, no. 4, 1991, pp. 127-240.{}

\item Gross, F. "Three-Dimensional Covariant Integral Equations for
Low-Energy Systems." Physics Review, vol. 186, no. 5, 1969, pp. 1448-1462.{}

\item Salpeter, E. E., and Bethe, H. A. "A Relativistic Equation for
Bound-State Problems." Physical Review, vol. 84, no. 6, 1951, pp.
1232-1242.{}

\item Itzykson, C., and Zuber, J.-B. Quantum Field Theory. McGraw-Hill,
1980.{}

\item Greiner, W., and Reinhardt, J. Quantum Electrodynamics. Springer,
1994.{}

\item Roberts, C. D., and Williams, A. G. "Dyson-Schwinger Equations and
Their Application to Hadronic Physics." Progress in Particle and Nuclear
Physics, vol. 33, 1994, pp. 477-575.{}

\item Maris, P., and Roberts, C. D. "Dyson-Schwinger Equations: A Tool for
Hadron Physics." International Journal of Modern Physics E, vol. 12, no. 2,
2003, pp. 297-365.{}

\item Fritzsch, H., Gell-Mann, M., and Leutwyler, H. "Advantages of the
Color Octet Gluon Picture." Physics Letters B, vol. 47, no. 4, 1973, pp.
365-368.{}

\item Politzer, H. D. "Asymptotic Freedom: An Approach to Strong
Interactions." Physics Reports, vol. 14, no. 4, 1974, pp. 129-180.{}

\item Fritzsch, H., and Gell-Mann, M. "Current Algebra: Quarks and What
Else?" eConf C720906V2, 1972, pp. 135-165.{}

\item Gross, D. J., and Wilczek, F. "Ultraviolet Behavior of Non-Abelian
Gauge Theories." Physical Review Letters, vol. 30, no. 26, 1973, pp.
1343-1346.{}

\item Politzer, H. D. "Reliable Perturbative Results for Strong
Interactions?" Physical Review Letters, vol. 30, no. 26, 1973, pp.
1346-1349.{}

\item Shifman, M. A., Vainshtein, A. I., and Zakharov, V. I. "QCD and
Resonance Physics. Theoretical Foundations." Nuclear Physics B, vol. 147,
no. 5-6, 1979, pp. 385-447.{}

\item Svetitsky, B., and Yaffe, L. G. "Critical Behavior at Finite
Temperature Confinement Transitions." Nuclear Physics B, vol. 210, no. 4,
1982, pp. 423-447.{}

\item Fukugita, M., and Ukawa, A. "Deconfining and Chiral Transitions of
Finite Temperature Quantum Chromodynamics in the Presence of Dynamical
Quarks." Physical Review Letters, vol. 57, no. 4, 1986, pp. 503-506.{}

\item Brambilla, N., et al. "QCD and Strongly Coupled Gauge Theories:
Challenges and Perspectives." The European Physical Journal C, vol. 74, no.
10, 2014, article 2981.

\item Wilson, K. G. "Non-Lagrangian Models of Current Algebra." Physical
Review, vol. 179, no. 5, 1969, pp. 1499-1512.{}

\item Wilson, K. G."Operator-Product Expansions and Anomalous Dimensions in
the Thirring Model." Physical Review D, vol. 2, no. 8, 1970, pp. 1473-1483.

\item Zimmermann, W. "Composite Operators in the Perturbation Theory of
Renormalizable Interactions." Annals of Physics, vol. 77, no. 2, 1973, pp.
570-601.{}

\item Shifman, M. A., Vainshtein, A. I., and Zakharov, V. I. "QCD and
Resonance Physics: Applications." Nuclear Physics B, vol. 147, no. 5-6,
1979, pp. 448-518. Cet article applique l'OPE \`{a} la chromodynamique
quantique (QCD) pour \'{e}tudier la physique des r\'{e}sonances et des \'{e}%
tats li\'{e}s.

\item Novikov, V. A., Shifman, M. A., Vainshtein, A. I., and Zakharov, V. I.
"Wilson's Operator Product Expansion: Can It Fail?" Nuclear Physics B, vol.
249, no. 3, 1985, pp. 445-471.{}

\item Les auteurs examinent les limites potentielles de l'OPE et discutent
de sa validit\'{e} dans diff\'{e}rentes situations physiques.{}

\item Braaten, E., and Nieto, A. "Effective Field Theory Approach to High
Temperature Thermodynamics." Physical Review D, vol. 51, no. 12, 1995, pp.
6990-7006.{}

\item Bagan, E., Steele, T. G., and Towner, M. D. "Operator Product
Expansion Beyond Perturbation Theory: Gluon Condensate Contributions to the
Heavy Quark Potential." Physical Review D, vol. 63, no. 3, 2001, article
034018.{}

\item Khodjamirian, A., and Ruckl, R. "QCD Sum Rules for Heavy Flavor
Physics." Advances in Ser. Direct. High Energy Physics, vol. 15, 1998, pp.
345-401.{}

\item Colangelo, P., and Khodjamirian, A. "QCD Sum Rules: A Modern
Perspective." At the Frontier of Particle Physics: Handbook of QCD, vol. 3,
2001, pp. 1495-1576.{}

\item Buchalla, G., and Cat\`{a}, O. "Effective Theory of a Dynamically
Broken Electroweak Standard Model at NLO." Journal of High Energy Physics,
vol. 2012, article 108.{}

\item Chetyrkin, K. G., and Maier, A. "Wilson Expansion of QCD Propagators
at Three Loops: Operators of Dimension Four and Five." Journal of High
Energy Physics, vol. 2010, article 125.

\item Wang, Zhi-Yong. "The (1,0)+(0,1) spinor description of the photon
field and its preliminary applications." arXiv preprint arXiv:1508.02321,
2015.

\item Wu, Xiang-Yao, et al. "Relativistic spinor equation of photon." arXiv
preprint arXiv:1401.0555, 2014.

\item Wang, Ruo Peng."The spinor field theory of the photon." arXiv preprint
arXiv:1109.3237, 2011.

\item Saito, Shinichi. "Spin of Photons: Nature of Polarisation." arXiv
preprint arXiv:2303.17112, 2023.

\item Su, Baoxia. "The spinor equation for the electromagnetic field." arXiv
preprint quant-ph/0011061, 2000.

\item Armour Jr, Rollin S. "Spin-1/2 Maxwell Fields." arXiv preprint
hep-th/0305084, 2003.

\item Chern, Tong. "Solving Maxwell's Equations." arXiv preprint
arXiv:2403.11181, 2024.

\item Duan, Yi-shi, Liu, Xin, et Fu, Li-bin. "Spinor Decomposition of SU(2)
Gauge Potential and The Spinor Structures of Chern-Simons and Chern
Density." Duan Yi-Shi et al 2003 Commun. Theor. Phys. 40 447.

\item Duan, Yi-shi, Ren, Ji-Rong, et Li, Ran. "Vector and Spinor
Decomposition of SU(2) Gauge Potential, Their Equivalence and Knot Structure
in SU(2) Chern-Simons Theory." Commun. Theor. Phys.47:875,2007.
\end{enumerate}

\pagebreak

\section*{Appendix 1: solutions of the saddle points equations and effective
theory}

We solve formally the saddle point equation for $S$. The results allow to
describe the effective field theory by replacing the results in the field
functionals.

\subsection*{Solution of the saddle point}

We study the minimization of the action functional for $\Psi _{I,\alpha
}^{\otimes k}$. 
\begin{equation*}
\exp \left( -S\left( \left\{ \Psi _{J,\alpha }^{\otimes l}\left(
U_{j}^{l}\right) \underset{l,k}{\otimes }\Psi _{I,\alpha }^{\otimes k}\left(
U_{i}^{k}\right) \right\} _{\alpha },\left\{ v_{\left\{ U_{i}^{k}\right\}
}\left( \Psi _{J,\alpha }^{\otimes k}\right) \otimes \Psi _{I,\alpha
}^{\otimes k}\right\} _{\left\{ U_{i}^{k}\right\} }\right) \right)
\end{equation*}%
The tensor products:%
\begin{equation*}
\Psi _{J}^{\otimes l}\left( U_{j}^{l}\right) \underset{l,k}{\otimes }\Psi
_{I}^{\otimes k}\left( U_{i}^{k}\right)
\end{equation*}%
and:%
\begin{equation*}
v_{\left\{ U_{i}^{k}\right\} }\left( \left\{ \Psi _{J}^{\otimes l}\right\}
\right) \otimes \Psi _{I}^{\otimes k}\left( U_{i}^{k}\right)
\end{equation*}%
will stand implicitely for:%
\begin{equation*}
\Psi _{J,\alpha }^{\otimes l}\left( U_{j}^{l}\right) \underset{l,k}{\otimes }%
\Psi _{I_{\alpha }}^{\otimes k}\left( U_{i}^{k}\right)
\end{equation*}%
and:%
\begin{equation*}
v_{U_{i}^{k}}\left\{ \Psi _{J,\alpha }^{\otimes k}\right\} \otimes \Psi
_{I,\alpha }^{\otimes k}\left( U_{i}^{k}\right)
\end{equation*}%
The indices $\alpha $ will be reintroduced later.

We will write the solutions of the equations:%
\begin{equation*}
\left\{ \Psi _{J}^{\otimes l}\left( U_{j}^{l}\right) \tbigoplus \Psi
_{I,0}^{\otimes k}\left( U_{i}^{k}\right) \right\} _{\substack{ l\leqslant m 
\\ l^{\prime }\leqslant m^{\prime }}}\equiv \left\{ \Psi _{I,0}^{\otimes
k}\left( U_{i}^{k}\right) \right\} _{\Psi _{J}^{\otimes l}\left(
U_{j}^{l}\right) }
\end{equation*}%
Writing the saddle point equation as:%
\begin{equation*}
\frac{\delta }{\delta \Psi _{I,\alpha }^{\otimes k}\left( U_{i}^{k}\right) }%
S\left( \left\{ \Psi _{J,\alpha }^{\otimes l}\left( U_{j}^{l}\right) 
\underset{l,k}{\otimes }\Psi _{I,\alpha }^{\otimes k}\left( U_{i}^{k}\right)
\right\} _{\alpha },\left\{ v_{\left\{ U_{i}^{k}\right\} }\left( \Psi
_{J,\alpha }^{\otimes k}\right) \otimes \Psi _{I,\alpha }^{\otimes
k}\right\} _{\left\{ U_{i}^{k}\right\} }\right) =0
\end{equation*}%
we assume that $S$ is a series in the variables:%
\begin{equation*}
\left\{ \Psi _{J,\alpha }^{\otimes l}\left( U_{j}^{l}\right) \underset{l,k}{%
\otimes }\Psi _{I,\alpha }^{\otimes k}\left( U_{i}^{k}\right) \right\}
_{\alpha }
\end{equation*}%
including a quadratic term in the $\Psi _{I,\alpha }^{\otimes k}\left(
U_{i}^{k}\right) $, and in the functionals:%
\begin{equation*}
\left\{ v_{\left\{ U_{i}^{k}\right\} }\left( \Psi _{J,\alpha }^{\otimes
k}\right) \otimes \Psi _{I,\alpha }^{\otimes k}\right\} _{\left\{
U_{i}^{k}\right\} }
\end{equation*}%
so that the saddle point equation for $\Psi _{I,\alpha }^{\otimes k}$ has
the form:%
\begin{eqnarray}
0 &=&\sum_{\substack{ s,\beta _{1},...,\beta _{s}  \\ t,\beta _{1}^{\prime
},...,\beta _{t}^{\prime }}}\tprod_{i}d\left(
U_{j}^{l_{i}}/f_{k_{i}l_{i}}\right) d\left( U_{i}^{k_{i}}\right)
\tprod_{i^{\prime }}d\left( U_{i^{\prime }}^{k_{i^{\prime }}}\right)
\label{S} \\
&&\times \mathcal{G}_{0}\left( U_{i}^{k},\left\{ U_{j,\beta
_{r}}^{l}/f_{kl},U_{i,\beta _{r}}^{k}\right\} ,\left\{ U_{i,\beta
_{r^{\prime }}^{\prime }}^{k}\right\} _{i^{\prime }},\left\{ \Psi _{J,\alpha
^{\prime }}^{\otimes l}\left( U_{j}^{l}\right) \underset{l,k}{\otimes }\Psi
_{I,\alpha ^{\prime }}^{\otimes k}\left( U_{i}^{k}\right) \right\} _{\alpha
^{\prime }\neq \alpha },\left\{ v_{\left\{ U_{i}^{k}\right\} }\left( \Psi
_{J,\alpha ^{\prime }}^{\otimes k}\right) \otimes \Psi _{I,\alpha ^{\prime
}}^{\otimes k}\right\} _{\left\{ U_{i}^{k}\right\} ,\alpha ^{\prime }\neq
\alpha }\right)  \notag \\
&&\times \left( \Psi _{J_{\alpha }}^{\otimes l}\left( U_{j,\beta
_{1}}^{l}/f_{kl},U_{i}^{k}\right) \tprod_{r=2}^{s}\Psi _{J_{\alpha
}}^{\otimes l}\left( U_{j,\beta _{r}}^{l}/f_{kl},U_{i,\beta _{r}}^{k}\right)
\Psi _{I,\alpha }^{\otimes k}\left( U_{i,\beta _{r}}^{k}\right)
\tprod_{r^{\prime }=1}^{t}v_{\left\{ U_{i,\beta _{r^{\prime }}^{\prime
}}^{k}\right\} }\left( \left\{ \Psi _{J,\alpha }^{\otimes l}\right\} \right)
\Psi _{I,\alpha }^{\otimes k}\left( U_{i,\beta _{r^{\prime }}^{\prime
}}^{k}\right) \right.  \notag \\
&&\left. +v_{\left\{ U_{i}^{k}\right\} }\left( \left\{ \Psi _{J,\alpha
}^{\otimes l}\right\} \right) \tprod_{r=1}^{s}\Psi _{J_{\alpha }}^{\otimes
l}\left( U_{j,\beta _{r}}^{l}/f_{kl},U_{i,\beta _{r}}^{k}\right) \Psi
_{I,\alpha }^{\otimes k}\left( U_{i,\beta _{r}}^{k}\right) \tprod_{r^{\prime
}=2}^{t}v_{\left\{ U_{i,\beta _{r^{\prime }}^{\prime }}^{k}\right\} }\left(
\left\{ \Psi _{J,\alpha }^{\otimes l}\right\} \right) \Psi _{I,\alpha
}^{\otimes k}\left( U_{i,\beta _{r^{\prime }}^{\prime }}^{k}\right) \right) 
\notag
\end{eqnarray}%
where $\mathcal{G}_{0}$ is a kernel depending on $\left\{ \Psi _{J,\alpha
^{\prime }}^{\otimes l}\left( U_{j}^{l}\right) \underset{l,k}{\otimes }\Psi
_{I,\alpha ^{\prime }}^{\otimes k}\left( U_{i}^{k}\right) \right\} _{\alpha
^{\prime }\neq \alpha }$. We assume that this kernel includes some delta
functions:%
\begin{equation*}
\delta \left( U_{i}^{k}-U_{i,\beta _{r}}^{k}\right)
\end{equation*}%
and:%
\begin{equation*}
\delta \left( U_{i}^{k}-U_{i,\beta _{r^{\prime }}^{\prime }}^{k}\right)
\end{equation*}%
to account for the possibility of some usual quadratic, or higher order,
terms in the action:%
\begin{equation*}
\int \left( \Psi _{J,\alpha }^{\otimes l}\left( U_{j}^{l}\right) \underset{%
l,k}{\otimes }\Psi _{I,\alpha }^{\otimes k}\left( U_{i}^{k}\right) \right)
^{n}dU_{j}^{l}dU_{i}^{k}/f_{kl}
\end{equation*}%
This equation (\ref{S}) can be solved recursively for each series expansion:

\begin{eqnarray*}
&&\Psi _{I,0,\alpha }^{\otimes k}\left( U_{i}^{k},\left\{ \Psi _{J}^{\otimes
l}\right\} \right) \\
&=&\sum_{\substack{ s,l_{1},...,l_{s},  \\ \alpha _{i},...,\alpha s}}%
\tprod_{i}d\left( U_{j}^{l_{i}}/f_{k_{i}l_{i}}\right) d\left(
U_{i}^{k_{i}}\right) \tprod_{i^{\prime }}d\left( U_{i^{\prime
}}^{k_{i^{\prime }}}\right) \\
&&\times \mathcal{K}_{0}\left( U_{i}^{k},\left\{
U_{i}^{k_{i}},U_{j}^{l_{i}}/f_{k_{i}l_{i}}\right\} _{i},\left\{
U_{i}^{k_{i^{\prime }}}\right\} _{i^{\prime }},\left\{ \alpha _{i}\right\}
\right) \tprod \Psi _{J_{\alpha _{i}}}^{\otimes l_{i}}\left(
U_{j}^{l_{i}}/f_{k_{i}l_{i}},U_{i}^{k_{i}}\right) \tprod v_{\left\{
U_{i^{\prime }}^{k_{i^{\prime }}}\right\} }\left( \left\{ \Psi _{J,\alpha
_{k}}^{\otimes k}\right\} \right)
\end{eqnarray*}%
We assume that $\mathcal{K}_{0}$ is independent from the set $\left\{ \alpha
_{i}\right\} $, that is:%
\begin{equation*}
\mathcal{K}_{0}\left( U_{i}^{k},\left\{
U_{i}^{k_{i}},U_{j}^{l_{i}}/f_{k_{i}l_{i}}\right\} _{i},\left\{
U_{i}^{k_{i^{\prime }}}\right\} _{i^{\prime }},\left\{ \alpha _{i}\right\}
\right) =\mathcal{K}_{0}\left( U_{i}^{k},\left\{
U_{i}^{k_{i}},U_{j}^{l_{i}}/f_{k_{i}l_{i}}\right\} _{i},\left\{
U_{i}^{k_{i^{\prime }}}\right\} _{i^{\prime }}\right)
\end{equation*}%
In the sequel, the realizations are understood. We also write:%
\begin{eqnarray}
&&\Psi _{I,0}^{\otimes k}\left( U_{i}^{k},\left\{ \Psi _{J}^{\otimes
l}\right\} \right)  \label{SDN} \\
&=&\sum_{s,l_{1},...,l_{s}}\int \tprod_{i}d\left(
U_{j}^{l_{i}}/f_{k_{i}l_{i}}\right) d\left( U_{i}^{k_{i}}\right)
\tprod_{i^{\prime }}d\left( U_{i^{\prime }}^{k_{i^{\prime }}}\right)  \notag
\\
&&\times \mathcal{K}_{0}^{v}\left( U_{i}^{k},\left\{
U_{i}^{k_{i}},U_{j}^{l_{i}}/f_{k_{i}l_{i}}\right\} _{i},\left\{
U_{i}^{k_{i^{\prime }}},U_{j}^{l_{i^{\prime }}}/f_{k_{i^{\prime
}}l_{i^{\prime }}}\right\} _{i^{\prime }}\right) \tprod_{i}\Psi
_{J}^{\otimes l_{i}}\left( U_{j}^{l_{i}}/f_{k_{i}l_{i}},U_{i}^{k_{i}}\right)
\tprod_{i^{\prime }}\Psi _{J}^{\otimes k}\left( U_{j}^{l_{i^{\prime
}}}/f_{k_{i^{\prime }}l_{i^{\prime }}},U_{i^{\prime }}^{k_{i^{\prime
}}}\right)  \notag
\end{eqnarray}%
where:%
\begin{equation}
\mathcal{K}_{0}^{v}\left( U_{i}^{k},\left\{
U_{i}^{k_{i}},U_{j}^{l_{i}}/f_{k_{i}l_{i}}\right\} _{i},\left\{
U_{i}^{k_{i^{\prime }}},U_{j}^{l_{i^{\prime }}}/f_{k_{i^{\prime
}}l_{i^{\prime }}}\right\} _{i^{\prime }}\right) =\mathcal{K}_{0}^{v}\left(
U_{i}^{k},\left\{ U_{i}^{k_{i}},U_{j}^{l_{i}}/f_{k_{i}l_{i}}\right\}
_{i}\right) \tprod_{i^{\prime }}v\left( U_{i}^{k_{i^{\prime
}}},U_{j}^{l_{i^{\prime }}}/f_{k_{i^{\prime }}l_{i^{\prime }}}\right)
\label{KV}
\end{equation}%
Changing the notation where $i,i^{\prime }$ is replaced by $i$, formula (\ref%
{SDN}) writes: 
\begin{eqnarray}
&&\Psi _{I,0}^{\otimes k}\left( U_{i}^{k},\left\{ \Psi _{J}^{\otimes
l}\right\} \right)  \label{PSR} \\
&=&\sum_{s,l_{i},...,l_{s}}\tprod_{i}d\left(
U_{j}^{l_{i}}/f_{k_{i}l_{i}}\right) d\left( U_{i}^{k_{i}}\right) \mathcal{K}%
_{0}^{v}\left( U_{i}^{k},\left\{
U_{i}^{k_{i}},U_{j}^{l_{i}}/f_{k_{i}l_{i}}\right\} _{i}\right)
\tprod_{i}\Psi _{J}^{\otimes l_{i}}\left(
U_{j}^{l_{i}}/f_{k_{i}l_{i}},U_{i}^{k_{i}}\right)  \notag
\end{eqnarray}%
\bigskip By changing variables in the functionals:%
\begin{eqnarray*}
&&\Psi _{I,0}^{\otimes k}\left( U_{i}^{k},\left\{ \Psi _{J}^{\otimes
l}\right\} \right) \\
&=&\sum_{s,l_{i},...,l_{s}}d\left( \left\{ U_{j}^{l_{i}}\right\}
_{l_{i}}\right) d\left( \left\{ U_{i}^{k_{i}}/f_{k_{i}l_{i}}\right\}
_{k_{i},l_{i}}\right) \tprod_{i}\Psi _{J}^{\otimes l_{i}}\left(
U_{j}^{l_{i}}\right) \mathcal{K}_{0}^{v}\left( U_{i}^{k},\left\{
U_{j}^{l_{i}}\right\} _{l_{i}},\left\{ U_{i}^{k_{i}}/f_{k_{i}l_{i}}\right\}
_{i}\right)
\end{eqnarray*}%
Reintroducing the realization indices this becomes:%
\begin{eqnarray}
&&\Psi _{I,0,\alpha }^{\otimes k}\left( U_{i}^{k},\left\{ \Psi _{J}^{\otimes
l}\right\} \right)  \label{BCG} \\
&=&\sum_{s,l_{i},...,l_{s}}\int d\left( \left\{ U_{j}^{l_{i}}\right\}
_{l_{i}}\right) d\left( \left\{ U_{i}^{k_{i}}/f_{k_{i}l_{i}}\right\}
_{k_{i},l_{i}}\right) \mathcal{K}_{0}^{v}\left( U_{i}^{k},\left\{
U_{j}^{l_{i}}\right\} _{l_{i}},\left\{ U_{i}^{k_{i}}/f_{k_{i}l_{i}}\right\}
_{i}\right) \left( \sum_{\left\{ \alpha _{i}^{\prime }\right\}
_{_{i\leqslant s}}}\tprod_{i}\Psi _{J,\alpha _{i}^{\prime }}^{\otimes
l_{i}}\left( U_{j}^{l_{i}}\right) \right)  \notag
\end{eqnarray}%
where:%
\begin{equation*}
\Psi _{J,\alpha }^{\otimes \sum_{i\leqslant n}l_{i}}\left(
U_{j}^{l_{i}}\right) =\sum_{\left\{ \alpha _{i}^{\prime }\right\}
_{_{i\leqslant s}}}\tprod_{i}\Psi _{J,\alpha _{i}^{\prime }}^{\otimes
l_{i}}\left( U_{j}^{l_{i}}\right)
\end{equation*}%
The dependency in $\alpha $ is justified by the fact that the action $S$ is
such that an element of $\left\{ \alpha _{i}^{\prime }\right\}
_{_{i\leqslant n}}$ is $\alpha $ or arises in the action in products
involving $\alpha $.

\subsection*{Projected functional}

Introducing (\ref{BCG}) into the functionals of the form:%
\begin{equation*}
\int g\left( v,U_{i}^{k}\right) v_{\left\{ U_{i}^{k}\right\} }\left( \left\{
\Psi _{J}^{\otimes l}\right\} \right) \Psi _{I}^{\otimes k}\left(
U_{i}^{k}\right) dvdU_{i}^{k}
\end{equation*}%
is performed first by evaluating these functionals on states $ev_{\left(
U_{i}^{l_{i}}\right) }\left( \left\{ \Psi _{J}^{\otimes l_{i}}\right\}
\right) $ that replaces $\Psi _{J}^{\otimes l_{i}}$ by its value $\Psi
_{J}^{\otimes l_{i}}\left( U_{i}^{l_{i}}\right) $. Then multiply by $v\left(
U_{i}^{k_{i}},U_{j}^{l_{i}}/f_{k_{i}l_{i}}\right) $. and integrate over $%
d\left( \left\{ U_{j}^{l_{i}}\right\} _{l_{i}}\right) $ to recover $%
v_{\left\{ U_{i}^{k}\right\} }\left( \left\{ \Psi _{J}^{\otimes l}\right\}
\right) $.

Practically this amounts to remove the integrals over $d\left( \left\{
U_{j}^{l_{i}}\right\} _{l_{i}}\right) $ in (\ref{BCG}) and consider: 
\begin{equation*}
\mathcal{K}_{0}^{v}\left( U_{i}^{k},\left\{
U_{i}^{k_{i}},U_{j}^{l_{i}}/f_{k_{i}l_{i}}\right\} _{i}\right)
\tprod_{i^{\prime }}v\left( U_{i}^{k_{i^{\prime }}},U_{j}^{l_{i^{\prime
}}}/f_{k_{i^{\prime }}l_{i^{\prime }}}\right)
\end{equation*}%
then, we replace the kernel in (\ref{KV}) by:%
\begin{eqnarray}
&&\mathcal{K}_{0}^{v}\left( U_{i}^{k},\left\{
U_{i}^{k_{i}},U_{j}^{l_{i}}/f_{k_{i}l_{i}}\right\} _{i},\left\{
U_{i}^{k_{i^{\prime }}},U_{j}^{l_{i^{\prime }}}/f_{k_{i^{\prime
}}l_{i^{\prime }}}\right\} _{i^{\prime }}\right)  \label{VK} \\
&\rightarrow &\mathcal{K}_{0}^{v}\left( U_{i}^{k},\left\{
U_{i}^{k_{i}},U_{j}^{l_{i}}/f_{k_{i}l_{i}}\right\} _{i},\left\{
U_{i}^{k_{i^{\prime }}},U_{j}^{l_{i^{\prime }}}/f_{k_{i^{\prime
}}l_{i^{\prime }}}\right\} _{i^{\prime }}\right) \tprod_{i}v\left(
U_{i}^{k_{i}},U_{j}^{l_{i}}/f_{k_{i}l_{i}}\right)  \notag \\
&=&\mathcal{K}_{0}^{v}\left( U_{i}^{k},\left\{
U_{i}^{k_{i}},U_{j}^{l_{i}}/f_{k_{i}l_{i}}\right\} _{i}\right)
\tprod_{i}v\left( U_{i}^{k_{i}},U_{j}^{l_{i}}/f_{k_{i}l_{i}}\right)
\tprod_{i^{\prime }}v\left( U_{i}^{k_{i^{\prime }}},U_{j}^{l_{i^{\prime
}}}/f_{k_{i^{\prime }}l_{i^{\prime }}}\right)  \notag
\end{eqnarray}%
and then resatblish integrals over $d\left( \left\{ U_{j}^{l_{i}}\right\}
_{l_{i}}\right) $.

Then the result is inserted in: 
\begin{equation*}
\int g\left( v,U_{i}^{k}\right) v_{\left\{ U_{i}^{k}\right\} }\left( \left\{
\Psi _{J}^{\otimes l}\right\} \right) \Psi _{I}^{\otimes k}\left(
U_{i}^{k}\right) dvdU_{i}^{k}
\end{equation*}

To write the result, we change the variables arising in the functionals. We
write:%
\begin{equation*}
U_{i}^{k}=\left( \left\{ U_{j}^{l}\right\} ,U^{k}/f_{kl}\right) 
\end{equation*}%
where $\left\{ U_{j}^{l}\right\} $ are\ coordinates in $U_{i}^{k}/\left(
U_{i}^{k}/f_{kl}\right) $. Then we replace:%
\begin{equation}
U_{i}^{k},\left\{ U_{i}^{k_{i}}/f_{k_{i}l_{i}},\right\} _{i}=\left( \left\{
U_{j}^{l}\right\} ,U_{i}^{k_{i}}/f_{k_{i}l_{i}}\right) ,\left\{
U_{i}^{k_{i}}/f_{k_{i}l_{i}}\right\} _{i}\rightarrow \left\{
U_{j}^{l}\right\} ,\left\{ U_{i}^{k_{i}}/f_{k_{i}l_{i}}\right\} _{i}
\label{HC}
\end{equation}

i.e. we include the variable $U^{k}/f_{kl}$ in the set indexed by $i$ (by
including a label $0$ for this variable). Moreover, using (\ref{VK}), we
rewrite:%
\begin{equation*}
\mathcal{K}_{0}^{v}\left( \left\{ U_{j}^{l_{i}}\right\} _{l_{i}},\left\{
U_{i}^{k_{i}}/f_{k_{i}l_{i}}\right\} _{i\leqslant n}\right) =\mathcal{K}%
_{0}\left( \left\{ U_{j}^{l_{i}}\right\} _{l_{i}},\left\{
U_{i}^{k_{i}}/f_{k_{i}l_{i}}\right\} _{i\leqslant n}\right) \tprod v\left(
U_{i}^{k_{i}},U_{j}^{l_{i}}/f_{k_{i}l_{i}}\right)
\end{equation*}%
and the functional becomes ultimately:%
\begin{eqnarray}
&&\int g\left( v,U_{i}^{k}\right) \mathcal{K}_{0}\left( \left\{
U_{j}^{l_{i}}\right\} _{l_{i}},\left\{ U_{i}^{k_{i}}/f_{k_{i}l_{i}}\right\}
_{i\leqslant n}\right) \tprod v\left(
U_{i}^{k_{i}},U_{j}^{l_{i}}/f_{k_{i}l_{i}}\right)  \label{PRC} \\
&&\times \sum_{\left\{ \alpha _{i}^{\prime }\right\} }\tprod_{i}\Psi
_{J,\left\{ \alpha _{i}^{\prime }\right\} }^{\otimes l_{i}}\left(
U_{j}^{l_{i}}\right) \tprod_{i}d\left\{
U_{j}^{l_{i}},U_{i}^{k_{i}}/f_{k_{i}l_{i}},\mathbf{\hat{\Lambda}}_{\alpha }^{%
\left[ k_{i}\right] }\left[ \Psi _{J},\nu \right] \right\}  \notag
\end{eqnarray}

\section*{Appendix 2: degeneracy}

To write the dependency of the saddle point solution we start with the
transformation parameters. We write:%
\begin{equation*}
\left\{ \mathbf{\hat{\Lambda}}_{\left\{ \alpha _{i\leqslant n}\right\} }^{%
\left[ \left\{ k_{i\leqslant n}\right\} \right] }\left[ \Psi _{J},\nu \right]
\right\} 
\end{equation*}%
for the set of parameters of the group $G_{k_{i},...,k_{n}}$ with varying $%
k_{i\leqslant n}$. The full\ set of parameters is:%
\begin{equation*}
\left( \left\{ \mathbf{\hat{\Lambda}}_{\left\{ \alpha _{i\leqslant
n}\right\} }^{\left[ \left\{ k_{i\leqslant n}\right\} \right] }\left[ \Psi
_{J},\nu \right] \right\} \right) _{n}
\end{equation*}%
fr $n$ running from $1$ to $\infty $.

We also note:%
\begin{equation*}
\left( \mathbf{\hat{\Lambda}}_{\left\{ \alpha _{i\leqslant n}\right\} }^{%
\left[ \left\{ k_{i\leqslant n}\right\} \right] }\left[ \Psi _{J},\nu \right]
\right) _{n}
\end{equation*}%
the parameters associated to an infinite given sequence $\left\{
k_{i}\right\} $. We also assume that the eigenvalues of symmetry parameters
satisfy functional relations:%
\begin{equation*}
h_{k_{n}}\left( \left\{ \mathbf{\hat{\Lambda}}_{\left\{ \alpha _{i\leqslant
n}\right\} }^{\left[ \left\{ k_{i\leqslant n}\right\} \right] }\left[ \Psi
_{J},U_{j}^{l},\nu \right] \right\} ,h_{p}\left( \left( \Psi _{J}\right)
,U_{j}^{l},\nu \right) \right) =0
\end{equation*}

We can write the vector kernels of transformations:%
\begin{eqnarray}
\left[ \mathbf{K}_{\left\{ \alpha _{i}\right\} _{i\leqslant n}}\left( \Psi
_{J}^{\otimes l}\right) \right] _{\left( U_{i}^{k_{i}}\right) ^{\prime
}}^{U_{i}^{k_{i}}} &=&\left[ \mathbf{K}_{i\leqslant n}\left( \Psi
_{J}^{\otimes l}\right) \right] _{\left( U_{i}^{k_{i}}\right) ^{\prime
}}^{U_{i}^{k_{i}}}  \label{gn1} \\
&=&\sum_{l_{i},\left\{ \alpha _{i}^{\prime }\right\} _{_{i\leqslant n}}}%
\mathbf{k}\left( \left\{ U_{j}^{l_{i}}\right\} _{i},U_{i}^{k_{i}},\left(
U_{i}^{k_{i}}\right) ^{\prime }\right) \tprod \Psi _{J,\alpha _{i}^{\prime
}}^{\otimes l_{i}}\left( U_{j}^{l_{i}}\right)  \notag \\
&=&\sum_{l_{i}}\mathbf{k}\left( \left\{ U_{j}^{l_{i}}\right\} _{i},\left\{
U_{i}^{k_{i}}\right\} ,\left\{ \left( U_{i}^{k_{i}}\right) ^{\prime
}\right\} _{i}\right) \sum_{\left\{ \alpha _{i}^{\prime }\right\}
_{_{i\leqslant n}}}\tprod \Psi _{J,\alpha _{i}^{\prime }}^{\otimes
l_{i}}\left( U_{j}^{l_{i}}\right)  \notag
\end{eqnarray}%
since the generators are the sum for all realizations arising in the action.
The kernel $\mathbf{k}$ has components dual to $\left\{ \mathbf{\hat{\Lambda}%
}_{\left\{ \alpha _{i\leqslant n}\right\} }^{\left[ \left\{ k_{i\leqslant
n}\right\} \right] }\left[ \Psi _{J},\nu \right] \right\} $. We note $%
\mathbf{L}_{_{i\leqslant n}}\left( \Psi _{J}^{\otimes l}\right) $ the
operator with kernel $\mathbf{K}_{_{i\leqslant n}}\left( \Psi _{J}^{\otimes
l}\right) $.

The solution is given by the group action:%
\begin{eqnarray}
&&\left\{ \Psi _{I,0,\alpha }^{\otimes k_{i}}\left( U_{i}^{k_{i}},\left(
\left\{ \mathbf{\hat{\Lambda}}_{\left\{ \alpha _{i\leqslant n}\right\} }^{%
\left[ \left\{ k_{i\leqslant n}\right\} \right] }\left[ \Psi _{J},\nu \right]
\right\} \right) _{n},\left\{ \Psi _{J}^{\otimes l}\right\} \right) \right\}
\label{TRN} \\
&=&\left\{ \exp \left( i\sum \mathbf{L}_{i\leqslant n}\left( \Psi
_{J}^{\otimes l}\right) .\left\{ \mathbf{\hat{\Lambda}}_{\left\{ \alpha
_{i\leqslant n}\right\} }^{\left[ \left\{ k_{i\leqslant n}\right\} \right] }%
\left[ \Psi _{J},\nu \right] \right\} \right) \Psi _{I,0,\alpha }^{\otimes
k_{i}}\left( U_{i}^{k_{i}},\left\{ \Psi _{J}^{\otimes l}\right\} \right)
\right\}  \notag
\end{eqnarray}%
The dependency in $\left( \Psi _{J}^{\otimes l}\right) $ kept implicit. To
isolate one of the solution:%
\begin{equation*}
\Psi _{I,0}^{\otimes k}\left( U_{i}^{k},\left( \left\{ \mathbf{\hat{\Lambda}}%
_{\alpha _{i}}^{\left[ k_{i}\right] }\left[ \Psi _{J},\nu \right] \right\}
_{i\leqslant n}\right) _{n},\left\{ \Psi _{J}^{\otimes l}\right\} \right)
\end{equation*}%
we write:%
\begin{eqnarray*}
&&\Psi _{I,0,\alpha }^{\otimes k}\left( U_{i}^{k},\left( \left\{ \mathbf{%
\hat{\Lambda}}_{\left\{ \alpha _{i\leqslant n}\right\} }^{\left[ \left\{
k,k_{i\leqslant n-1}\right\} \right] }\left[ \Psi _{J},\nu \right] \right\}
\right) _{n},\left\{ \Psi _{J}^{\otimes l}\right\} \right) \\
&=&\sum_{k_{i}}\int d\left( U_{i}^{k_{i}}\right) ^{\prime }\left[ \exp
\left( i\mathbf{L}\left( \Psi _{J}^{\otimes l}\right) .\left\{ \mathbf{\hat{%
\Lambda}}_{\mathbf{\alpha }}^{\left[ \mathbf{k}\right] }\left[ \Psi _{J},\nu %
\right] \right\} \right) \right] _{\left( U_{i}^{k_{i}}\right) ^{\prime
}}^{U_{i}^{k}}\Psi _{I,0}^{\otimes k}\left( \left( U_{i}^{k_{i}}\right)
^{\prime },\left\{ \Psi _{J}^{\otimes l}\right\} \right) \\
&=&\tprod\limits_{n}\sum_{k_{i\leqslant n}}\int d\left( U_{i}^{k_{i}}\right)
^{\prime }\left[ \exp \left( i\mathbf{L}_{i\leqslant n}\left( \Psi
_{J}^{\otimes l}\right) .\left\{ \mathbf{\hat{\Lambda}}_{\left\{ \alpha
_{i\leqslant n}\right\} }^{\left[ \left\{ k,k_{i\leqslant n-1}\right\} %
\right] }\left[ \Psi _{J},\nu \right] \right\} \right) \right] _{\left(
U_{i}^{k_{i}}\right) ^{\prime }}^{U_{i}^{k}}\Psi _{I,0,\alpha }^{\otimes
k}\left( \left( U_{i}^{k_{i}}\right) ^{\prime },\left\{ \Psi _{J}^{\otimes
l}\right\} \right)
\end{eqnarray*}%
and we will use the notatn $\left( \left\{ \mathbf{\hat{\Lambda}}_{\alpha }^{%
\left[ \left\{ k,k_{i\leqslant n-1}\right\} \right] }\left[ \Psi _{J},\nu %
\right] \right\} \right) _{n}$ for the set $\left( \left\{ \mathbf{\hat{%
\Lambda}}_{\left\{ \alpha _{i\leqslant n}\right\} }^{\left[ \left\{
k,k_{i\leqslant n-1}\right\} \right] }\left[ \Psi _{J},\nu \right] \right\}
\right) _{n}$ arising in the realization $\Psi _{I,\alpha ,0}^{\otimes k}$ .
This notation is valid, since the domain of the variables $\mathbf{\hat{%
\Lambda}}_{\alpha }^{\left[ k_{i}\right] }$ is the same for all
realizations. Consequently, the field writes:%
\begin{equation*}
\Psi _{I,0,\alpha }^{\otimes k}\left( U_{i}^{k},\left( \left\{ \mathbf{\hat{%
\Lambda}}_{\alpha }^{\left[ \left\{ k,k_{i\leqslant n-1}\right\} \right] }%
\left[ \Psi _{J},\nu \right] \right\} \right) _{n},\left\{ \Psi
_{J}^{\otimes l}\right\} \right)
\end{equation*}

\subsection*{Particular case}

For transformation groups that satisfies:%
\begin{equation*}
G_{k_{i},...,k_{n}}\subset G_{k_{i},...,k_{n+1}}
\end{equation*}%
we can assume that:%
\begin{equation*}
\mathbf{\hat{\Lambda}}_{\alpha }^{\left[ \left\{ k,k_{i\leqslant
n-1}\right\} \right] }\left[ \Psi _{J},\nu \right] \subset \mathbf{\hat{%
\Lambda}}_{\alpha }^{\left[ \left\{ k,k_{i\leqslant n}\right\} \right] }%
\left[ \Psi _{J},\nu \right]
\end{equation*}%
so that for given sequence $\left\{ k_{i}\right\} $:%
\begin{equation*}
\left( \mathbf{\hat{\Lambda}}_{\alpha }^{\left[ \left\{ k,k_{i\leqslant
n-1}\right\} \right] }\left[ \Psi _{J},\nu \right] \right) _{n}
\end{equation*}%
is an infinite dimension flag manifold, starting with $\mathbf{\hat{\Lambda}}%
_{\alpha }^{\left[ k\right] }\left[ \Psi _{J},\nu \right] $. The entire set
of parameters:%
\begin{equation*}
\left( \left\{ \mathbf{\hat{\Lambda}}_{\alpha }^{\left[ \left\{
k,k_{i\leqslant n-1}\right\} \right] }\left[ \Psi _{J},\nu \right] \right\}
\right) _{n}
\end{equation*}%
is thus an infinite number of flag manifolds, all starting at $\mathbf{\hat{%
\Lambda}}_{\alpha }^{\left[ k\right] }\left[ \Psi _{J},\nu \right] $.

To simplify we may assume that:%
\begin{equation*}
\mathbf{\hat{\Lambda}}_{\alpha }^{\left[ \left\{ k_{i\leqslant n}\right\} %
\right] }\left[ \Psi _{J},\nu \right] =\left\{ \mathbf{\hat{\Lambda}}%
_{\alpha }^{\left[ k_{i}\right] }\left[ \Psi _{J},\nu \right] \right\}
_{k_{1},..k_{n}}
\end{equation*}%
and that the groups parameters are constrained such that:%
\begin{equation*}
\left( \left\{ \mathbf{\hat{\Lambda}}_{\alpha }^{\left[ \left\{
k,k_{i\leqslant n-1}\right\} \right] }\left[ \Psi _{J},\nu \right] \right\}
\right) _{n}=\left\{ \mathbf{\hat{\Lambda}}_{\alpha }^{\left[ k\right] }%
\left[ \Psi _{J},\nu \right] ,\mathbf{\hat{\Lambda}}_{\alpha }^{\left[ k_{i}%
\right] }\left[ \Psi _{J},\nu \right] \right\} _{i}
\end{equation*}%
keeping in mind that this represents a distinguished set of points bearing
some properties of $U_{i}^{k}$. This cloud of points is a background space.
For such parameters, the constraints write:%
\begin{equation*}
h_{k_{n}}\left( \left\{ \mathbf{\hat{\Lambda}}_{\alpha _{i}}^{\left[ k_{i}%
\right] }\left[ \Psi _{J},U_{j}^{l},\nu \right] \right\} _{i\leqslant
n},h_{p}\left( \left( \Psi _{J}\right) ,U_{j}^{l},\nu \right) \right) =0
\end{equation*}

Note ultimately that in general: 
\begin{equation*}
\mathbf{\hat{\Lambda}}_{\alpha }^{\left[ k_{i}\right] }\left[ \Psi _{J},\nu %
\right] =\left( \mathbf{\hat{\Lambda}}_{\alpha }^{k_{i}}\left[ \Psi _{J},\nu %
\right] \right)
\end{equation*}

\subsection*{Restriction including constraints}

Since we have to consider functionals:%
\begin{equation*}
\Psi _{J}^{\otimes l}\left( U_{j}^{l}\right) \Psi _{I}^{\otimes k}\left(
U_{i}^{k}\right) \delta \left( f_{lk}\left( U_{j}^{l},U_{i}^{k}\right)
\right)
\end{equation*}%
we have to restrict $\Psi _{I}^{\otimes k}\left( U_{i}^{k}\right) $ to $%
\delta \left( f_{lk}\left( U_{j}^{l},U_{i}^{k}\right) \right) $:%
\begin{equation*}
\Psi _{I}^{\otimes k}\left( U_{i}^{k}\right) \rightarrow \Psi _{I}^{\otimes
k}\left( U_{i}^{k}\right) \delta \left( f_{lk}\left(
U_{j}^{l},U_{i}^{k}\right) \right)
\end{equation*}%
and the same applies to the $U_{i}^{k_{i}}$:%
\begin{equation*}
\Psi _{I}^{\otimes k_{i}}\left( U_{i}^{k_{i}}\right) \rightarrow \Psi
_{I}^{\otimes k_{i}}\left( U_{i}^{k_{i}}\right) \delta \left( \left\{
f_{l_{i}k_{i}}\left( U_{j}^{l_{i}},U_{i}^{k_{i}}\right) \right\} \right)
\end{equation*}%
The notation $\delta \left( \left\{ f_{l_{i}k_{i}}\left(
U_{j}^{l_{i}},U_{i}^{k_{i}}\right) \right\} \right) $ stands for the product:%
\begin{equation*}
\tprod\limits_{l_{i}}\delta \left( f_{l_{i}k_{i}}\left(
U_{j}^{l_{i}},U_{i}^{k_{i}}\right) \right)
\end{equation*}%
that is we take into account the whole constraints $f_{l_{i}k_{i}}\left(
U_{j}^{l_{i}},U_{i}^{k_{i}}\right) $ for $l_{i}$ variables.

This means that transformations (\ref{TRN}) have to be restricted to
projected fields:%
\begin{eqnarray}
&&\left\{ \Psi _{I,0,\alpha }^{\otimes k_{i}}\left( U_{i}^{k_{i}},\left(
\left\{ \mathbf{\hat{\Lambda}}_{\left\{ \alpha _{i\leqslant n}\right\} }^{%
\left[ \left\{ k_{i\leqslant n}\right\} \right] }\left[ \Psi _{J},\nu \right]
\right\} \right) _{n},\left\{ \Psi _{J}^{\otimes l}\right\} \right) \delta
\left( \left\{ f_{l_{i}k_{i}}\left( U_{j}^{l_{i}},U_{i}^{k_{i}}\right)
\right\} \right) \right\} \\
&=&\left\{ \exp \left( i\sum \mathbf{L}_{i\leqslant n}\left( \Psi
_{J}^{\otimes l}\right) .\left\{ \mathbf{\hat{\Lambda}}_{\left\{ \alpha
_{i\leqslant n}\right\} }^{\left[ \left\{ k_{i\leqslant n}\right\} \right] }%
\left[ \Psi _{J},\nu \right] \right\} \right) \Psi _{I,0,\alpha }^{\otimes
k_{i}}\left( U_{i}^{k_{i}},\left\{ \Psi _{J}^{\otimes l}\right\} \right)
\delta \left( \left\{ f_{l_{i}k_{i}}\left(
U_{j}^{l_{i}},U_{i}^{k_{i}}\right) \right\} \right) \right\}  \notag
\end{eqnarray}%
so that we consider transformations that commute with the projections $%
\delta \left( f_{lk}\left( U_{j}^{l},U_{i}^{k}\right) \right) $.

These constraints are not independent. Assuming the form:%
\begin{equation}
f_{k_{i}}\left( U_{i}^{k_{i}}\right) =f_{l_{i}}\left( U_{j}^{l_{i}}\right)
\label{NCR}
\end{equation}%
or more generally:%
\begin{equation}
f_{k_{i}}^{\left( \alpha \right) }\left( U_{i}^{k_{i}}\right)
=f_{l_{i}}\left( \left\{ U_{j}^{l_{i}}\right\} _{\alpha }\right)  \label{MCR}
\end{equation}%
for $\alpha $ running over finite set, such that $\cup _{\alpha }\left\{
U_{j}^{l_{i}}\right\} _{\alpha }=\left\{ U_{j}^{l_{i}}\right\} $, we can
replace:%
\begin{equation*}
\delta \left( \left\{ f_{l_{i}k_{i}}\left(
U_{j}^{l_{i}},U_{i}^{k_{i}}\right) \right\} \right) =\tprod\limits_{\alpha
}\delta \left( f_{l_{i}k_{i}}^{\left( \alpha \right) }\left(
U_{j}^{l_{i}},U_{i}^{k_{i}}\right) \right)
\end{equation*}%
and the rest of constraints are compatibility between the $U_{j}^{l_{i}}$.

Choosing $\alpha =1$, for simplicity, we have:%
\begin{equation}
\delta \left( \left\{ f_{l_{i}k_{i}}\left(
U_{j}^{l_{i}},U_{i}^{k_{i}}\right) \right\} \right) =\delta \left(
f_{l_{i}k_{i}}\left( U_{j}^{l_{i}},U_{i}^{k_{i}}\right) \right)  \label{SRC}
\end{equation}%
for some given $l_{i}$.

Due to the constraint, the parameters $\left\{ \mathbf{\hat{\Lambda}}%
_{\left\{ \alpha _{i\leqslant n}\right\} }^{\left[ \left\{ k_{i\leqslant
n}\right\} \right] }\left[ \Psi _{J},\nu \right] \right\} $ are restricted
to subsets:%
\begin{equation*}
\left\{ \mathbf{\hat{\Lambda}}_{\left\{ \alpha _{i\leqslant n}\right\} }^{%
\left[ \left\{ k_{i\leqslant n}\right\} \right] }\left[ \Psi _{J},\nu \right]
\right\} _{\delta \left( f_{l_{i}k_{i}}\left(
U_{j}^{l_{i}},U_{i}^{k_{i}}\right) \right) }\subset \left\{ \mathbf{\hat{%
\Lambda}}_{\left\{ \alpha _{i\leqslant n}\right\} }^{\left[ \left\{
k_{i\leqslant n}\right\} \right] }\left[ \Psi _{J},\nu \right] \right\}
\end{equation*}%
Practically, when $\Psi _{I,0}^{\otimes k_{i}}\left( U_{i}^{k_{i}}\right) $
is inserted in (\ref{PRC}) this subset of parameters is submitted in the $s$%
-th term of the sum to $\ s$ constraints due to the products:%
\begin{equation*}
\tprod_{i=1}^{s}\Psi _{J,\left\{ \alpha _{i}^{\prime }\right\} }^{\otimes
l_{i}}\left( U_{j}^{l_{i}}\right)
\end{equation*}%
and this corresponds to consider parameters of transformation that write:%
\begin{equation*}
\left\{ \mathbf{\hat{\Lambda}}_{\left\{ \alpha _{i\leqslant n}\right\} }^{%
\left[ \left\{ k_{i\leqslant n}\right\} \right] }\left[ \Psi _{J},\nu
,\left\{ U_{j}^{l_{i\leqslant s}}\right\} \right] \right\}
\end{equation*}%
where the constraints $\delta \left( f_{l_{i}k_{i}}\left(
U_{j}^{l_{i}},U_{i}^{k_{i}}\right) \right) $ have been solved by introducing
the dependency in $\left\{ U_{j}^{l_{i\leqslant s}}\right\} $ and by
choosing particular $l_{i}$ in $\left\{ U_{j}^{l_{i\leqslant s}}\right\} $
to implement the $s$ constraints (\ref{NCR}). In the text, given our
assumptions, we will replace:%
\begin{equation*}
\left\{ \mathbf{\hat{\Lambda}}_{\left\{ \alpha _{i\leqslant n}\right\} }^{%
\left[ \left\{ k_{i\leqslant n}\right\} \right] }\left[ \Psi _{J},\nu
,\left\{ U_{j}^{l_{i\leqslant s}}\right\} \right] \right\} \rightarrow
\left\{ \mathbf{\hat{\Lambda}}_{\alpha }^{\left[ \left\{ k_{i\leqslant
n}\right\} \right] }\left[ \Psi _{J},\nu ,\left\{ U_{j}^{l_{i}}\right\} %
\right] \right\}
\end{equation*}%
or:%
\begin{equation*}
\left\{ \mathbf{\hat{\Lambda}}_{\alpha }^{\left[ k,\left\{ k_{i\leqslant
n-1}\right\} \right] }\left[ \Psi _{J},\nu ,\left\{ U_{j}^{l_{i}}\right\} %
\right] \right\}
\end{equation*}%
if any $k$ is distinguished.

\section*{Appendix 3: Projected functional}

\subsection*{Disregarding constraint}

Without constraints in variables, including some degeneracy in solutions (%
\ref{BCG}) corresponds to consider parameters dependent kernels:%
\begin{equation}
\left\{ \mathcal{K}_{0}^{v}\left( \left\{ U_{j}^{l_{i}}\right\}
_{l_{i\leqslant s}},\left\{ U_{i}^{k_{i}}/f_{k_{i}l_{i}}\right\}
_{i\leqslant s},\left( \mathbf{\hat{\Lambda}}_{\alpha }^{\left[ k\right] }%
\left[ \Psi _{J},\nu ,\left\{ U_{j}^{l_{i\leqslant n}}\right\} \right]
,\left\{ \mathbf{\hat{\Lambda}}_{\alpha }^{\left[ \left\{ k_{i\leqslant
n}\right\} \right] }\left[ \Psi _{J},\nu \right] \right\} \right)
_{n}\right) \right\}  \label{KRN}
\end{equation}%
Linear combinations of degenerat background amounts to replace:%
\begin{equation*}
\mathcal{K}_{0}^{v}\left( U_{i}^{k},\left\{
U_{i}^{k_{i}},U_{j}^{l_{i}}/f_{k_{i}l_{i}}\right\} _{i}\right)
\end{equation*}%
in (\ref{PSR}) or (\ref{BCG}) by the following contribution:%
\begin{eqnarray}
&&\Psi _{\alpha }\left( \left( \mathbf{\hat{\Lambda}}_{\alpha }^{\left[ k%
\right] }\left[ \Psi _{J},\nu ,\left\{ U_{j}^{l_{i\leqslant n}}\right\} %
\right] ,\left\{ \mathbf{\hat{\Lambda}}_{\alpha }^{\left[ \left\{
k_{i\leqslant n}\right\} \right] }\left[ \Psi _{J},\nu \right] \right\}
\right) _{n},v\right)  \label{CNN} \\
&&\mathcal{K}_{0}^{v}\left( U_{i}^{k},\left\{ U_{j}^{l_{i}}\right\}
_{l_{i\leqslant s}},\left\{ U_{i}^{k_{i}}/f_{k_{i}l_{i}}\right\}
_{i\leqslant s},\left( \mathbf{\hat{\Lambda}}_{\alpha }^{\left[ k\right] }%
\left[ \Psi _{J},\nu ,\left\{ U_{j}^{l_{i\leqslant n}}\right\} \right]
,\left\{ \mathbf{\hat{\Lambda}}_{\alpha }^{\left[ \left\{ k_{i\leqslant
n}\right\} \right] }\left[ \Psi _{J},\nu \right] \right\} \right) _{n}\right)
\notag
\end{eqnarray}

We change variables arising in the functionals. We write:%
\begin{equation*}
U_{i}^{k}=\left( \left\{ U_{j}^{l}\right\} ,U^{k}/f_{kl}\right)
\end{equation*}%
where $\left\{ U_{j}^{l}\right\} $ are coordinates on $U_{i}^{k}/\left(
U_{i}^{k}/f_{kl}\right) $. Then we replace:%
\begin{eqnarray}
&&U_{i}^{k},\left\{ \mathbf{\hat{\Lambda}}_{\alpha }^{\left[ k\right] }\left[
\Psi _{J},\nu \right] \right\} _{i},\left\{
U_{i}^{k_{i}}/f_{k_{i}l_{i}},\left\{ \mathbf{\hat{\Lambda}}_{\alpha }^{\left[
\left\{ k_{i\leqslant n-1}\right\} \right] }\left[ \Psi _{J},\nu \right]
\right\} _{i}\right\} _{i}  \label{CH} \\
&=&\left( \left\{ U_{j}^{l}\right\} ,U_{i}^{k_{i}}/f_{k_{i}l_{i}}\right)
,\left\{ U_{i}^{k_{i}}/f_{k_{i}l_{i}},\left\{ \mathbf{\hat{\Lambda}}_{\alpha
}^{\left[ \left\{ k_{i\leqslant n}\right\} \right] }\left[ \Psi _{J},\nu %
\right] \right\} _{i}\right\} _{i}\rightarrow \left\{ U_{j}^{l}\right\}
,\left\{ U_{i}^{k_{i}}/f_{k_{i}l_{i}},\mathbf{\hat{\Lambda}}_{\alpha }^{%
\left[ \left\{ k_{i\leqslant n}\right\} \right] }\left[ \Psi _{J},\nu \right]
\right\} _{i}  \notag
\end{eqnarray}%
That is we include the variable $U^{k}/f_{kl}$ in the set indexed by $i$
(this is done by including a label $0$ for this variable). Similarly we
replace:

\begin{equation*}
\mathbf{\hat{\Lambda}}_{\alpha }^{\left[ k\right] }\left[ \Psi _{J},\nu
,\left\{ U_{j}^{l_{i\leqslant n}}\right\} \right] ,\left\{ \mathbf{\hat{%
\Lambda}}_{\alpha }^{\left[ \left\{ k_{i\leqslant n-1}\right\} \right] }%
\left[ \Psi _{J},\nu ,\left\{ U_{j}^{l_{i\leqslant n}}\right\} \right]
\right\} _{i}\rightarrow \left\{ \mathbf{\hat{\Lambda}}_{\alpha }^{\left[
\left\{ k,k_{i\leqslant n-1}\right\} \right] }\left[ \Psi _{J},\nu ,\left\{
U_{j}^{l_{i\leqslant n}}\right\} \right] \right\} _{i}
\end{equation*}%
Moreover, using (\ref{VK}), we rewrite the kernel as:%
\begin{eqnarray*}
&&\mathcal{K}_{0}^{v}\left( \left\{ U_{j}^{l_{i}}\right\} _{l_{i}},\left\{
U_{i}^{k_{i}}/f_{k_{i}l_{i}}\right\} _{i\leqslant n},\left( \left\{ \mathbf{%
\hat{\Lambda}}_{\alpha }^{\left[ \left\{ k,k_{i\leqslant n-1}\right\} \right]
}\left[ \Psi _{J},\nu \right] \right\} \right) _{n}\right)  \\
&=&\mathcal{K}_{0}\left( \left\{ U_{j}^{l_{i}}\right\} _{l_{i}},\left\{
U_{i}^{k_{i}}/f_{k_{i}l_{i}}\right\} _{i\leqslant n},\left( \left\{ \mathbf{%
\hat{\Lambda}}_{\alpha }^{\left[ \left\{ k,k_{i\leqslant n-1}\right\} \right]
}\left[ \Psi _{J},\nu \right] \right\} \right) _{n}\right) \tprod_{i}v\left(
U_{i}^{k_{i}},U_{j}^{l_{i}}/f_{k_{i}l_{i}}\right) 
\end{eqnarray*}%
where $S$ runs over the subset of $\left\{ 1,...,n\right\} $.

As a consequence, in a functional of tensor products:%
\begin{equation*}
\Psi _{J}^{\otimes l}\left( U_{j}^{l}\right) \underset{l,k}{\otimes }\Psi
_{I}^{\otimes k}\left( U_{i}^{k}\right)
\end{equation*}%
a copy for a single realization:%
\begin{equation*}
\int g\left( v,U_{i}^{k}\right) v_{\left\{ U_{i}^{k}\right\} }\left( \left\{
\Psi _{J}^{\otimes l}\right\} \right) \Psi _{I}^{\otimes k}\left(
U_{i}^{k}\right) dvdU_{i}^{k}
\end{equation*}%
is thus replaced by:%
\begin{eqnarray}
&&\int g\left( v,U_{i}^{k}\right) \mathcal{K}_{0}\left( \left\{
U_{j}^{l_{i}}\right\} _{l_{i}},\left\{ U_{i}^{k_{i}}/f_{k_{i}l_{i}}\right\}
_{i\leqslant n},\left( \left\{ \mathbf{\hat{\Lambda}}_{\alpha }^{\left[
\left\{ k,k_{i\leqslant n-1}\right\} \right] }\left[ \Psi _{J},\nu \right]
\right\} \right) _{n}\right) \tprod_{i}v\left(
U_{i}^{k_{i}},U_{j}^{l_{i}}/f_{k_{i}l_{i}}\right)  \label{PRF} \\
&&\times \sum_{\left\{ \alpha _{i}^{\prime }\right\} }\tprod_{i}\Psi
_{J,\left\{ \alpha _{i}^{\prime }\right\} }^{\otimes l_{i}}\left(
U_{j}^{l_{i}}\right) \Psi _{\alpha }\left( \left( \left\{ \mathbf{\hat{%
\Lambda}}_{\alpha }^{\left[ \left\{ k,k_{i\leqslant n-1}\right\} \right] }%
\left[ \Psi _{J},\nu \right] \right\} \right) _{n},v\right)
\tprod_{i}d\left\{ U_{j}^{l_{i}},U_{i}^{k_{i}}/f_{k_{i}l_{i}},\mathbf{\hat{%
\Lambda}}_{\alpha }^{\left[ k_{i}\right] }\left[ \Psi _{J},\nu \right]
\right\}  \notag
\end{eqnarray}

As in the core of the text, we write:%
\begin{equation*}
v_{\left\{ U_{i}^{k}\right\} }\left( \left\{ \Psi _{J,\alpha }^{\otimes
l}\right\} \right) =\int v\left( \left\{ U_{j}^{l}\right\}
,U_{i}^{l}/f_{kl}\right) \Psi _{J,\alpha }^{\otimes l}\left( U_{j}^{l}\right)
\end{equation*}%
and use our previous change of variable (\ref{CH}), to include $\left\{
U_{j}^{l}\right\} ,U_{i}^{l}/f_{kl}$ in the set indexed by $i$. Ultimately,
functionl (\ref{PRF}) rewrites:%
\begin{eqnarray}
&&\int g\left( v,U_{i}^{k}\right) \mathcal{K}_{0}\left( \left\{
U_{j}^{l_{i}}\right\} _{l_{i}},\left\{ U_{i}^{k_{i}}/f_{k_{i}l_{i}}\right\}
_{i\leqslant n},\left( \left\{ \mathbf{\hat{\Lambda}}_{\alpha }^{\left[
\left\{ k,k_{i\leqslant n-1}\right\} \right] }\left[ \Psi _{J},\nu \right]
\right\} \right) _{n}\right) \tprod_{i}v\left(
U_{i}^{k_{i}},U_{j}^{l_{i}}/f_{k_{i}l_{i}}\right)  \label{PR} \\
&&\times \Psi _{J,\alpha }^{\otimes l}\left( U_{j}^{l}\right) \sum_{\left\{
\alpha _{i}^{\prime }\right\} }\tprod_{i}\Psi _{J,\left\{ \alpha
_{i}^{\prime }\right\} }^{\otimes l_{i}}\left( U_{j}^{l_{i}}\right) \Psi
_{\alpha }\left( \left( \left\{ \mathbf{\hat{\Lambda}}_{\alpha }^{\left[
\left\{ k,k_{i\leqslant n-1}\right\} \right] }\left[ \Psi _{J},\nu \right]
\right\} \right) _{n},v\right)  \notag \\
&&\times \tprod_{i}d\left\{
U_{j}^{l_{i}},U_{i}^{k_{i}}/f_{k_{i}l_{i}}\right\} d\left( \left\{ \mathbf{%
\hat{\Lambda}}_{\alpha }^{\left[ \left\{ k,k_{i\leqslant n-1}\right\} \right]
}\left[ \Psi _{J},\nu \right] \right\} \right) _{n}  \notag
\end{eqnarray}

\subsection*{including constraints}

We include the constraints for parameters. We choose $n$ values of $l_{i}$
in $\left\{ U_{j}^{l_{i\leqslant n}}\right\} $ arising in (\ref{PR}) to
implement the $n$ constraints (\ref{NCR}). We then come back to the
derivation of (\ref{PRC}) and remove the integrals over $d\left( \left\{
U_{i}^{k_{i}}/f_{k_{i}l_{i}}\right\} _{l_{i}}\right) $ in (\ref{PRF}) to
write the functionals:%
\begin{eqnarray}
&&\int g\left( v,U_{i}^{k}\right) \mathcal{K}_{0}\left( \left\{
U_{j}^{l_{i}}\right\} _{l_{i}},\left\{ U_{i}^{k_{i}}/f_{k_{i}l_{i}}\right\}
_{i\leqslant n},\left( \left\{ \mathbf{\hat{\Lambda}}_{\alpha }^{\left[
\left\{ k,k_{i\leqslant n-1}\right\} \right] }\left[ \Psi _{J},\nu \right]
\right\} \right) _{n}\right) \tprod_{i}v\left(
U_{i}^{k_{i}},U_{j}^{l_{i}}/f_{k_{i}l_{i}}\right)  \\
&&\times \sum_{\left\{ \alpha _{i}^{\prime }\right\} }\tprod_{i}\Psi
_{J,\left\{ \alpha _{i}^{\prime }\right\} }^{\otimes l_{i}}\left(
U_{j}^{l_{i}}\right) \Psi _{\alpha }\left( \left( \left\{ \mathbf{\hat{%
\Lambda}}_{\alpha }^{\left[ \left\{ k,k_{i\leqslant n-1}\right\} \right] }%
\left[ \Psi _{J},\nu \right] \right\} \right) _{n},v\right)
\tprod_{i}d\left\{ U_{j}^{l_{i}},\mathbf{\hat{\Lambda}}_{\alpha }^{\left[
k_{i}\right] }\left[ \Psi _{J},\nu \right] \right\}   \notag
\end{eqnarray}%
This has projected the functional over the states defined by the set$\left\{
\tprod_{i}\Psi _{J,\left\{ \alpha _{i}^{\prime }\right\} }^{\otimes
l_{i}}\left( U_{j}^{l_{i}}\right) \right\} $. This implies the constraints (%
\ref{SRC}):%
\begin{equation*}
\delta \left( f_{l_{i}k_{i}}\left( U_{j}^{l_{i}},U_{i}^{k_{i}}\right)
\right) 
\end{equation*}%
and the restriction:%
\begin{equation*}
\Psi _{I}^{\otimes k_{i}}\left( U_{i}^{k_{i}}\right) \delta \left( \left\{
f_{l_{i}k_{i}}\left( U_{j}^{l_{i}},U_{i}^{k_{i}}\right) \right\} \right) 
\end{equation*}%
can be imposed by using the variables $U_{j}^{l_{i}}$, so that we write the
parameters dependency: 
\begin{equation*}
\left( \left\{ \mathbf{\hat{\Lambda}}_{\alpha }^{\left[ \left\{
k,k_{i\leqslant n-1}\right\} \right] }\left[ \Psi _{J},\nu \right] ,\left\{
U_{j}^{l_{i\leqslant n}}\right\} \right\} \right) _{n}
\end{equation*}%
and functional (\ref{PR}) rewrites:%
\begin{eqnarray}
&&\int g\left( v,U_{i}^{k}\right) \mathcal{K}_{0}\left( \left\{
U_{j}^{l_{i}}\right\} _{l_{i}},\left\{ U_{i}^{k_{i}}/f_{k_{i}l_{i}}\right\}
_{i\leqslant n},\left( \left\{ \mathbf{\hat{\Lambda}}_{\alpha }^{\left[
\left\{ k,k_{i\leqslant n-1}\right\} \right] }\left[ \Psi _{J},\nu ,\left\{
U_{j}^{l_{i\leqslant n}}\right\} \right] \right\} \right) _{n}\right)
\sum_{S,0\in S}\tprod_{i\in S}v\left(
U_{i}^{k_{i}},U_{j}^{l_{i}}/f_{k_{i}l_{i}}\right)   \label{PRT} \\
&&\times \Psi _{J,\alpha }^{\otimes l}\left( U_{j}^{l}\right) \sum_{\left\{
\alpha _{i}^{\prime }\right\} }\tprod_{i}\Psi _{J,\left\{ \alpha
_{i}^{\prime }\right\} }^{\otimes l_{i}}\left( U_{j}^{l_{i}}\right) \Psi
_{\alpha }\left( \left( \left\{ \mathbf{\hat{\Lambda}}_{\alpha }^{\left[
\left\{ k,k_{i\leqslant n-1}\right\} \right] }\left[ \Psi _{J},\nu ,\left\{
U_{j}^{l_{i\leqslant n}}\right\} \right] \right\} \right) _{n},v\right)  
\notag \\
&&\times \tprod_{i}d\left\{
U_{j}^{l_{i}},U_{i}^{k_{i}}/f_{k_{i}l_{i}}\right\} d\left( \left\{ \mathbf{%
\hat{\Lambda}}_{\alpha }^{\left[ \left\{ k,k_{i\leqslant n-1}\right\} \right]
}\left[ \Psi _{J},\nu ,\left\{ U_{j}^{l_{i\leqslant n}}\right\} \right]
\right\} \right) _{n}  \notag
\end{eqnarray}

We choose the form:%
\begin{equation*}
v\left( U_{i}^{k_{i}},U_{j}^{l_{i}}/f_{k_{i}l_{i}}\right) =v\left(
U_{j}^{l_{i}}\right) \delta \left( f_{k_{i}l_{i}}\left(
U_{i}^{k_{i}},U_{j}^{l_{i}}\right) \right)
\end{equation*}%
corresponding to current-current type interactions. As a consequence,
defining as before:%
\begin{equation*}
\Psi _{J,\alpha }^{\otimes \sum_{i\leqslant n}l_{i}}\left(
U_{j}^{l_{i}}\right) =\sum_{\left\{ \alpha _{i}^{\prime }\right\}
}\tprod_{i}\Psi _{J,\left\{ \alpha _{i}^{\prime }\right\} }^{\otimes
l_{i}}\left( U_{j}^{l_{i}}\right)
\end{equation*}%
as well as:%
\begin{eqnarray*}
&&\bar{g}\left( v,\left\{ U_{j}^{l_{i}},\right\} _{i\leqslant n},\left(
\left\{ \mathbf{\hat{\Lambda}}_{\alpha }^{\left[ \left\{ k,k_{i\leqslant
n-1}\right\} \right] }\left[ \Psi _{J},\nu ,\left\{ U_{j}^{l_{i\leqslant
n}}\right\} \right] \right\} \right) _{n}\right) \\
&=&\int g\left( v,U_{i}^{k}\right) \mathcal{K}_{0}\left( \left\{
U_{j}^{l_{i}}\right\} _{l_{i}},\left\{ U_{i}^{k_{i}}/f_{k_{i}l_{i}}\right\}
_{i\leqslant n},\left( \left\{ \mathbf{\hat{\Lambda}}_{\alpha }^{\left[
\left\{ k,k_{i\leqslant n-1}\right\} \right] }\left[ \Psi _{J},\nu ,\left\{
U_{j}^{l_{i\leqslant n}}\right\} \right] \right\} \right) _{n}\right)
dU_{i}^{k}/f_{kl}\tprod dU_{i}^{k_{i}}/f_{k_{i}l_{i}}
\end{eqnarray*}%
\bigskip

Formula (\ref{PRT}) is given by:%
\begin{eqnarray}
&&\int \bar{g}\left( v,\left\{ U_{j}^{l_{i}}\right\} _{i\leqslant n},\left(
\left\{ \mathbf{\hat{\Lambda}}_{\alpha }^{\left[ \left\{ k,k_{i\leqslant
n-1}\right\} \right] }\left[ \Psi _{J},\nu \right] \right\} \right)
_{n}\right)  \label{FNC} \\
&&\times \sum_{S,0\in S}\tprod_{i\in S}v\left( U_{j}^{l_{i}}\right) \Psi
_{J,\alpha }^{\otimes l}\left( U_{j}^{l}\right) \Psi _{J,\alpha }^{\otimes
\sum_{i\leqslant n}l_{i}}\left( U_{j}^{l_{i}}\right) \Psi _{\alpha }\left(
\left( \left\{ \mathbf{\hat{\Lambda}}_{\alpha }^{\left[ \left\{
k,k_{i\leqslant n-1}\right\} \right] }\left[ \Psi _{J},\nu \right] \right\}
\right) _{n},v\right)  \notag \\
&\equiv &\int \bar{g}\left( v,\left\{ U_{j}^{l_{i}},\mathbf{\hat{\Lambda}}%
_{\alpha }^{\left[ k_{i}\right] }\left[ \Psi _{J},\nu ,U_{j}^{l}\right]
\right\} _{i\leqslant n}\right) v\left( \left( U_{j}^{l_{i}}\right) \right)
\left( \Psi _{J,\alpha }^{\otimes \left( l+\sum_{i}l_{i}\right) }\left(
\left\{ U_{j}^{l_{i}}\right\} _{i},\left( \left\{ \mathbf{\hat{\Lambda}}%
_{\alpha }^{\left[ \left\{ k,k_{i\leqslant n-1}\right\} \right] }\left[ \Psi
_{J},\nu \right] \right\} \right) _{n},v\right) \right)  \notag
\end{eqnarray}%
with:%
\begin{eqnarray*}
&&\Psi _{J,\alpha }^{\otimes \left( l+\sum_{i}l_{i}\right) }\left( \left\{
U_{j}^{l_{i}}\right\} _{i},\left( \left\{ \mathbf{\hat{\Lambda}}_{\alpha }^{%
\left[ \left\{ k,k_{i\leqslant n-1}\right\} \right] }\left[ \Psi _{J},\nu %
\right] \right\} \right) _{n},v\right) \\
&=&\Psi _{J,\alpha }^{\otimes \left( l+\sum_{i}l_{i}\right) }\left(
U_{i}^{l},\left\{ U_{j}^{l_{i}}\right\} \right) \Psi _{\alpha }\left( \left(
\left\{ \mathbf{\hat{\Lambda}}_{\alpha }^{\left[ \left\{ k,k_{i\leqslant
n-1}\right\} \right] }\left[ \Psi _{J},\nu \right] \right\} \right)
_{n},v\right) \\
&=&\Psi _{J,\alpha }^{\otimes l}\left( U_{i}^{l}\right) \Psi _{J,\alpha
}^{\otimes \sum_{i\leqslant n}l_{i}}\left( U_{j}^{l_{i}}\right) \Psi
_{\alpha }\left( \left( \left\{ \mathbf{\hat{\Lambda}}_{\alpha }^{\left[
\left\{ k,k_{i\leqslant n-1}\right\} \right] }\left[ \Psi _{J},\nu \right]
\right\} \right) _{n},v\right)
\end{eqnarray*}%
and:%
\begin{equation*}
v\left( \left( U_{j}^{l_{i}}\right) \right) =\sum_{S,0\in S}\tprod_{i\in
S}v\left( U_{j}^{l_{i}}\right)
\end{equation*}

As a consequence, functional (\ref{FNC}) becomes:%
\begin{equation*}
\int \bar{g}\left( v,\left\{ U_{j}^{l_{i}},\mathbf{\hat{\Lambda}}_{\alpha }^{%
\left[ k_{i}\right] }\left[ \Psi _{J},\nu ,U_{j}^{l}\right] \right\}
_{i}\right) v\left( \left( U_{j}^{l_{i}}\right) \right) \left( \Psi
_{J,\alpha }^{\otimes \left( l+\sum_{i}l_{i}\right) }\left(
U_{j}^{l_{i}},\left( \left\{ \mathbf{\hat{\Lambda}}_{\alpha }^{\left[
\left\{ k,k_{i\leqslant n-1}\right\} \right] }\left[ \Psi _{J},\nu \right]
\right\} \right) _{n},v\right) \right)
\end{equation*}%
Setting $\bar{g}_{v}\rightarrow \bar{g}$ since the functionals are arbitrary
and the $v$ dependency can be absorbed in the definition of $\bar{g}$, this
yields:%
\begin{equation*}
\int v\left( \left( U_{j}^{l_{i}}\right) \right) \bar{g}\left( \left\{
U_{j}^{l_{i}}\right\} _{i},\left( \left\{ \mathbf{\hat{\Lambda}}_{\alpha }^{%
\left[ \left\{ k,k_{i\leqslant n-1}\right\} \right] }\left[ \Psi _{J},\nu %
\right] \right\} \right) _{n}\right) \Psi _{J,\alpha }^{\otimes \left(
\sum_{i}l_{i}\right) }\left( \left\{ U_{j}^{l_{i}}\right\} _{i},\left(
\left\{ \mathbf{\hat{\Lambda}}_{\alpha }^{\left[ \left\{ k,k_{i\leqslant
n-1}\right\} \right] }\left[ \Psi _{J},\nu \right] \right\} \right)
_{n},v\right)
\end{equation*}

In the case where:%
\begin{equation*}
\left( \left\{ \mathbf{\hat{\Lambda}}_{\alpha }^{\left[ \left\{
k,k_{i\leqslant n-1}\right\} \right] }\left[ \Psi _{J},\nu \right] \right\}
\right) _{n}=\left( \left\{ \mathbf{\hat{\Lambda}}^{\left[ \left\{
k,k_{i\leqslant n-1}\right\} \right] }\left[ \Psi _{J},\nu \right] \right\}
\right) _{n}
\end{equation*}%
are independent from $\alpha $, we can sum over the index and we have:%
\begin{eqnarray*}
&&\int v\left( \left( U_{j}^{l_{i}}\right) \right) \bar{g}\left( \left\{
U_{j}^{l_{i}}\right\} _{i},\left( \left\{ \mathbf{\hat{\Lambda}}^{\left[
\left\{ k,k_{i\leqslant n-1}\right\} \right] }\left[ \Psi _{J},\nu \right]
\right\} \right) _{n}\right) \sum_{\alpha }\Psi _{J,\alpha }^{\otimes \left(
\sum_{i}l_{i}\right) }\left( \left\{ U_{j}^{l_{i}}\right\} _{i},\left(
\left\{ \mathbf{\hat{\Lambda}}^{\left[ \left\{ k,k_{i\leqslant n-1}\right\} %
\right] }\left[ \Psi _{J},\nu \right] \right\} \right) _{n},v\right) \\
&\equiv &\int v\left( \left( U_{j}^{l_{i}}\right) \right) \bar{g}\left(
\left\{ U_{j}^{l_{i}}\right\} _{i},\left( \left\{ \mathbf{\hat{\Lambda}}^{%
\left[ \left\{ k,k_{i\leqslant n-1}\right\} \right] }\left[ \Psi _{J},\nu %
\right] \right\} \right) _{n}\right) \Psi _{J}^{\otimes \left(
\sum_{i}l_{i}\right) }\left( \left\{ U_{j}^{l_{i}}\right\} _{i},\left(
\left\{ \mathbf{\hat{\Lambda}}^{\left[ \left\{ k,k_{i\leqslant n-1}\right\} %
\right] }\left[ \Psi _{J},\nu \right] \right\} \right) _{n},v\right)
\end{eqnarray*}%
with:%
\begin{eqnarray*}
&&\Psi _{J}^{\otimes \left( \sum_{i}l_{i}\right) }\left( \left\{
U_{j}^{l_{i}},\right\} _{i},\left( \left\{ \mathbf{\hat{\Lambda}}^{\left[
\left\{ k,k_{i\leqslant n-1}\right\} \right] }\left[ \Psi _{J},\nu \right]
\right\} \right) _{n},v\right) \\
&=&\sum_{\alpha }\Psi _{J,\alpha }^{\otimes \left( \sum_{i}l_{i}\right)
}\left( \left\{ U_{j}^{l_{i}}\right\} _{i},\left( \left\{ \mathbf{\hat{%
\Lambda}}^{\left[ \left\{ k,k_{i\leqslant n-1}\right\} \right] }\left[ \Psi
_{J},\nu \right] \right\} \right) _{n},v\right)
\end{eqnarray*}

Note that if the dependency in:%
\begin{equation*}
\Psi _{J,\alpha }^{\otimes \left( \sum_{i}l_{i}\right) }\left( \left\{
U_{j}^{l_{i}}\right\} _{i},\left( \left\{ \mathbf{\hat{\Lambda}}^{\left[
\left\{ k,k_{i\leqslant n-1}\right\} \right] }\left[ \Psi _{J},\nu \right]
\right\} \right) _{n},v\right)
\end{equation*}%
have the form:%
\begin{equation*}
\int k_{v}\left( U_{j}^{l_{i}}\right) \Psi _{J,\alpha }^{\otimes \left(
\sum_{i}l_{i}\right) }\left( \left\{ U_{j}^{l_{i}}\right\} _{i},\left(
\left\{ \mathbf{\hat{\Lambda}}^{\left[ \left\{ k,k_{i\leqslant n-1}\right\} %
\right] }\left[ \Psi _{J},\nu \right] \right\} \right) _{n}\right)
\end{equation*}%
The functional becomes:%
\begin{equation*}
\int v\left( \left( U_{j}^{l_{i}}\right) \right) \bar{g}\left( \left\{
U_{j}^{l_{i}}\right\} _{i},\left( \left\{ \mathbf{\hat{\Lambda}}^{\left[
\left\{ k,k_{i\leqslant n-1}\right\} \right] }\left[ \Psi _{J},\nu \right]
\right\} \right) _{n}\right) \Psi _{J}^{\otimes \left( \sum_{i}l_{i}\right)
}\left( \left\{ U_{j}^{l_{i}}\right\} _{i},\left( \left\{ \mathbf{\hat{%
\Lambda}}^{\left[ \left\{ k,k_{i\leqslant n-1}\right\} \right] }\left[ \Psi
_{J},\nu \right] \right\} \right) _{n}\right)
\end{equation*}%
with $v\left( \left( U_{j}^{l_{i}}\right) \right) $ rescaled:%
\begin{equation*}
v\left( \left( U_{j}^{l_{i}}\right) \right) \rightarrow \tprod k_{v}\left(
U_{j}^{l_{i}}\right) \sum_{S,0\in S}\tprod_{i\in S}v\left(
U_{j}^{l_{i}}\right)
\end{equation*}

\subsubsection*{Remark}

If we choose for the functionals: 
\begin{equation*}
v_{\left\{ U_{i}^{k}\right\} }^{U_{j}^{l}/f_{lk}}\left\{ \Psi _{J}^{\otimes
l}\right\} =\Psi _{J}^{\otimes l}\left( U_{j}^{l}\right)
\end{equation*}%
that is a density which is a Dirac measure:%
\begin{equation*}
v\left( U_{j}^{l_{i}}\right) =\delta _{v\left( U_{j}^{l_{i}}\right) }
\end{equation*}%
we find:%
\begin{eqnarray*}
&&\int \bar{g}\left( \left\{ U_{j}^{l_{i}}\right\} _{i},\left( \left\{ 
\mathbf{\hat{\Lambda}}^{\left[ \left\{ k,k_{i\leqslant n-1}\right\} \right] }%
\left[ \Psi _{J},\nu \right] \right\} \right) _{n}\right) \Psi _{J}^{\otimes
\left( \sum_{i}l_{i}\right) }\left( \left\{ U_{j}^{l_{i}}\right\}
_{i},\left( \left\{ \mathbf{\hat{\Lambda}}^{\left[ \left\{ k,k_{i\leqslant
n-1}\right\} \right] }\left[ \Psi _{J},\nu \right] \right\} \right)
_{n},v\right) \\
&\equiv &F_{f,lin}\left( \Psi _{J}^{\otimes \sum_{i}l_{i}}\left( \left\{
U_{j}^{l_{i}}\right\} _{i},\left( \left\{ \mathbf{\hat{\Lambda}}^{\left[
\left\{ k,k_{i\leqslant n-1}\right\} \right] }\left[ \Psi _{J},\nu \right]
\right\} \right) _{n},v\right) \right)
\end{eqnarray*}

while if the $\mathbf{\hat{\Lambda}}_{\alpha }^{\left[ \left\{
k,k_{i\leqslant n-1}\right\} \right] }\left[ \Psi _{J},\nu \right] $ not
independent of the copies $\Psi _{J,\alpha }^{\otimes k_{p_{0}}}$,
identification is local.

\subsubsection*{Remark}

In (\ref{KRN}) and (\ref{CNT}), the kernel can be considered as
representation of the $G^{k,\left\{ k_{i\leqslant n-1}\right\} }$ with $%
n\leqslant s$, so that the kernel considered are rather:

\begin{equation}
\left\{ \mathcal{K}_{0}^{v}\left( \left\{ U_{j}^{l_{i}}\right\}
_{l_{i\leqslant s}},\left\{ U_{i}^{k_{i}}/f_{k_{i}l_{i}}\right\}
_{i\leqslant s},\left( \left\{ \mathbf{\hat{\Lambda}}_{\alpha }^{\left[
\left\{ k,k_{i\leqslant n-1}\right\} \right] }\left[ \Psi _{J},\nu \right]
\right\} \right) _{n\leqslant s}\right) \right\}
\end{equation}%
which corresponds to include the following contribution in (\ref{PSR}):%
\begin{equation}
\Psi _{\alpha }\left( \left( \left\{ \mathbf{\hat{\Lambda}}_{\alpha }^{\left[
\left\{ k,k_{i\leqslant n-1}\right\} \right] }\left[ \Psi _{J},\nu \right]
\right\} \right) _{n\leqslant s},v\right) \mathcal{K}_{0}^{v}\left( \left\{
U_{j}^{l_{i}}\right\} _{l_{i\leqslant s}},\left\{
U_{i}^{k_{i}}/f_{k_{i}l_{i}}\right\} _{i\leqslant s},\left( \left\{ \mathbf{%
\hat{\Lambda}}_{\alpha }^{\left[ \left\{ k,k_{i\leqslant n-1}\right\} \right]
}\left[ \Psi _{J},\nu \right] \right\} \right) _{n\leqslant s}\right)
\end{equation}%
and the field is:%
\begin{equation*}
\sum_{\alpha }\Psi _{J,\alpha }^{\otimes \left( \sum_{i=1}^{s}l_{i}\right)
}\left( \left\{ U_{j}^{l_{i}}\right\} _{i\leqslant s},\left( \left\{ \mathbf{%
\hat{\Lambda}}^{\left[ \left\{ k,k_{i\leqslant n-1}\right\} \right] }\left[
\Psi _{J},\nu \right] \right\} \right) _{n\leqslant s},v\right)
\end{equation*}

\section*{Appendix 4: invariance}

We start by assuming a projective invariance:%
\begin{equation*}
\int v\left( \left( U_{j}^{l_{i}}\right) \right) \frac{\delta \mathbf{\hat{%
\Lambda}}_{\alpha }^{\left[ \left\{ k,k_{i\leqslant n-1}\right\} \right] }%
\left[ \Psi _{J},\nu ,U_{j}^{l}\right] }{\delta \Psi _{J,\alpha }^{\otimes
l_{i}}\left( \left( U_{j}^{l_{i}}\right) ^{\prime }\right) }\nabla _{\mathbf{%
\hat{\Lambda}}_{\alpha }^{\left[ \left\{ k,k_{i\leqslant n-1}\right\} \right]
}}\Psi _{J,\alpha }^{\otimes \sum_{i}l_{i}}\left( \left\{
U_{j}^{l_{i}}\right\} _{i},\left( \left\{ \mathbf{\hat{\Lambda}}_{\alpha }^{%
\left[ \left\{ k,k_{i\leqslant n-1}\right\} \right] }\left[ \Psi _{J},\nu
,U_{j}^{l}\right] \right\} \right) _{n}\right)
\end{equation*}%
is not equal to $0$, but satisfies: 
\begin{eqnarray}
&&\int v\left( \left( U_{j}^{l_{i}}\right) \right) \frac{\delta \mathbf{\hat{%
\Lambda}}_{\alpha }^{\left[ \left\{ k,k_{i\leqslant n-1}\right\} \right] }%
\left[ \Psi _{J},\nu ,U_{j}^{l}\right] }{\delta \Psi _{J,\alpha }^{\otimes
l_{i}}\left( \left( U_{j}^{l_{i}}\right) ^{\prime }\right) }\nabla _{\mathbf{%
\hat{\Lambda}}_{\alpha }^{\left[ \left\{ k,k_{i\leqslant n-1}\right\} \right]
}}\Psi _{J,\alpha }^{\otimes \sum_{i}l_{i}}\left( \left\{
U_{j}^{l_{i}}\right\} _{i},\left( \left\{ \mathbf{\hat{\Lambda}}_{\alpha }^{%
\left[ \left\{ k,k_{i\leqslant n-1}\right\} \right] }\left[ \Psi _{J},\nu
,U_{j}^{l}\right] \right\} \right) _{n}\right)  \label{NVC} \\
&=&\int v\left( \left\{ \left( U_{j}^{l_{i}}\right) \right\} \right) \frac{%
\delta ^{\prime }V\left( \left\{ \Psi _{J,\alpha }^{\otimes
\sum_{i}l_{i}}\left( \left\{ U_{j}^{l_{i}}\right\} _{i},\left( \left\{ 
\mathbf{\hat{\Lambda}}_{\alpha }^{\left[ \left\{ k,k_{i\leqslant
n-1}\right\} \right] }\left[ \Psi _{J},\nu ,U_{j}^{l}\right] \right\}
\right) _{n}\right) \right\} \right) }{\delta \Psi _{J,\alpha }^{\otimes
l_{i}}\left( \left( U_{j}^{l_{i}}\right) ^{\prime },\left( \left\{ \mathbf{%
\hat{\Lambda}}_{\alpha }^{\left[ \left\{ k,k_{i\leqslant n-1}\right\} \right]
}\left[ \Psi _{J},\nu ,U_{j}^{l}\right] \right\} \right) _{n}\right) } 
\notag \\
&&\times \Psi _{J,\alpha }^{\otimes \sum_{i}l_{i}}\left( \left( \left\{
U_{j}^{l_{i}}\right\} _{i},\left( \left\{ \mathbf{\hat{\Lambda}}_{\alpha }^{%
\left[ \left\{ k,k_{i\leqslant n-1}\right\} \right] }\left[ \Psi _{J},\nu
,U_{j}^{l}\right] \right\} \right) _{n}\right) \right)  \notag
\end{eqnarray}%
and we show that we can define, at least locally, a field depending on
parameters independent from field variations.

Note that the gradients are not included in the right hand side since they
can be removed by redefinition of: 
\begin{equation*}
\Psi _{J,\alpha }^{\otimes \sum_{i}l_{i}}\left( \left( \left\{
U_{j}^{l_{i}}\right\} _{i},\left( \left\{ \mathbf{\hat{\Lambda}}_{\alpha }^{%
\left[ \left\{ k,k_{i\leqslant n-1}\right\} \right] }\left[ \Psi _{J},\nu
,U_{j}^{l}\right] \right\} \right) _{n}\right) \right)
\end{equation*}%
Functional $V$ is assumed to be defined as a series of functionals of the
type:%
\begin{eqnarray*}
&&V\left( \left\{ \Psi _{J,\alpha }^{\otimes \sum_{i}l_{i}}\left( \left\{
U_{j}^{l_{i}}\right\} _{i},\left( \left\{ \mathbf{\hat{\Lambda}}_{\alpha }^{%
\left[ \left\{ k,k_{i\leqslant n-1}\right\} \right] }\left[ \Psi _{J},\nu
,U_{j}^{l}\right] \right\} \right) _{n}\right) \right\} \right) \\
&=&\sum_{u}\tprod\limits_{p_{t}\leqslant u}\int \Psi _{J,\alpha
_{t},p_{t}}^{\otimes \sum_{i}l_{i}}\left( \left\{ \left\{
U_{j}^{l_{i}}\right\} _{i},\left( \left\{ \mathbf{\hat{\Lambda}}_{\alpha }^{%
\left[ \left\{ k,k_{i\leqslant n-1}\right\} \right] }\left[ \Psi _{J},\nu
,U_{j}^{l}\right] \right\} \right) _{n}\right\} _{I}\right) \\
&&\times A\left( \left\{ \left\{ U_{j}^{l_{i}}\right\} _{i},\left( \left\{ 
\mathbf{\hat{\Lambda}}_{\alpha }^{\left[ \left\{ k,k_{i\leqslant
n-1}\right\} \right] }\left[ \Psi _{J},\nu ,U_{j}^{l}\right] \right\}
\right) _{n}\right\} _{I}\right)
\end{eqnarray*}

Computing the variation:%
\begin{equation*}
\int v\left( \left( U_{j}^{l_{i}}\right) \right) \frac{\delta \Psi
_{J,\alpha }^{\otimes \sum_{i}l_{i}}\left( \left\{ U_{j}^{l_{i}}\right\}
_{i},\left( \left\{ \mathbf{\hat{\Lambda}}_{\alpha }^{\left[ \left\{
k,k_{i\leqslant n-1}\right\} \right] }\left[ \Psi _{J},\nu ,U_{j}^{l}\right]
\right\} \right) _{n}\right) }{\delta \Psi _{J,\alpha }^{\otimes
l_{i}}\left( \left( U_{j}^{l_{i}}\right) ^{\prime }\right) }
\end{equation*}%
\bigskip

and using assumption (\ref{NVC}) yields:%
\begin{eqnarray}
&&\int v\left( \left( U_{j}^{l_{i}}\right) \right) \frac{\delta \Psi
_{J,\alpha }^{\otimes \sum_{i}l_{i}}\left( \left\{ U_{j}^{l_{i}}\right\}
_{i},\left( \left\{ \mathbf{\hat{\Lambda}}_{\alpha }^{\left[ \left\{
k,k_{i\leqslant n-1}\right\} \right] }\left[ \Psi _{J},\nu ,U_{j}^{l}\right]
\right\} \right) _{n}\right) }{\delta \Psi _{J,\alpha }^{\otimes
l_{i}}\left( \left( U_{j}^{l_{i}}\right) ^{\prime }\right) }  \label{VRC} \\
&=&\int v\left( U_{i}^{l}\right) \frac{\delta ^{\prime }\Psi _{J\alpha
}^{\otimes \sum_{i}l_{i}}\left( \left\{ U_{j}^{l_{i}}\right\} _{i},\left(
\left\{ \mathbf{\hat{\Lambda}}_{\alpha }^{\left[ \left\{ k,k_{i\leqslant
n-1}\right\} \right] }\left[ \Psi _{J},\nu ,U_{j}^{l}\right] \right\}
\right) _{n},v\right) }{\delta \Psi _{J}^{\otimes l_{i}}\left( \left(
U_{j}^{l_{i}}\right) ^{\prime }\right) }  \notag \\
&&+\int v\left( \left\{ \left( U_{j}^{l_{i}}\right) \right\} \right) 
\underline{V}_{\alpha }\left( \left\{ \Psi _{J,\alpha }^{\otimes
\sum_{i}l_{i}}\left( \left\{ U_{j}^{l_{i}}\right\} _{i},\left( \left\{ 
\mathbf{\hat{\Lambda}}_{\alpha }^{\left[ \left\{ k,k_{i\leqslant
n-1}\right\} \right] }\left[ \Psi _{J},\nu ,U_{j}^{l}\right] \right\}
\right) _{n}\right) \right\} \right)  \notag \\
&&\times \Psi _{J,\alpha }^{\otimes \sum_{i}l_{i}}\left( \left( \left\{ 
\mathbf{\hat{\Lambda}}_{\alpha }^{\left[ \left\{ k,k_{i\leqslant
n-1}\right\} \right] }\left[ \Psi _{J},\nu ,U_{j}^{l}\right] \right\}
\right) _{n}\right)  \notag
\end{eqnarray}

with:%
\begin{eqnarray*}
&&\underline{V}_{\alpha }\left( \left\{ \Psi _{J,\alpha }^{\otimes
\sum_{i}l_{i}}\left( \left( \left\{ U_{j}^{l_{i}}\right\} _{i},\left(
\left\{ \mathbf{\hat{\Lambda}}_{\alpha }^{\left[ \left\{ k,k_{i\leqslant
n-1}\right\} \right] }\left[ \Psi _{J},\nu ,U_{j}^{l}\right] \right\}
\right) _{n}\right) \right) \right\} \right) \\
&=&\frac{\delta ^{\prime }V\left( \left\{ \Psi _{J,\alpha }^{\otimes
\sum_{i}l_{i}}\left( \left( \left\{ U_{j}^{l_{i}}\right\} _{i},\left(
\left\{ \mathbf{\hat{\Lambda}}_{\alpha }^{\left[ \left\{ k,k_{i\leqslant
n-1}\right\} \right] }\left[ \Psi _{J},\nu ,U_{j}^{l}\right] \right\}
\right) _{n}\right) \right) \right\} \right) }{\delta \Psi _{J,\alpha
}^{\otimes l_{i}}\left( \left( U_{j}^{l_{i}}\right) ^{\prime },\left(
\left\{ \mathbf{\hat{\Lambda}}_{\alpha }^{\left[ \left\{ k,k_{i\leqslant
n-1}\right\} \right] }\left[ \Psi _{J},\nu ,U_{j}^{l}\right] \right\}
\right) _{n}\right) }
\end{eqnarray*}%
and equation (\ref{VRC}) writes:

\begin{eqnarray*}
&&\int v\left( \left( U_{j}^{l_{i}}\right) \right) \frac{\delta \Psi
_{J,\alpha }^{\otimes \sum_{i}l_{i}}\left( \left\{ U_{j}^{l_{i}}\right\}
_{i},\left( \left\{ \mathbf{\hat{\Lambda}}_{\alpha }^{\left[ \left\{
k,k_{i\leqslant n-1}\right\} \right] }\left[ \Psi _{J},\nu ,U_{j}^{l}\right]
\right\} \right) _{n}\right) }{\delta \Psi _{J,\alpha }^{\otimes
l_{i}}\left( \left( U_{j}^{l_{i}}\right) ^{\prime }\right) } \\
&=&v\left( \left( U_{j}^{l_{i}}\right) ^{\prime }\right) \Psi _{J,\alpha
}^{\otimes \sum_{i}l_{i}}\left( \left\{ \left( U_{j}^{l_{i}}\right) ^{\prime
}\right\} _{i},\left( \left\{ \mathbf{\hat{\Lambda}}_{\alpha }^{\left[
\left\{ k,k_{i\leqslant n-1}\right\} \right] }\left[ \Psi _{J},\nu ,U_{j}^{l}%
\right] \right\} \right) _{n}\right) \\
&&+\int_{\left( U_{j}^{l_{i}}\right) /\left( U_{j}^{l_{i}}\right) ^{\prime
}}v\left( \left( U_{j}^{l_{i}}\right) \right) \underline{V}_{\alpha }\left(
\left\{ \Psi _{J,\alpha }^{\otimes \sum_{i}l_{i}}\left( \left\{
U_{j}^{l_{i}}\right\} _{i},\left( \left\{ \mathbf{\hat{\Lambda}}_{\alpha }^{%
\left[ \left\{ k,k_{i\leqslant n-1}\right\} \right] }\left[ \Psi _{J},\nu
,U_{j}^{l}\right] \right\} \right) _{n}\right) \right\} \right) \\
&&\times \Psi _{J,\alpha }^{\otimes \sum_{i}l_{i}}\left( \left( \left\{ 
\mathbf{\hat{\Lambda}}_{\alpha }^{\left[ \left\{ k,k_{i\leqslant
n-1}\right\} \right] }\left[ \Psi _{J},\nu ,U_{j}^{l}\right] \right\}
\right) _{n}\right)
\end{eqnarray*}%
This can be factored as:

\begin{eqnarray*}
&&\delta \Psi _{J,\alpha }^{\otimes \sum_{i}l_{i}}\left( \left\{
U_{j}^{l_{i}}\right\} _{i},\left( \left\{ \mathbf{\hat{\Lambda}}_{\alpha }^{%
\left[ \left\{ k,k_{i\leqslant n-1}\right\} \right] }\left[ \Psi _{J},\nu
,U_{j}^{l}\right] \right\} \right) _{n}\right) \\
&=&\int \left( \delta \left( U_{j}^{l_{i}}-\left( U_{j}^{l_{i}}\right)
^{\prime }\right) +\underline{V}_{\alpha }\left( \left\{ \Psi _{J,\alpha
}^{\otimes \sum_{i}l_{i}}\left( \left\{ U_{j}^{l_{i}}\right\} _{i},\left(
\left\{ \mathbf{\hat{\Lambda}}_{\alpha }^{\left[ \left\{ k,k_{i\leqslant
n-1}\right\} \right] }\left[ \Psi _{J},\nu ,U_{j}^{l}\right] \right\}
\right) _{n}\right) \right\} \right) \right) \\
&&\times \delta \Psi _{J,\alpha }^{\otimes l_{i}}\left( \left(
U_{j}^{l_{i}}\right) ^{\prime }\right) \Psi _{J,\alpha }^{\otimes
\sum_{i}l_{i}}\left( \left( \left\{ \mathbf{\hat{\Lambda}}_{\alpha }^{\left[
\left\{ k,k_{i\leqslant n-1}\right\} \right] }\left[ \Psi _{J},\nu ,U_{j}^{l}%
\right] \right\} \right) _{n}\right) \\
&=&\delta ^{\prime }\hat{\Psi}_{J,\alpha }^{\otimes \sum_{i}l_{i}}\left(
\left\{ U_{j}^{l_{i}}\right\} _{i},\left( \left\{ \mathbf{\hat{\Lambda}}%
_{\alpha }^{\left[ \left\{ k,k_{i\leqslant n-1}\right\} \right] }\left[ \Psi
_{J},\nu ,U_{j}^{l}\right] \right\} \right) _{n}\right)
\end{eqnarray*}%
where:%
\begin{eqnarray*}
&&\delta ^{\prime }\hat{\Psi}_{J,\alpha }^{\otimes \sum_{i}l_{i}}\left(
\left( U_{j}^{l_{i}}\right) \right) \\
&=&\int \left( \delta \left( U_{j}^{l_{i}}-\left( U_{j}^{l_{i}}\right)
^{\prime }\right) +\underline{V}_{\alpha }\left( \left\{ \Psi _{J,\alpha
}^{\otimes \sum_{i}l_{i}}\left( \left\{ U_{j}^{l_{i}}\right\} _{i},\left(
\left\{ \mathbf{\hat{\Lambda}}_{\alpha }^{\left[ \left\{ k,k_{i\leqslant
n-1}\right\} \right] }\left[ \Psi _{J},\nu ,U_{j}^{l}\right] \right\}
\right) _{n}\right) \right\} \right) \right) \delta \Psi _{J,\alpha
}^{\otimes l_{i}}\left( \left( U_{j}^{l_{i}}\right) ^{\prime }\right)
\end{eqnarray*}%
As a consequence, the initial variation of the field rewrites:%
\begin{eqnarray*}
&&\delta \Psi _{J,\alpha }^{\otimes \sum_{i}l_{i}}\left( \left\{
U_{j}^{l_{i}}\right\} _{i},\left( \left\{ \mathbf{\hat{\Lambda}}_{\alpha }^{%
\left[ \left\{ k,k_{i\leqslant n-1}\right\} \right] }\left[ \Psi _{J},\nu
,U_{j}^{l}\right] \right\} \right) _{n}\right) \\
&=&\delta ^{\prime }\hat{\Psi}_{J,\alpha }^{\otimes \sum_{i}l_{i}}\left(
\left( U_{j}^{l_{i}}\right) \right) \Psi _{J,\alpha }^{\otimes
\sum_{i}l_{i}}\left( \left( \left\{ \mathbf{\hat{\Lambda}}_{\alpha }^{\left[
\left\{ k,k_{i\leqslant n-1}\right\} \right] }\left[ \Psi _{J},\nu ,U_{j}^{l}%
\right] \right\} \right) _{n}\right)
\end{eqnarray*}%
which is performed at fixed $\left( \left\{ \mathbf{\hat{\Lambda}}_{\alpha
}^{\left[ \left\{ k,k_{i\leqslant n-1}\right\} \right] }\left[ \Psi _{J},\nu
,U_{j}^{l}\right] \right\} \right) _{n}$. Thus, even if $\Psi _{J,\alpha
}^{\otimes \sum_{i}l_{i}}$\ is not invariant, there is a related field $\hat{%
\Psi}_{J,\alpha }^{\otimes \sum_{i}l_{i}}$ that can be defined locally as a
function of an invariant family of parametrs $\mathbf{\hat{\Lambda}}_{\alpha
}^{\left[ \left\{ k,k_{i\leqslant n-1}\right\} \right] }$.

The solution:%
\begin{equation*}
\hat{\Psi}_{J,\alpha }^{\otimes \sum_{i}l_{i}}\left( \left\{
U_{j}^{l_{i}}\right\} ,\left( \left\{ \mathbf{\hat{\Lambda}}_{\alpha }^{%
\left[ \left\{ k,k_{i\leqslant n-1}\right\} \right] }\right\} \right)
_{n},v,\alpha \right) \delta \left( f\left( \left( \left\{ \mathbf{\hat{%
\Lambda}}_{\alpha }^{\left[ \left\{ k,k_{i\leqslant n-1}\right\} \right]
}\right\} \right) _{n},\left( \Psi _{J,0,\alpha }^{\otimes l_{i}},v\right)
\right) \right)
\end{equation*}%
satisfies the equation:%
\begin{eqnarray*}
&&\int \hat{v}\left( \left( U_{j}^{l_{i}}\right) \right) \frac{\delta
\left\{ \mathbf{\hat{\Lambda}}_{\alpha }^{\left[ \left\{ k,k_{i\leqslant
n-1}\right\} \right] }\left[ \Psi _{J},\nu ,U_{j}^{l}\right] \right\} }{%
\delta \Psi _{J,\alpha }^{\otimes l_{i}}\left( \left( U_{j}^{l_{i}}\right)
^{\prime }\right) }\nabla _{\mathbf{\hat{O}}}\hat{\Psi}_{J,\alpha }^{\otimes
\sum_{i}l_{i}}\left( \left\{ U_{j}^{l_{i}}\right\} _{i},\left( \left\{ 
\mathbf{\hat{\Lambda}}_{\alpha }^{\left[ \left\{ k,k_{i\leqslant
n-1}\right\} \right] }\left[ \Psi _{J},\nu ,U_{j}^{l}\right] \right\}
\right) _{n}\right) \\
&\simeq &\int v\left( \left\{ \left( U_{j}^{l_{i}}\right) \right\} \right) 
\frac{\delta ^{\prime }\hat{V}\left( \left\{ \hat{\Psi}_{J,\alpha }^{\otimes
\sum_{i}l_{i}}\left( \left\{ U_{j}^{l_{i}}\right\} _{i},\left( \left\{ 
\mathbf{\hat{\Lambda}}_{\alpha }^{\left[ \left\{ k,k_{i\leqslant
n-1}\right\} \right] }\left[ \Psi _{J},\nu ,U_{j}^{l}\right] \right\}
\right) _{n}\right) \right\} \right) }{\delta \Psi _{J,\alpha }^{\otimes
l_{i}}\left( \left\{ \left( U_{j}^{l_{i}}\right) ^{\prime },\mathbf{\hat{%
\Lambda}}_{\alpha }^{\left[ \left\{ k,k_{i\leqslant n-1}\right\} \right] }%
\left[ \Psi _{J},\nu ,U_{j}^{l}\right] \right\} \right) }
\end{eqnarray*}

with:%
\begin{equation*}
\hat{v}\left( \left( U_{j}^{l_{i}}\right) \right) =\int v\left( \left(
U_{j}^{l_{i}}\right) ^{\prime \prime }\right) \frac{\delta \Psi _{J,\alpha
}^{\otimes \sum_{i}l_{i}}\left( \left( U_{j}^{l_{i}}\right) ^{\prime \prime
}\right) }{\delta \hat{\Psi}_{J,\alpha }^{\otimes \sum_{i}l_{i}}\left(
U_{j}^{l_{i}}\right) }
\end{equation*}%
If the $\frac{\delta \mathbf{\hat{\Lambda}}_{\alpha }^{\left[ k_{i}\right] }%
\left[ \Psi _{J},\nu ,U_{j}^{l}\right] }{\delta \Psi _{J,\alpha }^{\otimes
l_{i}}\left( \left( U_{j}^{l_{i}}\right) ^{\prime }\right) }$ are
independent from realiztion, as before, the sum over realiztions verifies
the equation:%
\begin{eqnarray*}
&&\int \hat{v}\left( \left( U_{j}^{l_{i}}\right) \right) \frac{\delta 
\mathbf{\hat{\Lambda}}_{\alpha }^{\left[ k_{i}\right] }\left[ \Psi _{J},\nu
,U_{j}^{l}\right] }{\delta \Psi _{J,\alpha }^{\otimes l_{i}}\left( \left(
U_{j}^{l_{i}}\right) ^{\prime }\right) }\nabla _{\mathbf{\hat{O}}}\hat{\Psi}%
_{J}^{\otimes \sum_{i}l_{i}}\left( \left\{ U_{j}^{l_{i}}\right\} _{i},\left(
\left\{ \mathbf{\hat{\Lambda}}_{\alpha }^{\left[ \left\{ k,k_{i\leqslant
n-1}\right\} \right] }\left[ \Psi _{J},\nu ,U_{j}^{l}\right] \right\}
\right) _{n}\right) \\
&\simeq &v\left( \left\{ \left( U_{j}^{l_{i}}\right) \right\} \right) \frac{%
\delta ^{\prime }\hat{V}\left( \hat{\Psi}_{J}^{\otimes \sum_{i}l_{i}}\left(
\left\{ U_{j}^{l_{i}}\right\} _{i}\right) ,\left( \left\{ \mathbf{\hat{%
\Lambda}}_{\alpha }^{\left[ \left\{ k,k_{i\leqslant n-1}\right\} \right] }%
\left[ \Psi _{J},\nu ,U_{j}^{l}\right] \right\} \right) _{n}\right) }{\delta
\Psi _{J}^{\otimes l_{i}}\left( \left\{ \left( U_{j}^{l_{i}}\right) ^{\prime
},\mathbf{\hat{\Lambda}}_{\alpha }^{\left[ \left\{ k,k_{i\leqslant
n-1}\right\} \right] }\left[ \Psi _{J},\nu ,U_{j}^{l}\right] \right\}
\right) }
\end{eqnarray*}

\section*{Appendix 5: projection over eigenvalues of operators}

\subsection*{Projected functional for single eigenvalue}

Starting with the functional (\ref{VRF}):%
\begin{equation*}
\left\langle \Psi _{I}^{\otimes k}\left( U_{i}^{k}\right) \right\rangle _{%
\mathbf{\hat{\Lambda}}}=v_{I}^{\otimes k}\left( U_{i}^{k},\left\{ \Psi
_{J,\alpha }^{\otimes l_{i}}\right\} ,\underline{\mathbf{\hat{\Lambda}}}%
_{\infty ,\alpha }^{\left[ k\right] }\left[ \Psi _{J},\nu ,\left\{
U_{j}^{l_{i}}\right\} \right] \right) 
\end{equation*}%
computing these averages of the field ovr the space spanned by states $%
F_{0}^{\underline{\mathbf{\hat{\Lambda}}}}\left[ \left\{ \Psi _{I,\alpha
}^{\otimes k_{i}}\right\} ,\left\{ \Psi _{J,\alpha }^{\otimes l_{i}}\right\}
,v\right] $ needs to consider linear combinations over these states with
coefficients $\Psi \left( \underline{\mathbf{\hat{\Lambda}}}_{\infty ,\alpha
}^{\left[ k\right] }\left[ \Psi _{J},\nu ,\left\{ U_{j}^{l_{i}}\right\} %
\right] ,v\right) $. Averages $\left\langle \Psi _{I}^{\otimes k}\left(
U_{i}^{k}\right) \right\rangle _{\mathbf{\hat{\Lambda}}}$ are thus combined:%
\begin{eqnarray}
\left\langle \Psi _{I}^{\otimes k}\left( U_{i}^{k}\right) \right\rangle 
&\rightarrow &\int \Psi \left( \underline{\mathbf{\hat{\Lambda}}}_{\infty
,\alpha }^{\left[ k\right] }\left[ \Psi _{J},\nu ,\left\{
U_{j}^{l_{i}}\right\} \right] ,v\right) \left\langle \Psi _{I}^{\otimes
k}\left( U_{i}^{k}\right) \right\rangle _{\mathbf{\hat{\Lambda}}}d\underline{%
\mathbf{\hat{\Lambda}}}_{\infty ,\alpha }^{\left[ k\right] }\left[ \Psi
_{J},\nu ,\left\{ U_{j}^{l_{i}}\right\} \right]   \label{VGW} \\
&=&\int \Psi \left( \underline{\mathbf{\hat{\Lambda}}}_{\infty ,\alpha }^{%
\left[ k\right] }\left[ \Psi _{J},\nu ,\left\{ U_{j}^{l_{i}}\right\} \right]
,v\right) v_{I}^{\otimes k}\left( U_{i}^{k},\left\{ \Psi _{J,\alpha
}^{\otimes l_{i}}\right\} ,\underline{\mathbf{\hat{\Lambda}}}_{\infty
,\alpha }^{\left[ k\right] }\left[ \Psi _{J},\nu ,\left\{
U_{j}^{l_{i}}\right\} \right] \right) d\underline{\mathbf{\hat{\Lambda}}}%
_{\infty ,\alpha }^{\left[ k\right] }\left[ \Psi _{J},\nu ,\left\{
U_{j}^{l_{i}}\right\} \right]   \notag
\end{eqnarray}

We use that:%
\begin{equation*}
v_{I}^{\otimes k}\left( U_{i}^{k},\left\{ \Psi _{J,\alpha }^{\otimes
l_{i}}\right\} ,\underline{\mathbf{\hat{\Lambda}}}_{\infty ,\alpha }^{\left[
k\right] }\left[ \Psi _{J},\nu ,\left\{ U_{j}^{l_{i}}\right\} \right]
\right) 
\end{equation*}%
has an expansion:%
\begin{eqnarray*}
&&v_{I}^{\otimes k}\left( U_{i}^{k},\left\{ \Psi _{J,\alpha }^{\otimes
l_{i}}\right\} ,\underline{\mathbf{\hat{\Lambda}}}_{\infty ,\alpha }^{\left[
k\right] }\left[ \Psi _{J},\nu ,\left\{ U_{j}^{l_{i}}\right\} \right]
\right)  \\
&=&\sum_{s,l_{1},...,l_{s}}\int d\left( \left\{ U_{j}^{l_{i}}\right\}
_{l_{i}}\right) \tprod_{i}\Psi _{J}^{\otimes l_{i}}\left(
U_{j}^{l_{i}}\right) \mathcal{V}_{0}^{v}\left( U_{i}^{k},\left\{
U_{j}^{l_{i}}\right\} _{l_{i}},\underline{\mathbf{\hat{\Lambda}}}_{\infty
,\alpha }^{\left[ k\right] }\left[ \Psi _{J},\nu ,\left\{
U_{j}^{l_{i}}\right\} \right] \right) 
\end{eqnarray*}%
\bigskip As a consequence, replacing:%
\begin{equation*}
\Psi _{I}^{\otimes k}\left( U_{i}^{k}\right) \delta \left( f_{lk}\left(
U_{j}^{l},U_{i}^{k}\right) \right) 
\end{equation*}%
in the functional (\ref{FCL}) by its average (\ref{VGW}) and change
variables as before:%
\begin{eqnarray*}
U_{j}^{l},\left\{ U_{j}^{l_{i}}\right\} _{i} &\rightarrow &\left\{
U_{j}^{l_{i}}\right\} _{i} \\
\Psi _{J}^{\otimes l}\left( U_{j}^{l}\right) \tprod_{i}\Psi _{J}^{\otimes
l_{i}}\left( U_{j}^{l_{i}}\right)  &\rightarrow &\tprod_{i}\Psi
_{J}^{\otimes l_{i}}\left( U_{j}^{l_{i}}\right) 
\end{eqnarray*}%
leads to the functional, written for a single realization:%
\begin{equation*}
\int \bar{g}\left( \left\{ U_{j}^{l_{i}},\right\} ,\underline{\mathbf{\hat{%
\Lambda}}}_{\infty ,\alpha }^{\left[ k\right] }\left[ \Psi _{J},\nu ,\left\{
U_{j}^{l_{i}}\right\} \right] \right) \tprod_{i}\Psi _{J}^{\otimes
l_{i}}\left( U_{j}^{l_{i}}\right) \Psi \left( \underline{\mathbf{\hat{\Lambda%
}}}_{\infty ,\alpha }^{\left[ k\right] }\left[ \Psi _{J},\nu ,\left\{
U_{j}^{l_{i}}\right\} \right] ,v\right) 
\end{equation*}%
where:%
\begin{equation*}
\bar{g}\left( \left\{ U_{j}^{l_{i}},\underline{\mathbf{\hat{\Lambda}}}%
_{\alpha }^{\left[ k_{i}\right] }\left[ \Psi _{J},\nu ,U_{j}^{l}\right]
\right\} _{i}\right) =\int g\left( U_{i}^{k}/f_{kl},U_{j}^{l}\right) 
\mathcal{V}_{0}^{v}\left( U_{i}^{k},\left\{ U_{j}^{l_{i}}\right\} _{l_{i}},%
\underline{\mathbf{\hat{\Lambda}}}_{\infty ,\alpha }^{\left[ k\right] }\left[
\Psi _{J},\nu ,\left\{ U_{j}^{l_{i}}\right\} \right] \right)
dU_{i}^{k}/f_{kl}
\end{equation*}%
Introducing the eigenvalue explicit:%
\begin{eqnarray*}
\mathcal{V}_{0}^{v}\left( U_{i}^{k},\left\{ U_{j}^{l_{i}}\right\} _{l_{i}},%
\underline{\mathbf{\hat{\Lambda}}}_{\infty ,\alpha }^{\left[ k\right] }\left[
\Psi _{J},\nu ,\left\{ U_{j}^{l_{i}}\right\} \right] \right)  &\rightarrow &%
\mathcal{V}_{0}^{v}\left( U_{i}^{k},\left\{ U_{j}^{l_{i}}\right\} _{l_{i}},%
\underline{\mathbf{\hat{\Lambda}}}_{\infty ,\alpha }^{\left[ k\right] }\left[
\Psi _{J},\nu ,\left\{ U_{j}^{l_{i}}\right\} \right] ,\lambda \left( \Psi
_{J}^{\otimes k}\right) \right)  \\
\Psi \left( \underline{\mathbf{\hat{\Lambda}}}_{\infty ,\alpha }^{\left[ k%
\right] }\left[ \Psi _{J},\nu ,\left\{ U_{j}^{l_{i}}\right\} \right]
,v\right)  &\rightarrow &\Psi \left( \underline{\mathbf{\hat{\Lambda}}}%
_{\infty ,\alpha }^{\left[ k\right] }\left[ \Psi _{J},\nu ,\left\{
U_{j}^{l_{i}}\right\} \right] ,v,\lambda \left( \Psi _{J}^{\otimes k}\right)
\right) 
\end{eqnarray*}%
As before if the set: 
\begin{equation*}
\left\{ \underline{\mathbf{\hat{\Lambda}}}_{\infty ,\alpha }^{\left[ k\right]
}\left[ \Psi _{J},\nu ,\left\{ U_{j}^{l_{i}}\right\} \right] \right\}
=\left\{ \underline{\mathbf{\hat{\Lambda}}}_{\infty }^{\left[ k\right] }%
\left[ \Psi _{J},\nu ,\left\{ U_{j}^{l_{i}}\right\} \right] \right\} 
\end{equation*}%
does not depend on the realiztn, this becomes:%
\begin{eqnarray*}
&&F_{f,lin}\left( \left\{ \Psi _{J}^{\otimes l}\left( U_{j}^{l}\right)
\right\} _{l}\right)  \\
&=&\int \bar{g}\left( \left\{ U_{j}^{l_{i}}\right\} ,\underline{\mathbf{\hat{%
\Lambda}}}_{\infty }^{\left[ k\right] }\left[ \Psi _{J},\nu ,\left\{
U_{j}^{l_{i}}\right\} \right] ,\lambda \left( \Psi _{J}^{\otimes k}\right)
\right) \Psi _{J}^{\otimes \sum l_{i}}\left( \left\{ U_{j}^{l_{i}}\right\}
_{i},\underline{\mathbf{\hat{\Lambda}}}_{\infty ,\alpha }^{\left[ k\right] }%
\left[ \Psi _{J},\nu ,\left\{ U_{j}^{l_{i}}\right\} \right] ,\lambda \left(
\Psi _{J}^{\otimes k}\right) ,v\right) 
\end{eqnarray*}%
where:%
\begin{equation*}
\Psi _{J}^{\otimes \sum l_{i}}\left( \left\{ U_{j}^{l_{i}}\right\} ,%
\underline{\mathbf{\hat{\Lambda}}}_{\infty }^{\left[ k\right] }\left[ \Psi
_{J},\nu ,\left\{ U_{j}^{l_{i}}\right\} \right] ,\lambda \left( \Psi
_{J}^{\otimes k}\right) ,v\right) =\sum_{\alpha }\Psi _{J_{\alpha
}}^{\otimes \sum l_{i}}\left( \left\{ U_{j}^{l_{i}}\right\} _{i},\underline{%
\mathbf{\hat{\Lambda}}}_{\infty }^{\left[ k\right] }\left[ \Psi _{J},\nu
,\left\{ U_{j}^{l_{i}}\right\} \right] ,\lambda \left( \Psi _{J}^{\otimes
k}\right) ,v\right) 
\end{equation*}%
and:%
\begin{equation*}
\Psi _{J_{\alpha }}^{\otimes \sum l_{i}}\left( \left\{ U_{j}^{l_{i}}\right\}
,\underline{\mathbf{\hat{\Lambda}}}_{\infty }^{\left[ k\right] }\left[ \Psi
_{J},\nu ,\left\{ U_{j}^{l_{i}}\right\} \right] ,v,\lambda \left( \Psi
_{J}^{\otimes k}\right) \right) =\tprod_{i}\Psi _{J,\alpha }^{\otimes
l_{i}}\left( U_{j}^{l_{i}}\right) \Psi _{\alpha }\left( \underline{\mathbf{%
\hat{\Lambda}}}_{\infty }^{\left[ k\right] }\left[ \Psi _{J},\nu ,\left\{
U_{j}^{l_{i}}\right\} \right] ,\lambda \left( \Psi _{J}^{\otimes k}\right)
\right) 
\end{equation*}%
and:%
\begin{eqnarray*}
&&\bar{g}\left( \left\{ U_{j}^{l_{i}}\right\} ,\underline{\mathbf{\hat{%
\Lambda}}}_{\infty }^{\left[ k\right] }\left[ \Psi _{J},\nu ,\left\{
U_{j}^{l_{i}}\right\} \right] ,\lambda \left( \Psi _{J}^{\otimes k}\right)
\right)  \\
&=&\int g\left( U_{i}^{k}/f_{kl},U_{j}^{l}\right) \mathcal{V}_{0}^{v}\left(
U_{i}^{k},\left\{ U_{j}^{l_{i}}\right\} _{l_{i}},\underline{\mathbf{\hat{%
\Lambda}}}_{\infty }^{\left[ k\right] }\left[ \Psi _{J},\nu ,\left\{
U_{j}^{l_{i}}\right\} \right] ,\lambda \left( \Psi _{J}^{\otimes k}\right)
\right) dU_{i}^{k}/f_{kl}
\end{eqnarray*}

\subsection*{Average over several eigenspaces}

If we consider the average over several eigenspaces, we write:%
\begin{equation*}
\left( \left( \lambda ,\underline{\mathbf{\hat{\Lambda}}}_{\infty }^{\left[ k%
\right] }\right) \left[ \Psi _{J},\nu ,\left\{ U_{j}^{l_{i}}\right\} \right]
\right) 
\end{equation*}%
the variables:%
\begin{equation*}
\left( \lambda \left[ \Psi _{J},\nu ,\left\{ U_{j}^{l_{i}}\right\} \right] ,%
\underline{\mathbf{\hat{\Lambda}}}_{\infty }^{\left[ k\right] }\left[ \Psi
_{J},\nu ,\left\{ U_{j}^{l_{i}}\right\} \right] \right) 
\end{equation*}%
with the eigenvalues of the operator considered. $\lambda $ can be
multi-valued. The eigenstates write: 
\begin{equation*}
F\left[ \left[ \Psi _{J}^{\otimes l}\right] ,\left\{ \Psi ^{k}\right\}
,\left( \lambda ,\underline{\mathbf{\hat{\Lambda}}}_{\infty }^{\left[ k%
\right] }\right) \left[ \Psi _{J},\nu ,\left\{ U_{j}^{l_{i}}\right\} \right] %
\right] 
\end{equation*}%
and the average values in this state are:%
\begin{equation*}
\left\langle \Psi _{I}^{\otimes k}\left( U_{i}^{k}\right) \right\rangle
=v_{I}^{\otimes k}\left( U_{i}^{k},\left[ \Psi _{J}^{\otimes l}\right]
,\lambda ,\underline{\mathbf{\hat{\Lambda}}}_{\infty }^{\left[ k\right] }%
\left[ \Psi _{J},\nu ,\left\{ U_{j}^{l_{i}}\right\} \right] \right) 
\end{equation*}%
where:%
\begin{eqnarray*}
&&v_{I}^{\otimes k}\left( U_{i}^{k},\left[ \Psi _{J}^{\otimes l}\right]
,\left( \lambda ,\underline{\mathbf{\hat{\Lambda}}}_{\infty }^{\left[ k%
\right] }\right) \left[ \Psi _{J},\nu ,\left\{ U_{j}^{l_{i}}\right\} \right]
\right)  \\
&=&\int \Psi _{I}^{\otimes k}\left( U_{i}^{k}\right) \left\vert F_{0}\left[ %
\left[ \Psi _{J}^{\otimes l}\right] ,\left\{ \Psi _{J}^{\otimes l}\right\}
,\left( \lambda ,\underline{\mathbf{\hat{\Lambda}}}_{\infty }^{\left[ k%
\right] }\right) \left[ \Psi _{J},\nu ,\left\{ U_{j}^{l_{i}}\right\} \right] %
\right] \right\vert ^{2}\tprod \mathcal{D}\left\{ \Psi _{J}^{\otimes
l}\right\} 
\end{eqnarray*}%
and functional becomes:%
\begin{eqnarray*}
&&F_{f,lin}\left( \left\{ \Psi _{J,s_{p_{l^{\prime },0}}^{\otimes
k_{l^{\prime }}}}^{\otimes k_{l^{\prime }}}\left( \left( \left( U_{j}\right)
^{\left( p_{0},p_{l^{\prime }0}\right) }\right) ^{k_{l^{\prime }}}\right)
\right\} _{\left( p_{l^{\prime }0}\right) }\right)  \\
&=&\sum_{m}\sum_{_{D_{j}^{p_{0},p_{l},m}}}\sum_{m^{\prime
}}\sum_{D_{j,p_{u}}^{\left( p_{0},p_{l^{\prime }0},p_{l},p_{l^{\prime
}l}\right) ,m,m^{\prime }}}\int g\left( \left\{ \left[ p_{0},p_{l^{\prime }0}%
\right] \right\} _{l^{\prime }},\left\{ \left( \lambda ,\underline{\mathbf{%
\hat{\Lambda}}}_{\infty }^{\left[ k\right] }\right) \left[ \Psi _{J},\nu
,\left\{ U_{j}^{l_{i}}\right\} \right] \right\} _{\substack{ l\leqslant m \\ %
l^{\prime }\leqslant m^{\prime }}}\right)  \\
&&\times \Psi _{J,s_{p_{l^{\prime },0}}^{\otimes k_{l^{\prime }}}}^{\otimes
k_{l^{\prime }}}\left( \left[ p_{0},p_{l^{\prime }0}\right] ,\left\{ \left(
\lambda ,\underline{\mathbf{\hat{\Lambda}}}_{\infty }^{\left[ k\right]
}\right) \left[ \Psi _{J},\nu ,\left\{ U_{j}^{l_{i}}\right\} \right]
\right\} _{\substack{ l\leqslant m \\ l^{\prime }\leqslant m^{\prime }}}%
\right) d\left\{ \left[ p_{0},p_{l^{\prime }0}\right] \right\} _{0,l^{\prime
}}d\left( \lambda ,\underline{\mathbf{\hat{\Lambda}}}_{\infty }^{\left[ k%
\right] }\right) \left[ \Psi _{J},\nu ,\left\{ U_{j}^{l_{i}}\right\} \right] 
\end{eqnarray*}

\section*{Appendix 6: general form of the amplitudes}

The transport $P_{\lambda \lambda +\delta \lambda }$ is performed through
generator of translation operator:%
\begin{equation*}
\left( \Psi _{J}\left( U_{j}/\left[ U_{j}\right] ,\left\{ \underline{\mathbf{%
\hat{\Lambda}}}^{\left[ k_{i}\right] }\right\} ,\lambda +\delta \lambda
\right) -\Psi _{J}\left( U_{j}/\left[ U_{j}\right] ,\left\{ \underline{%
\mathbf{\hat{\Lambda}}}^{\left[ k_{i}\right] }\right\} ,\lambda \right)
\right) \frac{\delta }{\delta \Psi _{J}\left( U_{j}/\left[ U_{j}\right]
,\left\{ \underline{\mathbf{\hat{\Lambda}}}^{\left[ k_{i}\right] }\right\}
,\lambda \right) }
\end{equation*}

Since $\Psi _{J}\left( U_{j}/\left[ U_{j}\right] ,\left\{ \underline{\mathbf{%
\hat{\Lambda}}}^{\left[ k_{i}\right] }\right\} ,\lambda +\delta \lambda
\right) $ and $\Psi _{J}\left( U_{j}/\left[ U_{j}\right] ,\left\{ \underline{%
\mathbf{\hat{\Lambda}}}^{\left[ k_{i}\right] }\right\} ,\lambda \right) $ do
not act on the same space, their difference is not the derivative of an
operator $\Psi _{J}\left( U_{j}/\left[ U_{j}\right] ,\left\{ \underline{%
\mathbf{\hat{\Lambda}}}^{\left[ k_{i}\right] }\right\} ,\lambda \right) $
that can be defined for every $\lambda $. This derivative has to be
corrected to account for the change in spaces induced by the modification of 
$\lambda $.

Given the constraints:%
\begin{eqnarray*}
&&\left( \Psi _{J}\left( U_{j}/\left[ U_{j}\right] ,\left\{ \underline{%
\mathbf{\hat{\Lambda}}}^{\left[ k_{i}\right] }\right\} ,\lambda +\delta
\lambda \right) -\Psi _{J}\left( U_{j}/\left[ U_{j}\right] ,\left\{ 
\underline{\mathbf{\hat{\Lambda}}}^{\left[ k_{i}\right] }\right\} ,\lambda
\right) \right) \\
&=&\delta \lambda \left( \frac{\partial \Psi _{J}\left( U_{j}/\left[ U_{j}%
\right] ,\left\{ \underline{\mathbf{\hat{\Lambda}}}^{\left[ k_{i}\right]
}\right\} ,\lambda \right) }{\partial \lambda }+\left( \left( A_{\lambda
}\right) _{k}^{k^{\prime }}\left( U_{j}/\left[ U_{j}\right] \right) \Psi
_{J}\left( \left( U_{j}/\left[ U_{j}\right] \right) _{k^{\prime }},\left\{ 
\underline{\mathbf{\hat{\Lambda}}}^{\left[ k_{i}\right] }\right\} ,\lambda
\right) \right) _{k}\right) \\
&=&\delta \lambda \underline{\nabla }_{\lambda }\Psi _{J}\left( U_{j}/\left[
U_{j}\right] ,\left\{ \underline{\mathbf{\hat{\Lambda}}}^{\left[ k_{i}\right]
}\right\} ,\lambda \right)
\end{eqnarray*}%
\bigskip and the transport writes:%
\begin{equation*}
P_{\lambda \lambda +\delta \lambda }=\exp \left( \int i\delta \lambda 
\underline{\nabla }_{\lambda }\Psi _{J}\left( U_{j}/\left[ U_{j}\right]
,\left\{ \underline{\mathbf{\hat{\Lambda}}}^{\left[ k_{i}\right] }\right\}
,\lambda \right) \frac{\delta }{\delta \Psi _{J}\left( U_{j}/\left[ U_{j}%
\right] ,\left\{ \underline{\mathbf{\hat{\Lambda}}}^{\left[ k_{i}\right]
}\right\} ,\lambda \right) }\mathcal{D}\Psi _{J}\left( U_{j}/\left[ U_{j}%
\right] ,\left\{ \underline{\mathbf{\hat{\Lambda}}}^{\left[ k_{i}\right]
}\right\} ,\lambda \right) \right)
\end{equation*}

The matrix elements of:%
\begin{equation*}
\frac{\delta }{\delta \Psi _{J}\left( U_{j}/\left[ U_{j}\right] ,\left\{ 
\underline{\mathbf{\hat{\Lambda}}}^{\left[ k_{i}\right] }\right\} ,\lambda
\right) }
\end{equation*}%
are computed using dual basis, involving amplitudes of the form:%
\begin{equation*}
\exp \left( i\left( \Psi _{J}^{\prime }\left( U_{j}/\left[ U_{j}\right]
,\left\{ \underline{\mathbf{\hat{\Lambda}}}^{\left[ k_{i}\right] }\right\}
,\lambda \right) -\Psi _{J}\left( U_{j}/\left[ U_{j}\right] ,\left\{ 
\underline{\mathbf{\hat{\Lambda}}}^{\left[ k_{i}\right] }\right\} ,\lambda
\right) \right) \Pi _{J}\left( U_{j}/\left[ U_{j}\right] ,\left\{ \underline{%
\mathbf{\hat{\Lambda}}}^{\left[ k_{i}\right] }\right\} ,\lambda \right)
\right) 
\end{equation*}%
where $\Pi _{J}\left( U_{j}/\left[ U_{j}\right] ,\left\{ \underline{\mathbf{%
\hat{\Lambda}}}^{\left[ k_{i}\right] }\right\} ,\lambda \right) $ are
elements of dual basis. Using the parallel transport back to $\lambda
+\delta \lambda $ we can write:%
\begin{equation*}
\Psi _{J}^{\prime }\left( U_{j}/\left[ U_{j}\right] ,\left\{ \underline{%
\mathbf{\hat{\Lambda}}}^{\left[ k_{i}\right] }\right\} ,\lambda \right)
\rightarrow \Psi _{J}^{\prime }\left( U_{j}/\left[ U_{j}\right] ,\left\{ 
\underline{\mathbf{\hat{\Lambda}}}^{\left[ k_{i}\right] }\right\} ,\lambda
+\delta \lambda \right) 
\end{equation*}%
so that the matrices element involving $\frac{\delta }{\delta \Psi
_{J}\left( U_{j}/\left[ U_{j}\right] ,\left\{ \underline{\mathbf{\hat{\Lambda%
}}}^{\left[ k_{i}\right] }\right\} ,\lambda \right) }$ writes:%
\begin{equation*}
\exp \left( i\delta \lambda \underline{\nabla }_{\lambda }\Psi _{J}\left(
U_{j}/\left[ U_{j}\right] ,\left\{ \underline{\mathbf{\hat{\Lambda}}}^{\left[
k_{i}\right] }\right\} ,\lambda \right) \left( \Pi _{J}\left( U_{j}/\left[
U_{j}\right] ,\left\{ \underline{\mathbf{\hat{\Lambda}}}^{\left[ k_{i}\right]
}\right\} ,\lambda \right) \right) \right) 
\end{equation*}%
The matrix contribution of:%
\begin{equation*}
\exp \left( i\delta \lambda \underline{\nabla }_{\lambda }\Psi _{J}\left(
U_{j}/\left[ U_{j}\right] ,\left\{ \underline{\mathbf{\hat{\Lambda}}}^{\left[
k_{i}\right] }\right\} ,\lambda \right) \left( \Pi _{J}\left( U_{j}/\left[
U_{j}\right] ,\left\{ \underline{\mathbf{\hat{\Lambda}}}^{\left[ k_{i}\right]
}\right\} ,\lambda \right) \right) \right) \exp \left( iF\left( \Pi
_{J}\left( U_{j}/\left[ U_{j}\right] ,\left\{ \underline{\mathbf{\hat{\Lambda%
}}}^{\left[ k_{i}\right] }\right\} ,\lambda \right) \right) \right) 
\end{equation*}%
is then obtained by saddle point equation and we find:%
\begin{equation*}
\Pi _{J}\left( U_{j}/\left[ U_{j}\right] ,\left\{ \underline{\mathbf{\hat{%
\Lambda}}}^{\left[ k_{i}\right] }\right\} ,\lambda \right) =\underline{%
\nabla }_{\lambda }\Psi _{J}\left( U_{j}/\left[ U_{j}\right] ,\left\{ 
\underline{\mathbf{\hat{\Lambda}}}^{\left[ k_{i}\right] }\right\} ,\lambda
\right) +\text{something including field}
\end{equation*}%
Matrices elements involving only $\Psi _{J}\left( U_{j}/\left[ U_{j}\right]
,\left\{ \underline{\mathbf{\hat{\Lambda}}}^{\left[ k_{i}\right] }\right\}
,\lambda \right) $ \ Since the constraints are:%
\begin{equation*}
h_{k_{i}}\left( \left\{ \mathbf{\hat{\Lambda}}^{\left[ k_{i}\right] }\left[
\Psi _{J},U^{\left( j\right) }\right] \right\} _{i},h_{p}\left( \left( \Psi
_{J}\right) ,U^{\left( j\right) }\right) \right) =0
\end{equation*}%
non local terms write:%
\begin{eqnarray*}
&&\Psi _{J}\left( U_{j}/\left[ U_{j}\right] ,\left\{ \underline{\mathbf{\hat{%
\Lambda}}}^{\left[ k_{i}\right] }\right\} +\delta \left\{ \underline{\mathbf{%
\hat{\Lambda}}}^{\left[ k_{i}\right] }\right\} ,\lambda \right) -\Psi
_{J}\left( U_{j}/\left[ U_{j}\right] ,\left\{ \underline{\mathbf{\hat{\Lambda%
}}}^{\left[ k_{i}\right] }\right\} ,\lambda \right)  \\
&=&\delta \left\{ \underline{\mathbf{\hat{\Lambda}}}^{\left[ k_{i}\right]
}\right\} \left( \nabla _{\left\{ \underline{\mathbf{\hat{\Lambda}}}^{\left[
k_{i}\right] }\right\} }\Psi _{J}\left( U_{j}/\left[ U_{j}\right] ,\left\{ 
\underline{\mathbf{\hat{\Lambda}}}^{\left[ k_{i}\right] }\right\} ,\lambda
\right) +\left( \left( A_{\left\{ \underline{\mathbf{\hat{\Lambda}}}^{\left[
k_{i}\right] }\right\} }\right) _{k}^{k^{\prime }}\left( U_{j}/\left[ U_{j}%
\right] \right) \Psi _{J}\left( \left( U_{j}/\left[ U_{j}\right] \right)
_{k^{\prime }},\left\{ \underline{\mathbf{\hat{\Lambda}}}^{\left[ k_{i}%
\right] }\right\} ,\lambda \right) \right) _{k}\right)  \\
&=&\delta \left\{ \underline{\mathbf{\hat{\Lambda}}}^{\left[ k_{i}\right]
}\right\} \underline{\nabla }_{\left\{ \underline{\mathbf{\hat{\Lambda}}}^{%
\left[ k_{i}\right] }\right\} }\Psi _{J}\left( U_{j}/\left[ U_{j}\right]
,\left\{ \underline{\mathbf{\hat{\Lambda}}}^{\left[ k_{i}\right] }\right\}
,\lambda \right) 
\end{eqnarray*}%
and the matrices elements become:%
\begin{eqnarray*}
&&\left\langle \Psi _{J}\left( U_{j}/\left[ U_{j}\right] ,\left\{ \underline{%
\mathbf{\hat{\Lambda}}}^{\left[ k_{i}\right] }\right\} ,\lambda +\delta
\lambda \right) \right\vert \delta T_{\lambda \lambda +\delta \lambda
}\left\vert \Psi _{J}\left( U_{j}/\left[ U_{j}\right] ,\left\{ \underline{%
\mathbf{\hat{\Lambda}}}^{\left[ k_{i}\right] }\right\} ,\lambda \right)
\right\rangle  \\
&=&\left\langle \Psi _{J}\left( U_{j}/\left[ U_{j}\right] ,\left\{ 
\underline{\mathbf{\hat{\Lambda}}}^{\left[ k_{i}\right] }\right\} ,\lambda
+\delta \lambda \right) \right\vert \exp \left( i\delta \lambda S\left( \Psi
_{J}\right) \right) \left\vert \Psi _{J}\left( U_{j}/\left[ U_{j}\right]
,\left\{ \underline{\mathbf{\hat{\Lambda}}}^{\left[ k_{i}\right] }\right\}
,\lambda \right) \right\rangle 
\end{eqnarray*}%
Infinitesimally, this is generated by:%
\begin{eqnarray*}
&&S\left( \Psi _{J}\right)  \\
&=&S\left( \underline{\nabla }_{\lambda }\Psi _{J}\left( U_{j}/\left[ U_{j}%
\right] ,\left\{ \underline{\mathbf{\hat{\Lambda}}}^{\left[ k_{i}\right]
}\right\} ,\lambda \right) ,\underline{\nabla }_{\left\{ \underline{\mathbf{%
\hat{\Lambda}}}^{\left[ k_{i}\right] }\right\} }\Psi _{J}\left( U_{j}/\left[
U_{j}\right] ,\left\{ \underline{\mathbf{\hat{\Lambda}}}^{\left[ k_{i}\right]
}\right\} ,\lambda \right) ,\Psi _{J}\left( U_{j}/\left[ U_{j}\right]
,\left\{ \underline{\mathbf{\hat{\Lambda}}}^{\left[ k_{i}\right] }\right\}
,\lambda \right) \right) 
\end{eqnarray*}%
\bigskip 

\section*{Appendix 7 General formalism. Projected functional and effective
field}

\subsection*{Saddle point approach}

\subsubsection*{Projection of single subobject}

Starting with a generic functional involving only one of the $\left[
p_{l},p_{l^{\prime }l}\right] ^{k_{l^{\prime }}}$, and inserting the
solution of the saddle point equation yields:%
\begin{eqnarray*}
&&\int g\left( \left[ p_{l},p_{l^{\prime }l}\right] ^{k_{l^{\prime
}}},v\right) \times v_{\left\{ \left[ p_{l},p_{l^{\prime }l}\right]
^{k_{l^{\prime }}}\right\} }\left( \Psi _{J,s_{p_{l^{\prime },0}}^{\otimes
k_{l^{\prime }}}}^{\otimes k_{l^{\prime }}}\right)  \\
&&\times \Psi _{J,0,s_{p_{l^{\prime },l}}^{\otimes k_{l^{\prime
}}}}^{\otimes k_{l^{\prime }}}\left( \left[ p_{l},p_{l^{\prime }l}\right]
^{k_{l^{\prime }}},\left\{ \left[ p_{l_{1}},p_{l_{1}^{\prime }l_{1}}\right]
^{k_{l_{1}^{\prime }}}\right\} _{_{\substack{ l_{1}\leqslant m_{1} \\ %
l_{1}^{\prime }\leqslant m_{1}^{\prime }}}},v\right) \Psi \left( \left\{ %
\left[ 
\begin{array}{c}
\left( p_{l},p_{l^{\prime }l},k_{l^{\prime }}\right)  \\ 
\left[ \left\{ \Psi _{J}\left[ l^{\prime },0\right] \right\} ,v\right] 
\end{array}%
\right] \right\} ,v\right) dv
\end{eqnarray*}%
with $v_{\left\{ \left[ p_{l},p_{l^{\prime }l}\right] ^{k_{l^{\prime
}}}\right\} }\left( \Psi _{J,s_{p_{l^{\prime },0}}^{\otimes k_{l^{\prime
}}}}^{\otimes k_{l^{\prime }}}\right) $ was defined in (\ref{FC}):%
\begin{eqnarray*}
&&v_{\left[ p_{l},p_{l^{\prime }l}\right] ^{k_{l^{\prime }}}}\left( \Psi
_{J,s_{p_{l^{\prime },0}}^{\otimes k_{l^{\prime }}}}^{\otimes k_{l^{\prime
}}}\right)  \\
&=&\int v\left( \overline{\left[ p_{0},p_{l^{\prime }0}\right]
^{k_{l^{\prime }}}},\left\{ \left[ p_{l},p_{l^{\prime }l}\right]
^{k_{l^{\prime }}}\right\} \right) \times \Psi _{J,s_{p_{l^{\prime
},0}}^{\otimes k_{l^{\prime }}}}^{\otimes k_{l^{\prime }}}\left( \overline{%
\left[ p_{0},p_{l^{\prime }0}\right] ^{k_{l^{\prime }}}},\left\{ \left[
p_{l},p_{l^{\prime }l}\right] ^{k_{l^{\prime }}}\right\} \right) d\left( 
\overline{\left[ p_{0},p_{l^{\prime }0}\right] ^{k_{l^{\prime }}}}\right) 
\end{eqnarray*}%
Implicitly, in this section we write:%
\begin{equation*}
\left\{ \left[ 
\begin{array}{c}
\left( p_{l_{1}},p_{l_{1}^{\prime }l_{1}},k_{l_{1}^{\prime }}\right)  \\ 
\left[ \left\{ \Psi _{J}\left[ l^{\prime },0\right] \right\} ,v\right] 
\end{array}%
\right] \right\} _{_{\substack{ l\subset l_{1}\leqslant m_{1} \\ l^{\prime
}\subset l_{1}^{\prime }\leqslant m_{1}^{\prime }}}}=\left( \left[ 
\begin{array}{c}
\left( p_{l},p_{l^{\prime }l},k_{l^{\prime }}\right)  \\ 
\left[ \left\{ \Psi _{J}\left[ l^{\prime },0\right] \right\} ,v\right] 
\end{array}%
\right] \left\{ \left[ 
\begin{array}{c}
\left( p_{l_{1}},p_{l_{1}^{\prime }l_{1}},k_{l_{1}^{\prime }}\right)  \\ 
\left[ \left\{ \Psi _{J}\left[ l^{\prime },0\right] \right\} ,v\right] 
\end{array}%
\right] \right\} _{_{\substack{ l_{1}<m_{1} \\ l_{1}^{\prime }<m_{1}^{\prime
}}}}\right) 
\end{equation*}%
Expanding the solution leads to the projected functional:%
\begin{eqnarray*}
&&\sum_{_{\substack{ \left( s,s^{\prime },\left[ p,p^{\prime },k^{\prime }%
\right] \right)  \\ \left( m,m^{\prime },\left[ p,p^{\prime },k^{\prime }%
\right] \right) _{1}}}}\int g^{\mathcal{K}}\left( \left\{ \left[
p_{l_{1}},p_{l_{1}^{\prime }l_{1}}\right] ^{k_{l_{1}^{\prime }}}\right\} _{
_{\substack{ l_{1}\leqslant m_{1} \\ l_{1}^{\prime }\leqslant m_{1}^{\prime }
}}},v,\left\{ \left[ 
\begin{array}{c}
\left( p_{l_{1}},p_{l_{1}^{\prime }l_{1}},k_{l_{1}^{\prime }}\right)  \\ 
\left[ \left\{ \Psi _{J}\left[ l^{\prime },0\right] \right\} ,v\right] 
\end{array}%
\right] \right\} _{_{\substack{ l\subset l_{1}\leqslant m_{1} \\ l^{\prime
}\subset l_{1}^{\prime }\leqslant m_{1}^{\prime }}}}\right)  \\
&&\times \tprod_{\substack{ l\leqslant s \\ l_{i}^{\prime }\leqslant
s^{\prime }}}\Psi _{J,s_{p_{l_{i}^{\prime },0}}^{\otimes k_{l_{i}^{\prime
}}}}^{\otimes k_{l_{i}^{\prime }}}\left( \overline{\left[ p_{0},p_{l_{i}^{%
\prime }0}\right] ^{k_{l_{i}^{\prime }}}},\left\{ \left[
p_{l_{1}},p_{l_{1}^{\prime }l_{1}}\right] ^{k_{l_{1}^{\prime }}}\right\} _{_{%
\mathcal{P}}}\right) \tprod_{\mathcal{P\neq \emptyset }}v_{\left[
p_{l_{1}},p_{l_{1}^{\prime }l_{1}}\right] ^{k_{l_{1}^{\prime }}}}\left\{
\Psi _{J}^{\otimes k_{l_{i}^{\prime \prime }}}\right\}  \\
&&\times \Psi \left( \left\{ \left[ 
\begin{array}{c}
\left( p_{l},p_{l^{\prime }l},k_{l^{\prime }}\right)  \\ 
\left[ \left\{ \Psi _{J}\left[ l^{\prime },0\right] \right\} ,v\right] 
\end{array}%
\right] \right\} ,v\right) d\left( \left\{ \overline{\left[
p_{0},p_{l^{\prime }0}\right] ^{k_{l^{\prime }}}}\right\} _{\substack{ %
l\leqslant s \\ l^{\prime }\leqslant s^{\prime }}},\left\{ \left[
p_{l_{1}},p_{l_{1}^{\prime }l_{1}}\right] ^{k_{l_{1}^{\prime }}}\right\} 
_{\substack{ l_{1}\leqslant m_{1} \\ l_{1}^{\prime }\leqslant m_{1}^{\prime }
}}\right) dv
\end{eqnarray*}%
where:%
\begin{eqnarray*}
&&g^{\mathcal{K}}\left( \left\{ \left[ p_{l_{1}},p_{l_{1}^{\prime }l_{1}}%
\right] ^{k_{l_{1}^{\prime }}}\right\} _{_{\substack{ l_{1}\leqslant m_{1}
\\ l_{1}^{\prime }\leqslant m_{1}^{\prime }}}},\left\{ \overline{\left[
p_{0},p_{l^{\prime }0}\right] ^{k_{l^{\prime }}}}\right\} _{_{\substack{ %
l\leqslant s \\ l^{\prime }\leqslant s^{\prime }}}},v,\left\{ \left[ 
\begin{array}{c}
\left( p_{l_{1}},p_{l_{1}^{\prime }l_{1}},k_{l_{1}^{\prime }}\right)  \\ 
\left[ \left\{ \Psi _{J}\left[ l^{\prime },0\right] \right\} ,v\right] 
\end{array}%
\right] \right\} _{_{\substack{ l\subset l_{1}\leqslant m_{1} \\ l^{\prime
}\subset l_{1}^{\prime }\leqslant m_{1}^{\prime }}}}\right)  \\
&=&\sum_{l}g\left( \left[ p_{l},p_{l^{\prime }l}\right] ^{k_{l^{\prime
}}},v\right) \mathcal{K}_{0}\left( \left\{ \overline{\left[
p_{0},p_{l^{\prime }0}\right] ^{k_{l^{\prime }}}}\right\} _{_{\substack{ %
l\leqslant s \\ l^{\prime }\leqslant s^{\prime }}}},\left\{ \left[
p_{l_{1}},p_{l_{1}^{\prime }l_{1}}\right] ^{k_{l_{1}^{\prime }}}\right\} _{
_{\substack{ l_{1}\leqslant m_{1},l_{1}\neq l \\ l_{1}^{\prime }\leqslant
m_{1}^{\prime }}}},\left\{ \left[ 
\begin{array}{c}
\left( p_{l_{1}},p_{l_{1}^{\prime }l_{1}},k_{l_{1}^{\prime }}\right)  \\ 
\left[ \left\{ \Psi _{J}\left[ l^{\prime },0\right] \right\} ,v\right] 
\end{array}%
\right] \right\} _{_{\substack{ l\subset l_{1}\leqslant m_{1} \\ l^{\prime
}\subset l_{1}^{\prime }\leqslant m_{1}^{\prime }}}}\right) 
\end{eqnarray*}

and:%
\begin{eqnarray*}
&&\tprod_{_{_{\substack{ l_{2}\leqslant m_{2} \\ l_{2}^{\prime }\leqslant
m_{2}^{\prime }}}}}v_{\left[ p_{l_{2}},p_{l_{2}^{\prime }l_{2}}\right]
^{k_{l_{2}^{\prime }}}}\left\{ \Psi _{J}^{\otimes k_{l^{\prime }}}\right\} 
\\
&=&\tprod_{_{_{_{\substack{ l_{2}\leqslant m_{2} \\ l_{2}^{\prime }\leqslant
m_{2}^{\prime }}}}}}\int v\left( \overline{\left[ p_{0}^{\prime
},p_{l^{\prime }0}^{\prime }\right] ^{k_{l^{\prime }}^{\prime }}},\left\{ %
\left[ p_{l_{2}},p_{l_{2}^{\prime }l_{2}}\right] ^{k_{l_{2}^{\prime
}}}\right\} \right) \times \Psi _{J,s_{p_{l^{\prime },0}}^{\otimes
k_{l^{\prime }}}}^{\otimes k_{l^{\prime }}}\left( \overline{\left[
p_{0}^{\prime },p_{l^{\prime }0}^{\prime }\right] ^{k_{l^{\prime }}^{\prime
}}},\left\{ \left[ p_{l_{2}},p_{l_{2}^{\prime }l_{2}}\right]
^{k_{l_{2}^{\prime }}}\right\} \right) d\left( \overline{\left[
p_{0}^{\prime },p_{l^{\prime }0}^{\prime }\right] ^{k_{l^{\prime }}^{\prime
}}}\right) 
\end{eqnarray*}%
Gathering $p_{l_{1}}$ and $p_{l_{2}}\rightarrow p_{l_{1}}$, and $\left[
p_{0},p_{l^{\prime }0}\right] ^{k_{l^{\prime }}}$ nd $\overline{\left[
p_{0}^{\prime },p_{l^{\prime }0}^{\prime }\right] ^{k_{l^{\prime }}^{\prime
}}}$, the previous formula becomes:%
\begin{eqnarray*}
&&\sum_{_{\substack{ \left( s,s^{\prime },\left[ p,p^{\prime },k^{\prime }%
\right] \right)  \\ \left( m,m^{\prime },\left[ p,p^{\prime },k^{\prime }%
\right] \right) _{1}}}}\int \bar{g}^{\mathcal{K}}\left( \left\{ \left[
p_{l_{1}},p_{l_{1}^{\prime }l_{1}}\right] ^{k_{l_{1}^{\prime }}}\right\} _{
_{\substack{ l_{1}\leqslant m_{1} \\ l_{1}^{\prime }\leqslant m_{1}^{\prime }
}}},\left\{ \overline{\left[ p_{0},p_{l^{\prime }0}\right] ^{k_{l^{\prime }}}%
}\right\} _{\substack{ l\leqslant s \\ l^{\prime }\leqslant s^{\prime }}}%
,v,\left\{ \left[ 
\begin{array}{c}
\left( p_{l_{1}},p_{l_{1}^{\prime }l_{1}},k_{l_{1}^{\prime }}\right)  \\ 
\left[ \left\{ \Psi _{J}\left[ l^{\prime },0\right] \right\} ,v\right] 
\end{array}%
\right] \right\} _{_{\substack{ l\subset l_{1}\leqslant m_{1} \\ l^{\prime
}\subset l_{1}^{\prime }\leqslant m_{1}^{\prime }}}}\right)  \\
&&\times \tprod_{\substack{ l\leqslant s \\ l^{\prime }\leqslant s^{\prime }
}}\Psi _{J,s_{p_{l^{\prime },0}}^{\otimes k_{l_{i}^{\prime }}}}^{\otimes
k_{l^{\prime },i}}\left( \overline{\left[ p_{0},p_{l^{\prime }0}\right]
^{k_{l^{\prime }}}},\left\{ \left[ p_{l_{1}},p_{l_{1}^{\prime }l_{1}}\right]
^{k_{l_{1}^{\prime }}}\right\} _{\substack{ l_{1}\leqslant m_{1} \\ %
l_{1}^{\prime }\leqslant m_{1}^{\prime }}}\right) \Psi \left( \left\{ \left[ 
\begin{array}{c}
\left( p_{l},p_{l^{\prime }l},k_{l^{\prime }}\right)  \\ 
\left[ \left\{ \Psi _{J}\left[ l^{\prime },0\right] \right\} ,v\right] 
\end{array}%
\right] \right\} ,v\right)  \\
&&\times d\left( \left\{ \overline{\left[ p_{0},p_{l^{\prime }0}\right]
^{k_{l^{\prime }}}}\right\} _{\substack{ l\leqslant s \\ l^{\prime
}\leqslant s^{\prime }}},\left\{ \left[ p_{l_{1}},p_{l_{1}^{\prime }l_{1}}%
\right] ^{k_{l_{1}^{\prime }}}\right\} _{\substack{ l_{1}\leqslant m_{1} \\ %
l_{1}^{\prime }\leqslant m_{1}^{\prime }}}\right) dv
\end{eqnarray*}%
where:

\begin{eqnarray*}
&&\bar{g}^{\mathcal{K}}\left( \left\{ \left[ p_{l_{1}},p_{l_{1}^{\prime
}l_{1}}\right] ^{k_{l_{1}^{\prime }}}\right\} _{_{\substack{ l_{1}\leqslant
m_{1} \\ l_{1}^{\prime }\leqslant m_{1}^{\prime }}}},\left\{ \overline{\left[
p_{0},p_{l^{\prime }0}\right] ^{k_{l^{\prime }}}}\right\} _{\substack{ %
l\leqslant s \\ l^{\prime }\leqslant s^{\prime }}},v,\left\{ \left[ 
\begin{array}{c}
\left( p_{l_{1}},p_{l_{1}^{\prime }l_{1}},k_{l_{1}^{\prime }}\right)  \\ 
\left[ \left\{ \Psi _{J}\left[ l^{\prime },0\right] \right\} ,v\right] 
\end{array}%
\right] \right\} _{_{\substack{ l\subset l_{1}\leqslant m_{1} \\ l^{\prime
}\subset l_{1}^{\prime }\leqslant m_{1}^{\prime }}}}\right)  \\
&=&\sum_{\mathcal{P},\mathcal{P}_{1}}g^{\mathcal{K}}\left( \left\{ \left[
p_{l_{1}},p_{l_{1}^{\prime }l_{1}}\right] ^{k_{l_{1}^{\prime }}}\right\} _{_{%
\mathcal{P}_{1}}},\left\{ \left[ p_{0},p_{l^{\prime }0}\right]
^{k_{l^{\prime }}}\right\} _{\mathcal{P}},v,\left\{ \left[ 
\begin{array}{c}
\left( p_{l_{1}},p_{l_{1}^{\prime }l_{1}},k_{l_{1}^{\prime }}\right)  \\ 
\left[ \left\{ \Psi _{J}\left[ l^{\prime },0\right] \right\} ,v\right] 
\end{array}%
\right] \right\} _{_{\substack{ l\subset l_{1}\leqslant m_{1} \\ l^{\prime
}\subset l_{1}^{\prime }\leqslant m_{1}^{\prime }}}}\right)  \\
&&\times \tprod\limits_{\mathcal{P}^{c}}v\left( \left\{ \left[
p_{0},p_{l^{\prime }0}\right] ^{k_{l^{\prime }}}\right\} _{\mathcal{P}%
^{c}},\left\{ \left[ p_{l_{1}},p_{l_{1}^{\prime }l_{1}}\right]
^{k_{l_{1}^{\prime }}}\right\} _{_{\mathcal{P}_{1}^{c}}}\right) 
\end{eqnarray*}%
along with:%
\begin{equation*}
\left\{ \left[ p_{0},p_{l^{\prime }0}\right] ^{k_{l^{\prime }}}\right\} _{
_{\substack{ l_{1}\leqslant m_{1} \\ l_{1}^{\prime }\leqslant m_{1}^{\prime }
}}}=\left\{ \left\{ \left[ p_{0},p_{l^{\prime }0}\right] ^{k_{l^{\prime
}}}\right\} _{\mathcal{P}},\left\{ \left[ p_{0},p_{l^{\prime }0}\right]
^{k_{l^{\prime }}}\right\} _{\mathcal{P}^{c}}\right\} 
\end{equation*}%
and:%
\begin{equation*}
\left\{ \left[ p_{l_{1}},p_{l_{1}^{\prime }l_{1}}\right] ^{k_{l_{1}^{\prime
}}}\right\} _{_{\substack{ l_{1}\leqslant m_{1} \\ l_{1}^{\prime }\leqslant
m_{1}^{\prime }}}}=\left\{ \left\{ \left[ p_{l_{1}},p_{l_{1}^{\prime }l_{1}}%
\right] ^{k_{l_{1}^{\prime }}}\right\} _{_{\mathcal{P}_{1}}},\left\{ \left[
p_{l_{1}},p_{l_{1}^{\prime }l_{1}}\right] ^{k_{l_{1}^{\prime }}}\right\} _{_{%
\mathcal{P}_{1}^{c}}}\right\} 
\end{equation*}%
Changing the variables leads to rewrite the functional as:%
\begin{eqnarray*}
&&\sum_{_{\substack{ \left( s,s^{\prime },\left[ p,p^{\prime },k^{\prime }%
\right] \right)  \\ \left( m,m^{\prime },\left[ p,p^{\prime },k^{\prime }%
\right] \right) _{1}}}}\int \bar{g}^{\mathcal{K}}\left( \left\{ \overline{%
\left[ p_{l_{1}},p_{l_{1}^{\prime }l_{1}}\right] ^{k_{l_{1}^{\prime }}}}%
\right\} _{_{\substack{ l_{1}\leqslant m_{1} \\ l_{1}^{\prime }\leqslant
m_{1}^{\prime }}}},\left\{ \left[ p_{0},p_{l^{\prime }0}\right]
^{k_{l^{\prime }}}\right\} _{\substack{ l\leqslant s \\ l^{\prime }\leqslant
s^{\prime }}},v,\left\{ \left[ 
\begin{array}{c}
\left( p_{l_{1}},p_{l_{1}^{\prime }l_{1}},k_{l_{1}^{\prime }}\right)  \\ 
\left[ \left\{ \Psi _{J}\left[ l^{\prime },0\right] \right\} ,v\right] 
\end{array}%
\right] \right\} _{_{\substack{ l\subset l_{1}\leqslant m_{1} \\ l^{\prime
}\subset l_{1}^{\prime }\leqslant m_{1}^{\prime }}}}\right)  \\
&&\times \tprod\limits_{\mathcal{P}^{c}}v\left( \left\{ \left[
p_{0},p_{l^{\prime }0}\right] ^{k_{l^{\prime }}}\right\} _{\mathcal{P}%
^{c}},\left\{ \left[ p_{l_{1}},p_{l_{1}^{\prime }l_{1}}\right]
^{k_{l_{1}^{\prime }}}\right\} _{_{\mathcal{P}_{1}^{c}}}\right)  \\
&&\times \tprod_{\substack{ l\leqslant s \\ l^{\prime }\leqslant s^{\prime }
}}\Psi _{J,s_{p_{l^{\prime },0}}^{\otimes k_{l^{\prime }}}}^{\otimes
k_{l^{\prime }}}\left( \left[ p_{0},p_{l^{\prime }0}\right] ^{k_{l^{\prime
}}}\right) \Psi \left( \left\{ \left[ 
\begin{array}{c}
\left( p_{l},p_{l^{\prime }l},k_{l^{\prime }}\right)  \\ 
\left[ \left\{ \Psi _{J}\left[ l^{\prime },0\right] \right\} ,v\right] 
\end{array}%
\right] \right\} ,v\right) d\left( \left\{ \left[ p_{0},p_{l^{\prime }0}%
\right] ^{k_{l^{\prime }}}\right\} _{\substack{ l\leqslant s \\ l^{\prime
}\leqslant s^{\prime }}},\left\{ \overline{\left[ p_{l_{1}},p_{l_{1}^{\prime
}l_{1}}\right] ^{k_{l_{1}^{\prime }}}}\right\} _{\substack{ l_{1}\leqslant
m_{1} \\ l_{1}^{\prime }\leqslant m_{1}^{\prime }}}\right) dv
\end{eqnarray*}%
with:%
\begin{equation*}
\overline{\left[ p_{l_{1}},p_{l_{1}^{\prime }l_{1}}\right] ^{k_{l^{\prime }}}%
}=\overline{\left[ p_{l_{1}},p_{l_{1}^{\prime }l_{1}}\right] ^{k_{l^{\prime
}}}}/\tprod\limits_{p_{1,l},...,p_{m^{\prime },l}}f_{p_{1,l}...p_{m^{\prime
},l}}
\end{equation*}%
and the integration over the variables $\left\{ \overline{\left[
p_{l_{1}},p_{l_{1}^{\prime }l_{1}}\right] ^{k_{l_{1}^{\prime }}}}\right\} _{
_{\substack{ l_{1}\leqslant m_{1} \\ l_{1}^{\prime }\leqslant m_{1}^{\prime }
}}}$ leads to:%
\begin{eqnarray*}
&\equiv &\sum_{_{\substack{ \left( s,s^{\prime },\left[ p,p^{\prime
},k^{\prime }\right] \right)  \\ \left( m,m^{\prime },\left[ p,p^{\prime
},k^{\prime }\right] \right) _{1}}}}\int \bar{g}_{\nu }^{\mathcal{K}}\left(
\left\{ \left[ p_{0},p_{l^{\prime }0}\right] ^{k_{l^{\prime }}}\right\} 
_{\substack{ l\leqslant s \\ l^{\prime }\leqslant s^{\prime }}},v,\left\{ %
\left[ 
\begin{array}{c}
\left( p_{l_{1}},p_{l_{1}^{\prime }l_{1}},k_{l_{1}^{\prime }}\right)  \\ 
\left[ \left\{ \Psi _{J}\left[ l^{\prime },0\right] \right\} ,v\right] 
\end{array}%
\right] \right\} _{_{\substack{ l\subset l_{1}\leqslant m_{1} \\ l^{\prime
}\subset l_{1}^{\prime }\leqslant m_{1}^{\prime }}}}\right)  \\
&&\times \tprod_{\substack{ l\leqslant s \\ l^{\prime }\leqslant s^{\prime }
}}\Psi _{J,s_{p_{l^{\prime },0}}^{\otimes k_{l^{\prime }}}}^{\otimes
k_{l^{\prime }}}\left( \left[ p_{0},p_{l^{\prime }0}\right] ^{k_{l^{\prime
}}}\right) \Psi \left( \left\{ \left[ 
\begin{array}{c}
\left( p_{l},p_{l^{\prime }l},k_{l^{\prime }}\right)  \\ 
\left[ \left\{ \Psi _{J}\left[ l^{\prime },0\right] \right\} ,v\right] 
\end{array}%
\right] \right\} ,v\right) d\left( \left\{ \left[ p_{0},p_{l^{\prime }0}%
\right] ^{k_{l^{\prime }}}\right\} _{\substack{ l\leqslant s \\ l^{\prime
}\leqslant s^{\prime }}}\right) dv
\end{eqnarray*}%
where:%
\begin{eqnarray*}
&&\bar{g}_{\nu }^{\mathcal{K}}\left( \left\{ \overline{\left[
p_{l_{1}},p_{l_{1}^{\prime }l_{1}}\right] ^{k_{l_{1}^{\prime }}}}\right\} _{
_{\substack{ l_{1}\leqslant m_{1} \\ l_{1}^{\prime }\leqslant m_{1}^{\prime }
}}},\left\{ \left[ p_{0},p_{l^{\prime }0}\right] ^{k_{l^{\prime }}}\right\} 
_{\substack{ l\leqslant s \\ l^{\prime }\leqslant s^{\prime }}},v,\left\{ %
\left[ 
\begin{array}{c}
\left( p_{l_{1}},p_{l_{1}^{\prime }l_{1}},k_{l_{1}^{\prime }}\right)  \\ 
\left[ \left\{ \Psi _{J}\left[ l^{\prime },0\right] \right\} ,v\right] 
\end{array}%
\right] \right\} _{_{\substack{ l\subset l_{1}\leqslant m_{1} \\ l^{\prime
}\subset l_{1}^{\prime }\leqslant m_{1}^{\prime }}}}\right)  \\
&=&\int \bar{g}^{\mathcal{K}}\left( \left\{ \overline{\left[
p_{l_{1}},p_{l_{1}^{\prime }l_{1}}\right] ^{k_{l_{1}^{\prime }}}}\right\} _{
_{\substack{ l_{1}\leqslant m_{1} \\ l_{1}^{\prime }\leqslant m_{1}^{\prime }
}}},\left\{ \left[ p_{0},p_{l^{\prime }0}\right] ^{k_{l^{\prime }}}\right\} 
_{\substack{ l\leqslant s \\ l^{\prime }\leqslant s^{\prime }}},v,\left\{ %
\left[ 
\begin{array}{c}
\left( p_{l_{1}},p_{l_{1}^{\prime }l_{1}},k_{l_{1}^{\prime }}\right)  \\ 
\left[ \left\{ \Psi _{J}\left[ l^{\prime },0\right] \right\} ,v\right] 
\end{array}%
\right] \right\} _{_{\substack{ l\subset l_{1}\leqslant m_{1} \\ l^{\prime
}\subset l_{1}^{\prime }\leqslant m_{1}^{\prime }}}}\right)  \\
&&\times \tprod\limits_{\mathcal{P}^{c}}v\left( \left\{ \left[
p_{0},p_{l^{\prime }0}\right] ^{k_{l^{\prime }}}\right\} _{\mathcal{P}%
^{c}},\left\{ \left[ p_{l_{1}},p_{l_{1}^{\prime }l_{1}}\right]
^{k_{l_{1}^{\prime }}}\right\} _{_{\mathcal{P}_{1}^{c}}}\right) d\left(
\left\{ \overline{\left[ p_{l_{1}},p_{l_{1}^{\prime }l_{1}}\right]
^{k_{l_{1}^{\prime }}}}\right\} _{\substack{ l_{1}\leqslant m_{1} \\ %
l_{1}^{\prime }\leqslant m_{1}^{\prime }}}\right) 
\end{eqnarray*}

Ultimately, if we choose for the functional: 
\begin{equation*}
v_{\left\{ \left[ p_{l},p_{l^{\prime }l}\right] ^{k_{l^{\prime }}}\right\}
}=v_{\left[ p_{l},p_{l^{\prime }l}\right] ^{k_{l^{\prime }}},\alpha }^{%
\overline{\left[ p_{0},p_{l^{\prime }0}\right] ^{k_{l^{\prime }}}}}\left\{
\Psi _{J,\alpha ,s_{p_{l^{\prime },0}}^{\otimes k_{l^{\prime }}}}^{\otimes
k_{l^{\prime }}}\right\} 
\end{equation*}%
evaluated at $\Psi _{J,\alpha ,s_{p_{l^{\prime },0}}^{\otimes k_{l^{\prime
}}}}^{\otimes k_{l^{\prime }}}$ or equivalently if we consider subobjects
that are choosen as eigenstates of operators $\Lambda $ the functional
simplifies:%
\begin{eqnarray*}
&&\sum_{_{\substack{ \left( s,s^{\prime },\left[ p,p^{\prime },k^{\prime }%
\right] \right)  \\ \left( m,m^{\prime },\left[ p,p^{\prime },k^{\prime }%
\right] \right) _{1}}}}\int \bar{g}^{\mathcal{K}}\left( \left\{ \left[
p_{0},p_{l^{\prime }0}\right] ^{k_{l^{\prime }}}\right\} _{\substack{ %
l\leqslant s \\ l^{\prime }\leqslant s^{\prime }}},\left\{ \left[ 
\begin{array}{c}
\left( p_{l_{1}},p_{l_{1}^{\prime }l_{1}},k_{l_{1}^{\prime }}\right)  \\ 
\left[ \left\{ \Psi _{J}\left[ l^{\prime },0\right] \right\} ,v\right] ,%
\left[ p_{0},p_{l^{\prime }0}\right] ^{k_{l^{\prime }}}%
\end{array}%
\right] \right\} _{_{\substack{ l\subset l_{1}\leqslant m_{1} \\ l^{\prime
}\subset l_{1}^{\prime }\leqslant m_{1}^{\prime }}}}\right)  \\
&&\times \tprod_{\substack{ l\leqslant s \\ l^{\prime }\leqslant s^{\prime }
}}\Psi _{J,s_{p_{l^{\prime },0}}^{\otimes k_{l^{\prime }}}}^{\otimes
k_{l^{\prime }}}\left( \left[ p_{0},p_{l^{\prime }0}\right] ^{k_{l^{\prime
}}}\right) \Psi \left( \left\{ \left[ 
\begin{array}{c}
\left( p_{l},p_{l^{\prime }l},k_{l^{\prime }}\right)  \\ 
\left[ \left\{ \Psi _{J}\left[ l^{\prime },0\right] \right\} \right] ,\left[
p_{0},p_{l^{\prime }0}\right] ^{k_{l^{\prime }}}%
\end{array}%
\right] \right\} \right) d\left( \left\{ \left[ p_{0},p_{l^{\prime }0}\right]
^{k_{l^{\prime }}}\right\} _{\substack{ l\leqslant s \\ l^{\prime }\leqslant
s^{\prime }}}\right) 
\end{eqnarray*}%
with $\bar{g}^{\mathcal{K}}=\bar{g}_{1}^{\mathcal{K}}$.

Reintroducing the components $\alpha $ and summing over components yields
the functional:%
\begin{eqnarray*}
&&\sum_{_{\substack{ \left( s,s^{\prime },\left[ p,p^{\prime },k^{\prime }%
\right] \right)  \\ \left( m,m^{\prime },\left[ p,p^{\prime },k^{\prime }%
\right] \right) _{1}}}}\int \bar{g}^{\mathcal{K}}\left( \left\{ \left[
p_{0},p_{l^{\prime }0}\right] ^{k_{l^{\prime }}}\right\} _{\substack{ %
l\leqslant s \\ l^{\prime }\leqslant s^{\prime }}},\left\{ \left[ 
\begin{array}{c}
\left( p_{l_{1}},p_{l_{1}^{\prime }l_{1}},k_{l_{1}^{\prime }}\right)  \\ 
\left[ \left\{ \Psi _{J}\left[ l^{\prime },0\right] \right\} ,v\right] ,%
\left[ p_{0},p_{l^{\prime }0}\right] ^{k_{l^{\prime }}}%
\end{array}%
\right] \right\} _{_{\substack{ l\subset l_{1}\leqslant m_{1} \\ l^{\prime
}\subset l_{1}^{\prime }\leqslant m_{1}^{\prime }}}}\right)  \\
&&\times \tprod_{\substack{ l\leqslant s \\ l^{\prime }\leqslant s^{\prime }
}}\Psi _{J,s_{p_{l^{\prime },0}}^{\otimes k_{l^{\prime }}}}^{\otimes
k_{l^{\prime }}}\left( \left[ p_{0},p_{l^{\prime }0}\right] ^{k_{l^{\prime
}}},\left\{ \left[ 
\begin{array}{c}
\left( p_{l},p_{l^{\prime }l},k_{l^{\prime }}\right)  \\ 
\left[ \left\{ \Psi _{J}\left[ l^{\prime },0\right] \right\} \right] ,\left[
p_{0},p_{l^{\prime }0}\right] ^{k_{l^{\prime }}}%
\end{array}%
\right] \right\} \right) d\left( \left\{ \left[ p_{0},p_{l^{\prime }0}\right]
^{k_{l^{\prime }}}\right\} _{\substack{ l\leqslant s \\ l^{\prime }\leqslant
s^{\prime }}}\right) 
\end{eqnarray*}%
Now, we perform the sum over $\left( m,m^{\prime },\left[ p,p^{\prime
},k^{\prime }\right] \right) _{1}$ and define:%
\begin{eqnarray*}
&&\bar{g}^{\mathcal{K}}\left( \left\{ \left[ p_{0},p_{l^{\prime }0}\right]
^{k_{l^{\prime }}}\right\} _{\substack{ l\leqslant s \\ l^{\prime }\leqslant
s^{\prime }}},\left\{ \left[ 
\begin{array}{c}
\left( p_{l_{1}},p_{l_{1}^{\prime }l_{1}},k_{l_{1}^{\prime }}\right)  \\ 
\left[ \left\{ \Psi _{J}\left[ l^{\prime },0\right] \right\} ,v\right] ,%
\left[ p_{0},p_{l^{\prime }0}\right] ^{k_{l^{\prime }}}%
\end{array}%
\right] \right\} _{_{\substack{ l\subset l_{1}\leqslant m_{1} \\ l^{\prime
}\subset l_{1}^{\prime }\leqslant m_{1}^{\prime }}}}\right)  \\
&=&\sum_{\left( m,m^{\prime },\left[ p,p^{\prime },k^{\prime }\right]
\right) _{1}}\bar{g}^{\mathcal{K}}\left( \left\{ \left[ p_{0},p_{l^{\prime
}0}\right] ^{k_{l^{\prime }}}\right\} _{\substack{ l\leqslant s \\ l^{\prime
}\leqslant s^{\prime }}},\left\{ \left[ 
\begin{array}{c}
\left( p_{l_{1}},p_{l_{1}^{\prime }l_{1}},k_{l_{1}^{\prime }}\right)  \\ 
\left[ \left\{ \Psi _{J}\left[ l^{\prime },0\right] \right\} ,v\right] ,%
\left[ p_{0},p_{l^{\prime }0}\right] ^{k_{l^{\prime }}}%
\end{array}%
\right] \right\} _{_{\substack{ l\subset l_{1}\leqslant m_{1} \\ l^{\prime
}\subset l_{1}^{\prime }\leqslant m_{1}^{\prime }}}}\right) 
\end{eqnarray*}%
which leads to:%
\begin{eqnarray*}
&&\sum_{_{\left( s,s^{\prime },\left[ p,p^{\prime },k^{\prime }\right]
\right) }}\int \bar{g}^{\mathcal{K}}\left( \left\{ \left[ p_{0},p_{l^{\prime
}0}\right] ^{k_{l^{\prime }}}\right\} _{\substack{ l\leqslant s \\ l^{\prime
}\leqslant s^{\prime }}},\left\{ \left[ 
\begin{array}{c}
\left( p_{l},p_{l^{\prime }l},k_{l^{\prime }}\right)  \\ 
\left[ \left\{ \Psi _{J}\left[ l^{\prime },0\right] \right\} \right] ,\left[
p_{0},p_{l^{\prime }0}\right] ^{k_{l^{\prime }}}%
\end{array}%
\right] \right\} \right)  \\
&&\times \tprod_{\substack{ l\leqslant s \\ l^{\prime }\leqslant s^{\prime }
}}\Psi _{J,s_{p_{l^{\prime },0}}^{\otimes k_{l^{\prime }}}}^{\otimes
k_{l^{\prime }}}\left( \left[ p_{0},p_{l^{\prime }0}\right] ^{k_{l^{\prime
}}},\left\{ \left[ 
\begin{array}{c}
\left( p_{l},p_{l^{\prime }l},k_{l^{\prime }}\right)  \\ 
\left[ \left\{ \Psi _{J}\left[ l^{\prime },0\right] \right\} \right] ,\left[
p_{0},p_{l^{\prime }0}\right] ^{k_{l^{\prime }}}%
\end{array}%
\right] \right\} \right) d\left( \left\{ \left[ p_{0},p_{l^{\prime }0}\right]
^{k_{l^{\prime }}}\right\} _{\substack{ l\leqslant s \\ l^{\prime }\leqslant
s^{\prime }}}\right) 
\end{eqnarray*}

\subsubsection*{Projection of several subobjects}

when several subobjects are projected, the generalization is
straightforward. Defining: 
\begin{eqnarray*}
&&g^{\mathcal{K}}\left( \overline{\left[ p_{0},p_{l^{\prime }0}\right]
^{k_{l^{\prime }}}},\left\{ \left[ p_{l_{i,1}},p_{l_{i,1}^{\prime }l_{i,1}}%
\right] ^{k_{l_{i,1}^{\prime }}}\right\} _{_{\substack{ l_{i,1}\leqslant
m_{i,1}  \\ l_{i,1}^{\prime }\leqslant m_{i,1}^{\prime }}}},v,\left\{ \left[ 
\begin{array}{c}
\left( p_{l_{i,1}},p_{l_{i,1}^{\prime }l_{i,1}},k_{l_{i,1}^{\prime }}\right)
\\ 
\left[ \left\{ \Psi _{J}\left[ l^{\prime },0\right] \right\} ,v\right]%
\end{array}%
\right] \right\} _{_{\substack{ l_{i}\subset l_{i,1}\leqslant m_{i,1}  \\ %
l_{i}^{\prime }\subset l_{i,1}^{\prime }\leqslant m_{i,1}^{\prime }}}}\right)
\\
&=&\sum_{l}g\left( \left[ p_{l_{i}},p_{l_{i}^{\prime }l_{i}}\right]
^{k_{l_{i}^{\prime }}},v\right) \tprod\limits_{i}\mathcal{K}_{0}\left(
\left\{ \overline{\left[ p_{0},p_{l^{\prime }0}\right] ^{k_{l^{\prime }}}}%
\right\} _{_{\substack{ l\leqslant s  \\ l^{\prime }\leqslant s^{\prime }}}%
},\left\{ \left[ p_{l_{i,1}},p_{l_{i,1}^{\prime }l_{i,1}}\right]
^{k_{l_{i,1}^{\prime }}}\right\} _{_{\substack{ l_{i,1}\leqslant
m_{i,1},l_{i,1}\neq l_{i}  \\ l_{i,1}^{\prime }\leqslant m_{i,1}^{\prime }}}%
},v,\left\{ \left[ 
\begin{array}{c}
\left( p_{l_{i,1}},p_{l_{i,1}^{\prime }l_{i,1}},k_{l_{i,1}^{\prime }}\right)
\\ 
\left[ \left\{ \Psi _{J}\left[ l^{\prime },0\right] \right\} ,v\right]%
\end{array}%
\right] \right\} _{_{\substack{ l_{i}\subset l_{i,1}\leqslant m_{i,1}  \\ %
l_{i}^{\prime }\subset l_{i,1}^{\prime }\leqslant m_{i,1}^{\prime }}}}\right)
\end{eqnarray*}%
we obtain the formula presented in the text.

\subsection*{Projection over eigenstates of operator}

By changing variables in (\ref{FPT}), this expression writes:%
\begin{eqnarray*}
&&F_{f,lin}\left( \left\{ \Psi _{J,s_{p_{l^{\prime },0}}^{\otimes
k_{l^{\prime }}}}^{\otimes k_{l^{\prime }}}\left( \left[ p_{0},p_{l^{\prime
}0}\right] ^{k_{l^{\prime }}}\right) \right\} _{\left( p_{l^{\prime
}0}\right) }\right)  \\
&=&\sum_{m}\sum_{m^{\prime }}\int \sum_{_{\substack{ \left( s,s^{\prime },%
\left[ p,p^{\prime },k^{\prime }\right] \right)  \\ \left( m,m^{\prime },%
\left[ p,p^{\prime },k^{\prime }\right] \right) _{1}}}}d\left\{ \left[
p_{0},p_{l^{\prime }0}\right] ^{k_{l^{\prime }}}\right\} _{\substack{ %
l\leqslant s \\ l^{\prime }\leqslant s^{\prime }}}d\left\{ \left[ 
\begin{array}{c}
\left( p_{l},p_{l^{\prime }l},k_{l^{\prime }}\right)  \\ 
\left[ \left\{ \Psi _{J}\left[ l^{\prime },0\right] \right\} \right] ,\left[
p_{0},p_{l^{\prime }0}\right] ^{k_{l^{\prime }}}%
\end{array}%
\right] \right\} \bar{g}\left( \left\{ \left[ p_{0},p_{l^{\prime }0}\right]
^{k_{l^{\prime }}}\right\} _{_{\substack{ l\leqslant s \\ l^{\prime
}\leqslant s^{\prime }}}},\left\{ \left[ 
\begin{array}{c}
\left( p_{l_{1}},p_{l_{1}^{\prime }l_{1}},k_{l_{1}^{\prime }}\right)  \\ 
\left[ \left\{ \Psi _{J}\left[ l^{\prime },0\right] \right\} ,v\right] ,%
\left[ p_{0},p_{l^{\prime }0}\right] ^{k_{l^{\prime }}}%
\end{array}%
\right] \right\} _{_{_{\substack{ l\subset l_{1}\leqslant m_{1} \\ l^{\prime
}\subset l_{1}^{\prime }\leqslant m_{1}^{\prime }}}}}\right)  \\
&&\times \tprod_{\substack{ l\leqslant s \\ l^{\prime }\leqslant s^{\prime }
}}\Psi _{J,s_{p_{l^{\prime },0}}^{\otimes k_{l^{\prime }}}}^{\otimes
k_{l^{\prime }}}\left( \overline{\left[ p_{0},p_{l^{\prime }0}\right]
^{k_{l^{\prime }}}},\left\{ \left[ p_{l_{1}},p_{l_{1}^{\prime }l_{1}}\right]
^{k_{l_{1}^{\prime }}}\right\} _{_{\mathcal{P}}}\right) \tprod_{\mathcal{P}%
^{c}}v_{\left[ p_{l_{1}},p_{l_{1}^{\prime }l_{1}}\right] ^{k_{l_{1}^{\prime
}}}}\left\{ \Psi _{J}^{\otimes k_{l^{\prime }}}\right\} \Psi \left( \left\{ %
\left[ 
\begin{array}{c}
\left( p_{l},p_{l^{\prime }l},k_{l^{\prime }}\right)  \\ 
\left[ \left\{ \Psi _{J}\left[ l^{\prime },0\right] \right\} \right] ,\left[
p_{0},p_{l^{\prime }0}\right] ^{k_{l^{\prime }}}%
\end{array}%
\right] \right\} \right) 
\end{eqnarray*}%
\begin{eqnarray*}
&&\bar{g}\left( \left\{ \left[ p_{0},p_{l^{\prime }0}\right] ^{k_{l^{\prime
}}}\right\} _{_{\substack{ l\leqslant s \\ l^{\prime }\leqslant s^{\prime }}}%
},\left\{ \left[ 
\begin{array}{c}
\left( p_{l_{1}},p_{l_{1}^{\prime }l_{1}},k_{l_{1}^{\prime }}\right)  \\ 
\left[ \left\{ \Psi _{J}\left[ l^{\prime },0\right] \right\} ,v\right] ,%
\left[ p_{0},p_{l^{\prime }0}\right] ^{k_{l^{\prime }}}%
\end{array}%
\right] \right\} _{_{_{\substack{ l\subset l_{1}\leqslant m_{1} \\ l^{\prime
}\subset l_{1}^{\prime }\leqslant m_{1}^{\prime }}}}}\right)  \\
&=&\int g\left( \left\{ \overline{\left[ p_{l},p_{l^{\prime }l}\right] }%
\right\} _{l,l^{\prime }},\left\{ \left[ p_{0},p_{l^{\prime }0}\right]
\right\} _{l^{\prime }}\right)  \\
&&\tprod\limits_{l^{\prime }}\mathcal{V}_{0}\left( \left\{ \left[
p_{0},p_{l^{\prime }0}\right] ^{k_{l^{\prime }}}\right\} _{_{\substack{ %
l\leqslant s \\ l^{\prime }\leqslant s^{\prime }}}},\left\{ \overline{\left[
p_{l_{1}},p_{l_{1}^{\prime }l_{1}}\right] ^{k_{l_{1}^{\prime }}}}\right\} _{
_{\substack{ l_{1}\leqslant m_{1} \\ l_{1}^{\prime }\leqslant m_{1}^{\prime }
}}},\left\{ \left[ 
\begin{array}{c}
\left( p_{l_{1}},p_{l_{1}^{\prime }l_{1}},k_{l_{1}^{\prime }}\right)  \\ 
\left[ \left\{ \Psi _{J}\left[ l^{\prime },0\right] \right\} ,v\right] ,%
\left[ p_{0},p_{l^{\prime }0}\right] ^{k_{l^{\prime }}}%
\end{array}%
\right] \right\} _{_{_{\substack{ l\subset l_{1}\leqslant m_{1} \\ l^{\prime
}\subset l_{1}^{\prime }\leqslant m_{1}^{\prime }}}}}\right) d\overline{%
\left[ p_{l_{i}},p_{l_{i}^{\prime }l_{i}}\right] ^{k_{l_{i}^{\prime }}}}
\end{eqnarray*}%
Then, introducing the sum over copies yields:%
\begin{eqnarray*}
&&F_{f,lin}\left( \left\{ \Psi _{J,s_{p_{l^{\prime },0}}^{\otimes
k_{l^{\prime }}}}^{\otimes k_{l^{\prime }}}\left( \left[ p_{0},p_{l^{\prime
}0}\right] ^{k_{l^{\prime }}}\right) \right\} _{\left( p_{l^{\prime
}0}\right) }\right)  \\
&=&\sum_{m}\sum_{m^{\prime }}\int \sum_{\alpha }\sum_{_{\substack{ \left(
s,s^{\prime },\left[ p,p^{\prime },k^{\prime }\right] \right)  \\ \left(
m,m^{\prime },\left[ p,p^{\prime },k^{\prime }\right] \right) _{1}}}%
}d\left\{ \left[ p_{0},p_{l^{\prime }0}\right] ^{k_{l^{\prime }}}\right\} 
_{\substack{ l\leqslant s \\ l^{\prime }\leqslant s^{\prime }}}d\left\{ %
\left[ 
\begin{array}{c}
\left( p_{l},p_{l^{\prime }l},k_{l^{\prime }}\right)  \\ 
\left[ \left\{ \Psi _{J}\left[ l^{\prime },0\right] \right\} \right] ,\left[
p_{0},p_{l^{\prime }0}\right] ^{k_{l^{\prime }}}%
\end{array}%
\right] \right\}  \\
&&\times \bar{g}\left( \left\{ \left[ p_{0},p_{l^{\prime }0}\right]
^{k_{l^{\prime }}}\right\} _{_{\substack{ l\leqslant s \\ l^{\prime
}\leqslant s^{\prime }}}},\left\{ \left[ 
\begin{array}{c}
\left( p_{l_{1}},p_{l_{1}^{\prime }l_{1}},k_{l_{1}^{\prime }}\right)  \\ 
\left[ \left\{ \Psi _{J}\left[ l^{\prime },0\right] \right\} ,v\right] ,%
\left[ p_{0},p_{l^{\prime }0}\right] ^{k_{l^{\prime }}}%
\end{array}%
\right] \right\} _{_{\substack{ l\subset l_{1}\leqslant m_{1} \\ l^{\prime
}\subset l_{1}^{\prime }\leqslant m_{1}^{\prime }}}}\right)  \\
&&\times \tprod_{\substack{ l\leqslant s \\ l^{\prime }\leqslant s^{\prime }
}}\Psi _{J,\alpha ,s_{p_{l^{\prime },0}}^{\otimes k_{l^{\prime }}}}^{\otimes
k_{l^{\prime }}}\left( \overline{\left[ p_{0},p_{l^{\prime }0}\right]
^{k_{l^{\prime }}}},\left\{ \left[ p_{l_{1}},p_{l_{1}^{\prime }l_{1}}\right]
^{k_{l_{1}^{\prime }}}\right\} _{_{\mathcal{P}}}\right) \tprod_{\mathcal{P}%
^{c}}v_{\left[ p_{l_{1}},p_{l_{1}^{\prime }l_{1}}\right] ^{k_{l_{1}^{\prime
}}}}\left\{ \Psi _{J}^{\otimes k_{l^{\prime }}}\right\} \Psi _{\alpha
}\left( \left\{ \left[ 
\begin{array}{c}
\left( p_{l},p_{l^{\prime }l},k_{l^{\prime }}\right)  \\ 
\left[ \left\{ \Psi _{J}\left[ l^{\prime },0\right] \right\} \right] ,\left[
p_{0},p_{l^{\prime }0}\right] ^{k_{l^{\prime }}}%
\end{array}%
\right] \right\} \right) 
\end{eqnarray*}%
As before if the set $\left\{ \left\{ \left[ 
\begin{array}{c}
\left[ p_{l_{i}},p_{l_{i}^{\prime }l_{i}},k_{l_{i}^{\prime }}\right]  \\ 
\left[ \left\{ \Psi _{J}\left[ l^{\prime },0\right] \right\} ,v\right] 
\end{array}%
\right] \right\} \right\} $ is independent of the realization, the previous
formula becomes:%
\begin{eqnarray*}
&&F_{f,lin}\left( \left\{ \Psi _{J,s_{p_{l^{\prime },0}}^{\otimes
k_{l^{\prime }}}}^{\otimes k_{l^{\prime }}}\left( \left[ p_{0},p_{l^{\prime
}0}\right] ^{k_{l^{\prime }}}\right) \right\} _{\left( p_{l^{\prime
}0}\right) }\right)  \\
&=&\sum_{m}\sum_{m^{\prime }}\int \sum_{_{\substack{ \left( s,s^{\prime },%
\left[ p,p^{\prime },k^{\prime }\right] \right)  \\ \left( m,m^{\prime },%
\left[ p,p^{\prime },k^{\prime }\right] \right) _{1}}}}d\left\{ \left[
p_{0},p_{l^{\prime }0}\right] ^{k_{l^{\prime }}}\right\} _{\substack{ %
l\leqslant s \\ l^{\prime }\leqslant s^{\prime }}}d\left\{ \left[ 
\begin{array}{c}
\left( p_{l},p_{l^{\prime }l},k_{l^{\prime }}\right)  \\ 
\left[ \left\{ \Psi _{J}\left[ l^{\prime },0\right] \right\} \right] ,\left[
p_{0},p_{l^{\prime }0}\right] ^{k_{l^{\prime }}}%
\end{array}%
\right] \right\}  \\
&&\times \bar{g}\left( \left\{ \left[ p_{0},p_{l^{\prime }0}\right]
^{k_{l^{\prime }}}\right\} _{_{\substack{ l\leqslant s \\ l^{\prime
}\leqslant s^{\prime }}}},\left\{ \left[ 
\begin{array}{c}
\left( p_{l_{1}},p_{l_{1}^{\prime }l_{1}},k_{l_{1}^{\prime }}\right)  \\ 
\left[ \left\{ \Psi _{J}\left[ l^{\prime },0\right] \right\} ,v\right] ,%
\left[ p_{0},p_{l^{\prime }0}\right] ^{k_{l^{\prime }}}%
\end{array}%
\right] \right\} _{_{\substack{ l\subset l_{1}\leqslant m_{1} \\ l^{\prime
}\subset l_{1}^{\prime }\leqslant m_{1}^{\prime }}}}\right) \Psi
_{J,\tprod\limits_{l^{\prime }}s_{p_{l^{\prime },0}}^{\otimes k_{l^{\prime
}}}}^{\otimes \sum_{l^{\prime }}k_{l^{\prime }}}\left(
\tprod\limits_{l^{\prime }}\left[ p_{0},p_{l^{\prime }0}\right]
^{k_{l^{\prime },i}},\left\{ \left[ 
\begin{array}{c}
\left( p_{l},p_{l^{\prime }l},k_{l^{\prime }}\right)  \\ 
\left[ \left\{ \Psi _{J}\left[ l^{\prime },0\right] \right\} \right] ,\left[
p_{0},p_{l^{\prime }0}\right] ^{k_{l^{\prime }}}%
\end{array}%
\right] \right\} \right) 
\end{eqnarray*}%
where:%
\begin{eqnarray*}
&&\Psi _{J,\tprod\limits_{l^{\prime }}s_{p_{l^{\prime },0}}^{\otimes
k_{l^{\prime }}}}^{\otimes \sum_{l^{\prime }}k_{l^{\prime }}}\left(
\tprod\limits_{l^{\prime }}\left[ p_{0},p_{l^{\prime }0}\right]
^{k_{l^{\prime },i}},\left\{ \left[ 
\begin{array}{c}
\left( p_{l},p_{l^{\prime }l},k_{l^{\prime }}\right)  \\ 
\left[ \left\{ \Psi _{J}\left[ l^{\prime },0\right] \right\} \right] ,\left[
p_{0},p_{l^{\prime }0}\right] ^{k_{l^{\prime }}}%
\end{array}%
\right] \right\} \right)  \\
&=&\sum_{\mathcal{P},\alpha }\tprod_{\substack{ l\leqslant s \\ l^{\prime
}\leqslant s^{\prime }}}\Psi _{J,\alpha ,s_{p_{l^{\prime },0}}^{\otimes
k_{l^{\prime }}}}^{\otimes k_{l^{\prime }}}\left( \overline{\left[
p_{0},p_{l^{\prime }0}\right] ^{k_{l^{\prime }}}},\left\{ \left[
p_{l_{1}},p_{l_{1}^{\prime }l_{1}}\right] ^{k_{l_{1}^{\prime }}}\right\} _{_{%
\mathcal{P}}}\right)  \\
&&\times \tprod_{\mathcal{P}^{c}}v_{\left[ p_{l_{1}},p_{l_{1}^{\prime }l_{1}}%
\right] ^{k_{l_{1}^{\prime }}}}\left\{ \Psi _{J}^{\otimes k_{l^{\prime
}}}\right\} \Psi _{\alpha }\left( \left\{ \left[ 
\begin{array}{c}
\left( p_{l},p_{l^{\prime }l},k_{l^{\prime }}\right)  \\ 
\left[ \left\{ \Psi _{J}\left[ l^{\prime },0\right] \right\} \right] ,\left[
p_{0},p_{l^{\prime }0}\right] ^{k_{l^{\prime }}}%
\end{array}%
\right] \right\} \right) 
\end{eqnarray*}

\subsection*{Sequences of sets of subobjects and parameters}

The effective field (\ref{FS}) and their functionals (\ref{Fs}) where
defined for a sequence of two maps $\left[ p_{l_{i}},p_{l_{i}^{\prime }l_{i}}%
\right] $ and tensor power of states $k_{l_{i}^{\prime }}$:%
\begin{equation}
\tprod_{i}\Psi _{J,\alpha _{i},s_{p_{\eta _{i^{\prime }}^{\prime }\eta
_{i}}}^{\otimes k_{\eta _{i^{\prime }}^{\prime }\eta _{i}}}}^{\otimes
k_{\eta _{i^{\prime }}^{\prime }\eta _{i}}}\left( \left[ p_{\eta
_{i}},p_{\eta _{i^{\prime }}^{\prime }\eta _{i}}\right] ^{k_{\eta
_{i^{\prime }}^{\prime }\eta _{i}}},\left\{ \left[ 
\begin{array}{c}
\left( p_{l_{i}},p_{l_{i}^{\prime }l_{i}},k_{l_{i}^{\prime }}\right)  \\ 
\left[ \left\{ \Psi _{J,\alpha }\left[ p_{\eta },p_{\eta ^{\prime }\eta }%
\right] \right\} ,v\right] 
\end{array}%
\right] \right\} ,v_{\eta }\right) 
\end{equation}%
and the system was described by several field $\Psi _{J,s_{p_{l^{\prime
},\eta }}^{\otimes k_{l^{\prime }\eta }}}^{\otimes k_{l^{\prime }\eta }}$
while the functionls are describd b (\ref{Fs}).

Considering rather general sequences:%
\begin{equation*}
\left[ p_{l_{i}},p_{l_{i}^{\prime }l_{i}},...p_{...l_{i}^{\prime \prime
}l_{i}^{\prime }l_{i}},..\right]
\end{equation*}%
that is:%
\begin{equation}
\left[ p_{\left( \mathbf{l}_{i}\right) _{1}},p_{\left( \mathbf{l}_{i}\right)
_{2}},...p_{\left( \mathbf{l}_{i}\right) _{n}},...\right]
\end{equation}%
where the $\left( \mathbf{l}_{i}\right) _{n}$ are the $n$ first terms of any
infinite sequnce:%
\begin{equation*}
\left( \mathbf{l}_{i}\right) _{\infty }
\end{equation*}%
We can write (\ref{Ms}) as:%
\begin{equation*}
\left[ \mathbf{p}_{\left( \mathbf{l}_{i}\right) _{\infty }}\right]
\end{equation*}%
representing in infinite sequence of subobject. We can also consider the
sequenc for any "rank $k$" subobjc:%
\begin{equation*}
\left[ \mathbf{p}_{\left( \mathbf{l}_{i}\right) _{\geqslant k,\infty }}%
\right]
\end{equation*}%
starting t $k$ and partial seqnc:%
\begin{equation*}
\left[ \mathbf{p}_{\left( \mathbf{l}_{i}\right) _{<k}}\right]
\end{equation*}%
ending at $k-1$, and:%
\begin{equation*}
\left[ \mathbf{p}_{\left( \mathbf{l}_{i}\right) _{\geqslant l,<k}}\right]
\end{equation*}%
starting at $l$ and endg at $k-1$, so that:%
\begin{eqnarray*}
\left[ \mathbf{p}_{\left( \mathbf{l}_{i}\right) _{\infty }}\right] &=&\left[ 
\mathbf{p}_{\left( \mathbf{l}_{i}\right) _{\geqslant k,\infty }}\right] %
\left[ \mathbf{p}_{\left( \mathbf{l}_{i}\right) _{<k}}\right] \\
&=&\left[ \mathbf{p}_{\left( \mathbf{l}_{i}\right) _{\geqslant k,\infty }}%
\right] \left[ \mathbf{p}_{\left( \mathbf{l}_{i}\right) _{\geqslant l,<k}}%
\right] \left[ \mathbf{p}_{\left( \mathbf{l}_{i}\right) _{<l}}\right]
\end{eqnarray*}

Note that the sequencs may end at different object indexd by $\eta $ leadn
to define:%
\begin{equation*}
\left[ \mathbf{p}_{\left( \eta _{i},\mathbf{l}_{i}\right) _{\infty }}\right]
,\left[ \mathbf{p}_{\left( \eta _{i},\mathbf{l}_{i}\right) _{<k}}\right] ,%
\left[ \mathbf{p}_{\left( \mathbf{l}_{i}\right) _{\eta ,\geqslant k,\infty }}%
\right] ,\left[ \mathbf{p}_{\left( \mathbf{l}_{i}\right) _{\geqslant l,<k}}%
\right]
\end{equation*}%
satisfying:%
\begin{eqnarray*}
\left[ \mathbf{p}_{\left( \eta _{i},\mathbf{l}_{i}\right) _{\infty }}\right]
&=&\left[ \mathbf{p}_{\left( \mathbf{l}_{i}\right) _{\geqslant k,\infty }}%
\right] \left[ \mathbf{p}_{\left( \eta _{i},\mathbf{l}_{i}\right) _{<k}}%
\right] \\
&=&\left[ \mathbf{p}_{\left( \mathbf{l}_{i}\right) _{\geqslant k,\infty }}%
\right] \left[ \mathbf{p}_{\left( \mathbf{l}_{i}\right) _{\geqslant l,<k}}%
\right] \left[ \mathbf{p}_{\left( \eta _{i},\mathbf{l}_{i}\right) _{<l}}%
\right]
\end{eqnarray*}

Assuming that sequencs in decompositn are finite, we shld replac:%
\begin{equation*}
\left[ p_{\eta _{i}},p_{\eta _{i^{\prime }}^{\prime }\eta _{i}}\right]
^{k_{\eta _{i^{\prime }}^{\prime }\eta _{i}}}\rightarrow \left[ \mathbf{p}%
_{\left( \eta _{i},\mathbf{l}_{i}\right) _{<k}}\right] ^{\mathbf{k}_{\mathbf{%
p}_{\left( \eta _{i},\mathbf{l}_{i}\right) _{<k}}}}
\end{equation*}%
to denote the tensor powers intervening in the decompostn. The components of
the vector $\mathbf{k}_{\mathbf{p}_{\left( \eta _{i},\mathbf{l}_{i}\right)
_{<k}}}$ are the tensor powers of the parameters $\left[ \mathbf{p}_{\left(
\eta _{i},\mathbf{l}_{i}\right) _{<k}}\right] $ arisin in the decompositon
of the initial fld.

This leads to the fld:%
\begin{equation}
\tprod_{i}\Psi _{J,,s_{\left[ \mathbf{p}_{\left( \eta _{i},\mathbf{l}%
_{i}\right) _{<k}}\right] }^{\otimes \mathbf{k}_{\mathbf{p}_{\left( \eta
_{i},\mathbf{l}_{i}\right) _{<k}}}}}^{\otimes \mathbf{k}_{\mathbf{p}_{\left(
\eta _{i},\mathbf{l}_{i}\right) _{<k}}}}\left( \left[ \mathbf{p}_{\left(
\eta _{i},\mathbf{l}_{i}\right) _{<k}}\right] ^{\mathbf{k}_{\mathbf{p}%
_{\left( \eta _{i},\mathbf{l}_{i}\right) _{<k}}}},\left\{ \left[ 
\begin{array}{c}
\left( \left[ \mathbf{p}_{\left( \eta _{i},\mathbf{l}_{i}\right) _{<k}}%
\right] ,\mathbf{k}_{\mathbf{p}_{\left( \eta _{i},\mathbf{l}_{i}\right)
_{<k}}}\right) \\ 
\left[ \left\{ \Psi _{J,\alpha }\left[ \mathbf{p}_{\left( \eta _{i},\mathbf{l%
}_{i}\right) _{<k}}\right] \right\} ,v\right]%
\end{array}%
\right] \right\} _{i},v_{\eta }\right)
\end{equation}%
and sbbjct flds f rnk $l$:%
\begin{equation}
\tprod_{i}\Psi _{J,,s_{\left[ \mathbf{p}_{\left( \eta _{i},\mathbf{l}%
_{i}\right) _{\geqslant l,<k}}\right] }^{\otimes \mathbf{k}_{\mathbf{p}%
_{\left( \eta _{i},\mathbf{l}_{i}\right) _{\geqslant l,<k}}}}}^{\otimes 
\mathbf{k}_{\mathbf{p}_{\left( \eta _{i},\mathbf{l}_{i}\right) _{\geqslant
l,<k}}}}\left( \left[ \mathbf{p}_{\left( \eta _{i},\mathbf{l}_{i}\right)
_{\geqslant l,<k}}\right] ^{\mathbf{k}_{\mathbf{p}_{\left( \eta _{i},\mathbf{%
l}_{i}\right) _{\geqslant l,<k}}}},\left\{ \left[ 
\begin{array}{c}
\left( \left[ \mathbf{p}_{\left( \eta _{i},\mathbf{l}_{i}\right) _{\geqslant
l,<k}}\right] ,\mathbf{k}_{\mathbf{p}_{\left( \eta _{i},\mathbf{l}%
_{i}\right) _{\geqslant l,<k}}}\right) \\ 
\left[ \left\{ \Psi _{J,\alpha }\left[ \mathbf{p}_{\left( \eta _{i},\mathbf{l%
}_{i}\right) _{\geqslant l,<k}}\right] \right\} ,v\right]%
\end{array}%
\right] \right\} _{i},,v_{\eta }\right)
\end{equation}%
with functionl:%
\begin{eqnarray}
&&\sum_{_{\left\{ \left( s,s^{\prime },\left[ \mathbf{p}_{\left( \eta _{i},%
\mathbf{l}_{i}\right) _{<k}}\right] \right) \right\} }}\int \bar{g}^{%
\mathcal{K}}\left( \left[ \left\{ \left[ \mathbf{p}_{\left( \eta _{i},%
\mathbf{l}_{i}\right) _{\geqslant l,<k}}\right] ^{\mathbf{k}_{\mathbf{p}%
_{\left( \eta _{i},\mathbf{l}_{i}\right) _{\geqslant l,<k}}}}\right\} 
_{\substack{ i\leqslant s  \\ i^{\prime }\leqslant s^{\prime }}}\right]
,\left\{ \left[ 
\begin{array}{c}
\left( \left[ \mathbf{p}_{\left( \eta _{i},\mathbf{l}_{i}\right) _{\geqslant
l,<k}}\right] ,\mathbf{k}_{\mathbf{p}_{\left( \eta _{i},\mathbf{l}%
_{i}\right) _{\geqslant l,<k}}}\right) \\ 
\left[ \left\{ \Psi _{J,\alpha }\left[ \mathbf{p}_{\left( \eta _{i},\mathbf{l%
}_{i}\right) _{\geqslant l,<k}}\right] \right\} ,v\right]%
\end{array}%
\right] \right\} _{i},\left[ v\right] \right) \\
&&\times \Psi _{J,\tprod\limits_{i^{\prime }}s_{p_{\eta _{i^{\prime
}}^{\prime }\eta _{i}}}^{\otimes k_{\eta _{i^{\prime }}^{\prime }\eta
_{i}}}}^{\otimes \sum_{i^{\prime }}k_{\eta _{i^{\prime }}^{\prime }\eta
_{i}}}\left( \left[ \left\{ \left[ \mathbf{p}_{\left( \eta _{i},\mathbf{l}%
_{i}\right) _{\geqslant l,<k}}\right] ^{\mathbf{k}_{\mathbf{p}_{\left( \eta
_{i},\mathbf{l}_{i}\right) _{\geqslant l,<k}}}}\right\} _{\substack{ %
i\leqslant s  \\ i^{\prime }\leqslant s^{\prime }}}\right] ,\left\{ \left[ 
\begin{array}{c}
\left( \left[ \mathbf{p}_{\left( \eta _{i},\mathbf{l}_{i}\right) _{\geqslant
l,<k}}\right] ,\mathbf{k}_{\mathbf{p}_{\left( \eta _{i},\mathbf{l}%
_{i}\right) _{\geqslant l,<k}}}\right) \\ 
\left[ \left\{ \Psi _{J,\alpha }\left[ \mathbf{p}_{\left( \eta _{i},\mathbf{l%
}_{i}\right) _{\geqslant l,<k}}\right] \right\} ,v\right]%
\end{array}%
\right] \right\} _{i},\left[ v\right] \right)  \notag \\
&&\times d\left( \left\{ \left[ 
\begin{array}{c}
\left( \left[ \mathbf{p}_{\left( \eta _{i},\mathbf{l}_{i}\right) _{\geqslant
l,<k}}\right] ,\mathbf{k}_{\mathbf{p}_{\left( \eta _{i},\mathbf{l}%
_{i}\right) _{\geqslant l,<k}}}\right) \\ 
\left[ \left\{ \Psi _{J,\alpha }\left[ \mathbf{p}_{\left( \eta _{i},\mathbf{l%
}_{i}\right) _{\geqslant l,<k}}\right] \right\} ,v\right]%
\end{array}%
\right] \right\} _{ii}\right) d\left( \left\{ \left[ p_{\eta _{i}},p_{\eta
_{i^{\prime }}^{\prime }\eta _{i}}\right] ^{k_{\eta _{i^{\prime }}^{\prime
}\eta _{i}}}\right\} _{\substack{ l\leqslant s  \\ l^{\prime }\leqslant
s^{\prime }}}\right)  \notag
\end{eqnarray}%
Th cas of objct functnl being understood for $l=0$, which amounts to replac $%
\left( \eta _{i},\mathbf{l}_{i}\right) _{\geqslant 0,<k}\rightarrow \left(
\eta _{i},\mathbf{l}_{i}\right) _{<k}$.

Thus the spaces of parametrs consists of sequencs of parametrs, each of them
defind b sbbjcts:%
\begin{equation*}
\left\{ \left\{ \left[ 
\begin{array}{c}
\left( \left[ \mathbf{p}_{\left( \eta _{i},\mathbf{l}_{i}\right) _{<k}}%
\right] ,\mathbf{k}_{\mathbf{p}_{\left( \eta _{i},\mathbf{l}_{i}\right)
_{<k}}}\right) \\ 
\left[ \left\{ \Psi _{J,\alpha }\left[ \mathbf{p}_{\left( \eta _{i},\mathbf{l%
}_{i}\right) _{<k}}\right] \right\} ,v\right]%
\end{array}%
\right] \right\} _{i},\cup _{l}\left\{ \left[ 
\begin{array}{c}
\left( \left[ \mathbf{p}_{\left( \eta _{i},\mathbf{l}_{i}\right) _{\geqslant
l,<k}}\right] ,\mathbf{k}_{\mathbf{p}_{\left( \eta _{i},\mathbf{l}%
_{i}\right) _{\geqslant l,<k}}}\right) \\ 
\left[ \left\{ \Psi _{J,\alpha }\left[ \mathbf{p}_{\left( \eta _{i},\mathbf{l%
}_{i}\right) _{\geqslant l,<k}}\right] \right\} ,v\right]%
\end{array}%
\right] \right\} _{i}\right\}
\end{equation*}%
Th trnctd fnctnls crrspnd t dcmpstns rsng nl sng sbbjcts strtng frm sm rnk.
Ths mpls sm prstnc f th sbbjct t lrgr scl.

The equations fr subobjct flds becoms (which for the bjct correspndn t $l=0$%
):%
\begin{eqnarray}
&&0=\int v\left( \left[ \left\{ \left[ \mathbf{p}_{\left( \eta _{i},\mathbf{l%
}_{i}\right) _{\geqslant l,<k}}\right] ^{\mathbf{k}_{\mathbf{p}_{\left( \eta
_{i},\mathbf{l}_{i}\right) _{\geqslant l,<k}}}}\right\} _{\substack{ %
i\leqslant s  \\ i^{\prime }\leqslant s^{\prime }}}\right] \right) \frac{%
\delta \left\{ \left[ 
\begin{array}{c}
\left( \left[ \mathbf{p}_{\left( \eta _{i},\mathbf{l}_{i}\right) _{\geqslant
l,<k}}\right] ,\mathbf{k}_{\mathbf{p}_{\left( \eta _{i},\mathbf{l}%
_{i}\right) _{\geqslant l,<k}}}\right) \\ 
\left[ \left\{ \Psi _{J,\alpha }\left[ \mathbf{p}_{\left( \eta _{i},\mathbf{l%
}_{i}\right) _{\geqslant l,<k}}\right] \right\} ,v\right]%
\end{array}%
\right] \right\} }{\delta \Psi _{J,s_{\mathbf{p}_{\left( \eta _{i},\mathbf{l}%
_{i}\right) }}^{\otimes \mathbf{k}_{\mathbf{p}_{\left( \eta _{i},\mathbf{l}%
_{i}\right) _{\geqslant l,<k}}}}}^{\otimes \sum_{i^{\prime }}\mathbf{k}_{%
\mathbf{p}_{\left( \eta _{i},\mathbf{l}_{i}\right) _{\geqslant
l,<k}}}}\left( \left[ \mathbf{p}_{\left( \eta _{i},\mathbf{l}_{i}\right)
_{\geqslant l,<k}}\right] ^{\mathbf{k}_{\mathbf{p}_{\left( \eta _{i},\mathbf{%
l}_{i}\right) _{\geqslant l,<k}}}}\right) } \\
&&\times \frac{\nabla \Psi _{J,\tprod\limits_{i^{\prime }}s_{p_{\eta
_{i^{\prime }}^{\prime }\eta _{i}}}^{\otimes k_{\eta _{i^{\prime }}^{\prime
}\eta _{i}}}}^{\otimes \sum_{i^{\prime }}\mathbf{k}_{\mathbf{p}_{\left( \eta
_{i},\mathbf{l}_{i}\right) _{\geqslant l,<k}}}}\left( \left\{ \left[ \left\{ %
\left[ \mathbf{p}_{\left( \eta _{i},\mathbf{l}_{i}\right) _{\geqslant l,<k}}%
\right] ^{\mathbf{k}_{\mathbf{p}_{\left( \eta _{i},\mathbf{l}_{i}\right)
_{\geqslant l,<k}}}}\right\} _{\substack{ i\leqslant s  \\ i^{\prime
}\leqslant s^{\prime }}}\right] \right\} _{_{\substack{ i\leqslant s  \\ %
i^{\prime }\leqslant s^{\prime }}}},\left\{ \left[ 
\begin{array}{c}
\left( \left[ \mathbf{p}_{\left( \eta _{i},\mathbf{l}_{i}\right) _{\geqslant
l,<k}}\right] ,\mathbf{k}_{\mathbf{p}_{\left( \eta _{i},\mathbf{l}%
_{i}\right) _{\geqslant l,<k}}}\right) \\ 
\left[ \left\{ \Psi _{J,\alpha }\left[ \mathbf{p}_{\left( \eta _{i},\mathbf{l%
}_{i}\right) _{\geqslant l,<k}}\right] \right\} ,v\right]%
\end{array}%
\right] \right\} ,v\right) }{\nabla \left\{ \left[ 
\begin{array}{c}
\left( \left[ \mathbf{p}_{\left( \eta _{i},\mathbf{l}_{i}\right) _{\geqslant
l,<k}}\right] ,\mathbf{k}_{\mathbf{p}_{\left( \eta _{i},\mathbf{l}%
_{i}\right) _{\geqslant l,<k}}}\right) \\ 
\left[ \left\{ \Psi _{J,\alpha }\left[ \mathbf{p}_{\left( \eta _{i},\mathbf{l%
}_{i}\right) _{\geqslant l,<k}}\right] \right\} ,v\right]%
\end{array}%
\right] \right\} }  \notag
\end{eqnarray}

\section*{Appendix 8. \textbf{Exemple}}

An example can be detailed by considering a system involving some boson +
fermions system, classically described by the hamiltonian (the free fermion
part is omitted):%
\begin{equation*}
H=\int A\left( k\right) k^{2}A\left( -k\right) +\bar{\psi}\left(
k_{1}\right) A\left( k\right) \gamma \psi \left( -k_{1}-k\right)
\end{equation*}%
or by the alternative form: 
\begin{equation*}
\int A\left( k\right) k^{2}A\left( -k\right) +\int \bar{\psi}\left(
k_{1}\right) A\left( k\right) \gamma \psi \left( -k_{2}\right) \delta \left(
k+k_{1}-k_{2}\right)
\end{equation*}%
We consider this system defined by two spaces of states, one for bosons and
one for fermions constrained by momentum conservation. We will rewrite the
constraint in a form suitable for the present formalism and then project the
state space for bosons along eigenspaces depending on the fermion degrees of
freedom. This corresponds to the case where one type of states, i.e., the
bosons (corresponding above to the states depending on $U^{\left( i\right) }$%
), is projected onto some subspace to produce an effective theory for the
fermions (the subspace parameterized by $U^{\left( j\right) }$). However,
due to constraints, this projection also depends on the states parameterized
by $U^{\left( j\right) }$.

The application of our procedure follows several steps. First, rewriting the
constraints in terms of operators. Then decomposing the operator along which
we want to project in terms of eigenvalues of fermion degrees of freedom.
Then, given this decomposition, project each partial Hamiltonian in this
decomposition onto the lowest boson eigenstate.

\subsection*{Constraint}

First, introducing:%
\begin{equation*}
\bar{\psi}\left( k_{1}\right) A\left( k\right) \gamma \psi \left(
-k_{2}\right) =A\left( k\right) B\left( k_{1}-k_{2}\right)
\end{equation*}%
whose commutation relation with momentum operators are given by:%
\begin{eqnarray*}
\left[ K_{a}^{\left( k\right) },A\left( \hat{k}\right) \right] &=&kA\left(
k\right) \delta \left( k-\hat{k}\right) \\
\left[ K_{b}^{\left( k^{\prime }\right) },B\left( \hat{k}^{\prime }\right) %
\right] &=&k^{\prime }B\left( k^{\prime }\right) \delta \left( k^{\prime }-%
\hat{k}^{\prime }\right)
\end{eqnarray*}%
by imposing the constraints:%
\begin{equation*}
\left[ K_{a}^{\left( k\right) },A\left( k\right) \right] =-\left[
K_{b}^{\left( k^{\prime }\right) },B\left( k^{\prime }\right) \right]
\end{equation*}%
we recover the initial form including $\delta \left( k+k_{1}-k_{2}\right) $.
In the basis $\left\vert \lambda ,k\right\rangle _{A}$ for $A\left( k\right) 
$, $\left\vert \lambda ,k^{\prime }\right\rangle _{B}$ for $B\left(
k^{\prime }\right) $\ the states satisfying the constraint are:%
\begin{eqnarray*}
&&\dprod \left\vert \lambda _{1},k_{1}\right\rangle _{A}\left\vert \lambda
_{2},-k_{1}\right\rangle _{A}...\left\vert \lambda _{l},k_{l}\right\rangle
_{A}\left\vert \mu _{1},k_{1}\right\rangle _{B}\left\vert \mu
_{2},-k_{1}\right\rangle _{B}...\left\vert \mu _{l},-k_{l}\right\rangle _{B}
\\
&&\times \dprod_{\hat{k}\neq k_{i},\hat{k}^{\prime }\neq -k_{i}}\left\vert 0,%
\hat{k}\right\rangle _{A}...\left\vert 0,\hat{k}^{\prime }\right\rangle
_{B}...
\end{eqnarray*}

\subsection*{Decomposition of the hamiltonian in fermions eigenstates}

To project the boson states on some eigenspaces, we have to decompose the
interaction terms in a way that is diagonal in the fermion degrees of
freedom. We write $\left\vert S^{\left( j\right) }\right\rangle $ the common
eigenvectors of: 
\begin{equation*}
\int \bar{\psi}\left( k_{1}\right) \gamma \psi \left( -k_{1}-k\right) dk_{1}
\end{equation*}

and rewrite the Hamiltonian for $A$:%
\begin{eqnarray*}
H_{A} &=&\int ka_{k}^{+}a_{k}^{-}+\left[ \left\langle S^{\left( j\right)
}\right\vert \int \bar{\psi}\left( k_{1}\right) \gamma \psi \left(
-k_{1}-k\right) dk_{1}\left\vert S^{\left( j\right) }\right\rangle \right]
a_{k}^{+}+\left[ \left\langle S^{\left( j\right) }\right\vert \int \bar{\psi}%
\left( k_{1}\right) \gamma \psi \left( -k_{1}+k\right) dk_{1}\left\vert
S^{\left( j\right) }\right\rangle \right] a_{k}^{-} \\
&=&\int k\left( a_{k}^{+}+\frac{1}{k}\left[ \left\langle S^{\left( j\right)
}\right\vert \int \bar{\psi}\left( k_{1}\right) \gamma \psi \left(
-k_{1}-k\right) dk_{1}\left\vert S^{\left( j\right) }\right\rangle \right]
\right) \left( a_{k}^{-}+\frac{1}{k}\left[ \left\langle S^{\left( j\right)
}\right\vert \int \bar{\psi}\left( k_{1}\right) \gamma \psi \left(
-k_{1}+k\right) dk_{1}\left\vert S^{\left( j\right) }\right\rangle \right]
\right) \\
&&-\frac{1}{k}\left\vert \left\langle S^{\left( j\right) }\right\vert \int 
\bar{\psi}\left( k_{1}\right) \gamma \psi \left( -k_{1}-k\right)
dk_{1}\left\vert S^{\left( j\right) }\right\rangle \right\vert ^{2}
\end{eqnarray*}%
For a given state $S^{\left( j\right) }$, this can be written $H_{A}\left(
S^{\left( j\right) }\right) $ and this corresponds to the operator $\hat{H}%
_{i}\left( U^{\left( j\right) }\right) $ in th decomposition (\ref{P}). This
hamiltonian depends on the fermion degrees of freedom.

\subsection*{Boson eigenstates}

The eigenstates for $H_{A}\left( \left\vert S^{\left( j\right)
}\right\rangle \right) $ are, for gvn state $\left\vert S^{\left( j\right)
}\right\rangle $: 
\begin{eqnarray*}
\dprod \left( a_{k}^{+}+\frac{1}{k}\left\vert \left\langle S^{\left(
j\right) }\right\vert \int \bar{\psi}\left( k_{1}\right) \gamma \psi \left(
-k_{1}-k\right) dk_{1}\left\vert S^{\left( j\right) }\right\rangle
\right\vert \right) \left\vert \hat{0}\right\rangle  &=&\dprod \hat{a}%
_{k}^{+}\left\vert \hat{0}_{k}\left( S^{\left( j\right) }\right)
\right\rangle  \\
\left\vert \hat{0}_{k}\left( S^{\left( j\right) }\right) \right\rangle 
&=&\exp \left( -\frac{c_{k}^{2}\left( S^{\left( j\right) }\right) }{2}%
\right) \exp \left( -c_{k}\left( S^{\left( j\right) }\right)
a_{k}^{+}\right) \left\vert 0\right\rangle  \\
c_{k}\left( S^{\left( j\right) }\right)  &=&\frac{1}{k}\left\langle
S^{\left( j\right) }\right\vert \int \bar{\psi}\left( k_{1}\right) \gamma
\psi \left( -k_{1}+k\right) dk_{1}\left\vert S^{\left( j\right)
}\right\rangle 
\end{eqnarray*}%
where we define:%
\begin{equation*}
\left\vert \hat{0}_{k}\left( S^{\left( j\right) }\right) \right\rangle
\equiv \left\vert k,S^{\left( j\right) }\right\rangle 
\end{equation*}%
Then, the product of state includes the constraint:%
\begin{equation*}
\left\vert \hat{0}\left( k,S^{\left( j\right) }\right) \right\rangle
\left\vert \hat{0}\left( k_{1}+k_{2},S^{\left( j\right) }\right)
\right\rangle \delta \left( k_{1}+k_{2}-k\right) 
\end{equation*}

\subsection*{Decomposition in bosons eigenstates}

The interaction part:%
\begin{equation*}
\bar{\psi}\left( k_{1}\right) \left( A\left( k\right) \gamma \right) \psi
\left( -k_{1}-k\right)
\end{equation*}%
can then be rewritten decomposing the boson states according to the basis $%
\left\vert N_{k}\right\rangle \left\langle N_{k}\right\vert $ with:%
\begin{equation*}
N_{k}=\left( a_{k}^{+}+\frac{1}{k}\left\vert \left\langle S^{\left( j\right)
}\right\vert \int \bar{\psi}\left( k_{1}\right) \gamma \psi \left(
-k_{1}-k\right) dk_{1}\left\vert S^{\left( j\right) }\right\rangle
\right\vert \right) \left( a_{k}^{-}+\frac{1}{k}\left\vert \left\langle
S^{\left( j\right) }\right\vert \int \bar{\psi}\left( k_{1}\right) \gamma
\psi \left( -k_{1}-k\right) dk_{1}\left\vert S^{\left( j\right)
}\right\rangle \right\vert \right)
\end{equation*}

We have:%
\begin{eqnarray*}
&&\bar{\psi}\left( k_{1}\right) \left( A\left( k\right) \gamma \right) \psi
\left( -k_{1}-k\right) \\
&=&\int \sum_{i}\left\vert S^{\left( j\right) }\right\rangle
\sum_{N_{k},N_{k}^{\prime }}\left\langle S^{\left( j\right) }\right\vert
\left\vert N_{k}\right\rangle \left\langle N_{k}\right\vert A\left( k\right)
\left\vert N_{k}^{\prime }\right\rangle \left\langle N_{k}^{\prime
}\right\vert \int \bar{\psi}\left( k_{1}\right) \gamma \psi \left(
-k_{1}-k\right) dk_{1}\left\vert N_{k}^{\prime }\right\rangle \left\vert
S^{\left( j\right) }\right\rangle \left\langle S^{\left( j\right)
}\right\vert dk
\end{eqnarray*}

\subsection*{Projection on lowest boson eigenstate}

Now we project:%
\begin{equation*}
\sum_{N_{k},N_{k}^{\prime }}\left\vert N_{k}\right\rangle \left\langle
N_{k}\right\vert A\left( k\right) \left\vert N_{k}^{\prime }\right\rangle
\left\langle N_{k}^{\prime }\right\vert
\end{equation*}%
on the lowest boson eigenstate:%
\begin{equation*}
\left\vert \hat{0}_{k}\left( S^{\left( j\right) }\right) \right\rangle
\left\langle \hat{0}_{k}\left( S^{\left( j\right) }\right) \right\vert
\end{equation*}%
so that the field $A\left( k\right) $ is replaced by: 
\begin{eqnarray*}
&&\left\vert N_{k}\right\rangle \left\langle N_{k}\right\vert A\left(
k\right) \left\vert N_{k}^{\prime }\right\rangle \left\langle N_{k}^{\prime
}\right\vert \\
&\rightarrow &\left\vert \hat{0}_{k}\left( S^{\left( j\right) }\right)
\right\rangle \sum_{N_{k},N_{k}^{\prime }}\left\langle \hat{0}_{k}\left(
S^{\left( j\right) }\right) \right\vert \left\vert N_{k}\right\rangle
\left\langle N_{k}\right\vert A\left( k\right) \left\vert N_{k}^{\prime
}\right\rangle \left\langle N_{k}^{\prime }\right\vert \left\vert \hat{0}%
_{k}\left( S^{\left( j\right) }\right) \right\rangle \left\langle \hat{0}%
_{k}\left( S^{\left( j\right) }\right) \right\vert
\end{eqnarray*}%
Gvn that:%
\begin{equation*}
\left\langle N_{k^{\prime }}\right\vert A\left( k\right) _{1}\left\vert
N_{k}\right\rangle
\end{equation*}%
is non nul for $N_{k^{\prime }}=N_{k}\pm 1$, we can use that;%
\begin{equation*}
\left\langle N_{k^{\prime }}\right\vert A\left( k\right) _{1}\left\vert
N_{k}\right\rangle =\sqrt{n+1}\delta _{k^{\prime },k}
\end{equation*}%
to rewrite:%
\begin{eqnarray*}
&&\sum_{N_{k},N_{k}^{\prime }}\left\langle \hat{0}_{k}\left( S^{\left(
j\right) }\right) \right\vert \left\vert N_{k}\right\rangle \left\langle
N_{k}\right\vert A\left( k\right) \left\vert N_{k}^{\prime }\right\rangle
\left\langle N_{k}^{\prime }\right\vert \left\vert \hat{0}_{k}\left(
S^{\left( j\right) }\right) \right\rangle \\
&=&\sum_{n}\exp \left( -\frac{c_{k}^{2}\left( S^{\left( j\right) }\right) }{2%
}\right) c_{k}^{n+1}\left( S^{\left( j\right) }\right) \frac{1}{\sqrt{\left(
n+1\right) !}}\sqrt{n+1}\frac{1}{\sqrt{n!}}c_{k}^{n}\left( S^{\left(
j\right) }\right) \exp \left( -\frac{c_{k}^{2}\left( S^{\left( j\right)
}\right) }{2}\right) \\
&=&\sum_{n}\exp \left( -\frac{c_{k}^{2}\left( S^{\left( j\right) }\right) }{2%
}\right) \exp \left( -\frac{c_{k}^{2}\left( S^{\left( j\right) }\right) }{2}%
\right) \exp \left( c_{k}^{2}\left( S^{\left( j\right) }\right) \right)
c_{k}\left( S^{\left( j\right) }\right) \\
&=&c_{k}\left( S^{\left( j\right) }\right)
\end{eqnarray*}%
As a consequence, the projection of the interaction part along the lowest
eigenstates for $A$ writes:%
\begin{eqnarray*}
&&\bar{\psi}\left( k_{1}\right) \left( A\left( k\right) \gamma \right) \psi
\left( -k_{1}-k\right) \\
&=&\int \sum_{i}\left\vert S^{\left( j\right) }\right\rangle
\sum_{N_{k},N_{k}^{\prime }}\left\langle S^{\left( j\right) }\right\vert
\left\vert N_{k}\right\rangle \left\langle N_{k}\right\vert A\left( k\right)
\left\vert N_{k}^{\prime }\right\rangle \left\langle N_{k}^{\prime
}\right\vert \int \bar{\psi}\left( k_{1}\right) \gamma \psi \left(
-k_{1}-k\right) dk_{1}\left\vert N_{k}^{\prime }\right\rangle \left\vert
S^{\left( j\right) }\right\rangle \left\langle S^{\left( j\right)
}\right\vert dk \\
&\rightarrow &\int \sum_{i}\left\vert S^{\left( j\right) }\right\rangle
\left\vert \hat{0}_{k}\left( S^{\left( j\right) }\right) \right\rangle
c_{k}\left( S^{\left( j\right) }\right) \left\langle \hat{0}_{k}\left(
S^{\left( j\right) }\right) \right\vert \left\langle S^{\left( j\right)
}\right\vert \int \bar{\psi}\left( k_{1}\right) \gamma \psi \left(
-k_{1}-k\right) dk_{1}\left\vert S^{\left( j\right) }\right\rangle
\left\langle S^{\left( j\right) }\right\vert dk \\
&=&\int \sum_{i}\left\vert S^{\left( j\right) }\right\rangle \left\vert \hat{%
0}_{k}\left( S^{\left( j\right) }\right) \right\rangle c_{k}\left( S^{\left(
j\right) }\right) \left\langle S^{\left( j\right) }\right\vert \int \bar{\psi%
}\left( k_{1}\right) \gamma \psi \left( -k_{1}-k\right) dk_{1}\left\vert
S^{\left( j\right) }\right\rangle \left\langle \hat{0}_{k}\left( S^{\left(
j\right) }\right) \right\vert \left\langle S^{\left( j\right) }\right\vert dk
\\
&=&\int \sum_{i}\left\vert S^{\left( j\right) }\right\rangle \left\vert \hat{%
0}_{k}\left( S^{\left( j\right) }\right) \right\rangle \frac{1}{k}%
\left\langle S^{\left( j\right) }\right\vert \int \bar{\psi}\left(
k_{1}\right) \gamma \psi \left( -k_{1}+k\right) dk_{1}\left\vert S^{\left(
j\right) }\right\rangle \\
&&\times \left\langle S^{\left( j\right) }\right\vert \int \bar{\psi}\left(
k_{1}\right) \gamma \psi \left( -k_{1}-k\right) dk_{1}\left\vert S^{\left(
j\right) }\right\rangle \left\langle \hat{0}_{k}\left( S^{\left( j\right)
}\right) \right\vert \left\langle S^{\left( j\right) }\right\vert dk
\end{eqnarray*}%
If: 
\begin{equation*}
\dprod\limits_{k}\left\vert \hat{0}_{k}\left( S^{\left( j\right) }\right)
\right\rangle \simeq \left\vert \hat{0}\right\rangle
\end{equation*}%
the projection reduces to:%
\begin{eqnarray*}
&&\bar{\psi}\left( k_{1}\right) \left( A\left( k\right) \gamma \right) \psi
\left( -k_{1}-k\right) \\
&\rightarrow &\left\vert \hat{0}\right\rangle \int \left( \int \bar{\psi}%
\left( k_{1}\right) \gamma \psi \left( -k_{1}+k\right) dk_{1}\right) \frac{1%
}{k}\left( \int \bar{\psi}\left( k_{1}\right) \gamma \psi \left(
-k_{1}-k\right) dk_{1}\right) dk\left\langle \hat{0}\right\vert
\end{eqnarray*}

Note that the projection on the $\left\vert \hat{0}_{k}\left( S^{\left(
j\right) }\right) \right\rangle $ leads to effective amplitudes: 
\begin{equation*}
\int \sum_{i}\left\vert S^{\left( j\right) }\right\rangle \left\langle
S^{\left( j\right) }\right\vert \int \bar{\psi}\left( k_{1}\right) \gamma
\psi \left( -k_{1}+k\right) dk_{1}\frac{1}{k}\int \bar{\psi}\left(
k_{1}\right) \gamma \psi \left( -k_{1}-k\right) dk_{1}\left\vert S^{\left(
j\right) }\right\rangle \left\langle S^{\left( j\right) }\right\vert dk
\end{equation*}%
that is, to an effective self interaction term:%
\begin{equation*}
\int \left( \int \bar{\psi}\left( k_{1}\right) \gamma \psi \left(
-k_{1}+k\right) dk_{1}\frac{1}{k}\int \bar{\psi}\left( k_{1}\right) \gamma
\psi \left( -k_{1}-k\right) dk_{1}\right) dk
\end{equation*}

\subsection*{Remark 1}

\bigskip The results corresponds in term of path integral to integrate the
boson degrees of freedom:%
\begin{eqnarray*}
&&\int \exp \left( -\int \left( A\left( k\right) k^{2}A\left( -k\right) +%
\bar{\psi}\left( k_{1}\right) \left( a_{k}^{+}\gamma \right) \psi \left(
-k_{1}-k\right) +\bar{\psi}\left( k_{1}\right) \left( a_{k}^{-}\gamma
\right) \psi \left( -k_{1}+k\right) \right) \right) \mathcal{D}A\left(
k\right) \\
&=&\exp \left( -\int \left( \int \bar{\psi}\left( k_{1}\right) \gamma \psi
\left( -k_{1}+k\right) dk_{1}\right) \frac{1}{k}\left( \int \bar{\psi}\left(
k_{1}\right) \gamma \psi \left( -k_{1}-k\right) dk_{1}\right) \right)
\end{eqnarray*}%
with boundary conditions $A\left( k\right) +\bar{\psi}\left( k_{1}\right)
\gamma \psi \left( -k_{1}-k\right) \rightarrow 0$: which projects on the $%
\left\vert \hat{0}_{k}\left( S^{\left( j\right) }\right) \right\rangle $.

\subsection*{\textbf{Remark 2:}}

We can rewrite the constraints in terms of conditions on parameter spaces.
In this example, states:%
\begin{equation*}
\dprod\limits_{k}\left\vert \hat{0}_{k}\left( S^{\left( j\right) }\right)
\right\rangle \left\vert S^{\left( j\right) }\right\rangle
\end{equation*}%
can be parametrized as:%
\begin{equation*}
\dprod\limits_{k}\left\vert \hat{0}_{k}\left( \lambda _{k}\right)
\right\rangle \left\vert S^{\left( j\right) }\right\rangle \delta \left(
\lambda _{k}-\left\langle S^{\left( j\right) }\right\vert \int \bar{\psi}%
\left( k_{1}\right) \gamma \psi \left( -k_{1}+k\right) dk_{1}\left\vert
S^{\left( j\right) }\right\rangle \right)
\end{equation*}%
where states $\left\vert \hat{0}_{k}\left( \lambda _{k}\right) \right\rangle 
$ are defined by:%
\begin{equation*}
\left\vert \hat{0}_{k}\left( \lambda _{k}\right) \right\rangle =\exp \left( -%
\frac{\lambda _{k}^{2}}{2k^{2}}\right) \exp \left( -\frac{\lambda _{k}}{k}%
a_{k}^{+}\right) \left\vert 0\right\rangle
\end{equation*}%
The $\int \bar{\psi}\left( k_{1}\right) \gamma \psi \left( -k_{1}+k\right)
dk_{1}$ commute and the $\left\vert S^{\left( j\right) }\right\rangle $ are
thus parametrized as:%
\begin{equation*}
\left\vert \left\{ \mu _{k}\right\} _{k\in 
\mathbb{R}
}\right\rangle
\end{equation*}%
so that states become:%
\begin{equation*}
\dprod\limits_{k}\left\vert \hat{0}_{k}\left( \lambda _{k}\right)
\right\rangle \left\vert \left\{ \mu _{k}\right\} _{k\in 
\mathbb{R}
}\right\rangle \delta \left( \lambda _{k}-\mu _{k}\right)
=\dprod\limits_{k}\left\vert \hat{0}_{k}\left( \lambda _{k}\right)
\right\rangle \left\vert \left\{ \lambda _{k}\right\} _{k\in 
\mathbb{R}
}\right\rangle
\end{equation*}%
with the condition that $\left\vert \left\{ \lambda _{k}\right\} _{k\in 
\mathbb{R}
}\right\rangle =0$ for\ non admissible values of the $\left\{ \lambda
_{k}\right\} $. The eingenstates of $\int \bar{\psi}\left( k_{1}\right)
\gamma \psi \left( -k_{1}+k\right) dk_{1}$\ and $A\left( k\right) $ are
constrained by $\delta $ condition.

\section*{Appendix 9. \textbf{Formulation in terms of constrained operators
and description of} \textbf{States}:}

This section describes more precisely the constraints in the initial
parameters spaces as constraints between operators. We also study the
consequences of the operator constraints on the states. We provide an
example in the end.

\subsection*{Projection operator and Constraints in terms of operators}

Starting with operators $\left( U_{j}\right) H_{i}\left( U_{j}\right) $
acting on the states spaces over the parameter space $U_{i}$ and depending
on the $U_{j}$ parameters, we assume they can be written as combinations of
some operator $\mathbf{\hat{\Lambda}}_{i}\left( U_{j}\right) $ similar to
some band hamiltonin:%
\begin{equation}
\left( U_{j}\right) H_{i}\left( U_{j}\right) =H_{i}\left( U_{j},\mathbf{\hat{%
\Lambda}}_{i}\left( U_{j}\right) \right)  \label{frma}
\end{equation}%
where $\mathbf{\hat{\Lambda}}_{i}\left( U^{\left( j\right) }\right) $ is a
combination of the operators $\mathbf{U}^{\left( i\right) }$ of
multiplication by sum of tensor products of $U_{i}$ and $\Pi ^{\left(
i\right) }\left( U^{\left( i\right) }\right) $ is conjugate to the $\mathbf{U%
}^{\left( i\right) }$. The combination depends on $\left( U^{\left( j\right)
}\right) $. Acting on the $U_{i}$ states, prtrs $\mathbf{\hat{\Lambda}}%
_{i}\left( U_{j}\right) $ belong the the space of operators on the larger
states space spanned by the $\left\vert \hat{U}^{\left( i\right) }\mid \hat{U%
}^{\left( j\right) }\right\rangle $. We aim at writing the constraint
between $\hat{U}^{\left( i\right) }$ and $\hat{U}^{\left( j\right) }$
degrees of freedom as a relation between the $\mathbf{\hat{\Lambda}}%
_{i}\left( U_{j}\right) $ and some operators acting on the state space
corresponding to the $\hat{U}^{\left( j\right) }$.

We start by considering by some functions, possibly milti-valued, $%
A^{i}\left( U^{\left( j\right) }\right) $ satifisfying:%
\begin{equation*}
F\left( A^{i},U^{\left( j\right) }\right) =0
\end{equation*}%
with $F$ a vector value function of $\left( A^{i},U^{\left( j\right)
}\right) $.

This can be written as an equation operators with eigenvalues $%
A^{i},U^{\left( j\right) }$ when operatrs commute:%
\begin{equation*}
F\left( \mathbf{A}^{i},\mathbf{U}^{\left( j\right) }\right) =0
\end{equation*}%
The commutation condition is satisfied if we assume that $\mathbf{A}^{i}$ is
a function of the unconstrained generators $\mathbf{\Lambda }_{i}$ and
conjugate $\Pi _{\mathbf{\Lambda }_{i}}^{i}$. This is an identity for the
matrices elements. The constraints become in terms of operators: 
\begin{equation*}
F\left( \mathbf{A}^{i}\left( \left\{ \mathbf{\Lambda }_{i}\right\} ,\left\{
\Pi _{\mathbf{\Lambda }_{i}}^{i}\right\} \right) ,\mathbf{U}^{\left(
j\right) }\right) \equiv F\left( \left\{ \mathbf{\Lambda }_{i}\right\}
,\left\{ \Pi _{\mathbf{\Lambda }_{i}}^{i}\right\} ,\mathbf{U}^{\left(
j\right) }\right) =0
\end{equation*}%
with $\Pi ^{i}$ conjugate to the $\mathbf{\Lambda }_{i}$ and $\mathbf{U}%
^{\left( j\right) }$ defined as multplication over the states $\left\{
\left\vert U_{j}\right\rangle \right\} _{\lambda }$. This allows to rewrite
some combination $A^{j}$ of the eignvalues of $\mathbf{U}^{\left( j\right) }$
as functions of the eigenvals of the $\left( \mathbf{\Lambda }_{i}\right) $.
This expression is not local since $\left\{ \mathbf{\Lambda }_{i}\right\}
,\left\{ \Pi _{\mathbf{\Lambda }_{i}}^{i}\right\} $ do not commute.

The states for gvn $\lambda $ can then be written: 
\begin{equation*}
\left\{ \left\vert \Sigma \left( \lambda ,\mathbf{\Lambda }_{i}\right)
\right\rangle _{\times F\left( \mathbf{\Lambda }_{i},\Pi ^{i},U_{j}\right)
}\left\vert U_{j}\right\rangle \right\} _{\lambda }
\end{equation*}%
As an example, the states $\mathbf{A}^{i}\left( \mathbf{\Lambda }_{i},\Pi
^{i}\right) \left\vert 0\right\rangle $ can be assumed to be series $%
\sum_{\alpha ,\beta }\left( \mathbf{\Lambda }_{i}\right) ^{\alpha }\left(
\Pi ^{i}\right) ^{\beta }$ acting on some vacuum $\left\vert 0\right\rangle $
with $\mathbf{\Lambda }_{i}=\left( \mathbf{\Lambda }_{i}\right) ^{+}+\left( 
\mathbf{\Lambda }_{i}\right) $, $\Pi ^{i}=\left( \mathbf{\Lambda }%
_{i}\right) ^{+}-\left( \mathbf{\Lambda }_{i}\right) ^{-}$, and we have:%
\begin{equation*}
\mathbf{A}^{i}\left( \mathbf{\Lambda }_{i},\Pi ^{i}\right) \left\vert
0\right\rangle =\sum_{m}a_{m}\left( \left( \mathbf{\Lambda }_{i}\right)
^{+}\right) ^{m}\left\vert 0\right\rangle
\end{equation*}%
and the eigenvalues of $A_{i}\left( \mathbf{\Lambda }_{i},\Pi ^{i}\right) $
are sums of tensor powers of eigenvalues $\left( \mathbf{\Lambda }%
_{i}\right) ^{\otimes m}$.

The constrt written in terms of eignvals: 
\begin{equation*}
F\left( A^{i},U^{\left( j\right) }\right) =0
\end{equation*}%
translate in writting $p$ coordinates of states $\left\vert U^{\left(
j\right) }\right\rangle $ by functionals $h$ of series $\sum_{m}b_{m}^{j}%
\left( \mathbf{\Lambda }_{i}\right) ^{\otimes m}$ and $\left\vert U^{\left(
j\right) }\right\rangle =\left\vert h\left( \sum_{m}b_{m}^{j}\left( \mathbf{%
\Lambda }_{i}\right) ^{\otimes m}\right) ,U^{\left( j/p\right)
}\right\rangle $ where $U^{\left( j/p\right) }$ describes the remaining
degrees of freedom.

The states can then be written as:%
\begin{equation}
\left\vert \lambda \left( U_{j}\right) ,\left\{ \mathbf{\Lambda }%
_{i}\right\} \right\rangle \left\vert h\left( \left[ \left\{ \mathbf{\Lambda 
}_{i}\right\} \right] \right) ,U^{\left( j/p\right) }\right\rangle
\label{CTV}
\end{equation}%
an infinite number of coordinates $\Lambda _{i}$ of $\mathbf{\Lambda }_{i}$
are involvd in the series expansion of this state. More precisely, an
infinite sequence consistng of an increasing number of points arises in the
expansion. This is similar to the flag manifold described in the first part.
Note that we also recover the form of states divided into parameter
dependent states and remaining degrees of freedom.

\bigskip Similar to some covariant formulation, we can also consider the
states:%
\begin{equation*}
\left\vert \Sigma \left( \lambda \left( U_{j}\right) ,\left\{ \mathbf{%
\Lambda }_{i}\right\} \right) \right\rangle \left\vert h\left( \left[
\left\{ \mathbf{\Lambda }_{i}\right\} \right] \right) ,U^{\left( j/p\right)
}\right\rangle
\end{equation*}%
In the sequel, we will assume that the constraint is global so that we can
write the $\left( U_{j}\right) H_{i}\left( U_{j}\right) $ as:%
\begin{equation}
H_{i}\left[ U^{\left( j\right) },\left\{ \mathbf{\Lambda }_{i}\right\} %
\right] \delta \left( f^{\left( p_{j}\right) }\left( \left\{ \mathbf{\Lambda 
}_{i}\right\} ,A^{\left( j\right) }\right) \right)  \label{frmb}
\end{equation}%
where the $\left\{ \mathbf{\Lambda }_{i}\right\} $ is a set of unconstrained
operators. The constraint has to be understood for the matrices elements of
the $f^{\left( p_{j}\right) }\left( \left\{ \mathbf{\Lambda }_{i}\right\}
,A^{\left( j\right) }\right) $ in a tensor basis of eigenvectors of $\left\{ 
\mathbf{\Lambda }_{i}\right\} $ and $A^{\left( j\right) }$. \ The expansion
of $f^{\left( p_{j}\right) }\left( \left\{ \mathbf{\Lambda }_{i}\right\}
,A^{\left( j\right) }\right) $ is non local and involves infinite number of
eigenvlues $\left\{ \mathbf{\Lambda }_{i}\right\} $ through integrls.

The eigenstates write:%
\begin{equation*}
\left\vert \lambda \left( U_{j}\right) ,\left\{ \mathbf{\Lambda }%
_{i}\right\} \right\rangle \left\vert h\left( \left[ \left\{ \mathbf{\Lambda 
}_{i}\right\} \right] \right) ,U^{\left( j/p\right) }\right\rangle
\end{equation*}%
and the $\left\{ \mathbf{\Lambda }_{i}\right\} $ rzpresent degeneracies. Or
covarianntly:%
\begin{equation*}
\left\vert \Sigma \left( \lambda \left( U_{j}\right) ,\left\{ \mathbf{%
\Lambda }_{i}\right\} \right) \right\rangle \left\vert h\left( \left[
\left\{ \mathbf{\Lambda }_{i}\right\} \right] \right) ,U^{\left( j/p\right)
}\right\rangle
\end{equation*}%
\textbf{Remark: }For given $U^{\left( j\right) }$ the $\Sigma \left( \lambda
\left( U_{j}\right) ,\left\{ \mathbf{\Lambda }_{i}\right\} \right) $ are
eigenstates of commuting set of operators: $\left\{ \mathbf{\Lambda }%
_{i}\right\} \left( U^{\left( j\right) }\right) $: 
\begin{equation*}
\left( \mathbf{\Lambda }_{i}^{\left( 1\right) }\mathbf{\Lambda }_{i}^{\left(
2\right) }-\mathbf{\Lambda }_{i}^{\left( 2\right) }\mathbf{\Lambda }%
_{i}^{\left( 1\right) }\right) \left\vert \Sigma \left( \lambda \left(
U_{j}\right) ,\left\{ \mathbf{\Lambda }_{i}\right\} \right) \right\rangle =0
\end{equation*}

\subsection*{\textbf{Description of the constraint}}

Using (\ref{frmb}):%
\begin{equation*}
H_{i}\left[ U^{\left( j\right) },\mathbf{A}^{\left( i\right) }\right] \delta
\left( f^{\left( p_{j}\right) }\left( \mathbf{U}^{\left( j\right) },\mathbf{A%
}^{\left( i\right) }\right) \right)
\end{equation*}%
there are $p_{j}$ operators $\mathbf{A}^{\left( i\right) }$ combination of
some components of $\mathbf{\Lambda }_{i}$ and conjugate $\Pi _{\mathbf{%
\Lambda }_{i}}^{i}$. Function $f$ with $p_{j}$ components constrains the $%
A^{\left( i\right) }$ and $U^{\left( j\right) }$. When no confusion arises, $%
p_{j}$ is written $p$. The index corresponds to the number of contact pont $%
U^{\left( j\right) }$, $U^{\left( i\right) }$. There are $p_{j}$ eigenvalues
of $\mathbf{A}^{\left( i\right) }$ that can be expressed as functionals of
the $\left\{ \mathbf{\Lambda }_{i}\right\} $, eigenvalues of $\left\{ 
\mathbf{\hat{\Lambda}}_{i}\right\} $ (see (\ref{CTV})). Introducing the
corresponding $p_{j}$ eigenstates $\left\vert a^{\left( i\right)
}\right\rangle $ of the operators $\mathbf{A}^{\left( i\right) }$: 
\begin{equation*}
\left\vert U^{\left( j\right) }\right\rangle \simeq Vect\left\{ \left\vert
a^{\left( i\right) }\right\rangle \right\} \left\vert U^{\left( j\right)
/p}\right\rangle
\end{equation*}%
The state $\left\vert U^{\left( j\right) /p}\right\rangle $ denotes the
remaining independent degrees of freedom. The $\left\vert a^{\left( i\right)
}\right\rangle $ generate a subspace defined by parameters $U^{\left(
j\right) p}$. 
\begin{equation*}
\left\vert U^{\left( j\right) }\right\rangle \simeq \left\vert U^{\left(
j\right) p}\right\rangle \left\vert U^{\left( j\right) /p}\right\rangle
\end{equation*}%
After projection on some eigenstates for $\lambda ^{\left( i\right) }\left(
U^{\left( j\right) }\right) ,\left\{ \mathbf{\Lambda }_{i}\right\} $: 
\begin{equation}
\left\vert \lambda ^{\left( i\right) }\left( U^{\left( j\right) }\right)
,\left\{ \mathbf{\Lambda }_{i}\right\} \right\rangle \delta \left( f\left(
\left\{ \mathbf{\Lambda }_{i}\right\} ,a^{\left( j\right) }\right) \right)
\left\vert U^{\left( j\right) }\right\rangle \simeq \left\vert \lambda
^{\left( i\right) }\left( U^{\left( j\right) }\right) ,\left\{ \mathbf{%
\Lambda }_{i}\right\} \right\rangle \left\vert h\left( \left\{ \left\{ 
\mathbf{\Lambda }_{i}\right\} \right\} ,U^{\left( j\right) /p}\right)
\right\rangle \left\vert U^{\left( j\right) /p}\right\rangle  \label{frm1}
\end{equation}%
The function $h\left( \left\{ \mathbf{\Lambda }_{i}\right\} ,U^{\left(
j\right) /p}\right) $ with $p$ components. The dependency in $U^{\left(
j\right) /p}$ will be considered implicit and we write $h\left( \left\{ 
\mathbf{\Lambda }_{i}\right\} \right) $.

The states $\left\vert \lambda ^{\left( i\right) }\left( U^{\left( j\right)
}\right) ,\left\{ \mathbf{\Lambda }_{i}\right\} \right\rangle \left\vert
h\left( \left\{ \mathbf{\Lambda }_{i}\right\} \right) \right\rangle $
combine with coefficient $H\left( h\left( \left\{ \mathbf{\Lambda }%
_{i}\right\} \right) \right) $ to produce also states: 
\begin{equation*}
\int H\left( h\left( \left\{ \mathbf{\Lambda }_{i}\right\} \right) \right)
\left\vert \lambda ^{\left( i\right) },\left\{ \mathbf{\Lambda }_{i}\right\}
\right\rangle \left\vert h\left( \left\{ \mathbf{\Lambda }_{i}\right\}
\right) \right\rangle
\end{equation*}%
that correspond to "wave functions" $\left\vert H\left( \left\{ \mathbf{%
\Lambda }_{i}\right\} \right) \right\rangle $.

We can assume that the eignvalues $\left\{ \mathbf{\Lambda }_{i}\right\} $
can be divided into $\left\{ \mathbf{\Lambda }_{i}\right\} _{p}$ and $%
\left\{ \mathbf{\Lambda }_{i}\right\} _{/p}$ so that: 
\begin{equation*}
\left\vert \lambda ^{\left( i\right) }\left( U^{\left( j\right) }\right)
,\left\{ \mathbf{\Lambda }_{i}\right\} \right\rangle \left\vert h\left(
\left\{ \mathbf{\Lambda }_{i}\right\} ,U^{\left( j\right) /p}\right)
\right\rangle \left\vert U^{\left( j\right) /p}\right\rangle =\left\vert
\lambda ^{\left( i\right) }\left( U^{\left( j\right) }\right) ,\left\{ 
\mathbf{\Lambda }_{i}\right\} _{/p}\right\rangle \left\vert h\left( \left\{ 
\mathbf{\Lambda }_{i}\right\} _{p},U^{\left( j\right) /p}\right)
\right\rangle \left\vert U^{\left( j\right) /p}\right\rangle
\end{equation*}

\subsection*{\textbf{Exemple}}

Assume the eigenstates of $\mathbf{A}^{\left( i\right) }$ 
\begin{equation*}
\left\vert A^{\left( i\right) }\right\rangle =\left\vert \left( a^{\left(
i\right) },n_{a^{\left( i\right) }},\mu _{a^{\left( i\right) }}\right)
\right\rangle
\end{equation*}

Considerng the particular form of (\ref{frmb}):%
\begin{eqnarray*}
&&H_{i}\left[ U^{\left( j\right) },O^{\left( i\right) }\right] \delta \left(
f^{\left( p_{j}\right) }\left( U^{\left( j\right) },A^{\left( i\right)
}\right) \right) \\
&=&\sum \left( \left( \left\{ \mathbf{\Lambda }_{i}\left( U^{\left( j\right)
}\right) \right\} \right) ^{2}-\alpha ^{i}\left( U^{\left( j\right) }\right)
\left\{ \mathbf{\Lambda }_{i}\left( U^{\left( j\right) }\right) \right\}
\right) ^{2} \\
&&\times \delta \left( f^{\left( p_{j}\right) }\left( U^{\left( j\right)
},A^{\left( i\right) }\right) \right) +\alpha \left( U^{\left( j\right)
}\right)
\end{eqnarray*}%
where:%
\begin{eqnarray*}
\alpha ^{i}\left( U^{\left( j\right) }\right) &=&\left\langle A^{\left(
i\right) }\right\vert \delta \left( N^{i}-N^{j}\right) \left\vert A^{\left(
i\right) }\right\rangle \delta \left( f^{\left( p_{j}\right) }\left(
U^{\left( j\right) },A^{\left( i\right) }\right) \right) \\
\alpha \left( U^{\left( j\right) }\right) &=&\left\langle A^{\left( i\right)
}\right\vert N^{j}\left\vert A^{\left( j\right) }\right\rangle
\end{eqnarray*}%
and $h\left( \left\{ \mathbf{\Lambda }_{i}\left( U^{\left( j\right) }\right)
\right\} \right) =h_{j}\left( \mathbf{\Lambda }_{i}\right) $

\begin{eqnarray*}
\left\vert h\left( \left\{ \mathbf{\Lambda }_{i}\left( U^{\left( j\right)
}\right) \right\} \right) \right\rangle &=&\left\vert h\left( \left\{ 
\mathbf{\Lambda }_{i}\left( U^{\left( j\right) }\right) \right\}
,n_{a^{\left( j\right) }},\mu _{a^{\left( j\right) }}\right) \right\rangle \\
&\equiv &\left\vert \left\{ \mathbf{\Lambda }_{i}\right\} ,n_{\bar{o}%
^{\left( i\right) }},\mu _{\bar{o}^{\left( i\right) }}\right\rangle \\
&\equiv &\left( A_{\mu _{\bar{o}^{\left( i\right) }}}^{+}\right) ^{n_{\bar{o}%
^{\left( i\right) }}}\left( \left\{ \mathbf{\Lambda }_{i}\right\} \right)
\left\vert 0\right\rangle
\end{eqnarray*}%
and for several $\left\{ \mathbf{\Lambda }_{i}\left( U^{\left( j\right)
}\right) \right\} $%
\begin{eqnarray*}
\left\vert h\left( \left\{ \mathbf{\Lambda }_{i}\right\} \right)
\right\rangle &=&\left\vert h\left( \left\{ \mathbf{\Lambda }_{i}\right\}
,n_{a^{\left( j\right) }},\mu _{a^{\left( j\right) }}\right) \right\rangle \\
&\equiv &\left\vert \left\{ \mathbf{\Lambda }_{i}\right\} ,n_{\bar{o}%
^{\left( i\right) }},\mu _{\bar{o}^{\left( i\right) }}\right\rangle \\
&\equiv &\left( A_{\mu _{\left\{ \mathbf{\Lambda }_{i}\right\} }}^{+}\right)
^{n_{\left\{ \mathbf{\Lambda }_{i}\right\} }}\left( \left\{ \mathbf{\Lambda }%
_{i}\right\} \right) \left\vert 0\right\rangle
\end{eqnarray*}

\end{document}